\documentclass[]{aa}
\usepackage{graphicx}
\usepackage{txfonts}
\usepackage{longtable}
\usepackage[]{natbib}
\bibpunct{(}{)}{;}{a}{}{,} % to follow the A&A style
\def\lax    {\ifmmode{_<\atop^{\sim}}\else{${_<\atop^{\sim}}$}\fi}
\def\gax    {\ifmmode{_>\atop^{\sim}}\else{${_>\atop^{\sim}}$}\fi}
\def\kms    {\ifmmode{{\rm ~km~s}^{-1}}\else{~km~s$^{-1}$}\fi}

\def\arcmper  {\ifmmode \rlap.{' }\else $\rlap{.}' $\fi}
\def\arcs   {$^{\prime\prime}$}
\def\arcsper  {\ifmmode \rlap.{'' }\else $\rlap{.}'' $\fi}
\def\arcsgper  {\ifmmode \rlap.^{s }\else $\rlap{.}^s $\fi}
\def\deg      {\ifmmode^\circ\else$^\circ$\fi}     %Degree sign

\def\hper     {\ifmmode \rlap.^{h}\else $\rlap{.}^h$\fi}

\def\lsun {L$_{\odot}$}
\def\m1       {$^{-1}$}

\def\mper     {\ifmmode \buildrel m\over . \else $\buildrel m\over .$\fi}
\def\msun {M$_{\odot}$}

\def\dopt{$D_{\rm 25}$}
\def\dopttwo{$D_{\rm 25}^2$}
\def\>           {$>$}
\def\<           {$<$}

\def\simlt       {\lower.5ex\hbox{$\; \buildrel < \over \sim \;$}}
\def\simgt       {\lower.5ex\hbox{$\; \buildrel > \over \sim \;$}}

\newcommand{\mhtwo}{$M_{\rm H_2}$}
\newcommand{\mhtwocenter}{$M_{\rm H_2,center}$}

\newcommand{\mhtwomapextra}{$M_{\rm H_2, map, extra}$}
\newcommand{\mhi}{$M_{\rm HI}$}
\newcommand{\mstar}{$M_{*}$}
\newcommand{\lsunk}{$L_{\rm k,\odot}$}
\newcommand{\lb}{$L_{\rm B}$}
\newcommand{\lk}{$L_{\rm K}$}
\newcommand{\lfir}{$L_{\rm FIR}$}
\newcommand{\lir}{$L_{\rm IR}$}
\newcommand{\taudep}{$\tau_{\rm dep}$}

\begin{document}
   \title{The AMIGA sample of isolated galaxies}
   \subtitle{IX. Molecular gas properties \dag \thanks{
  \dag\ Full Tables~1, 4 and 5  
 are only available in electronic form at the CDS via anonymous ftp to {\tt cdsarc.u-
strasbg.fr (130.79.128.5)} or via {\tt http://cdsweb.u-strasbg.fr/cgi-bin/qcat?J
/A+A/} and from {\tt http://amiga.iaa.es/}. 
}}
\titlerunning{The AMIGA sample of isolated galaxies. X }

   \author{U. Lisenfeld\inst{1,2}
          \and
          D. Espada\inst{3,4}
          \and
          L. Verdes-Montenegro\inst{3}
          \and
          N. Kuno\inst{5}
          \and
          S. Leon\inst{6}
           \and 
           J. Sabater\inst{7}
          \and
          N. Sato\inst{8}
           \and
          J. Sulentic\inst{3}
          \and
           S. Verley\inst{1}
           \and
           M. S. Yun\inst{9}
               }
               
%  \offprints{U. Lisenfeld}

   \institute{
       Departamento de F\'\i sica Te\'orica y del Cosmos,
         Universidad de Granada, 18071 Granada, Spain, \email{ute@ugr.es}
          \and
         Instituto Carlos I de F\'{\i}sica Te\'orica y Computacional, Universidad de Granada, Spain
         \and
Instituto de Astrof\'{\i}sica de Andaluc\'{\i}a (CSIC) Apdo. 3004, 18080 Granada, Spain
%              \email{lourdes@iaa.es}
        \and
National Astronomical Observatory of Japan, 2-21-1 Osawa, Mitaka, Tokyo 181-8588, Japan
         \and
Nobeyama Radio Observatory, Minamimaki, Minamisaku, Nagano 384-1805, Japan 
 \and
 Joint Alma Observatory/ESO, Las Condes, Santiago, Chile
\and
Institute for Astronomy, University of Edinburgh, Edinburgh EH9 3HJ, UK
\and 
Student Center for Independent Research in the Science, Wakayama University, 930 Sakaedani, Wakayama, 640-8510, Japan
\and
Department of Astronomy, University of Massachusetts, Amherst, MA 01003, USA 
            }

   \date{Received ; accepted }

% \abstract{}{}{}{}{} 
% 5 {} token are mandatory
 
  \abstract
  {}
%   {The effect of the environment on the molecular gas content of a galaxy has been a  controversial and yet unsolved question.}
 {We characterize the molecular gas content (ISM cold phase) using CO
emission of a redshift-limited subsample  of  isolated galaxies from the AMIGA
(Analysis of the interstellar Medium of Isolated GAlaxies) project in order to
provide a comparison
sample for studies of galaxies in different environments.}
 {We present the $^{12}$CO(1-0) data for 273 AMIGA galaxies, most of them ($n=186$) from our own
 observations   with the IRAM 30m and the FCRAO 14m telescopes and the rest from the literature.
 We constructed a redshift-limited sample containing galaxies with 1500\kms\ $<$ v $<$5000\kms\  and excluded
 objects with morphological evidence of possible interaction. 
 This sample ($n=173$)
 is the basis for our statistical analysis. 
It contains galaxies with molecular gas masses, \mhtwo , in the range of $\sim 10^{8} - 10^{10}$ \msun .
%and is $\sim$ 80\% complete down to m$_{\rm B,corr} < 13.9$ mag. 
 It is dominated, both in absolute number and  in
 detection rate, by spiral galaxies of type $T=3-5$ (Sb-Sc).
  Most galaxies were observed with a single pointing towards their centers. Therefore, we performed an extrapolation to the total molecular gas mass expected in the
   entire disk based on the assumption of an exponential distribution.
    We then studied the relationships between \mhtwo\ and  other galactic properties  (\lb , \dopttwo, \lk , \lfir , and \mhi) .}
 {We find correlations between \mhtwo\ and \lb ,   \dopttwo,  \lk , and  \lfir .  The tightest correlation of \mhtwo\ holds with \lfir\ and,
 for $T=3-5$, with \lk , and the poorest    with
 \dopttwo .The 
correlations with  \lfir\  and \lk\ are very close to linearity. The correlation with \lb\  is nonlinear
so that \mhtwo/\lb\ increases with \lb .  
%We define a deficiency parameter,
 %based on these correlations, that can be used to search for possible enhancements 
 %or deficiencies of the molecular gas in other samples. 
 The molecular and the atomic gas masses of our sample show no strong correlation.   
  We find a low mean value, log(\mhtwo/\mhi) = -0.7 (for $T=3-5$), and a strong decrease in
  this ratio with morphological type.
 The molecular gas column density and the surface density of the star formation rate (the Kennicutt-Schmidt law)
 show a tight correlation with a rough unity slope.
  We compare the relations of  \mhtwo\ with  \lb\  and \lk\   found for AMIGA galaxies 
  to samples of interacting galaxies from the literature and find an indication
 for an enhancement of the molecular gas in interacting galaxies of up to 0.2-0.3 dex.
}
  {}
%   {The correlations with other galactic parameters
 %  that we presented can be used to study
 % deviations of the molecular gas content in other  samples.}

     \maketitle

%
%________________________________________________________________

\section{Introduction}
\label{Sect:Introduction}

A major and longlasting debate in astronomy 
involves the relative roles of ``nature''
and ``nurture'' in galaxy formation and evolution 
\citep[e.g.][]{1976ApJS...32..171S,1978ApJ...219...46L,1985MNRAS.214...87J,1987ApJ...320...49B}. 
%(e.g. Sulentic 1976, Larson \& Tinsley 1978, Joseph \& Wright 1985, Bushhouse 1987). 
Although it is  broadly accepted that
galaxy evolution strongly depends on the environment,
the quantitative effect of ``nurture'' on certain galactic properties
is still a matter of debate. 

The molecular gas content  is  an important quantity of a galaxy because it is directly
related to its capacity for star formation (SF).
We still need  to determine, however, how the environment affects the amount of the molecular gas.
Galaxies in clusters \citep[][Scott et al. in prep.]{1989ApJ...344..171K,1997A&A...327..522B} 
%(Kenney \& Young 1989, Boselli et al. 1997) 
and groups \citep{1998ApJ...497...89V,1998A&A...330...37L}
%(Verdes-Montenegro et al. 1998, Leon et al. 1998) 
seem
to have a normal molecular gas content, even though they can be highly deficient in atomic gas.
On the other hand, some authors
\citep{1993A&A...269....7B,1994A&A...281..725C,2004A&A...422..941C} 
%Combes et al. (1994) and Braine \& Combes (1993) 
find an enhanced molecular gas content in interacting
galaxies, in contrast to 
the results of \citet{1997ApJ...490..166P},
who concluded that the molecular gas content
is not affected by interaction in strongly interacting pairs or Virgo cluster galaxies.

To clarify the role played by the environment, a well-defined
sample of isolated galaxies is needed to serve as  a 
zero level for studies dealing with the effect of interactions.
Most previous studies investigating  the   properties of molecular gas in  
isolated and interacting galaxies
 \citep{1988ApJ...334..613S,1993A&A...272..123S,1997A&A...327..522B,2001PASJ...53..713N,2003ApJS..145..259H,2009AJ....137.4670L}
have generally not defined any very clear criterion for isolation. 
%(Solomon \& Sage 1988; Sage 1993; Boselli et al. 1997; 
%Leroy et al. 2009, Helfer et al. 2003,  Nishiyama et al. 2001).
%
%Only the sample of 
%\citet{2003A&A...411..381S} 
%%Sauty et al. (2003) 
%presented CO observations of well-defined isolated galaxies, some  of which are
%included in the present study.
\citet{1997ApJ...490..166P} carried out a CO study comparing isolated and interacting galaxies.
Their sample of isolated galaxies is composed of 68 galaxies from various
sources, selected in a much less rigorous way than the present study and biased towards
infrared-luminous objects.
The only survey explicitly focusing on isolated  galaxies, and in particular
on galaxies from the Catalogue of Isolated Galaxies, is the one by  \citet{2003A&A...411..381S}.  They present the
CO data of 99 optically-selected spiral galaxies with recession velocities up to 14000\kms\
and briefly compare the properties of the molecular gas
mass to the blue luminosity and atomic gas mass. A detailed analysis of the properties of that
sample is, however, not presented there.
{The largest previous CO survey was the FCRAO Extragalactic CO  survey \citep{1995ApJS...98..219Y} observing
$\sim$ 300  nearby galaxies.  The major difference with respect to  the present study is that it 
 did not consider the isolation of the galaxies as a criterion. Furthermore, it  only contained 
bright galaxies (either m$_{\rm B, corr} < 13$ mag, or $F_{60 \mu m} > 5$ Jy  or $F_{100 \mu m} > 100$ Jy)
whereas our samples also  includes fainter objects. }

The project AMIGA 
\citep[`Analysis of the interstellar Medium of Isolated GAlaxies",][]{2005A&A...436..443V}
%("Analysis of the interstellar Medium of Isolated GAlaxies",
%Verdes-Montenegro et al. 2005) 
was started to provide such a reference sample
by characterizing the properties of the interstellar medium (ISM) and star formation (SF)
in isolated galaxies in the local Universe. 
It is based on the Catalogue of Isolated Galaxies
\citep[CIG][]{1973SoSAO...8....3K}
%(Karatchentseva 1973) 
which is composed of 1050 galaxies located in the Northern hemisphere.
The AMIGA project is  presented in  \citet{2005A&A...436..443V}.
A considerable amount of work has been done since then in order 
to refine the sample. This work includes
the revision of all CIG positions  \citep{2003A&A...411..391L},
the determination of   POSS2-based morphologies  and the identification of 
 galaxies showing signs of possible interaction \citep{2006A&A...449..937S} and
the reevaluation and quantification of the 
degree of isolation  \citep{2007A&A...472..121V,2007A&A...470..505V}.
The results of this project 
consistently find that  the AMIGA galaxies have the lowest SF activity as well as the lowest 
presence of Active Galactic Nuclei (AGN)  in the local Universe.
This is obtained both from the far-infrared (FIR) luminosity derived from IRAS data \citep{2007A&A...462..507L},
 and  the radio continumm emission \citep{2008A&A...485..475L}, which are both SF tracers.
 The rate of AGN candidates, derived from IRAS colors and radio continuum emission
 of the AMIGA galaxies is lowest compared to similar studies from the literature \citep{2008A&A...486...73S}. 
 Optical photometric analysis of Sb-Sc galaxies in the AMIGA sample showed that most galaxies
 have  pseudo-bulges instead of classical bulges, and 
 a comparison with samples of spiral galaxies selected without isolation criteria revealed
 that the isolated galaxies tend to host larger bars, are more symmetric, less concentrated and less clumpy
 \citep{2008MNRAS.390..881D}.
 These findings strongly support that the AMIGA sample represents the most isolated galaxies in the local Universe
where secular evolution is dominant. 
Espada et al. (in prep.) study the HI content and 
\citep{espada11a} found the smallest fraction of asymmetric HI profiles in the AMIGA sample
when compared with any sample yet studied.

The revised AMIGA sample is 
reasonably complete ($ \sim$ 80-95\%) down  to m$_{\rm B,corr} \le15.0$ mag 
\citep{2005A&A...436..443V} 
%(Verdes-Montenegro et al. 2005) 
and it is currently one of the largest sample of nearby
isolated galaxies in the Northern hemisphere. It consists of galaxies whose 
structure and evolution have been driven largely or entirely by 
internal rather than by external forces 
%(i.e. as close to pure nature  as exists in the local Universe) 
at least during the last
 3 Gyr \citep{2005A&A...436..443V}.
The data are being released and periodically updated at http:// amiga.iaa.es 
where a Virtual Observatory 
interface with different query modes has been implemented. 

In the present paper we present and analyze 
CO observations of a redshift-limited subsample of this catalogue.
The goal is to characterize the properties of the molecular gas, traced by CO,
of isolated galaxies and to provide a reference sample for studies investigating
the role of the environment.

 \section{The sample}
\label{sample}

 For a study of the molecular gas content we had to restrict the number of galaxies, since
 observation of the entire optically complete sample ($n\sim 700$) required too much
 telescope time.
  We chose to build  a redshift-limited subsample  by
 selecting galaxies with recession velocities in the range of  $1500-5000$ \kms . 
The completeness limit of of the AMIGA sample of 15 mag
 corresponds to blue luminosity of log10(\lb/\lsun) = 8.55  and  log10(\lb/\lsun) = 9.60 
 at the distances derived  for these velocities with a Hubble constant of 75 km s$^{-1}$ Mpc$^{-1}$.
 The range was chosen in order to avoid (i) very nearby galaxies for which
the condition of isolation is not reliable \citep{2007A&A...472..121V}
and (ii) distant galaxies which are difficult
to detect in CO.  The restriction in velocity 
provides us with a sample probing a defined volume in space.

 There are  278  galaxies in this velocity range in the CIG. {We have CO data for 
 201 of these objects, mostly from our own observations  (180 galaxies) with the
 30m telescope of the Instituto de Radioastronom\'\i a Milim\'etrica (IRAM)
 at the Pico Veleta and with the 14m Five College Radio Astronomical Observatory (FCRAO),
 and the rest from the literature.
 We then
excluded those galaxies that were identified by us in a visual inspection of optical images
as having signs of a  possible  present or past interaction (see description of Table~\ref{general-data}  for more
details on the criteria). 
%($n=39$) , obtaining a sample of 234 galaxies to which we refer as our total CO sample.
%We have CO data for 201 galaxies  in the velocity range between 1500 and 5000 \kms .  
%After excluding these  objects  we have 
This leaves us with
 173 isolated galaxies with CO data in the velocity range between 1500 and 5000 \kms .
% This sample is $\sim$ 80\% complete down to m$_{\rm B,corr} < 13.9$ mag.
 We refer to this sample as
the redshift-limited CO sample and we will use it for the statistical analysis throughout this paper.

%-----------------------------------------------------------------------------------
\begin{table*}
\caption{General data for the total CO sample. }
\begin{center}
\begin{tabular}{lccccccccc} 
\hline
CIG & Dist & Vel & D$_{\rm 25}$ & $i$ &T(RC3) & Inter &  log(\lb ) & log(\lfir ) &  log(\lk )   \\
      & [Mpc]  &  [\kms ] &  [$\arcmin$] &   [$^\circ$] &           &           &  [\lsun ] & [\lsun ] &  [\lsunk ] \\
(1) & (2) & (3) &(4) & (5) & (6) & (7)  & (8) & (9) & (10) \\
\hline
     1   &  96   & 7299   &  1.39   &  65   &   5   &   1   & 10.57     & 10.28   & 11.23  \\
     4   &  31   & 2310   &  3.29   &  84   &   5   &   0   & 10.36     & 10.08   & 10.98  \\
     6   &  61   & 4528   &  0.70   &  65   &   7   &   1   &  9.80     &  9.82   & 10.21  \\
    10   &  63   & 4613   &  1.05   &  63   &   5   &   0   &  9.78 & $<$  9.44   & 10.09  \\
... &  ... &... & ... & ... & ... & ...  & ...  & ... \\
\hline
\end{tabular}

The full table is available in electronic form at the CDS and from http://amiga.iaa.es. 
\end{center}
\label{general-data}
\end{table*}
%-----------------------------------------------------------------------------------

Additionally, we have CO data for 72 galaxies outside this velocity range.
Six galaxies are  from our own observations (with velocities between 5000 and 5500 \kms) and the rest is from the literature.
Thus, in total, we have  $^{12}$CO(1-0) data for  273 CIG galaxies.  We refer to this
sample as the total CO sample, and list the corresponding data in 
Tables~\ref{general-data} and \ref{mol_mass}, but we do not use it for any statistical analysis.}

Recently, an update of 
the basic properties of the galaxies in the AMIGA sample was
 carried out  for the blue magnitude, optical isophotal diameter
$D_{25}$,
 velocity, and morphology and interaction degree based on higher
 resolution images (from SDSS or our own images). 
  The details are
described in Espada et al. (in prep.).
In order to provide a self-contained data set for the present
paper, we list in Table~\ref{general-data}  the basic data relevant for the
total CO sample. 
The columns are:

\begin{enumerate}
\item CIG: Entry number in the Catalogue of Isolated Galaxies (CIG).
\item Dist: Distance in Mpc, based on $H_0 = 75$ \kms\ Mpc$^{-1}$.
\item Vel: Recession velocity in \kms .
\item $D_{25}$: Isophotal optical major  diameter at the isophotal level
25 mag arcsec$^{-1}$ in the B-band, \dopt , in arcmin  from the Lyon Extragalactic Database (LEDA).
\item $i$:  Inclination from LEDA. 
\item T(RC3): Morphological type T(RC3) as determined by our morphological revision,
given in the RC3 numerical scale \citep{1991trcb.book.....D}.

\item Inter: Code for morphological signs of a possible interaction. 

 0: No signs of large distortions which cannot be clearly explained by e.g. dust.

1: At least one of the following signs of possible interaction:
asymmetric, lopsided, 
%(Verley et al 2007),
distorted,
confirmed pair in the Nasa Extragalactic Database (NED) or
in \citet{2007A&A...470..505V}, 
%Verley et al. 2007), 
integral sign shape, warp,
tidal features (tail, bridge, shell).

2:  Merger-like morphology or superposition of two galaxies. 

\item log(\lb ): Decimal logarithm of the blue luminosity, \lb ,  in units of solar bolometric luminosities.
We calculated \lb ,  from the flux, $f_B$,
as \lb = $\nu f_B$, where $\nu$ is the central frequency of the blue band.
 
\item log(\lfir ): Decimal logarithm of the far-infrared luminosity in units of solar bolometric luminosities,  taken from
\citet{2007A&A...462..507L} and adapted to the revised distances used in the present paper.
{\lfir\ is   is computed from the IRAS fluxes at 60 and 100 $\mu$m, $F_{\rm 60}$ and  $F_{\rm 100})$,
as  $\log(L_{\rm FIR}/L_{\sun}) = \log(FIR) + 2 \log(D) + 19.495$,
where $D$ is distance in Mpc and 
$FIR = 1.26 \times 10^{-14} (2.58F_{60} + F_{100})$ W\,m$^{-2}$
\citep{1988ApJS...68..151H}. }

%(taken from Lisenfeld et al. 2007, adapted
%to the revised distances used in the present paper) 

\item log(\lk ): 
{Decimal logarithm of the  luminosity in the K-band, in units of the solar luminosity in the $K_{\rm S}$-band
($L_{\rm K,\odot} = 5.0735 \times 10^{32}$ erg s$^{-1}$), calculated from the extrapolated magnitude in the $K_S$ (2.17 $\mu$m)
band from the 2MASS Extended Source Catalogue \citep{2000AJ....119.2498J}.
The magnitudes were available for 250 galaxies of our sample.
We calculated the $K_{\rm S}$ luminosity, \lk ,  from the total (extrapolated) $K_{\rm S}$ flux, $f_{\rm K}$,
as $L_{\rm K} = \nu f_{\rm K}(\nu)$, where $\nu$ is the central frequency of the $K_{\rm S}$-band.
\lk\ is a good measure of the total stellar mass.
% \mstar which can be estimated by
%adopting a mass-to-luminosity ratio
%of \msun/$L_{K,\odot} = 0.66$ \citep{2001MNRAS.326..255C}   % 1.32 for Salpeter
%(including a factor of 0.5) (Bell et al. 2003, Kauffmann et al 2003)  to change from the Salpeter to the %\citet{2001MNRAS.322..231K}  %Kroupa  
%Initial Mass Function (IMF)). 
}
\end{enumerate}

%----------------------------------
\begin{figure}[h!]
\includegraphics[width=6.5cm,angle=270]{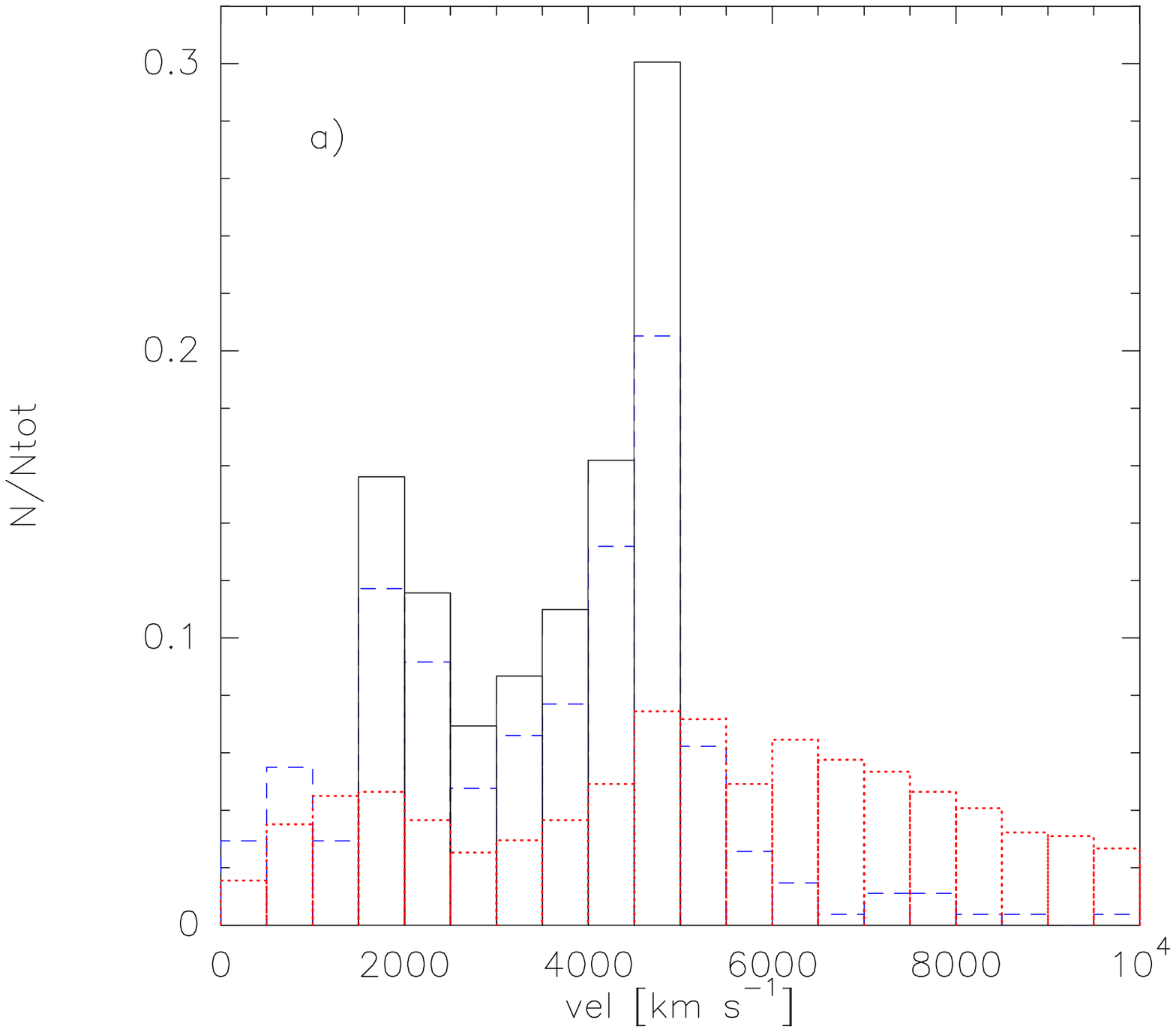} 
\includegraphics[width=6.5cm,angle=270]{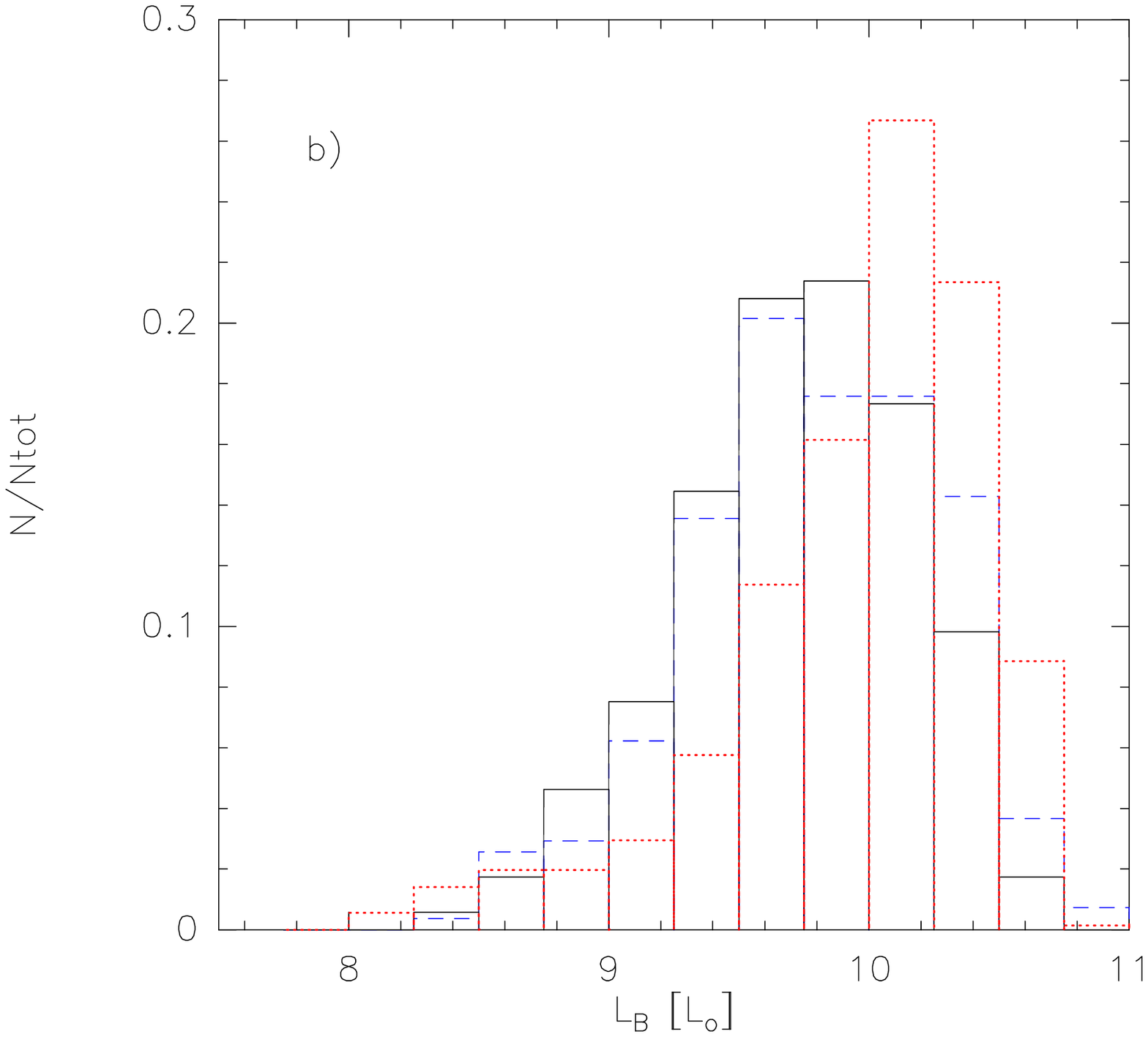}
\includegraphics[width=6.5cm,angle=270]{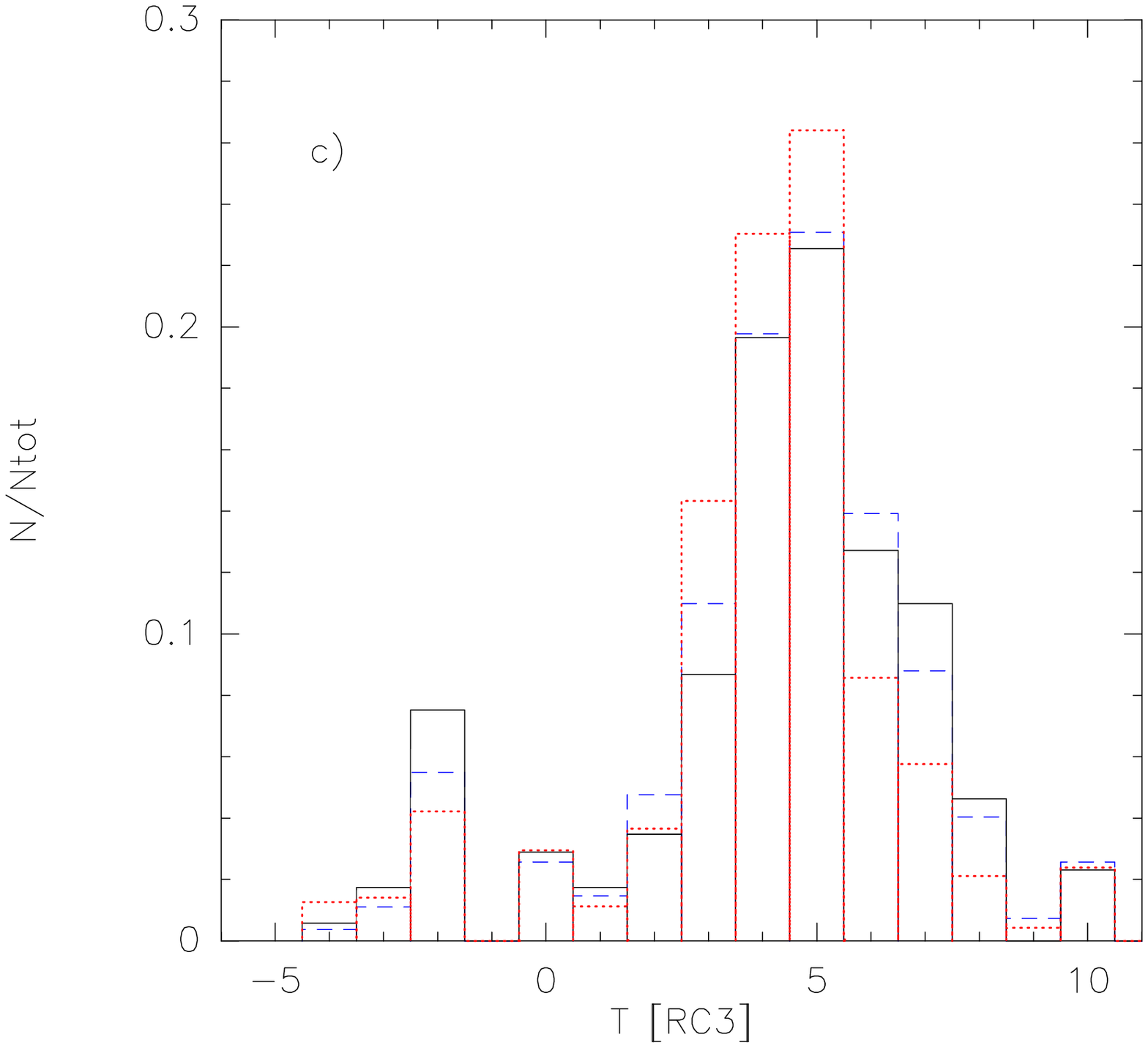}
\caption{Basic properties (normalized distributions) of the redshift-restricted CO sample ($n=173$, full black line),
total CO  sample, {excluding possibly interacting objects} ($n=234$, dashed blue line) 
 as well as the optically complete sample \citep[$n=712$,][dotted red line]{2005A&A...436..443V}.
 a) Recession velocity,  b) optical luminosity, log(\lb) and  c) morphological type  (T(RC3), as in the RC3 catalogue).}
\label{compare_opt_sample}
\end{figure}
%----------------------------------

In Fig.\ref{compare_opt_sample} we present some of the basic characteristics
of the CO samples and compare them to those of
the  optically complete sample \citep{2005A&A...436..443V}.
The latter   ($n=712$) is composed of CIG galaxies  with
$m_{\rm B}$ in the range of 11 - 15 mag
and is  80 - 95 \% complete.
The main difference between the CO and the optically complete sample is
the larger spread in velocity of the latter (Fig.\ref{compare_opt_sample}a). The CO samples are at a lower
velocity, especially the redshift-restricted sample. This leads to a slightly lower 
optical luminosity (Fig.\ref{compare_opt_sample}b) of the CO samples 
since  the number of luminous objects is higher at larger
distances due to the  Malmquist bias.

The distribution of the  morphological
types (Fig.\ref{compare_opt_sample}c) is very similar in all the three samples.
All samples are dominated by spiral galaxies.
The relative number of early type galaxies ($T=(-5) -  0$) is 8\% for the
optically complete sample and 13\% for the CO redshift-restricted sample.
The distribution is peaked around galaxies with types $T =3-5$ (63\% for the
optically complete sample and 51\% for the CO redshift-restricted sample).

%-------------------------------------------------------------------------------
\section{CO observations and analysis}

We carried out  CO(1-0) observations with the FCRAO  and
with the IRAM 30m telescope.  
 % and 9 at Nobeyama. 
We observed galaxies with isophotal diameters
%\dopt $<$ 45\arcs were observed with the Nobeyama 45m telescope, 
%those with45\arcs $<
  \dopt$ <100$\arcsec\  at  the 30m telescope  and galaxies with
   \dopt$ \ge100$\arcsec\  at 
 the 14m FCRAO telescope.  In this way we tried  to optimize the agreement between
 beam size and optical diameters and minimize the fraction of missing flux in the observations
done with  a single pointing.
 We observed 100 galaxies at the FCRAO radio-telescope and 101 at the IRAM 30m
 telescope. 
 In order to check the consistency of 
the results  we observed 15 galaxies at 
both telescopes.

\subsection{Observations}

\subsubsection{IRAM 30m telescope}

We observed the $^{12}$CO(1--0) line at 115\,GHz 
with the IRAM 30-meter telescope on Pico Veleta  using the dual
polarization receivers A100 and B100, together with the 512
$\times$ 1 MHz filterbanks.
%A total of 170 hours of observing time was spent between November 2002 and January 2004.
%
%Parallely, the $^{12}$CO(2--1) line was observed, but the 
%data quality was generally poor, so that the data will not be shown here.  
%{Check the data still to see whether it is usable.}
%
The observations were done in wobbler switching mode with a 
wobbler throw of 120\arcs\ in azimuthal direction.
%The beam size of this antenna at 115GHz is 22\arcsec, which corresponds to 
%3.5 kpc at a distance of 2500 km s$^{-1}$, assuming a morphological 
%constant H$_{0}$ = 75 \kms Mpc$^{-1}$.
Pointing was
monitored on nearby quasars every 60 -- 90  minutes.
%the rms offset being \approx xxxxx \arcs.
% --> This will be difficult to check
%%The observations were generally carried out in medium-quality  weather due to the
%%fact that they were part of pooled observations in which the excellent
%%weather was dedicated to more weather demanding projects.
% The galaxies were grouped into seven ranges of velocity with a width of 500 km s$^{-1}$ to avoid 
% losing time for tuning. 
The integration time on source was typically 0.5 to 1.5 hours. 
The mean  system temperatures was 320\,K 
on the $T_{\rm A}^*$ scale.  
All CO spectra and intensities are
presented on the main beam temperature scale ($T_{\rm mb}$) which is
defined as $T_{\rm mb} = (F_{\rm eff}/B_{\rm eff})\times T_{\rm A}^*$.
The IRAM forward
efficiency, $F_{\rm eff}$, was 0.95  and
the beam efficiency, $B_{\rm eff}$, 0.75.
The peak intensities of our sources ranged between about 10 and 80 mK ($T_{\rm mb}$).

Most galaxies were observed at the central position with a single pointing.
The galaxies with the strongest emission at the center 
(19 galaxies) were mapped along the major axis with 
a  spacing of 15\arcsec , until  a root mean square  (rms) noise of about
3 mK was reached for a velocity resolution of 10.6 \kms .

\subsubsection{FCRAO 14m telescope}

The observations at the FCRAO were done with the receiver SEQUOIA, a $4\times 4$
pixel array operating from 85 to 115 GHz. We used the so called ``beam switching'' mode, 
in which the telescope switches position between the source and a reference position
89.2\arcsec\ apart in azimuth.
%The information in one pixel when it is ON source and the signal in the opposite pixel when it is OFF source, 
Two of the pixels in the array alternated between the ON- and the OFF-position,
doubling in this way  the effective integration time.

The typical observing time per object was about 2 hours. 
The pointing was checked  between the observations of different sources using a 
nearby quasar.
The mean system temperatures  was 380\,K (on the  $T_{\rm A}^*$ scale).
All CO spectra and luminosities are
presented on the main beam temperature scale ($T_{\rm mb}$) which is
defined as $T_{\rm mb} = T_{\rm A}^*/B_{\rm eff}$. 
The main beam efficiency is
$B_{\rm eff}=0.45$. 
We observed each galaxy with one pointing at their central position.
The peak intensities of our sources ranged between about 10 and 80 mK ($T_{\rm mb}$).

\subsubsection{Literature data}

%--------------------------------------------------------------------------------------

\begin{table}
\caption{List of papers from which  the CO data have been compiled. }
\begin{center}
\begin{tabular}{cllr} 
\hline
Bibl. code & Article               & Telescope$^1$       &  N \\ \hline\hline
1 & This work & 1, 2 & 189 \\
2& {\citet{1995ApJS...98..219Y}} &  2, 5           & 25 \\
3&  {\citet{1993A&A...269....7B}} &   1               & 11 \\
4&   {\citet{2003A&A...411..381S}} & 1, 3, 5 & 99 \\
5& {\citet{1996A&AS..115..439E}} & 3, 4    & 15 \\
%%%6&  Solomon \& Sage (1988)    & 2, 5             & 7  \\  No data in final sample
6& {\citet{1993A&A...272..123S}} &   5        & 7 \\
%7&  {\citet{1996A&AS..118...47C}} &   1, 3         & 6\\
\hline 
%8&He93& \citet{2003ApJS..145..259H}   &   BIMA           & 9\\   
\end{tabular}
\end{center}

$^1$The  codes for the telescopes are listed in Table~\ref{telescopes}.
\label{literature}
\end{table}

%--------------------------------------------------------------------------------------

\begin{table}
\caption{Main parameters of the used radio-telescopes.}
\begin{center}
\begin{tabular}{clccc} 
\hline
 Code & Radio-Telescope  &Diameter&$\theta_{HPBW}$ (115 GHz) & Jy/K$^1$      \\ \hline\hline
% 1    &  Nobeyama       &45m     & 10$\arcsec$ & 4.2  \\                           %3 these are the codes  
 1    &  IRAM           &30m     & 21$\arcsec$      & 5  \\                     %2  
 2    &  FCRAO          &14m     &  45$\arcsec$     & 20  \\                       %1  
 3    & SEST            &15m     &   43$\arcsec$    & 19  \\                       %32 
 4    & Onsala          &20m     &   33$\arcsec$    & 12  \\               %8  
 5    & NRAO Kitt-Peak            &12m     &   55$\arcsec$    & 32  \\\hline                 %33 
\end{tabular}

$^1$ The conversion factor from  T$_{\rm mb}$  in K to flux in Jy.
\end{center}
\label{telescopes}
\end{table}
%--------------------------------------------------------------------------------------

We furthermore searched the
 literature  and found CO(1-0) data for 131 objects, 87 of them had not been observed by us.
 We list the references for these data in  Table~\ref{literature}. 
 %Note that the total number
%of observations from the literature is higher than 131, because 
Some galaxies were observed at several telescopes.  
Table~\ref{telescopes}  provides information about  the telescopes used in the different
surveys: 
 the antenna size (column 3), the half power beam width (HPBW) (column 4) 
 and the conversion factor Jy/K (on the  T$_{\rm mb}$ scale) at 115 GHz  (column 5).
 % Sauty data:
%We included their data in our study (62 galaxies of the total CO sample, and
%26 galaxies in the redshift-limited CO sample) for those galaxies not observed
%by us and without  better data from other observations in the literature. 

\subsection{Data reduction}

The data from both telescopes were reduced in the standard way using the CLASS software
in the GILDAS package\footnote{http://www.iram.fr/IRAMFR/GILDAS}.
The data reduction consisted in dismissing poor scans, flagging bad channels,
subtracting a baseline  and averaging the spectra for the
same object and position. 
In most cases a constant baseline was subtracted and only in a few cases 
the subtraction of a  linear baseline was required.
%The spectra were then smoothed in velocity to a 
%resolution of 10.6 \kms\ for the 30m data.

%%%%%%%%%%%%%%%%%%%%%%%%%%%%%%%%%%%%%
\subsection{Spectra and integrated intensities}

The CO(1--0) profiles of the detections and  tentative detections 
observed by us at the IRAM 30m and  FCRAO 14m are shown 
in Appendix A
(Fig.~\ref{spectra_iram} and Fig.~\ref{spectra_fcrao}, respectively).
The spectral shapes observed are very diverse.
Both single and double peaked lines are present and the line widths span 
a wide range.

The velocity integrated  intensity, 
$I_{\rm CO} =\int T_{\rm mb} {\rm ~dv}$ (in K km s$^{-1}$),
was calculated from the spectra with a velocity  resolution of 10.6 \kms\ for the IRAM spectra,
respectively 13.1 \kms\ for the FCRAO spectra
by summing up all channels with significant emission.
Its error was calculated as:
\begin{displaymath}
{\rm error~}(I_{\rm CO}) = \sigma (W_{\rm CO} ~\delta V_{\rm CO})^{1/2}~ [{\rm K~km~s^{-1}}],
\end{displaymath} 
where $\sigma$ is the rms noise of the spectrum, $W_{\rm CO}$ is the CO line width, and 
$\delta V_{\rm CO}$ is the spectral resolution.
For undetected galaxies, we calculate a 3$\sigma$ upper limit, assuming a line width of 
300 \kms ,  as:

\begin{displaymath}
{I_{u.l.}({\rm CO})=3 \times \sigma (300 \delta V_{\rm CO})^{1/2} ~ [{\rm K~km~s^{-1}}]}.
\end{displaymath}

In Table~\ref{COdata} we list  the following items:

\begin{enumerate}
\item Entry number in the Catalogue of Isolated Galaxies (CIG). An asterisk added to the number means that
the detection is marginal. In our statistical analysis we treat marginal detections as upper limits.
\item Off. $\alpha$:  RA offset  from the center  in arcsec.
\item Off. $\delta$: Declination offset  from the center in arcsec.
\item rms: root mean square noise in mK for a velocity resolution of 10.6 \kms\ (IRAM), respectively
13.1 \kms\ (FCRAO).
\item $I_{\rm CO}$: velocity integrated CO line temperature $\int$ $T_{\rm mb} {\rm ~dv}$, in K km s$^{-1}$, and its  error.
\item V$_{\rm CO}$: mean velocity of the CO line, in km s$^{-1}$.
\item W$_{\rm CO}$: zero line-width of the CO spectrum, in km s$^{-1}$.
\item Tel: radio-telescope code, as listed in Table~\ref{telescopes}.
\end{enumerate}

%***********************************************************
 \begin{table*}
 \caption{Velocity integrated $^{12}$CO(1-0) line intensities, mean velocity  and line widths for the CIG galaxies 
 observed by us.}  
 \begin{tabular}{cccccccc} 
 \hline
 CIG & Off. $\alpha$ & Off. $\delta$  & rms & I$_{\rm CO}$ &  V$_{\rm CO}$ & W$_{\rm CO}$ & Tel. \\  
  &  [\arcsec ] &  [\arcsec ] &  [mK]  &  [K km s$^{-1}$]  & [\kms ] & [\kms ] & \\
(1)&(2)&(3)&(4)&(5)&(6)&(7)&(8)\\
\hline
    10$^*$ &    0  &   0  &  3.47  &  0.50  $\pm$  0.15      &    4980  &     180 &  1 \\
    27      &   0  &   0  &  4.21  &  3.54  $\pm$  0.23      &    4586  &     287 &  1 \\
    27      &  13  &   8  &  7.21  &  $<$  1.22   & -  & -  &  1 \\
    29$^*$ &    0  &   0  &  5.24  &  0.49  $\pm$  0.16      &    4162  &      92 &  1 \\
   ...  &    ..   &   ...  &  ...  & ...      &    ...  &  ...      &   \\
   \hline
\label{COdata}
\end{tabular}

The full table is available in electronic form at the CDS and from http://amiga.iaa.es.

\end{table*}

%__________________________________________________________________

\subsection{Comparison between  IRAM and  FCRAO data}

In order to 
check the relative calibration between the IRAM 30m  and the
FCRAO 14m telescope and to guarantee that these two data sets are comparable,
  we observed 15 galaxies
with both telescopes. We expect a ratio of the velocity integrated 
intensities of $I_{\rm CO-IRAM}/I_{\rm CO-FCRAO}=1$  for 
emission  homogeneously filling the beams,
 and $I_{\rm CO-IRAM}/I_{\rm CO-FCRAO}=(\Theta_{\rm  FCRAO}/\Theta_{\rm  IRAM})^2 = 4.5$,
 where  $\Theta_{\rm  FCRAO}$ and $\Theta_{\rm  IRAM}$  are the FWHM of the respective beams,
 for a point-like emission.

Four galaxies  (CIG 66, 181, 281 and 330) 
were detected at both telescopes. The  ratios of $I_{\rm CO-IRAM}/I_{\rm CO-FCRAO}$ 
range between 1.1 and 2.3,
consistent with the value expected for slightly concentrated emission.
Six galaxies were detected at IRAM, but only tentatively detected 
(CIG 176, CIG 355) or
undetected (CIG 217, CIG 561, CIG 609, CIG 622) 
at the FCRAO. The lower limit for $I_{\rm CO-IRAM}/I_{\rm CO-FCRAO}$ 
in five cases was between 0.94 and 3.6, consistent with the expected range of
values.
For CIG 217 this value is higher  ($I_{\rm CO-IRAM}/I_{\rm CO-FCRAO}=6.1$) than the 
theoretical upper limit. Since the detection at IRAM has a high signal-to-noise ratio, the most likely reason
is an underestimate of the upper limit of the 
FCRAO data.
There is one object with a detection at the FCRAO and only a tentative detection
at IRAM (CIG 433), and one with a nondetection at IRAM (CIG 268).
The ratio of the intensities in both cases is   $I_{\rm CO-IRAM}/I_{\rm CO-FCRAO}=0.3$, indicating an
underestimate in the IRAM data. 
In the remaining three cases,  both observations  were either no detection or tentative detections.

We conclude that  there is very good agreement between the detected values
at both telescopes,   and in most cases (with the exception of three galaxies)  also for  objects only detected at
one telescope. This gives us confidence that the calibration of the two data sets
is consistent.

%--------------------------------------------------------------------------------------------------
 
\subsection{Calculation of the molecular gas mass}
 \label{molecular-gas-mass}

The molecular gas mass (\mhtwo ) is calculated using
a Galactic conversion factor of  
N(H$_2$)/I$_{CO} = 2. 0\times 10^{20}$ cm $^{-2}$(K km s$^{-1}$)$^{-1}$
\citep[e.g.][]{1986ApJ...309..326D}
yielding:

\begin{equation}
M_{\rm H_2} [M_\odot] = 75 I_{\rm CO} D^2 \Omega ,
\label{cal_mol_mass}
\end{equation}
\noindent where 
$I_{\rm CO}$ is the velocity integrated CO line intensity in K \kms, $D$ is the distance in Mpc and $\Omega$ is the area covered
by the observations in arcsec$^2$ (i.e. $\Omega=1.13 \Theta_B^2$ 
for a single pointing with  a Gaussian beam of FWHM $\Theta_B$). 
We do not include the mass of heavy metal (mostly helium) in the
molecular gas mass. 
%{It might be better to include helium fraction. This can be easily changed.}

Most of our objects were observed at the central position in a single pointing since the mapping
of the entire galaxy would have been too time-consuming. We therefore might have missed
part of the CO emission for galaxies where  the emission is more extended than the beam.
This fraction  depends on the galaxy size, inclination and on the telescope beams. 
It is thus necessary to correct for this loss, and we do this by extrapolating
\mhtwo\ observed in the central beam to the total mass in the galaxy.
In the next subsection we explain how we carried out this correction.

%--------------------------------------------------------------------------------------------------
 
\subsubsection{Aperture correction}
 \label{aperature-correction}
 
% \paragraph{CO distribution in galaxies}

In order to  apply an aperture correction, we need to predict the 
distribution of the CO emission. 
CO maps of nearby spiral galaxies
\citep{2001PASJ...53..757N,2001ApJ...561..218R,2008AJ....136.2782L} 
%(Nishiyama, Nakai \& Kuno 2001; Regan et al. 2001; Leroy et al. 2008)
have shown that  the radial distribution of $I_{CO}(r)$  in galaxies
can be well described by an exponential function with a scale length $r_e$:

\begin{equation}
I_{CO}(r) = I_{0}  \exp(-r/r_{\rm e}). 
\label{exp_disk}
\end{equation}

The CO scale length, $r_{\rm e}$, is well correlated and similar to the optical exponential scale length 
\citep{2001ApJ...561..218R,2008AJ....136.2782L}. %(Regan et al. 2001; Leroy et al. 2008).
It also correlates, although less tightly, with
the optical  radius at the 25mag isophote, $r_{\rm 25}$.
\citet{2008AJ....136.2782L}  %Leroy et al. (2008)
derived for spiral galaxies from
the THINGS survey 
a mean value of  $\alpha = r_{\rm e}/r_{\rm 25} =0.2$.
We derived the same mean value for $\alpha$ from  the data of \citet{2001ApJ...561..218R} %Regan et al. (2001)
for 15 spiral galaxies observed
in  the BIMA Survey of Nearby Galaxies (BIMA-SONG)  and from the data of 
\citet{2001PASJ...53..713N} %Nishiyama et al. (2001) 
for 25 spiral galaxies observed with
 the Nobeyama 45m telescope.
We also used the data of \citet{1995ApJS...98..219Y}, %Young et al. (1995) 
who studied the molecular gas content and distribution  in a sample of
300 nearby galaxies, to derive the $r_{\rm e}$.
% They found an exponential distribution to be a good
%fit for about half of the objects. 
They found a mean ratio between 
 the effective CO diameter, $D_{\rm CO}$,  the
 diameter  within which 70\% of the CO emission is situated,
 and the optical diameter of $D_{\rm CO}/D_{\rm 25} = 0.5$.
For an  exponential
distribution one can derive that  $D_{\rm CO} \times 0.5 = 2.5 r_e$. 
Thus, their data also yield $r_e/r_{\rm 25} = \frac{D_{\rm CO}\times 0.5/2.5}{D_{\rm 25}\times 0.5} = 
 \frac{0.5}{2.5} =0.2$.

Finally, we use the data of the 19 galaxies (all of them with morphological type $T \ge 2$)
mapped along the major axis with the IRAM 30m telescope as a further test.
Although our data is  not sufficiently detailed  to  fit  the radial distribution (we have only 3-5  detected spectra
along the major axis), we can use it to
test whether (i)  an exponential distribution is a reasonable 
description of the CO distribution, (ii) 
the scale length derived by other studies is in agreement with our data, and (iii) the
predictions for the extrapolated \mhtwo\ are in agreement
with our mapped values.  

For the first two tests,
we fitted an exponential distribution independently to each  side of the CO distribution along the major axis. 
From the 38 resulting fits, there were only   six cases where no exponential fit could be applied within the error
bars. In  five cases we derived  $\alpha \ge 0.4$, in seven cases  $0.4 < \alpha < 0.25$, 
in three cases  $\alpha < 0.15$ and in  17 cases, the majority,
 $0.15 \le \alpha \le 0.25$. Thus, our data 
are in general consistent with the value of $\alpha$ found by
other studies. %, although they also confirm the relatively large scatter.

In summary, we conclude that an exponential distribution of the molecular
gas distribution with $\alpha = r_e/r_{\rm 25} = 0.2$ is a good approximation based
on the CO maps  for nearby spiral galaxies available up to date and also consistent
with our data.
We adopt the same value for both spiral galaxies and early type
galaxies.
In early-type galaxies, the molecular gas extent is much less known.
However, for our study, this uncertainty is not  important because
the  number of objects with $T\le 0$  is low  ($n =23$, with  eight detections) and
we focus on our results and conclusions 
on  spiral galaxies, in particular of types $T=3-5$, which dominate the sample.
%where the molecular gas distribution 
%is similar (M. Bureau, private communication).

%paragraph{Aperture-corrected molecular gas mass}

We now use these results  to calculate the aperture correction which 
we define as the ratio between
the total (extrapolated) molecular gas mass \mhtwo, and the molecular gas mass in the
central pointing, \mhtwocenter ,

\begin{equation}
f_{\rm ap} = M_{\rm H_2} /M_{\rm H_2,center} .
\label{corr_factor}
\end{equation}

The total molecular gas mass is calculated  
by spatially integrating $I_{\rm CO} (r)$  from Eq.~\ref{exp_disk}
and  using Eq.~\ref{cal_mol_mass}.
This yields :
\begin{eqnarray}
M_{\rm H_2} [M_\odot]  &=&  75 D^2  \int_0^\infty I(r) 2\pi dr \\ \nonumber
&=&  75 D^2  \int_0^\infty I_{0}  \exp(-r/r_e)   2\pi dr
 =75 D^2  I_0 2 \pi r_e^2 .
\end{eqnarray}

Similarly, we calculate \mhtwocenter\ by convolving the exponential 
CO intensity distribution with a Gaussian beam. This yields:

\begin{eqnarray}
\label{extra_form}
 & M_{\rm H_2, center} [M_\odot]   =  
 75 D^2 4 I_0 \int_0^\infty dx  \int_0^\infty dy  \\ \nonumber
 & \exp\left(-\ln(2)\left[\left(\frac{2x}{\Theta_B}\right)^2 +
\left(\frac{2y\cos(i)}{\Theta_B}\right)^2\right] \right)
 \exp\left(-\frac{\sqrt{x^2+y^2}}{r_e}\right),
\end{eqnarray}
where $i$ is the inclination of the disk.
% and $\Theta_B$ the FWHM of the Gaussian beam.
The integration of Eq.~\ref{extra_form} 
is carried out numerically. 

\begin{figure}
\centering
\includegraphics[width=7cm,angle=0]{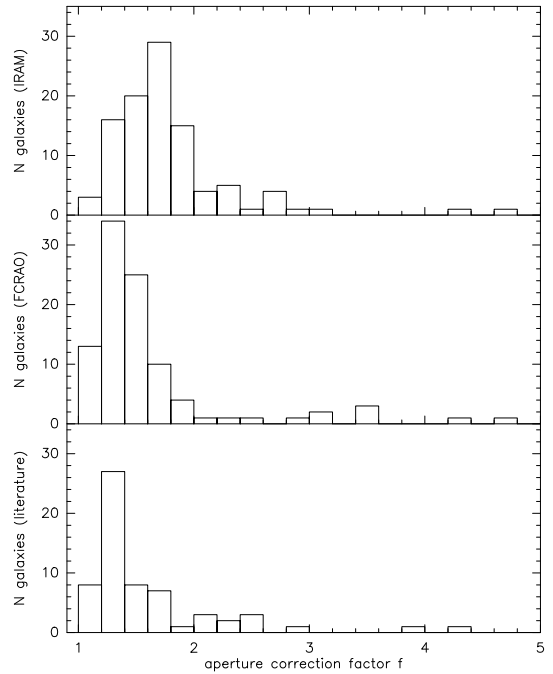}
\caption{Histogram of the aperture correction factor, $f_{\rm ap}$, for galaxies
observed by us with the IRAM 30m telescope (upper panel),
with FCRAO (middle panel) and for galaxies taken from the literature (lower panel).
}
\label{histo_extra}
\end{figure}

Thus, the correction factor, $f_{\rm ap} = $\mhtwo/\mhtwocenter ,
depends on the ratio of the scale length and the beam size, $r_e/\Theta_B$, as well as  the galaxy inclination $i$.
%The scale length is derived for each galaxy from $r_{25}$, as described above, with $r_e = \alpha  r_{25}$.
Fig. \ref{histo_extra}  shows the distribution of $f_{\rm ap}$ for the galaxies in our sample.
The correction factors are generally low:
81\%  of the galaxies have $f_{\rm ap}<2$, and 92\%  $f_{\rm ap}<3$.
Only nine galaxies have a correction factor above 5. All of them are nearby
(v $< 1000$\kms) galaxies with a large angular size (between 4\arcmin and 20\arcmin)
that are not included in our redshift-limited sample.
 
%*************************************************
%*************************************************

\begin{figure}
\centering
\includegraphics[width=4.6cm,angle=270]{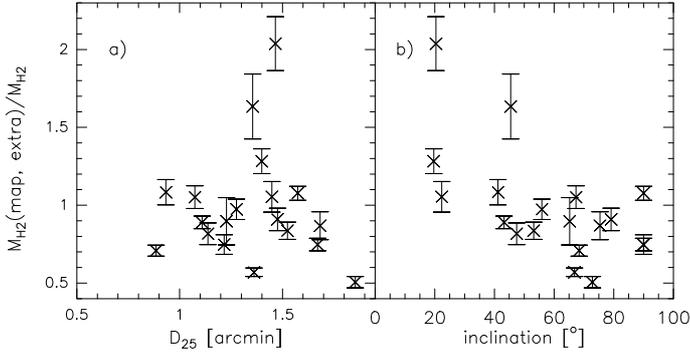}
\caption{{Ratio between the  molecular gas mass,  mapped along the major axis with the IRAM 30m telescope, and extrapolated to the
entire disk assuming azimuthal symmetry (\mhtwomapextra ) and the molecular gas mass extrapolated from the
central pointing as explained in Sect. 3.5.1} (\mhtwo )
as a function of  diameter   (a) 
and galaxy inclination (b). 
%In the panels c) and d) we show the 
%corresponding values, but have extrapolated the mapped molecular gas mass
%azimuthally, assuming radial symmetry and taking into account the inclination.
}
\label{fig:comp_extra_mapped}
\end{figure}

%\clearpage
In order to carry out test (iii) %with respect to the validity of this extrapolation 
 we compared for the 19 galaxies, that were mapped along their major axis with
the IRAM 30m telescope, the extrapolated \mhtwo\ to
 {the mapped molecular gas mass  which we 
  extrapolated  to the mass in the entire disk  by assuming azimuthal symmetry  (\mhtwomapextra ).
 Fig. \ref{fig:comp_extra_mapped}  presents the ratio  of  these
 two masses as a function
 of \dopt\ and $i$.
For most objects,  both masses agree reasonably well,  with ratios  ranging between 0.5 and 1.3.
There are only two outliers, with ratios around 2. These galaxies (CIG 84 and CIG 28) have
 a very flat CO distribution along the major axis so that our extrapolation, assuming an exponentially
 decreasing distribution, underestimates the true amount of molecular gas.  
  No trend with neither  \dopt\  nor the inclination is seen, showing that no apparent bias is   introduced
 by the aperture correction. 
 The mean mass ratio  is   $1.0$, with a standard
 deviation of 0.3.}

% The slightly lower \mhtwo\ derived from the mapping is
% most likely due to the fact that the mapping was performed along the major axis, so that some
% emission was missed. 
% When we correct for this effect assuming azimuthal symmetry, we get
% a mean of \mhtwo/\mhtwomap\ = 1.0 (standard deviation of 0.3).
 
 %This hypothesis is supported by the fact that most of the galaxies with
 %($i  \sim 90^\circ$)  have a ratio close to 1. 

% There is a  weak trend with inclination in the sense that
%  low inclination galaxies have a lower mapped-to-extrapolated molecular gas mass ratio.
 % This trend is expected since the mapping was  done only along the major
 %   axis. Therefore only in edge-on galaxies ($i  \sim 90^\circ$)  the mapping observations
 %   are expected to measure the total molecular
 %   gas content.   For high inclinations the mass ratio approaches 1, supporting 
 %   the correctness of our aperture correction.
%
 % In order to test this and correct for the missing emission, we extrapolated in the lower
 % Irow (panel c) and d)) the mapped value  to 
 % I to the total molecular gas mass in the disk, assuming
 % Ithat the distribution is axisymmetric and taking into account the inclination of the disk.
 % IThe ratio between both molecular gas mass are now closer to 1 
 % Iand no trend with inclination is seen
 % IThe good agreement between the observed and extrapolated values  and the lack
 % Iof trend of the ratio with diameter or inclination
 % Istrongly supports the correctness of our  that the extrapolation according to eq. (xxx).
 
%*************************************************
%*************************************************

%*************************************************

\section{Results}
 \label{results}
 
\subsection{Molecular gas content}

In Table ~\ref{mol_mass} we list \mhtwo\ for the individual galaxies.
If observations from different references
were available for a single galaxy, we inspected the spectra  and
discarded those of poorer quality. In case of similarly good data
we gave preference to mapped observations or to  
observations from the telescope with a larger beam in order to 
avoid flux loss.
For the 44 objects that had data from the literature and were also observed by us, our data was in
general of
better quality, with the exception of three objects
(CIG 512, CIG 604, CIG 626).
The columns in Table ~\ref{mol_mass} are:

\begin{enumerate}
\item  CIG: Entry number in the Catalogue of Isolated Galaxies (CIG).
\item log(\mhtwo) center: Decimal logarithm of \mhtwo\ towards the central pointing
in solar masses,  derived from Eq. ~\ref{cal_mol_mass}.
An asterix  denotes tentative detections that are
treated as upper limits in the statistical analysis. 
%We list this quantity mainly for reference, because
%in our analysis we always use the extrapolated molecular mass.
\item log(\mhtwo) mapped: Decimal logarithm of the mapped \mhtwo\, in solar masses,
calculated in the following way: for the
data from the literature, the angular separation between the individual pointings was always larger than
the beam size so that the total \mhtwo\ could be calculated as the sum
of the individual pointings. For our own observations with the IRAM 30m telescope,
 the spacing between
the individual pointing was 15'', which is smaller than the
FWHM of the beam (21\arcsec) so that in this case we had to take  the overlap
of the individual pointings into account. We calculated the mapped \mhtwo\
from Eq. ~\ref{cal_mol_mass}  where   $I_{\rm CO}$  is taken as the mean value of the
different pointings, and  $\Omega$  is the total area covered by the
mapping, approximated as  $21\arcsec\times 36\arcsec$, $21\arcsec\times 51\arcsec$, $21\arcsec\times 66\arcsec$,
$21\arcsec\times 81\arcsec$ for 2, 3, 4, and 5 pointings, respectively.
%We  corrected for this overlap after 
% summing the individual pointings. The correction factor was
%given, for a map consisting of $n$ individual, consecutive  pointings,
% by the ratio between the area covered by the mapping and $n$  times 
% the solid angle  $\Omega$ of the beam .
%The correction factor was 0.76, 0.71, 0.69 and 0.68  for 2, 3, 4, and 5 pointings, respectively.
%
%If no mapping was carried out, the
%entry is 0 {maybe change this to "-"}.
\item  log(\mhtwo) extrapol.: Total (extrapolated) \mhtwo\ in solar masses, 
calculated as described in  Section 3.5.1. An asterisk  denotes tentative detections that are
treated as upper limits in the statistical analysis.
\item Tel.: Radio-telescope code, as  in Table~\ref{telescopes}.
\item Ref.: Bibliographic code, as in Table~\ref{literature}.

\end{enumerate}

%--------------------------------------------------------------------------------------------------
\begin{table}
\caption{Molecular gas mass}
\begin{center}
\begin{tabular}{lccccc} 
\hline
CIG &  \multicolumn{3}{c}{log(M$_{\rm H_2}$) [\msun\ ] } &Tel. & Ref. \\
 &   center & mapped & extrapol. & & \\
(1) & (2) & (3) & (4) & (5) & (6)  \\
\hline
     1   &    9.47   &     0   &  9.58   &   3   &  4  \\
     4   &     8.96   &  9.08   &  9.26   &   2   &  2  \\
     6   &   $<$8.14   &     0   &  $<$8.14   &   5   &  4  \\
    10   &    7.88$*$   &     0   &  8.13$*$   &   1   &  1  \\
    .... &    ..... &  ..... &  ..... &  ..... &  .....  \\
\hline

\end{tabular}
%{make  detection code consistent}
\end{center}

The full table is available in electronic form at the CDS and from http://amiga.iaa.es.
\label{mol_mass}
\end{table}
%--------------------------------------------------------------------------------------------------

%--------------------------------------------------------------------------------------------------
\begin{figure}
\begin{center}
\includegraphics[width=7cm,angle=270]{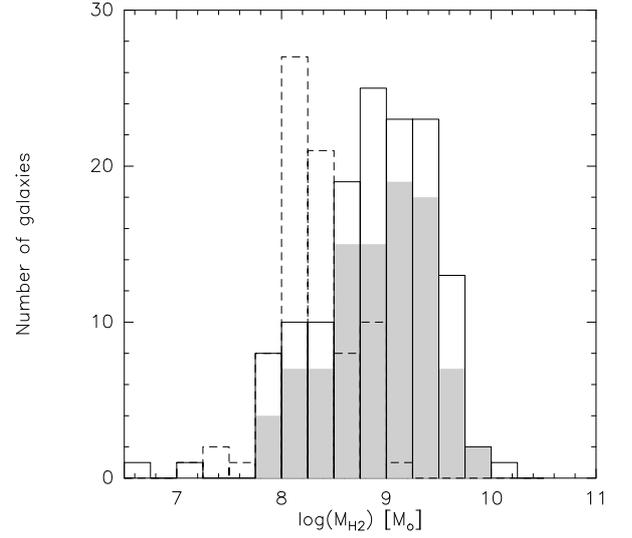}
\caption{Histogram  of the molecular gas mass, \mhtwo . Galaxies flagged as potentially 
interacting were excluded. The full line corresponds to all detections ($n = 131$),  the gray filled histogram shows detections
in the redshift limited sample ($n = 89$) and the dashed lines shows tentative or nondetections in the
redshift-limited sample ($n = 84$). }
\label{histo-mh2}
\end{center}
\end{figure}
%--------------------------------------------------------------------------------------------------

%--------------------------------------------------------------------------------------------------
\begin{figure}
\begin{center}
\includegraphics[width=7cm,angle=270]{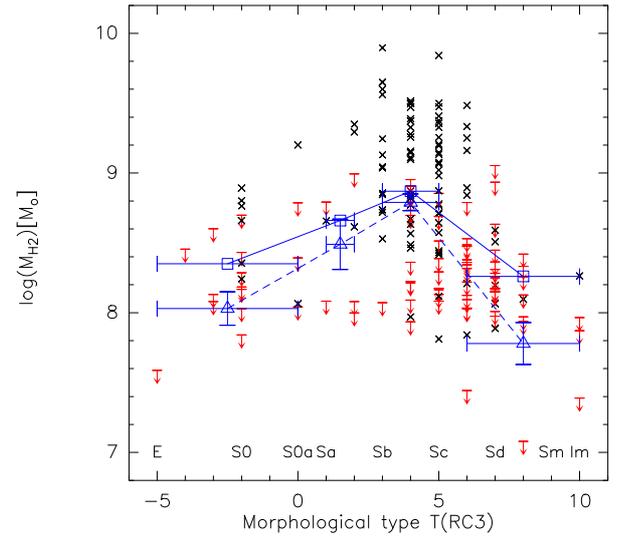}
\caption{The molecular gas mass for the redshift-limited sample 
as a function of morphological type. 
Triangles  %(from  Table~\ref{tab1_average}) 
denote
the mean value and its error for a range of morphological types and squares the median value,
as listed in   Table~\ref{tab1_average}. The error bars in the x-direction denote the range of
morphological types over which the mean and median have been taken.
}
\label{mh2-hubble}
\end{center}
\end{figure}
%--------------------------------------------------------------------------------------------------

Fig.~\ref{histo-mh2} shows the distribution of log(\mhtwo)\footnote{
Here, and in the following, we always use to the extrapolated molecular gas mass in our analysis and 
denote it \mhtwo\ for simplicity.},
for detections in the total and redshift-limited CO samples and for 
nondetections in the latter.  The distributions  in the total and in the redshift-limited samples are
very similar. The mean \mhtwo\  of 
 the redshift-limited sample is   log(\mhtwo) = 8.30 $\pm$ 0.08
 (see Table~\ref{tab1_average}), calculated with the package ASURV that takes into
 account the upper limits. Here and throughout the paper
 we use  the ASURV\footnote{
 Astronomy Survival Analysis (ASURV) Rev. 1.1 is a generalized statistical package that implements the methods
 presented by \citet{1985ApJ...293..192F}     and  \citet{1986ApJ...306..490I},  and is described in detail in  
\citet{1990ApJ...364..104I}   and   \citet{1992ASPC...25..245L}.
 }
package which applies  survival analysis in the presence of upper or lower limits and  calculates the mean value based
on the Kaplan-Meier estimator. 
%The median value we present for comparison are obtained by treating the upper limits as detections.

 Fig.~\ref{mh2-hubble} shows the distribution of the molecular gas mass as a function of the morphological
 type. The mean and median 
 values are listed in Table~\ref{tab1_average}.
 Our sample is dominated by spiral galaxies of type  $T=3-5$ (Sb-Sc).  Not only is the total number
 of objects greatest in this range, but also the detection rate. Therefore, we can
 derive the most reliable results for these types. Both for earlier and for later types,
 the detection rates are very low, making a detailed analysis difficult. 
The molecular gas mass is largest
 for spiral galaxies of  $T=3-5$, and decreases 
 both for earlier and later types. 
 There are eight early-type galaxies,  of type S0 and S0a,  with detections in \mhtwo , and five
 of them have unusually high molecular gas masses in the range of those for spiral galaxies
 (CIG 332, CIG 481, CIG 498, CIG 733 and CIG 1015).

%%%%%%%%%%%%%%%%%%%%%%%%%%%%%%%%
\subsection{Relation of  \mhtwo\ to other parameters}

In the following we investigate the relations between \mhtwo\ and 
\lb  , $D_{25}$, \lfir , \lk\  and  \mhi . 
The first two quantities (\lb  , $D_{25}$) were chosen because they are  in general
available for any galaxy and are therefore useful  to predict
the expected \mhtwo . \lfir\  is very closely related
to \mhtwo\ because of their common relation to
SF.  { \lk\ is dominated by the emission of low-mass stars which are}
the result of the long-term SF history of an object and determine the gravitational potential
which influences the SF  activity.
 Finally, we compare \mhtwo\ to \mhi\
in order to derive the molecular-to-atomic gas mass ratio as a function of morphological
type.

The results of the regression analysis as well as the Spearman rho correlation coefficient derived in this section are listed
 in  Table~\ref{regression}, and the mean and median 
 values of the  ratios {for different subsamples} are in Table~\ref{tab1_average}.
  We use the package  ASURV to calculate the bisector regression line applying the Schmitt's binning method
 \citep{1985ApJ...293..178S}
 as the only method offered by ASURV able to deal with censored data  in  both the dependent and the 
 independent variable\footnote{
 The Schmitt method was partially reimplemented and wrapped into Python. It can be found at
 http://amiga.iaa.es/software/python-asurv.
 }.
 
% We list both the parameters of the bisector fit and the parameters of the fit obtained by treating 
% \mhtwo\ as the dependent parameter. {MORE}.

%%%%%%%%%%%%%%%%%%%%%%%%%%%%%%%%
\subsubsection{Optical luminosity}

%*************************************************
\begin{figure}
\includegraphics[width=8cm,angle=0]{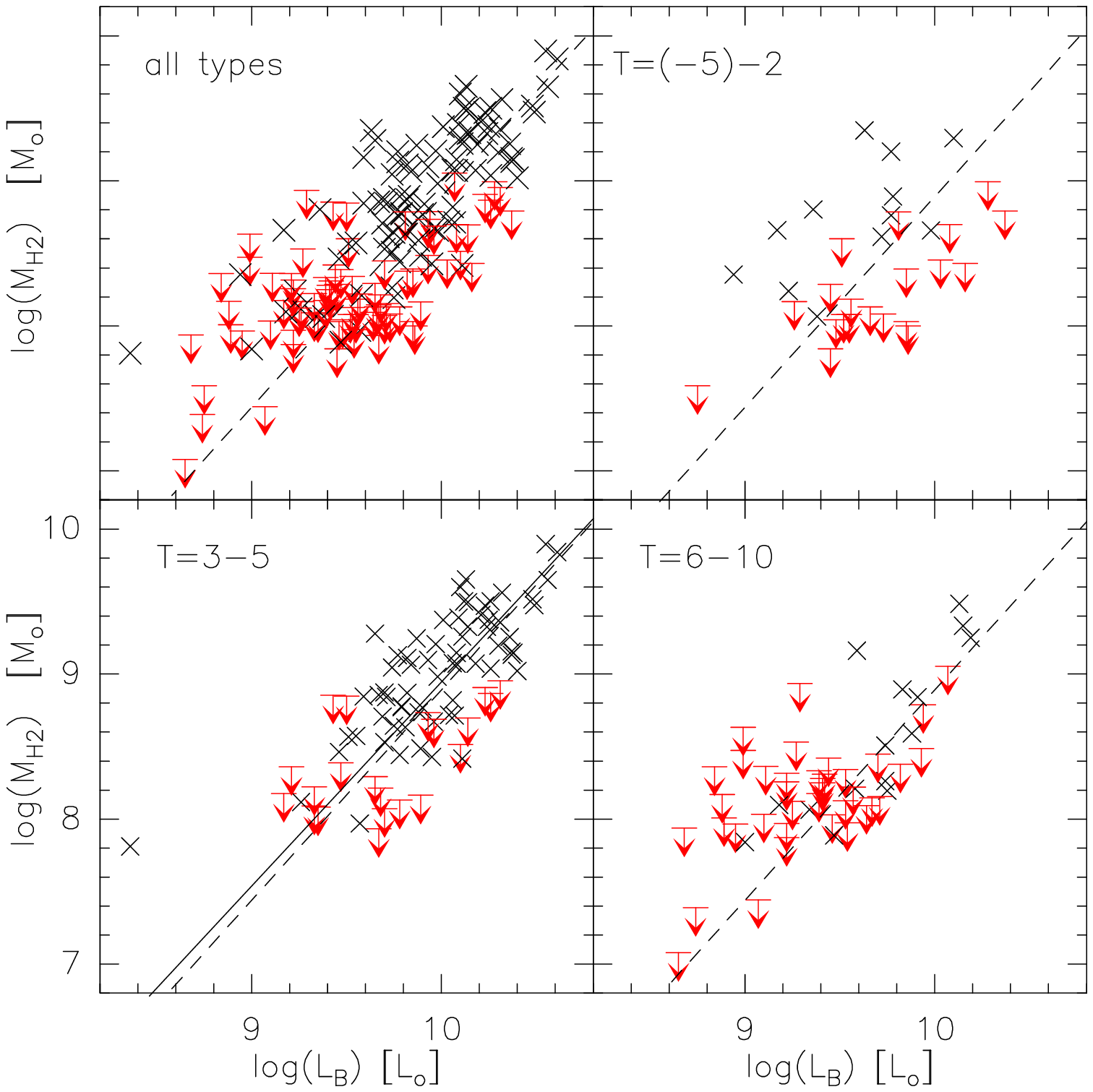}
\caption{ \mhtwo\ vs. \lb\ for the redshift-limited CO sample
($n=173$), including all morphological types, and for groups of 
different morphological types. 
The dashed line gives
the best  fit bisector (derived with ASURV) for all morphological types and  the solid line is the best fit
for the $T =3-5$ sample.
}
\label{mh2-lb_types}
\end{figure}
%*************************************************

%*************************************************
\begin{figure}
\includegraphics[width=7cm,angle=270]{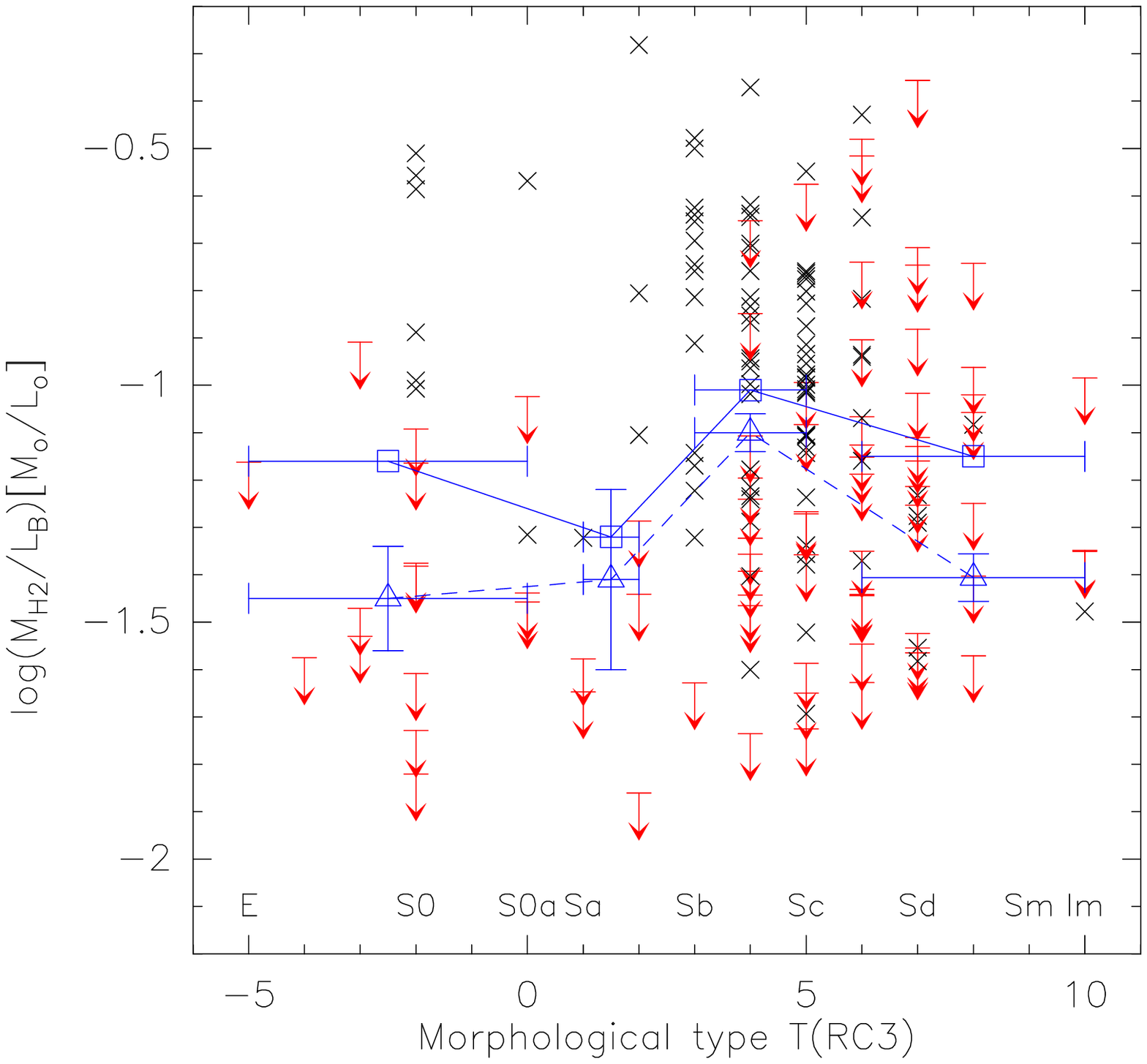}
\caption{ The ratio between \lb  and \mhtwo\ 
as a function of morphological type. 
Triangles denote the
mean value and its error for a range of morphological types %(listed in Table~\ref{tab1_average})  
and squares  denote
the median values, 
%calculated treating upper limits as detections, 
as listed in   Table~\ref{tab1_average}. The error bars in the x-direction denote the range of
morphological types over which the mean and median have been taken.}
\label{mh2-lb-hubble}
\end{figure}
%*************************************************

%*************************************************
\begin{figure}
\includegraphics[width=6cm,angle=270]{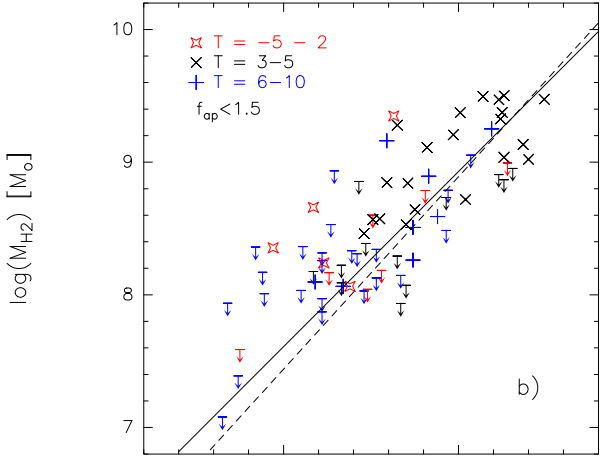}
\caption{ \mhtwo\ vs. \lb\ for galaxies in the redshift-limited CO sample
with an aperture correction factor of less
than $f_{\rm ap} < 1.5$  ($n=78$).
The solid line is the best fit bisector line derived with ASURV
for this restricted sample, and the dashed lines gives, for comparison,
the fit for the entire redshift-limited CO sample.
}
\label{mh2-lb_small_aperture}
\end{figure}
%*************************************************

%*************************************************
%begin{figure}
%\includegraphics[width=6cm,angle=270]{mh2_extra_lb_diff_t.ps}
%\includegraphics[width=6cm,angle=270]{mh2_extra_lb_diff_3_t_small_corr.ps}
%\includegraphics[width=6.85cm,angle=270]{mh2_extra_lb_t_3_5.ps}
%\caption{ \mhtwo\ vs. \lb\ for the redshift-limited CO sample
%($n=173$).
%Panel a) shows  the sample with different codings for different groups of morphological types,
%panel b)
%only for galaxies with an aperture correction factor of less
%than $f_{\rm ap} < 1.5$  ($n=78$), and panel c)
%only galaxies with morphological type $T = 3-5$ (Sb-Sc).
%The solid line is the best fit bisector line derived with ASURV
%for the respective sample, and the dashed lines gives, for comparison,
%the fit for the entire redshift-limited CO sample from panel a) (see Table~\ref{regression}).
%}
%\label{mh2-lb_all}
%\end{figure}
%*************************************************

Fig.~\ref{mh2-lb_types} shows the relation between \mhtwo\ and \lb ,
for the the entire redshift-limited sample and for subsamples of different morphological types,
together with the  the best-fit bisector regression lines.
 The high number of upper limits
for galaxies of type $T=(-5) - 2$ and $T=6-10$ impedes the calculation of a reliable regression line for
these individual subgroups.  However, we note that there is no apparent deviation 
 from the mean best-fit regression for all morphological types for these two groups.
A slope considerably larger than 1 is found for the correlation between \mhtwo\ and \lb ,
in agreement with  \citet{1997ApJ...490..166P}.
%in the sense that  \mhtwo\  increases faster than  \lb . 
This means that the ratio  \mhtwo/\lb\  tends to increase with \lb . {
Even though the scatter around the best-fit is large, so that no strong correlation exists between
\mhtwo/\lb\ and \lb , we confirm the variation of \mhtwo/\lb\
by finding that  the mean values  is indeed higher  for high \lb\  than for low \lb :
 for galaxies (all morphological types)  with \lb $< 10^{10}$ \lsun\ ($n=123$ galaxies,  $n_{\rm}= 68$ upper limits)  we obtain
 \mhtwo/\lb =$-1.36\pm 0.05$
and for galaxies with \lb $\ge 10^{10}$ \lsun\ ($n/n_{\rm} = $50/11)   \mhtwo/\lb = $-1.10\pm 0.06$. 
The corresponding numbers for the $T=3-5$ subsample are
$-1.16\pm 0.06$ (\lb $< 10^{10}$ \lsun , $n/n_{\rm} = $48/16) and
$-1.04\pm 0.05$ (\lb $\ge 10^{10}$ \lsun , $n/n_{\rm} = $40/5).
This trend 
has to be taken into account when using this ratio as an indicator of
an enhancement of \mhtwo .}

The ratio between  \mhtwo\ and \lb\ is shown in Fig.~\ref{mh2-lb-hubble}
as a function of morphological type. The values are listed in Table~\ref{tab1_average}.
The ratio is highest for galaxies of type $T=3-5$. {It is however remarkable that  galaxies of type S0 and S0a
with detections in CO have much higher values of \mhtwo/\lb ,  in the range of 
spiral galaxies of type $T = 3-5$, than early-type galaxies with nondetection in CO.}

In order to check  whether the extrapolation of the molecular gas mass to the
entire disk has introduced any biases, we show in Fig.~\ref{mh2-lb_small_aperture} the relation between 
\mhtwo\ and   \lb\ only for  galaxies with
$f_{\rm ap} < 1.5$. No significant difference  compared to the entire redshift-limited CO sample can be seen, 
and the best-fit regression 
coefficients are the same within the errors
 (Table~\ref{regression}).
 We found this good agreement between
 entire redshift-limited CO sample and the subsample of
galaxies with $f_{\rm ap} < 1.5$ for all correlations studied.
Therefore,
in the following subsections we do not show the $f_{\rm ap} < 1.5$ correlation separately, but
we list in Tables~\ref{tab1_average} and  \ref{regression} the mean values and the regression coefficients 
for this  subsample.

\begin{table*}
      \caption{Mean and median values.}
\begin{tabular}{lccccccccc}
\hline
Sample &   $ <$ log(\mhtwo)$>$  &  $< \log(\frac{M_{\rm H2}}{L_{\rm B}})>$ &  $< \log(\frac{M_{\rm H2}}{D_{\rm 25}^2})>$  & $ < \log(\frac{M_{\rm H2}}{L_{\rm K}})>$  &  
 $<\log(\frac{M_{\rm H2}}{M_{\rm HI}})>$   & 
 $<\log(\frac{L_{\rm FIR}}{M_{\rm H2}}) >$ &$<$log(SFE)$>$\\
  & [\msun] &  [\msun/\lsun] & [\msun/kpc$^2$] & [\msun/\lsunk] &  & [\lsun/\msun] & [yr$^{-1}$] \\
   & median &median &median &median &   median &   median &   median\\
   &  $n/n_{\rm up}$ &   $n/n_{\rm up}$ &  $n/n_{\rm up}$ &   $n/n_{\rm up}$ &  $n/n_{\rm up}$ &  $n/n_{\rm up}$ &  $n/n_{\rm up}$ \\
(1) & (2) & (3) & (4) & (5) & (6) & (7)  & (8) \\
\hline
{E-Im ($T$=(-5) - 10)}                &   8.30$\pm$0.08 &  -1.28$\pm$0.04 & 6.00$\pm$0.05 &  -1.87$\pm$0.05 &  -0.94$\pm$ 0.05 & 0.72$\pm$0.03 & -8.94\\
                                 &              8.61                  &  -1-10& 6.22              &           -1.58                          & -0.73                                  & 0.64& -9.02\\
                                  & 173/79 &173/79 &  173/79 &157/65& 153/63 & 97/22 &  97/22 \\
{E-Im }($f_{\rm ap}<$ 1.5) &  8.30$\pm$0.12 &  -1.29$\pm$0.06 &  6.06$\pm$0.07 &   -1.80$\pm$0.06 & -0.96$\pm$0.08 &  0.80$\pm$0.06&-8.86\\
                                     & 8.51                    &      -1.07       &   6.29   & -1.55                        &  -0.76                       &  0.69  & -8.97 \\
                                  & 76/42 & 76/42 & 76/42 & 65/33  & 66/33 & 39/12 &  39/12 \\
\hline
E-S0a ($T$=(-5) - 0) &  8.03$\pm$0.12 & -1.45$\pm$0.11 & 5.79$\pm$0.13& -2.32$\pm$0.10 &  -0.46$\pm$0.21& 0.95$\pm$0.08 & -8.71\\
                                 &              8.35     &     -1.16          &    6.27         &       -2.06                &   -0.25                       &  0.94& -8.72 \\
                                 & 23/15& 23/15& 23/15&  22/15  &8/3& 6/0 & 6/0 \\
Sa - Sab ($T$=1-2)  & 8.49$\pm$0.18 &-1.41$\pm$0.19 &5.76$\pm$0.20 &  -2.27$\pm$0.19 & -0.58$\pm$0.21 &0.69$\pm$0.06&-8.97 \\
                                         & 8.66              &   -1.32         &   5.89        &         -2.06                 &   -0.53                         &  0.63   & -9.03\\
                                         &9/5  & 9/5      &9/5     &9/5      & 8/4 &  6/3 & 6/3 \\
Sb-Sc ($T$= 3-5)  &  8.79$\pm$0.06& -1.10$\pm$0.04 &6.26$\pm$0.05 &  -1.66$\pm$0.04  &  -0.72 $\pm$0.06 &  0.63$\pm$0.03& -9.03\\
                               &                8.87   & -1.01   &  6.37      &        -1.58                           &   -0.60                      &   0.62  & -9.04\\
                               & 88/21  & 88/21  & 88/21  & 87/20 & 85/19  & 68/10 & 68/10 \\
Scd-Im ($T$= 6-10)  &   7.78$\pm$0.15 & -1.41$\pm$0.05 & 5.79$\pm$0.05& -1.62$\pm$0.06&  -1.38$\pm$0.07 &  0.99$\pm$0.09 &-8.67\\
                                                  & 8.26   &         -1.15        &   6.01        &        -1.37            &        -1.07                        &   0.63 &-9.03\\
	   & 53/38 & 53/38 & 53/38 & 39/25 &52/37 & 17/9 &17/9 \\

%--------------------------------------------------  INVERTED
%\begin{table*}
%      \caption{Mean and median values.}
%      %of \mhtwo , \mhtwo/\mhi , \mhtwo/\lk , \lfir/\mhtwo and SFE.}
%\begin{tabular}{c|ccccccc}
%\hline
%   & {E-Im}   & {E-Im }($f_{\rm ap}<$ 1.5) & E-S0a  & Sa- Sab & Sb-Sc   & Sb-Sc \\
%  &  ($T$=-5 - 10)      &    ($T$=-5 - 10)  & ($T$=-5 - 0) &   ($T$= 1-2) & ($T$= 3-5) & ($T$= 3-5)   \\
%\hline
%$ <$ log(\mhtwo)$>$  & 8.30$\pm$0.08 &  8.03$\pm$0.12 & 8.03$\pm$0.12 & 8.49$\pm$0.18 & 8.79 $\pm$0.06 & 7.78$\pm$0.15  \\
%median                         &  8.61     &  8.51  & 8.35   &   8.66   & 8.87   & 8.26   \\
%$n/n_{\rm up}$          & 173/79  &  76/42  & 23/15 & 9/5  & 88/21  & 53/38 \\ 
%$< \log(\frac{M_{\rm H2}}{L_{\rm B}})>$   \\
%median   &   \\
%$n/n_{\rm up}$  &    \\ 
%$< \log(\frac{M_{\rm H2}}{D_{\rm 25}^2})>$  &\\
%median   &   \\
%$n/n_{\rm up}$  &    \\ 
%$< \log(\frac{M_{\rm H2}}{L_{\rm K}})>$  &\\
%median   &   \\
%$n/n_{\rm up}$  &    \\ 
%$<\log(\frac{M_{\rm H2}}{M_{\rm HI}})>$  &\\
%median   &   \\
%$n/n_{\rm up}$  &    \\ 
% $<\log(\frac{L_{\rm FIR}}{M_{\rm H2}}) >$ & \\
%median   &   \\
%  $<$log(SFE)$>$\\
%median   &   \\
%$n/n_{\rm up}$  &    \\ 
%
\hline
\hline
\end{tabular}
\label{tab1_average}

%In the total number of galaxies column 6, we excluded 20 galaxies of the total sample.
%16 galaxies have  both non-detection in  \mhi\ and \mhtwo .  A further 4 galaxies have 
% no useful values for \mhi (either
%non-detection or  tentative detection or were not observed  in HI).
%They are not included either  because   ASURV cannot
%caculated mean values for samples including both upper and lower limits.
  Mean value and its error calculated with the program ASURV for different subsamples,
 and (below)  median value (calculated treating the upper limits as detections).
For the mean  and median 
 value of
log(\mhtwo/\lfir) and log(\mhtwo/\mhi) only galaxies with detection in \lfir\  or \mhi , respectively,
were taken into account, since
ASURV cannot handle both upper and lower limits. 
Below:  Total number of galaxies and number of upper limits in \mhtwo\  taken
into account for the means and median.
%
%$^{(1)}$ 152 is the total number for \mhtwocenter/\mhi , and 153 for
%\mhtwo/\mhi\ . The difference is due to CIG 96 for which no
%CO was detected at the central position but at positions offset from the center. Therefore 
% $M_{\rm mapped}$ could be determined which was
%adopted to be the same as \mhtwo\ in this case.
\end{table*}
%--------------------------

%-----------------------------------------------------------------------------------------------------
\begin{table*}
\caption{Correlation analysis
 \label{regression}}
\begin{center}
\begin{tabular}{lccccccc}
\hline
Sample       &  $n$ & $n_{up}$ & slope   &   intercept   & slope   &   intercept  & corr. coeff   \\
   &  & & (bisector) & (bisector) & (\mhtwo\  dep.) &  (\mhtwo\  dep.)  &  $r$ \\
(1) & (2) & (3) &(4) & (5) & (6 ) & (7)  & (8) \\
 \\ \hline
 \mhtwo\ vs. \lb \\
 \hline 
{E-Im ($T$=(-5) - 10)}                  & 173   &  79   & 1.45$\pm$ 0.08 &  -5.61$\pm$0.77 & 1.12$\pm$ 0.08 &  -2.43$\pm$0.83 & 0.66 \\  
{E-Im} ($f_{\rm ap}<1.5)^*$ & 76 & 42 & 1.32$\pm$0.10 &  -4.27$\pm$0.94 & 1.06$\pm$0.10 &  -1.79$\pm$0.98 & 0.63\\
Sb-Sc ($T=3-5$ )    &   88 &  21 &1.41$\pm$0.10 & -5.12$\pm$0.97 & 1.06$\pm$0.10 &  -1.69$\pm$1.02 &0.65\\
 \hline 
 \mhtwo\ vs. $D_{\rm 25}^2$ \\ 
 \hline 
{E-Im ($T$=(-5) - 10)}                  &  173  &  79  & 1.41$\pm$0.05 & 5.10$\pm$0.50 & 0.88$\pm$0.09 &  6.36$\pm$0.23 &0.52 \\  
{E-Im} ($f_{\rm ap}<1.5)^*$ & 76 & 42 & 1.32$\pm$0.08 & 5.47$\pm$0.71 & 0.91$\pm$0.11 & 6.38$\pm$0.26 & 0.47\\
Sb-Sc (3-5)    &  88 & 21 &1.31$\pm$0.09 & 5.50$\pm$0.93 & 0.79$\pm$0.10 &6.82$\pm$0.27 & 0.53\\
 \hline 
  \mhtwo\ vs. \lk \\ 
 \hline 
{ E-Im ($T$=(-5) - 10) }                 &  157 &  65 & 1.05$\pm$0.07 &  {-2.27$\pm$0.72} & 0.72$\pm$0.07 &   {1.13$\pm$0.74} &  0.64 \\  
{E-Im } ($f_{\rm ap}<1.5)^*$ & 65& 33& 0.99$\pm$0.09 &  {-1.54$\pm$0.96} & 0.69$\pm$0.09 &  {1.41$\pm$0.95} &0.66 \\
Sb-Sc ($T=3-5$)    & 87 & 20 &1.18$\pm$0.07 &  -3.49$\pm$0.78 &0.95$\pm$0.08 &  {-1.14$\pm$0.85} & 0.74\\
\hline 
 \mhtwo\ vs. \lfir  \\ 
 \hline 
{ E-Im ($T$=(-5) - 10) }      & 172  & 97   & 1.16$\pm$0.08 &  -2.14$\pm$0.72 & 0.98$\pm$0.06 &  -0.46$\pm$0.61 & 0.80 \\  
{E-Im} ($f_{\rm ap}<1.5)^*$ & 75& 48 & 1.15$\pm$0.13 & -2.09$\pm$1.23 & 0.89$\pm$0.11 & 0.28$\pm$1.04 & 0.71 \\
Sb-Sc ($T=3-5$)  & 88   & 30 & 1.04$\pm$0.06&  -1.00$\pm$0.55& 0.90$\pm$0.06&  0.29$\pm$ 0.58 & 0.82 \\
\hline 
$\Sigma_{\rm SFR}$ vs. $\Sigma_{\rm H_2}$    \\ 
 \hline 
{ E-Im ($T$=(-5) - 10) }       & 172  & 97   & 0.89$\pm$0.07 &  -2.78$\pm$0.20 & 0.82$\pm$0.07 &  2.25$\pm$0.20 & 0.65\\  
{E-Im} ($f_{\rm ap}<1.5)^*$ & 75  & 48 & 1.00$\pm$0.14 & -2.77$\pm$0.38 & 0.57$\pm$0.10 & 1.53$\pm$0.31 & 0.56 \\
Sb-Sc ($T=3-5$)  & 88   & 30 & 0.88$\pm$0.07&  -2.82$\pm$0.21& 0.84$\pm$0.08&  2.39$\pm$ 0.22 &0.71 \\
\hline
\end{tabular}
\end{center}

The entries are:
Column 1: Subsample considered. 
Column 2: Total number of galaxies in the respective samples. 
Column 3: Number of galaxies with upper limits in \mhtwo\  and/or \lfir . 
Column 4: Bisector slope and its error of the best-fit regression line derived with the Schmitt binning method in the ASURV package. The slope and intercept are defined as
 log(A) = intercept + log(B) $\times$ slope, 
where $A$ is \mhtwo\ or $\Sigma_{\rm H_2}$  and $B$ is 
 \lb , \lk,  \lfir , (\dopt)$^2$ or  $\Sigma_{\rm SFR}$, respectively.
Column 5: Bisector intercept and its error.
Column 6 and 7: Slope and intercept and their errors of the best-fit regression line derived with the Schmitt binning method in the ASURV package adopting \lb, \lfir\ , \lk\,  $D_{\rm 25}^2$ or  $\Sigma_{\rm SFR}$ as independent variable. 
Column 8: Spearman's rho correlation coefficient, calculated with ASURV.

$*$ Only galaxies for which the aperture correction factor, $f_{\rm ap}$ (see Sect. 3.5.1), is less than 1.5.

\end{table*}

%%%%%%%%%%%%%%%%%%%%%%%%%%%%%%%%
\subsubsection{Optical isophotal diameter}

\begin{figure}
\begin{center}
\includegraphics[width=8cm,angle=0]{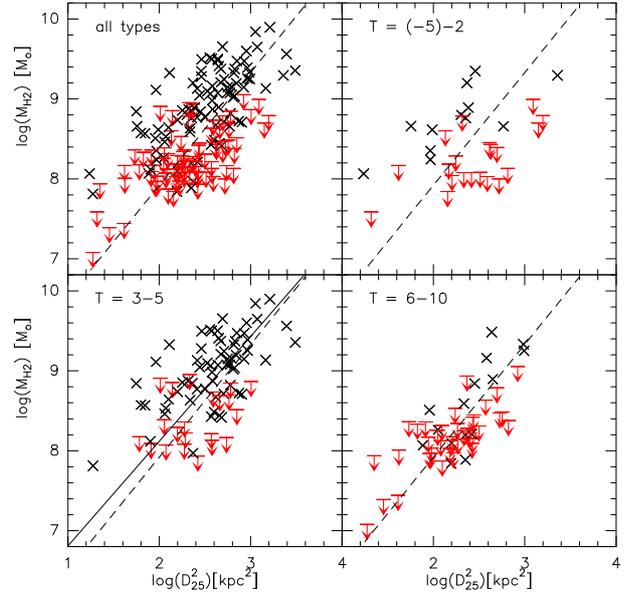}
\caption{
 \mhtwo\ vs. square of the optical isophotal diameter, \dopttwo\ , 
%for galaxies with morphological type $T=3-5$ of the  redshift-limited CO sample.  
 for the redshift-limited CO sample
($n=173$), including all morphological types, and for groups of 
different morphological types. 
The dashed line gives
the best  fit bisector (derived with ASURV) for all morphological types and  the solid line is the best fit
for the $T =3-5$ sample (see Table~\ref{regression}).
}
\label{mh2-diam-all}
\end{center}
\end{figure}

\begin{figure}
\includegraphics[width=7cm,angle=270]{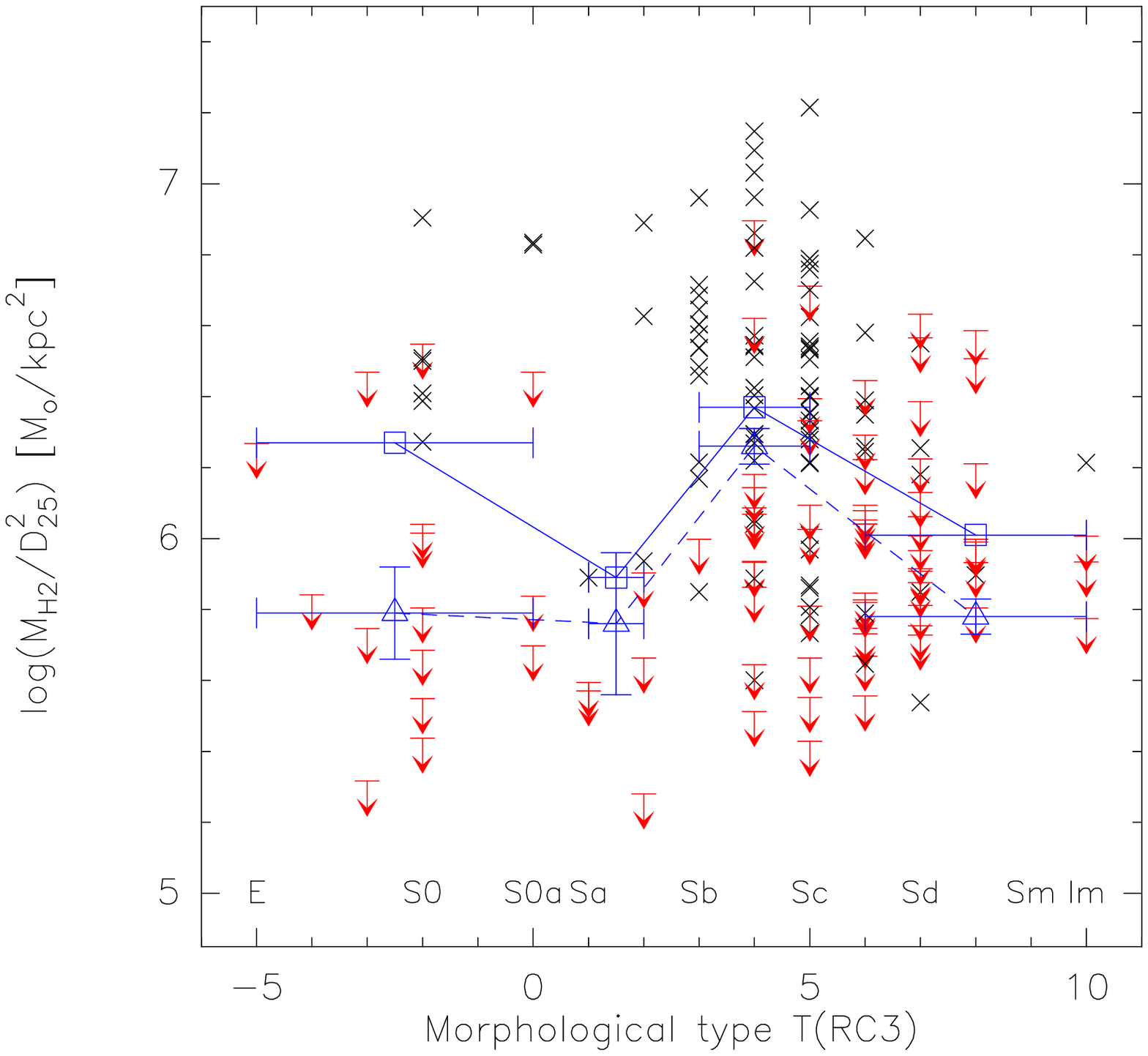}
\caption{The ratio between \mhtwo\  and  \dopttwo\  for the redshift-limted sample 
as a function of morphological type. 
Triangles denote the
mean value and its error for a range of morphological types %(listed in Table~\ref{tab1_average})  
and squares
the median values, 
%calculated treating upper limits as detections, 
as listed in   Table~\ref{tab1_average}. The error bars in the x-direction show the range of
morphological types over which the mean and median have been taken.}
\label{mh2-over-diam-hubble}
\end{figure}

{Fig.~\ref{mh2-diam-all} shows the relation between \mhtwo\ and \dopttwo,
for the  entire redshift-limited sample and for subsamples of different morphological types,
together with the  best-fit bisector regression lines. For early-type galaxies ($T\le 0$) only a very
poor correlation is visible, whereas galaxies of type $T=6-10$ seem to follow the same correlation as those of type $T=3-5$.

The correlation between \mhtwo\ and \dopttwo, has the lowest correlation coefficient ($r\sim 0.5$) among those considered in this
 paper (Table~\ref{regression}).
Although the bisector slope is formally larger than 1, we do not find a variation
of \mhtwo/\dopttwo\ with increasing optical diameter. This shows that for this poor correlation the regression
slope has to be taken with caution.

Fig.~\ref{mh2-over-diam-hubble} shows the ratio \mhtwo/\dopttwo\
as a function of morphological type. The values are listed in Table~\ref{tab1_average}.
The ratio is highest for spiral galaxies of
type $T=3-5$.
Similarly to  \mhtwo/\lb,  S0 and S0a galaxies with detections in CO have high values of  \mhtwo/\dopttwo , similar
to those of  spiral galaxies of type $T = 3-5$,   whereas  \mhtwo/\dopttwo\ of the nondetections is much lower.}

%*************************************************
%*************************************************

%*************************************************
%*************************************************

%%%%%%%%%%%%%%%%%%%%%%%%%%%%%%%%
\subsubsection{Luminosity in the K-band}

%-----------------------------------------------------
\begin{figure}
\includegraphics[width=8.cm,angle =0]{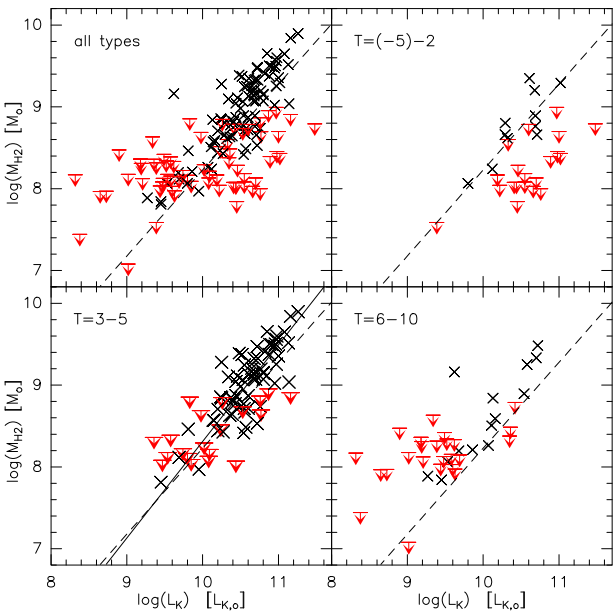}
\caption{The relation between \mhtwo\ and \lk\  
%for morphological type $T=3-5$ in the redshift-limited CO sample. 
for the redshift-limited CO sample
($n=173$), including all morphological types, and for groups of 
different morphological types. 
The dashed line gives
the best  fit bisector (derived with ASURV) for all morphological types and  the solid line is the best fit
for the $T =3-5$ sample (see Table~\ref{regression}).
}
\label{mh2_extra_mstar}
\end{figure}
%-----------------------------------------------------

{Fig.~\ref{mh2_extra_mstar} shows the relation between \mhtwo\ and \lk,
for the entire redshift-limited sample and for subsamples of different morphological types,
together with the  the best-fit bisector regression lines. 
The relation   is close to linear. 
The correlation is very good ($r=0.73$) for spiral galaxies of type $T=3-5$.
The distribution of the emission from spiral galaxies of later types ($T=6-10$) is consistent with
this correlation. 
However,  early-type galaxies ($T\le0$) show a very poor correlation. 
Only the objects with CO detections follow the same
correlation  as spiral galaxies whereas most galaxies with CO nondetections have
upper limits for \mhtwo\  that lie considerably below it.}

%
% The coefficients of the linear regression
% for the three samples considered (entire redshift-limited CO sample, galaxies
%with $f_{\rm ap} < 1.5$ and galaxies of morphological type $T=3-5$) are identical within the errors.
%Therefore, the  ratio between both quantities is roughly
%constant for the entire luminosity range and a study of this ratio
%as a function of morphological type is meaningful.

\begin{figure}
\centerline{\includegraphics[width=7cm,angle=270]{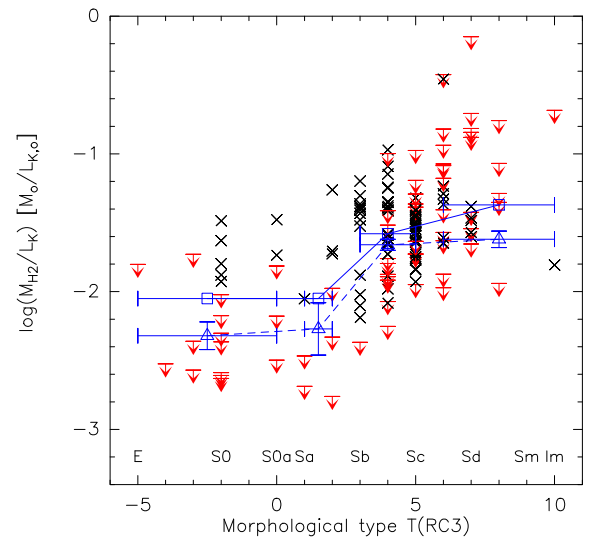}\quad
}
\caption{The ratio between \mhtwo\ and \lk\  as a function of morphological type. 
Triangles denote the
mean value and its error for a range of morphological types %(listed in Table~\ref{tab1_average})  
and squares  denote
the median values, calculated treating upper limits as detections,
 as listed in   Table~\ref{tab1_average}. The error bars in the x-direction denote the range of
morphological types over which the mean and median have been taken.
}
\label{mh2_extra_over_mstar}
\end{figure}

The ratio  \mhtwo/\lk\ is shown in 
Fig.~\ref{mh2_extra_over_mstar}.
It is lowest for early-type galaxies (up to $T=2$), and increases for later types
by a factor 3-5.
From $T=3$ on, the ratio is approximately constant and does not show the decrease it shows for
\mhtwo/\lb\ and \mhtwo/\dopttwo .
Early-type (E+S0s) galaxies have  a lower mean  molecular gas mass per \lk\
 than spiral galaxies. This is caused by galaxies not detected in CO, whereas, as seen before,
S0 and S0a detected in CO 
have   higher values of \mhtwo/\lk , in the same range 
as spiral galaxies.

%-----------------------------------------------------
%-----------------------------------------------------

%%%%%%%%%%%%%%%%%%%%%%%%%%%%%%%%
\subsubsection{FIR luminosity,  SF  rate and efficiency}

%---------------------------------------------------------------------------------
\begin{figure}
%\begin{center}
%\includegraphics[width=7cm,angle=270]{mh2_extra_lfir_t_3_5.ps}
\includegraphics[width=8cm,angle=0]{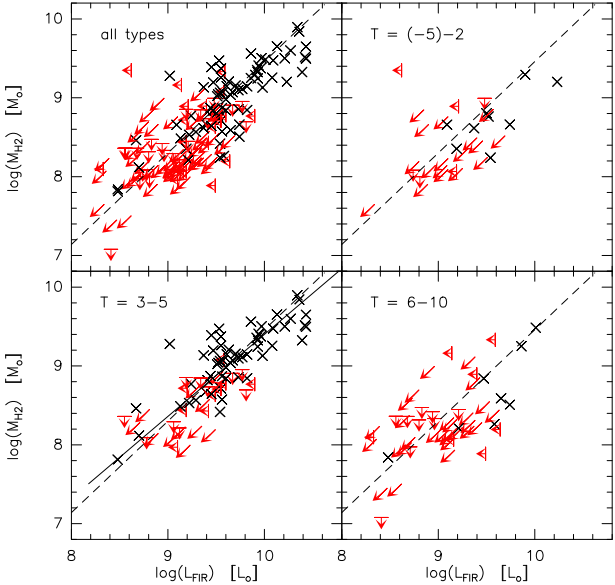}
\caption{\mhtwo\ vs. \lfir\ for  
%galaxies with morphological type $T=3-5$ of the  redshift-limted sample.  
 for the redshift-limited CO sample
($n=173$), including all morphological types, and for groups of 
different morphological types. 
The dashed line gives
the best  fit bisector (derived with ASURV) for all morphological types and  the solid line is the best fit
for the $T =3-5$ sample (see Table~\ref{regression}).
}
\label{mh2-lfir-all}
%\end{center}
\end{figure}
%---------------------------------------------------------------------------------

A good correlation is known to exist 
between  \mhtwo\ and 
\lfir\ \citep[e.g.][]{1991ARA&A..29..581Y} % (e.g. Young \& Scoville 1992)
because both quantities are directly related to
SF:  the molecular gas as the fuel for SF and  \lfir\ as a tracer for SF based on  the heating of the
dust by newly born stars.
{Fig.~\ref{mh2-lfir-all} shows the relation between \mhtwo\ and \lfir,
for the entire redshift-limited sample and for subsamples of different morphological types,
together with the  best-fit bisector regression lines. 
We find a good correlations with a roughly  linear slope (Table~\ref{regression}).
For early ($T\le0$) and late ($T=6-10$) type galaxies, the distribution of   \mhtwo\ vs. \lfir\ is
consistent with the correlation found for galaxies of type $T=3-5$.}
%As in preceeding subsection, we do not show the
%results for earlier and later morphological-types because of the high number of non-dections
%which make a linear regression unreliable. 
%  The coefficients of the linear regression
% for the three samples (entire redshift-limited sample,
%$f_{\rm ap} < 1.5$ and T=3-5) are identical within the errors.

%---------------------------------------------------------------------------------
\begin{figure}
\begin{center}
\includegraphics[width=7cm,angle=270]{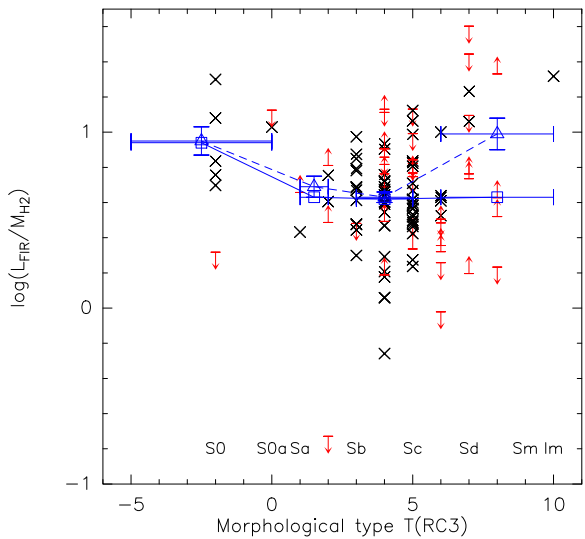}
\caption{The ratio between \lfir\  and \mhtwo\ 
as a function of morphological type. Galaxies with
upper limits at both wavelengths are excluded. 
Triangles  denote the
mean value and its error for a range of morphological types %(listed in Table~\ref{tab1_average})  
and square signs denote
the median values, 
%calculated treating upper limits as detections, 
as listed in   Table~\ref{tab1_average}. The error bars in the x-direction show the range of
morphological types over which the mean and median have been taken.
}
\label{mh2-over-lfir-morphological}
\end{center}
\end{figure}
%---------------------------------------------------------------------------------

Fig.~\ref{mh2-over-lfir-morphological} shows the ratio  \lfir/\mhtwo\   as a function of morphological type.
Early-type galaxies have a higher value,  possibly due to
a large fraction of their FIR emission 
not being heated by young stars  but by  the general interstellar radiation field.
For late-type spirals ($T\ge 6$) the mean ratio increases again, but  the low number of
values and detections make any firm conclusion difficult.  

{These values and trends for \lfir/\mhtwo\ are consistent with earlier results
of \citet{1996AJ....112.1903Y} who studied them in a sample of 120 galaxies included
in the FCRAO survey. They used a higher value of N(H$_2$)/I$_{CO} = 2. 8\times 10^{20}$ cm$^{-2}$ (K km s$^{-1}$)$^{-1}$, 
but also define the FIR luminosity in the range between $1-500\, \mu$m which is, according to their prescription,
between $\sim 0.1-0.2$ dex higher.
Since both  differences roughly compensate, their values of $L_{\rm IR}$/\mhtwo\ are comparable to our
values \lfir/\mhtwo .
They find  very similar values of   \lfir/\mhtwo\   for different morphological types as we do.
They obtain value  between $0.55\pm 0.08$ and $0.61\pm 0.06$  
 for morphological types $T=3-5$  ($n = 45$),
higher values (between $0.70\pm 0.13$ and $1.53\pm 0.24$)  for later type galaxies  ($T=6-10, n=19$), and,
for earlier spiral types ($T=1-2, n=14$),   mean values ($0.65\pm 0.20$ and $0.53\pm 0.09$) that are
similar to galaxies of type $T=3-5$.  Their sample only includes three galaxies of type S0-S0a and no elliptical galaxies, so that we cannot compare earlier
types.}

%They find  a mean value of \lfir/\mhtwo\ between
%$0.68\pm0.07$, $0.81\pm0.06$ and  $0.72\pm0.09$  (after adapting to our CO-to-H$_2$ conversion factor)
% for isolated, field and Virgo Sbc-Sc galaxies, respectively.
%The values for isolated and Virgo galaxies are  very close to the value that we obtain  for Sb-Sc galaxies (Tab.~\ref{tab1_average}).

% NOTES:
% Look at their TAb. 3 and Sect. 3.4 for definition of Lir:
% They say that it gives the fluxbetween 1 and 500 mu. Their formula is very similar to ours, but it included a factor C (to
%correct flux below 40 and above 120 mu) which depends on f60/f100. It is tabulated in
% Catalogued Galaxies in the IRAS Survey (1985), Tab. B.1 which I cannot find.
% I derived Lfir with my formula for the N157 and 253 (bright) and NGC 925 (less bright) and derived
% 0.16, 0.13 dex LESS for the first two and 0.20 less for the last.  I will assume a value in 
% but is the same as ours,  
%
% The x-factor which is 2.8e20 (see TAb. 3 or Sect. 3.2). Compared to ours this gives a 0.146dex lower MH2.
%
% They give the values for  Lir/MH2  in Tab. 5.
%
% The difference in their definition of LIR and MH2 roughly compensates.

{ \lfir\ is a  good tracer for the star formation rate (SFR) due to two reasons: 
(i) young stars are formed in dense regions where dust opacities are high and (ii) the
dust extinction curve peaks in the ultraviolet so that the radiation from young, massive stars
is preferentially absorbed. Therefore,   \lfir\ as a good SFR tracer especially for actively
SF galaxies.  It has to be used with some caution in galaxies with a low
SFR where dust heated  from old stars  can contribute  to \lfir , or in galaxies
with a low metallicity and thus a low dust opacity \citep[e.g.][]{2003ApJ...586..794B}.

Keeping these limitations in mind,} we use the formula of \citet{1998ARA&A..36..189K}  % Kennicutt (1998)
to calculate the SFR:

\begin{equation}
{\rm SFR (M_\odot yr^{-1}) = 4.5 \times 10^{-44} L_{\rm IR} \, ({\rm erg\, yr^{-1}}}), 
\end{equation}
where L$_{\rm IR}$ is the total FIR luminosity in the range 8 -1000 $\mu$m. {This
formula assumes a Salpeter Initial Mass Function. We convert this to a
value based on the \citet{2001MNRAS.322..231K} IMF by deviding by a factor 1.59 
\citep{2008AJ....136.2782L}.}
In our analysis we use  \lfir ,  calculated following the formula given
by \citet{1988ApJS...68..151H}, % Helou et al. (1988) 
which estimates the FIR emission in the wavelength range
 of 42--122.5 $\mu$m. We  estimate L$_{\rm IR}$ from \lfir\ 
using the result of \citet{2003ApJ...586..794B} % Bell (2003) 
 that on average L$_{\rm IR} \sim 2 \times$ \lfir\   
for a heterogenous sample of normal and starbursting galaxies.
Adopting this factor and the conversion to the Kroupa IMF we calculate
the SFR from the \lfir\ as:

\begin{eqnarray}
\label{sfr_lfir}
{\rm SFR  (M_\odot yr^{-1})} &=& 4.5  \times 2 \times  \frac{1}{1.59} 10^{-44}  \,  {\rm L_{FIR}  (erg\, yr^{-1})} \\  \nonumber
& = & 2.2 \times 10^{-10}  {\rm L_{FIR}} (L_\odot).
\end{eqnarray}

We derive an average SFR for galaxies of type $T=3-5$ of  $\sim 0.7$ \msun yr$^{-1}$.
The ratio \lfir/\mhtwo\ is proportional
to the star formation efficiency (SFE), defined here as the ratio between SFR and \mhtwo . 
The mean value of the SFE for galaxies of type $T=3-5$ is $10^{-9}$ yr$^{-1}$
(Table~\ref{tab1_average}).

%%%%%%%%%%%%%%%%%%%%%%%%%%%%%%%%
\subsubsection{Atomic gas mass}

%---------------------------------------------------------
\begin{figure}
\includegraphics[width=8cm,angle=0]{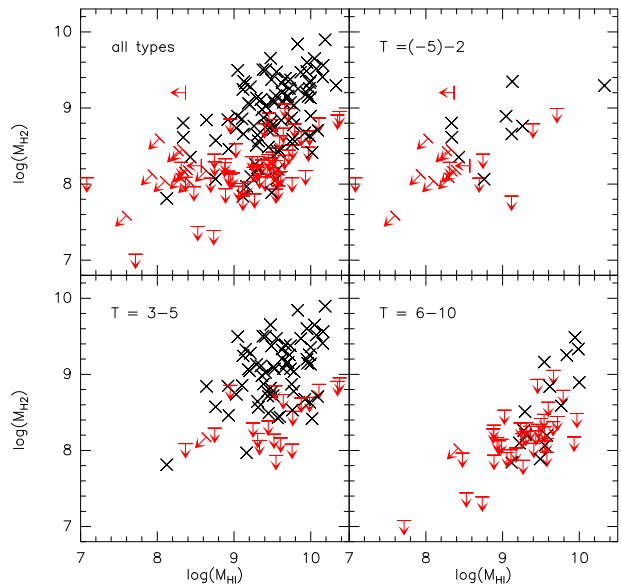}
\caption{The relation between \mhtwo\ and \mhi\ for groups of different morphological types.
}
\label{mh2_extra_mhi}
\end{figure}
%---------------------------------------------------------

%---------------------------------------------------------
\begin{figure}
\includegraphics[width=7cm,angle=270]{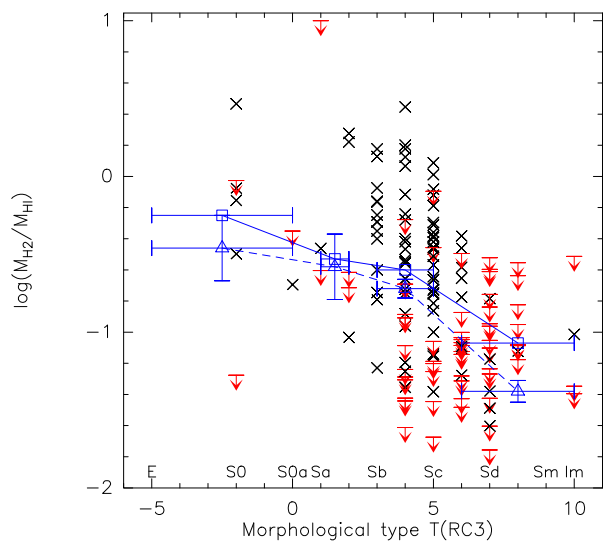}
\caption{The \mhtwo/\mhi\  ratio as a function of the morphological type. 
Triangles denote the
mean value and  its error for a range of morphological types %(listed in Table~\ref{tab1_average})  
and squares denote
the median values, 
%calculated treating upper limits as detection, 
as listed in   Table~\ref{tab1_average}. The error bars in the x-direction denote the range of
morphological types over which the mean and median have been taken.
}
\label{mh2_extra_over_mhi}
%On the right we
%show \mhtwo\ calculated only for the central pointing (for comparison with studies
%done in similar way) and on the right the values calculated for the
%\mhtwo .}
\end{figure}
%---------------------------------------------------------

%The ratio between \mhtwo\ and \mhi\ as a function of morphological type
%has been a controversy (see e.g. Casoli et al. 1998). 
Fig. \ref{mh2_extra_mhi} shows the relation between \mhtwo\ and \mhi\ for different
groups of morphological types. 
{The correlation with \mhi\ is much poorer than
the other correlations considered here, with a Spearman correlation coefficients of 
$r = 0.44, 0.44, 0.29, 0.57$ for the entire redshift-limited sample,   $T=(-5) - 2$,  $T=3-5$ and  $T=6-10$,
respectively.   In most galaxies \mhi\  is higher than  \mhtwo .}
Some differences can be seen for the different morphological types:
Early type galaxies ($T=-5 - 2)$ have a high number of upper limits in both
\mhtwo\ and \mhi .  The upper limits mainly populate the low \mhi\ and \mhtwo\ part of the
diagram, whereas the detections have values of \mhtwo\ and \mhi\ comparable
to those of spiral galaxies.
Late-type galaxies ($T= 6 - 10)$ are shifted towards higher \mhi\  ($\simgt 10^9$ \msun )
but lower ($ \simlt 10^9$ \msun ) \mhtwo\ with a large
number of upper limits in the latter.  Galaxies of morphological types  $T=3-5$
 have both  high atomic gas  and
molecular gas masses  ($\simgt 10^9$ \msun ) with a low number of upper limits in \mhtwo\ and
almost none in \mhi .

Fig.~\ref{mh2_extra_over_mhi}  displays the ratio \mhtwo/\mhi\
as a function of morphological type, showing a strong variation with $T$.
The highest values are found for early-type galaxies, up to 
$T = 0$ (albeit with a high uncertainty due to the high number of upper limits).
For later  types, \mhtwo/\mhi\ is decreasing strongly.
%In the right panes, we show \mhtwocenter/\mhi\ which is used for comparison with
%literature data. The tendency  is basically the same, but the individual values are
%slightly lower. For both cases 
The mean ratio \mhtwo/\mhi\ is significantly lower than 1 for
all morphological types.

%The ratio of \mhtwo/\mhi\ is correlated with \lfir/\lb\ (Fig. xxx), the highest
%values being found for the FIR-brightest galaxies.

%%%%%%%%%%%%%%%%%%%%%%%%%%%%%%%%
\subsection{Expected molecular gas content in a galaxy}

We showed in Sect. 4.2 that good correlations exist   between \mhtwo\  and 
other parameters  of a galaxy  (\lb , \lk\ and \lfir)
and a somewhat poorer correlation with \dopttwo .
In Table~\ref{regression} we list, apart from the linear regression parameters, also the 
Spearman's rho correlation coefficient {and in Table~\ref{tab1_average} the different
ratios depending on the morphological types.
All these relations  can be used to study differences in the molecular gas content
 of other sample, like e.g. interacting galaxies,
with respect to isolated galaxies.

The best correlation exists between
\mhtwo\  and \lfir ($r \ge 0.8$). \lfir\ is thus a very reliable way for predicting the expected molecular gas content in
a galaxy. However, \lfir\ might not be a good parameter  when searching for 
variations of \mhtwo\ in  interacting galaxies because \lfir\ itself, tracing the SFR, is
easily affected in such an environment.
A very good, and roughly linear, correlation also exists  between \mhtwo\  and \lk\ for galaxies of type $T=3-5$ ($r = 0.73$).
The luminosity in the K-band as a measure of the total stellar mass is less affected by recent events than
\lfir\ or \lb\ and thus a good normalization parameter when searching for changes in \mhtwo.
This correlation is, however,  poor for early type galaxies ($T\le 2$), where the undetected   objects have
ratios \mhtwo/\lk\ well below the values for spiral galaxies with $T\ge3$
(see  Fig.\ref{mh2_extra_mstar} and    Table~\ref{tab1_average}) so that for those types \lk\ is not recommened
as a measure of the expected  \mhtwo.
There is also   a good correlation between \mhtwo\ and \lb\  ($r\sim0.65$). 
%Here, also E+S0s galaxies seem to follow the same trend
%between \mhtwo\ and \lb\ as spiral galaxies (see Fig.~\ref{mh2-lb_types}).
The ratio \mhtwo/\lb\ is however
not constant, but increases with \lb\ which has to taken into account in any comparison.
Finally, the poorest correlation ($r\sim 0.5$) exists with the isophotal diameter which 
is not  a very reliable parameter for predicting \mhtwo.

Apart from using the ratio of \mhtwo\ to another parameter, }
we can use the correlations, defined by the linear regression parameters
listed in Table~\ref{regression},   for  predicting the expected \mhtwo\ and
determine  whether
a deficiency or an excess of  \mhtwo\ 
exists in an object  in comparison to isolated galaxies. 

We define the deficiency in \mhtwo\ in an analogous way
to the definition in \mhi\ of \citet[e.g.][]{1984AJ.....89..758H}:

\begin{equation}
{\rm Def}(M_{\rm H_2}) = \log(M_{\rm H_2, predicted}) - \log(M_{\rm H_2, observed}).
\end{equation}

Note that in this definition, a positive value of  Def(\mhtwo ) means
a deficiency and a negative value means an excess.
We can derive the predicted \mhtwo ,
$M_{\rm H_2, predicted}$, from any of the parameters, $X$, 
(\lb , \dopttwo , \lk\ or \lfir) as:

\begin{equation}
\log(M_{\rm H_2,predicted})  = intercept + slope \times \log(X),
\end{equation}
where {\it intercept} and {\it slope} are the parameters of the best fit listed in Table~\ref{regression}.
{These  definitions allow us to directly take  the effect of a 
nonlinearity of a correlation into account. %between \mhtwo\ and another parameter.

In Table~\ref{regression} we have given the values for two types of regressions:
The bisector fit and the fit obtained by taking \mhtwo\  as the dependent variable, O(Y$|$X), and
minimizing the distance of the \mhtwo\ measurements from the best-fit regression line.
The regression parameters are different for both methods because of the scatter in the
data and the difference is larger for poorer correlations.
%This  difference  allows to estimate the range uncertainty  in the method. 
The O(Y$|$X) fit is the appropiate regression for predicting \mhtwo\ from \lb , whereas the bisector
regression is the best estimate for the underlying correlation between two parameters
\citep{1990ApJ...364..104I}.
For a sample covering the same luminosity range,  O(Y$|$X)
is the best way for predicting the expected \mhtwo. For samples with a different 
luminosity range, however, the bisector fit is better since it provides a more
reliable extrapolation.
}

%This way of finding a possible deviation of \mhtwo\ is  more precise than
%using the ratio  \mhtwo/$X$  especially in the cases of $X=$\lb\ or  \dopttwo\ where
%the correlation with \mhtwo\ is strongly non-linear.
% In these cases
%a high ratio does not necessarily indicate an enhanced molecular gas mass,
% but it  can simply shows that the corresponding galaxy
%is more luminous or larger. 

%%%%%%%%%%%%%%%%%%%%%%%%%%%%%%%%%%%%%%%%
\subsection{The Kennicutt-Schmidt law}

\begin{figure}
\includegraphics[width=7cm,angle =270]{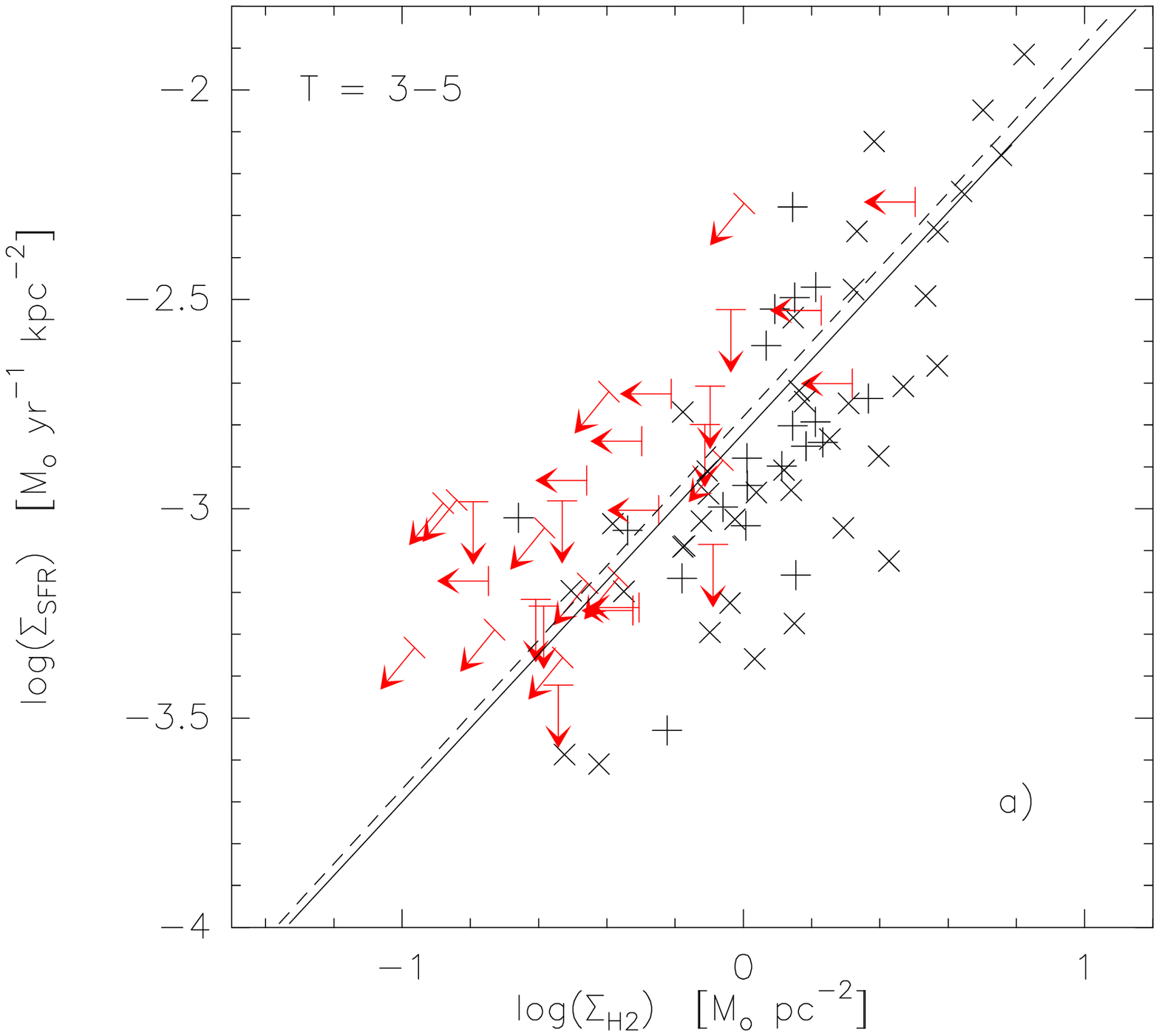}
\includegraphics[width=7cm,angle =270]{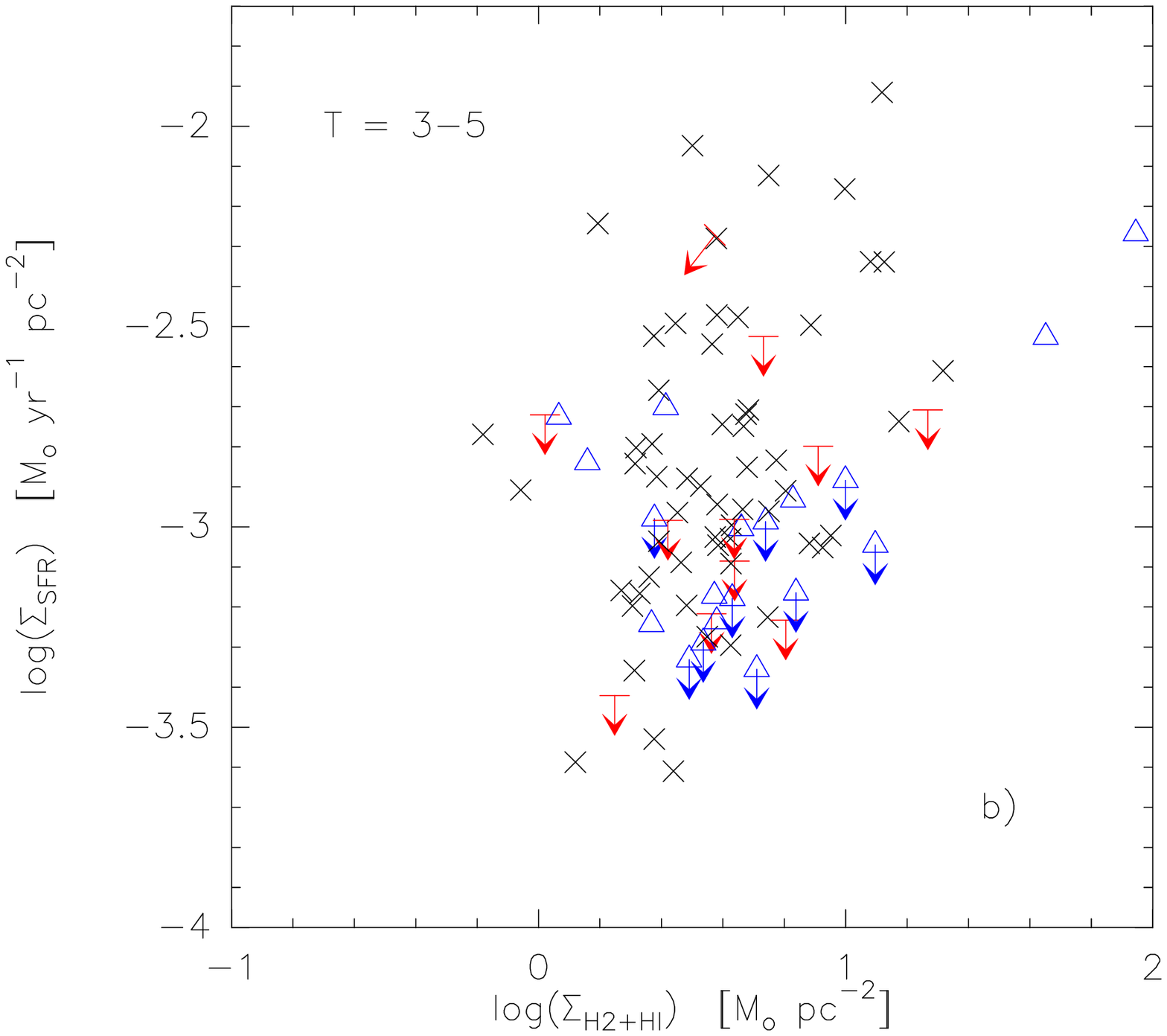}
\caption{{\it Panel a)}: The relation between the surface density of the molecular gas and the SFR per area
for morphological types $T=3-5$, calculated by dividing the molecular gas mass and the
SFR derived from \lfir\  by the surface of the galaxies,  $ \pi (D_{\rm 25}/2)^2$. The  full line shows the bisector best fit for
this sample, and the dashed line the best fit to the sample of all morphological types.
{\it Panel b:} The same relation with the total (atomic + molecular) gas surface density.  The blue triangles show the
galaxies with either an upper limit in HI or in CO. The gas surface density for these galaxies has a lower
and an upper limit which for the galaxies in this figure are very close together (within the size of the triangle).
}
\label{schmidt-kennicutt}
\end{figure}

We used our data to calculate the Kennicutt-Schmidt law, i.e. the relation between the 
disk-averaged gas column  density and the disk-averaged
SFR per area, $\Sigma_{\rm SFR}$.
Fig.~\ref{schmidt-kennicutt}a shows the relation with the molecular gas column density, $\Sigma_{\rm H2}$, and
Fig.~\ref{schmidt-kennicutt}b with the total (molecular+atomic) gas column density, $\Sigma_{\rm H2+HI}$.
The surface
densities were calculated by dividing the SFR, respectively the (extrapolated) molecular or total gas mass,
by the galaxy surface $\pi$\dopttwo\ /4.
A clear correlation exists with the molecular gas, but none with the total gas, showing a lack
of correlation with the atomic
gas column density. 
Our findings  are in agreement with previous results based on spatially
resolved analysis
\citep{2002ApJ...569..157W,2008AJ....136.2846B,2010A&A...510A..64V},
 showing that the SFR is strongly related to
molecular gas only.
The best-fit regression yields $\Sigma_{\rm SFR} \propto \Sigma_{\rm H_2}^{0.89\pm 0.07}$
(Table~\ref{regression}). This slope, close to unity, is again consistent with the results
of \citet{2002ApJ...569..157W} and \citet{2008AJ....136.2846B} for a spatially resolved  analysis.

%%%%%%%%%%%%%%%%%%%%%%%%%%%%%%%%
\section{Discussion}

We compare our data to that of other samples of ``normal"  galaxies
and also search for differences in samples of   interacting galaxies. 
The samples of normal galaxies that we consider are made of not obviously
interacting galaxies, which are however selected without  a clear isolation criteria,
and  some of these samples contain, e.g.  several cluster galaxies.
In all these comparisons we have adjusted the molecular gas masses
to our definition, i.e. same conversion factor and no consideration of the helium
fraction. 
%Furthermore, we adapt the values of the SFRs and \lk\ to the
%Salpeter IMF that we use in this paper. For this, we multiply the values derived
%by a Kroupa IMF by a factor 1.6 \citep{2008AJ....136.2782L},  and the Chabrier IMF by 1.7 \citep{2011arXiv1104.0019S} .

\subsection{Comparison to studies of normal galaxies}

\subsubsection{Relation between molecular gas, FIR and blue luminosity } 

The  nonlinear relation between \mhtwo\ and \lb\ has been
found by other groups as well \citep[see][and references therein]{1997ApJ...490..166P},
with similar slopes as found by us. Their study was based on a smaller number ($n=68$) of galaxies
selected with a less rigorous  criterion with respect to the environment.
%In Fig.\ref{compare_mh2_casasola_perea}  we compare the fit obtained by
%us to that of  \citet{1997ApJ...490..166P}. Our value for the predicted \mhtwo\ at a given \lb\ is 
%slightly lower than the value of the fit by \citet{1997ApJ...490..166P}, even though we
%have extrapolated \mhtwo\ to the total galaxy. This  indicates 
%that our sample indeed shows a lower
They discussed the cause of the nonlinearity and conclude that the most likely
reason is extinction affecting \lb\ and increasing with galaxy luminosity. They
predicted that the relation between \mhtwo\ and luminosities at longer wavelengths
should be more linear. This prediction is confirmed by the nearly linear relation 
found between \mhtwo\ and \lk\ 
in our analysis.

%% We find for MH2 independent variable: slope 1|1.93 = 0.52 with Lb,
%% and 1|1.35 = 0.74+-0.06 for Lfir
%%

{We compared our mean value of \mhtwo/\lb\ for galaxies of type $T=3-5$ to results
for a large sample of normal galaxies, studied by \citet{2003A&A...405....5B}.
They searched the literature for galaxies with data for their ISM properties, excluding
galaxies with a known pecularity (interacting, disturbed, galaxies with polar rings
or counterrotation) and with active galactic nuclei. Their sample includes
177 galaxies of type $T=3-5$ with CO(1-0) data (160 detections and 17 upper limits) with
values of  log(\lb) between 9 and 11.
They derive a mean value (adapted to our convention for the calculation of the \mhtwo\
and \lb ) of \mhtwo/\lb = $-0.92\pm0.04$ for these galaxies. When restricting the range of \lb\ to log(\lb) = 10 - 10.6,
the mean value is $-0.82\pm0.05$, about 0.2 dex higher than the corresponding value for the 
AMIGA sample (see Tab.~\ref{compare_other_samples}).
 showing that AMIGA galaxies have a lower
molecular gas content. When comparing their values for \mhtwo/\dopttwo\ a
similar difference is found.}

The relation between \mhtwo\ and \lfir\ in nonstarburst galaxies has been found to be
close to linear in other studies  \citep[e.g.][]{1997ApJ...490..166P,2004ApJ...606..271G},
 in agreement with our results.

\subsubsection{Gas depletion time}

We derived in Sect. 4.2.4 the SFE (defined as the SFR/\mhtwo ). 
The SFE is directly related to the molecular gas depletion time, \taudep , by
\taudep = SFE$^{-1}$. The gas depletion time for our sample
has a mean value of log(\taudep) = 8.9 yr for the entire redshift-limited sample and
log(\taudep) = 9 yr for galaxies of type $T=3-5$ (see Table~\ref{tab1_average}),
with a spread of values roughly ranging between  log(\taudep) = 8.5 yr  and
 log(\taudep) = 9.5 yr (see Fig.~\ref{mh2-over-lfir-morphological}).

This value can be compared to those found in recent surveys.
\citet{2011ApJ...730L..13B}  derived a mean gas depletion time of 2.35 Gyr from spatially resolved observations of 30 nearby galaxies in the 
HERACLES survey. This values  includes a helium fraction of a factor 1.36, 
thus giving \taudep = 1.7 Gyr without helium.   They furthermore showed that this value is consistent with 
a wide range of molecular gas depletions times from the literature, albeit with
a large standard deviation of 0.23 dex. This value is only slightly higher than our value for isolated galaxies.
The small difference could be due to the fact that our value is global, and thus might encompass
some FIR emission  not directly associated to SF from the outskirts of the galaxies, whereas the
value from \citet{2011ApJ...730L..13B} is from  a spatially resolved study.

\citet{2011arXiv1104.0019S}   studied the molecular gas depletion time, \taudep ,
 for a volume limited
sample of 222 galaxies with $0.025 < z < 0.05$ observed in the COLD GASS survey. They 
found values for the gas depletion time
in the range of  roughly log(\taudep) = 8.6 yr and 9.5 yr, with a mean value
of  \taudep = 1 Gyr 
{(for a \citet{2003PASP..115..763C}  
IMF which gives a mass comparable to within $\sim$ 10\%  to the
Kroupa IMF). }
Both the range and the mean agree very well with our values.

They furthermore found a good correlation of \taudep\ with \mstar\ and with
the specific SFR, sSFR = SFR/\mstar . We tested both correlations with our data.
We did not find a similarly good correlation with \lk (which is in a good approximation proportional
to the stellar mass \mstar ) although
our sample covered a similar range of stellar masses, up to several $10^{11}$ \msun .
Taking into account all morphological types, we found evidence for a weak trend
(correlation coefficient $r=0.25$) which completely disappeared when restricting
the sample to $T=3-5$ ($r=0.08$).

We could confirm with our data the existence of  a correlation between \taudep\ and sSFR.
This correlation is however expected because both \taudep\ and the sSFR depend on the
SFR, and  \lk\ and \mhtwo\ show a good 
correlation. Thus, the respective ratios are expected to correlate.

\subsubsection{The molecular-to-atomic gas ratio}

\begin{figure}
\begin{center}
\includegraphics[width=7cm,angle=270]{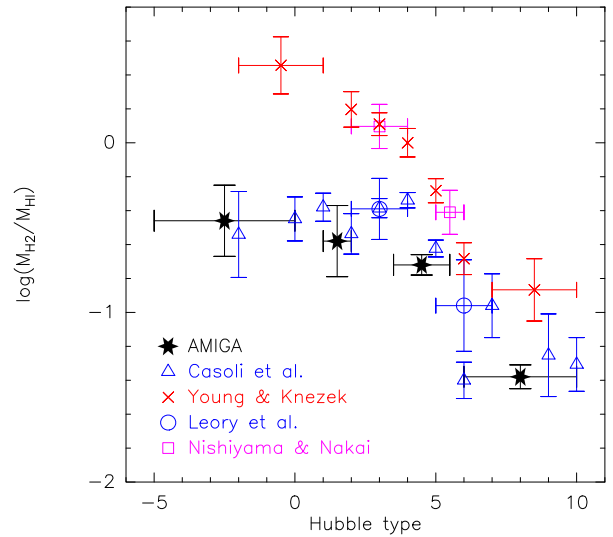}
\caption{Comparison of our values for \mhtwo/\mhi\  (black stars) to those
from the literature: \citet{1998A&A...331..451C} (blue triangles),
\citet{1989ApJ...347L..55Y} (red crosses), \citet{2008AJ....136.2782L} (blue circles) and
\citet{2001PASJ...53..713N} (magenta squares). 
The molecular gas mass have all been adapted to common conversion factor 
of $2\times 10^{20}$cm$^{-2}$ and no consideration of the helium mass.
}
\label{mh2_mhi_compare_young_casoli}
\end{center}
\end{figure}

The value of  \mhtwo/\mhi\ in galaxies as a function of 
morphological type has been  controversial. 
In a  large survey of spiral galaxies, \citet{1995ApJS...98..219Y} % Young et al. (1995)
mapped a sample of about 300 galaxies with the FCRAO telescope. 
\citet{1989ApJ...347L..55Y} %Young \& Knezek 
studied the
dependence of \mhtwo\ on \mhi\ for that sample.
%, the blue luminositiy and the optical diameter. 
They derived a continous decrease of
 \mhtwo/\mhi\  from early to late-type galaxies with
mean  values  above 1 for early types.
\citet{1998A&A...331..451C} % Casoli et al. (1998) 
studied the molecular gas properties of a large ($n=528$), heterogenous
 sample of galaxies,
composed of data from the literature and their own observations.
Their data  consisted both of mapped galaxies and of objects
where only the central position had been observed.
In agreement with \citet{1995ApJS...98..219Y}, they found a decrease in  \mhtwo/\mhi\  from early to
late-type galaxies, but obtained  much lower values for  \mhtwo/\mhi ,
especially for early-type spirals.
In Fig.~\ref{mh2_mhi_compare_young_casoli}  we compare our results to these two studies 
and furthermore  to  the THINGS sample \citep{2008AJ....136.2782L} and 
the Nobeyama survey \citep{2001PASJ...53..713N}. 
\citet{2003A&A...405....5B} (not included in the Fig.~\ref{mh2_mhi_compare_young_casoli} ) 
provide values for a sample of 427 normal galaxies 
 in agreement with \citet{1998A&A...331..451C} for $T>0$
and higher values (by 0.5-1 dex) for $T\le 0$.

The data from all studies show  a  decrease of the  \mhtwo/\mhi\ 
gas mass towards late morphological  types. A pronounced step in the ratio
takes place at  $T \sim 6$.  The low values found for late-type galaxies
could be due to two effects: (i) late-type galaxies are 
richer in HI  \citep[][]{1984AJ.....89..758H} % (e.g. Haynes \& Giovanelli 1984)  
or  (ii) they {have a lower molecular gas content.
A comparison of \mhi\ and \mhtwo\  to the blue luminosity or the optical diameter
shows that both effects take place: The ratio \mhi/\lb\ (respectively  \mhi/\dopttwo) increases
for types $T\ge 6$ by $\sim 0.2-0.3$ dex whereas \mhtwo/\lb (respectively  \mhtwo/\dopttwo) decreases
by $\sim 0.3-0.5$ dex \citep{1989ApJ...347L..55Y,1998A&A...331..451C,2003A&A...405....5B}.
Our data show the same behaviour.
The strong decrease of \mhtwo/\lb\ and  \mhtwo/\dopttwo\ could be due to a real decrease in
the molecular gas mass or due to the fact that late-type galaxies tend
to have lower metallicities so that we probably underestimate the true \mhtwo\
 by using the Galactic conversion factor.
}

% Comments: For late types
% Bettoni: MH2/LB lower by about 0.5, MHI/Lb higher by about 0.3 (dex) similar for norm. with d25)
% Young& Knezek: MH2/LB lower by about 0.3-0.5, MHI/Lb higher by about 0.2 0.3 (dex) (similar for norm. with d25)
% Casoli: MH2/d25**2 lower by about 0.5, MHI/d25**2 higher by about  0.2(dex) 

Although the general trends 
in the data sets are the same, with a pronounced and continous decrease of 
\mhtwo/\mhi\ from early to late-type  galaxies, there are considerable differences between the
different samples (see Fig.~\ref{mh2_mhi_compare_young_casoli}).
The differences are up to an order of magnitude for elliptical, lenticular and early-type spiral galaxies (up to
$T \sim 3$), and less (up to 0.4 dex) for later type spirals.
There is in general a good agreement between the values found by Casoli et al.
and ours, the match being very good
for early-type galaxies (up to T= 0) whereas for later-type galaxies our
values lie somewhat below.
The results from the THINGS survey \citep{2008AJ....136.2782L} % (Leory et al. 2008) 
are also in agreement with the data of \citeauthor{1998A&A...331..451C}, and
only slightly higher than our values.
The results by \citet{1989ApJ...347L..55Y} % Young \& Knezek 
and \citet{2001PASJ...53..713N} % Nishiyama \& Nakai (2001) 
give much higher values than the other surveys.
This could be due to the sample selection. In the sample of 
\citet{1989ApJ...347L..55Y} % Young \& Knezek 
FIR or optically bright galaxies were selected, so that
a high CO emission can be expected. Furthermore, the \citet{1989ApJ...347L..55Y} % Young \& Knezek
survey contains objects from the Virgo Cluster, and cluster galaxies are
known to be HI deficient on average. Cluster galaxies were excluded in
\citeauthor{1998A&A...331..451C}, % Casoli et al., 
and are not  present in our sample either.
A selection effect can also account for the difference between the
samples of \citeauthor{2001PASJ...53..713N}, \citeauthor{2008AJ....136.2782L} % Nishiyama \& Nakai, Leroy 
and our data. The samples differ in the
mean \lfir/\lb\ with the \citeauthor{2001PASJ...53..713N} % Nishiyama \& Nakai 
sample being the FIR brightest with a mean
 log(\lfir/\lb ) = $0.10\pm0.07$ whereas the  \citeauthor{2008AJ....136.2782L} % Leroy 
and our sample show lower values
(log(\lfir/\lb )$= -0.51\pm0.06$, respectively log(\lfir/\lb )$= -0.49\pm0.03$). Since 
galaxies with a higher log(\lfir/\lb) tend to have a higher
\mhtwo/\mhi , the observed trend is understandable.

A noticeable point is that the mean \mhtwo/\mhi\  of the AMIGA
sample represents for all morphological types the lower limit of all samples.
For late-types, the AMIGA galaxies have the lowest \mhtwo/\mhi ,
even though the \mhtwo\ is extrapolated unlike other
studies that are usually based on a  central pointing only.
%Given the heterogeneity  of the comparison samples and methods
%for the derivation of the molecular gas mass
%(central pointing, full mapping or  major axis mapping) , it is difficult
%to assess the significance of this lowest value. 
{This is in line with the low molecular gas content that we found in comparison to the
sample of \citet{2003A&A...405....5B}. }
%
%Without any further analysis, it is impossible to conclude whether this
%low value is due to a lower molecular gas content, or a higher atomic gas
%content compared to other, less isolated galaxies.
%The second option seems however plausible, as interactions tend
%to affect the (extended) atomic gas more effectively than the molecular
%gas which is much more concentrated towards the central region of galaxies. 
%
%In principle, we
%expect isolated galaxies to have a low \mhtwo/\mhi\
%due to the low present SF activity \citep{2007A&A...462..507L}.
%%the low degree of evolution that they have experiences in their life. 
%We can state that the present result is consistent with this picture.

%*****************************************************************
\subsection{Comparison to studies of interacting galaxies}

\subsubsection{Is the molecular gas content enhanced in interacting galaxies?}

Several studies in the past have concluded that  \mhtwo\ in 
interacting galaxies  is enhanced \citep{ 1993A&A...269....7B,2004A&A...422..941C,1994A&A...281..725C}, based on 
a higher  value of \mhtwo/\lb\ or \mhtwo/\dopttwo\   compared to noninteracting galaxies.
However, we found  that  {the ratio \mhtwo/\lb\ increases with \lb\ and this trend has
to be taken into account when comparing isolated and interacting samples.}
%Since interacting galaxies are
%generally more luminous, this trend} alone might be able to explain the enhanced value
%of  \mhtwo/\lb .
Indeed, \citet{1997ApJ...490..166P}  found  no difference 
 in the {\it correlation} between \mhtwo\ and  \lb\  for samples of
isolated, strongly and weakly perturbed galaxies. 
%We confirm their finding
%when comparing their sample to ........

We use 
the sample of  \citet{2004A&A...422..941C} of interacting galaxies to search for a possible 
excess in \mhtwo\ with respect to AMIGA galaxies. 
Their sample includes 153 galaxies
with molecular gas data from different sources. \citeauthor{2004A&A...422..941C} found  that  the
mean ratio of \mhtwo/\lb\ for spiral galaxies was, depending on the morphological
type,  between about 0.2 and 1.0 dex  higher than for a sample of 427 noninteracting galaxies 
from \citet{2003A&A...405....5B} and concluded that
\mhtwo\ was enhanced in interacting galaxies.
However, since their sample of interacting galaxies is on average 0.5 mag brighter, a higher
\mhtwo/\lb\ is already expected  due to the higher luminosity. 
%Using the slope of the regression
%between \mhtwo\ and \lb\ from Table\ref{regression}) we expect a
%\mhtwo/\lb\ to be higher by a factor of $0.5 \times 0.4 \times 1.4 =  0.28$ for the brighter, interacting sample.
%
We furthermore include the samples of strongly and weakly perturbed galaxies from \citet{1997ApJ...490..166P} 
in this test.
The weakly perturbed sample has 43 galaxies and  includes classes 1, 2, and 3 of \citet{1988ApJ...334..613S}
and class 2 objects from the luminous IRAS sample of \citet{1991ApJ...370..158S}. The 
strongly perturbed sample has 35 galaxies and includes interaction class 4 of 
 \citet{1988ApJ...334..613S}, interaction classes 3 and 4 of \citet{1991ApJ...370..158S}
 and closely interacting pairs from \citet{1994A&A...281..725C}.

{We  searched for a possible enhancement of \mhtwo\ in comparison to  \lb\ and to \lk\ in these samples, 
 both by applying the deficiency parameter and by comparing
the ratios. 
The mean values are listed in Table~\ref{compare_other_samples}.
We note  that the interacting samples are more
luminous (by about 0.5 dex) in both \lb\ and \lk\ than the AMIGA sample so that we have  to extrapolate
the relations found for the AMIGA galaxies  to higher luminosities.
For the calculation of  the deficiency parameter we therefore use the bisector fit as the
best fitting method for extrapolations. We use the fits for the AMIGA $T=3-5$ subsample to compare to the
weakly and strongly perturbed samples
WPER and SPER, since  spiral galaxies are the better comparison  for these actively star-forming objects.
For the  \citeauthor{2004A&A...422..941C} sample information about the morphological types are
available and we are able to do the analysis both
for the entire sample  and for $T=3-5$, using the values for the corresponding morphological types in AMIGA.
When comparing the ratio \mhtwo/\lb\ we restrict the galaxies that we consider to the same range
of \lb\ in order to avoid effects caused by  the nonlinearity in the correlation. For the roughly linear  \mhtwo/\lk\ ratio this 
is not necessary (we checked that no difference in  \mhtwo/\lk\ was present for low and high \lk\ in the AMIGA sample).
}

\begin{figure}
\includegraphics[width=7.cm,angle=270]{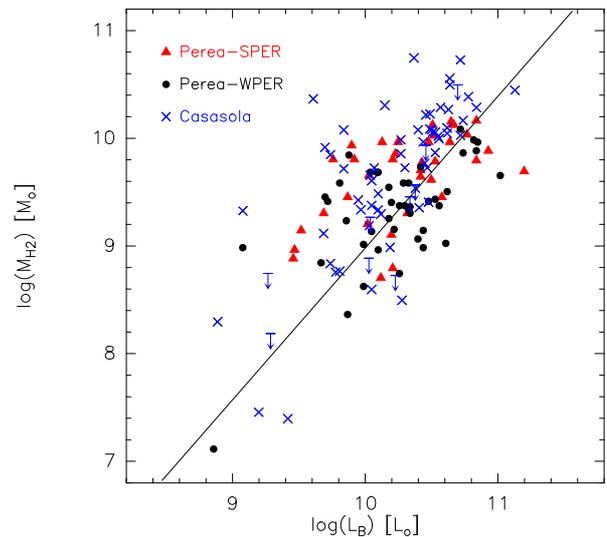}
\caption{The molecular gas mass vs. \lb\ for different sample of interacting galaxies:
Galaxies of type $T=3-5$ from the sample of \citet{2004A&A...422..941C} 
 and a sample of weakly  (WPER) and strongly (SPER) perturbed galaxies from
 \citet{1997ApJ...490..166P}.
 We adapted both the molecular gas masses and \lb\ to our definition.
The full line is the regression fit obtained for the AMIGA $T=3-5$
sample  from Table~\ref{regression}.
}
\label{compare_mh2_lb_casasola_perea}
\end{figure}

\begin{figure}
\includegraphics[width=7.cm,angle=270]{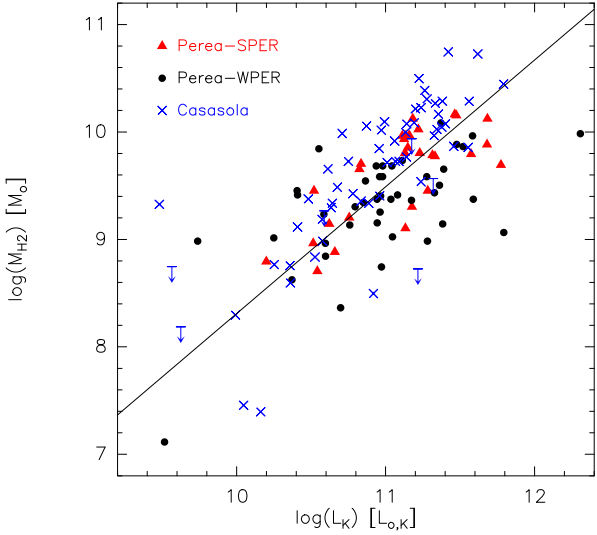}
\caption{The molecular gas mass (adapted to our conversion factor) vs. \lk\ for the same samples as in 
Fig.~\ref{compare_mh2_lb_casasola_perea} .
The full line is the regression fit obtained for the AMIGA $T=3-5$
sample  from Table~\ref{regression}.
}
\label{compare_mh2_lk_casasola_perea}
\end{figure}

Fig.\ref{compare_mh2_lb_casasola_perea}  shows \mhtwo\ vs. \lb\ for the  three samples
 compared to the bisector fit of the AMIGA $T=3-5$ sample.
%We adapted the molecular gas mass of Casasola et al. (2004) to our conversion factor and the no helium fraction
%{change later}. 
{An excess of  \mhtwo\ is visible for both the Casasola $T=3-5$ and SPER sample.
This is confirmed by the mean value of the deficiency (Table~\ref{compare_other_samples}),
%In order to quantify this further, 
%we  calculated the  molecular gas deficiency  for each galaxy  from its
%\lb , following Eqs. (8) and (9) and the values for the linear (bisector) regression fit for the AMIGA sample
%from Table~\ref{regression}.  
%
%We did this first the full sample using the regression line of AMIGA for the full sample ($n = 153, n_{\rm up} = 35$),.
%Since the morphological distribution of the sample is similar as ours (see their 
%Fig. 1) with a small fraction of early type galaxies ($T\le 0$) of about 15\% , this
%is a reasonable comparison. We derive a mean value for the molecular gas deficiency of 
%def(\mhtwo) = $0.30 \pm 0.07$, showing no excess but rather a small deficiency of the molecular gas.
%We also calculated the mean deficiency only for spiral galaxies of morphological type $T=3-5$ ($n = 69, n_{\rm up} = 9$),
%using the corresponding regression fit for the AMIGA sample, and obtained
%def(\mhtwo) = $-0.08 \pm 0.07$.
both for the   \citeauthor{2004A&A...422..941C} $T=3-5$ and the SPER sample a clear excess (i.e. a negative 
value of the deficiency) of \mhtwo\ is found.
This excess in the SPER sample was not found in  \citet{1997ApJ...490..166P} 
due to the different regression line they used.
For the  WPER  sample, on the other hand,
no deficiency nor excess is  found.
% however, it is not very significant, given the small number of galaxies
%and the large standard deviation of 0.5 of the distribution
%of def(\mhtwo ).
When comparing the ratio \mhtwo/\lb\ we confirm the higher values for both the Casasola and the SPER sample, with 
mean values of about 0.3 dex higher than for  the AMIGA $T=3-5$ sample.  
%In this comparison we have calculated  \mhtwo/\lb\ for the same range of \lb\  both for the AMIGA
%and  interacting samples in order not to be affected by the dependence on luminosity.
The value of  \mhtwo/\lb\ for the WPER sample is compatible with AMIGA within the errors.
The entire (i.e. all morphological types)   \citeauthor{2004A&A...422..941C} sample 
has a  value  of  \mhtwo/\lb\ compatible with AMIGA, showing the importance of taking  the morphological type into account.

When doing a similar analysis based on \lk\ (see Fig.~\ref{compare_mh2_lk_casasola_perea}), 
we find again no indications for an enhancement in \mhtwo\
for the  WPER sample nor for the total Casasola sample, neither from the deficiency nor from the ratio. 
There is  a clear excess in \mhtwo\ for galaxies of type $T=3-5$ from the Casasola sample, quantified both
in the ratio,  which is about 0.3 dex higher for  $T=3-5$ AMIGA galaxies,
and in the deficiency parameter. 
Also the SPER sample shows a higher  \mhtwo/\lk\ ratio than the AMIGA sample, however, no indication
of an \mhtwo\ excess from the deficiency parameter. 
%The deficiency
%parameter with respect to the best-fit regression for all morphological types gives
%a deficiency of -0.25, indicating an excess of \mhtwo\ and showing that for this small sample size and a the practically linear correlation
%with \lk\, the ratio \mhtwo/\lk\ is probably the most robust estimator.
The discrepancy between the two indicators (mean and deficiency) is probably due to the small
sample size and the small amount of the molecular gas excess. We suggest to rely in this case
on \mhtwo/\lk\  as the more robust indicator.

In summary, by comparing the molecular gas mass to \lk\ and \lb\ and using both the 
deficiency and the ratios \mhtwo/\lk\ and \mhtwo/\lb\   we found the clearest evidence for  an
enhancement  of \mhtwo\ in the Casasola sample of interacting galaxies of type $T=3-5$, which is
the largest comparison sample in our study. Evidence for an enhancement was also found 
 for the Perea sample of strongly perturbed interacting galaxies (SPER).
No evidence for any enhancement was found for the Perea sample of weakly perturbed interacting sample (WPER).
%For the sample of  strongly perturbed interacting
%galaxies (SPER sample) we found evidence of  enhancement based on \lb\, and weaker
%evidence base on  \lk . 
Based on the present data, it is not entirely clear where the differences between these
samples come from. 
A possible reason could be a lower degree of 
interaction in the WPER sample.
We would  like to point out the importance of matching the luminosity range
when comparing parameters with a nonlinear correlation, as \mhtwo/\lb , and the
importance of comparing the same morphological types, since there are generally 
large difference between early type and spiral galaxies, in particular for \mhtwo/\lk .}

What influence has the fact that the molecular gas masses in these samples are  not aperture corrected as in the
AMIGA sample? 
We do not expect this effect to be very important in interacting galaxies where
the SF and thus the molecular gas usually tend to be more concentrated to the central regions.
However, we can make an estimate of the importance of this effect
for the \citeauthor{2004A&A...422..941C}  sample. 
% It is not possible to reproduce easily with which telescope the CO(1-0) were taken
%and whether the galaxies were mapped or not. Thus, it is difficult to carry out an aperture correction
%aposteriori,  but we can give an estimate for the importance of this effect.
 The {median} angular diameter of the galaxies {of type $T=3-5$} in
this  sample is 90\arcsec . %and no relation with \lb\  can be seen.
If we assume that the galaxies were observed
only at the central position with a  beam size of 50\arcsec\ (the typical beam size
of the radio telescopes used in surveys) the predicted
aperture correction according to our prescription is between a factor of 1.2 and 1.4 (for
edge-on and face-on galaxies, respectively). Thus, in this case the total molecular gas content
would be 0.08-0.15 dex higher. 
%Still, no  molecular gas excess with respect
%to the AMIGA sample would be present for this sample.
{Thus, if the molecular gas in the interacting galaxies of the Casasola sample are distributed in a similar
way as in isolated galaxies, this difference would strengthen the finding  that  \mhtwo\ is enhanced in this sample
of interacting galaxies. It also shows the importance of mapping the molecular
gas  in galaxies in order to be able to compare different samples in a relyable way.}

%************************************************ORIGINAL
\begin{table*}
      \caption{Deficiencies of \mhtwo , \lfir\ and \lfir/\mhtwo\ in interacting sample and AMIGA}
\begin{tabular}{lcccccc}
\hline
\hline
%(1) & (2) & (3) & (4) & (5)  \\
   & \multicolumn{2}{c}{AMIGA} & \multicolumn{2}{c}{Casasola et al. (2004)}  & \multicolumn{2}{c}{Perea et al. (1997)} \\
  &  all types  & $T=3-5$ & all types  & $T=3-5$ &  WPER & SPER \\
\hline
$<$ def(\mhtwo\ )$>$ (\lb , bisec.)  & 0.07$\pm$0.04 & 0.07$\pm$0.04 & 0.13$\pm$0.07 & -0.31$\pm$0.07 & -0.03$\pm$0.07 &  -0.25$\pm$0.08 \\
%$<$ def(\mhtwo\ )$>$ ( \lb , \mhtwo dep.)  & 0.02$\pm$0.04 &  -0.01$\pm$0.04 & -0.196$\pm$0.07 & -0.181$\pm$0.07 & -0.30$\pm$0.07 & -0.50$\pm$0.07  \\ 
$n/n_{\rm up}$      &  173/79 & 88/21 & 153/35 & 68/9 & 43/0 & 35/0 \\
$<$log(\mhtwo/\lb)$>$  (for $10^{10}< $\lb $\ge 10^{10.6}$)    &    -1.10$\pm$0.06 &-1.04$\pm$0.05 & -1.12$\pm$0.08 & -0.71$\pm$0.08 & -0.93$\pm$0.06 & -0.71$\pm$0.08 \\
$n/n_{\rm up}$ &  50/11 & 40/5 & 85/13 & 39/6 & 24/0 & 19/0  \\
\hline
%$<$ def(\mhtwo\ )$>$ (\lk , bisec.) & -0.14$\pm$0.19 &  0.05$\pm$0.04 & 0.03$\pm$0.08 & -0.16$\pm$0.06 & -0.07$\pm$0.07 & -0.25$\pm$0.06 \\  Values for 
%       Wper and Sper based on all T AMIGA fit
$<$ def(\mhtwo\ )$>$ (\lk , bisec.) & -0.14$\pm$0.19 &  0.05$\pm$0.04 & 0.03$\pm$0.08 & -0.16$\pm$0.06 & 0.13$\pm$0.07 & -0.02$\pm$0.05 \\   
$<$log(\mhtwo/\lk)$>$   & -1.87$\pm$0.05 & -1.66$\pm$0.06& -1.70$\pm$0.08 &-1.37$\pm$0.07 & -1.65$\pm$0.07 &-1.47$\pm$0.05 \\
                  $n/n_{\rm up}$ & 173/79 & 88/21 & 132/29 & 60/6  & 43/0 & 32/0 \\
                  \hline
$<$def(\lfir)$>$ &   -0.02$\pm$0.04 & -0.05$\pm$0.04 & -0.39$\pm$0.02&-0.44$\pm$0.03&  -0.48$\pm$0.08&  -1.07$\pm$0.09 \\
     $n/n_{\rm up}$   & 172/75 & 88/20 &628/0  &340/0 &43/0 &35/0  \\
$<$log(\lfir/\mhtwo) $>$     & 0.72$\pm$0.03  &   0.63$\pm$0.03 &  0.89$\pm$0.05 &   0.75$\pm$0.05   & 0.98$\pm$0.07 & 1.33$\pm$0.05\\
     $n/n_{\rm up}$   & 97/22 & 68/10 & 105/15  &51/4 &43/0 &35/0  \\
\hline
\hline
\label{compare_other_samples}
\end{tabular}
\end{table*}
%*****************************************************************

%************************************************ORIGINAL
%\begin{table*}
%      \caption{Deficiencies of \mhtwo\ , \lfir\ and \lfir/\mhtwo\ in interacting sample and AMIGA}
%\begin{tabular}{lcccccc}
%\hline
%\hline
%%(1) & (2) & (3) & (4) & (5)  \\
%Sample & $n/n_{\rm up}$ &  $<$ def(\mhtwo\ )$>$  & $n/n_{\rm up}$ & $<$def(\lfir)$>$ & $n/n_{\rm up}$  &log(\lfir/\mhtwo) \\
%\hline
%Casasola et al. (2004), all           & 153/35 & 0.28$\pm$0.07  &  628/0  & -0.05$\pm$0.02& 105/15  &  0.89$\pm$0.05\\
%Casasola et al. (2004), $T=3-5$ & 68/9& -0.04$\pm$0.07  &  340/0 &  -0.05$\pm$0.03 & 51/4 & 0.69$\pm$0.05 \\
%Perea et al. (1997), WPER & 43/0 &   -0.11$\pm$0.08& 43/0  &  -0.44$\pm$0.08& 43/0 & 0.98$\pm$0.07 \\
%Perea et al. (1997), SPER &38/3 &  -0.27$\pm$0.09 & 35/0   &  -0.99$\pm$0.10 &35/0 &  1.33$\pm$0.05\\
%\hline
%AMIGA , all   & 173/79 & 0.07$\pm$0.04  &  172/75  & -0.02$\pm$0.04 & 97/22  &  0.72$\pm$0.03\\
%AMIGA , $T=3-5$    & 88/21 & 0.01$\pm$0.04  &  88/20  & -0.05$\pm$0.04 & 68/10  &  0.63$\pm$0.03\\
%\hline
%\hline
%\label{compare_other_samples}
%\end{tabular}
%\end{table*}
%*****************************************************************

%*****************************************************************
\subsubsection{Is \lfir\ and the SFE enhanced  in interacting galaxies?}

\begin{figure}
\includegraphics[width=7.cm,angle=270]{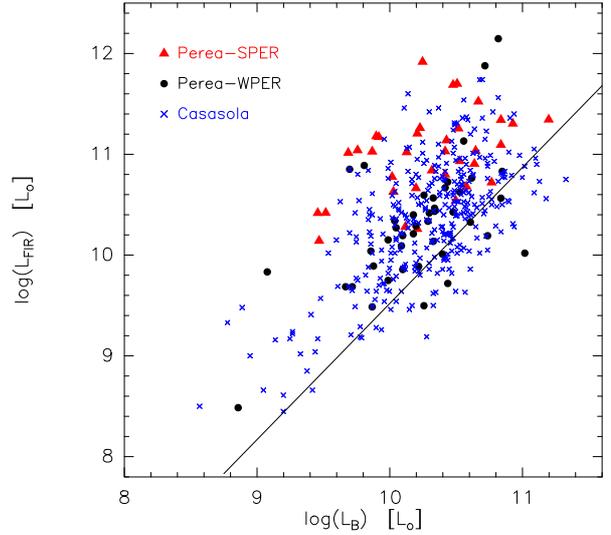}
\caption{The FIR luminosity vs. blue luminosity for different samples of interacting galaxies:
 \citet{2004A&A...422..941C} (only spiral galaxies with type $T=3-5$ are shown, in order not to overload the plot),
 and a sample of weakly  (WPER) and strongly (SPER) perturbed galaxies from
\citet{1997ApJ...490..166P}.
The full line is the regression fit obtained for the AMIGA $T=3-5$
sample from \citet{2007A&A...462..507L}.
}
\label{compare_lfir_casasola_perea}
\end{figure}

We use the same samples as in the previous subsection to  look for a possible enhancement of 
\lfir\ in interacting galaxies.  In Fig.~\ref{compare_lfir_casasola_perea} 
\lfir\ is compared to the blue luminosity. We include the 
best-fit regression line found for the AMIGA $T=3-5$ sample presented in
\citet[][]{2007A&A...462..507L} (Table 6,  log(\lfir) = 1.35$\times$  log(\lb)-3.98).
We see a clear excess of \lfir\ with respect to this regression line
 for the three samples, the strongest for the sample of strongly perturbed galaxies (SPER).
The excess can be quantified
by calculating the \lfir\  deficiency, defined in an anologous way as the \mhtwo\ deficiency.
The values are listed in Table~\ref{compare_other_samples}
and confirm the visual impressions of an excess in \lfir\ of about an order of 
magintude for the SPER sample.

The large excess in \lfir\ together with a smaller (or no) excess 
in \mhtwo\  results in a higher value of  SFE $\propto$ \lfir/\mhtwo\
compared to the AMIGA sample for the strongly and weakly perturbed samples.
  Comparing these values  to 
the results for the AMIGA sample  we find an increase  of about a factor 5 for
the strongly perturbed sample and of about 2 for the weakly perturbed sample.
{The SFE in the Casasola sample is similar to the AMIGA sample.}

For higher infrared luminosities,  the value of 
 \lfir/\mhtwo\  is known to increase strongly. \citet{1991ApJ...370..158S} studied the molecular gas 
%
% Sanders used LIR(8-1000), we adopt that this is about LFIR(40-120), X-factor of 3e20
%
content in  luminous infrared galaxies, ranging from \lir\ = $10^{10}$ \lsun\  to several $10^{12}$ \lsun ,
and showed that the ratio \lir/\mhtwo\ increases strongly with IR luminosity and with the degree of 
interaction. Whereas for isolated, low-luminosity (\lir $< 10^{11}$ \lsun)  galaxies they found values
of  \lir/\mhtwo\  similar to ours,  \lir/\mhtwo\  increases by a factor of about 10 for galaxies
with \lir $\sim 10^{12}$ \lsun\ which are mostly advanced mergers.
Similarly high values of \lfir/\mhtwo $\sim 50$ have been found by \citet{1997ApJ...478..144S} for ULIRGs.
%
% They find a median value of LFIR/L'CO =160, for a conversion factor of 3e20 which
% corresponds to LFIT/MH2 = 160/(5*2/3) = 48
%
%Taking into account that the molecular gas mass in ULIRGs is overestimated
%by a factor of 3-5 due to the different physical conditions of the molecular gas
%in galaxies \citep{1997ApJ...478..144S,1998ApJ...507..615D},
%the true  ratio of \lfir/\mhtwo\ is even higher.

\section{Conclusions and Summary}
\label{conclusion}

We presented   molecular gas masses, based on CO observations,
for a sample  of 273 isolated galaxies and we performed a statistical analysis for 
a redshift-limited sample of 173 isolated galaxies with recession velocities 
between 1500 and 5000 \kms .  The observations covered in most cases only the
central position of the galaxies. In order to correct for the missing molecular gas
mass outside the observed area, we derived and applied an aperture correction
assuming an exponentially decreasing CO disk. 
We obtained the following results:

\begin{enumerate}

\item We compared the molecular gas mass to different  parameters
(\lb , \lk, \dopttwo\ and \lfir ) in order to characterize
the relations followed by isolated galaxies
and establish a baseline that can be used to find possible deviations for interacting
galaxies. We concentrated our analysis on the morphological types $T=3-5$ (Sb-Sc)
which represents the bulk of our sample (51\% of the galaxies are of these types) 
where also the detection rate  of CO (74\%) was highest.

\item  We found  good  correlations with these parameters, 
roughly linear in the case of  \lk\  and  \lfir\ and  nonlinear for 
 \lb\ and \dopttwo\ . The tightest correlations are with \lfir\ and, for spiral galaxies with $T=3-5$,
 with \lk , whereas the correlation with \dopttwo\ is the poorest.
{Due to the  nonlinearity of the  correlation,  the ratio \mhtwo/ \lb\ %, respectively \mhtwo/\dopttwo ,   
changes with \lb\ which has to be taken into account when comparing it to other samples.}
%is no good indicators to test for a possible deficit or enhancement of the molecular gas.
We describe a deficiency parameter,
defined in analogy to the deficiency parameter for  the atomic gas as
the difference between the logarithm of the expected molecular gas mass and
the logarithm of the observed molecular gas mass. The expected molecular
gas mass can be calculated from any of the parameters studied by us
(\lb , \dopt , \lk , \lfir ) using the  correlation coefficients listed
in Table~\ref{regression}.

\item We applied these relations and the resulting expressions to three samples  from the literature
\citep{1997ApJ...490..166P,2004A&A...422..941C}. {For the sample of Casasola et al. and the sample
of strongly interacting galaxies of Perea et al.,  we  found  clear
evidence for an enhancement of \mhtwo\ in comparison to \lb\ and \lk ,
while for  a sample of weakly interacting galaxies  from Perea et al. no difference with respect to the AMIGA sample was found. A possible reason for this
difference could be a higher degree of interaction in the first two samples.}
% \lfir , a tracer for the SFR,
% was significantly enhanced in all three samples, resulting in a higher value for the star formation efficiency
%(defined as SFR/molecular gas mass)  in the sample of Perea et al., whereas for 
%the Casasola sample no significant AMIGA.}

\item We derived a mean molecular gas depletion time, {\taudep\ (defined as \mhtwo/SFR),
of log(\taudep) = 9.0 yr  for spiral galaxies ($T=3-5$) and a slightly lower value of 
log(\taudep) = 8.9 yr  for all morphological types (both values for a Kroupa IMF)}, in reasonable
agreement with other studies of nearby galaxies \citep{2011ApJ...730L..13B,2011arXiv1104.0019S}.

\item No good correlation was found between \mhtwo\ and \mhi .
The ratio between the molecular and the atomic gas mass
decreases significantly from early to late-type galaxies, with a difference of up to a factor of 10.
The ratio \mhtwo /\mhi\ of the AMIGA galaxies is well below 1 for all morphological types, with a mean value
of log(\mhtwo/\mhi) = -0.72 for galaxies of type $T=3-5$.
We compared our values to those of other noninteracting samples and found that
the AMIGA galaxies had the lowest values for all spiral galaxies.
% It is unclear whether
%this low ratio indicates a relatively low molecular gas mass or a high atomic gas mass.

\item We used our data to compare the disk averaged  surface densities of the molecular,  $\Sigma_{\rm H_2}$,
and molecular+atomic gas, $\Sigma_{\rm H_2+HI}$,
to those of the SFR, $\Sigma_{\rm SFR}$. We found a good correlation between the logarithms of $\Sigma_{\rm H_2}$ 
 and $\Sigma_{\rm SFR}$, with a slope  close to 1.  No correlation with of $\Sigma_{\rm SFR}$
with  $\Sigma_{\rm H_2+HI}$ was found.

\end{enumerate}

\begin{acknowledgements}
We would like to thank the referee for useful comments and suggestions.
This work has
been supported by the research projects  AYA2008-06181-C02 and
 AYA2007-67625-C02-02  from the Spanish Ministerio de Ciencia y
Educaci\'on and the Junta
de Andaluc\'\i a (Spain) grants P08-FQM-4205, FQM-0108 and TIC-114.
DE was supported by a Marie Curie International Fellowship within the 6th European Community Framework Programme (MOIF-CT-2006-40298).
UL  warmly thanks IPAC (Caltech),  where this work was finished during a sabbatical stay,  for their hospitality.
We also thank K.C. Xu for helpful discussion.
This work is   based on observations with the  Instituto de Radioastronomia Milim\'etrica IRAM 30m and the
  Five College Radio Astronomy (FCRAO) 14m. IRAM is supported by INSU/CNRS (France), MPG (Germany, and IGN (Spain).
  The FCRAO is supported by NSF grant AST 0838222.
  We made use of the Nasa Extragalactic Database (NED) and of the Lyon Extragalactic Database (LEDA).
\end{acknowledgements}

\bibliography{cig-co-ref}

\appendix

\section{CO Spectra} 

Fig.~\ref{spectra_iram} shows the CO(1--0) profiles of the detections and  tentative detections 
observed by us at the IRAM 30m telescope and  Fig.~\ref{spectra_fcrao} those observed at the FCRAO 14m 
telescope.

%%%% IRAM spectra
 
 \begin{figure*} 
\centerline{\includegraphics[width=3cm,angle=270]{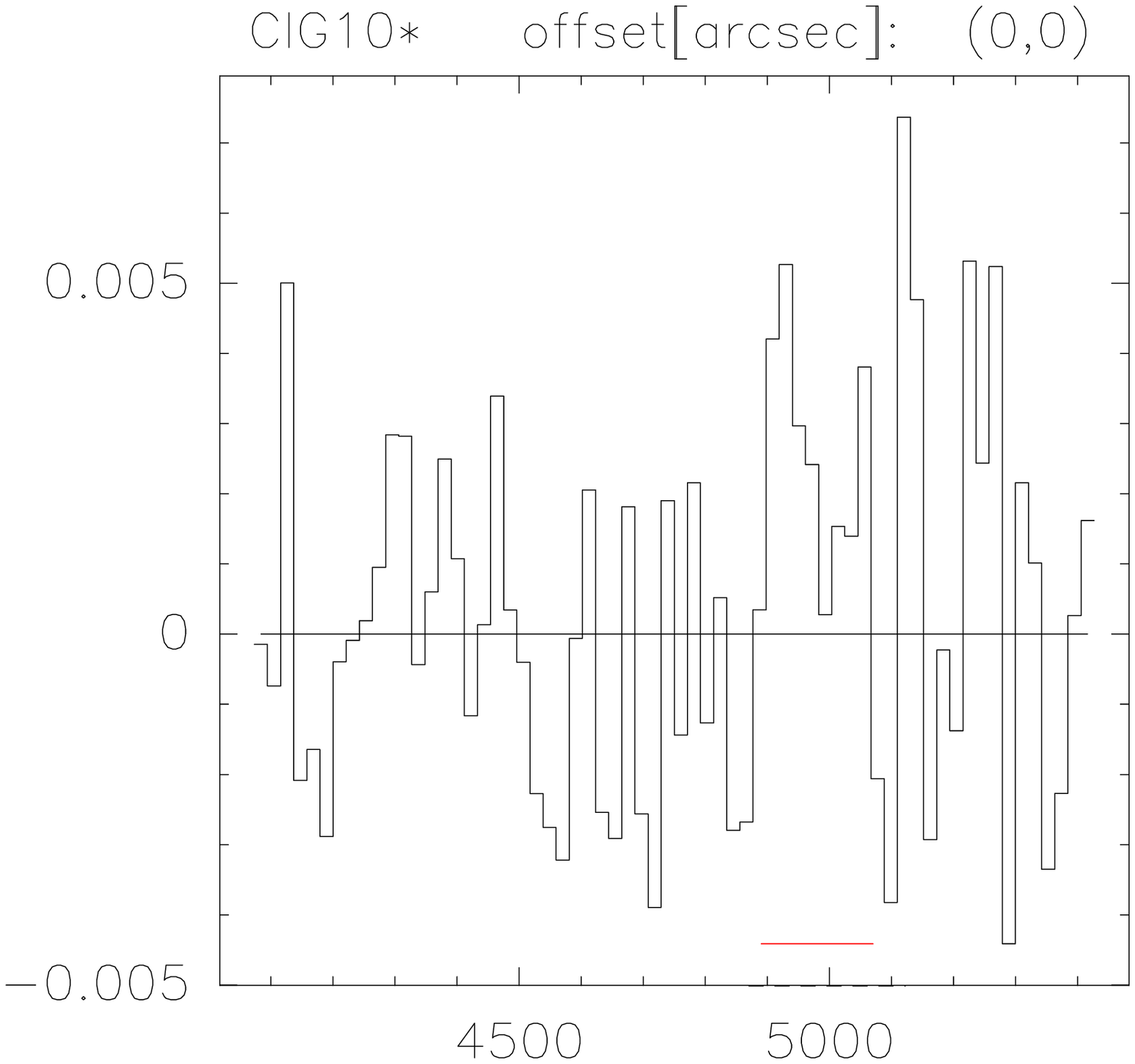} \quad 
\includegraphics[width=3cm,angle=270]{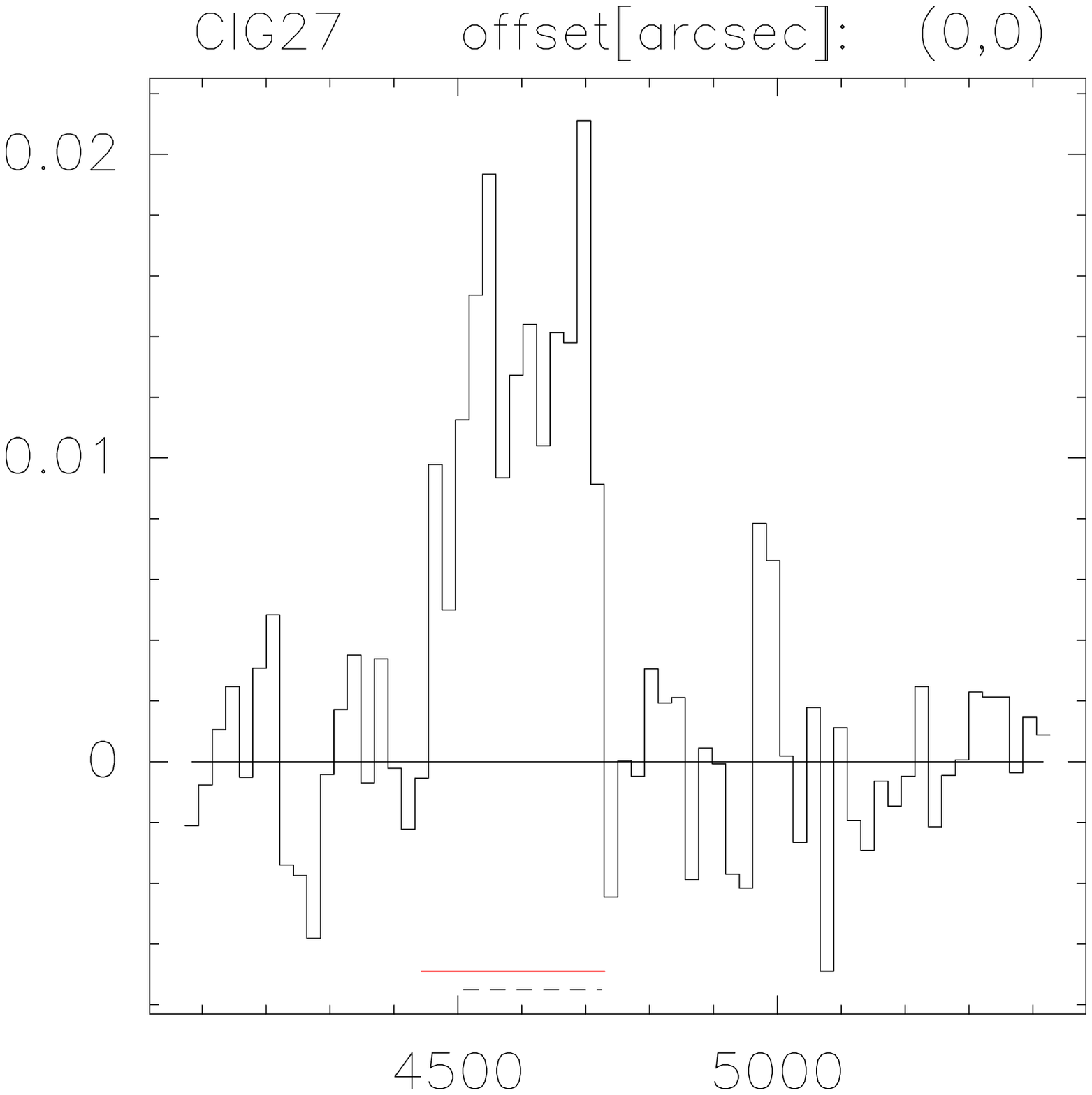}\quad 
\includegraphics[width=3cm,angle=270]{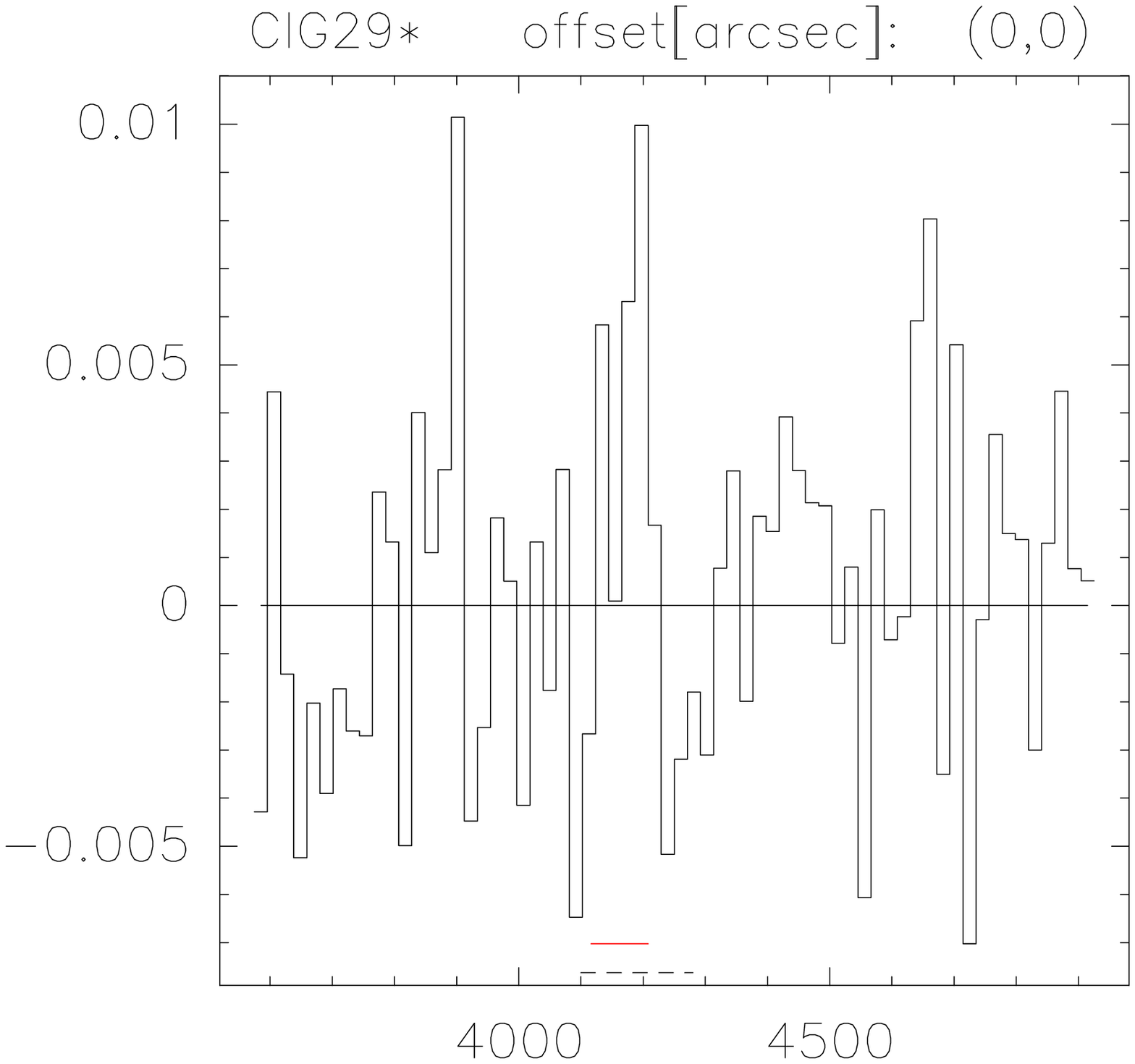}\quad 
\includegraphics[width=3cm,angle=270]{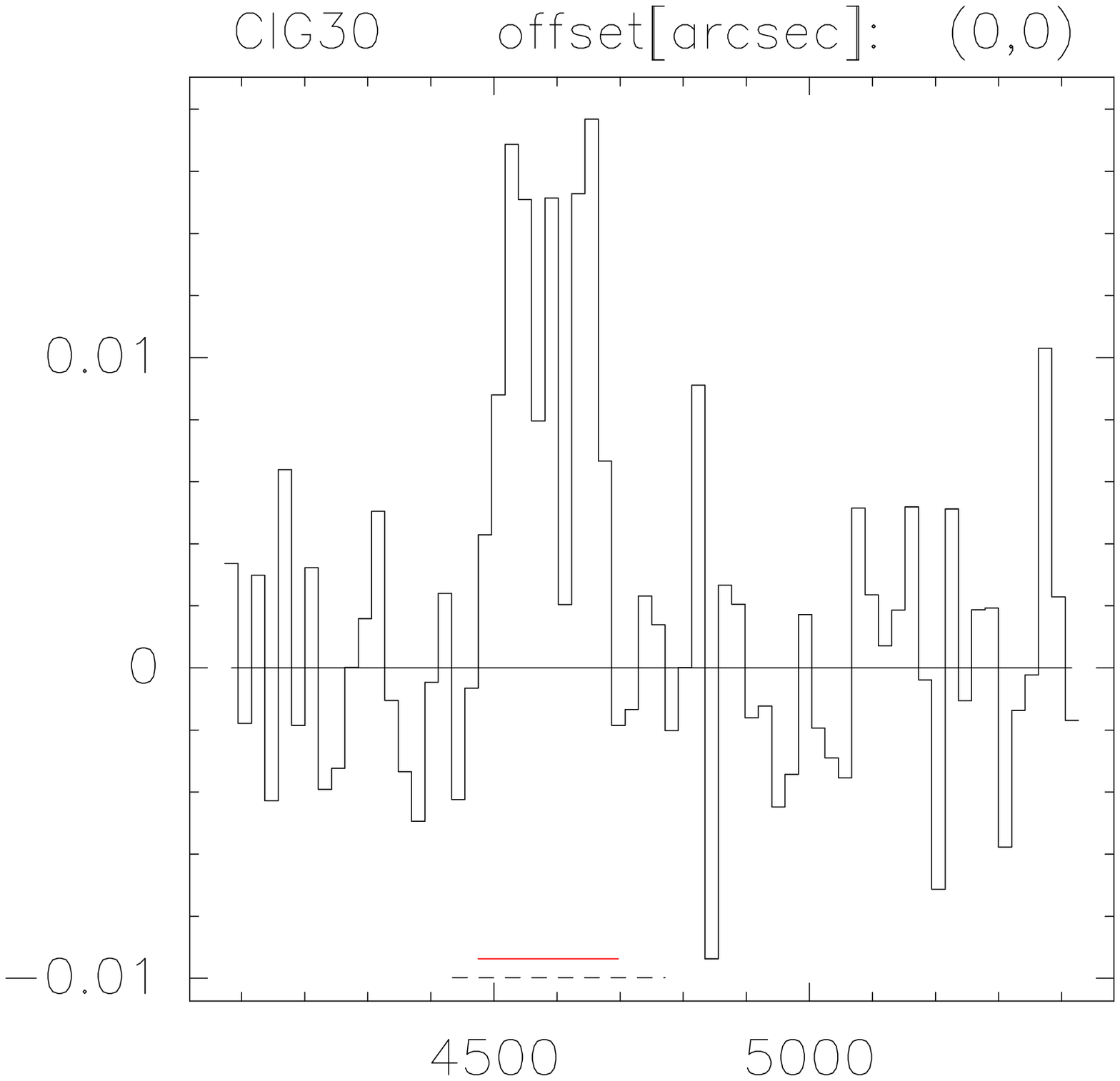}\quad 
\includegraphics[width=3cm,angle=270]{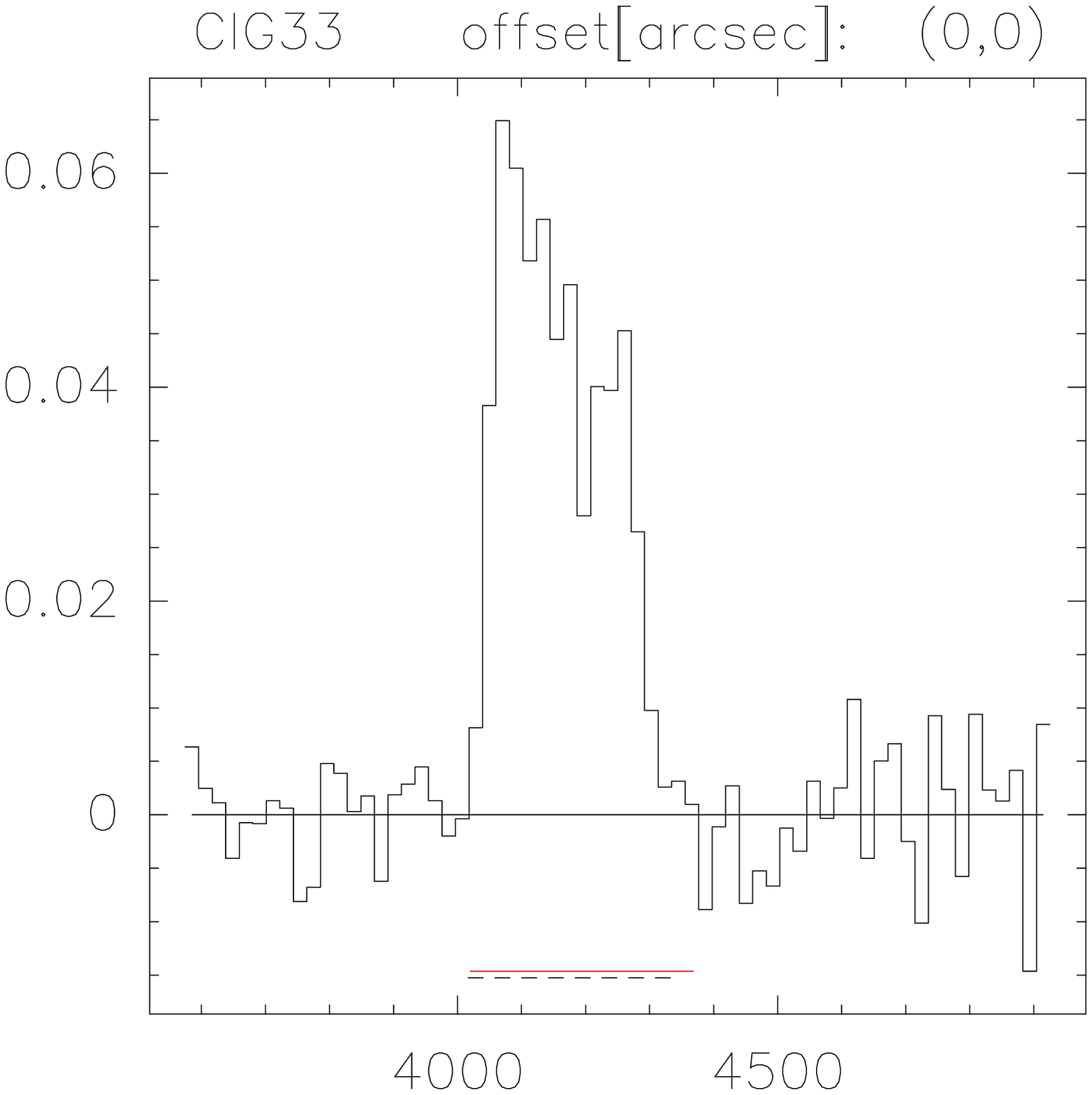}} 
\centerline{\includegraphics[width=3cm,angle=270]{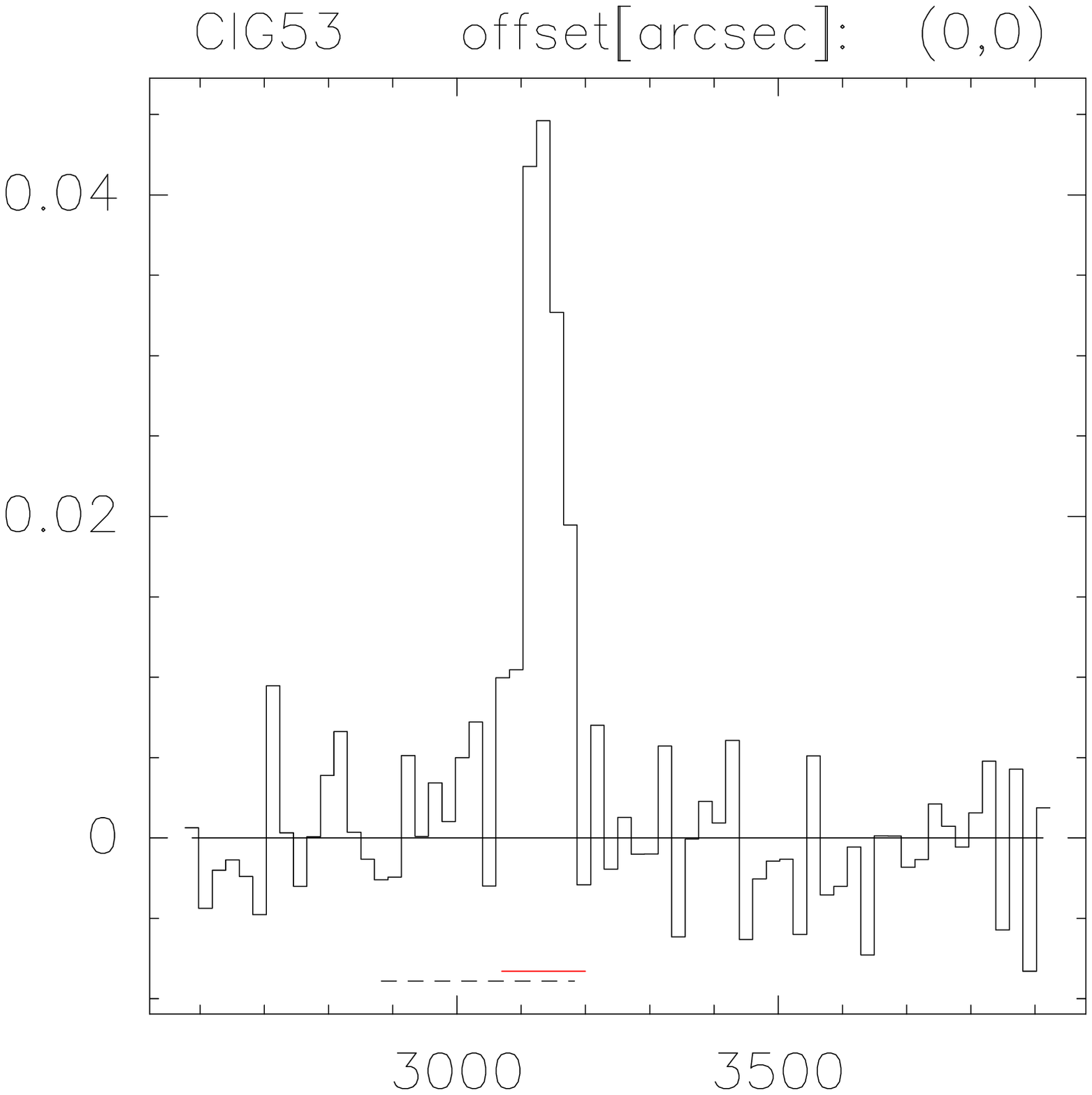} \quad 
\includegraphics[width=3cm,angle=270]{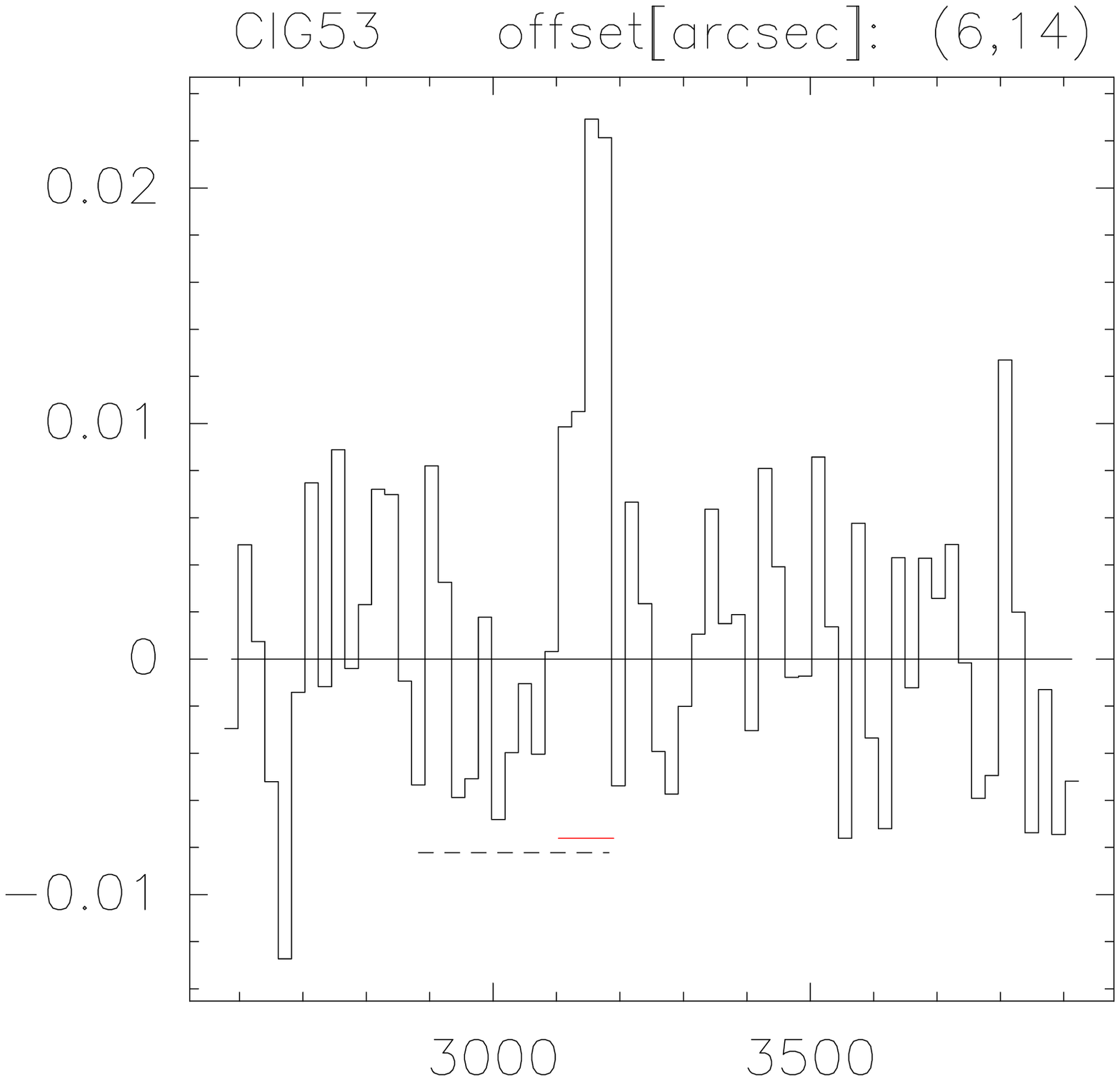}\quad 
\includegraphics[width=3cm,angle=270]{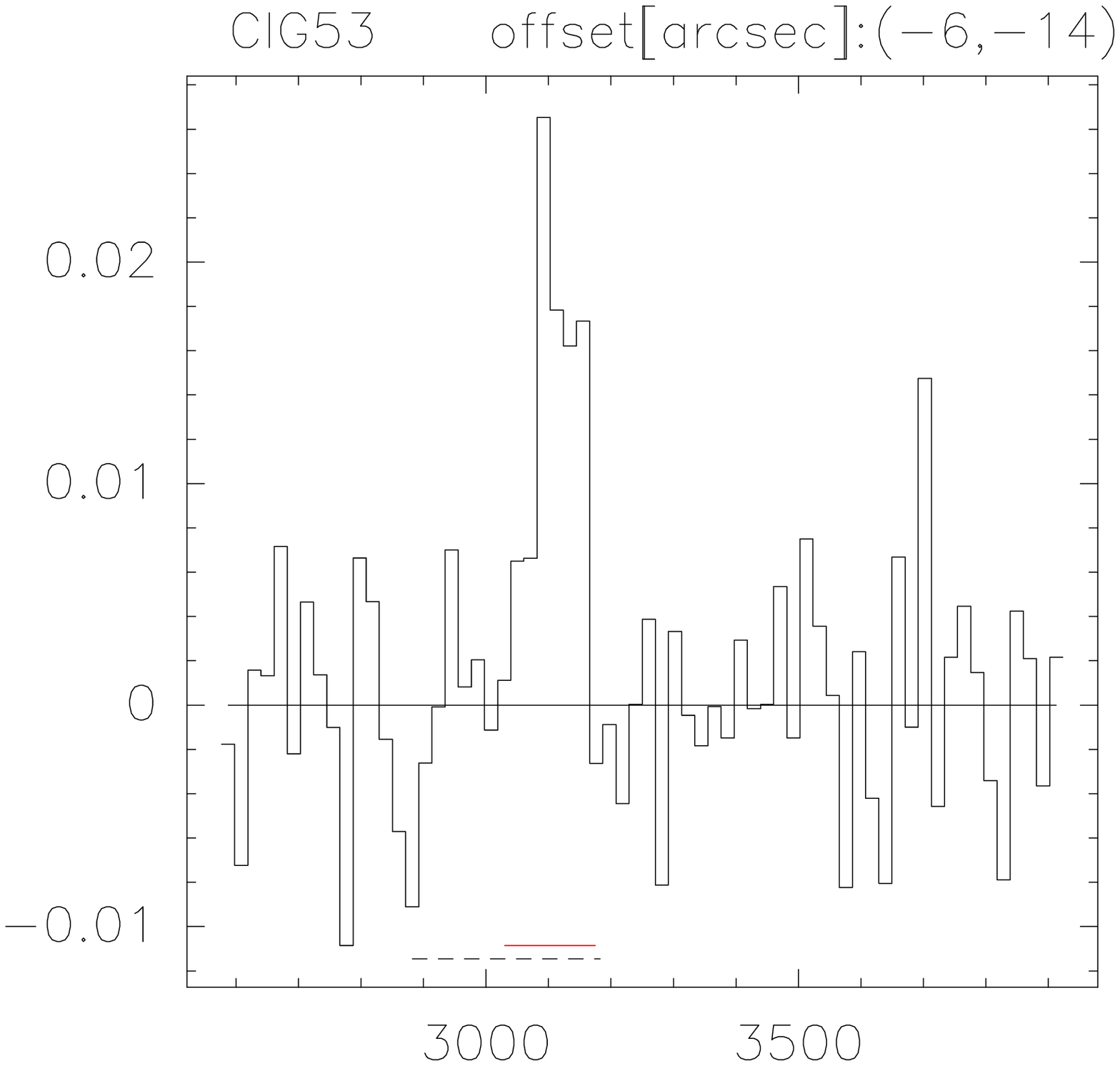}\quad 
\includegraphics[width=3cm,angle=270]{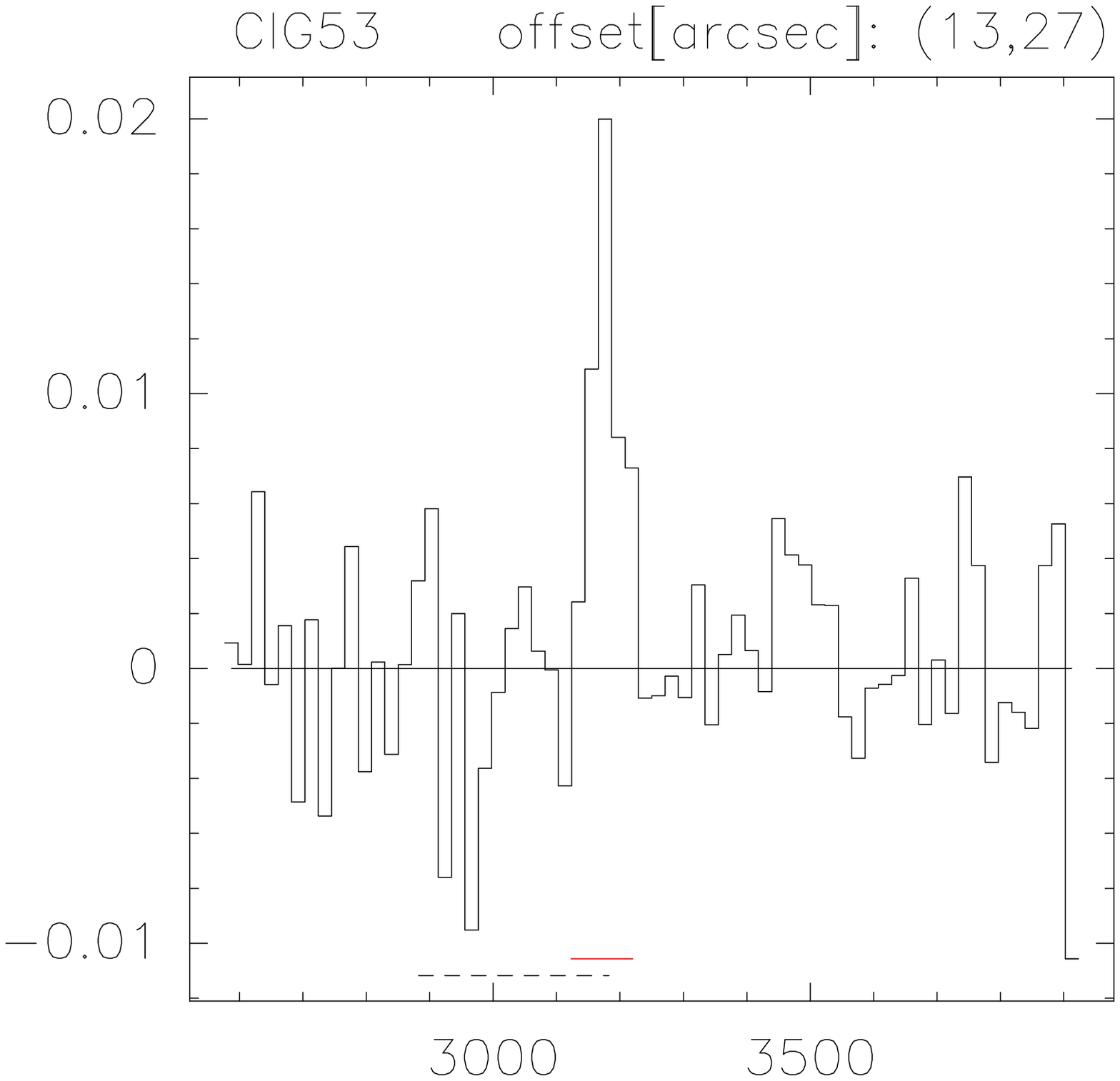}\quad 
\includegraphics[width=3cm,angle=270]{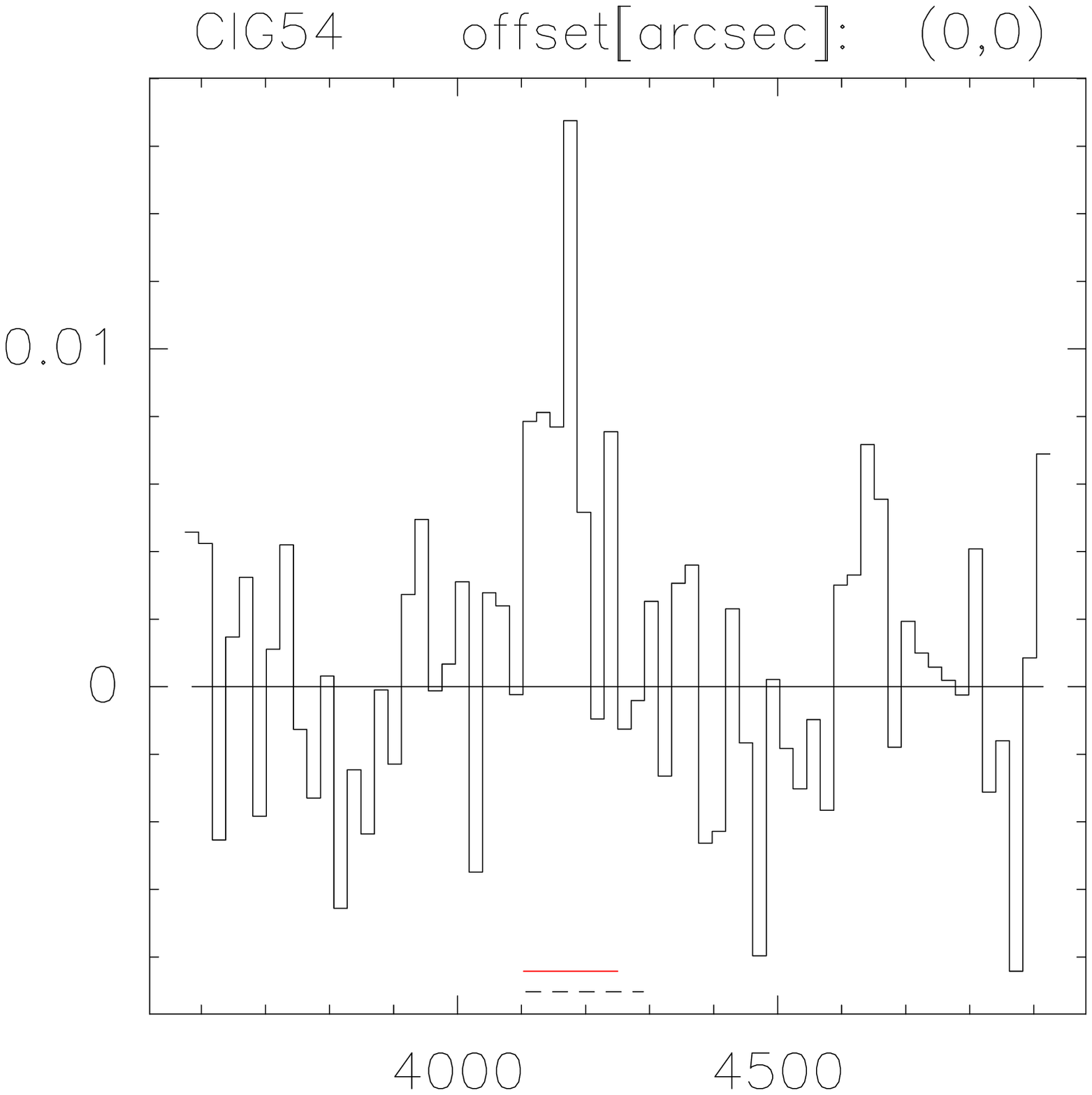}} 
\centerline{\includegraphics[width=3cm,angle=270]{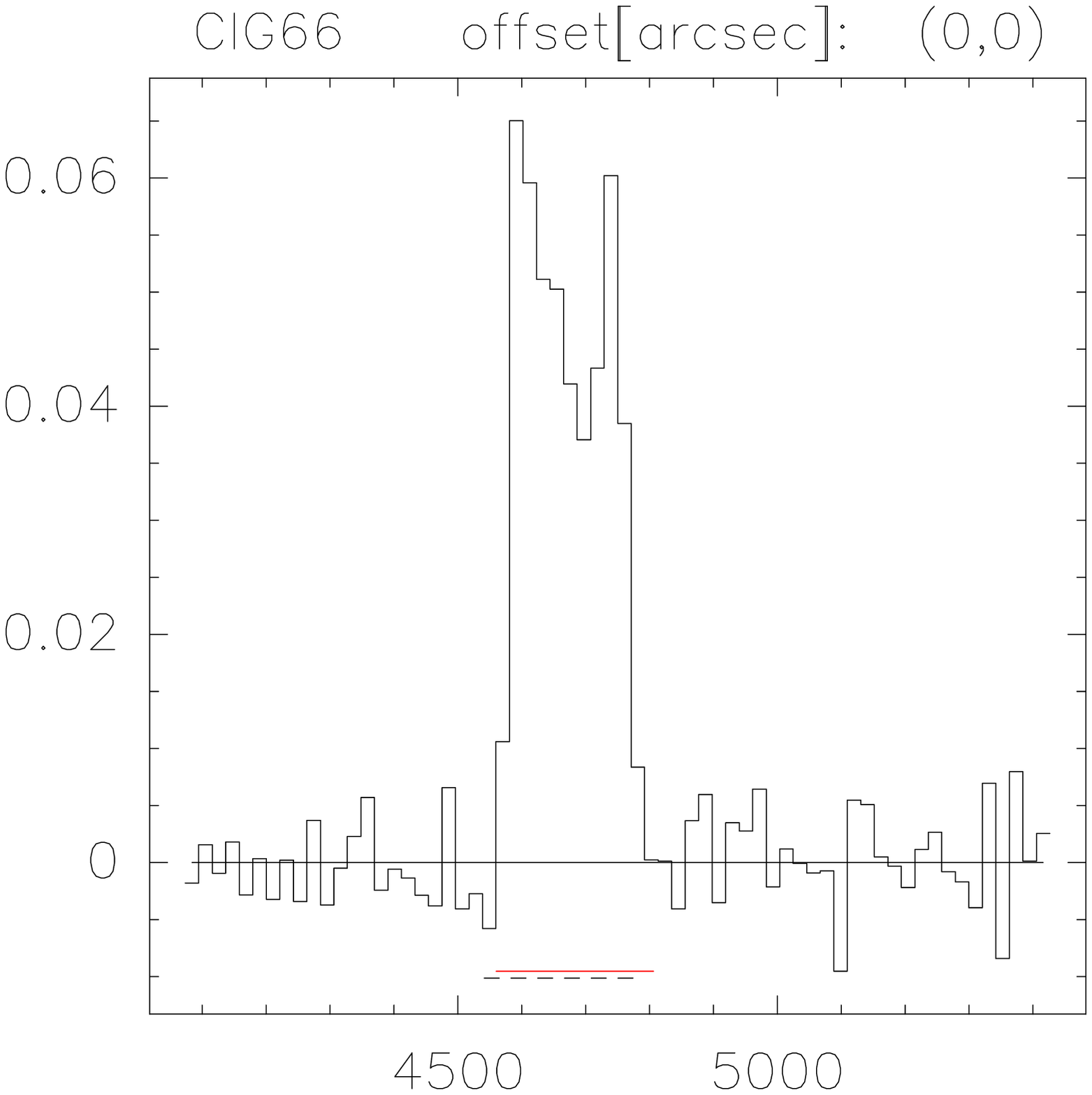} \quad 
\includegraphics[width=3cm,angle=270]{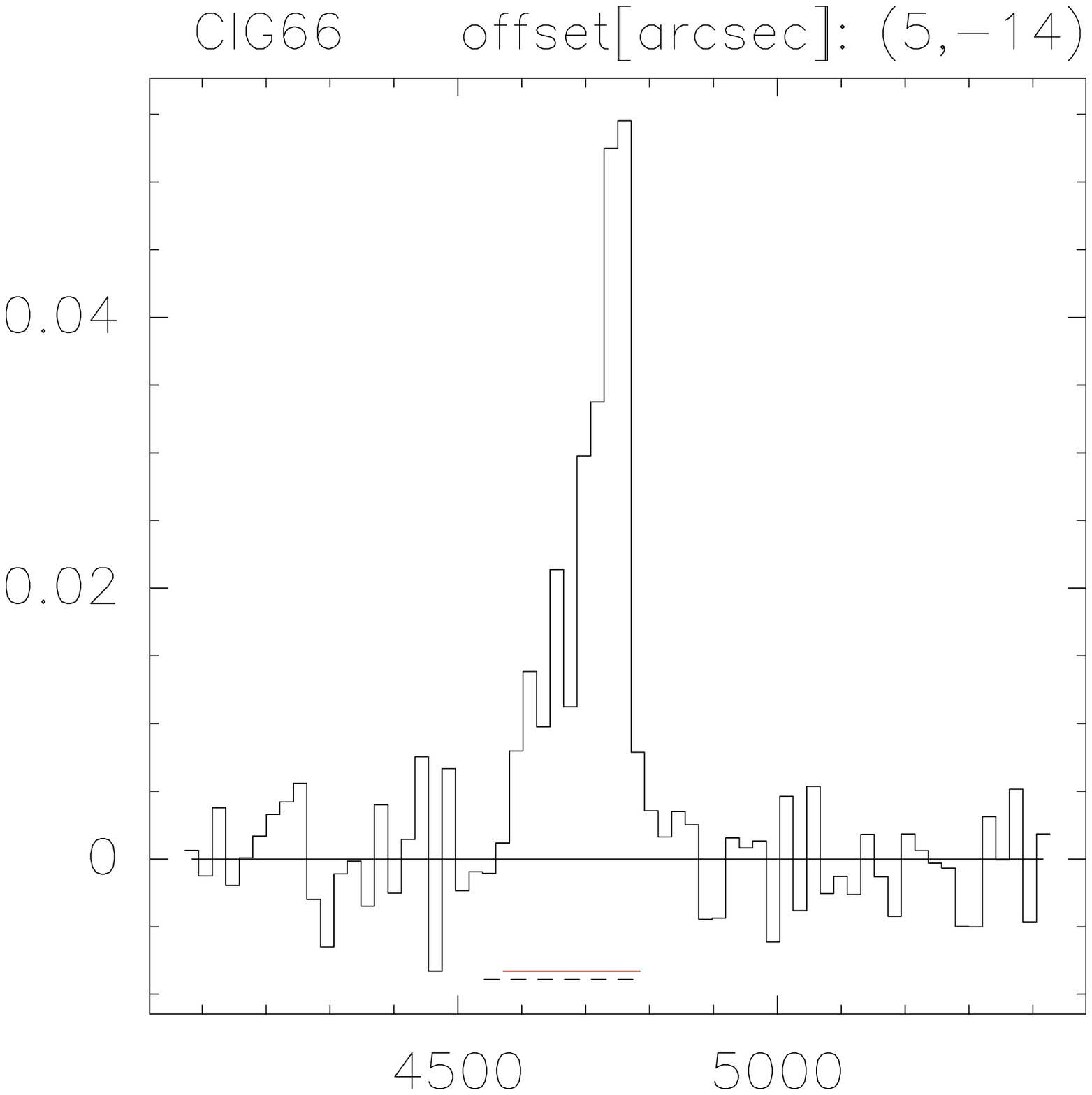}\quad 
\includegraphics[width=3cm,angle=270]{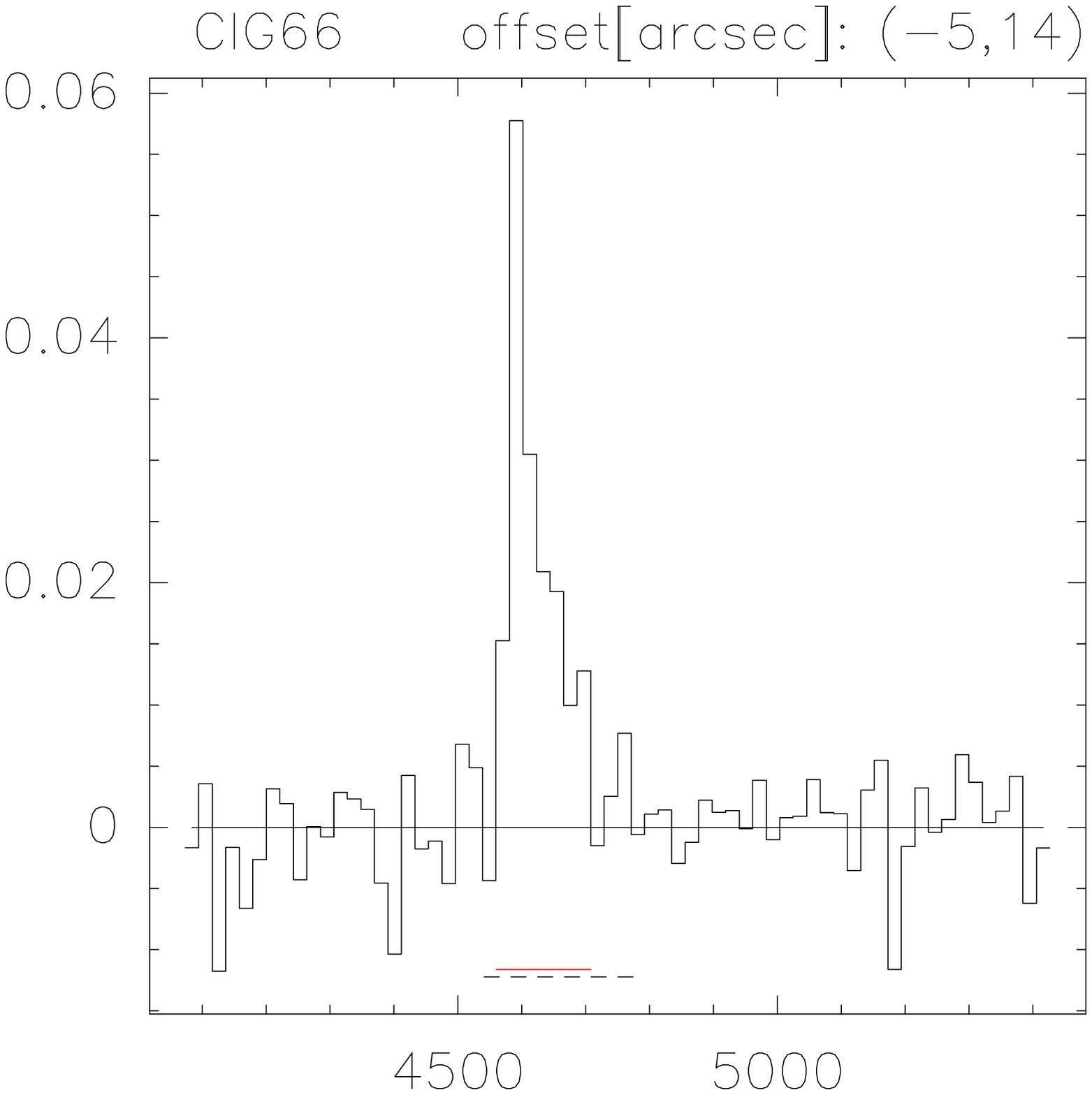}\quad 
\includegraphics[width=3cm,angle=270]{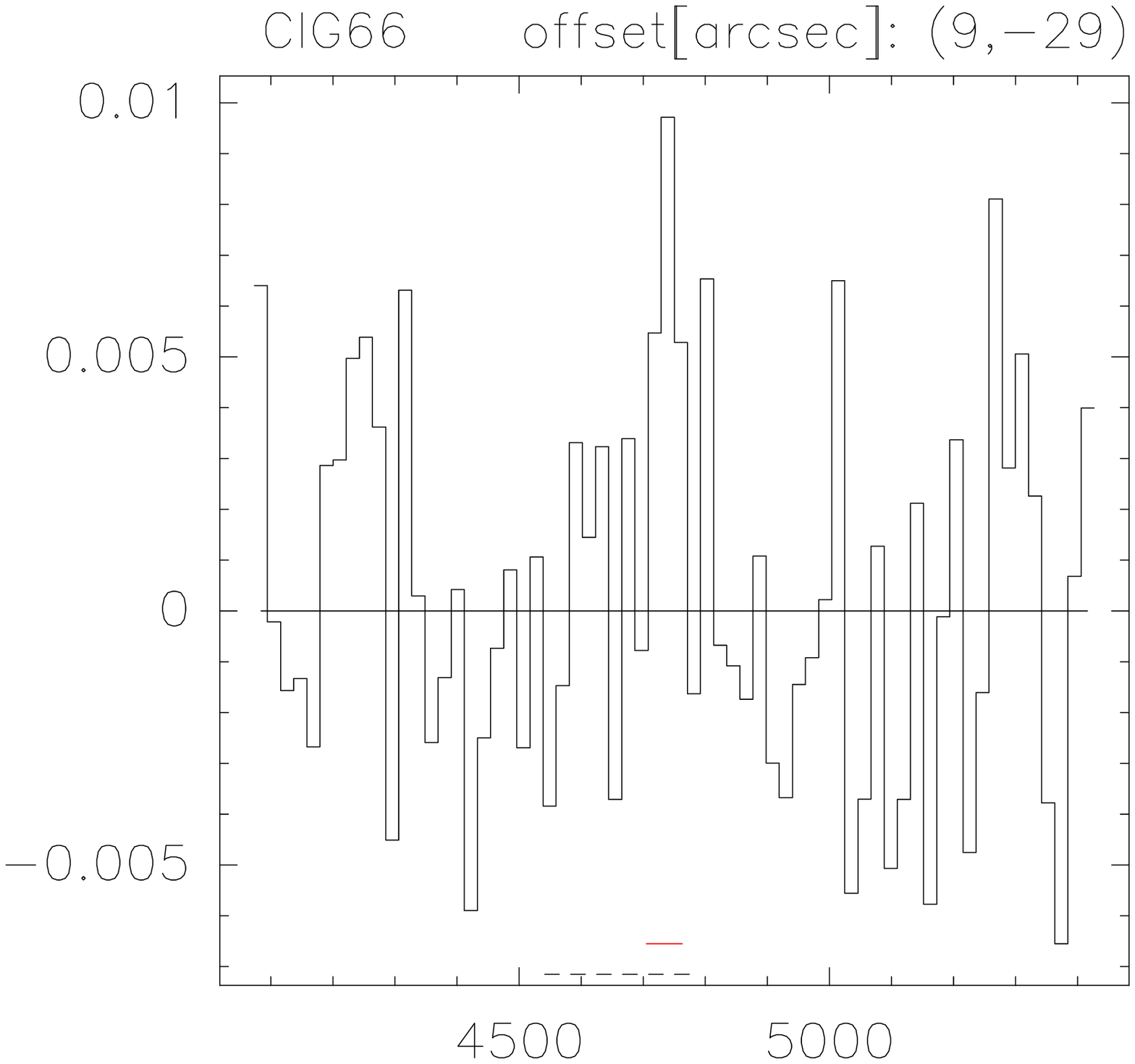}\quad 
\includegraphics[width=3cm,angle=270]{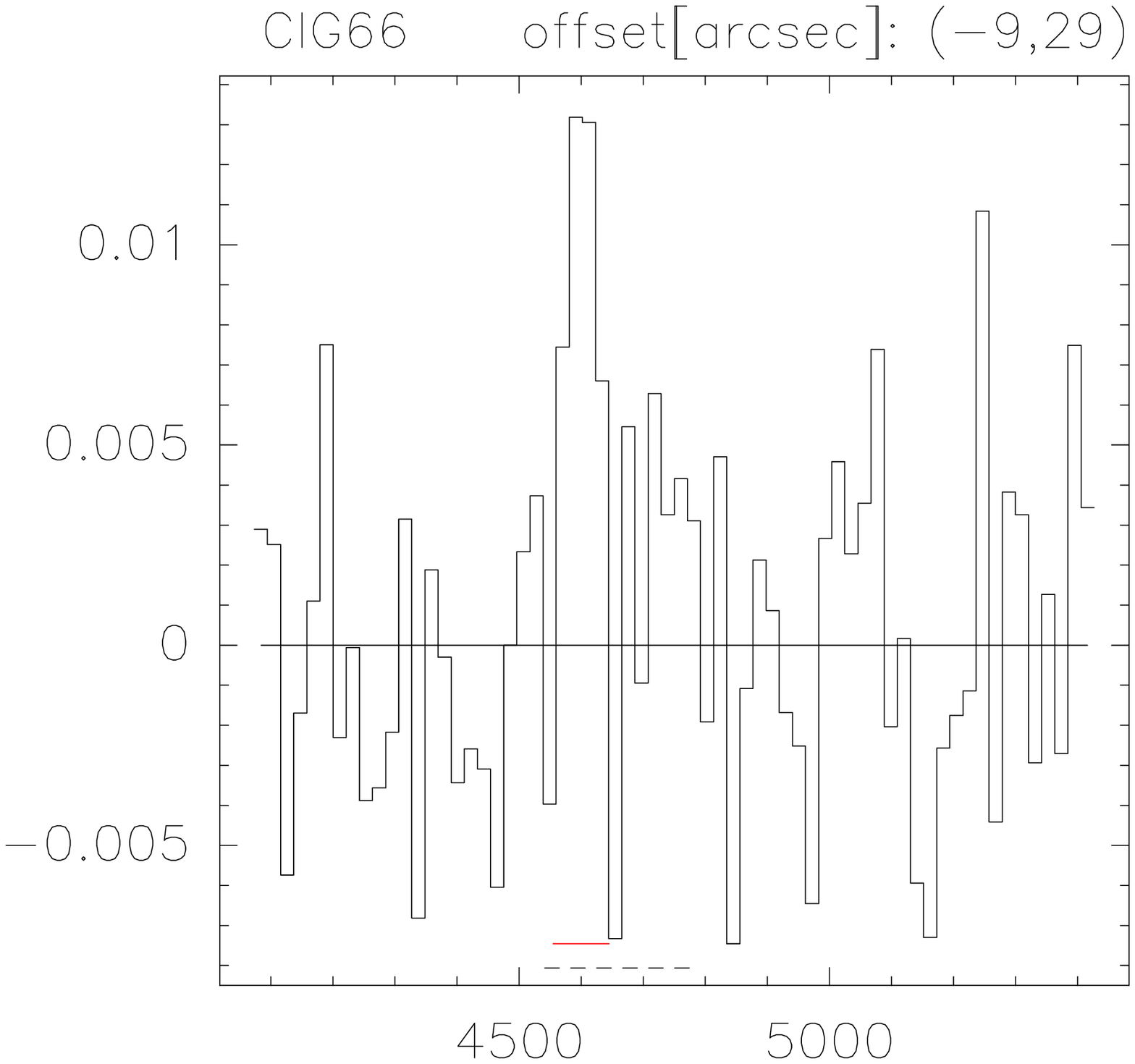}} 
\centerline{\includegraphics[width=3cm,angle=270]{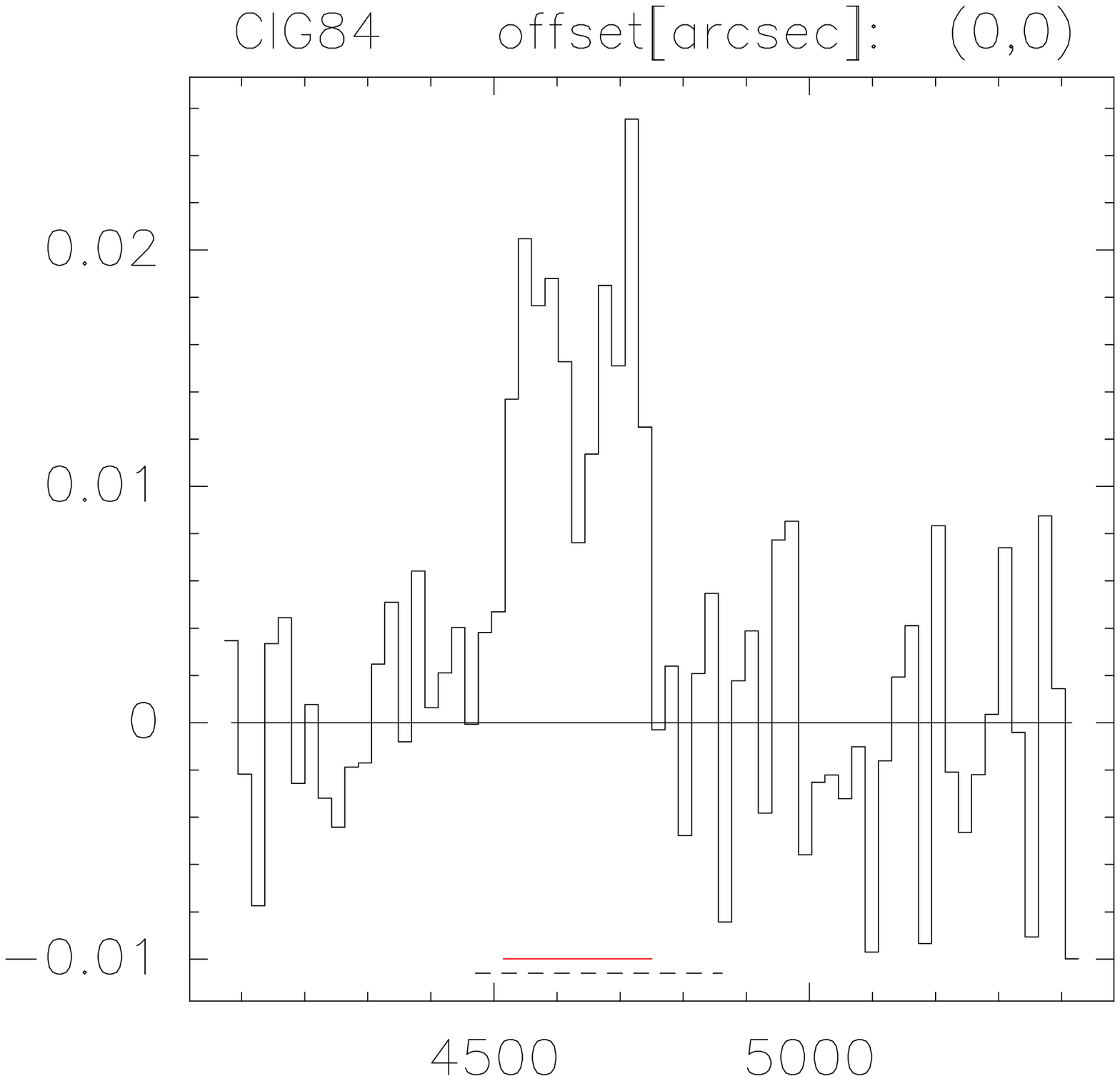} \quad 
\includegraphics[width=3cm,angle=270]{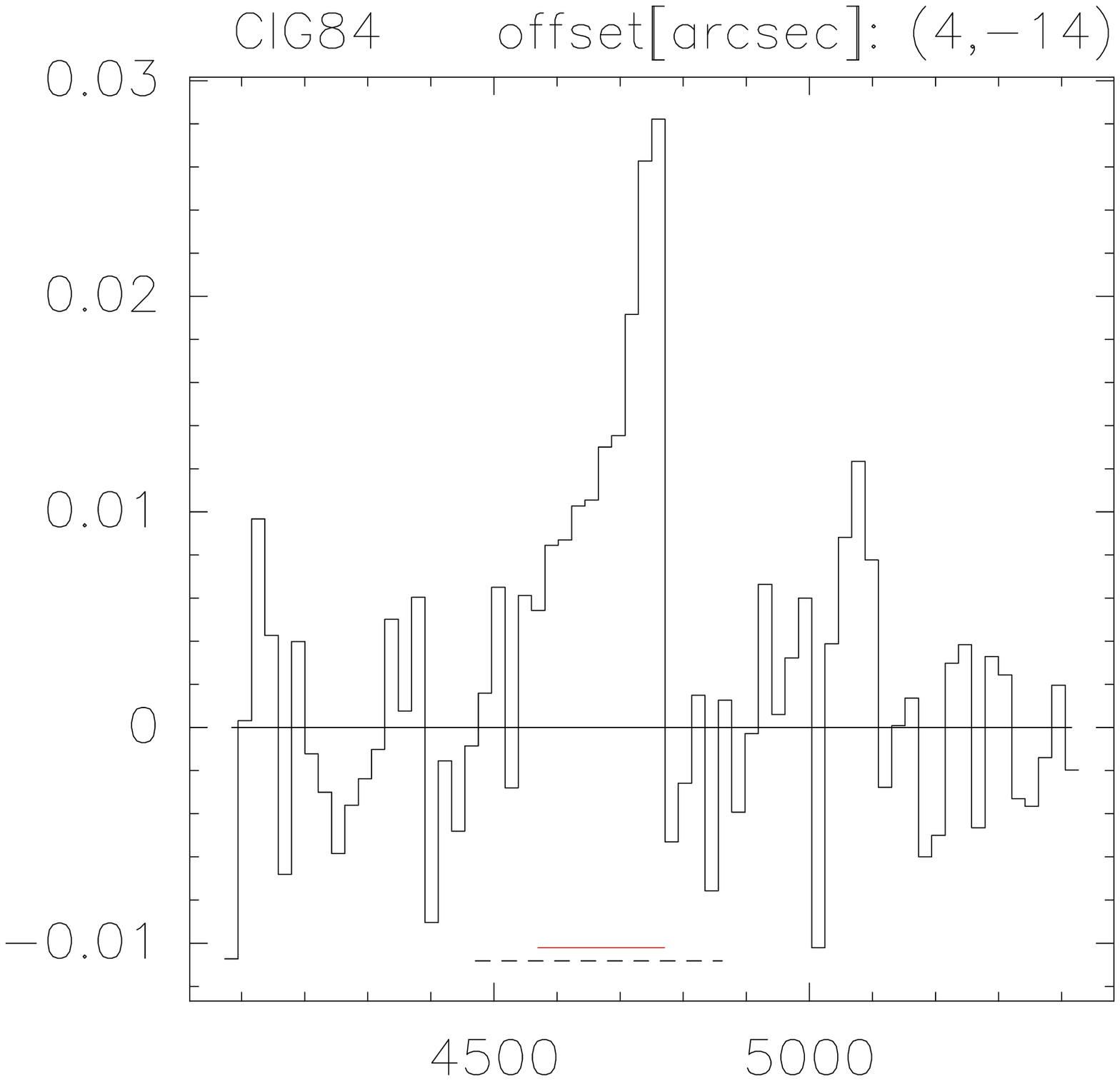}\quad 
\includegraphics[width=3cm,angle=270]{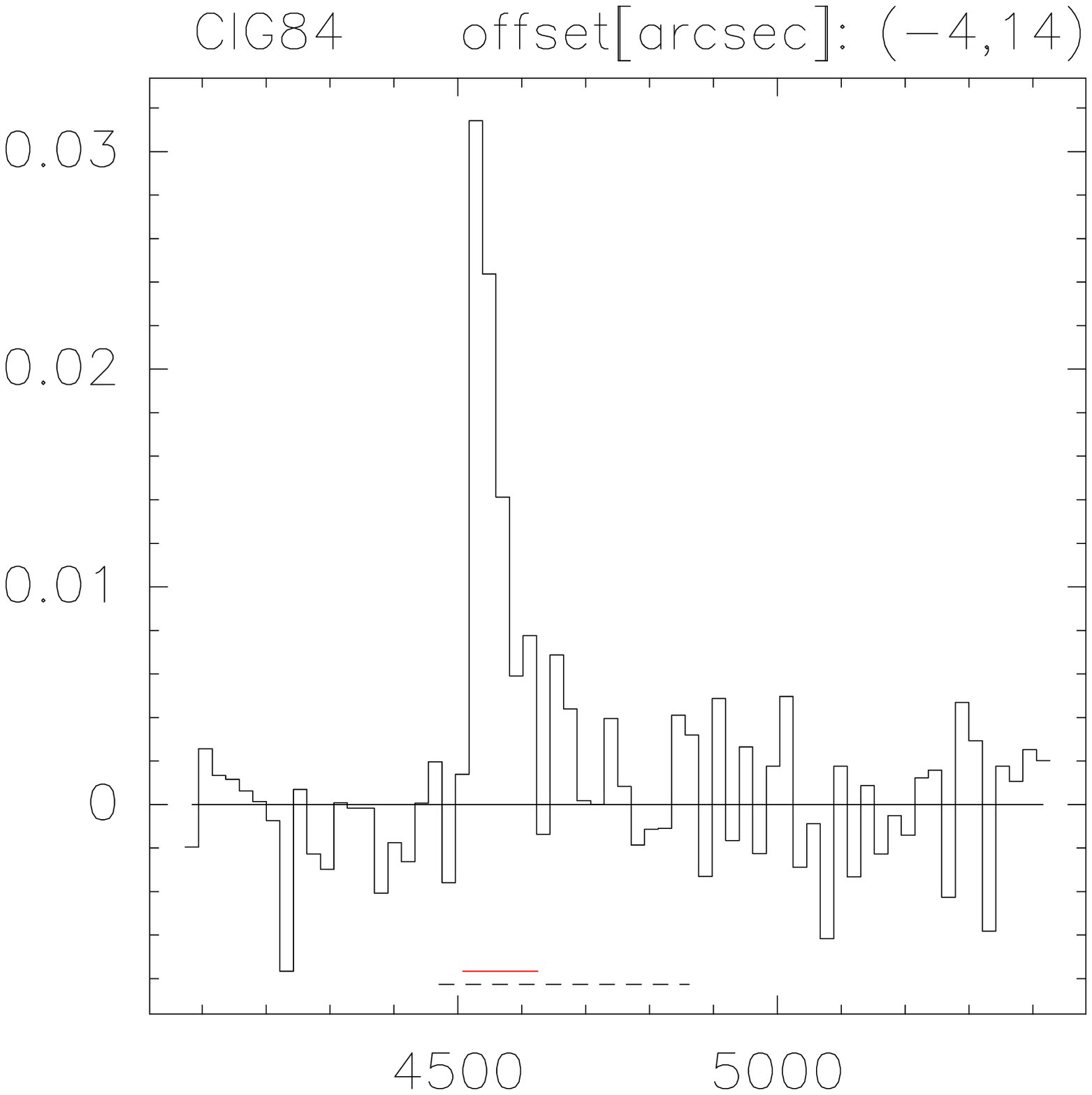}\quad 
\includegraphics[width=3cm,angle=270]{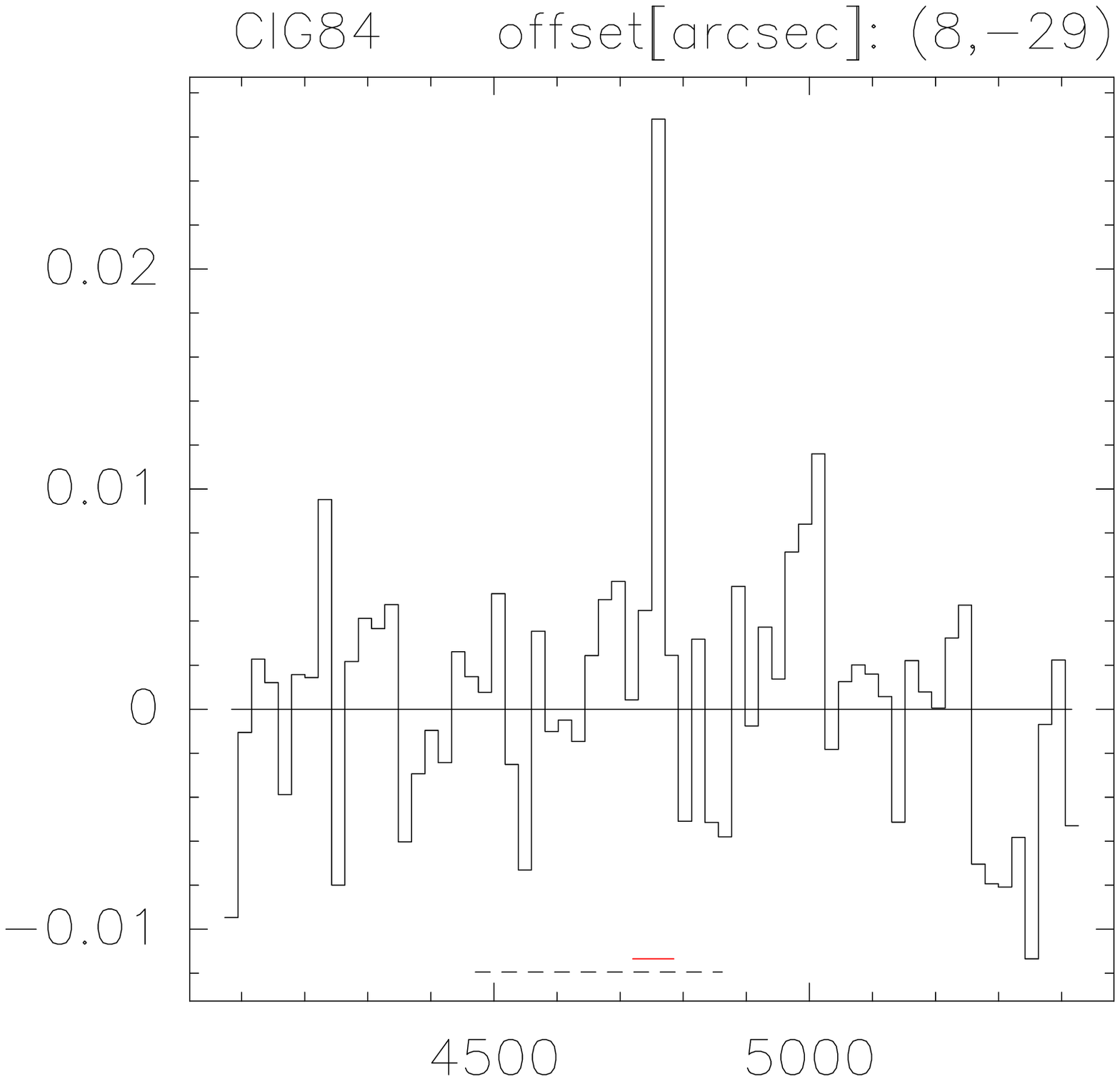}\quad 
\includegraphics[width=3cm,angle=270]{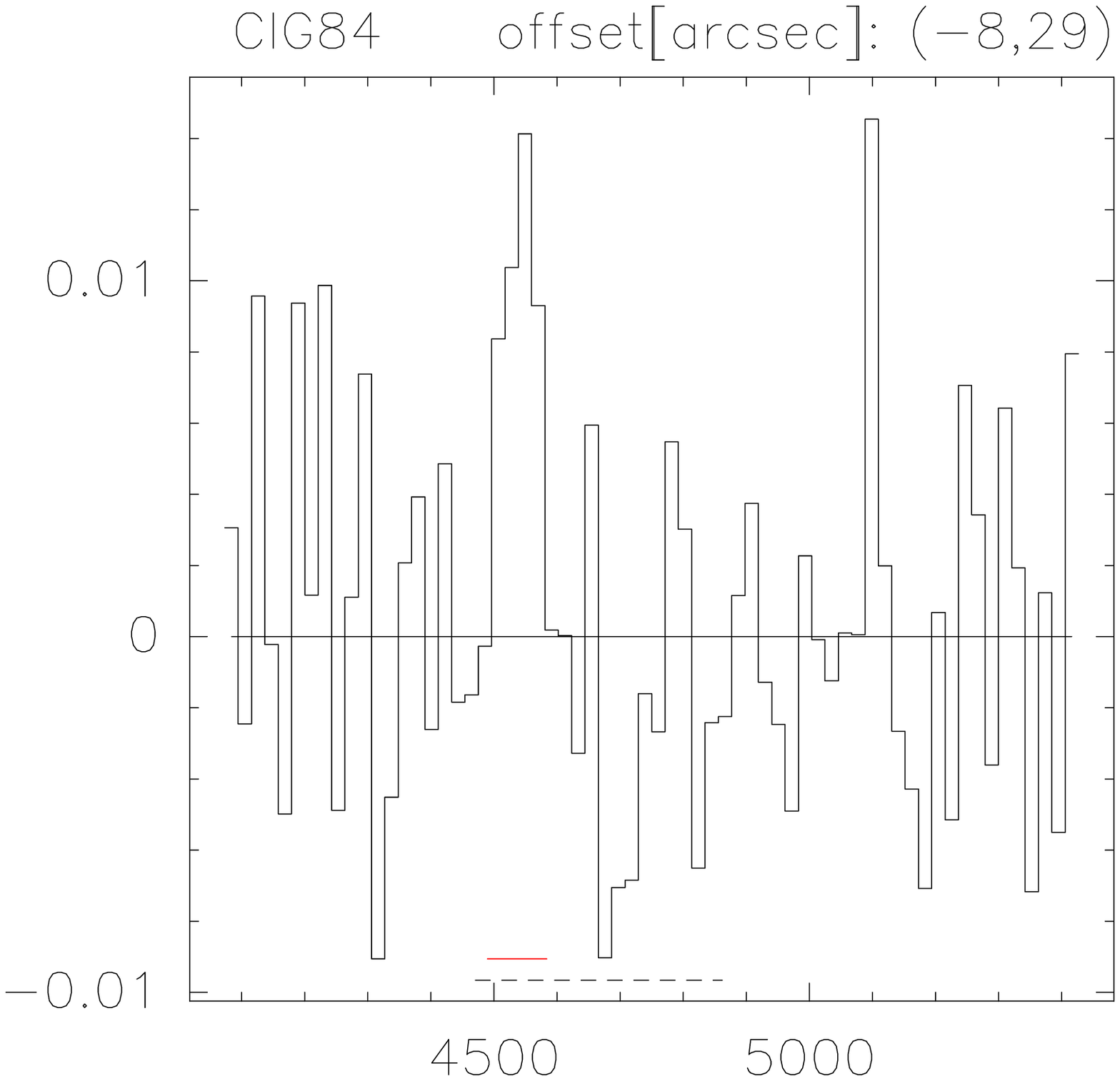}} 
\centerline{\includegraphics[width=3cm,angle=270]{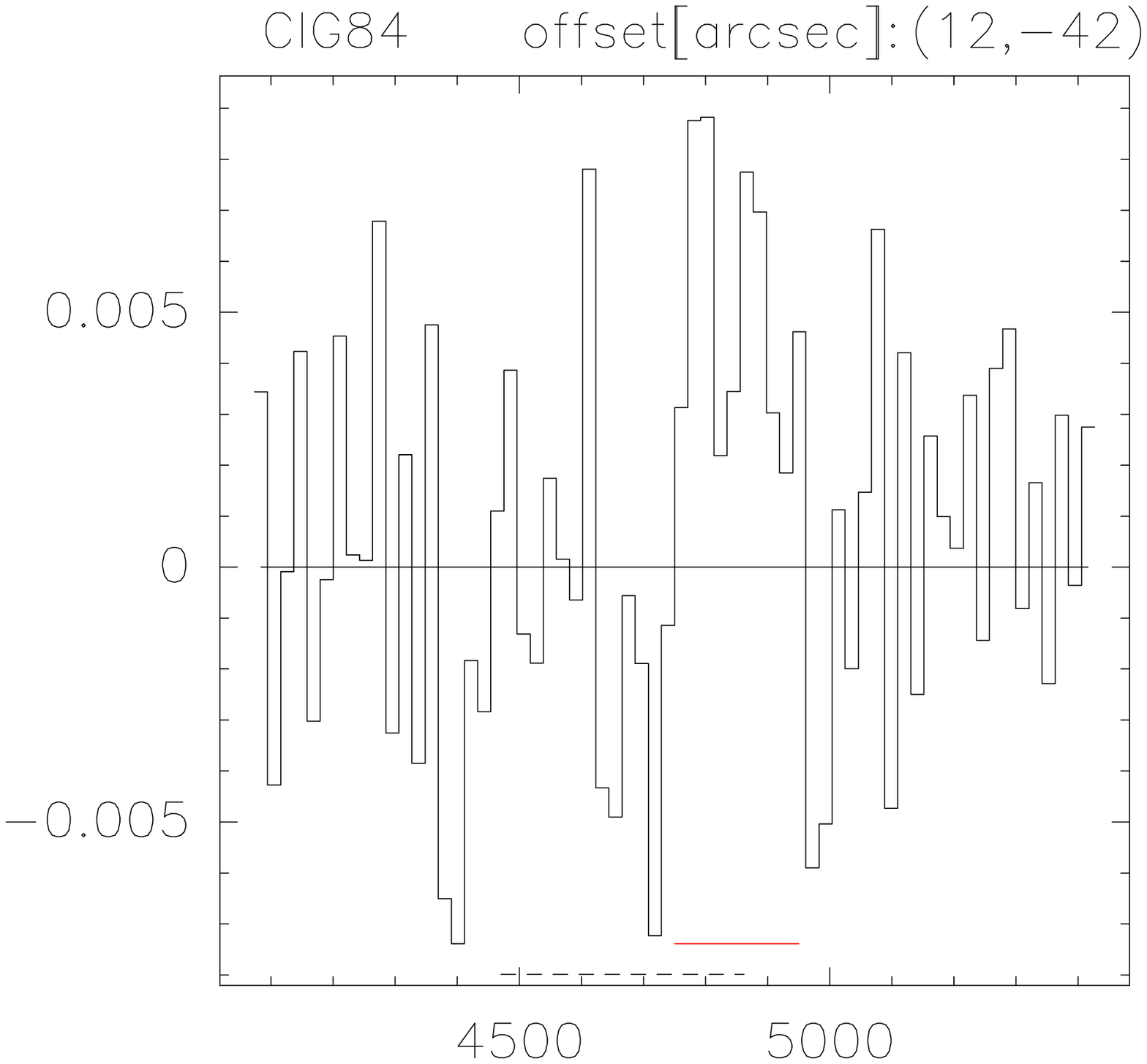} \quad 
\includegraphics[width=3cm,angle=270]{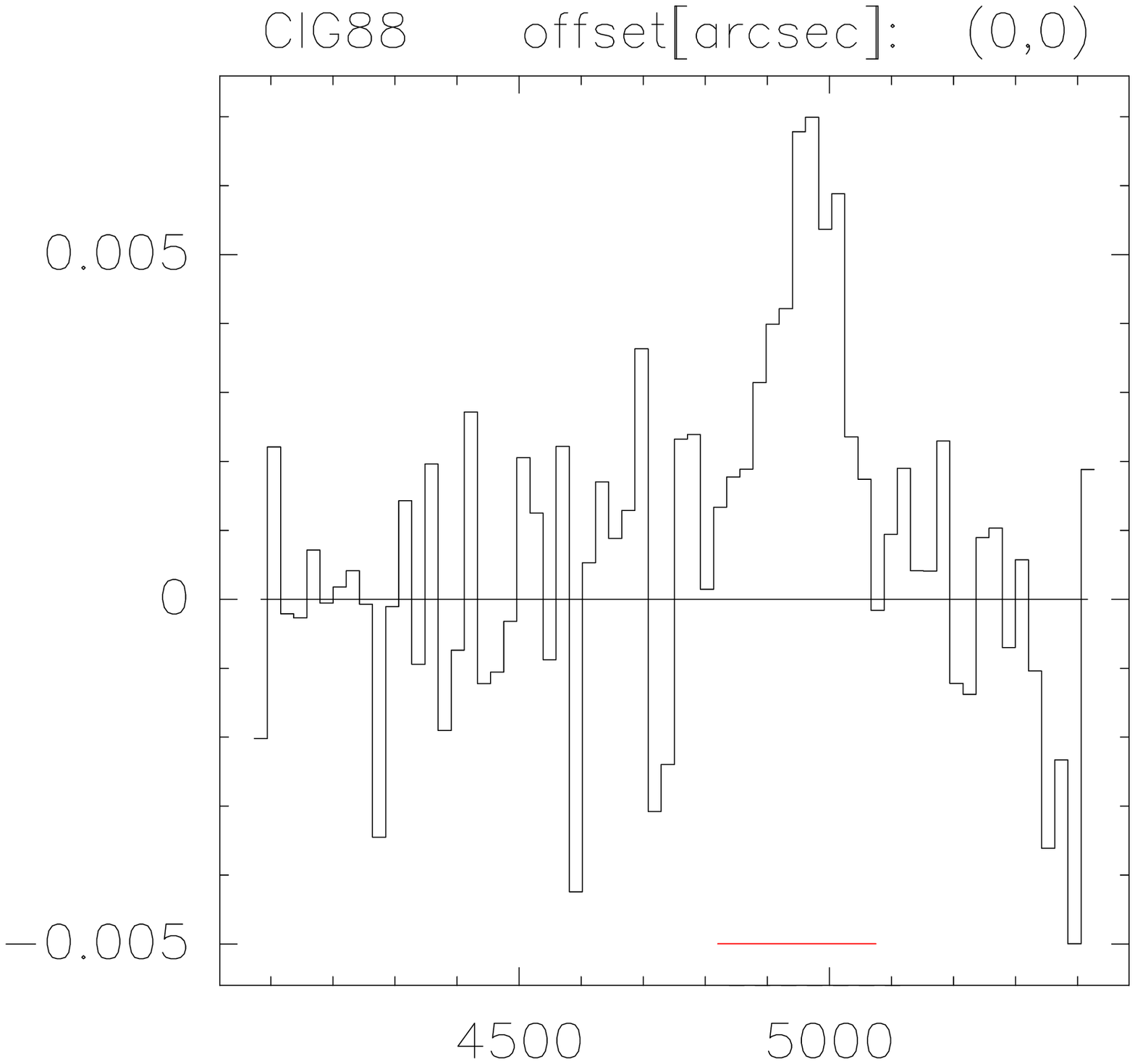}\quad 
\includegraphics[width=3cm,angle=270]{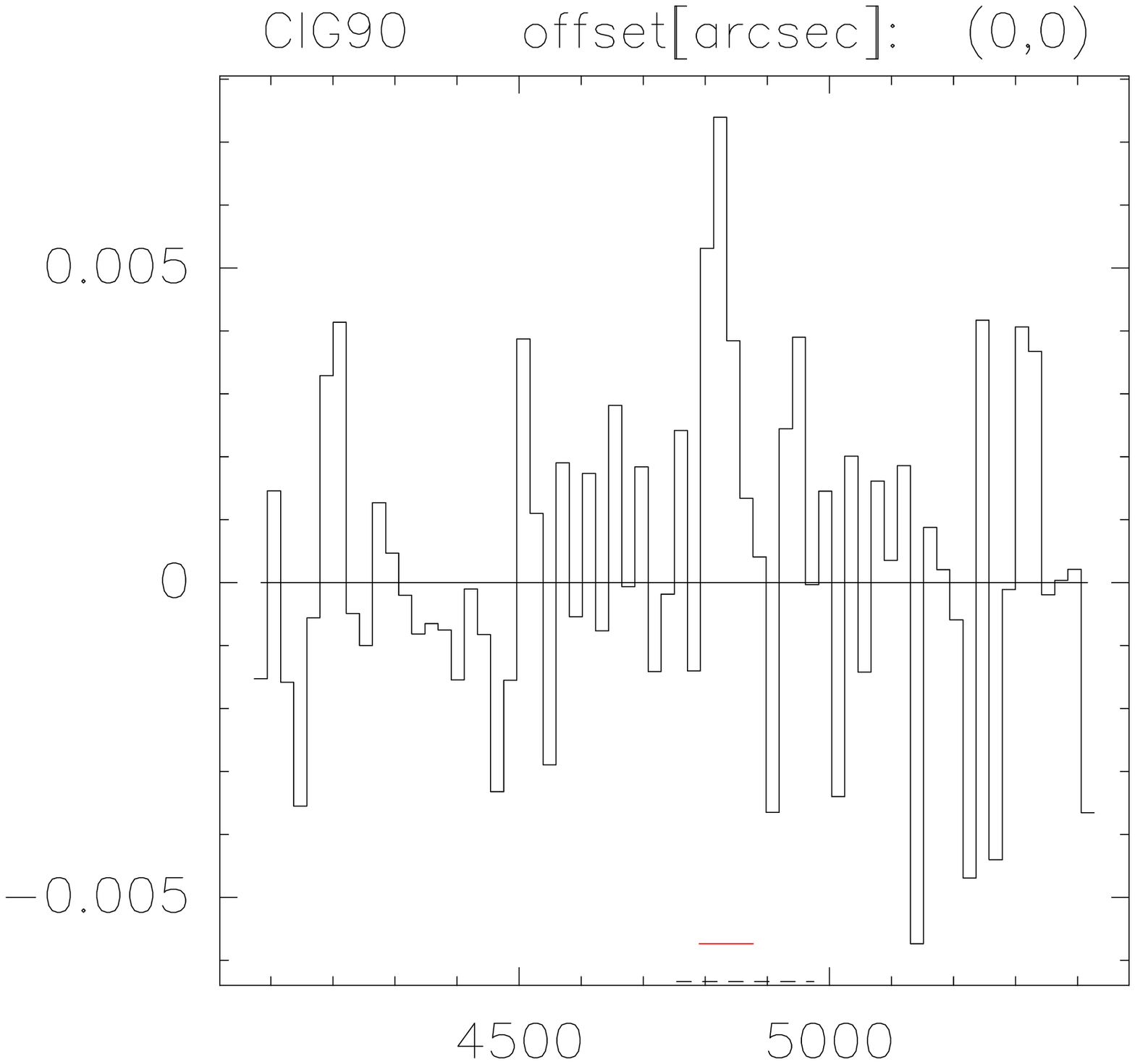}\quad 
\includegraphics[width=3cm,angle=270]{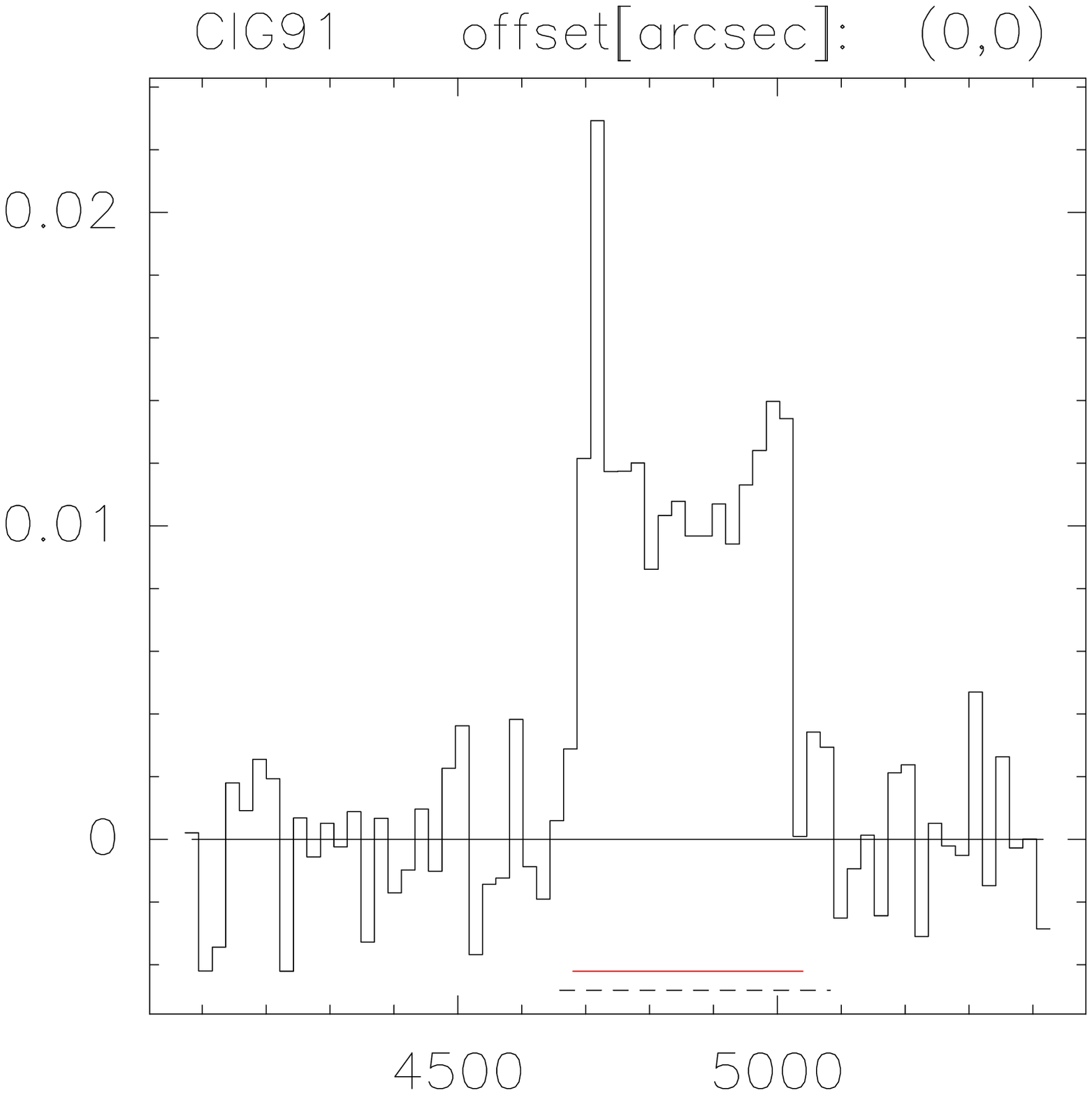}\quad 
\includegraphics[width=3cm,angle=270]{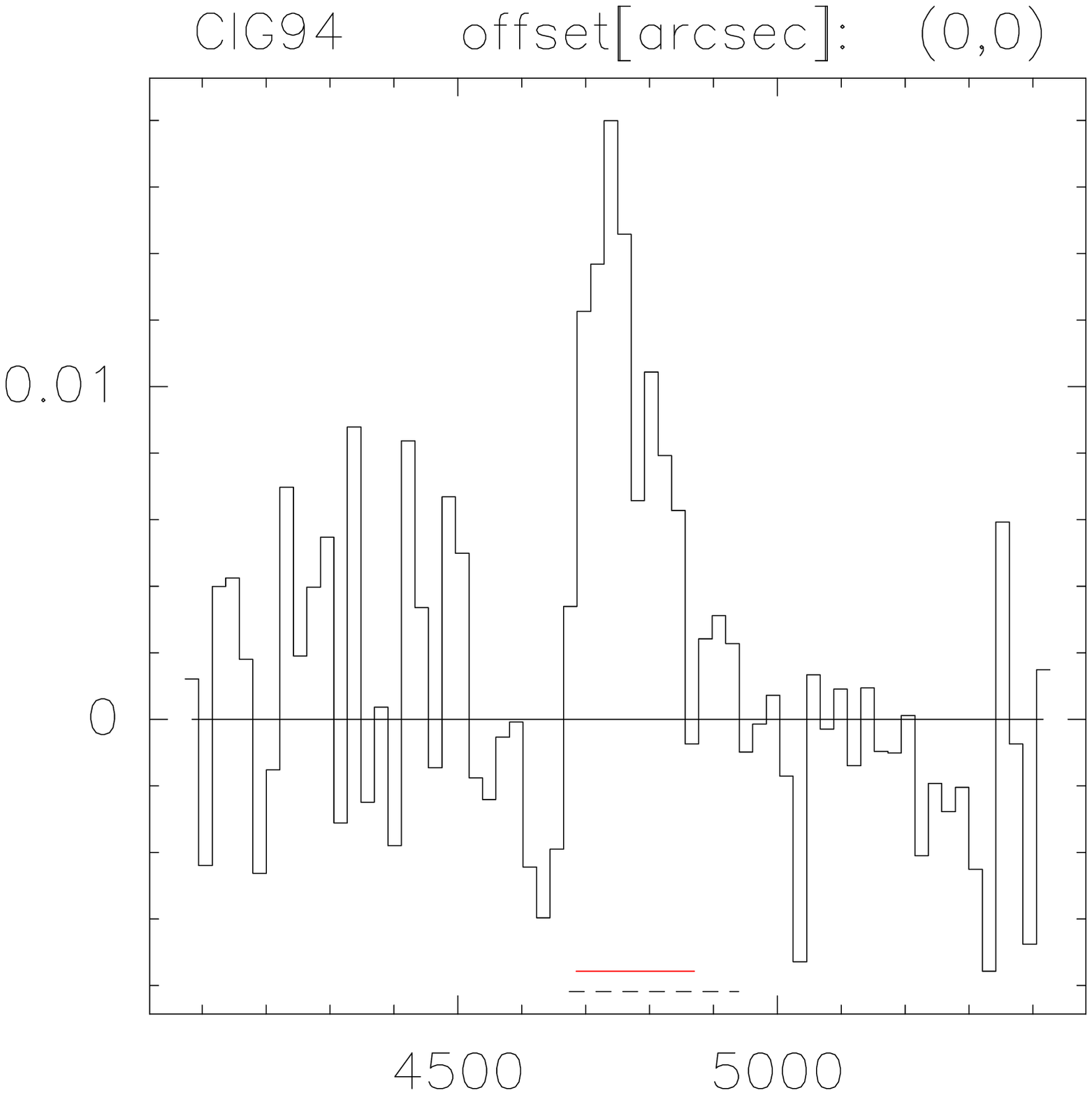}} 
\centerline{\includegraphics[width=3cm,angle=270]{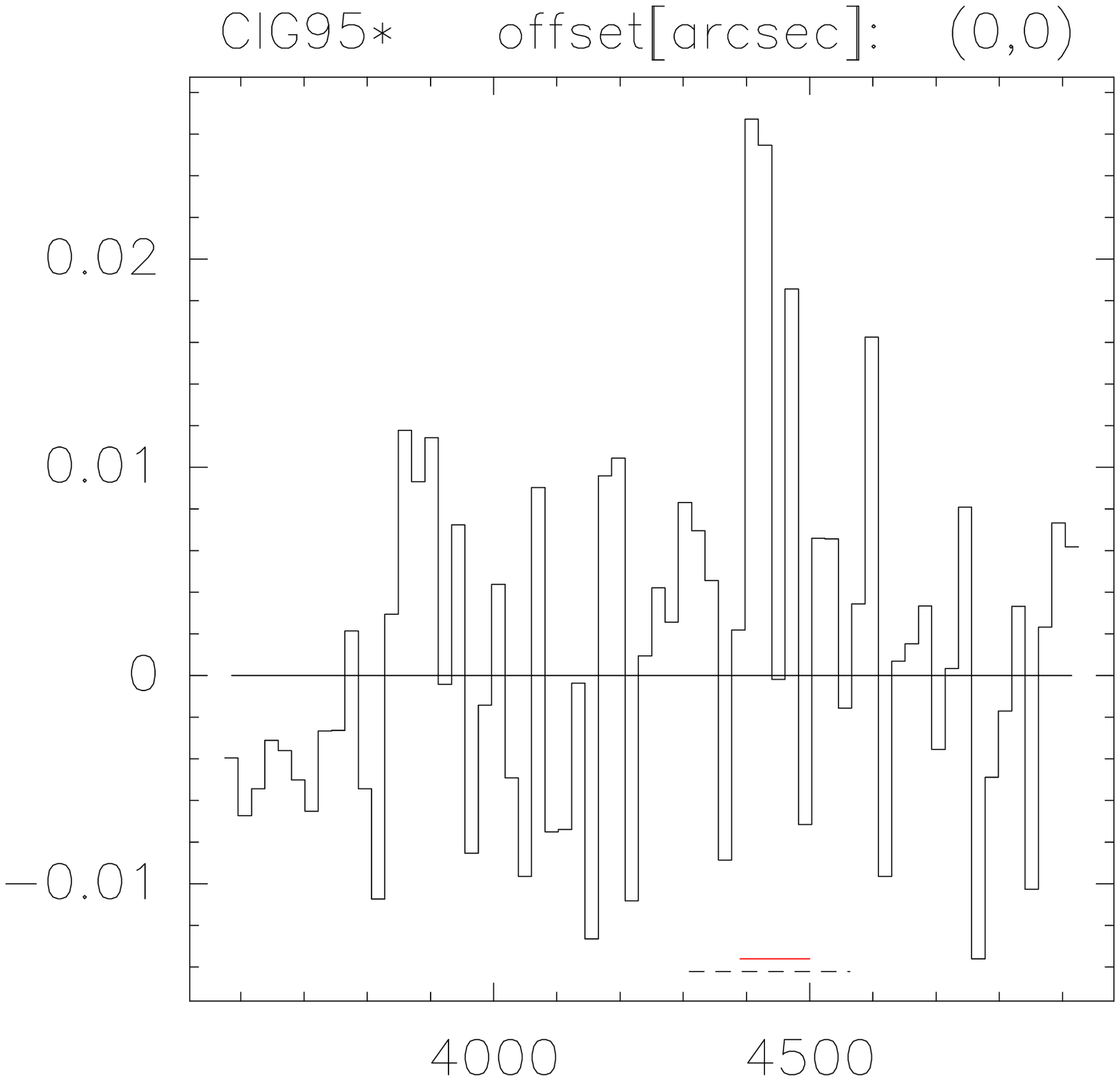} \quad 
\includegraphics[width=3cm,angle=270]{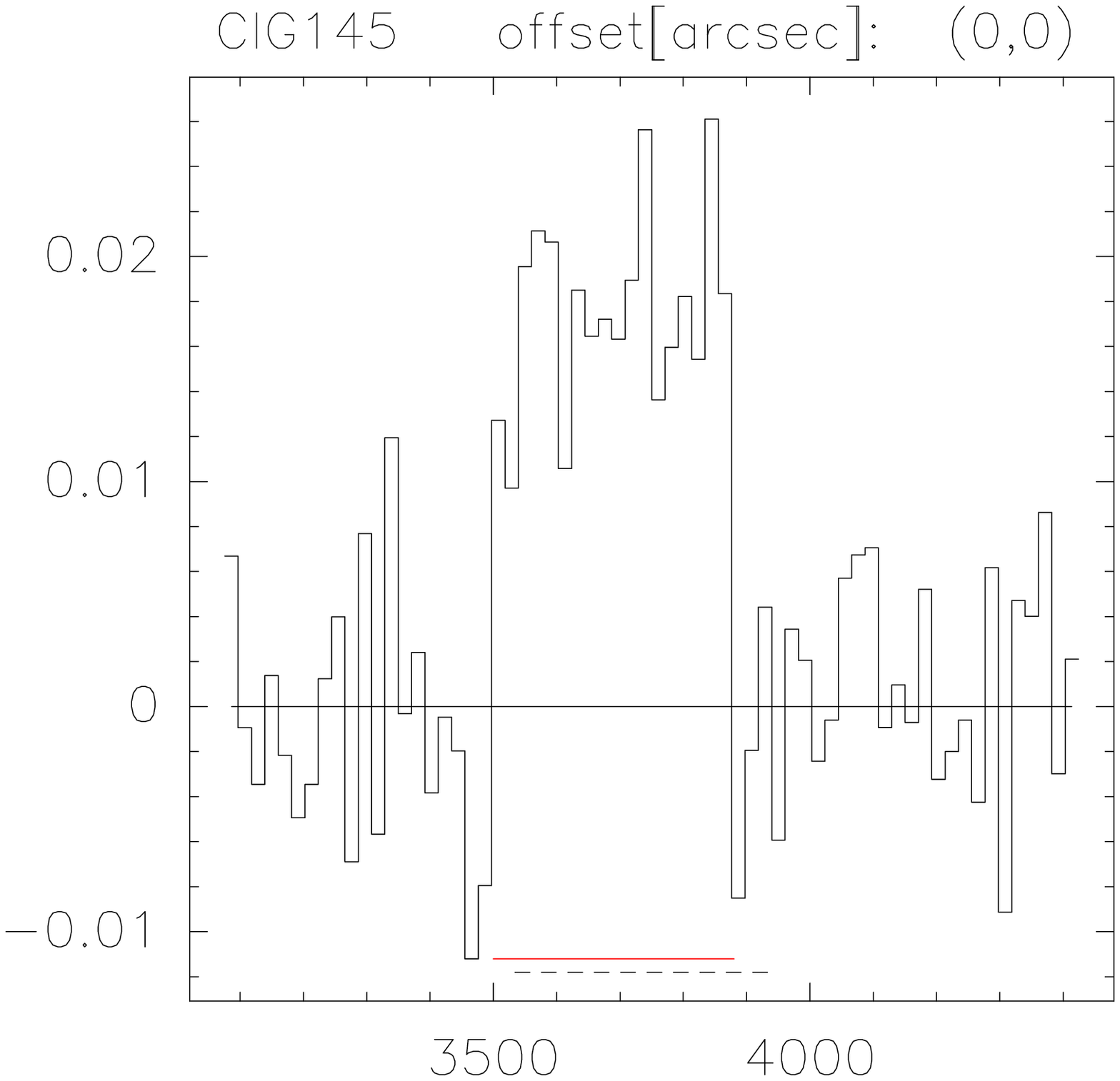}\quad 
\includegraphics[width=3cm,angle=270]{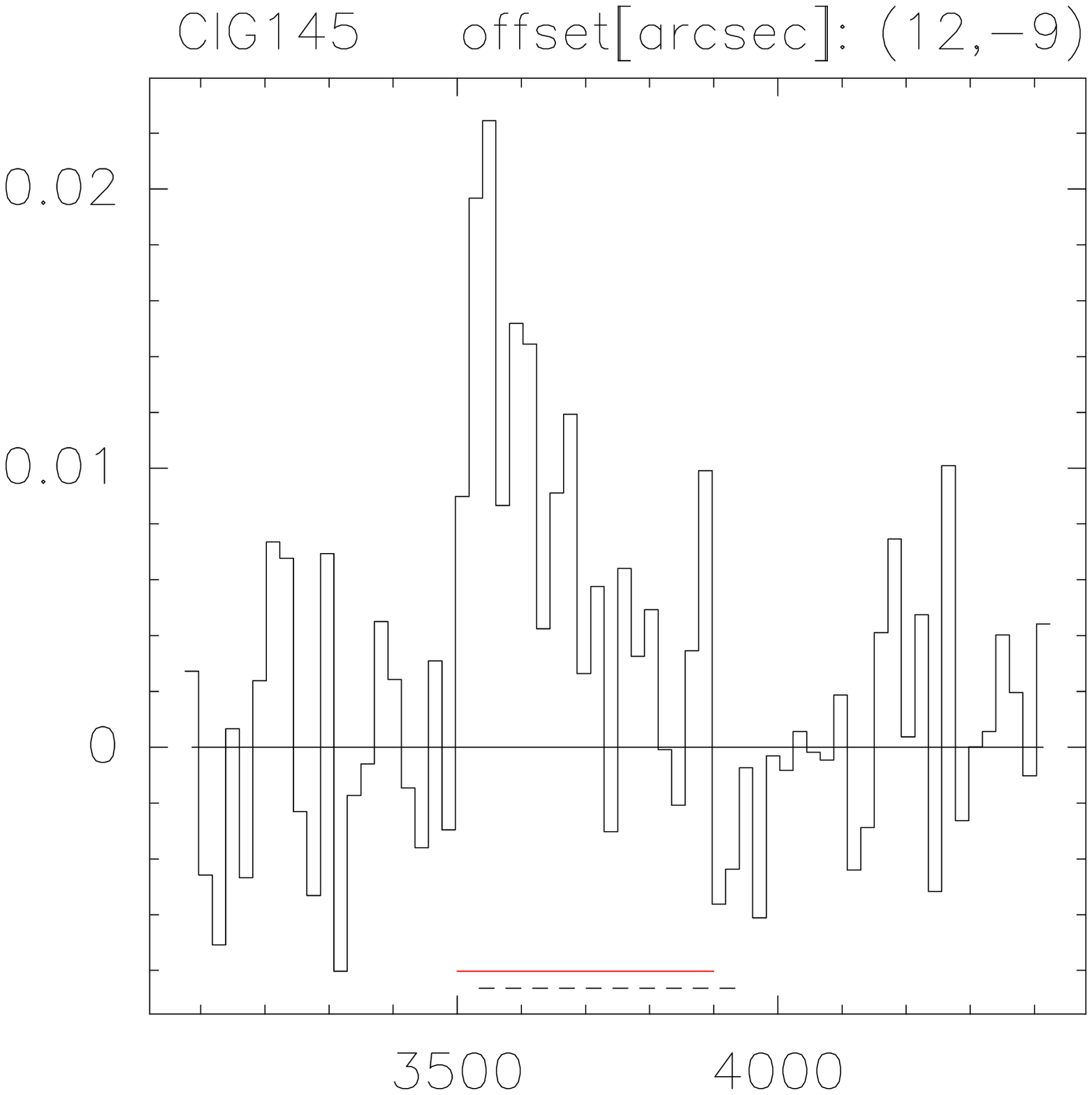}\quad 
\includegraphics[width=3cm,angle=270]{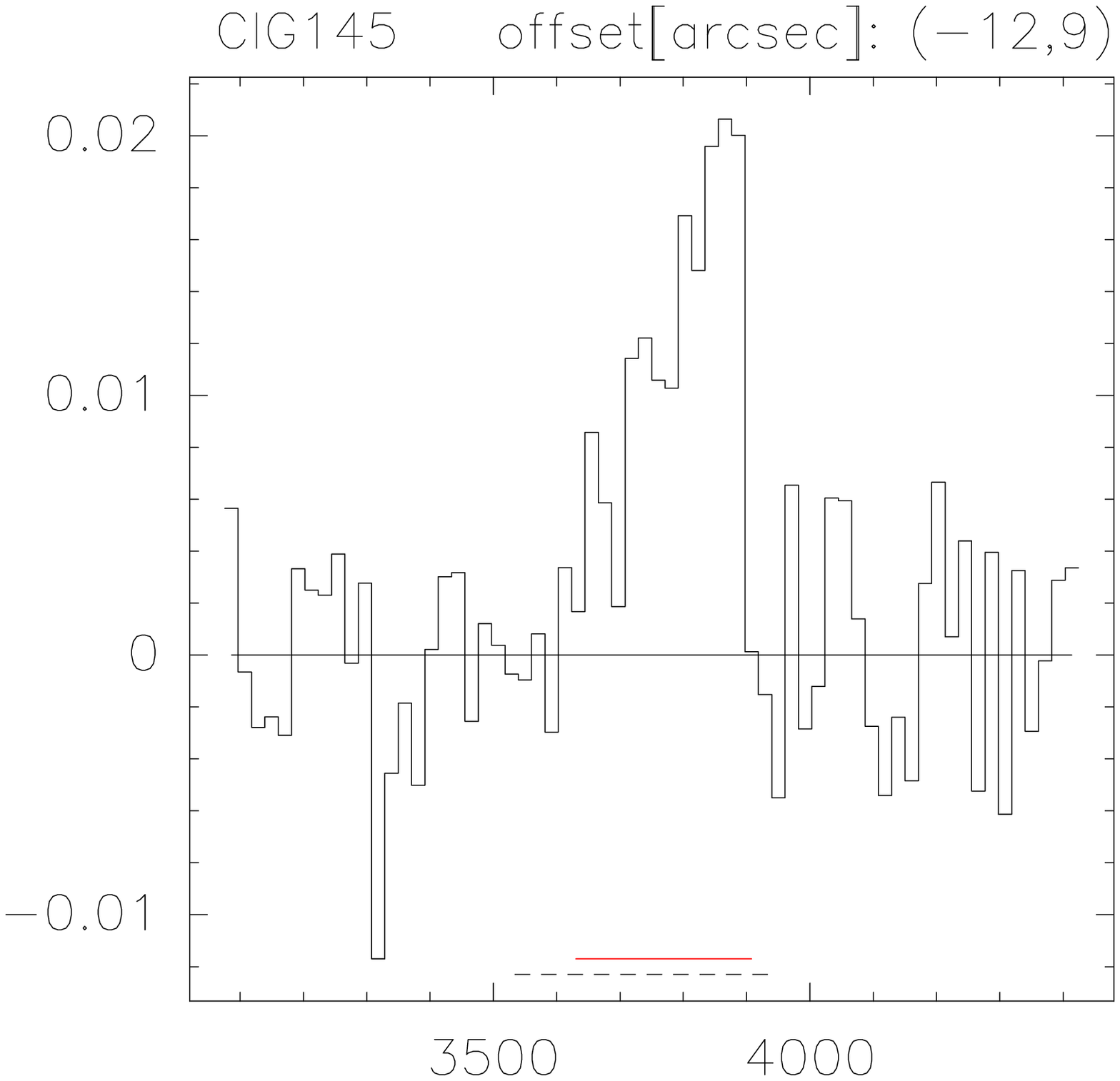}\quad 
\includegraphics[width=3cm,angle=270]{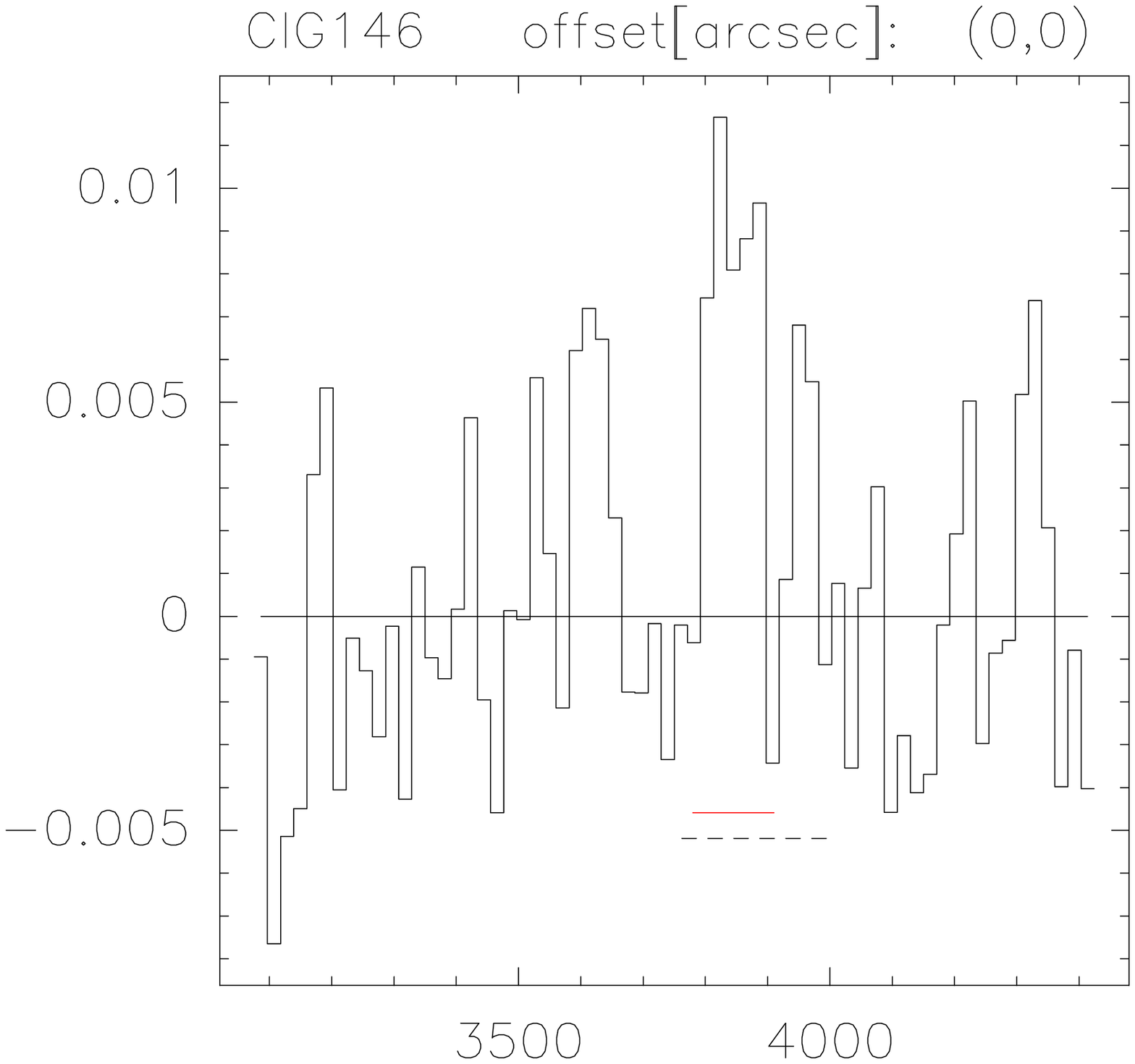}} 
\centerline{\includegraphics[width=3.3cm,angle=270]{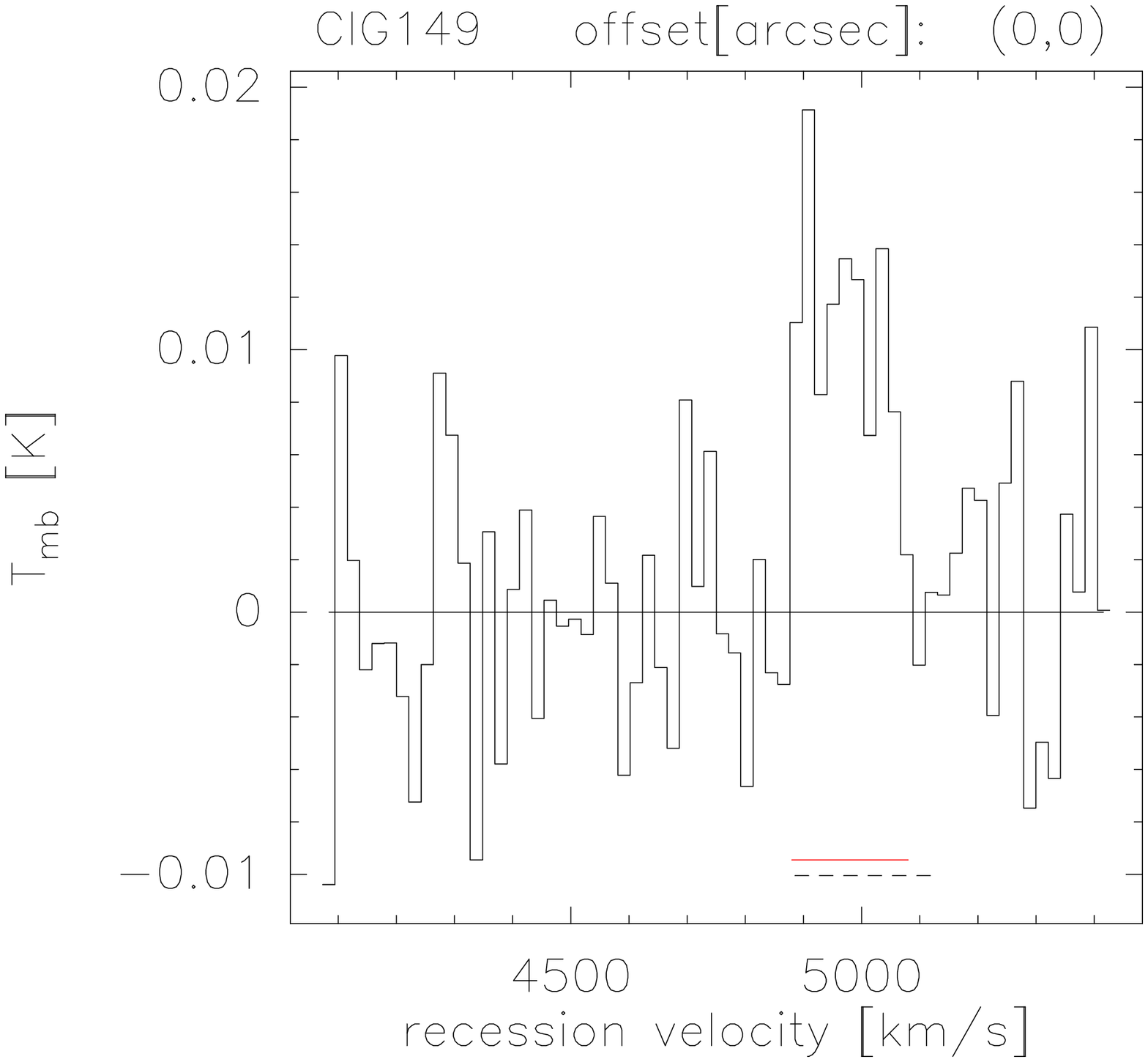} \quad 
\includegraphics[width=3cm,angle=270]{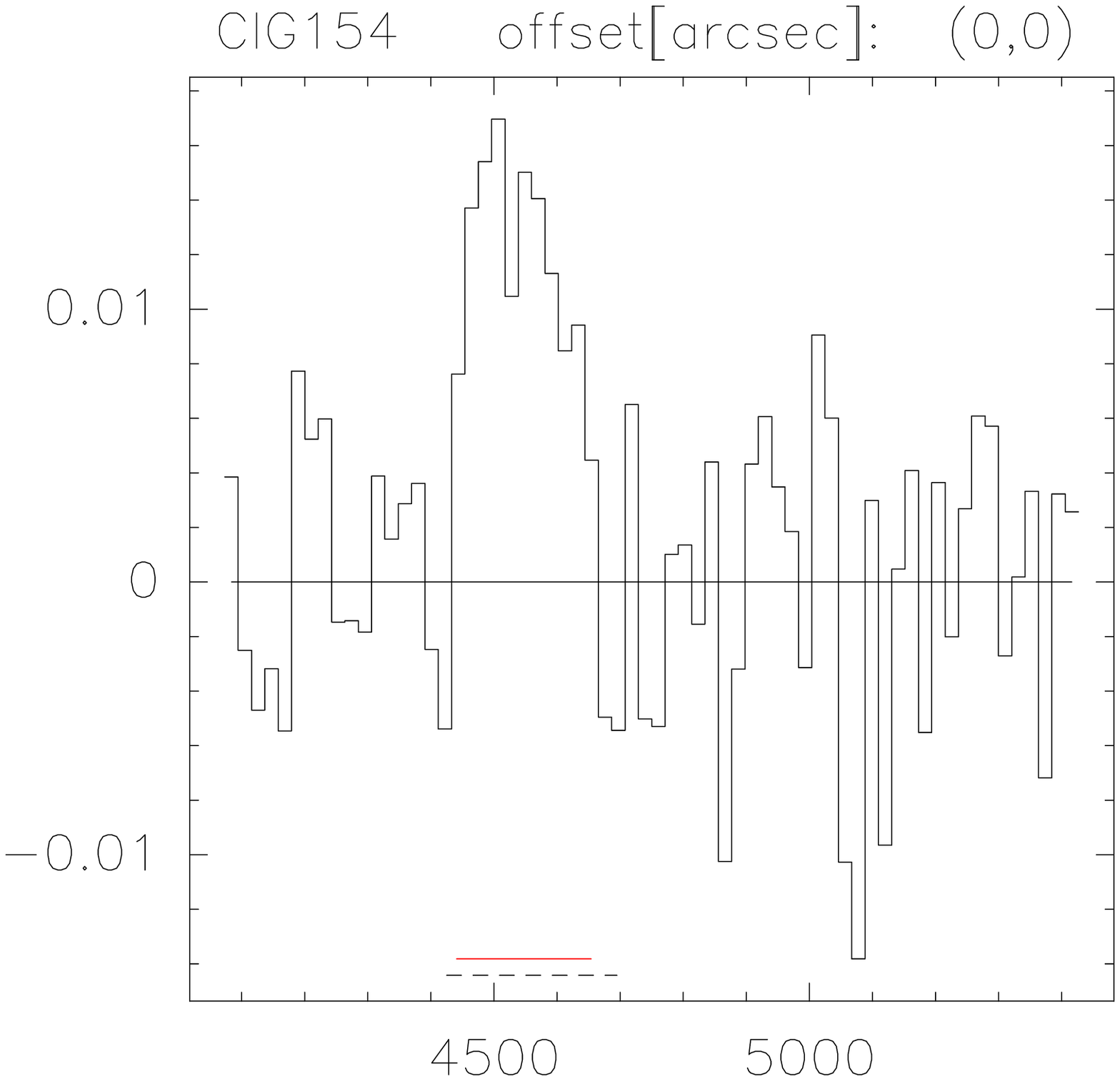}\quad 
\includegraphics[width=3cm,angle=270]{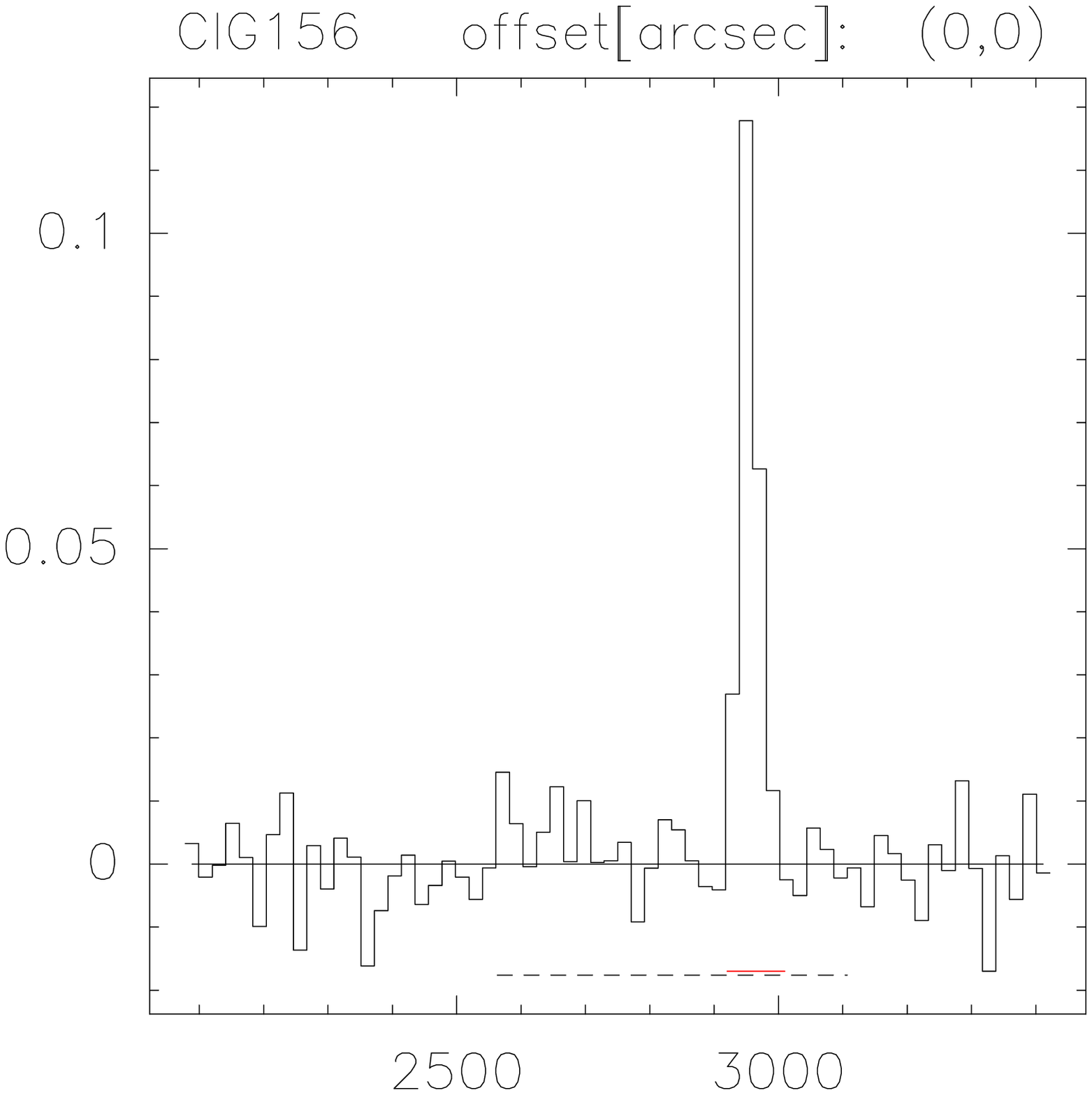}\quad 
\includegraphics[width=3cm,angle=270]{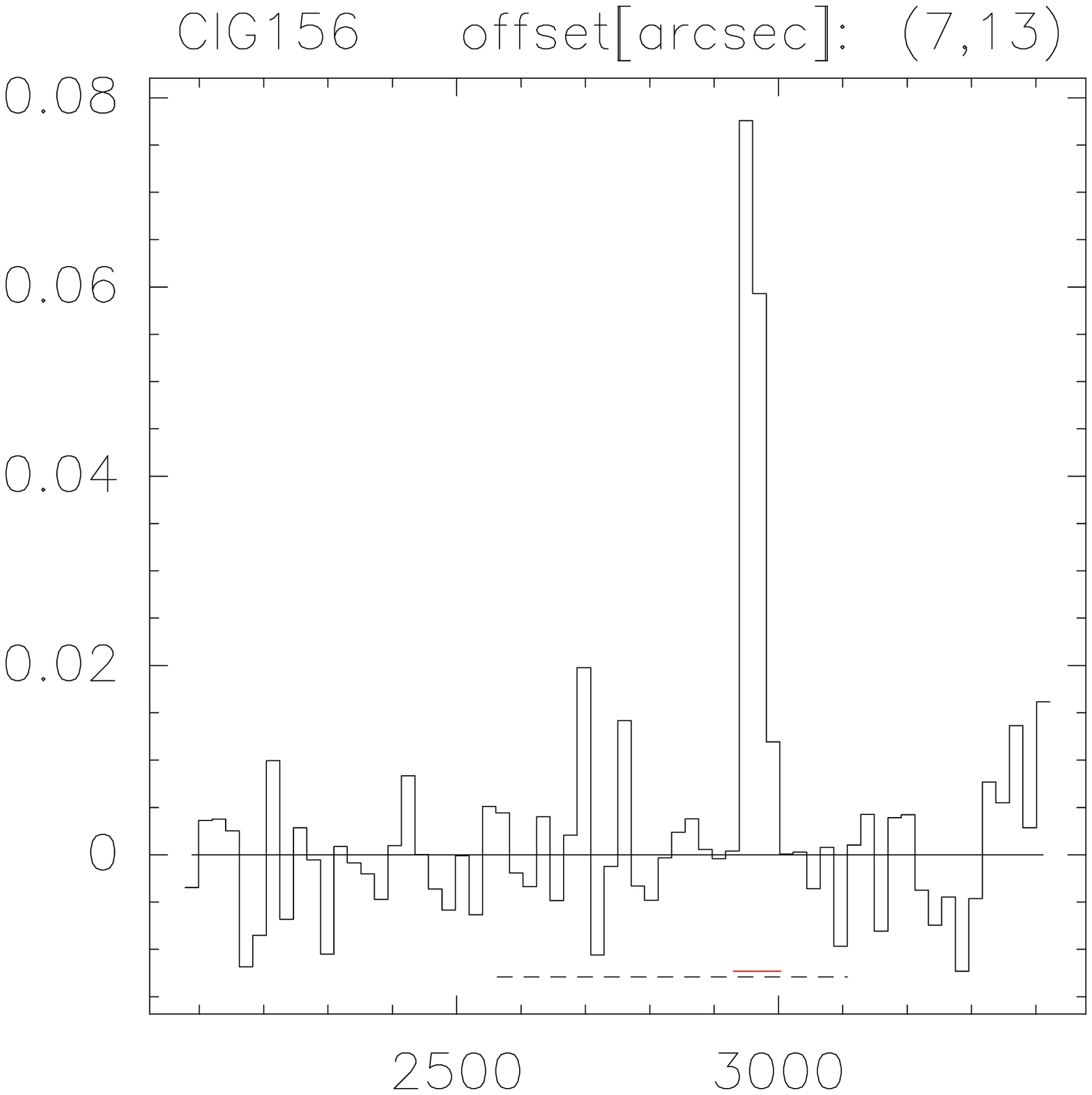}\quad 
\includegraphics[width=3cm,angle=270]{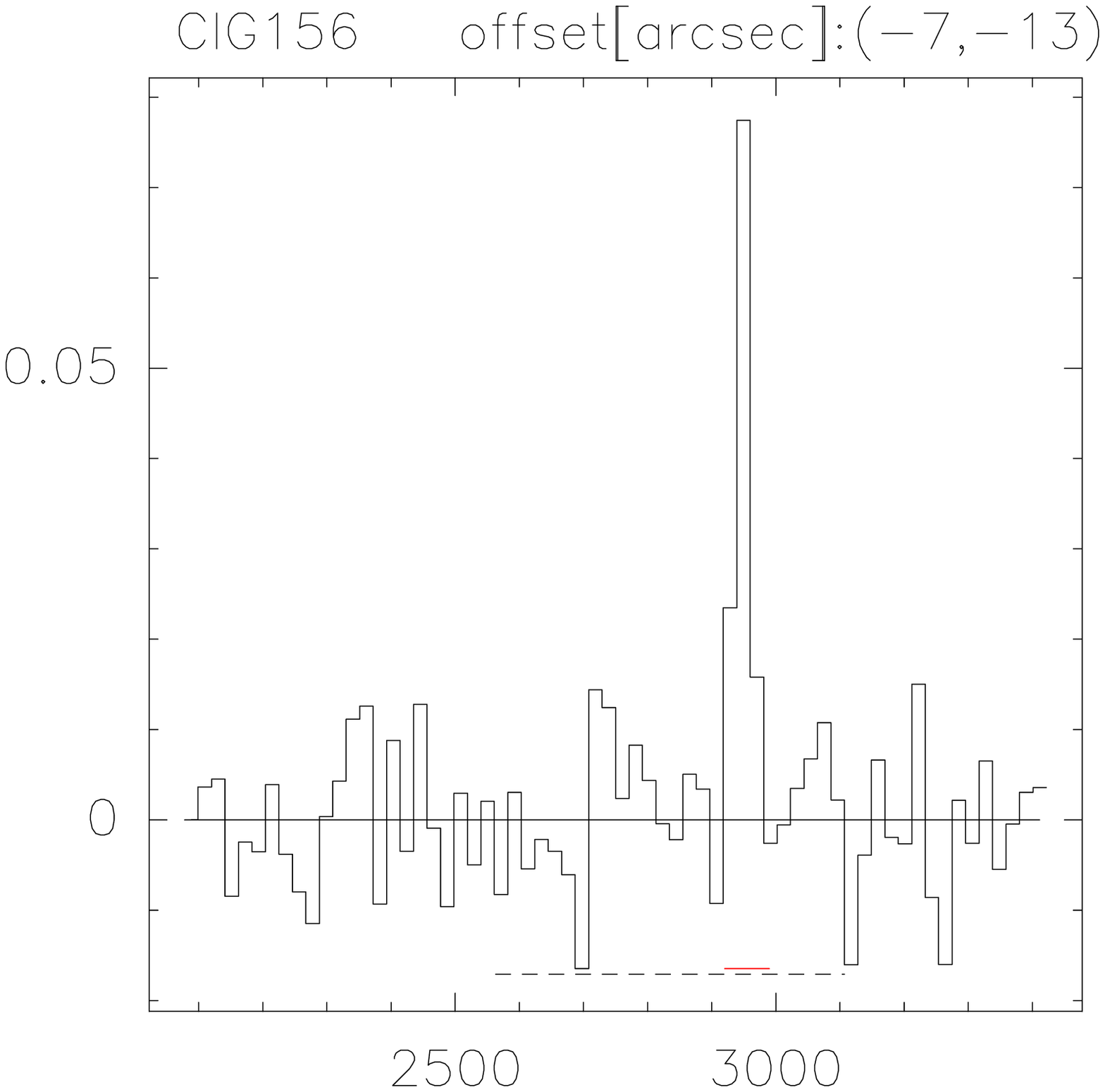}} 
\caption{CO(1-0) spectra for the  galaxies detected with the IRAM 30m telescope. The x-axis represents
 the recession velocity in \kms\ and the y-axis the main beam brightness temperature  $T_{\rm mb}$  in K. The spectral resolution is
20.8 \kms\ in most cases except for some individual galaxies for which
a higher or lower resolution was required to clearly show the line.
The full (red) line segment shows the line width of the CO line adopted for the 
determination of the velocity  integrated intensity. The dashed (black) line segment
is the HI  line width at 20\% peak level, W$_{20}$.
An asterisk next to the name indicates a marginal detection.} 
\label{spectra_iram} 
\end{figure*} 

%\clearpage
  
\begin{figure*} 
\centerline{\includegraphics[width=3cm,angle=270]{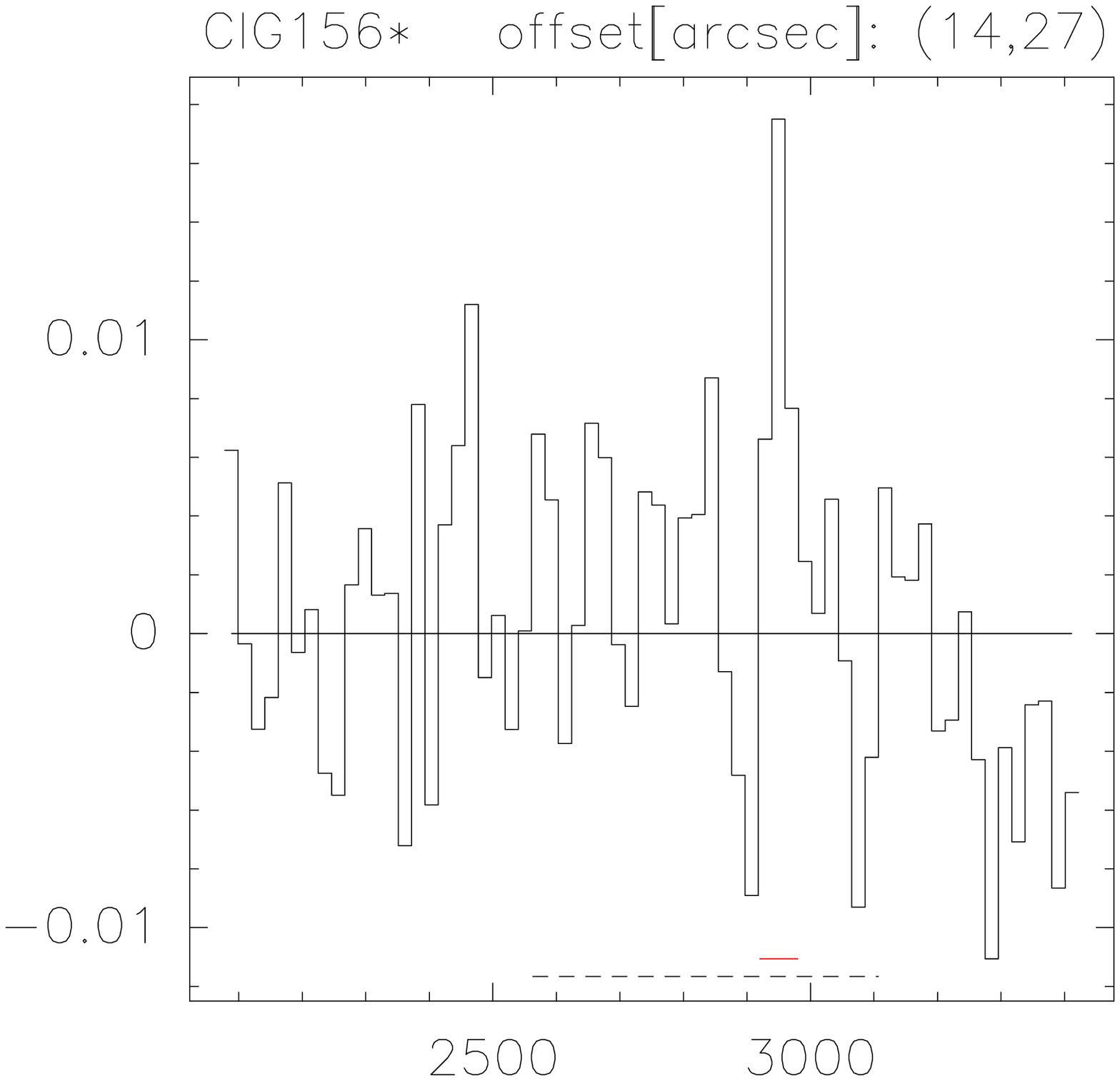} \quad 
\includegraphics[width=3cm,angle=270]{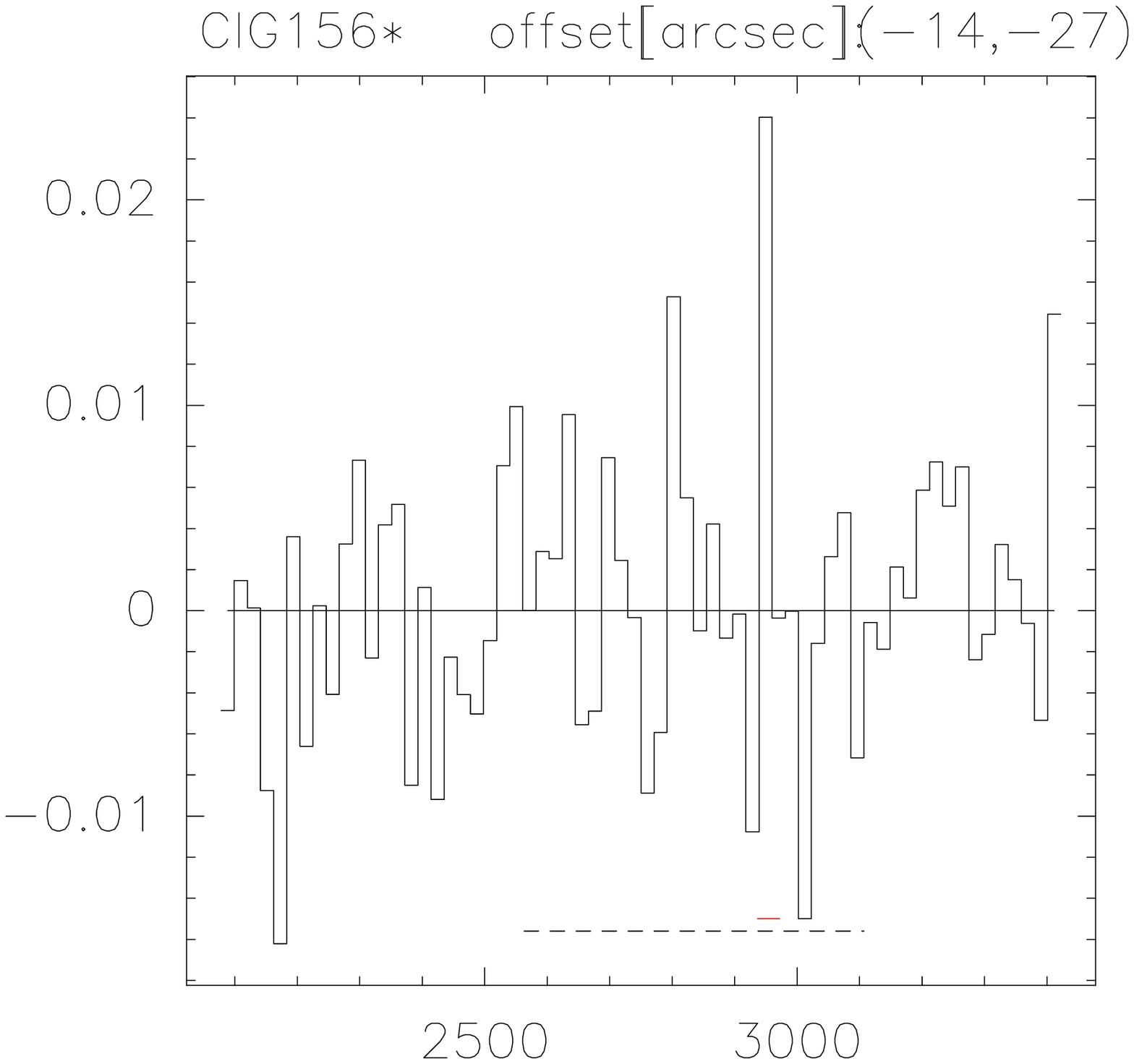}\quad 
\includegraphics[width=3cm,angle=270]{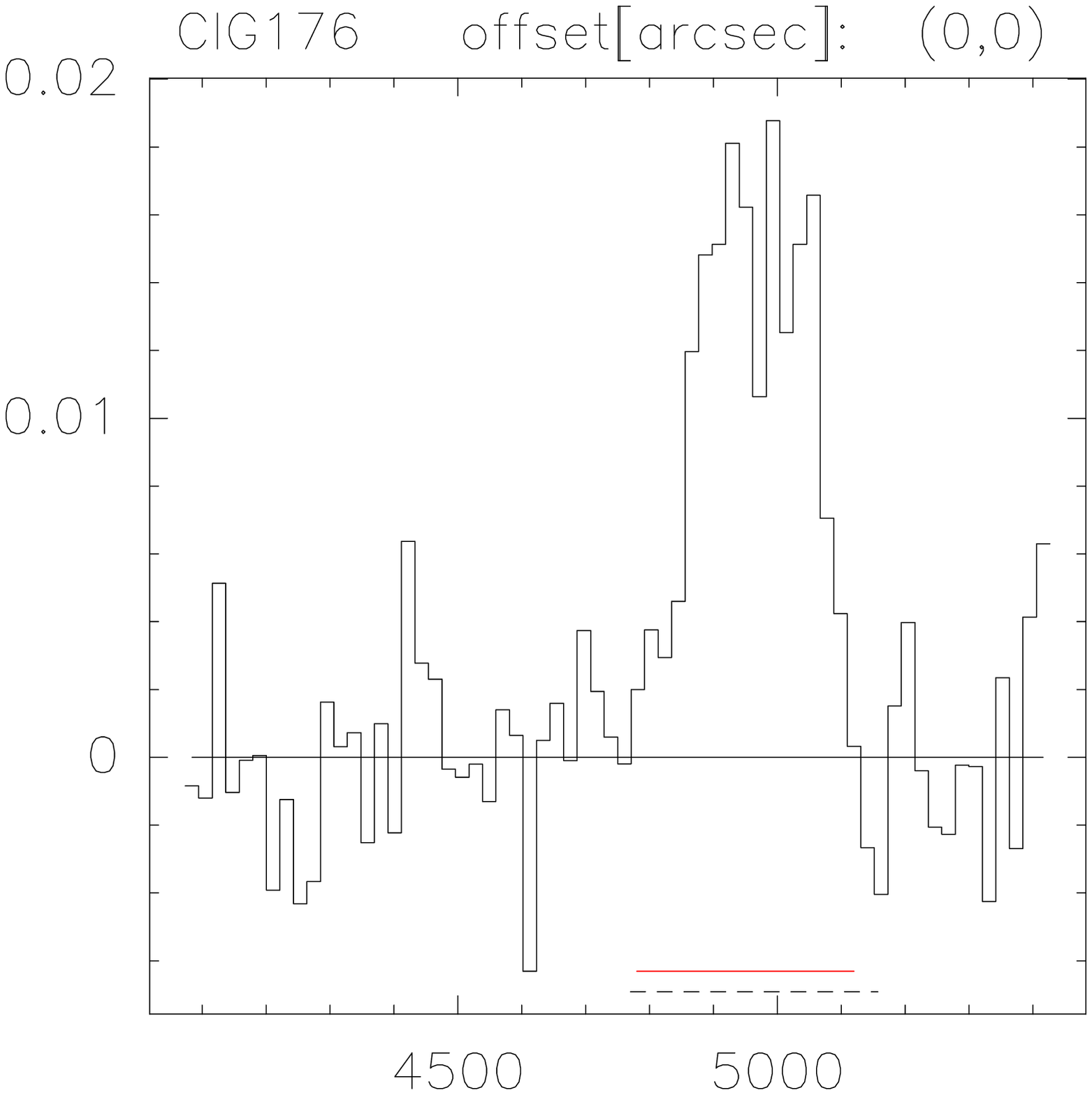}\quad 
\includegraphics[width=3cm,angle=270]{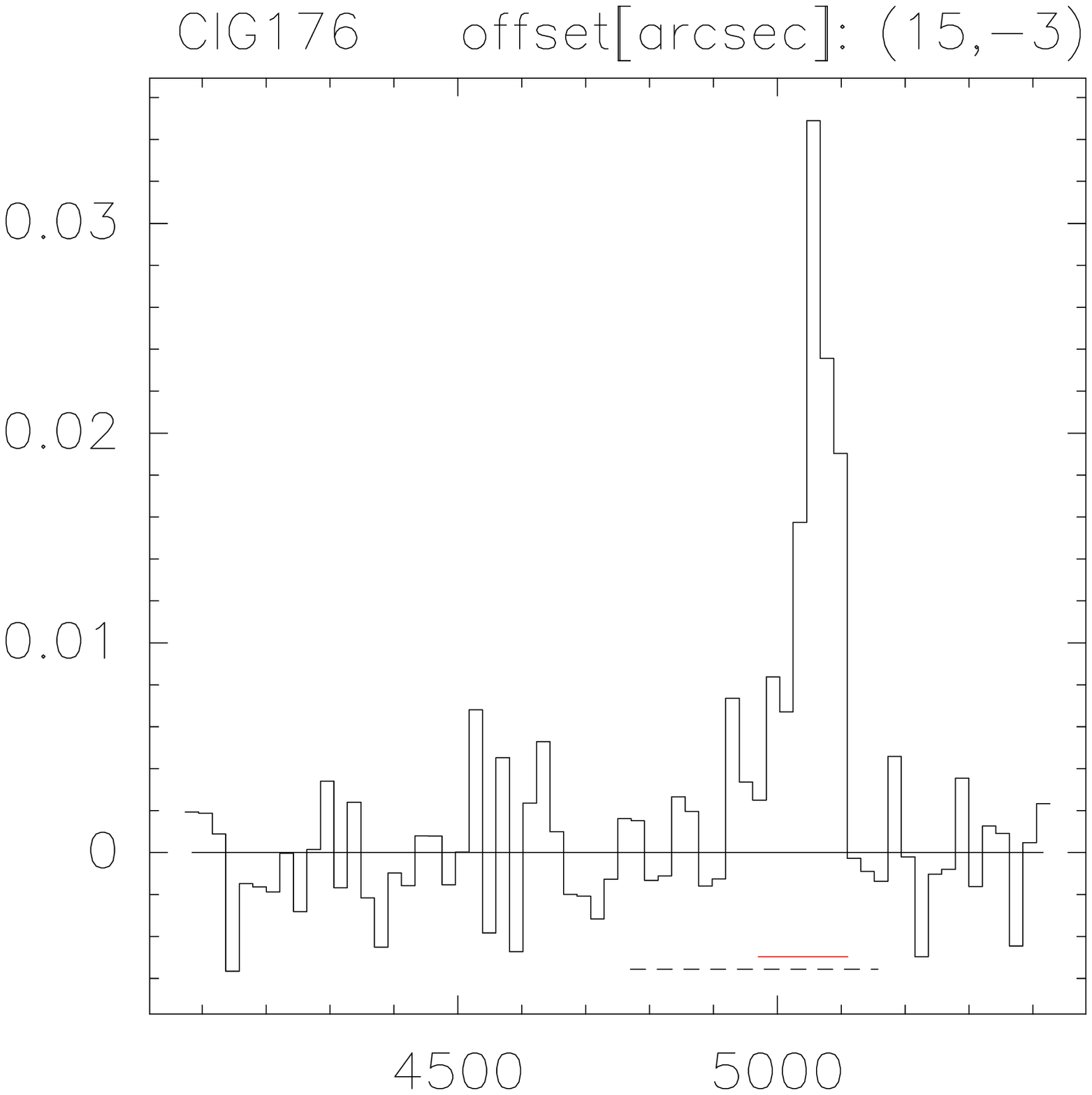}\quad 
\includegraphics[width=3cm,angle=270]{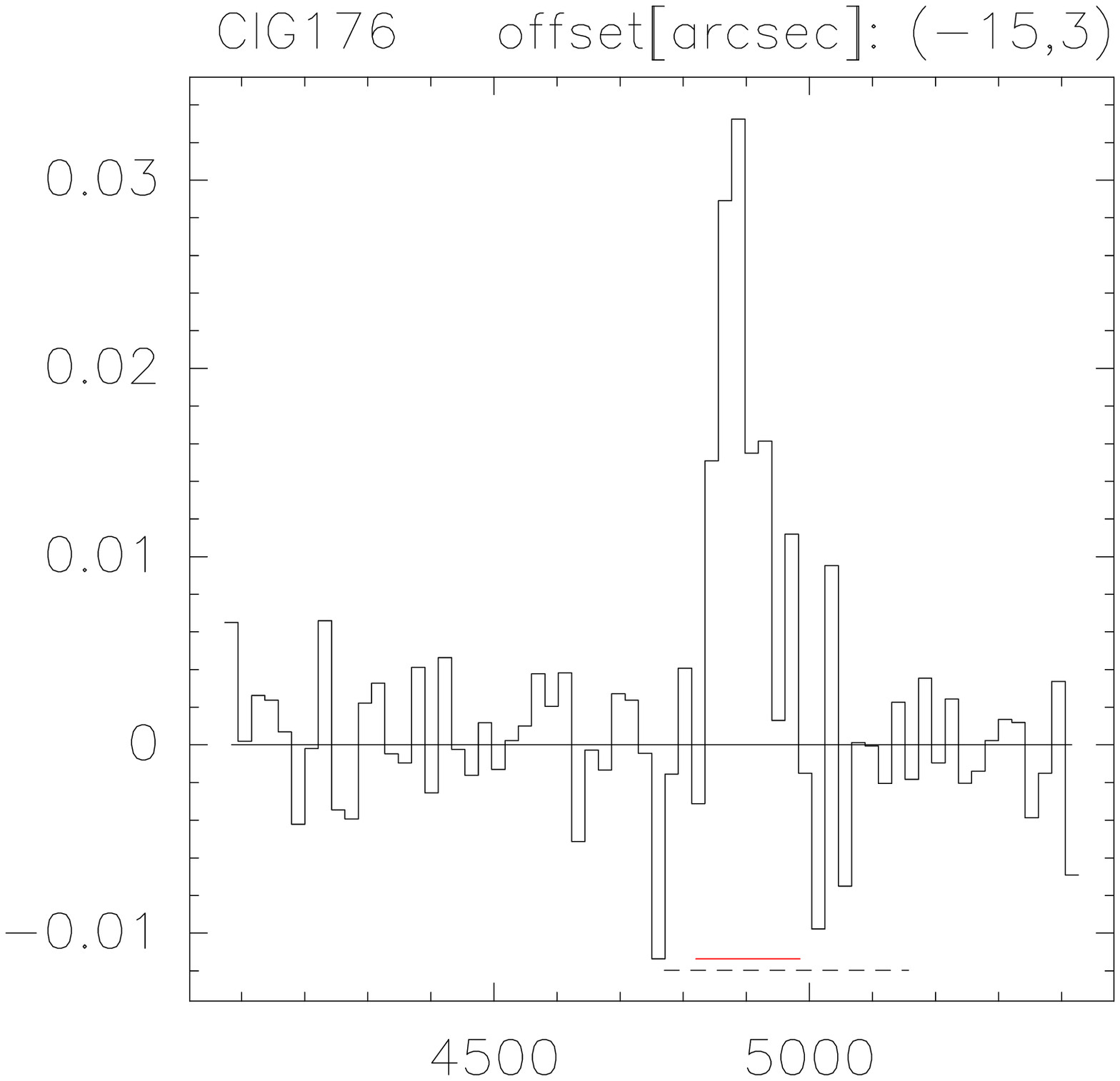}} 
\centerline{\includegraphics[width=3cm,angle=270]{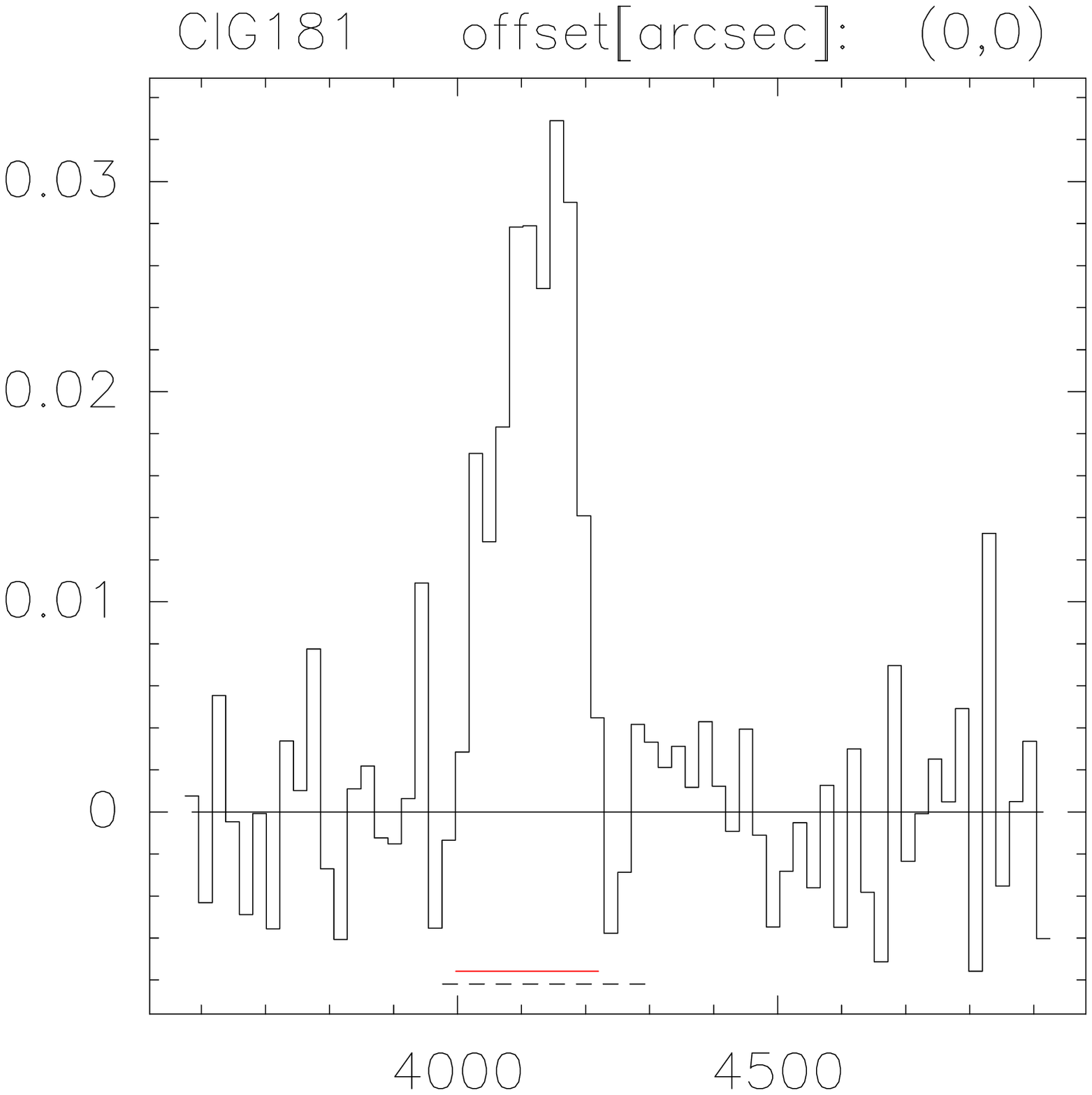} \quad 
\includegraphics[width=3cm,angle=270]{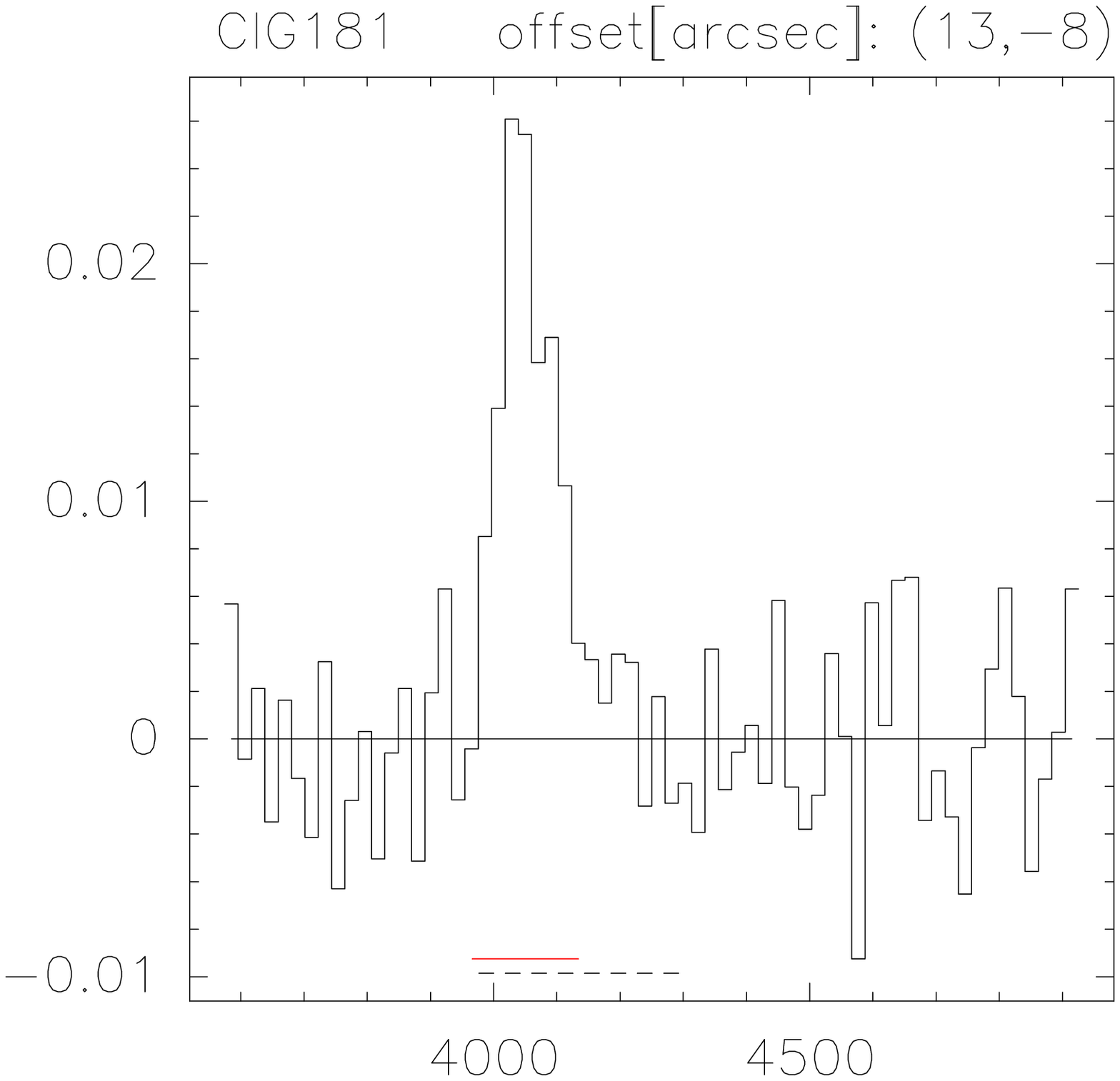}\quad 
\includegraphics[width=3cm,angle=270]{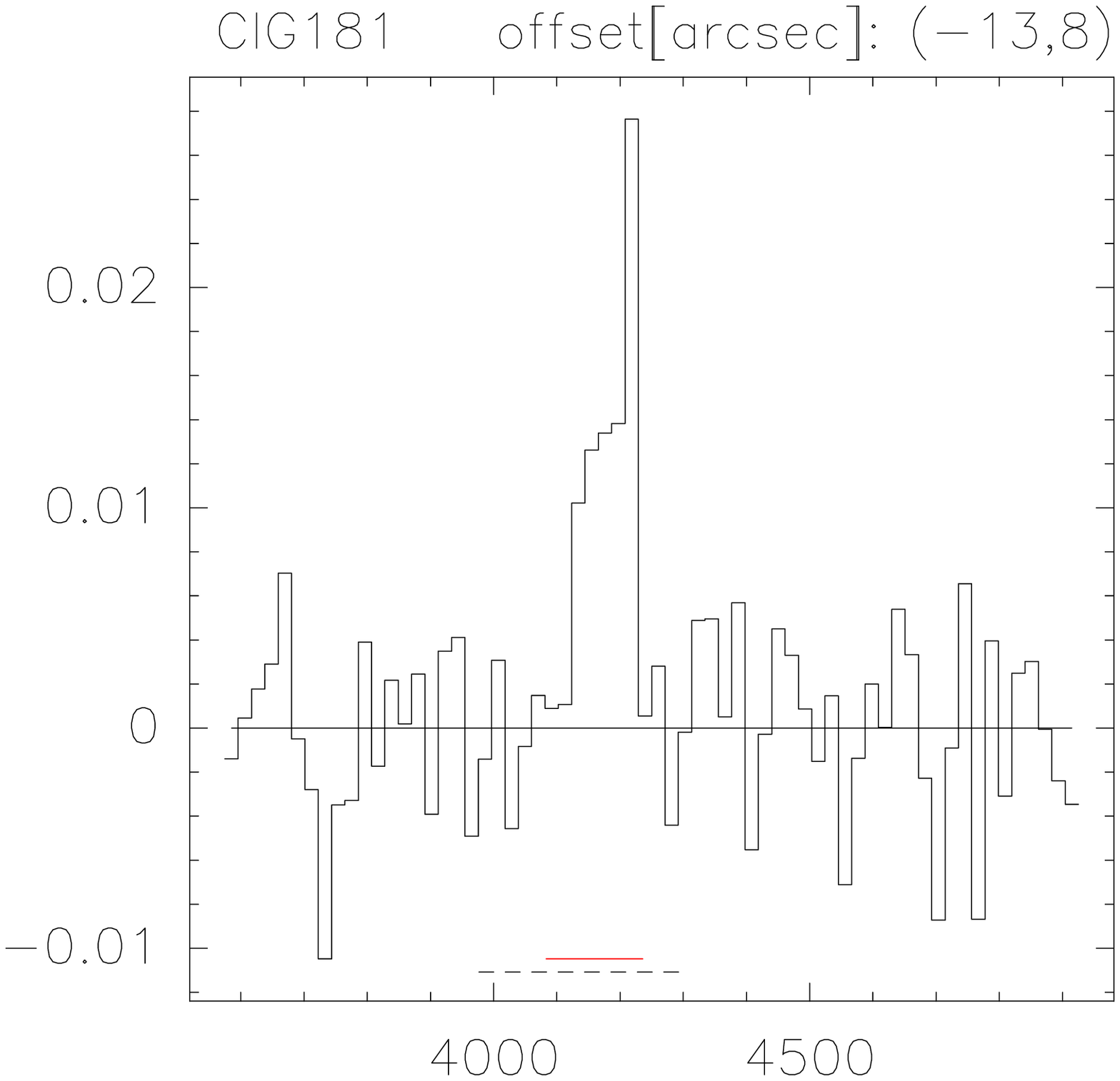}\quad 
\includegraphics[width=3cm,angle=270]{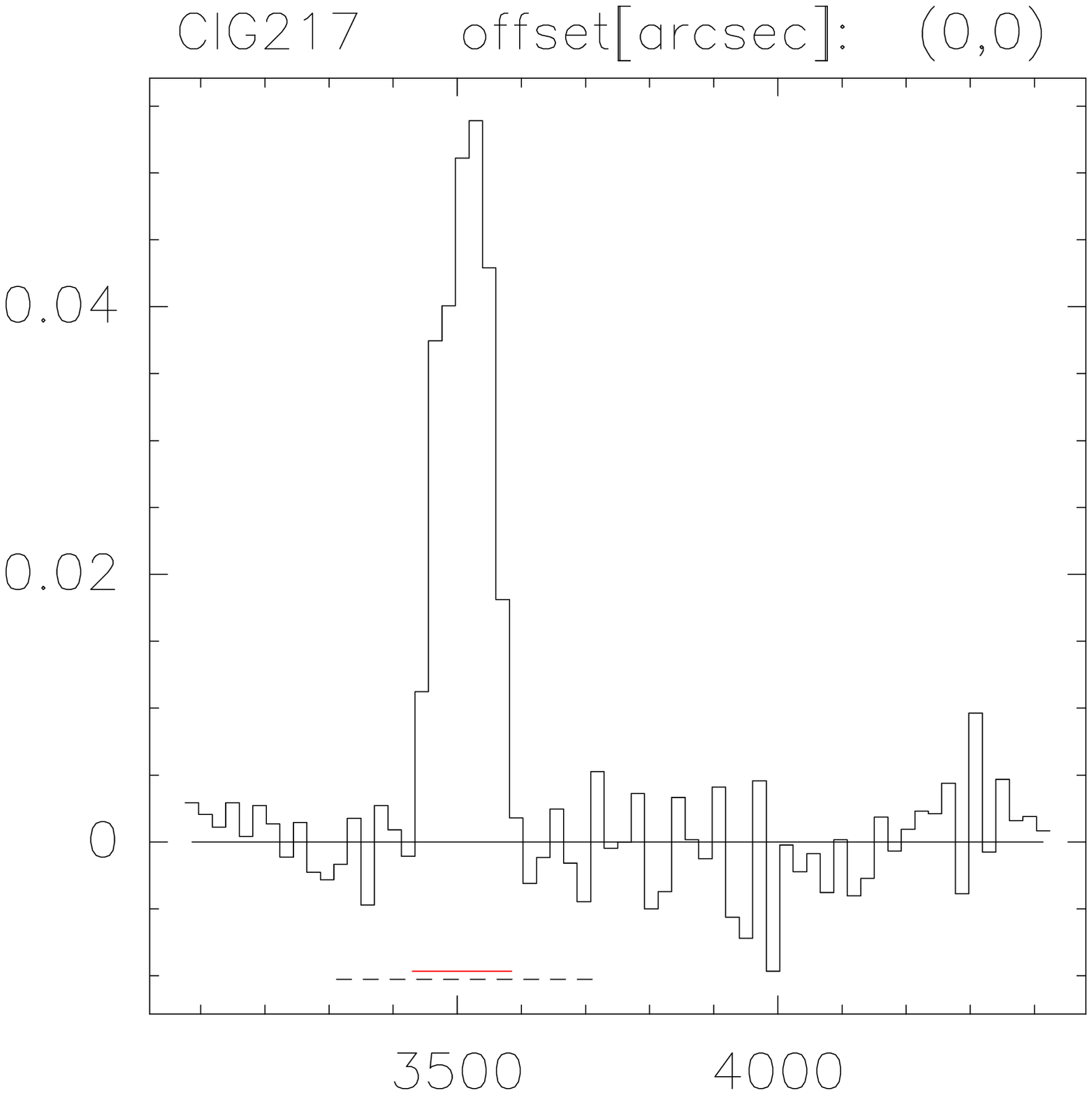}\quad 
\includegraphics[width=3cm,angle=270]{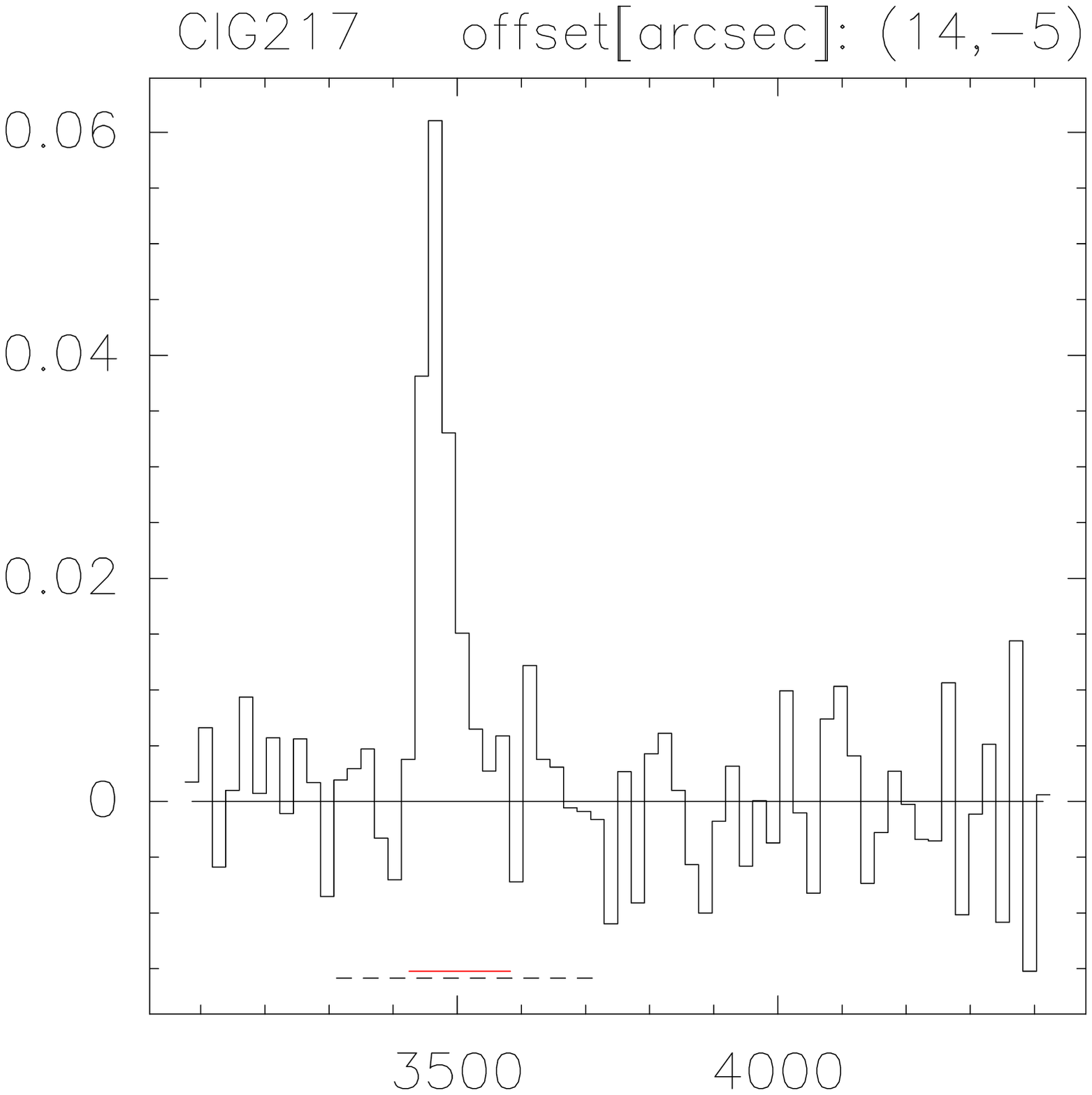}} 
\centerline{\includegraphics[width=3cm,angle=270]{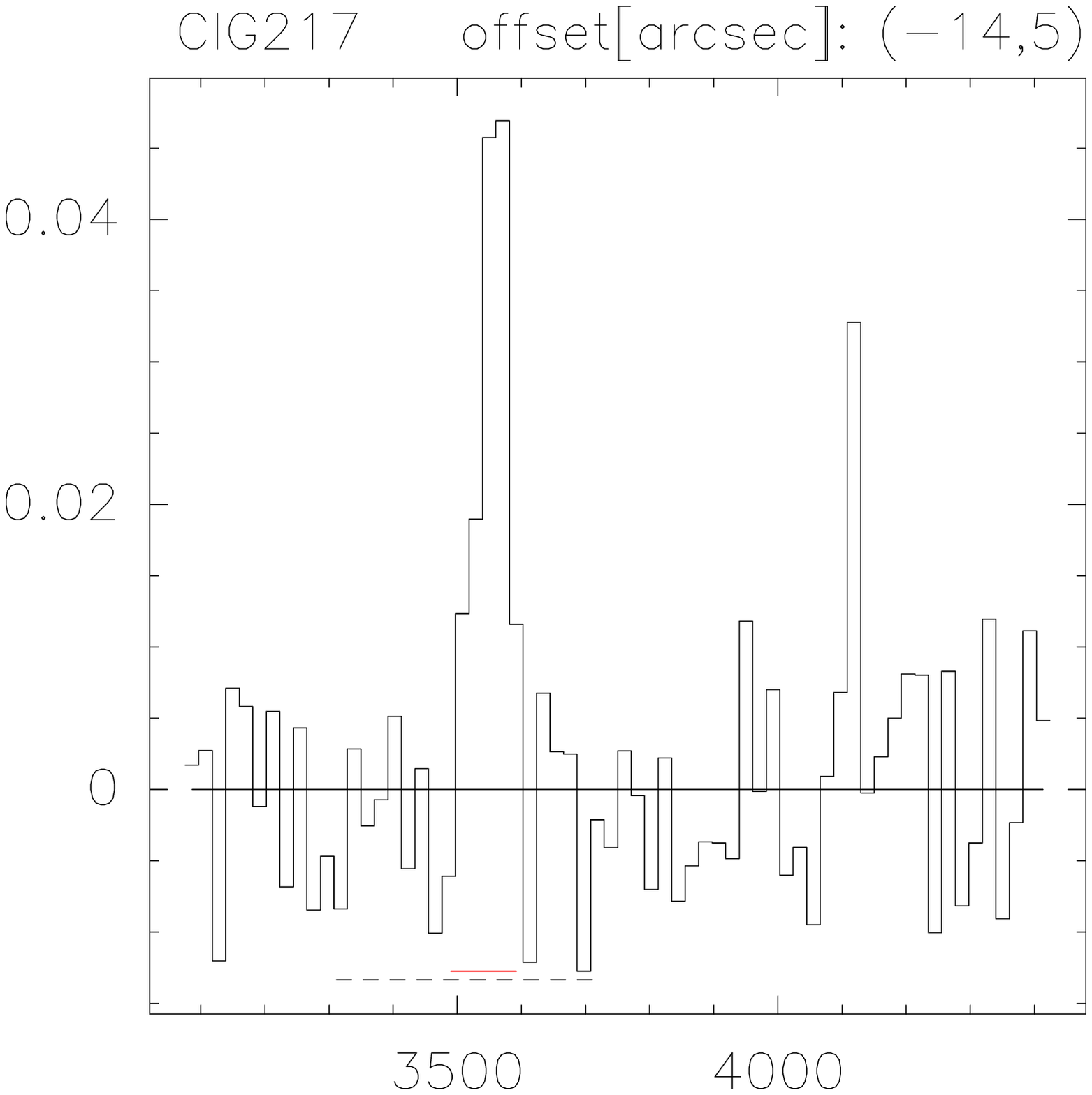} \quad 
\includegraphics[width=3cm,angle=270]{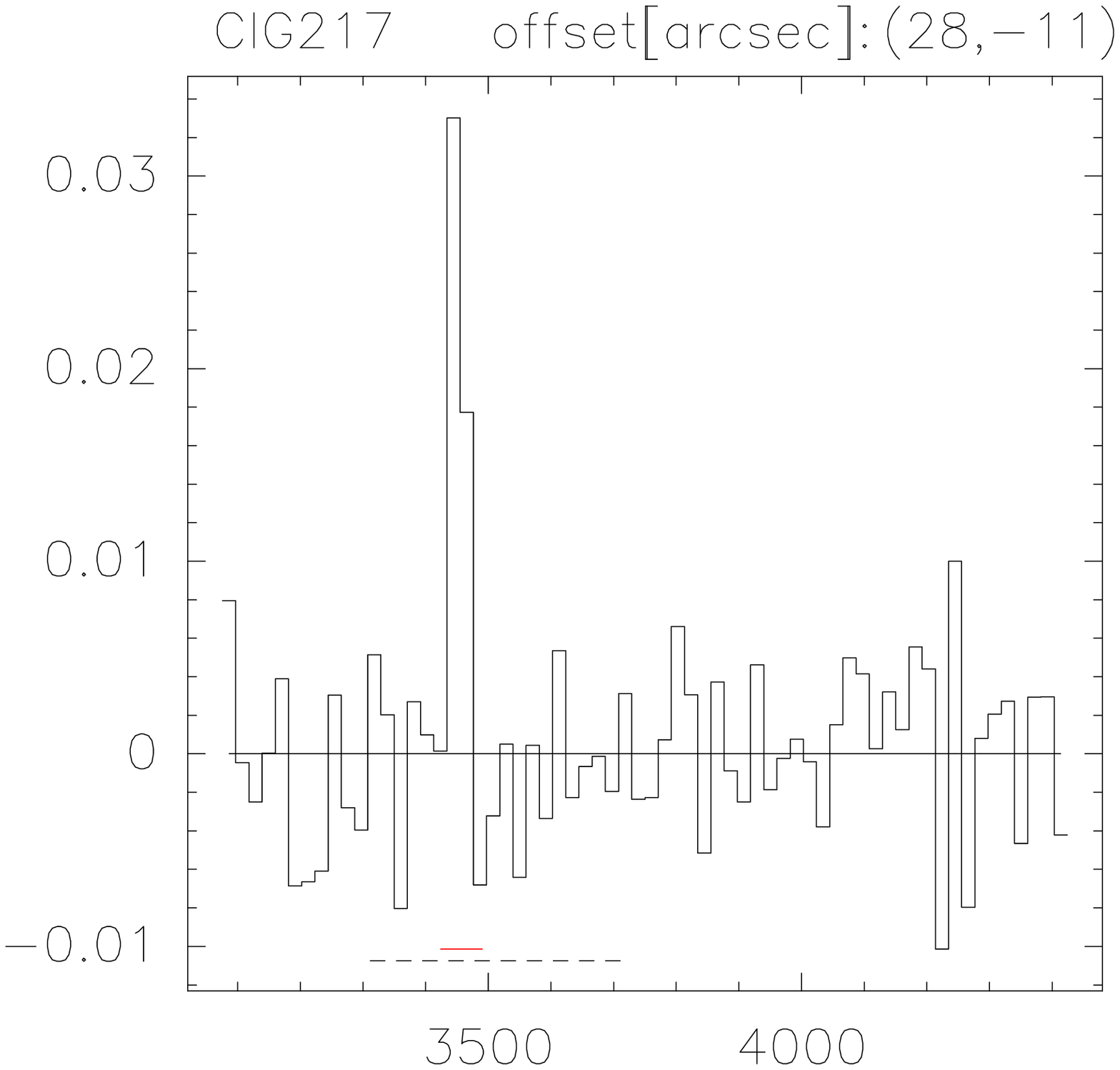}\quad 
\includegraphics[width=3cm,angle=270]{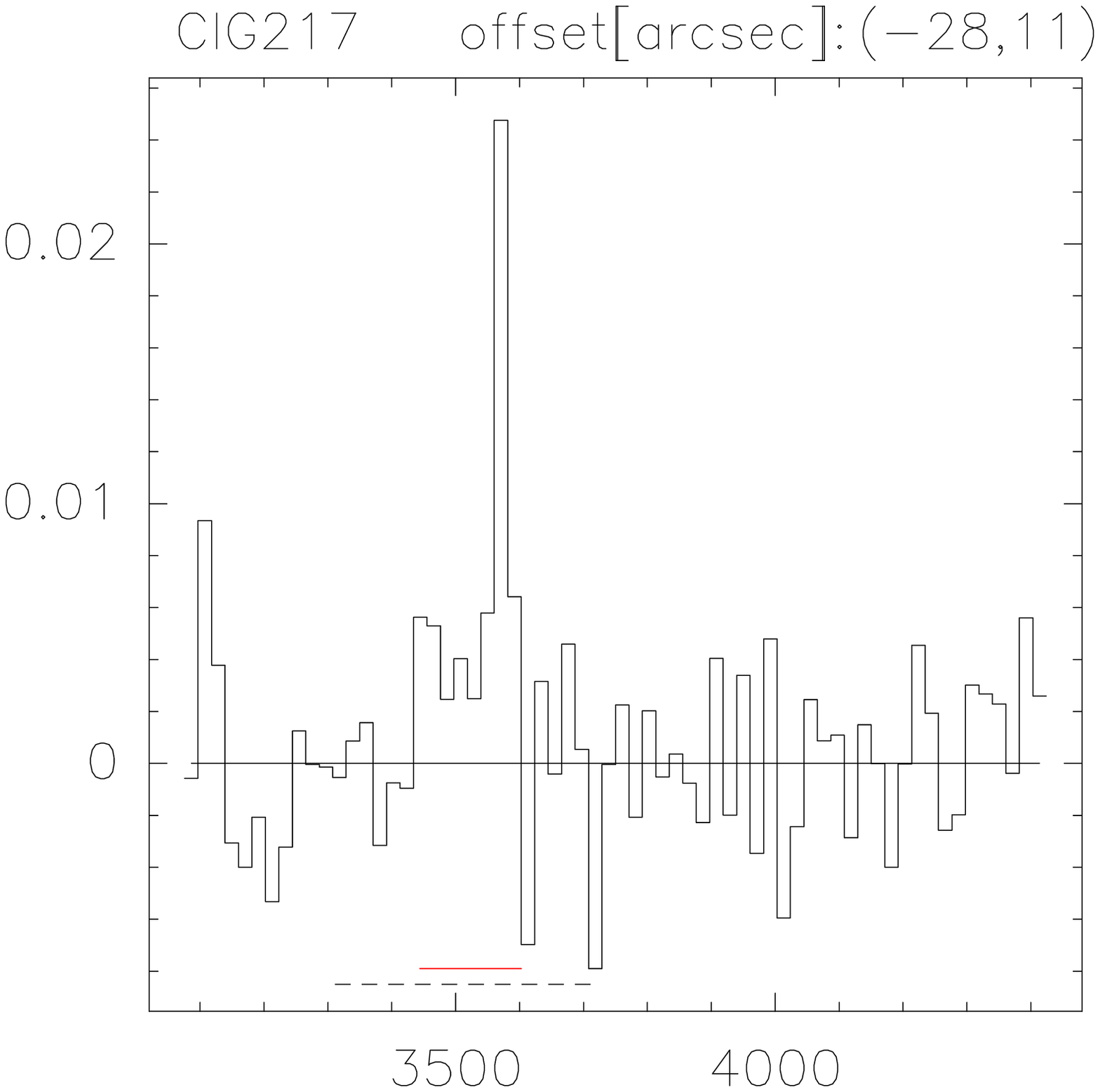}\quad 
\includegraphics[width=3cm,angle=270]{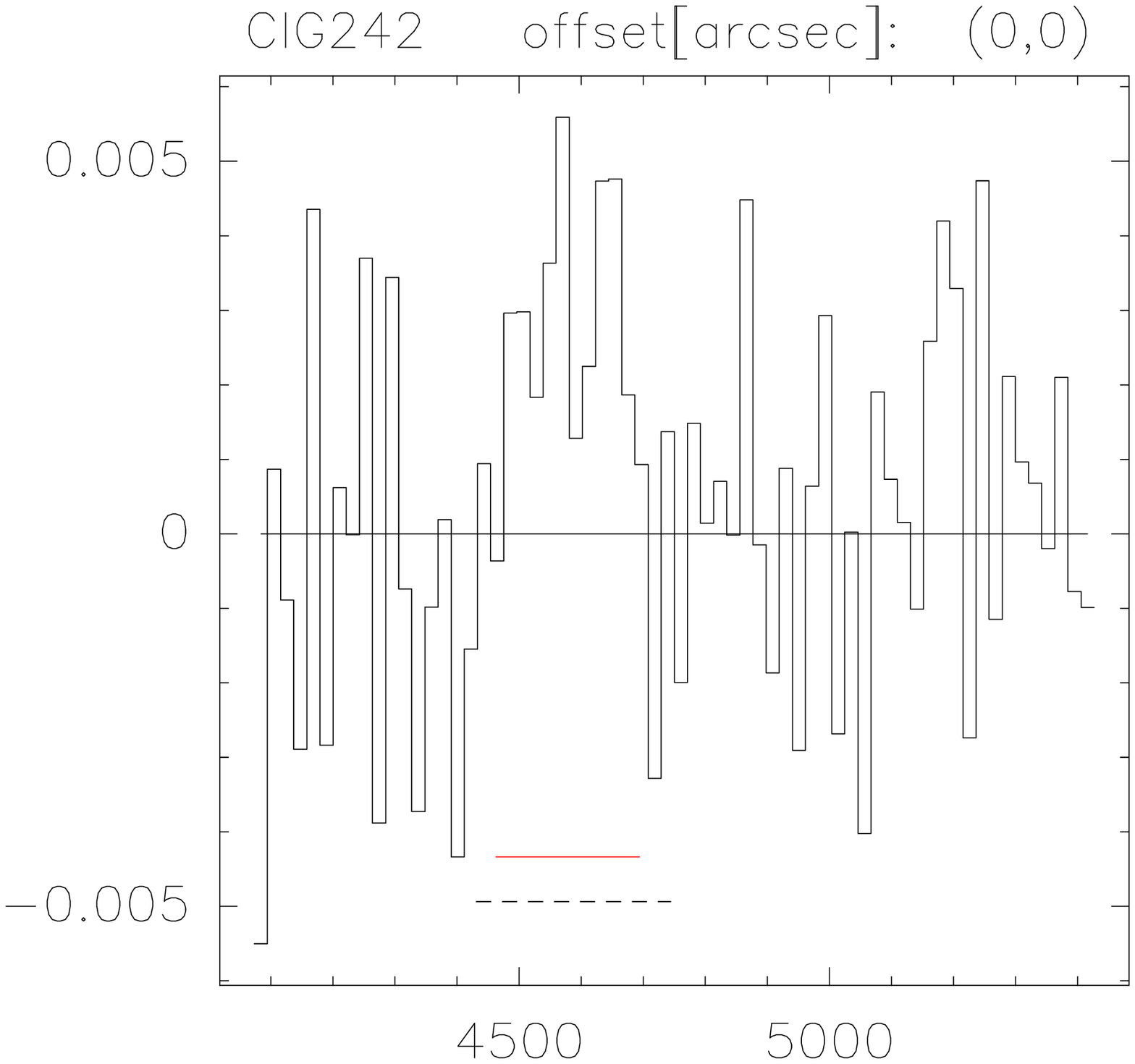}\quad 
\includegraphics[width=3cm,angle=270]{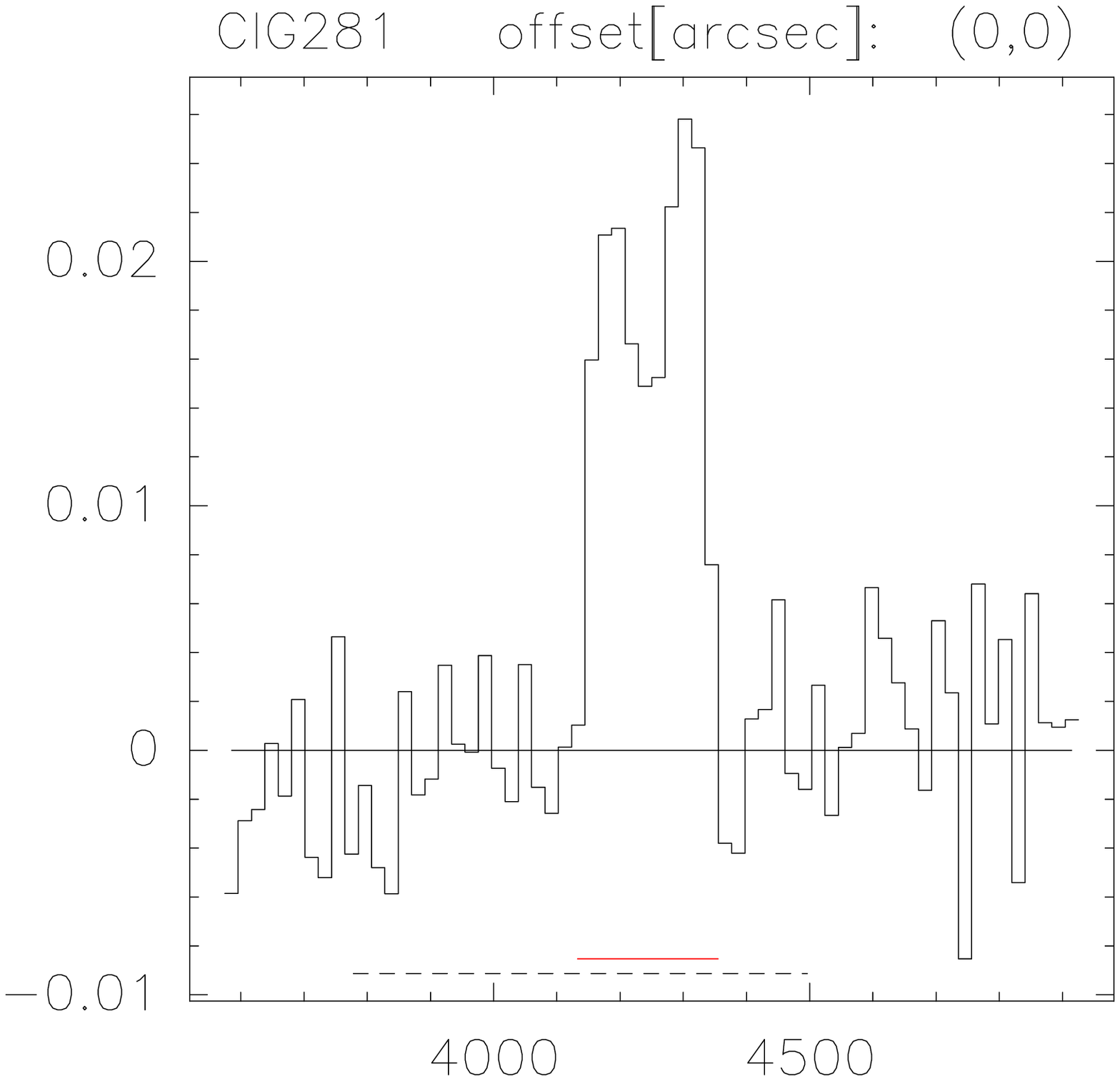}} 
\centerline{\includegraphics[width=3cm,angle=270]{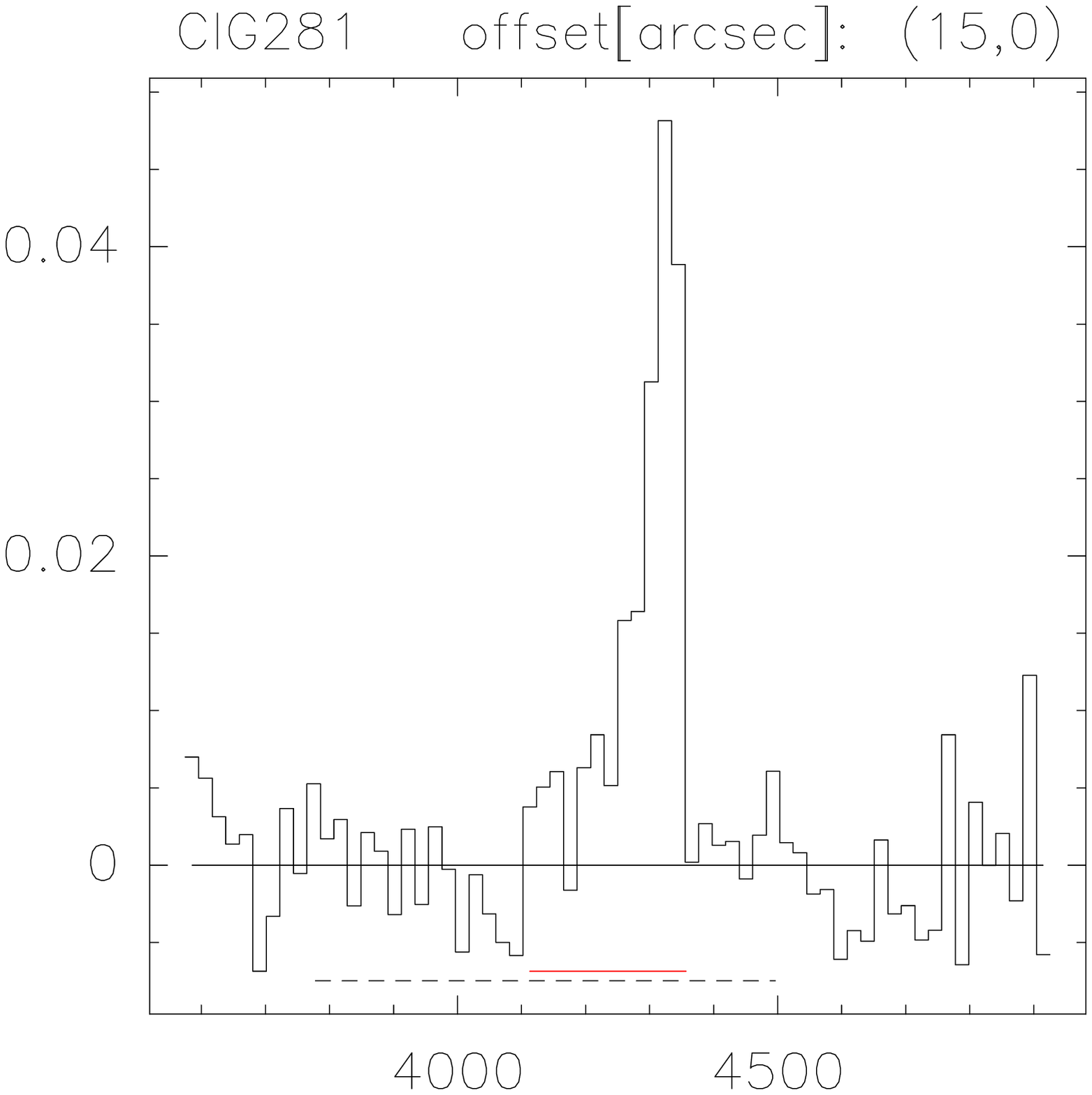} \quad 
\includegraphics[width=3cm,angle=270]{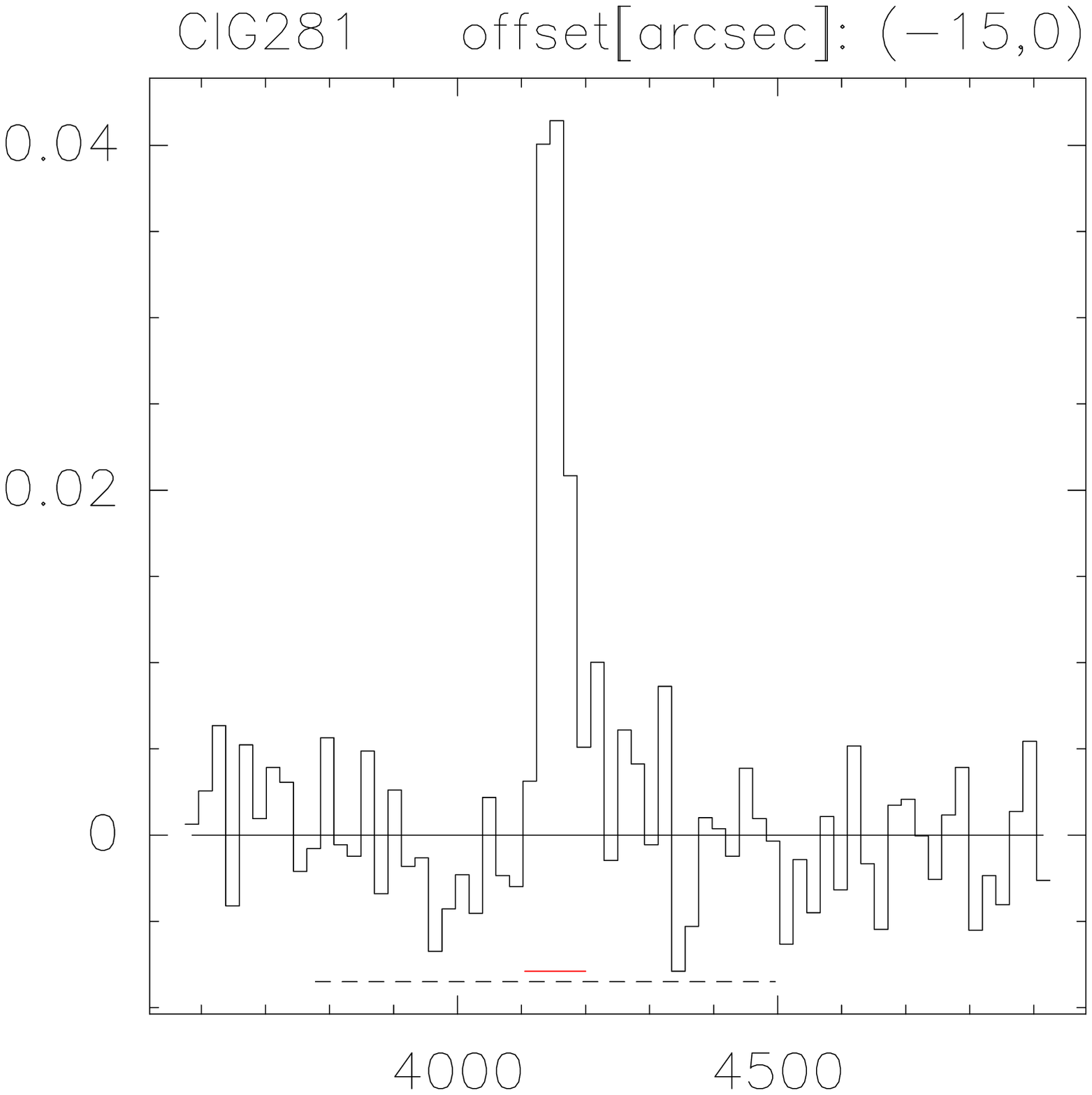}\quad 
\includegraphics[width=3cm,angle=270]{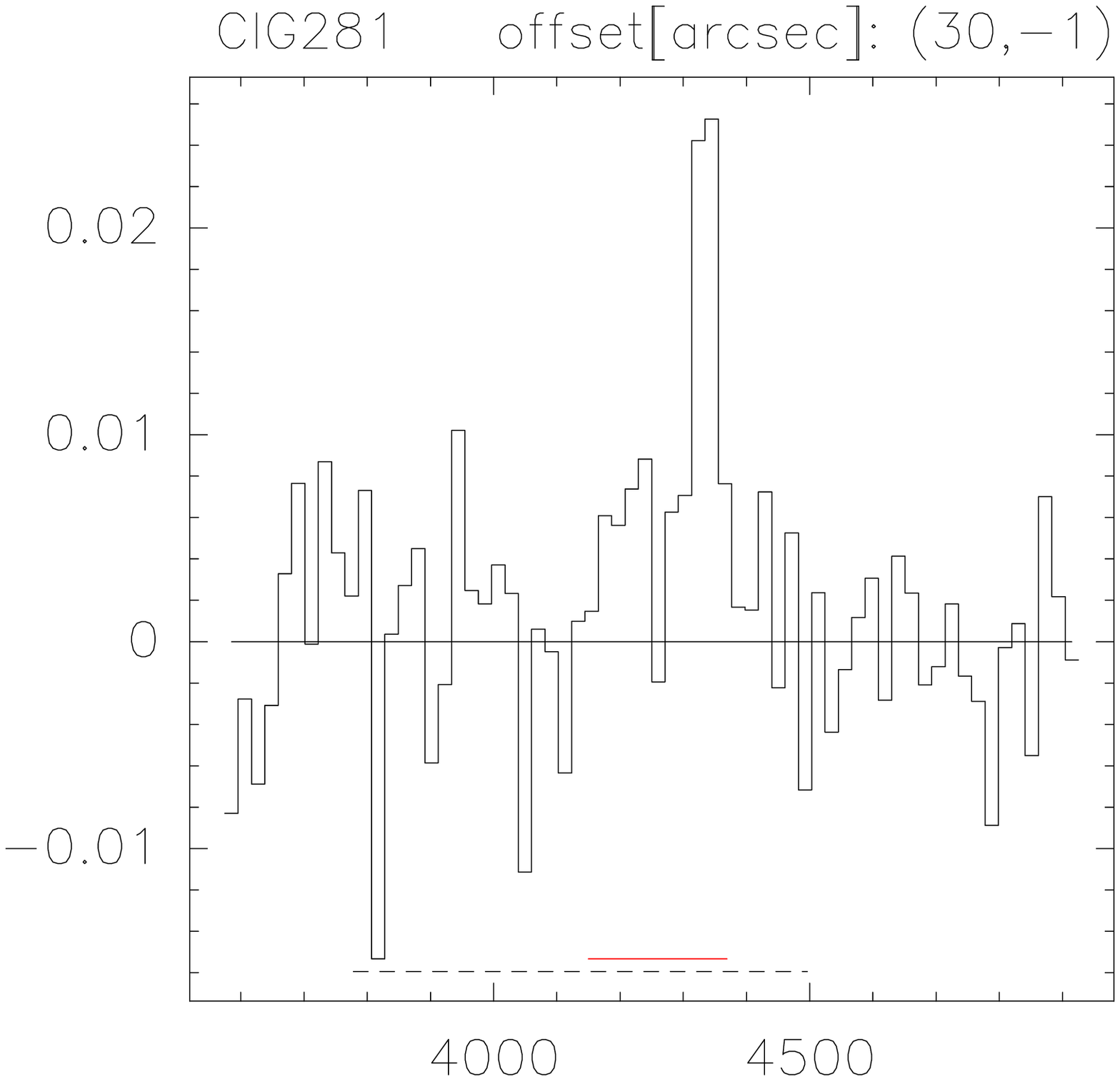}\quad 
\includegraphics[width=3cm,angle=270]{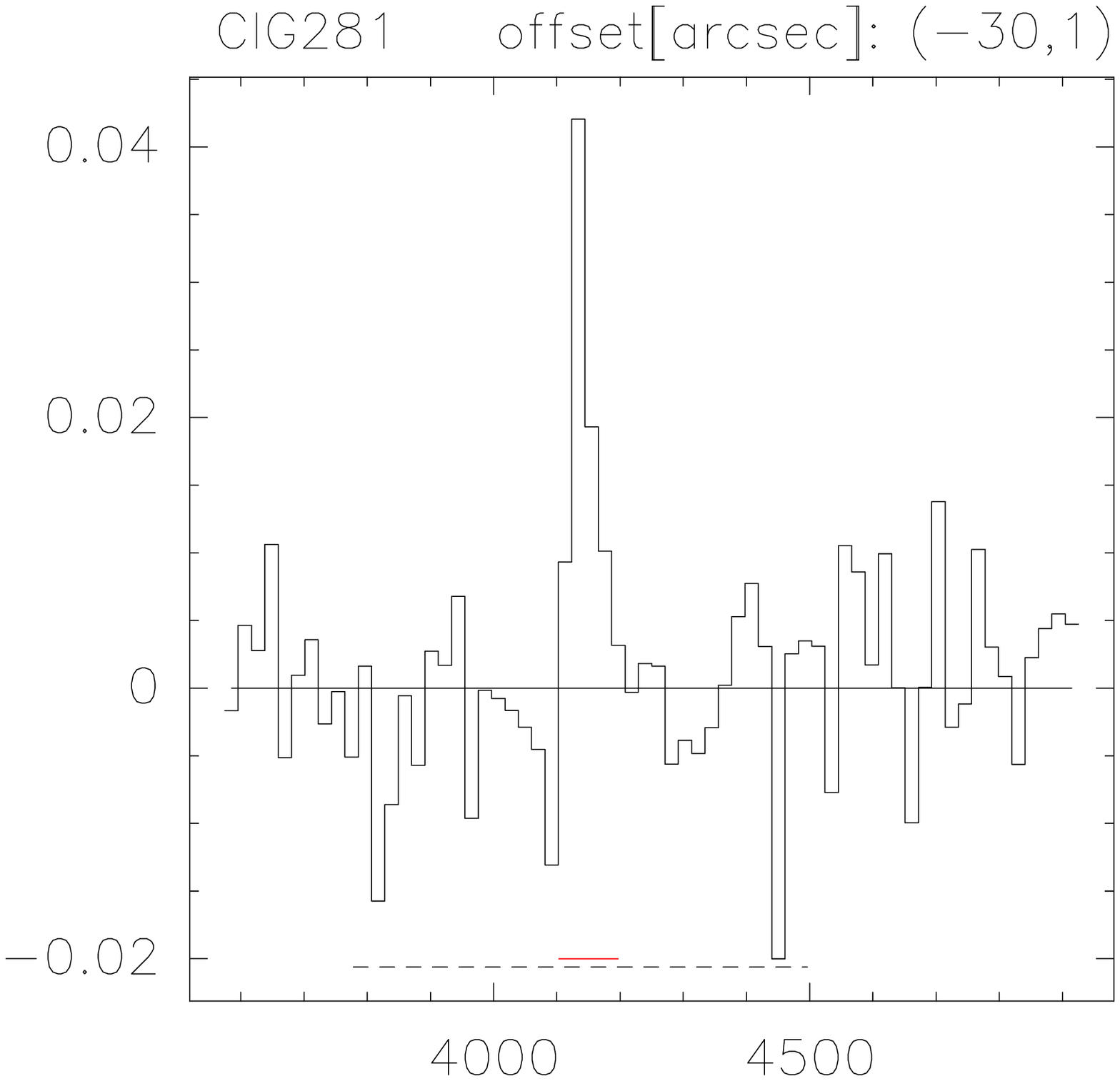}\quad 
\includegraphics[width=3cm,angle=270]{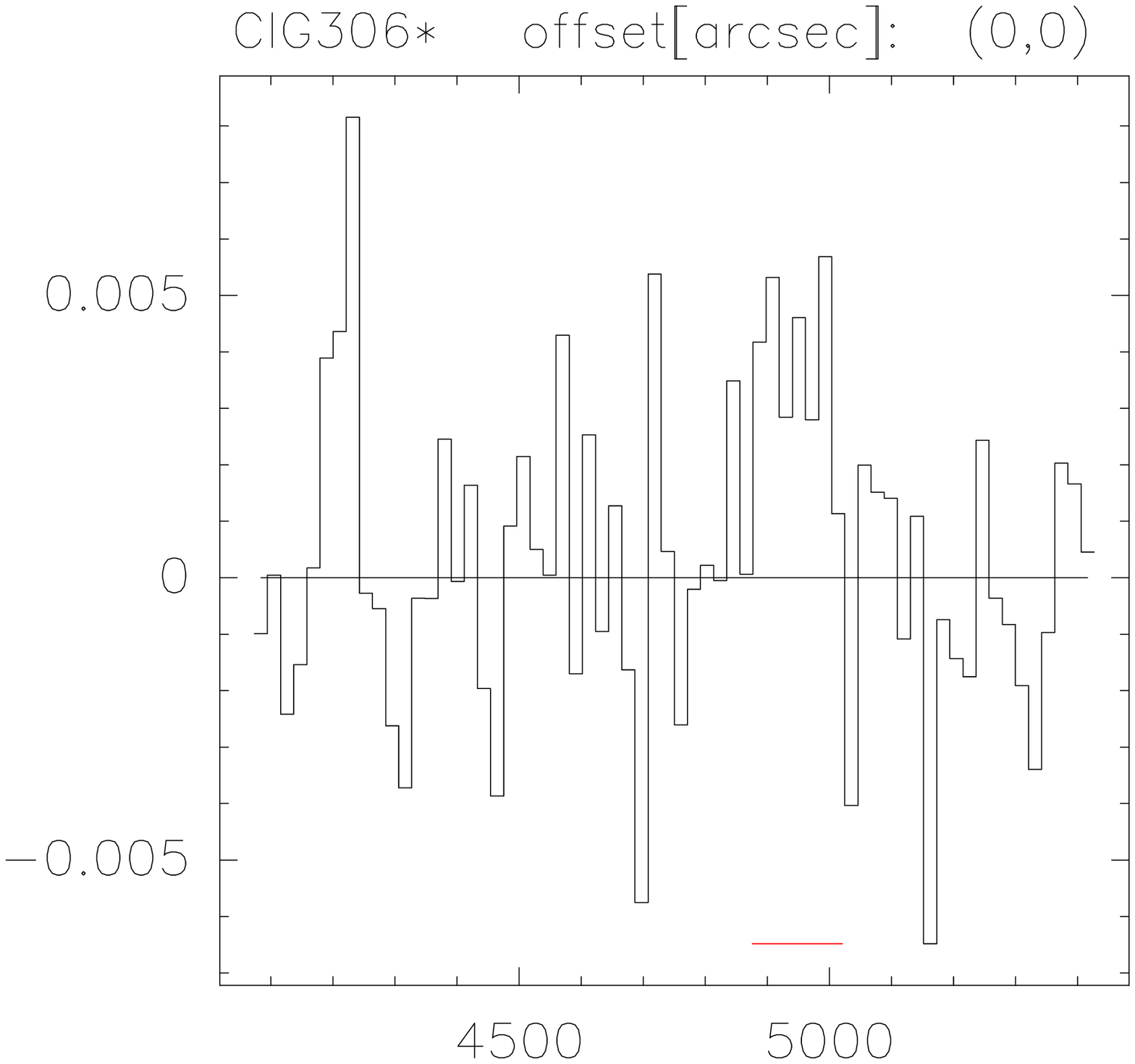}} 
\centerline{\includegraphics[width=3cm,angle=270]{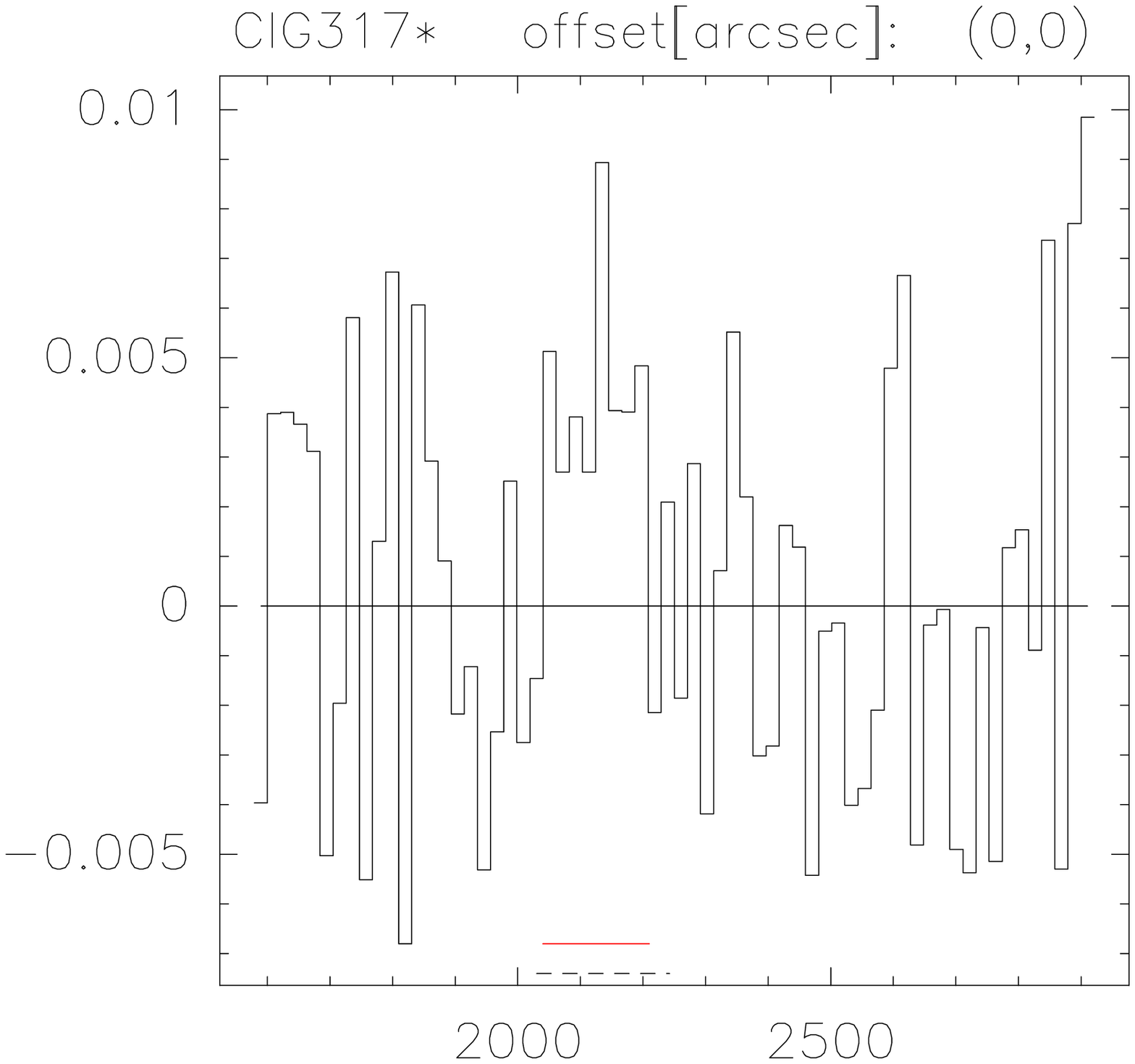} \quad 
\includegraphics[width=3cm,angle=270]{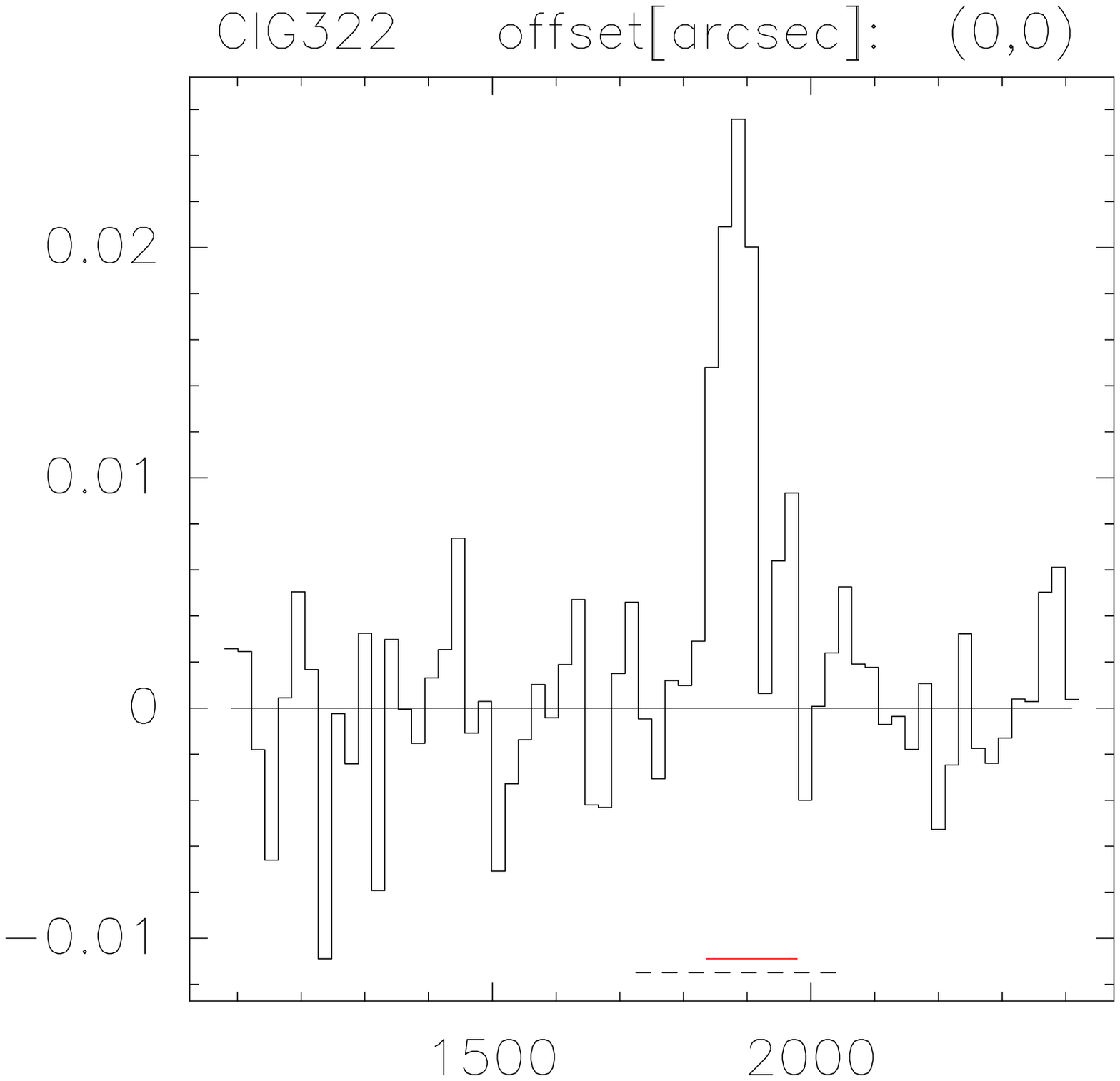}\quad 
\includegraphics[width=3cm,angle=270]{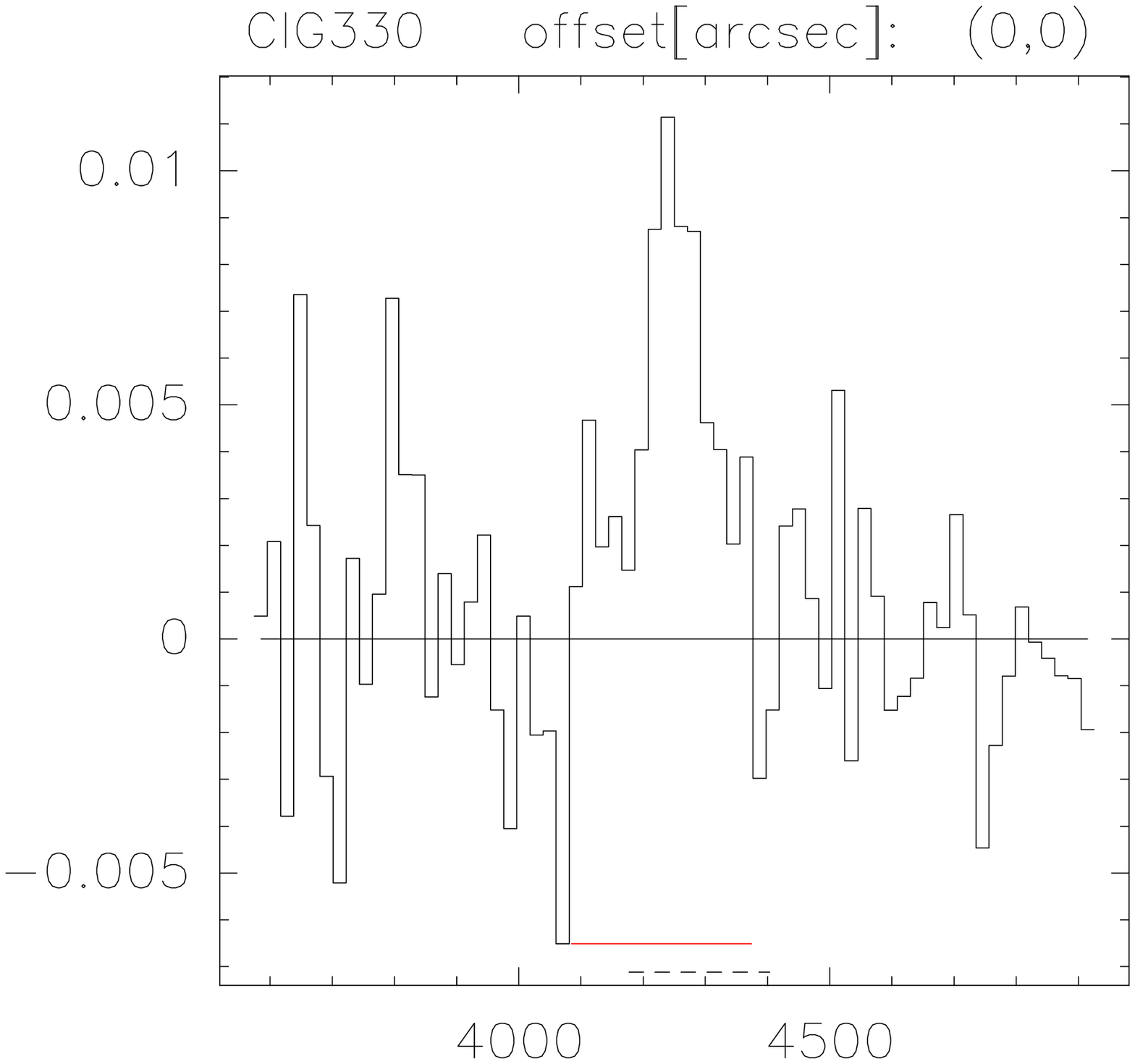}\quad 
\includegraphics[width=3cm,angle=270]{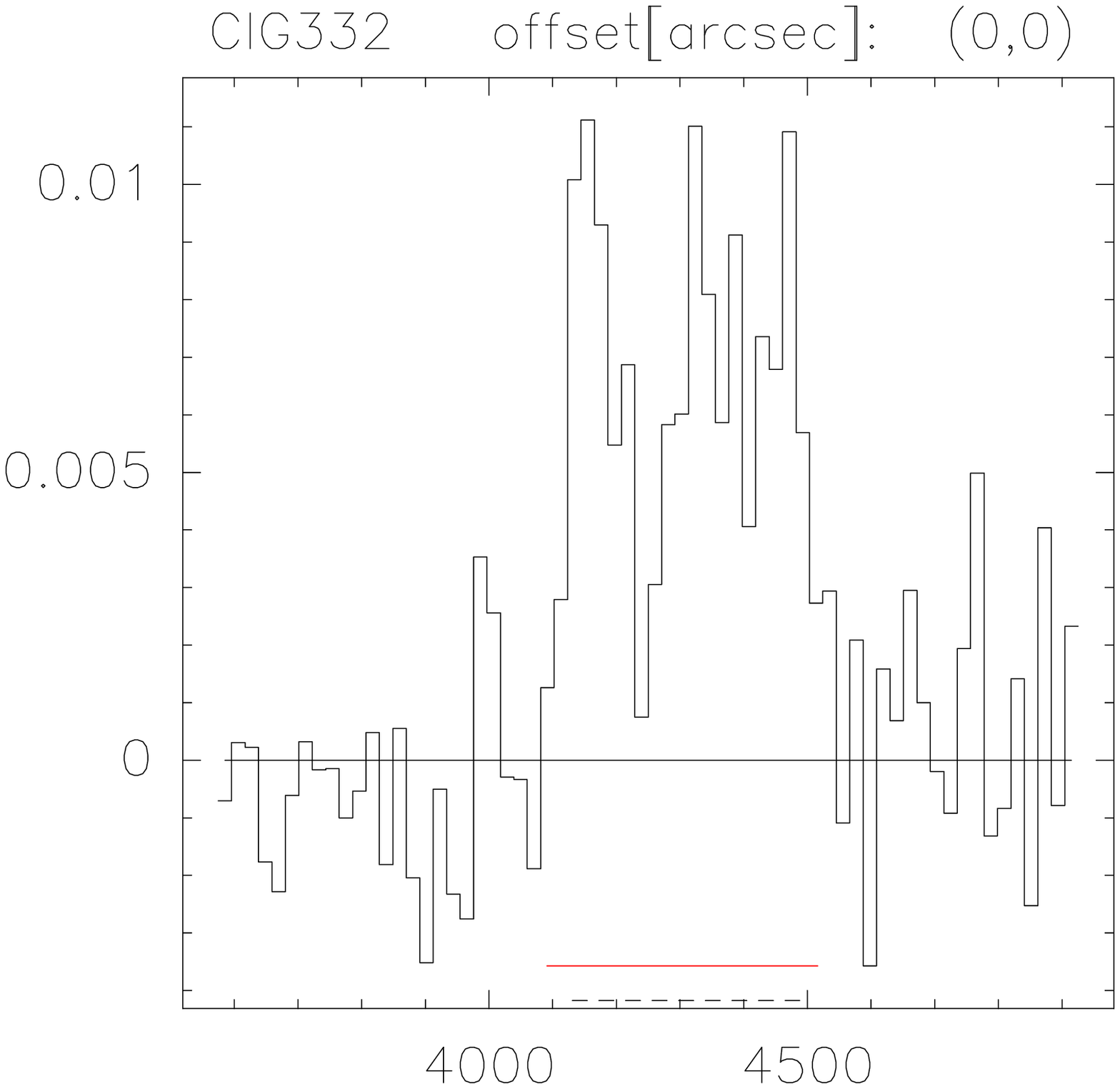}\quad 
\includegraphics[width=3cm,angle=270]{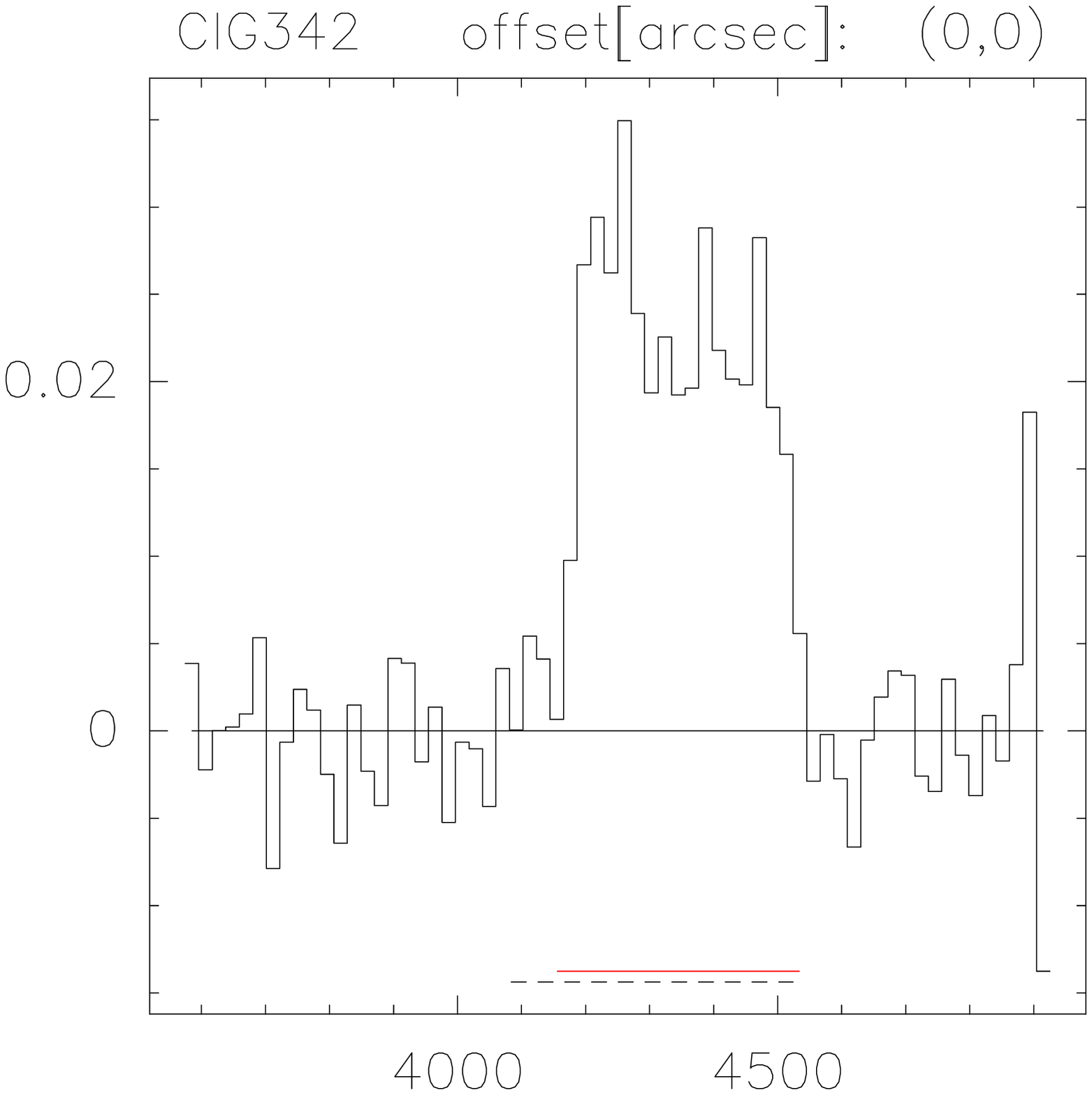}} 
\centerline{\includegraphics[width=3cm,angle=270]{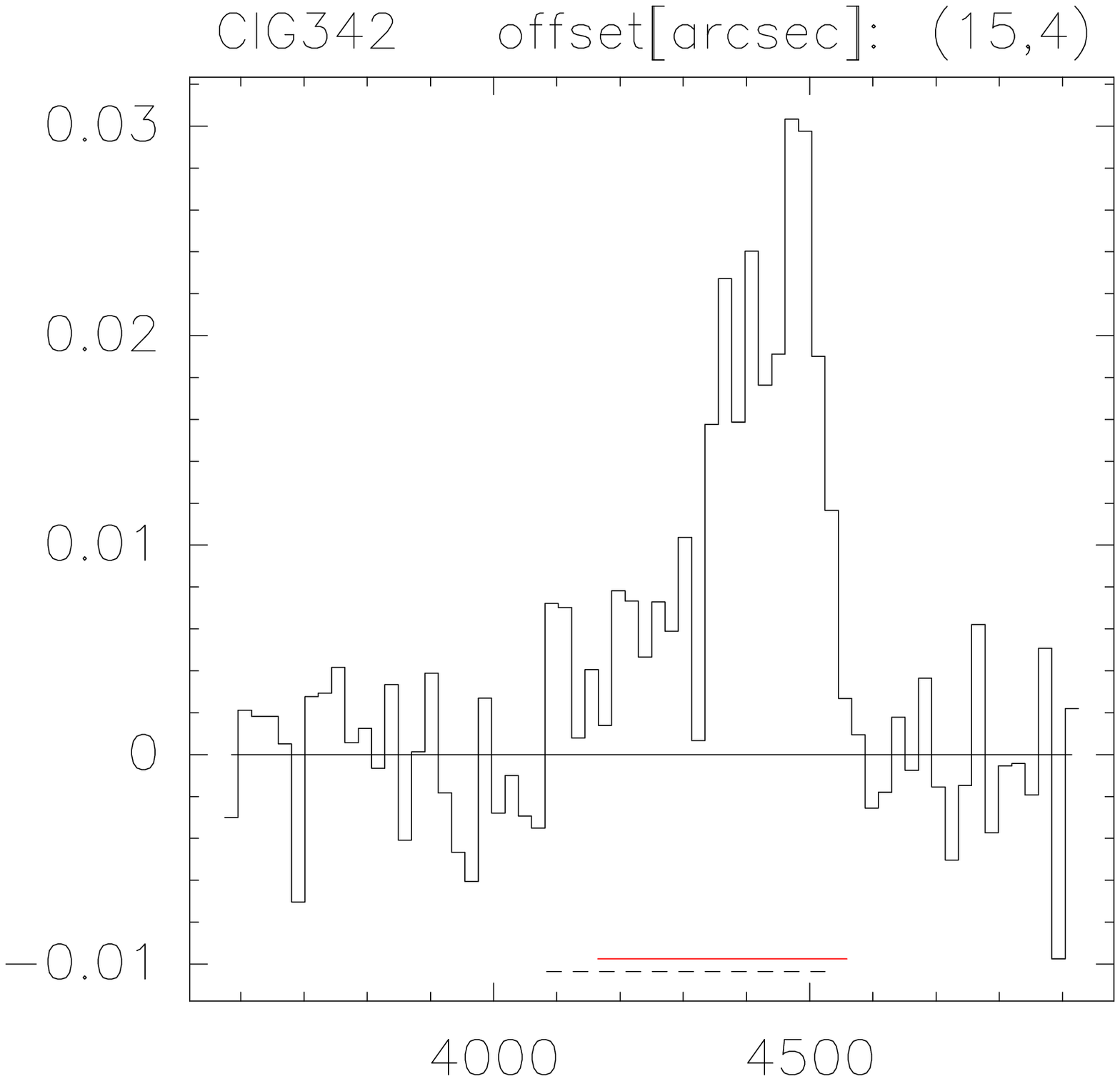} \quad 
\includegraphics[width=3cm,angle=270]{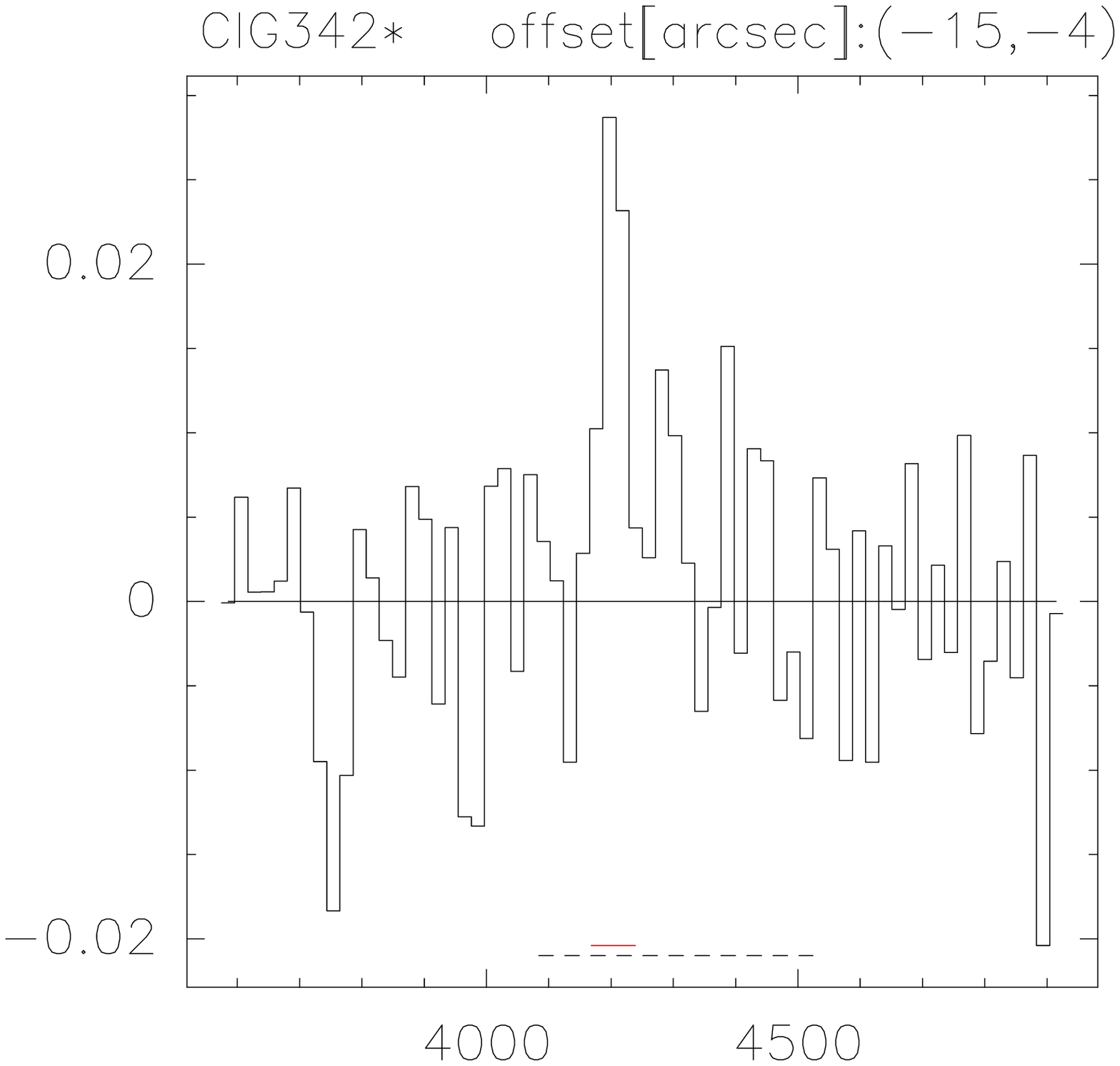}\quad 
\includegraphics[width=3cm,angle=270]{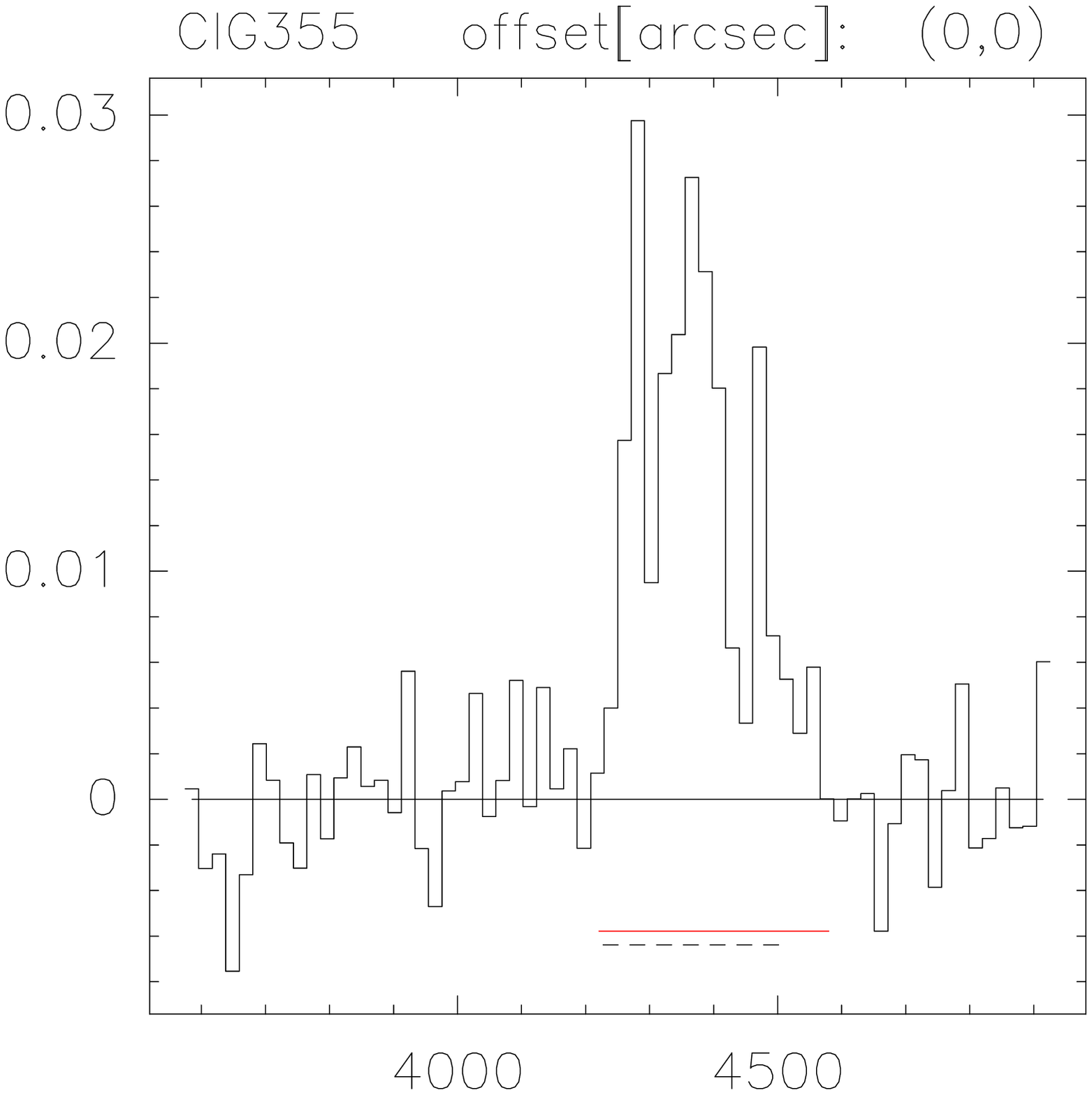}\quad 
\includegraphics[width=3cm,angle=270]{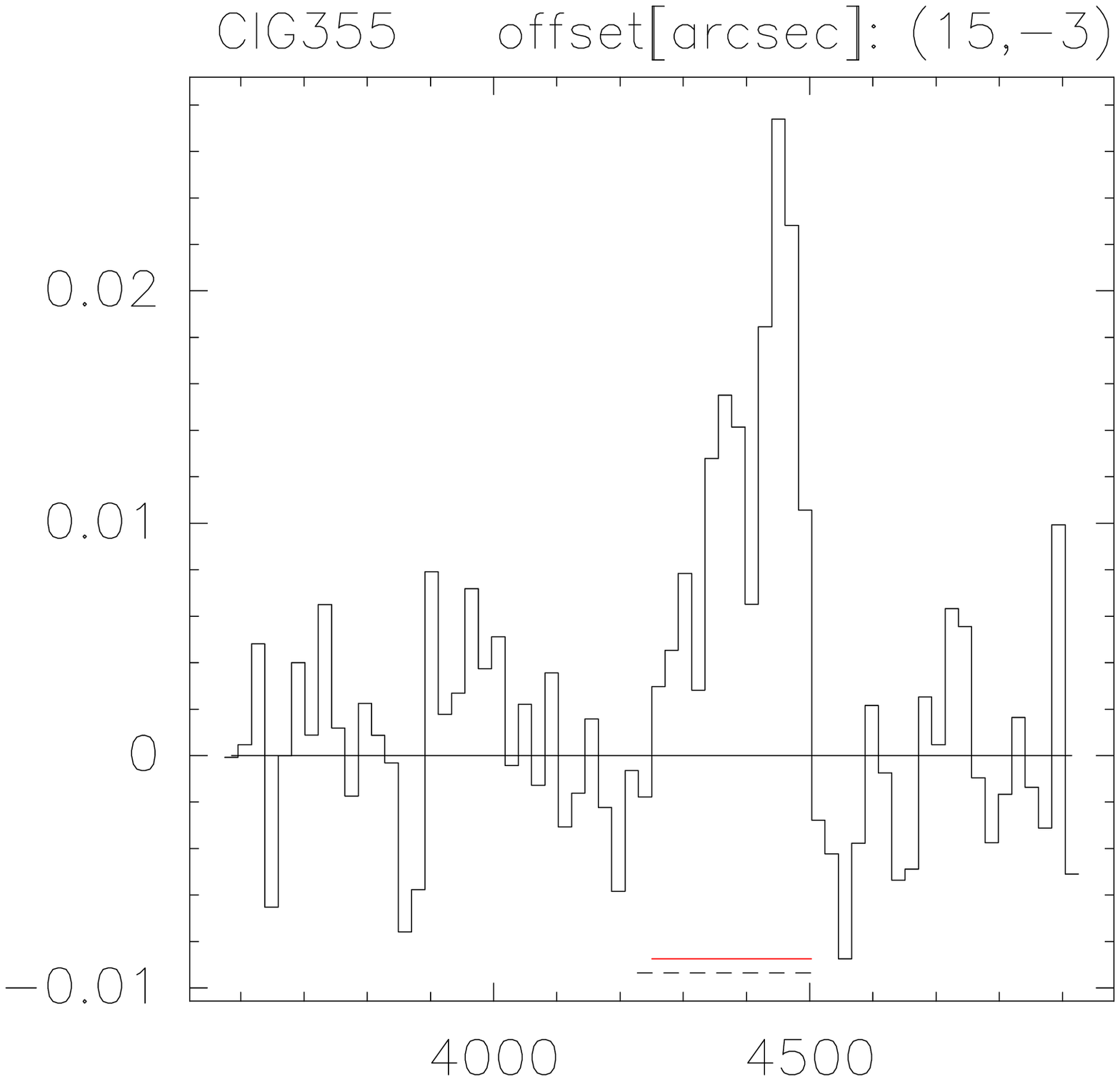}\quad 
\includegraphics[width=3cm,angle=270]{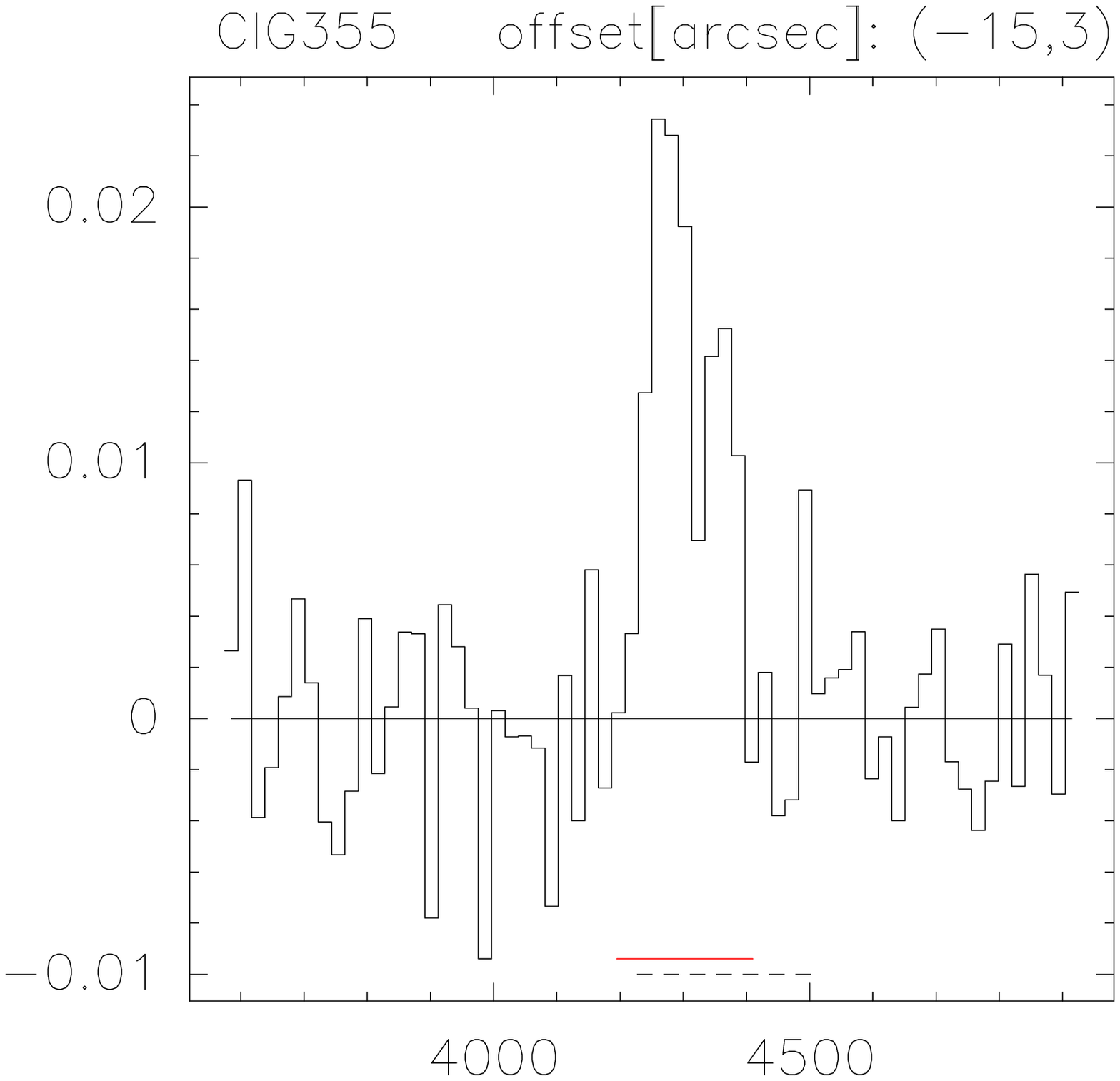}} 
\centerline{\includegraphics[width=3.3cm,angle=270]{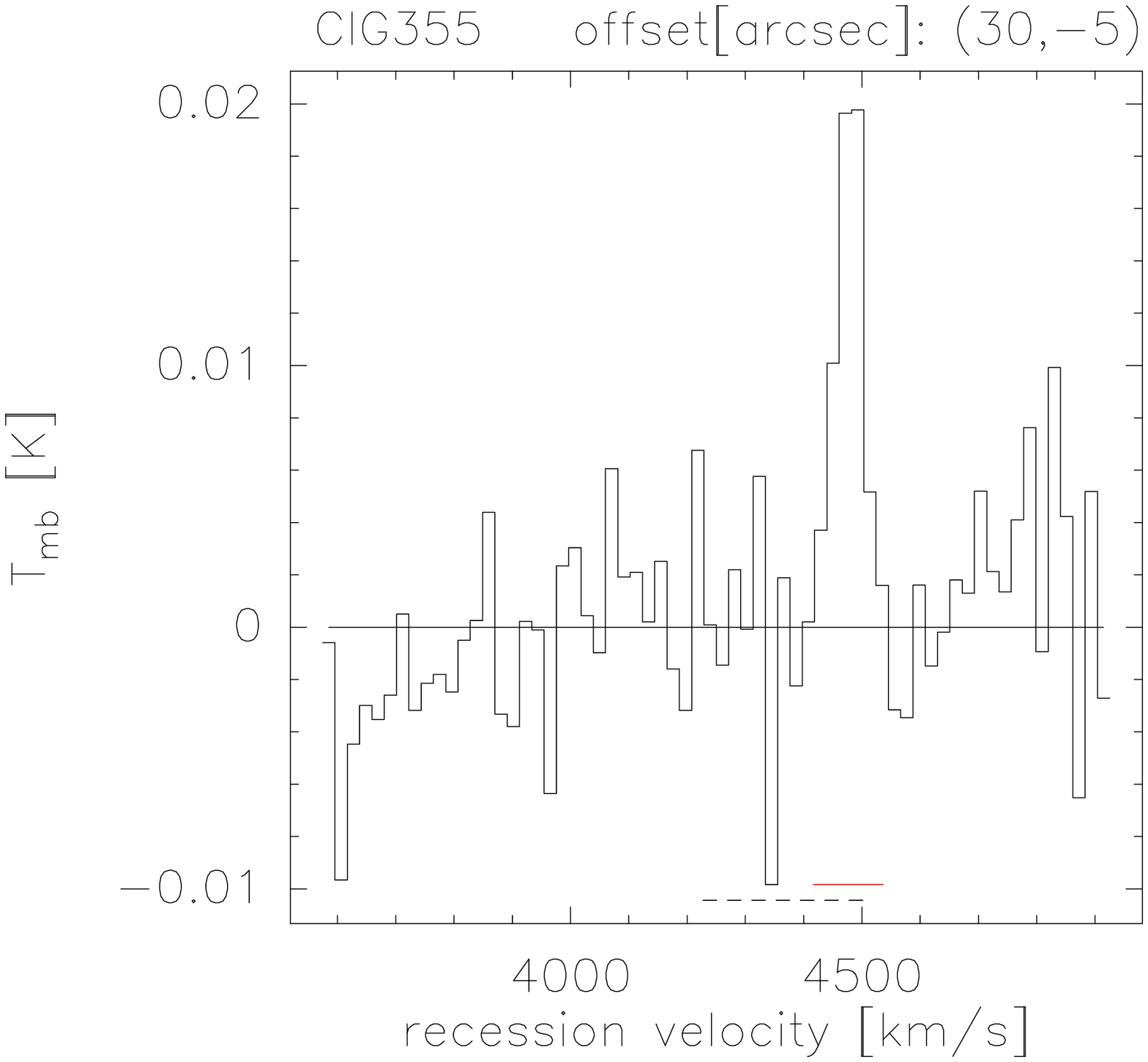} \quad 
\includegraphics[width=3cm,angle=270]{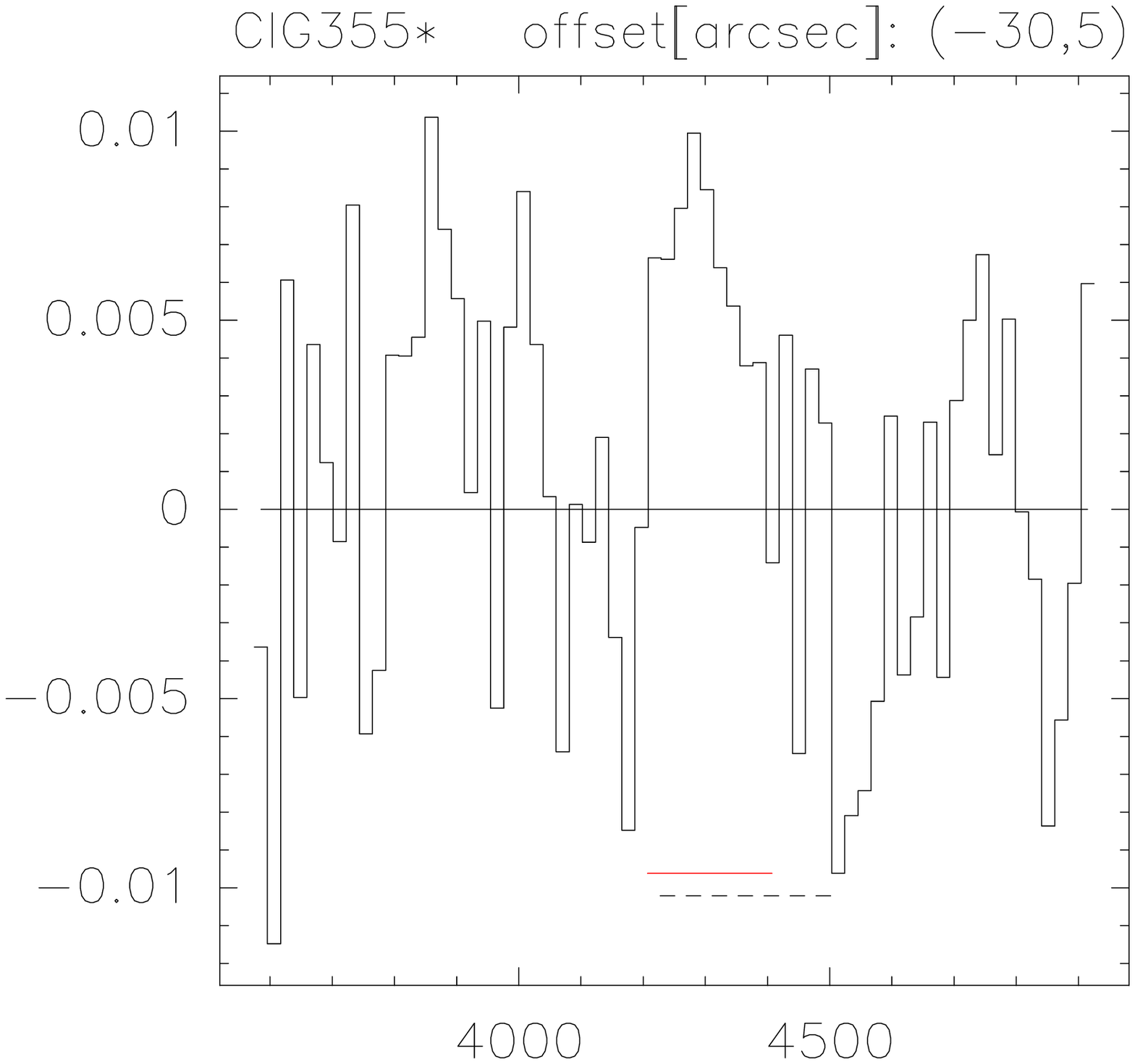}\quad 
\includegraphics[width=3cm,angle=270]{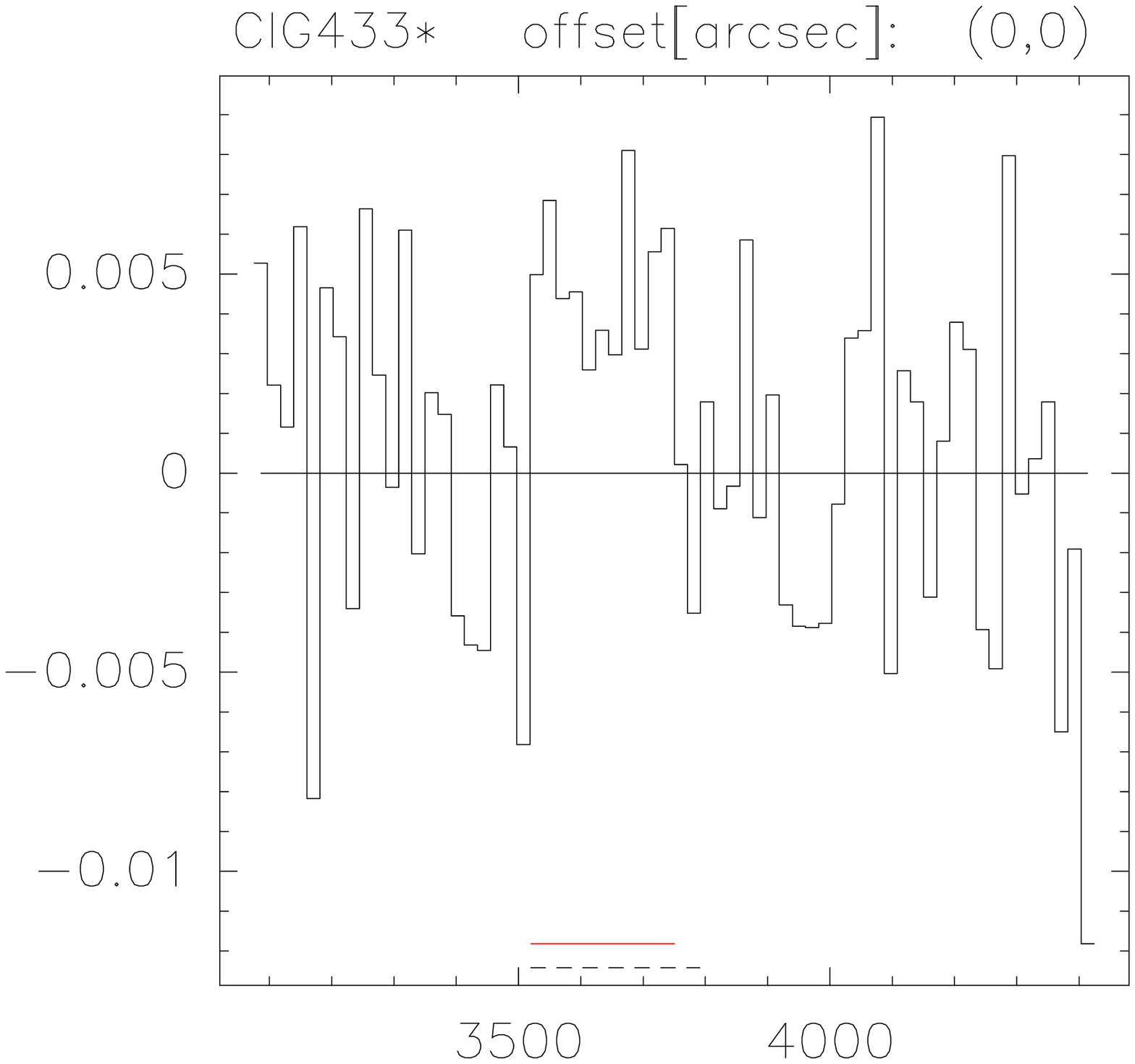}\quad 
\includegraphics[width=3cm,angle=270]{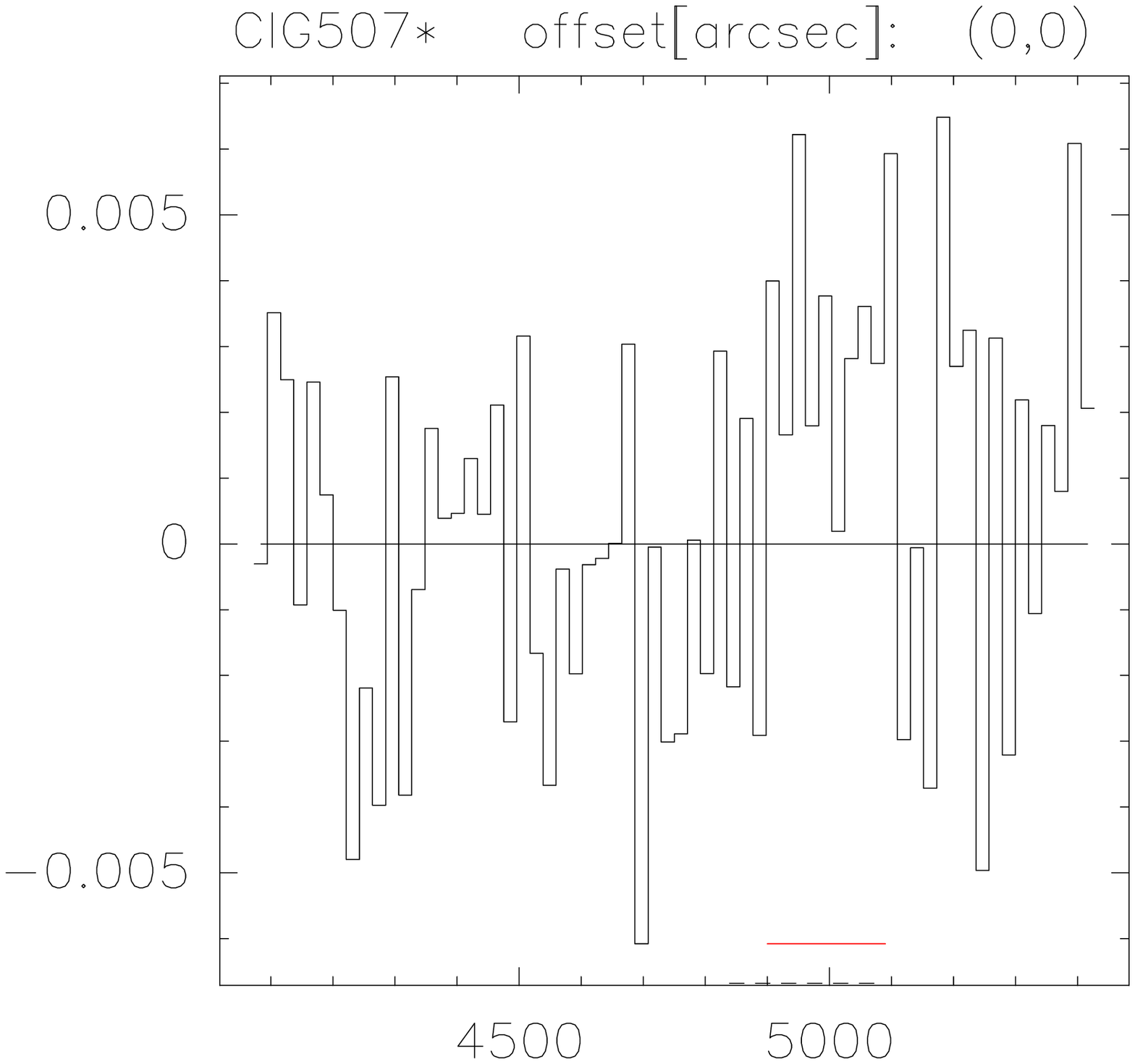}\quad 
\includegraphics[width=3cm,angle=270]{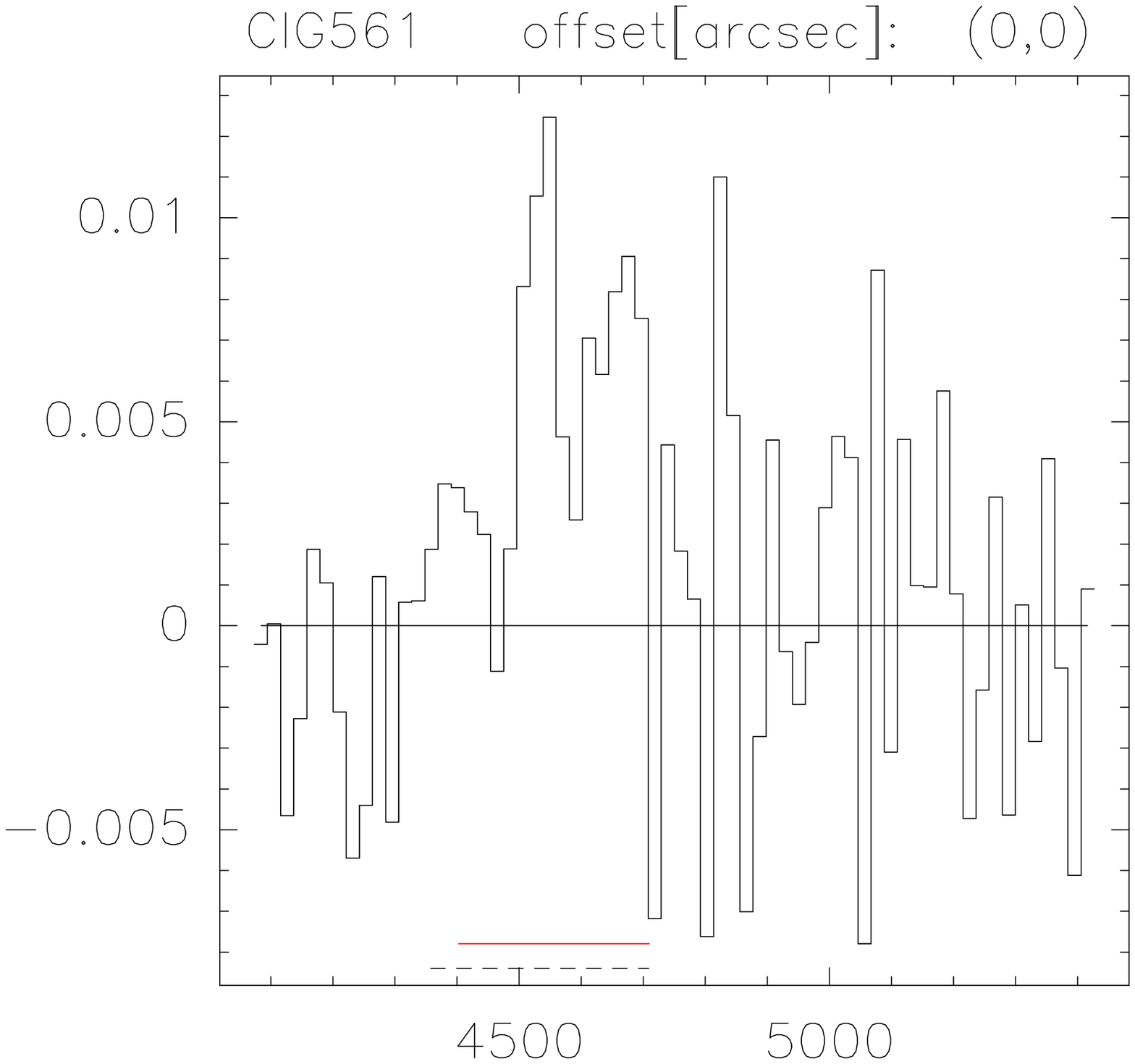}} 
\addtocounter{figure}{-1} 
\caption{(continued)} 
\end{figure*} 
  
\begin{figure*} 
\centerline{\includegraphics[width=3cm,angle=270]{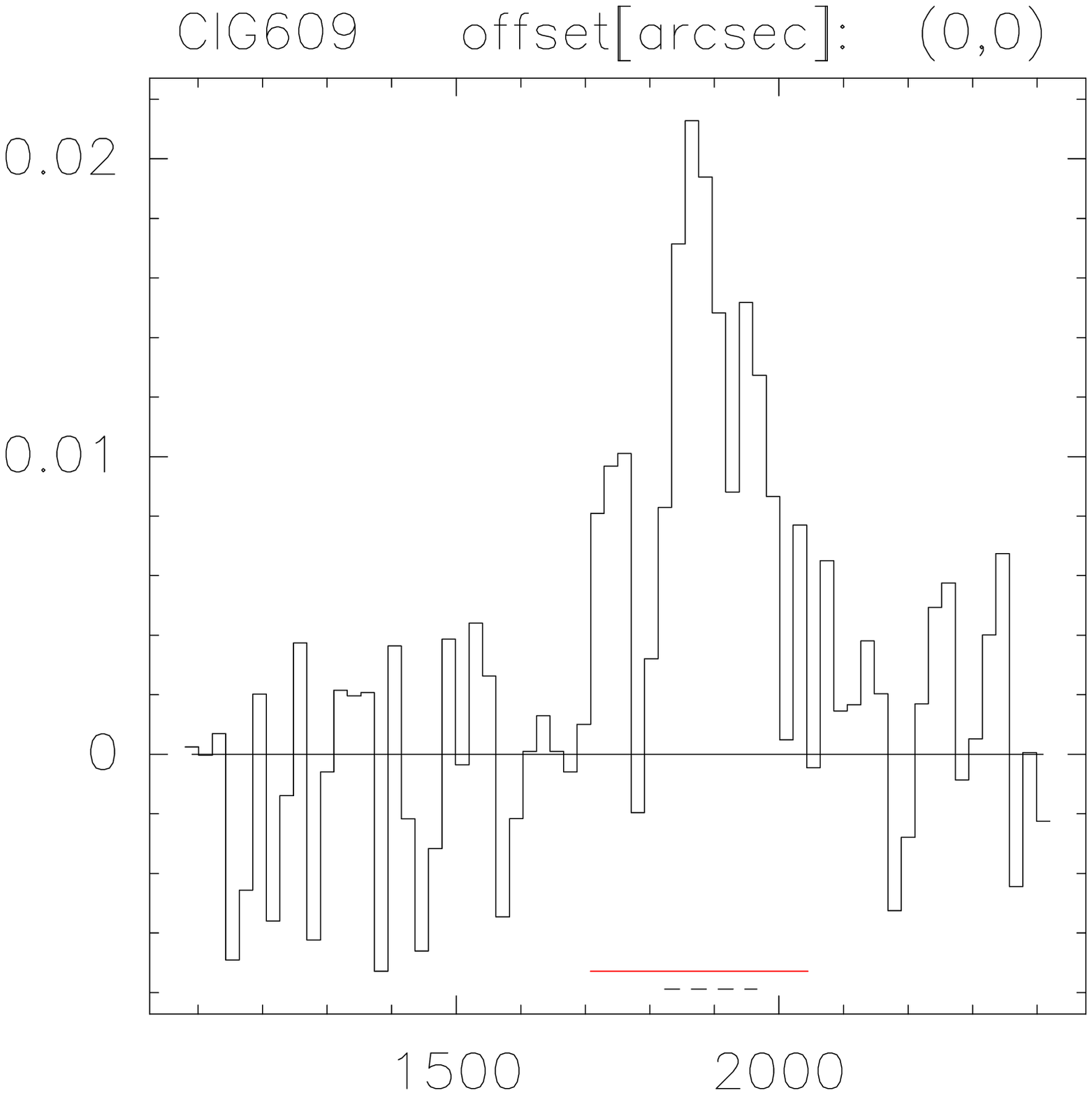} \quad 
\includegraphics[width=3cm,angle=270]{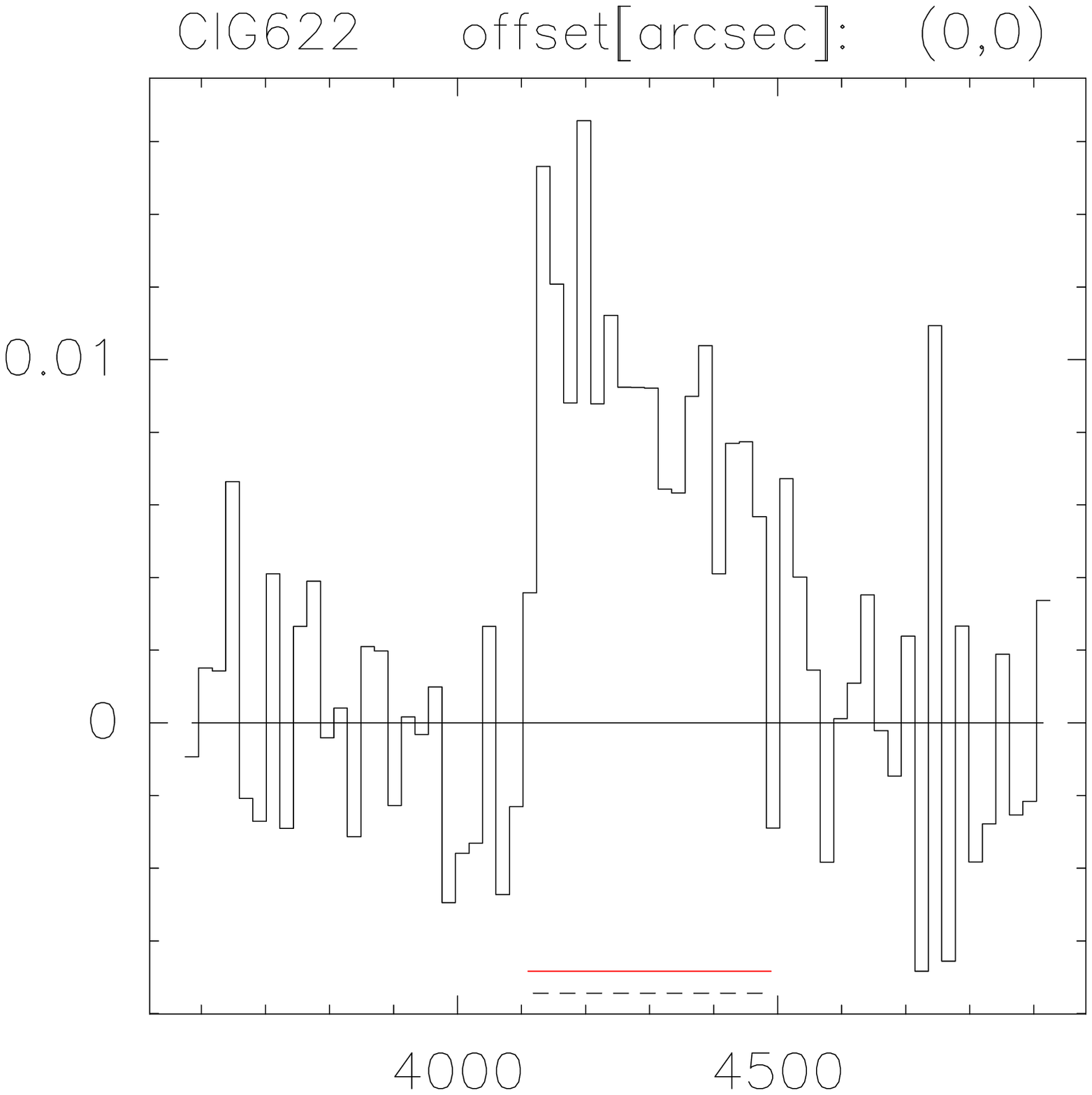}\quad 
\includegraphics[width=3cm,angle=270]{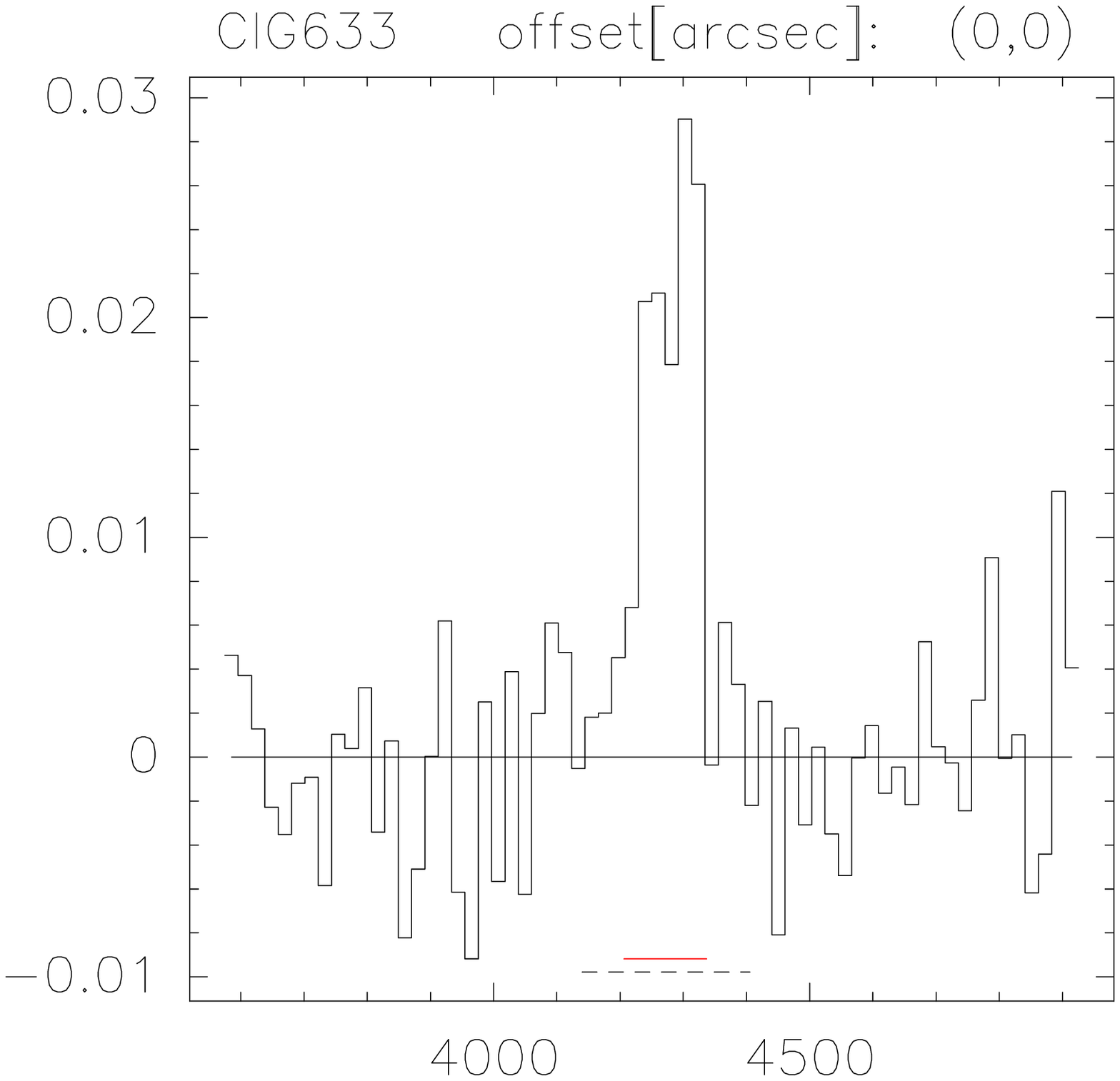}\quad 
\includegraphics[width=3cm,angle=270]{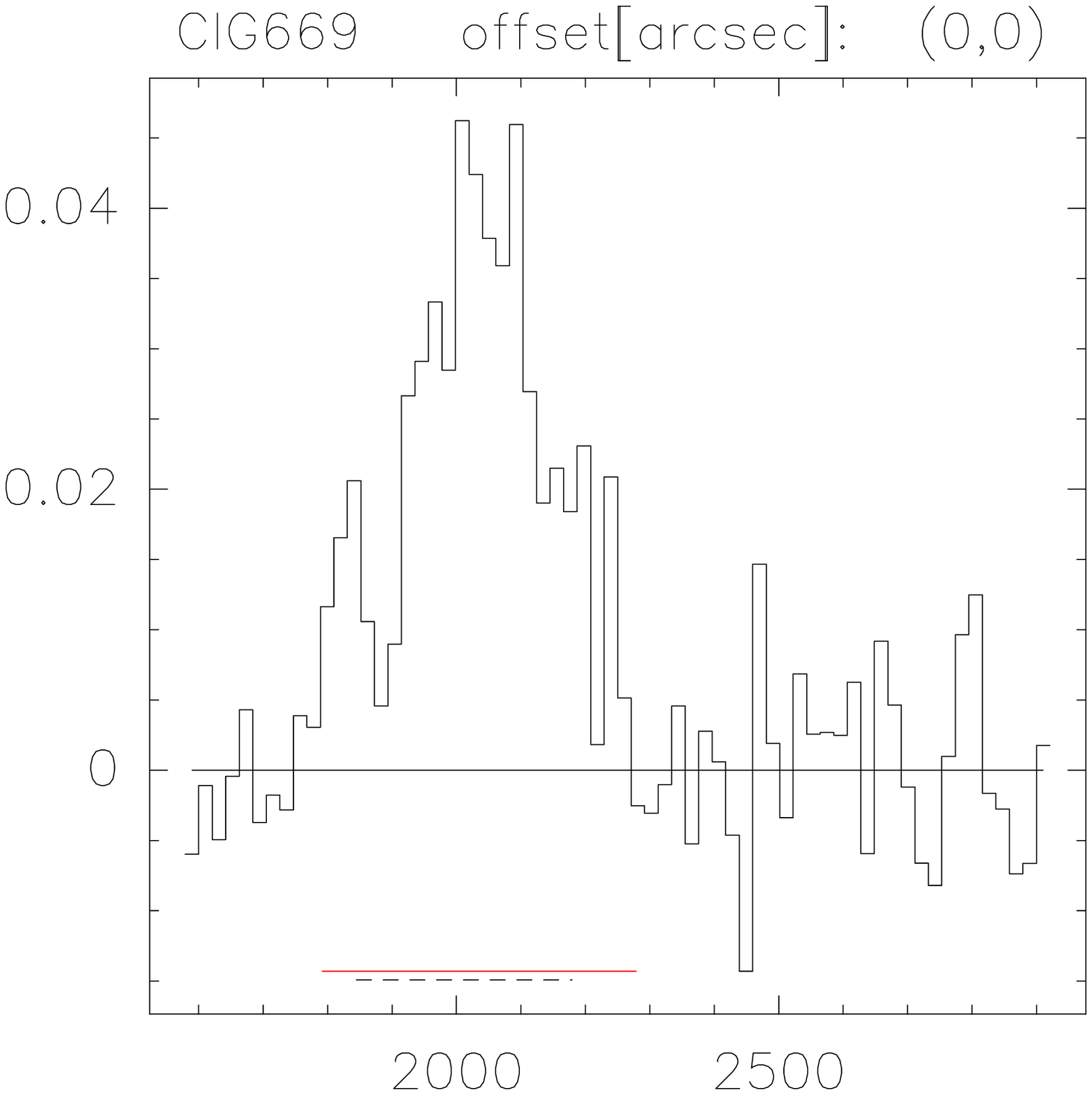}\quad 
\includegraphics[width=3cm,angle=270]{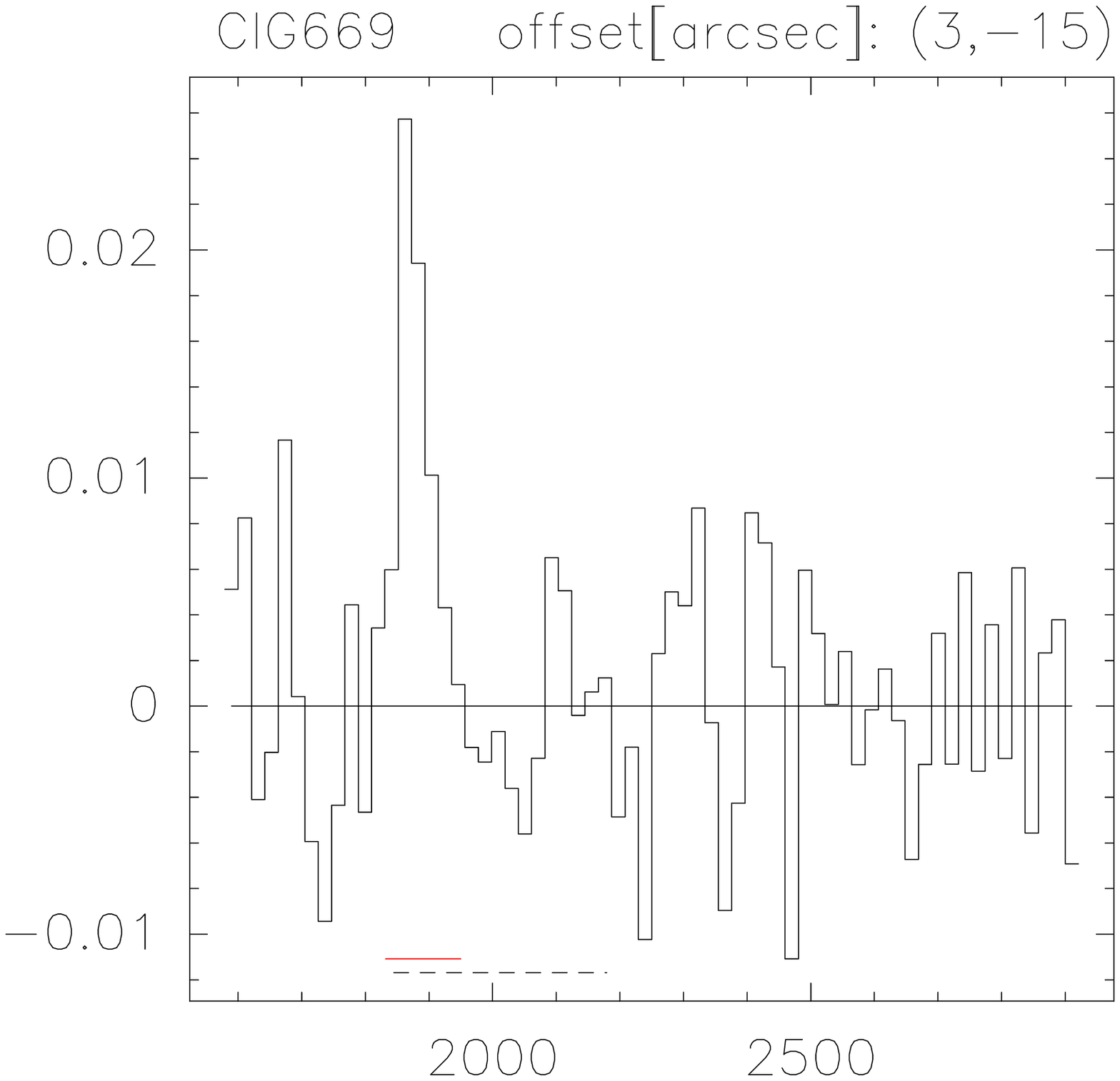}} 
\centerline{\includegraphics[width=3cm,angle=270]{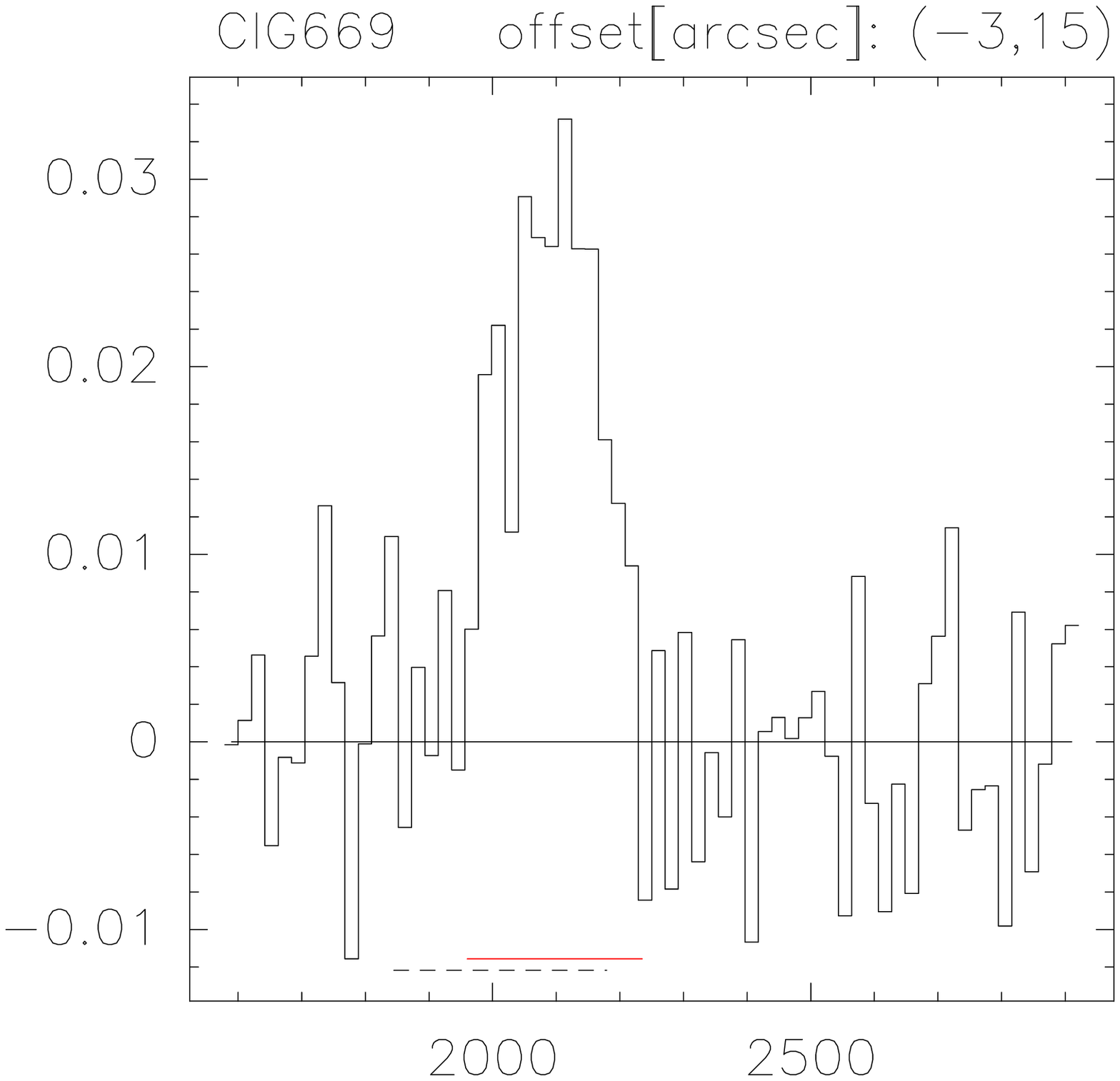} \quad 
\includegraphics[width=3cm,angle=270]{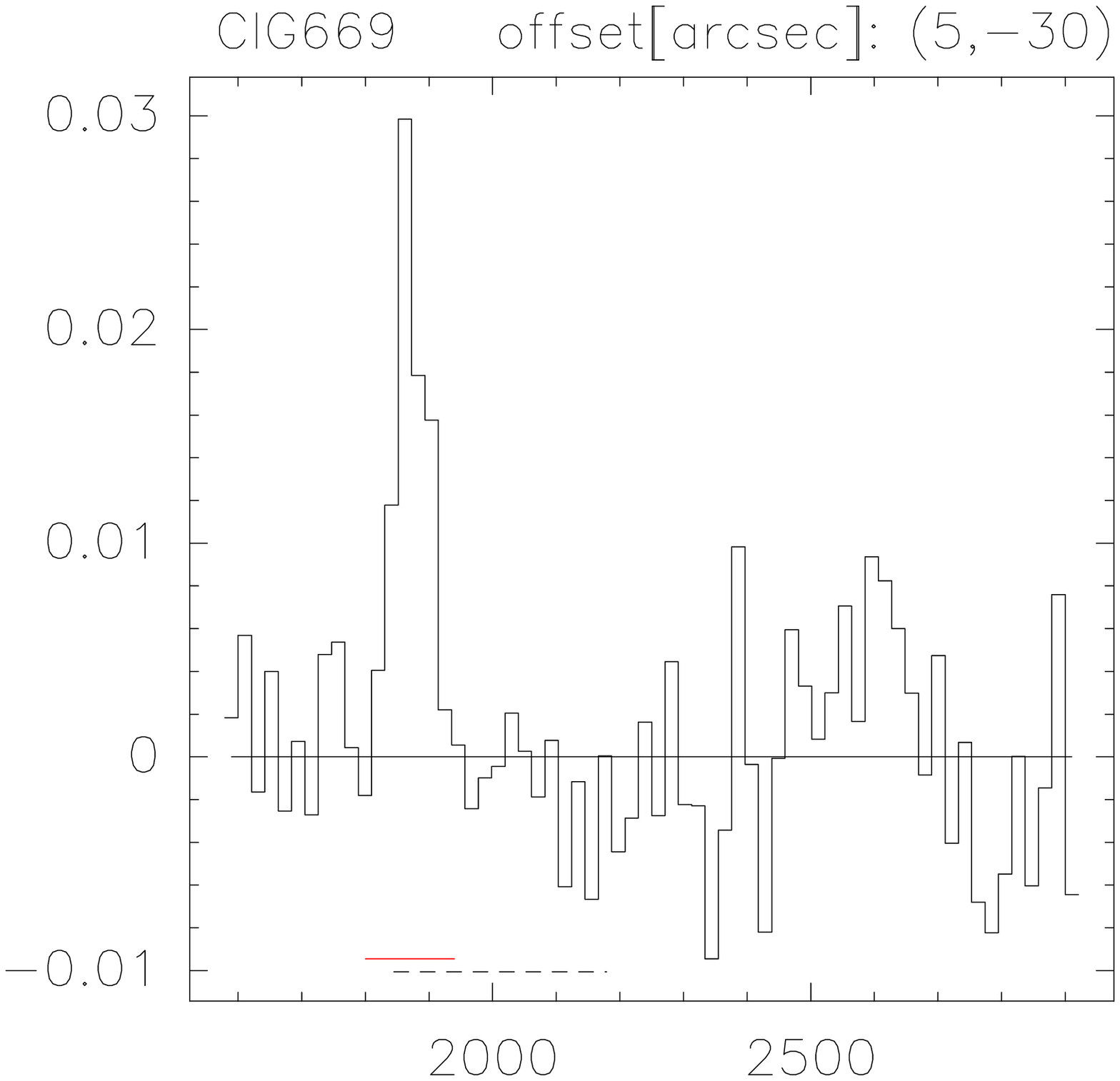}\quad 
\includegraphics[width=3cm,angle=270]{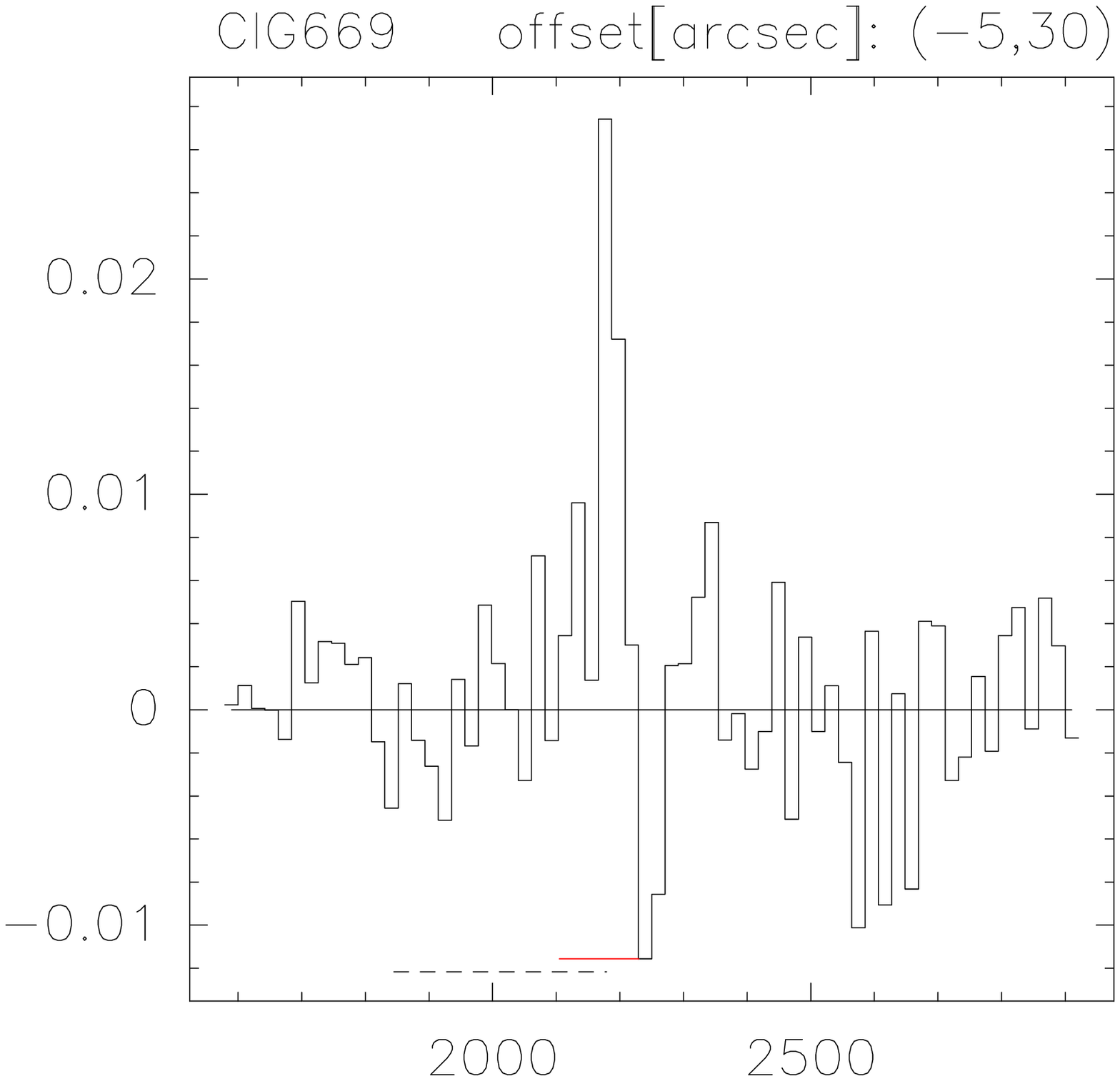}\quad 
\includegraphics[width=3cm,angle=270]{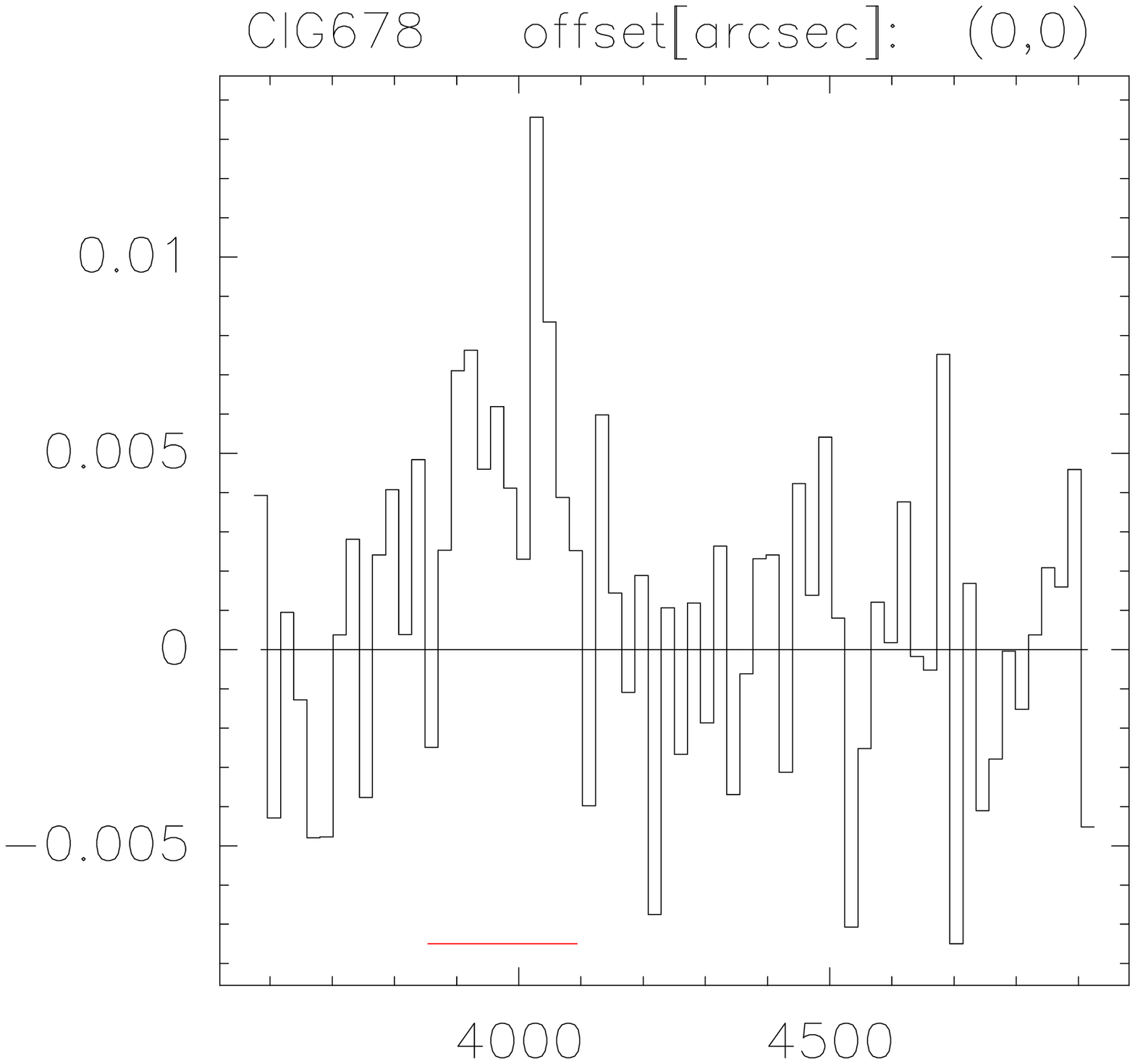}\quad 
\includegraphics[width=3cm,angle=270]{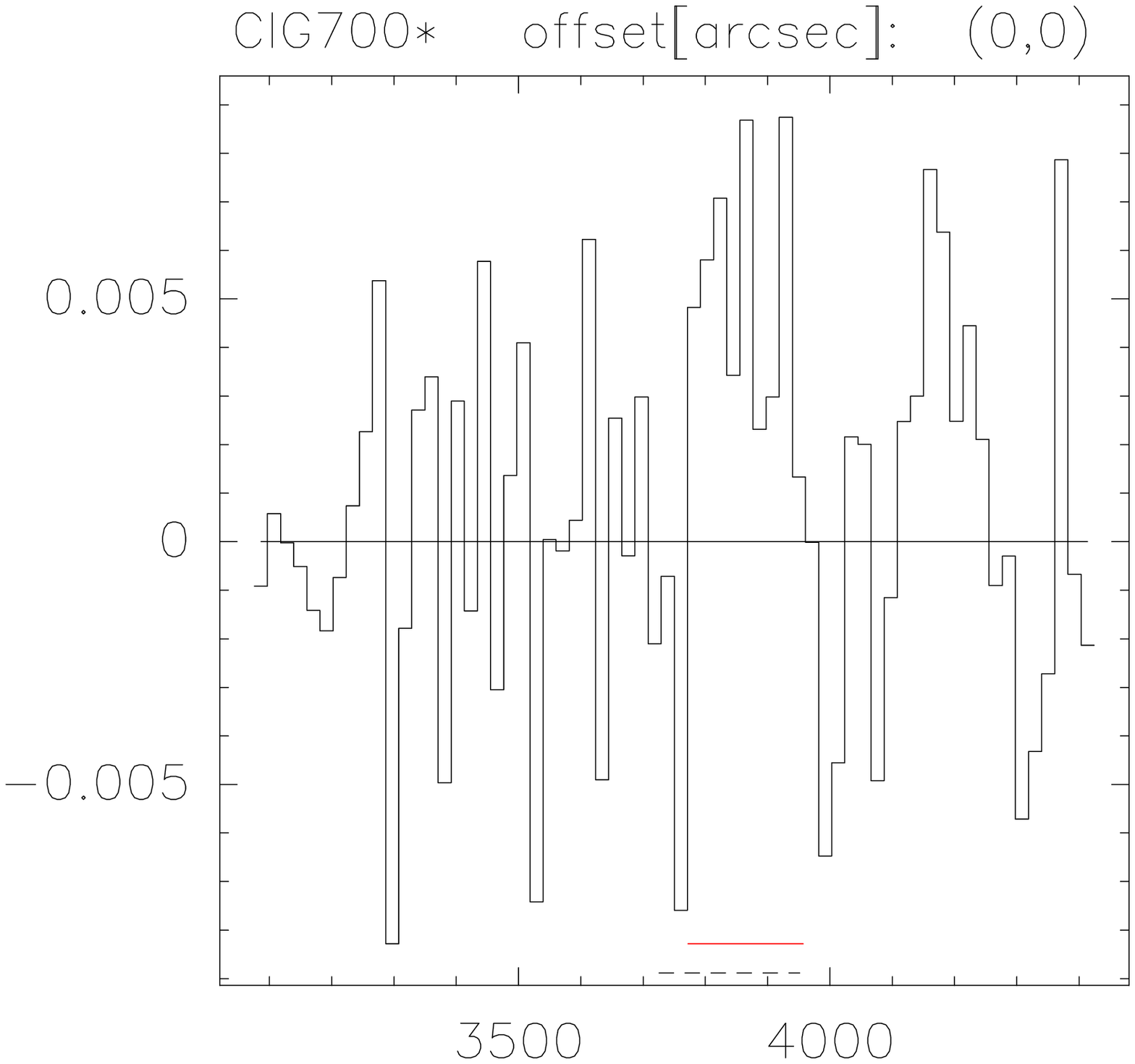}} 
\centerline{\includegraphics[width=3cm,angle=270]{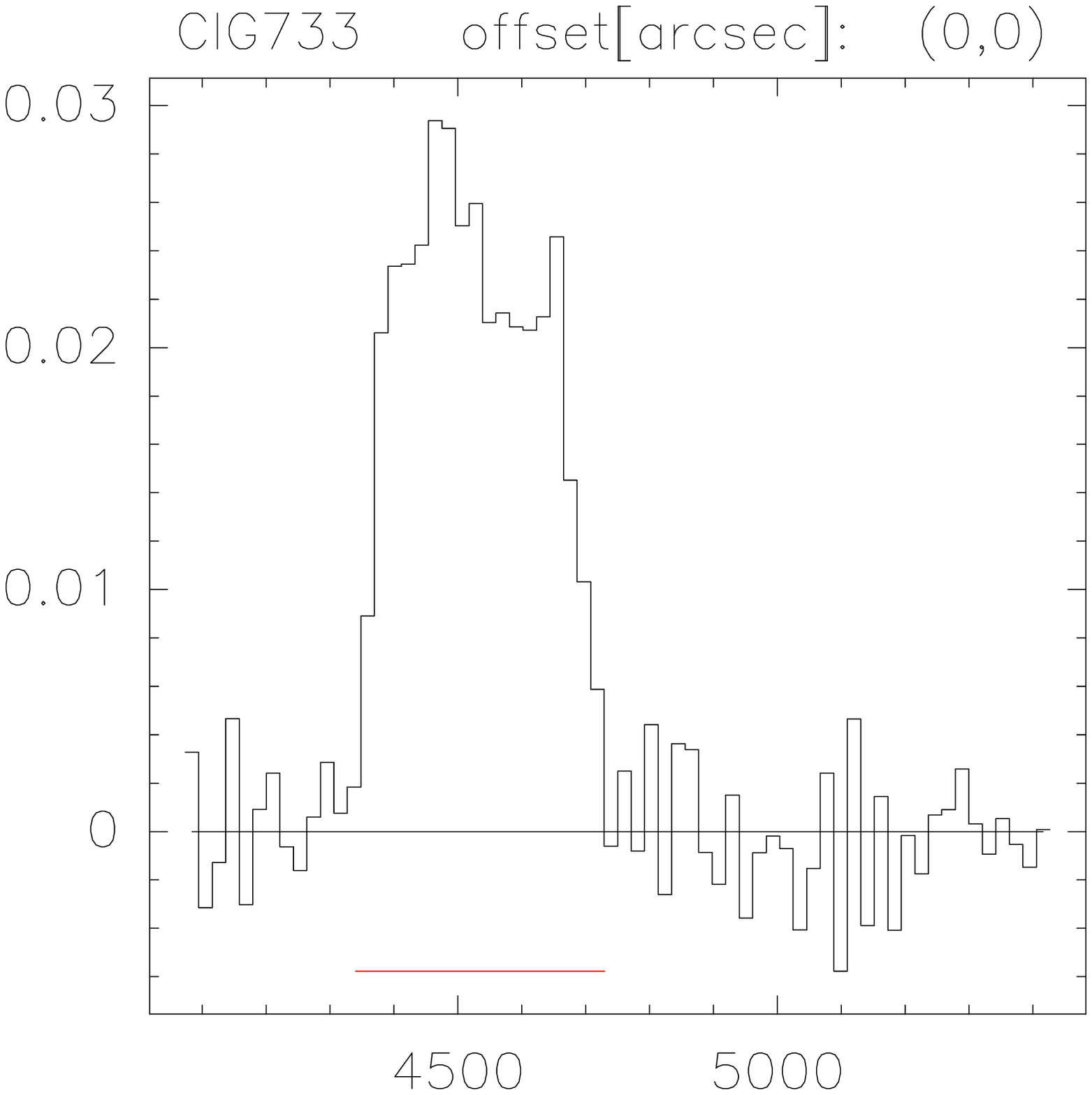} \quad 
\includegraphics[width=3cm,angle=270]{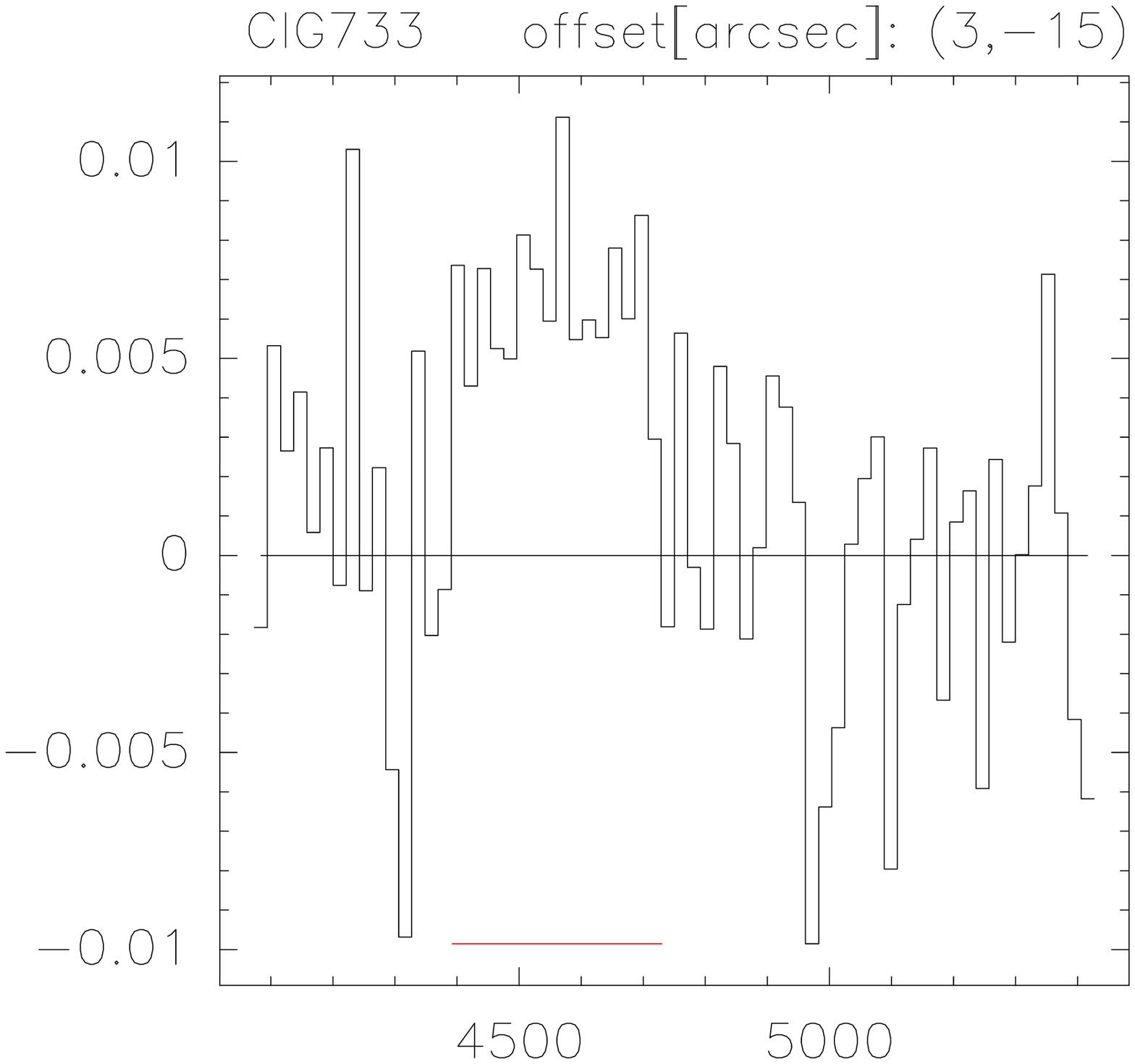}\quad 
\includegraphics[width=3cm,angle=270]{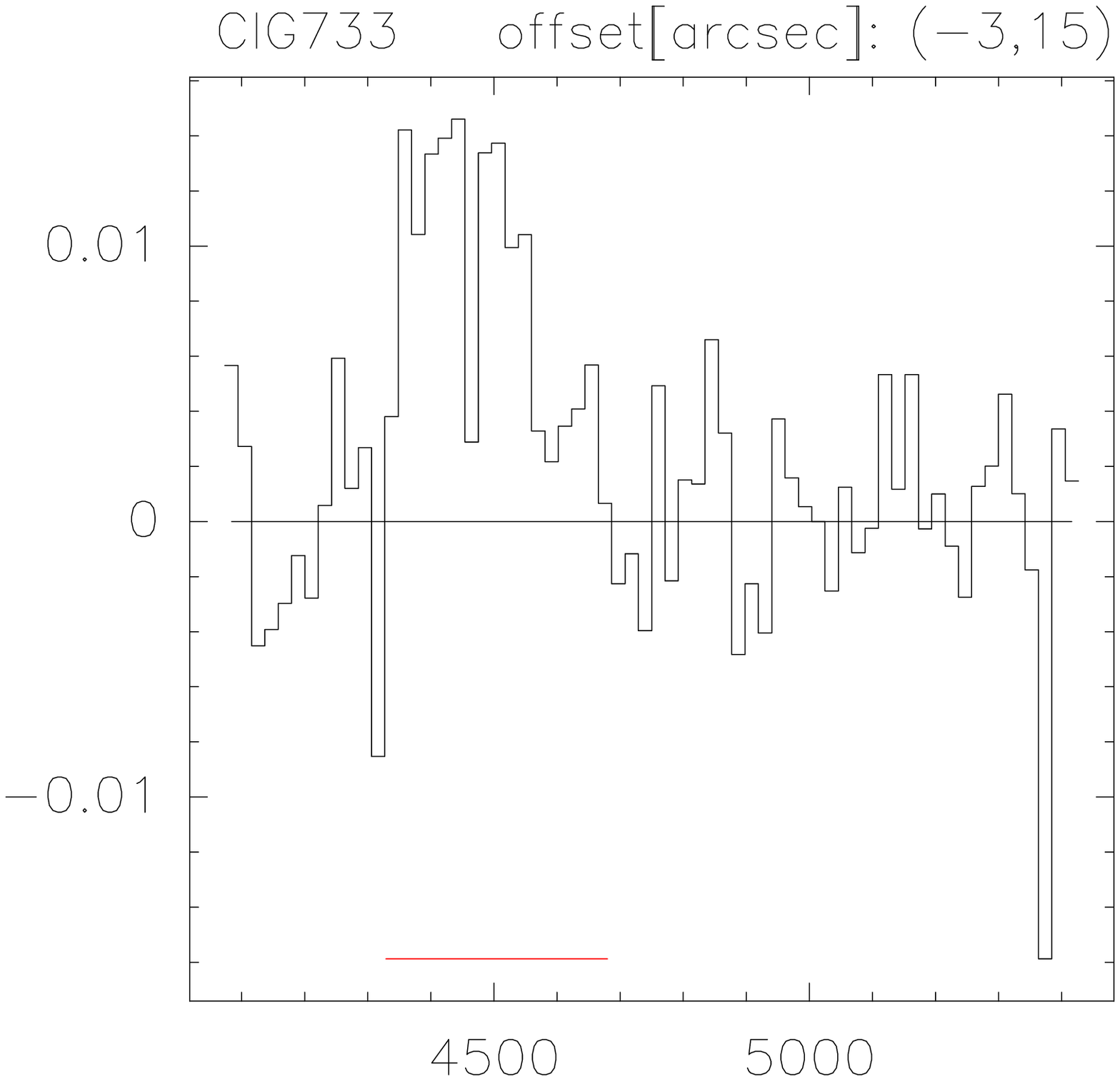}\quad 
\includegraphics[width=3cm,angle=270]{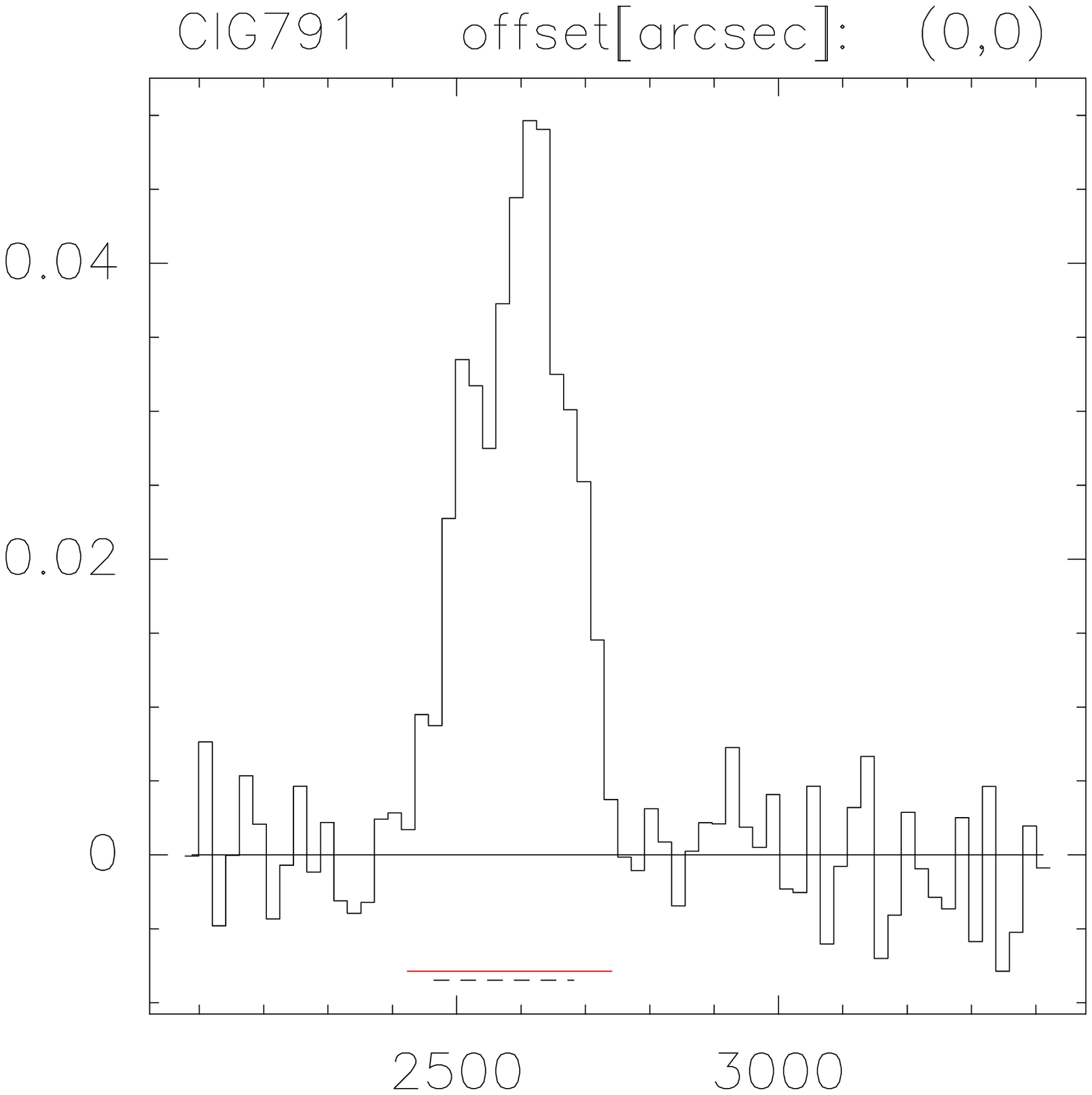}\quad 
\includegraphics[width=3cm,angle=270]{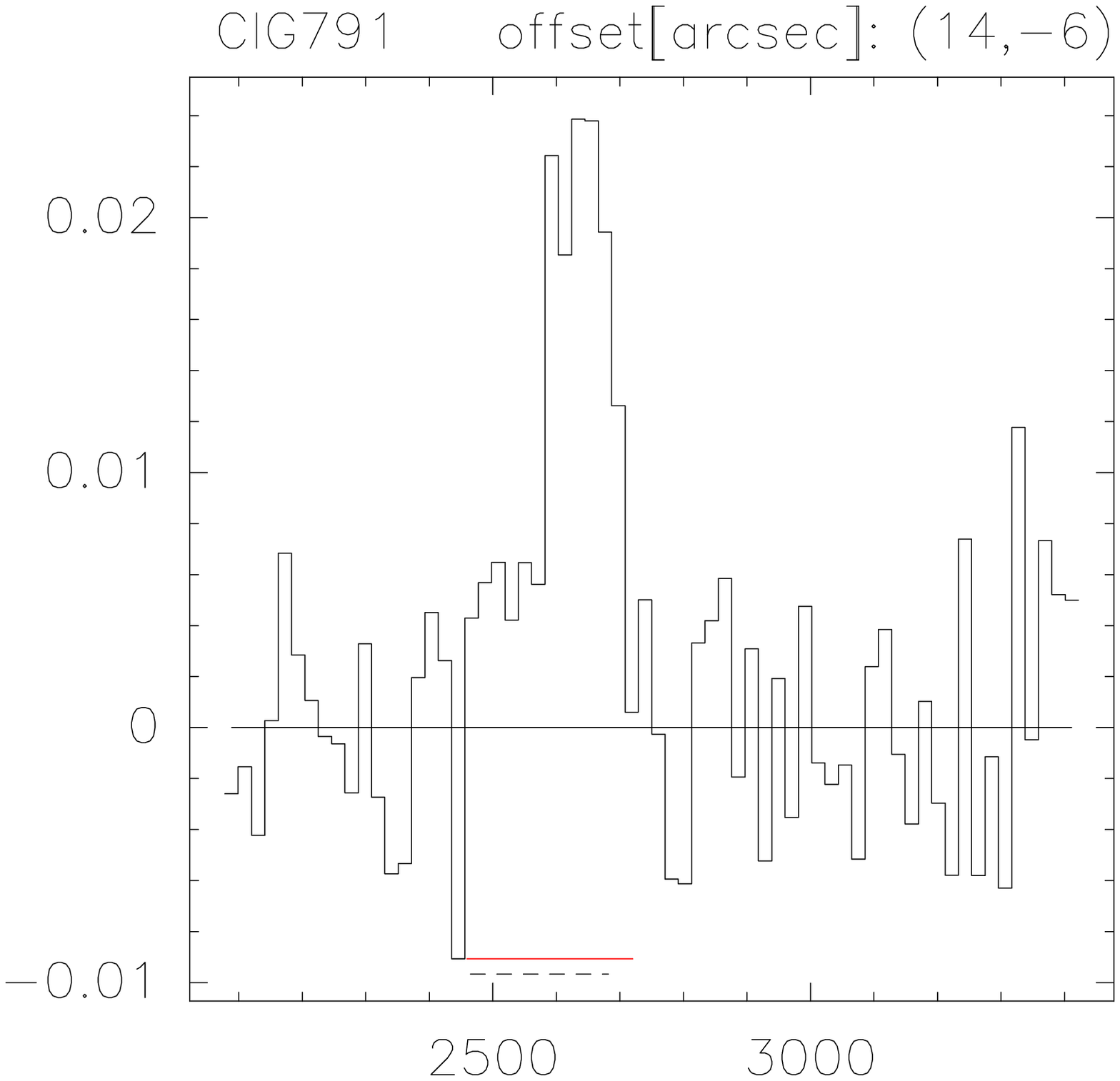}} 
\centerline{\includegraphics[width=3cm,angle=270]{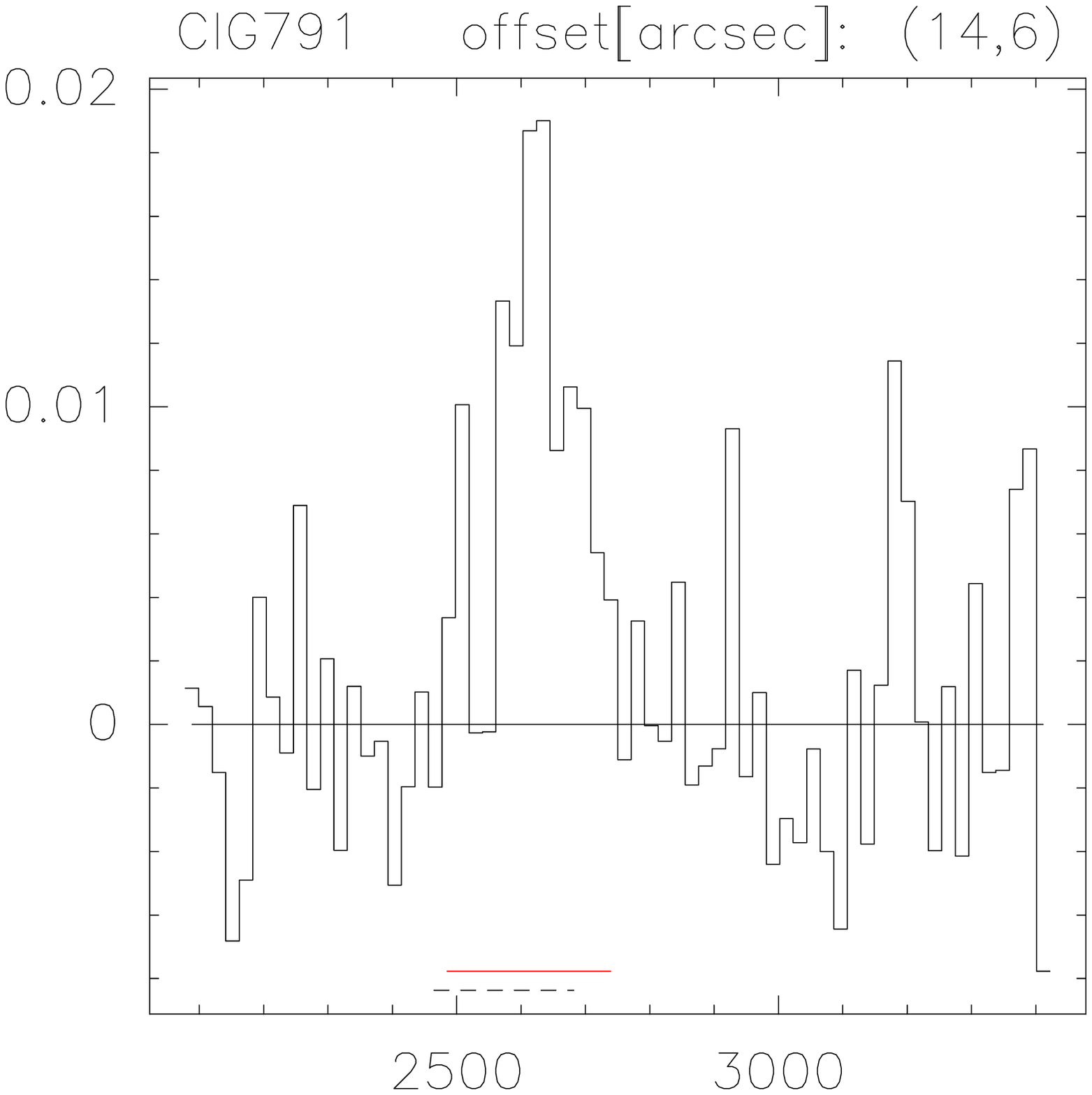} \quad 
\includegraphics[width=3cm,angle=270]{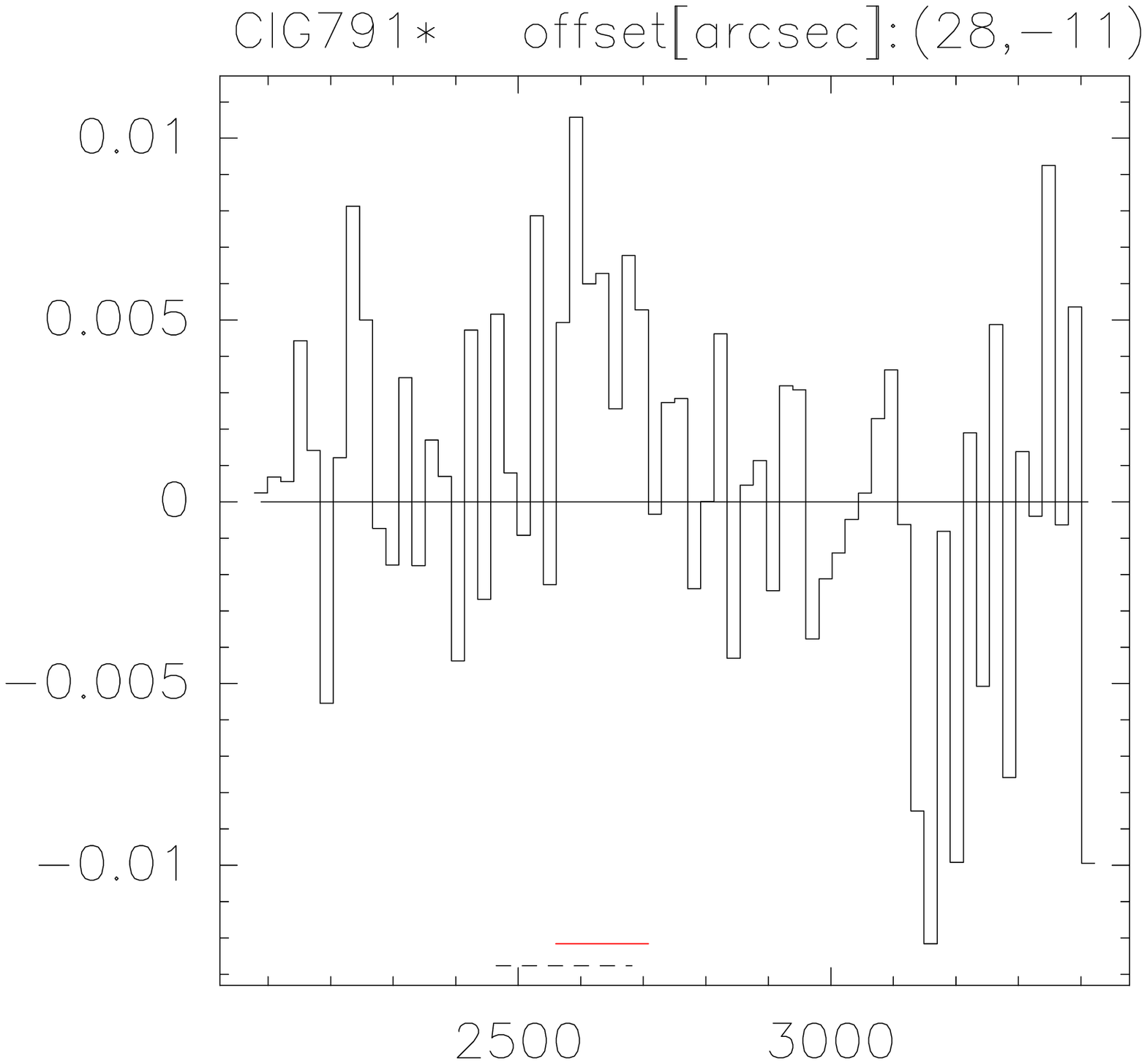}\quad 
\includegraphics[width=3cm,angle=270]{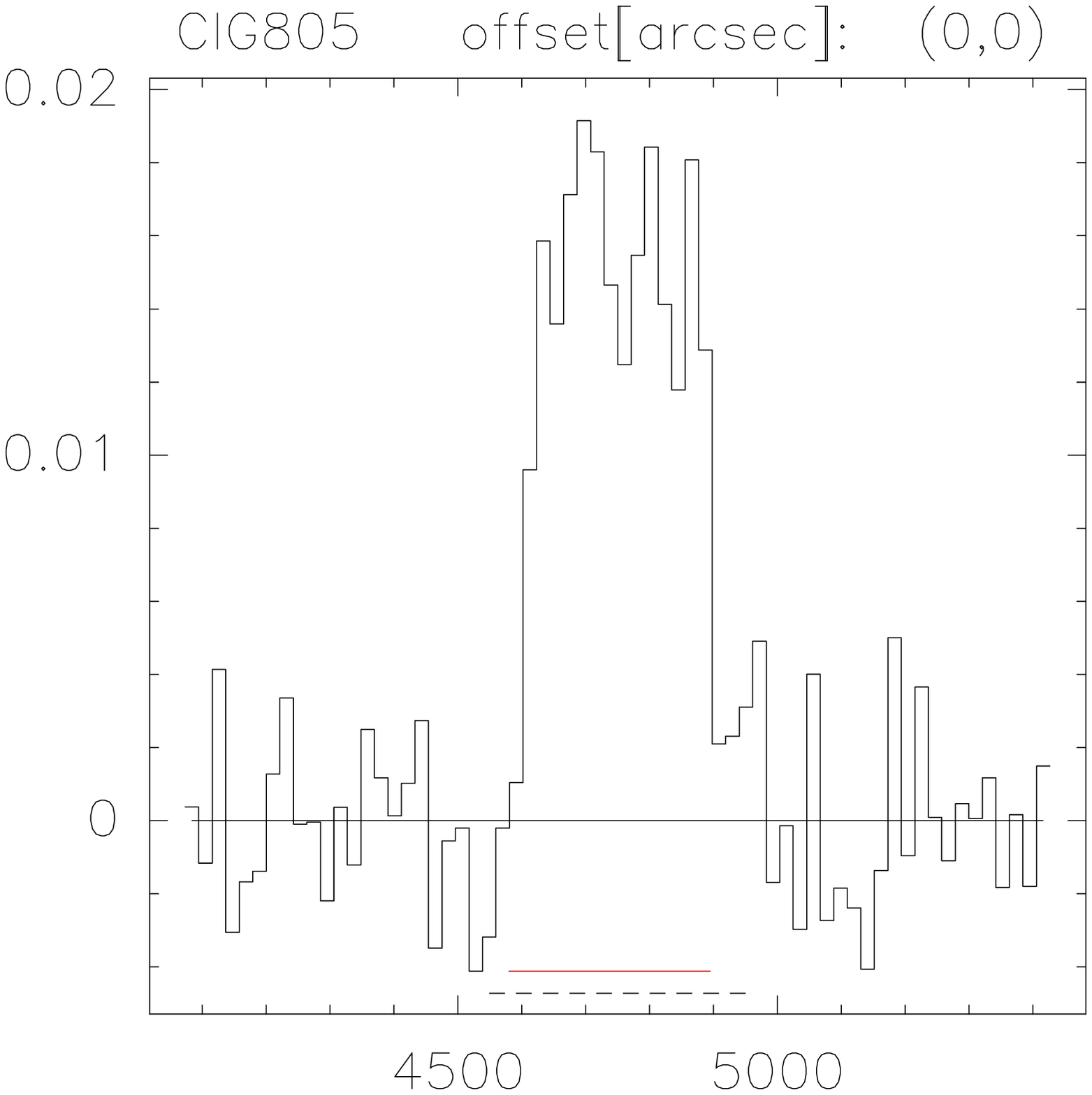}\quad 
\includegraphics[width=3cm,angle=270]{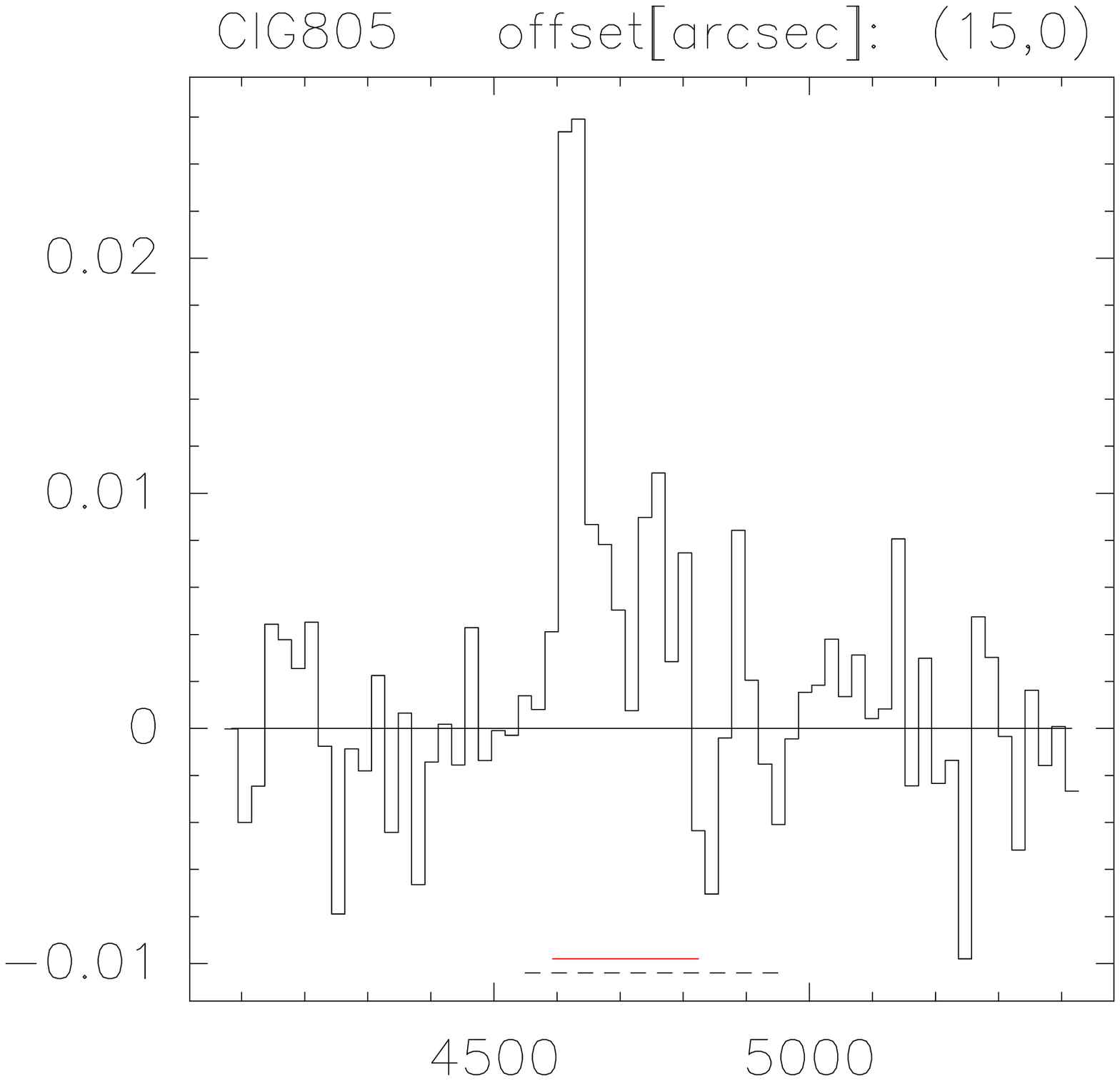}\quad 
\includegraphics[width=3cm,angle=270]{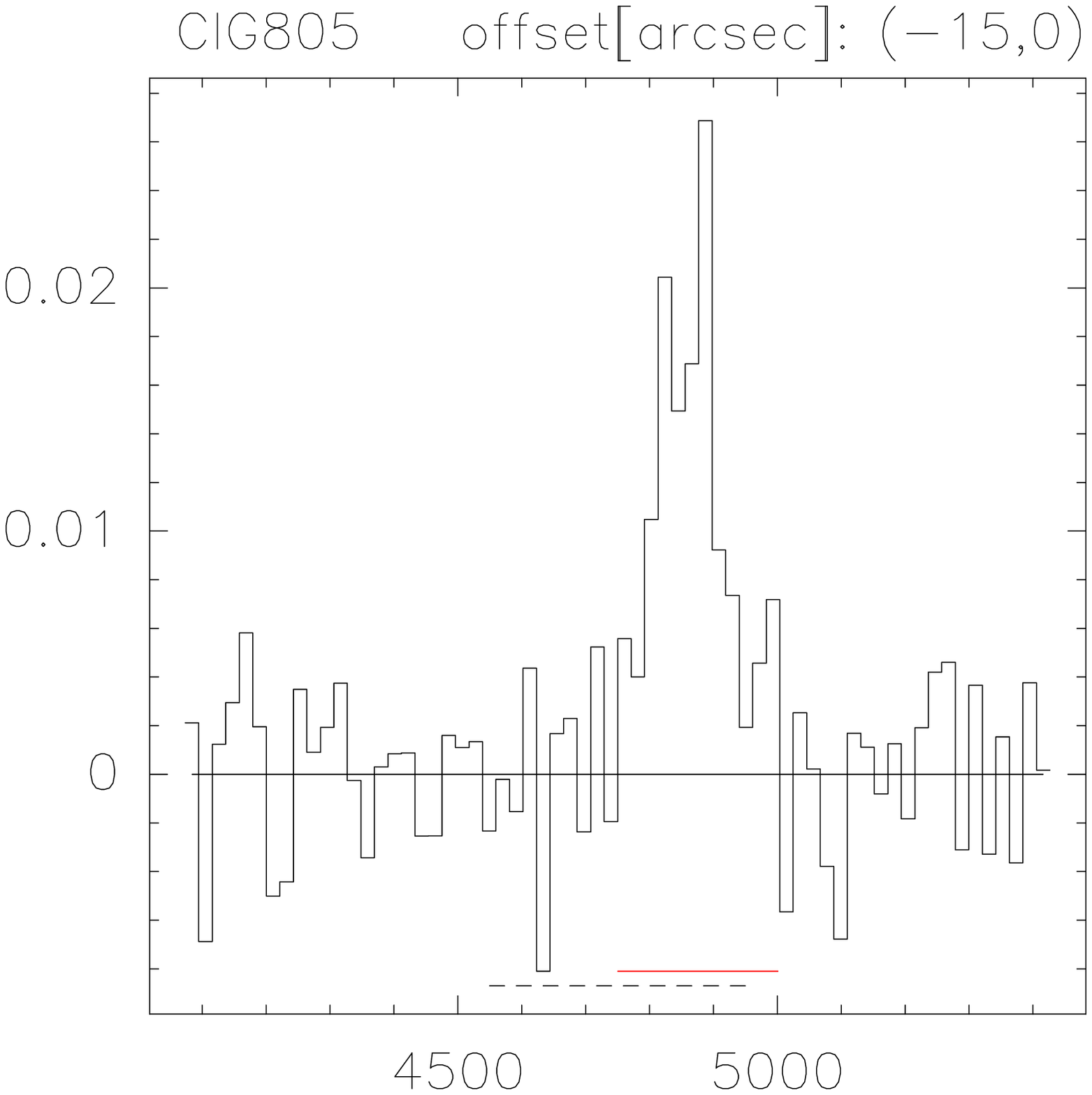}} 
\centerline{\includegraphics[width=3cm,angle=270]{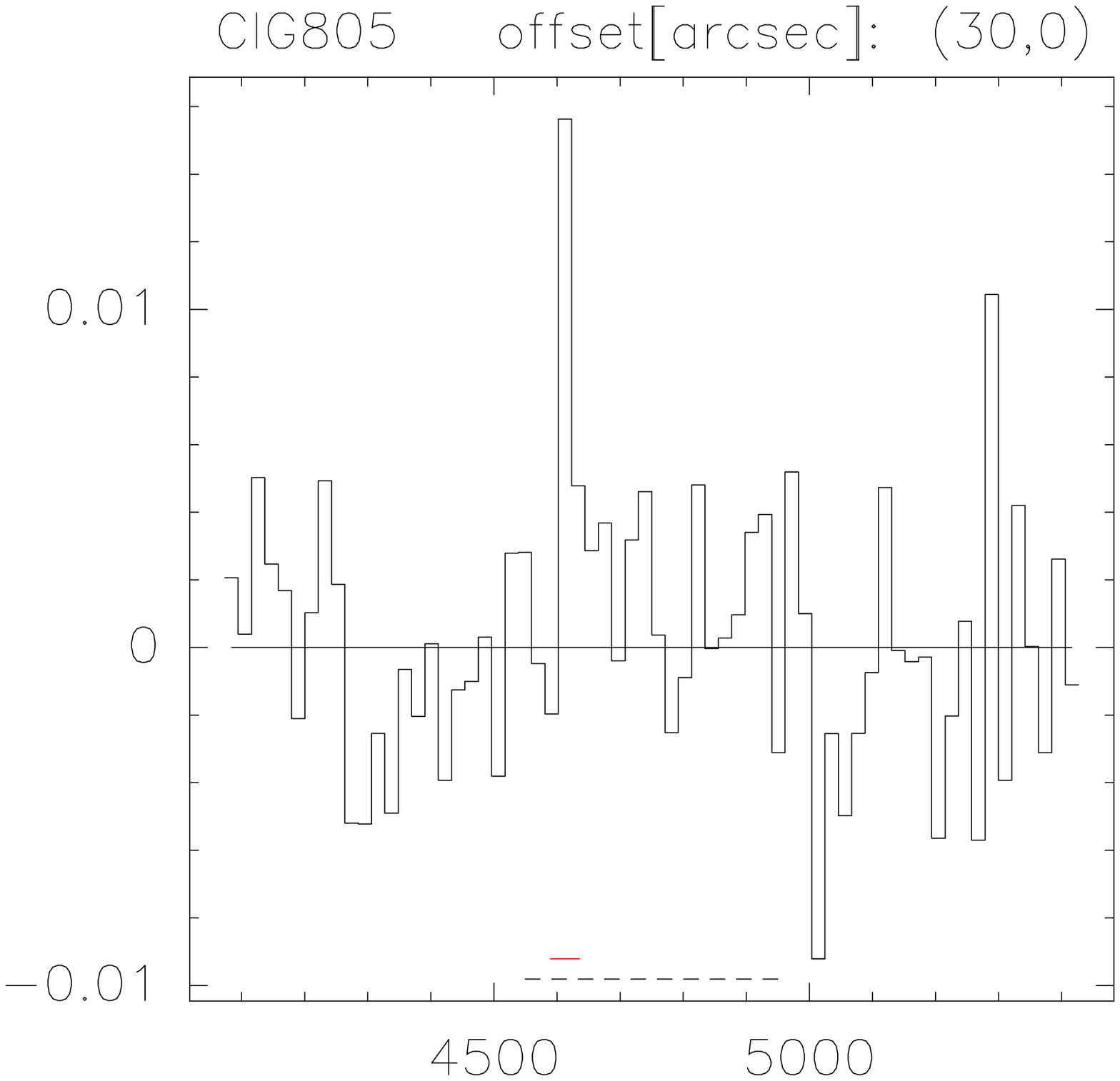} \quad 
\includegraphics[width=3cm,angle=270]{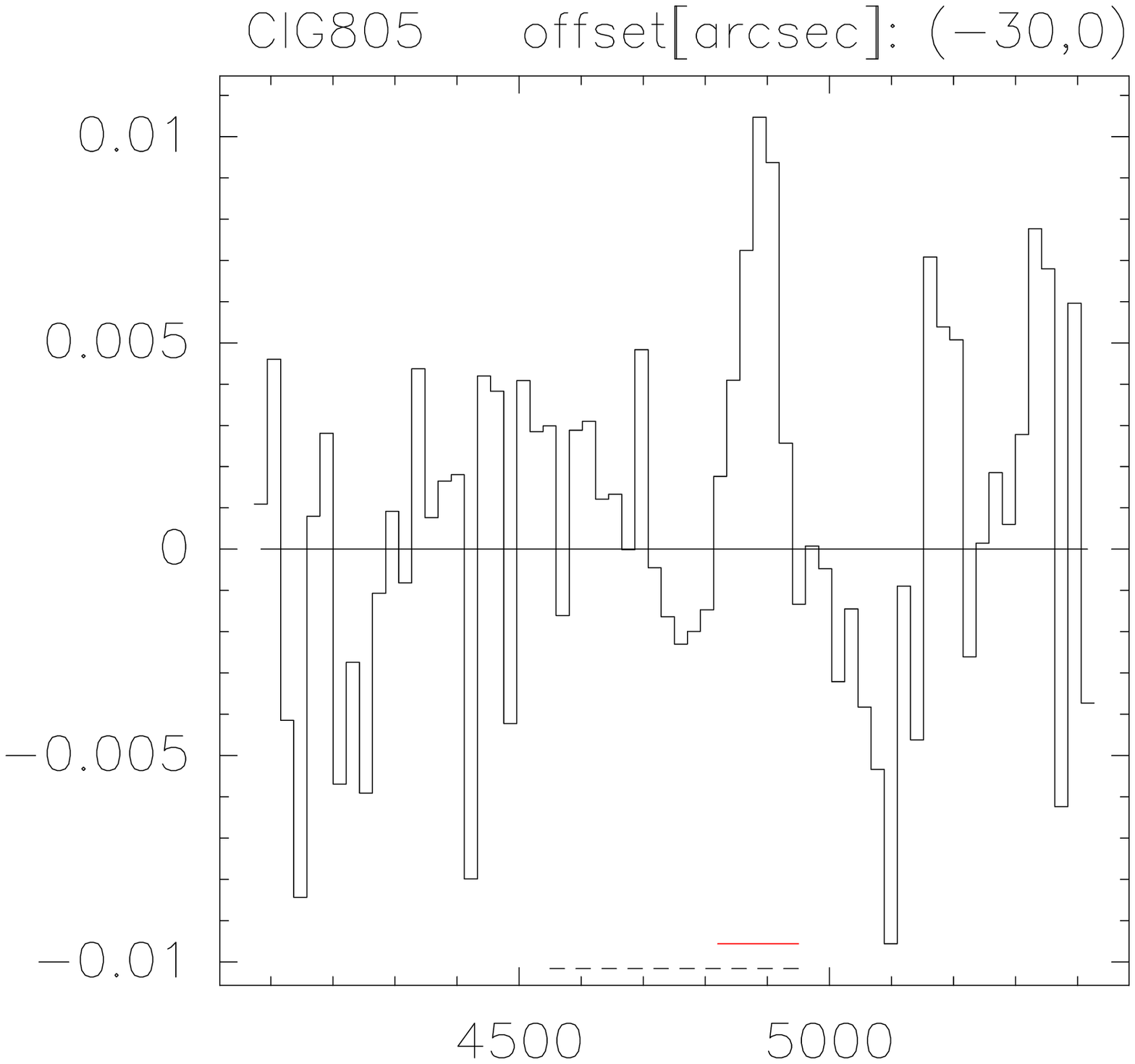}\quad 
\includegraphics[width=3cm,angle=270]{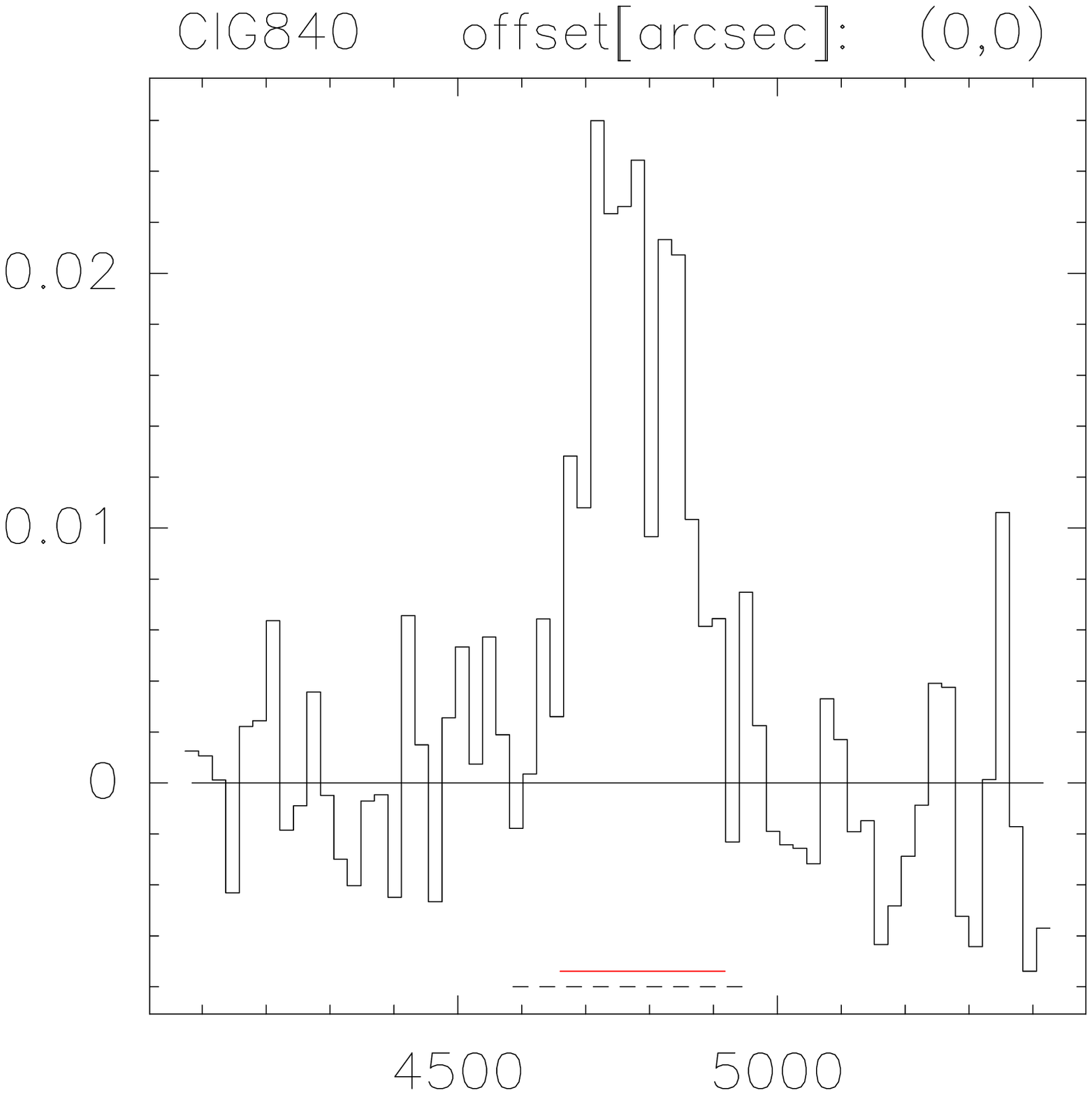}\quad 
\includegraphics[width=3cm,angle=270]{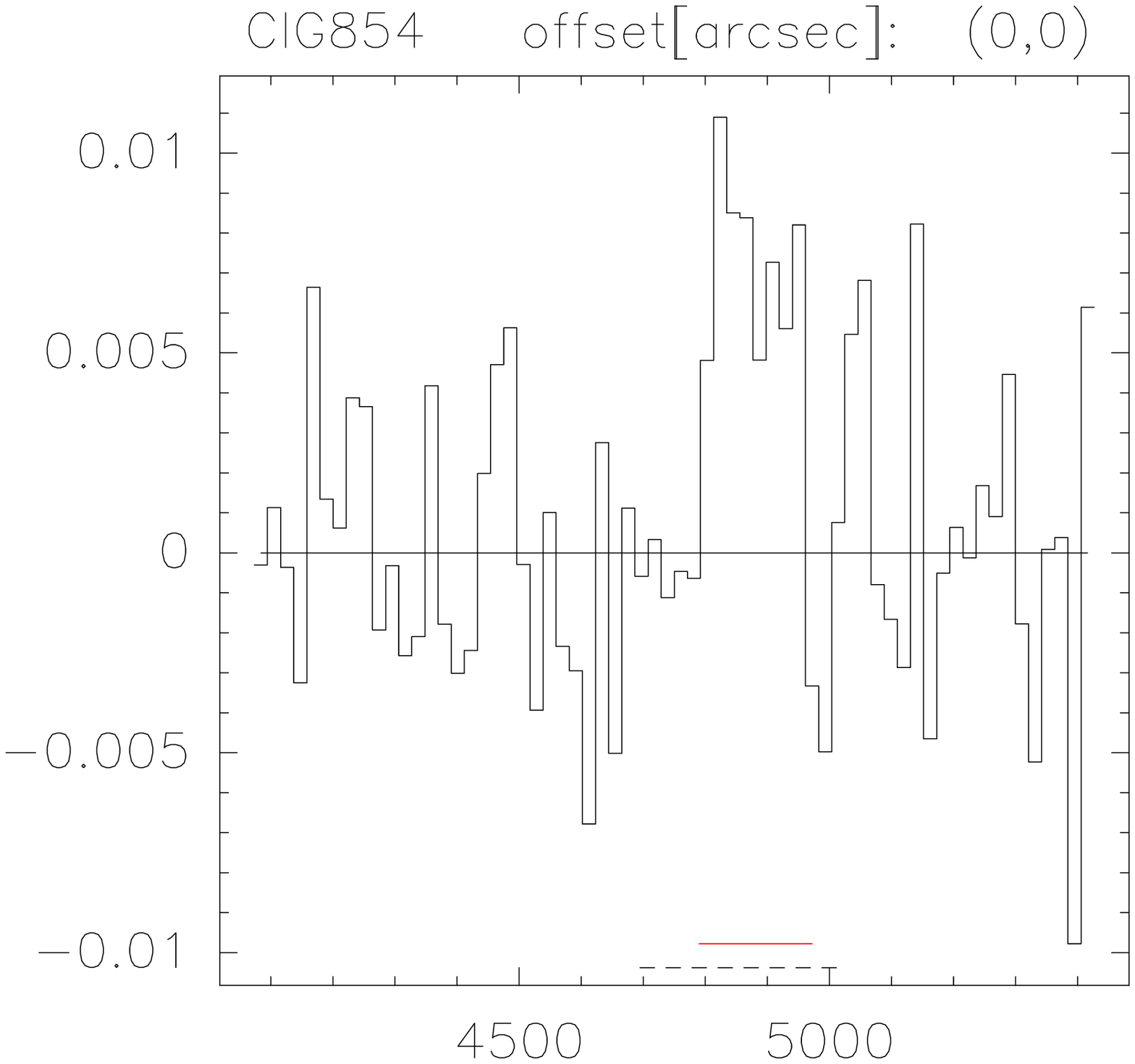}\quad 
\includegraphics[width=3cm,angle=270]{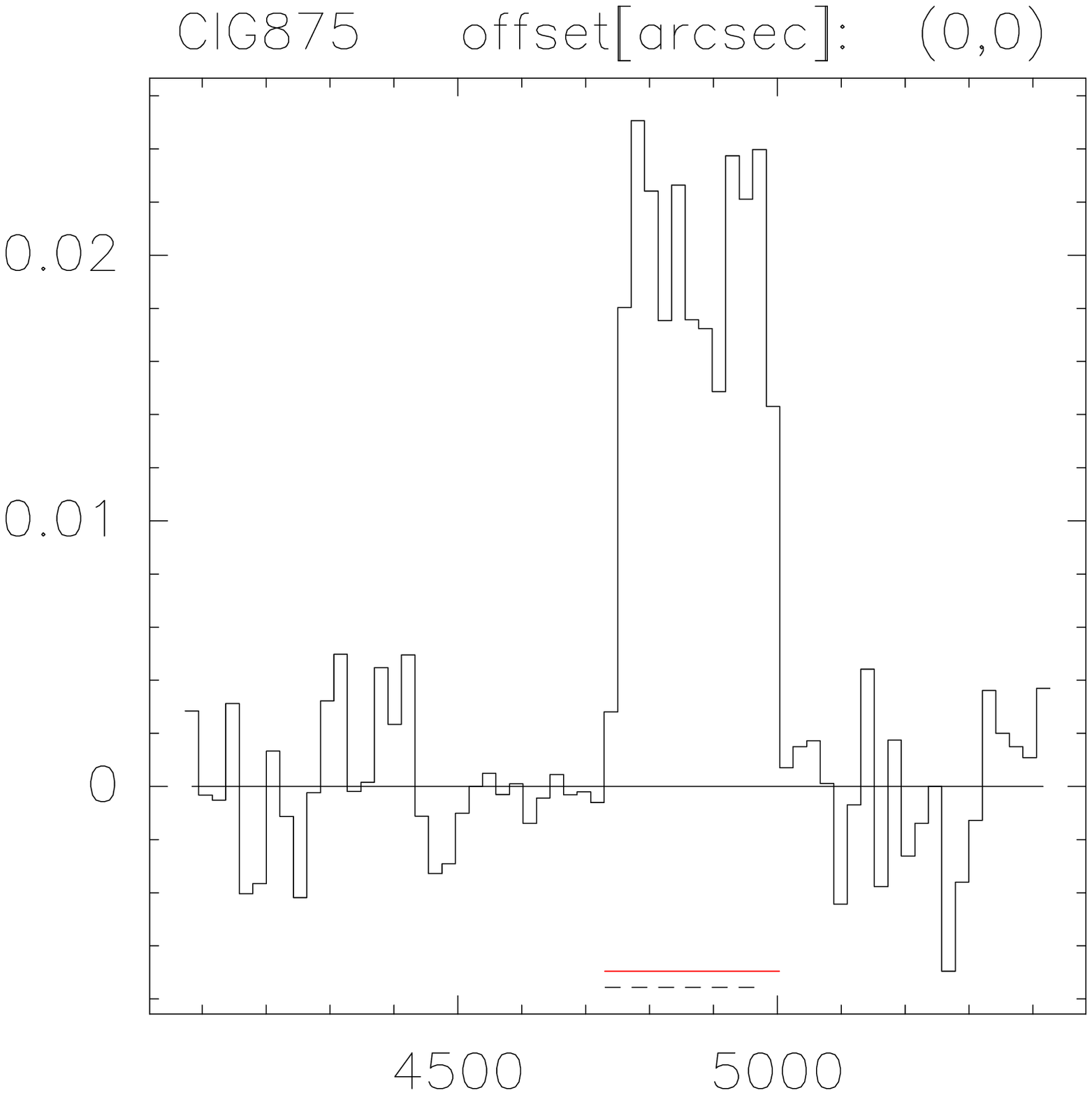}} 
\centerline{\includegraphics[width=3cm,angle=270]{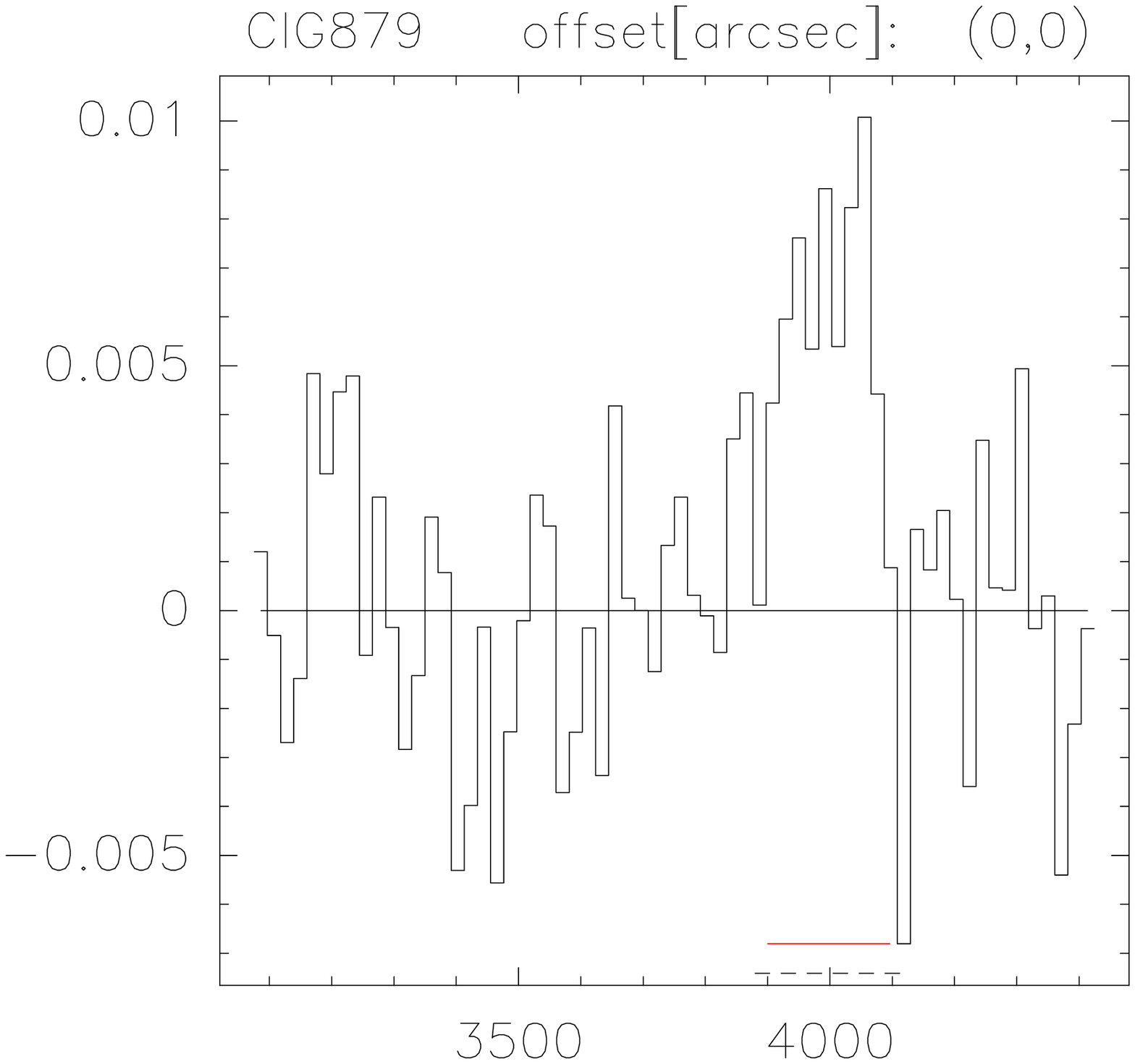} \quad 
\includegraphics[width=3cm,angle=270]{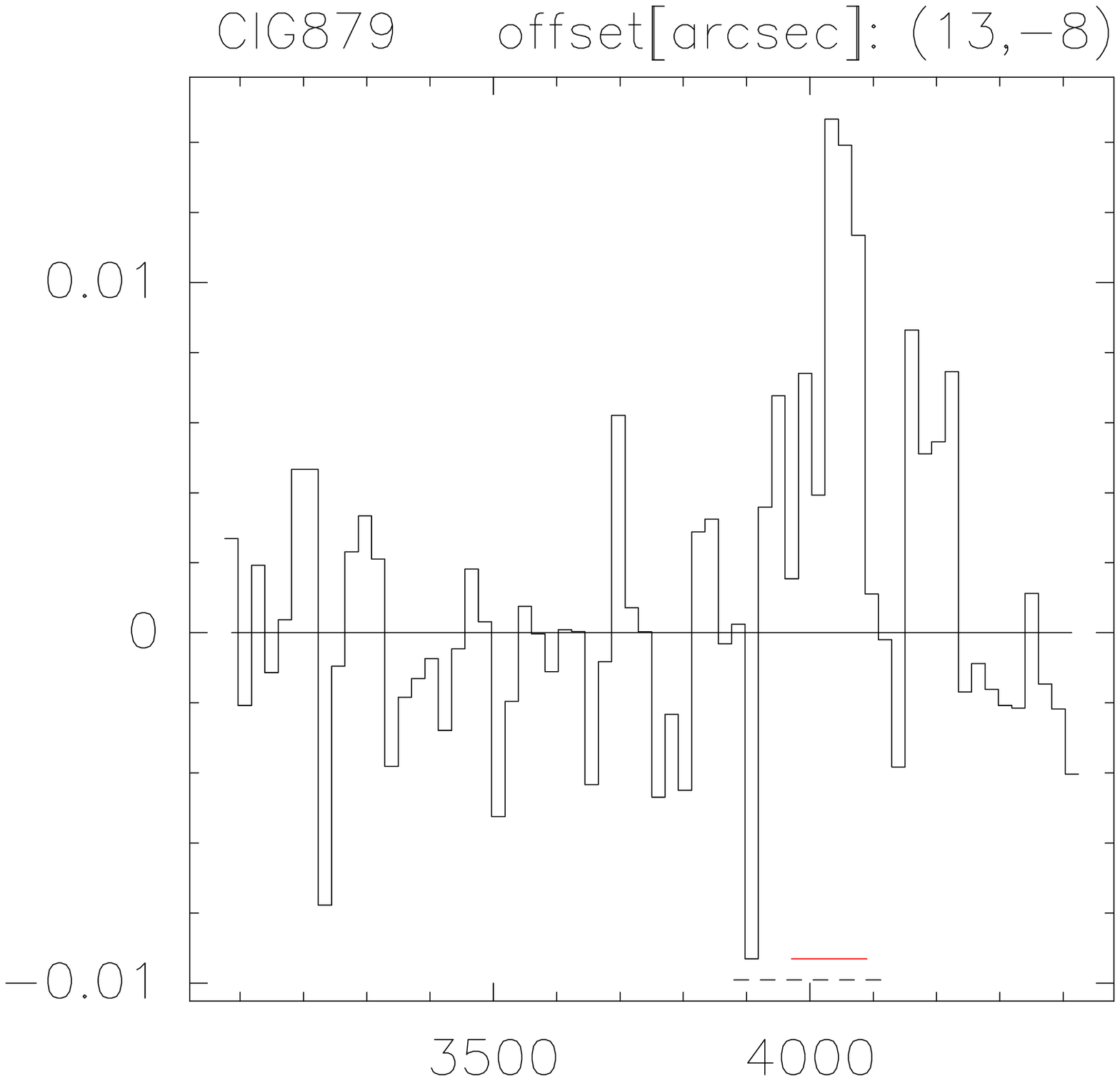}\quad 
\includegraphics[width=3cm,angle=270]{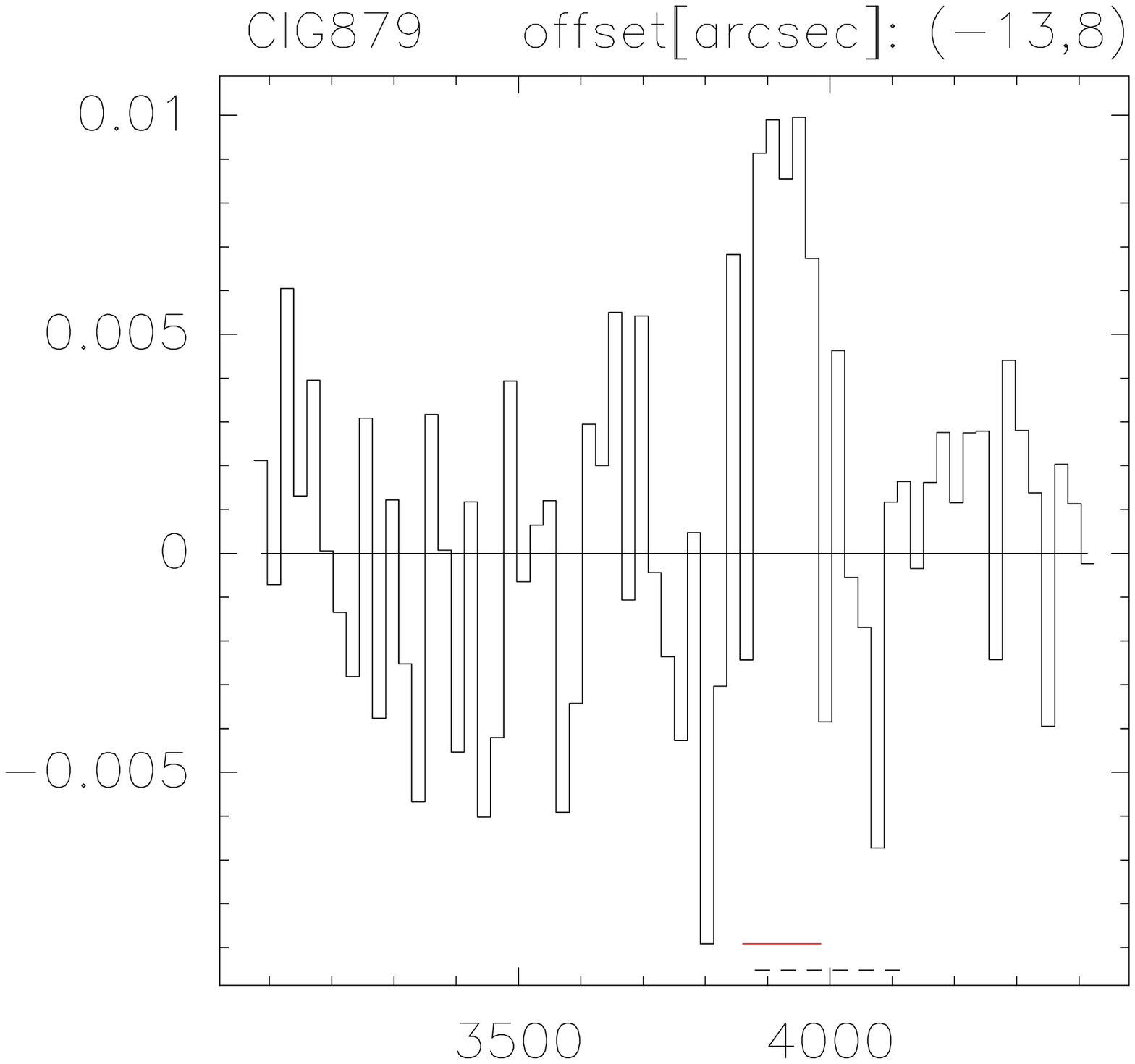}\quad 
\includegraphics[width=3cm,angle=270]{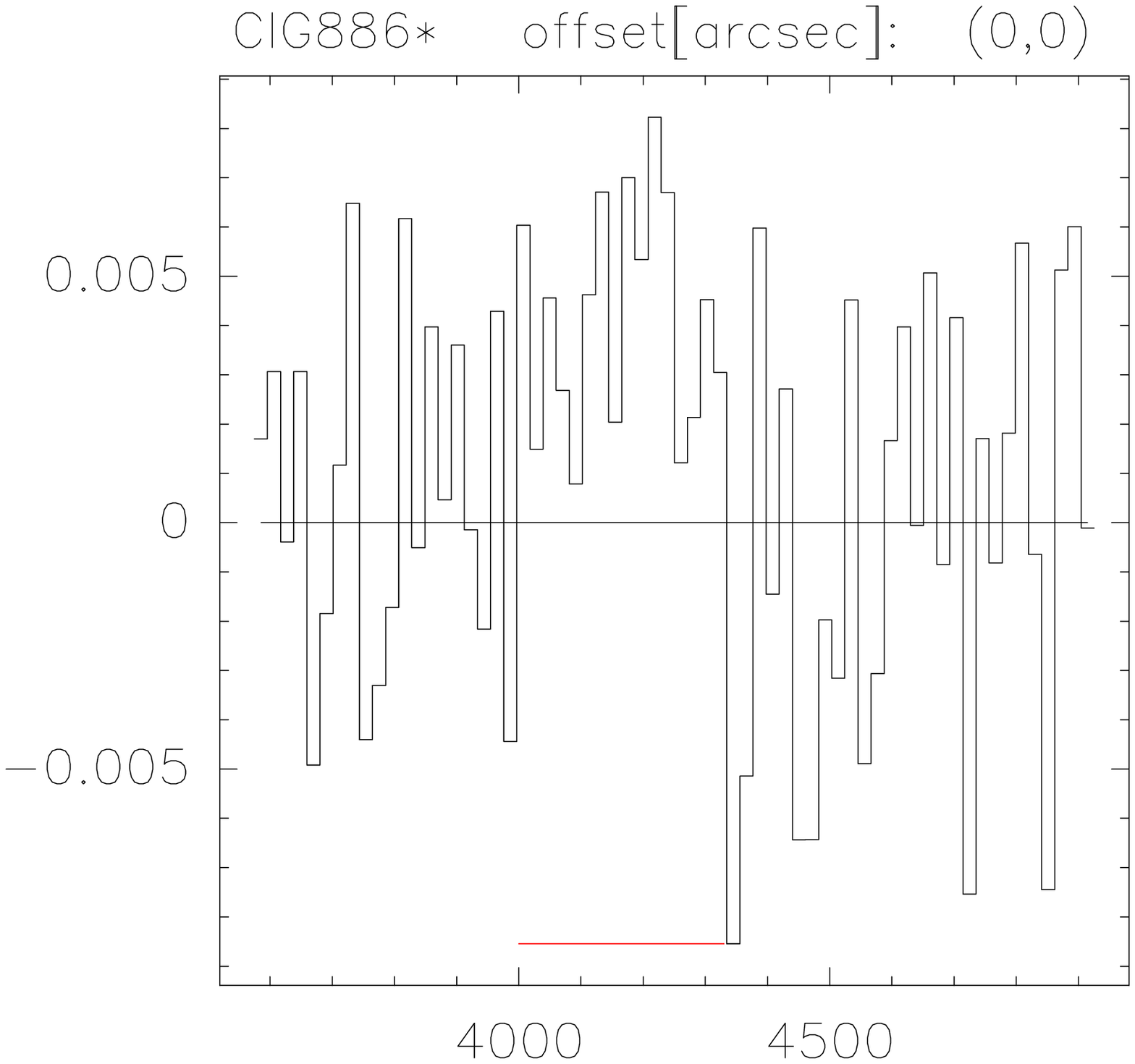}\quad 
\includegraphics[width=3cm,angle=270]{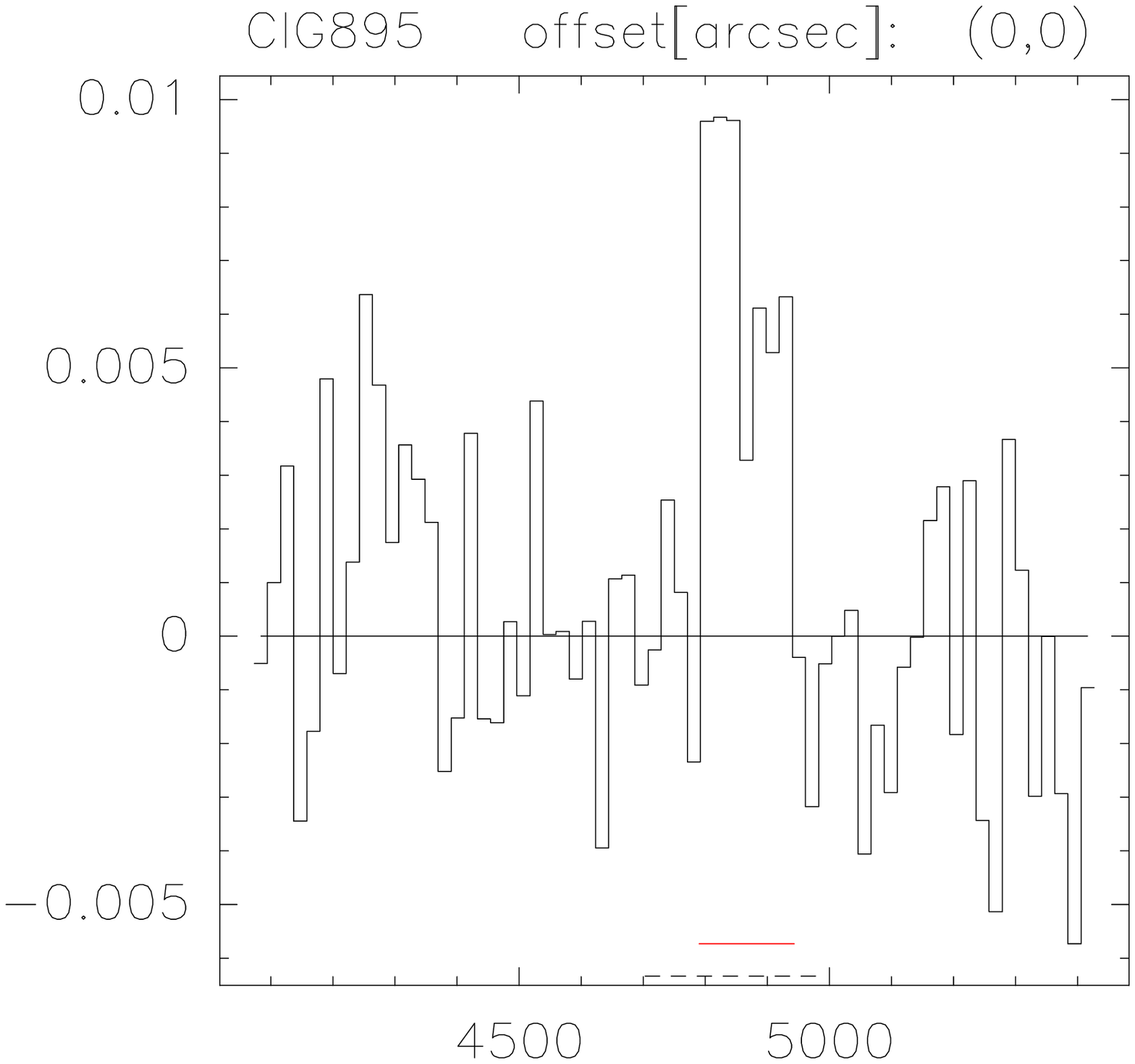}} 
\centerline{\includegraphics[width=3.3cm,angle=270]{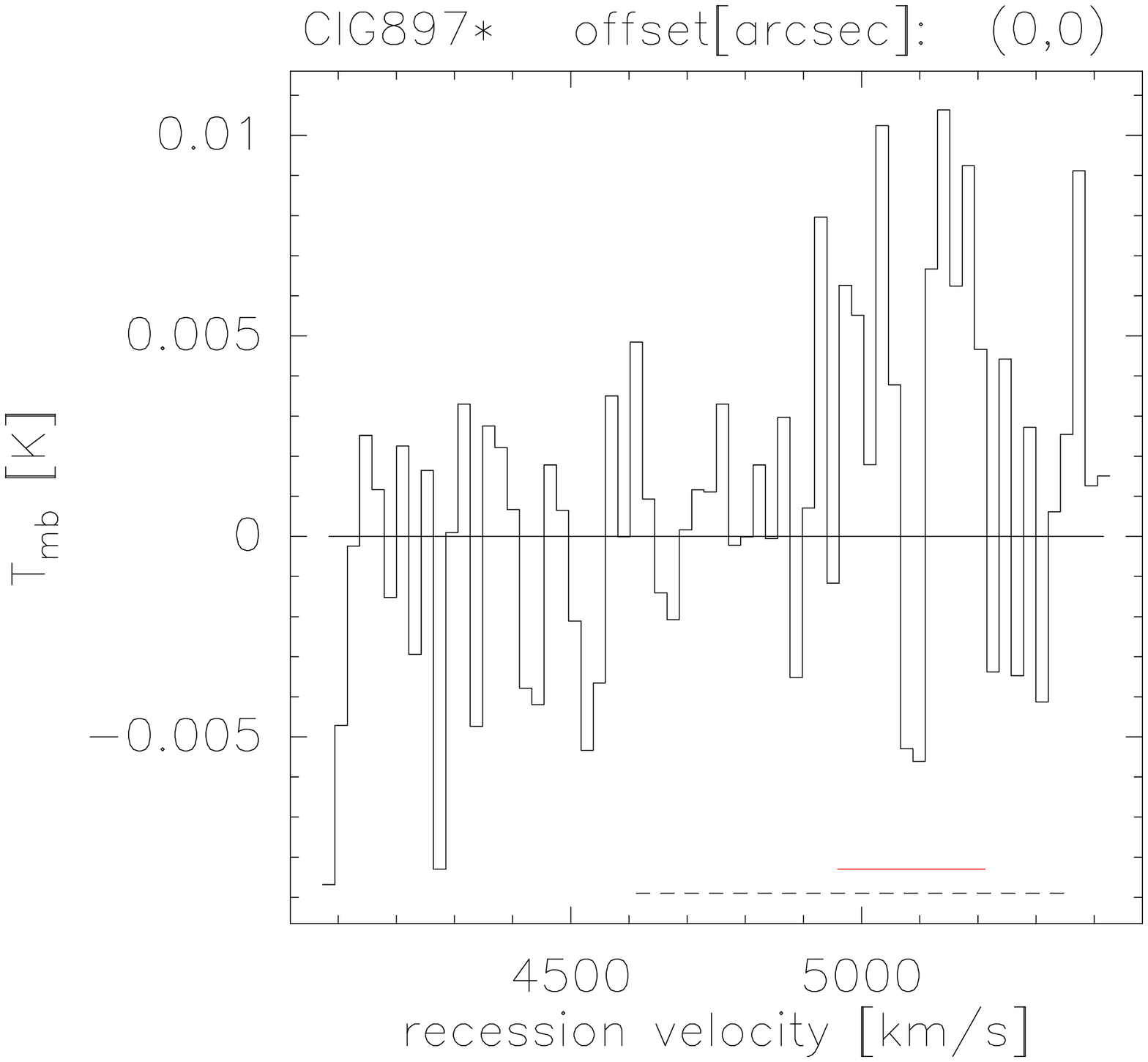} \quad 
\includegraphics[width=3cm,angle=270]{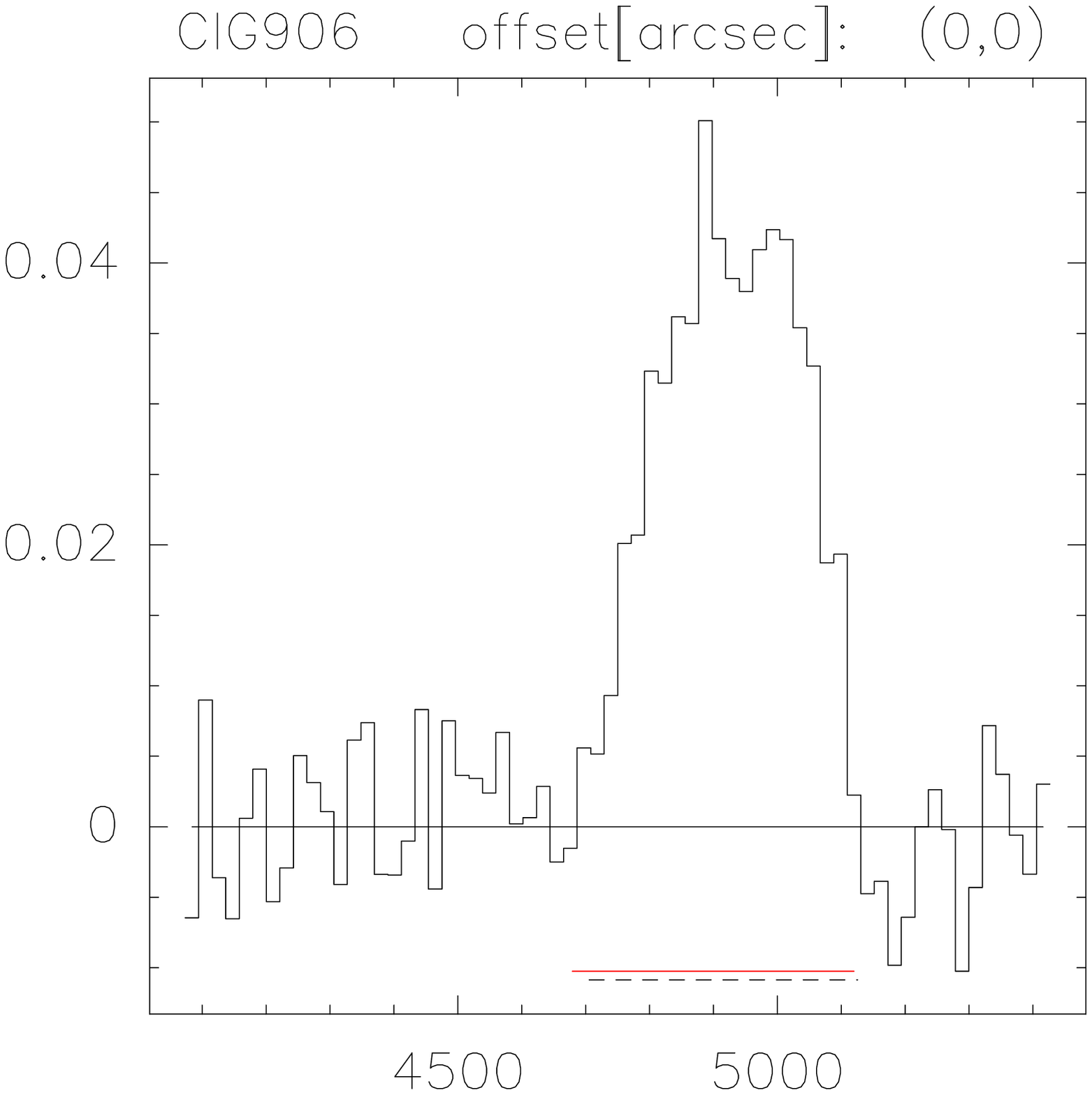}\quad 
\includegraphics[width=3cm,angle=270]{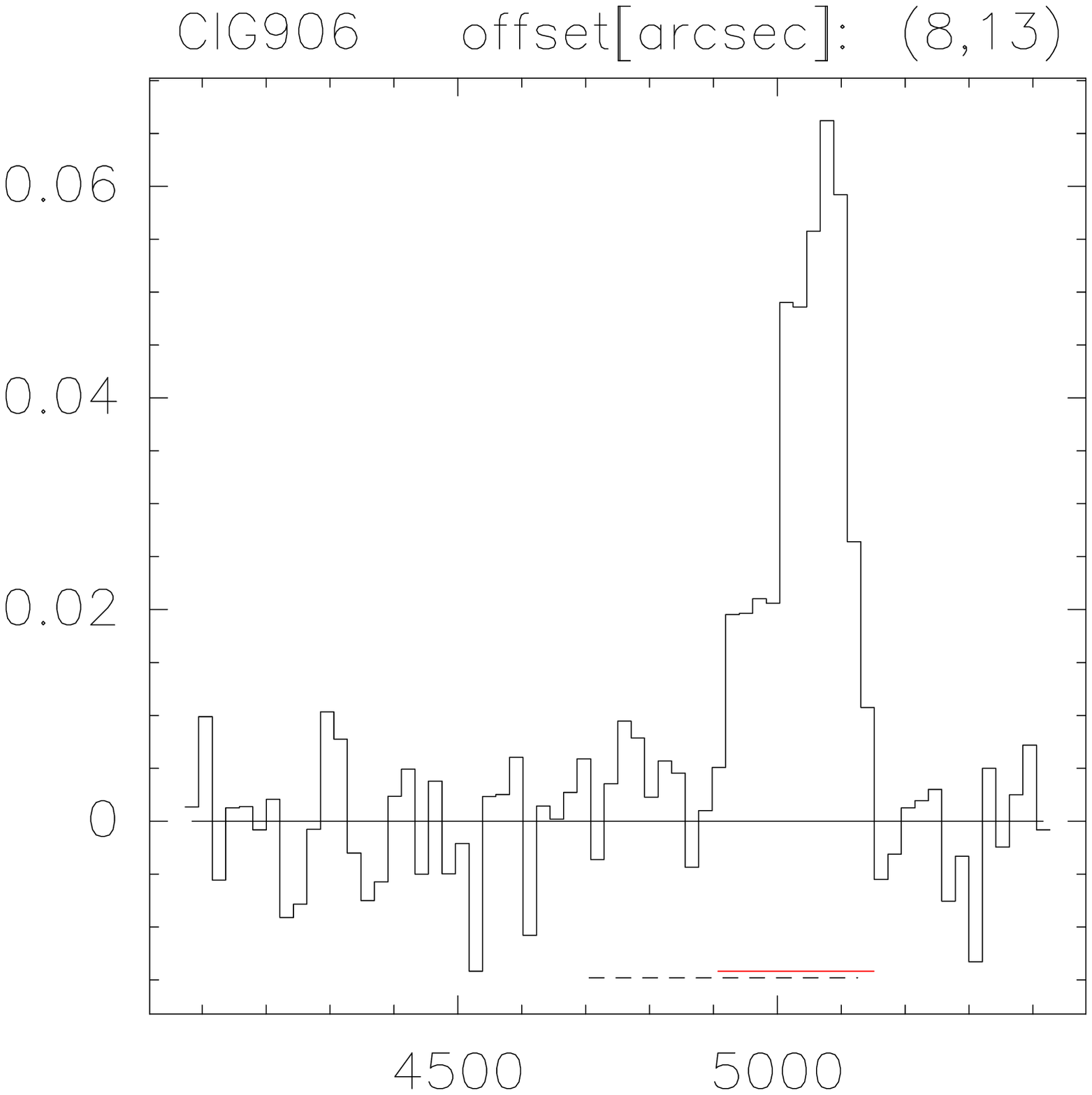}\quad 
\includegraphics[width=3cm,angle=270]{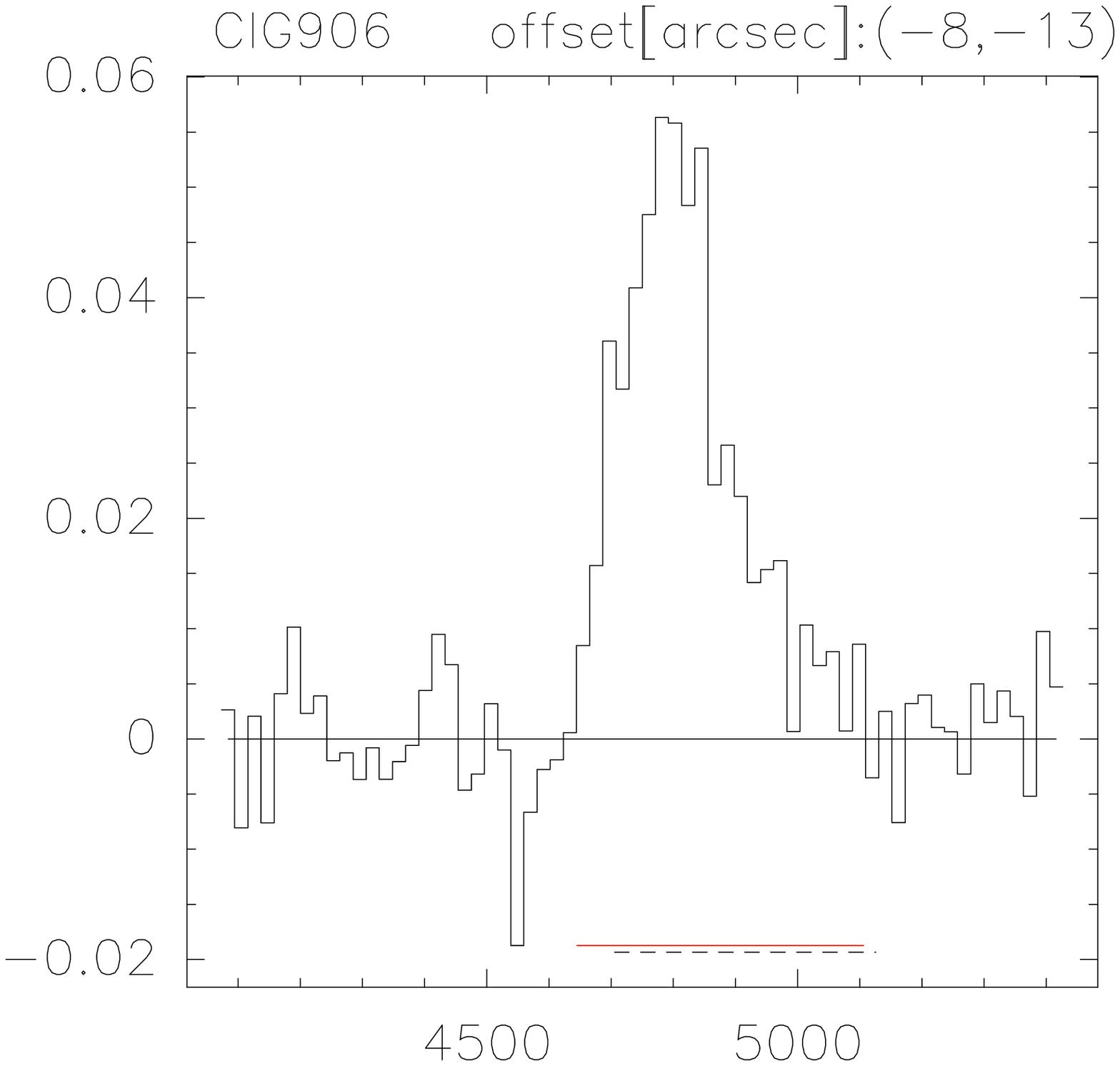}\quad 
\includegraphics[width=3cm,angle=270]{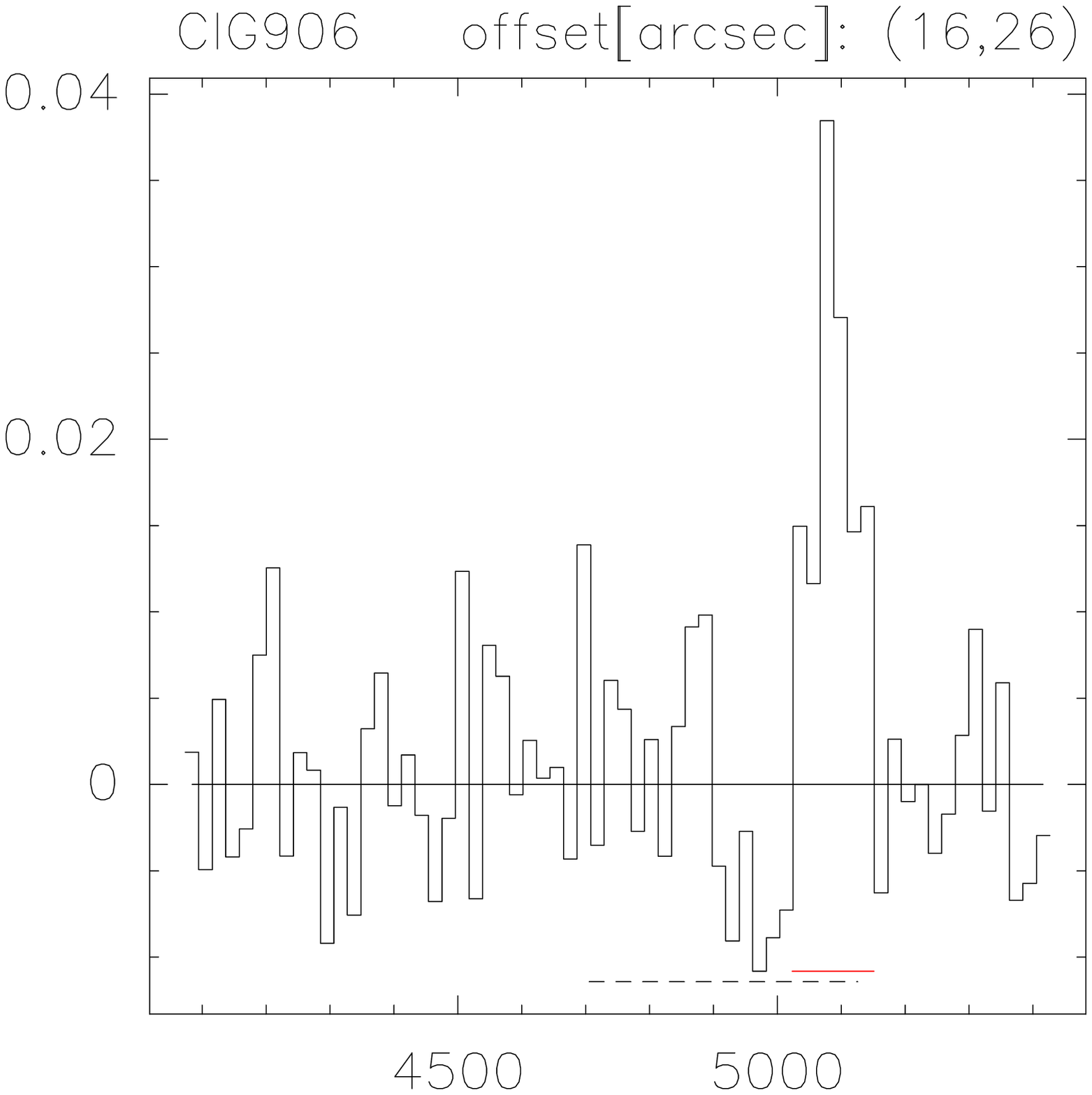}} 
\addtocounter{figure}{-1} 
\caption{(continued)} 
\end{figure*} 
  
\begin{figure*} 
\centerline{\includegraphics[width=3cm,angle=270]{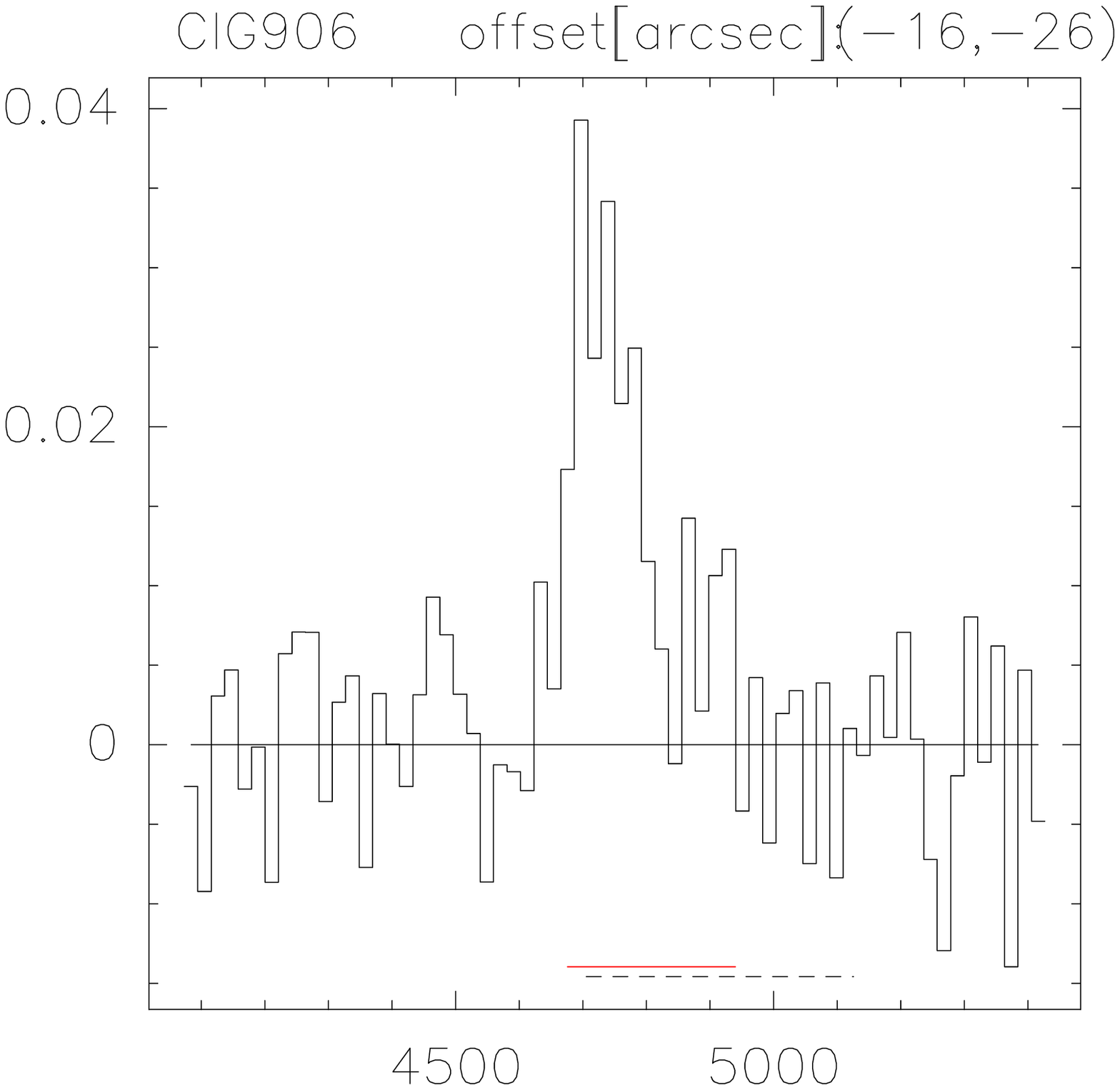} \quad 
\includegraphics[width=3cm,angle=270]{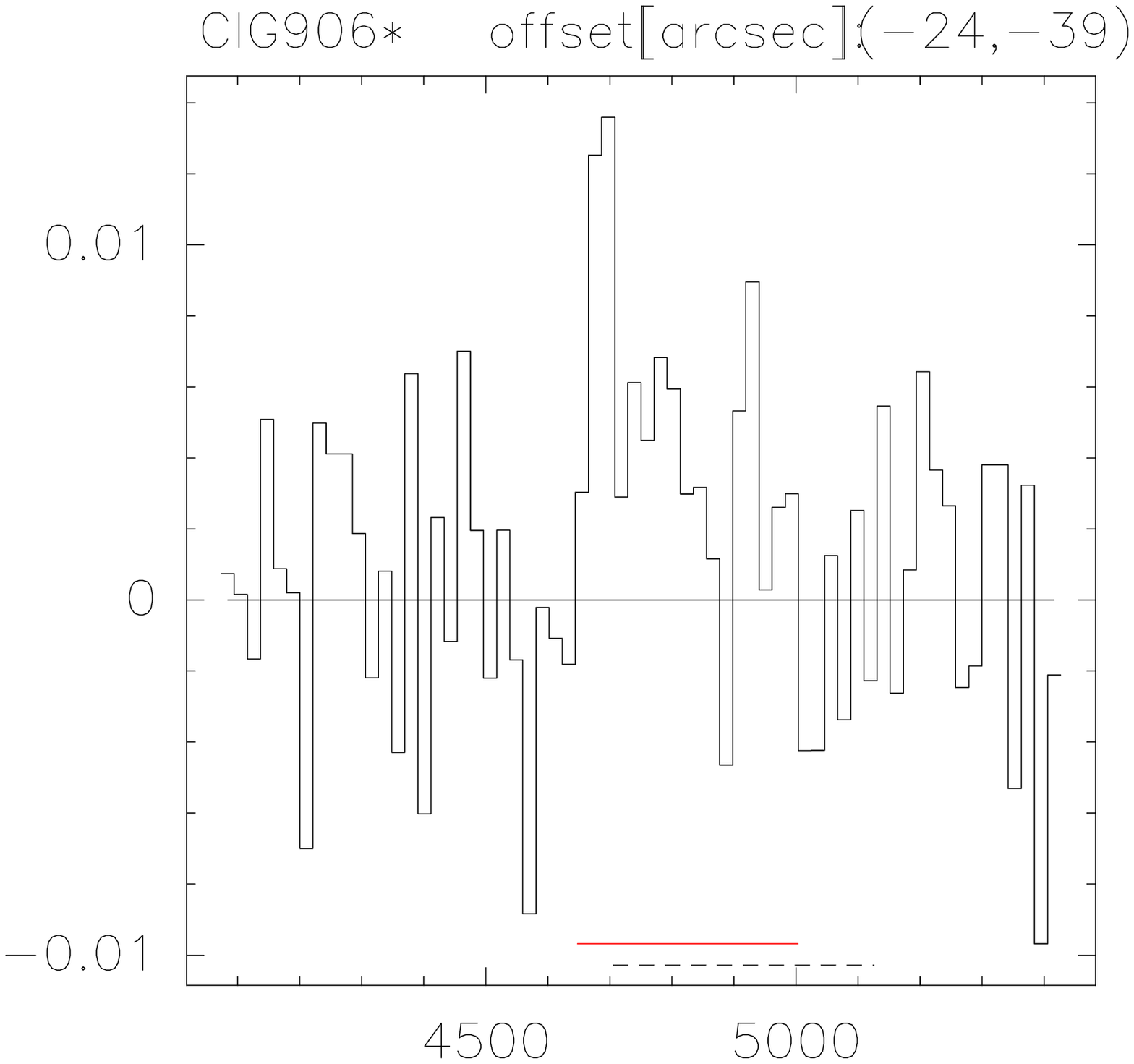}\quad 
\includegraphics[width=3cm,angle=270]{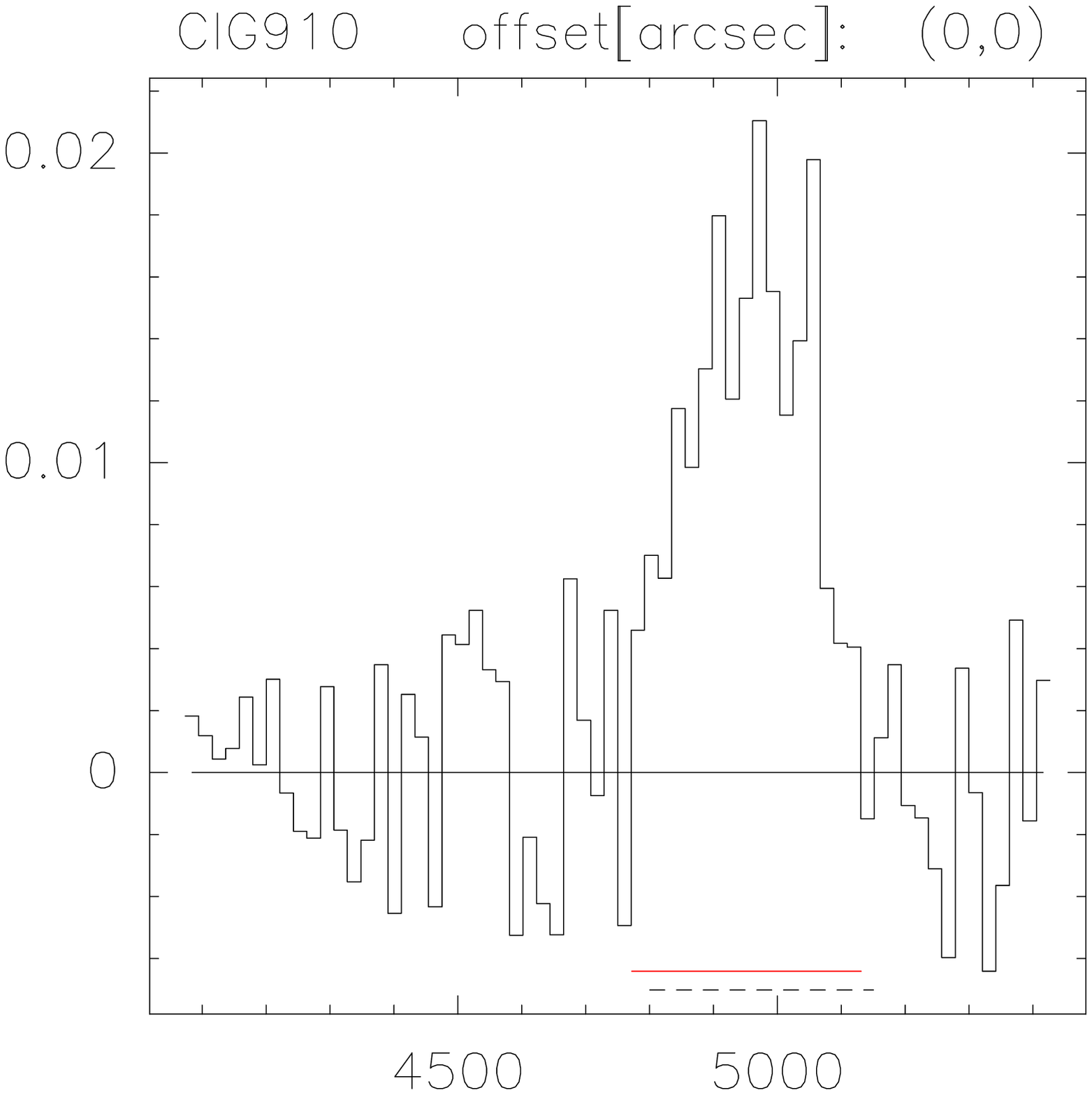}\quad 
\includegraphics[width=3cm,angle=270]{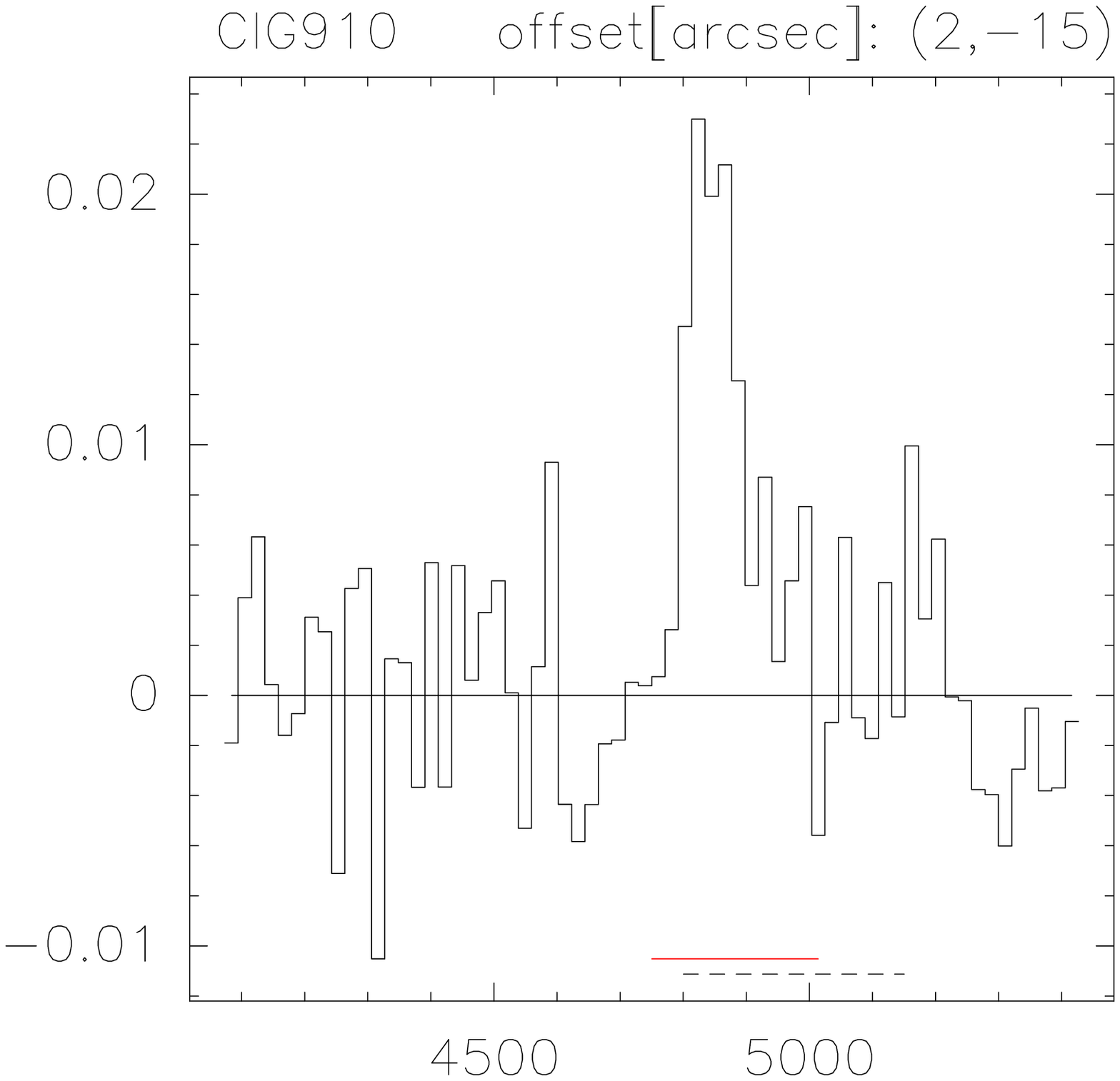}\quad 
\includegraphics[width=3cm,angle=270]{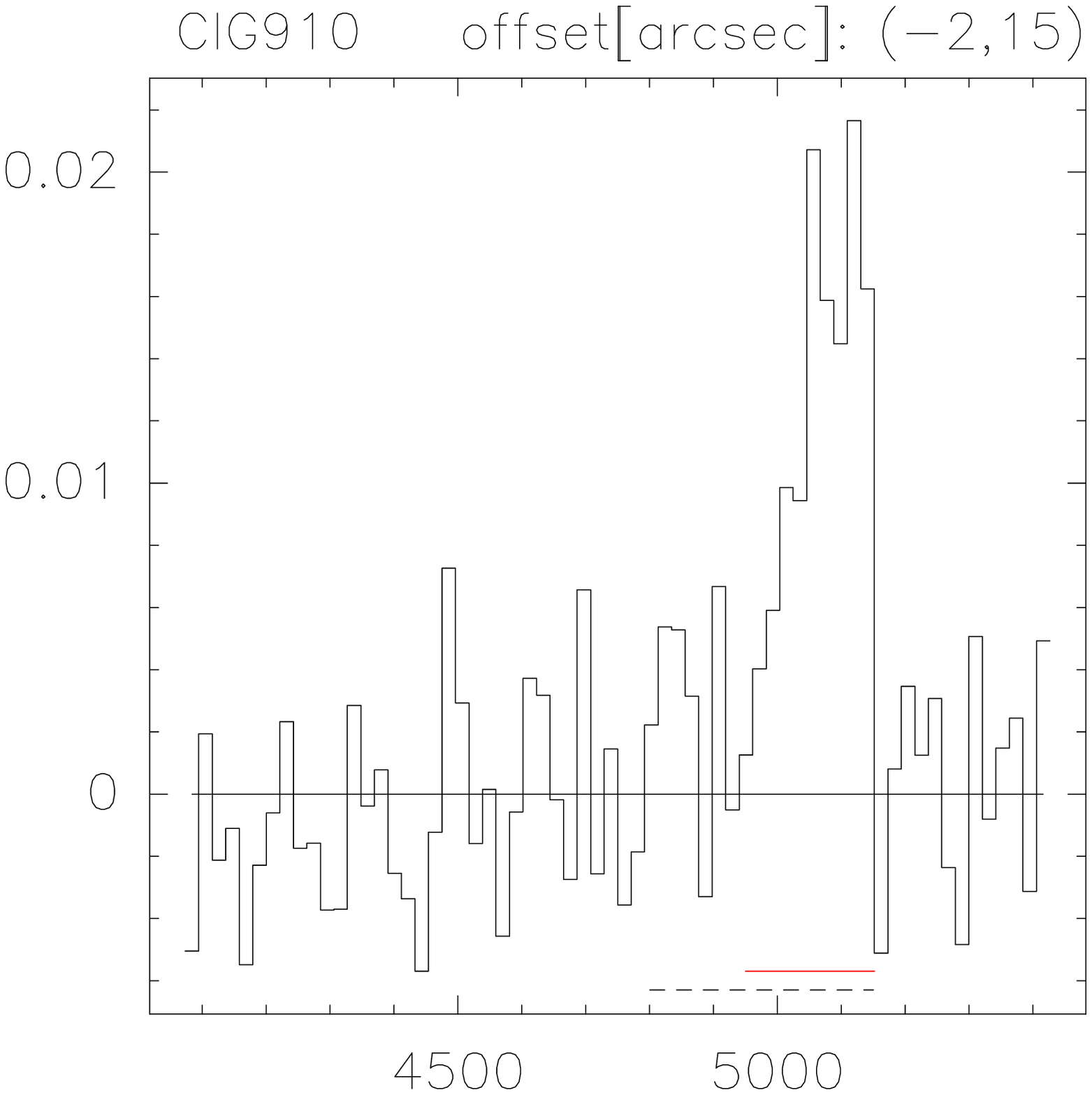}} 
\centerline{\includegraphics[width=3cm,angle=270]{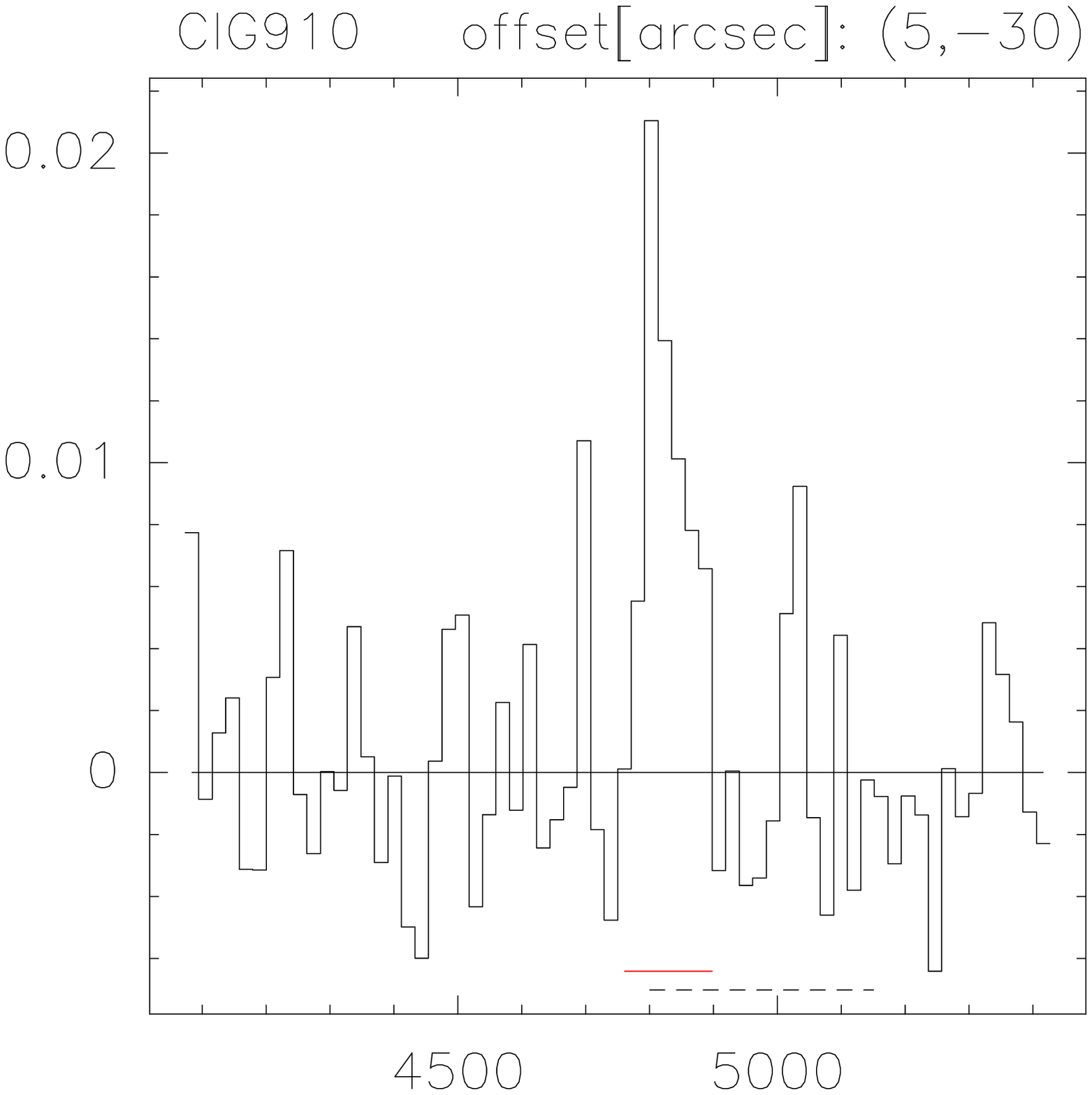} \quad 
\includegraphics[width=3cm,angle=270]{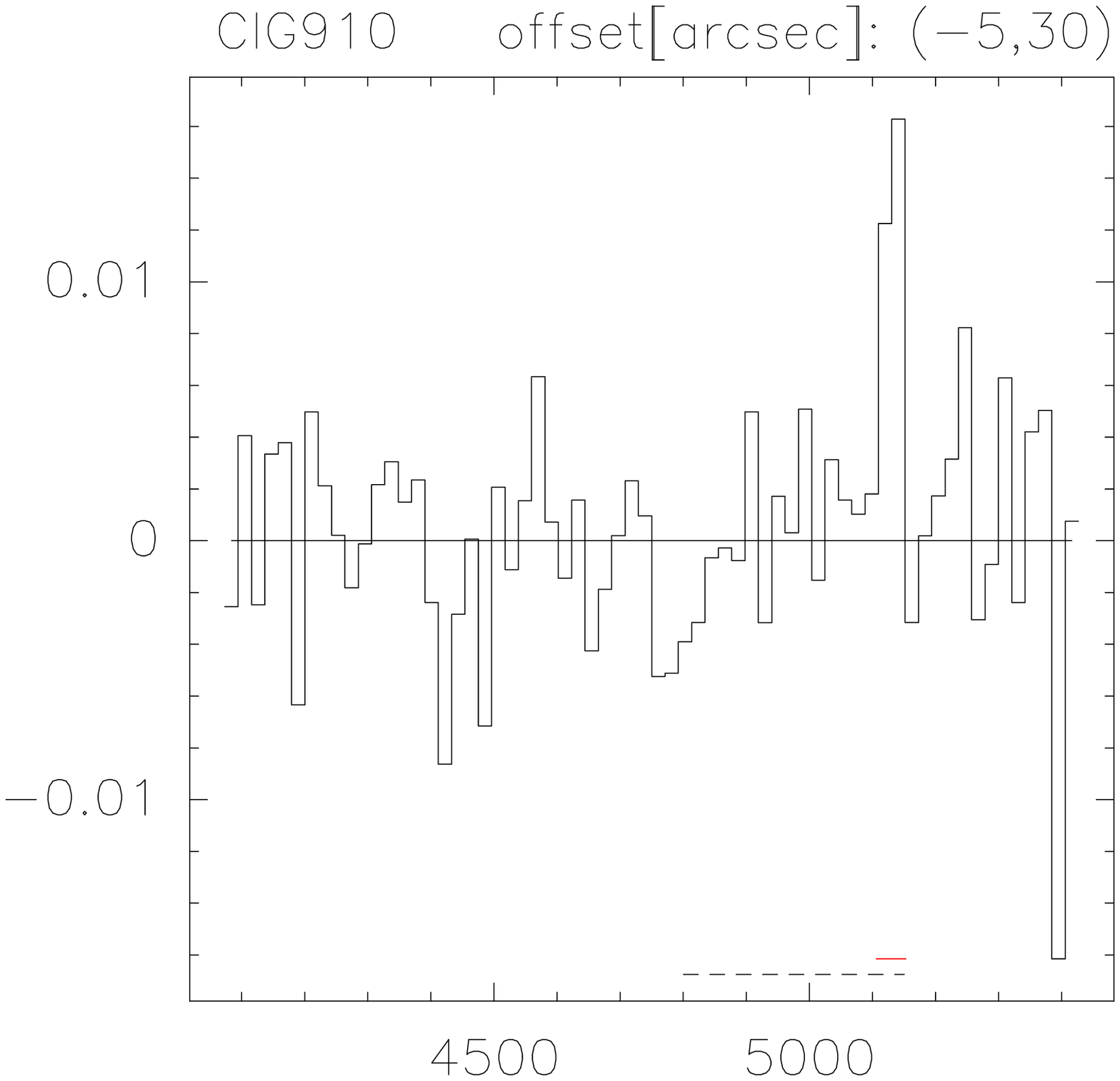}\quad 
\includegraphics[width=3cm,angle=270]{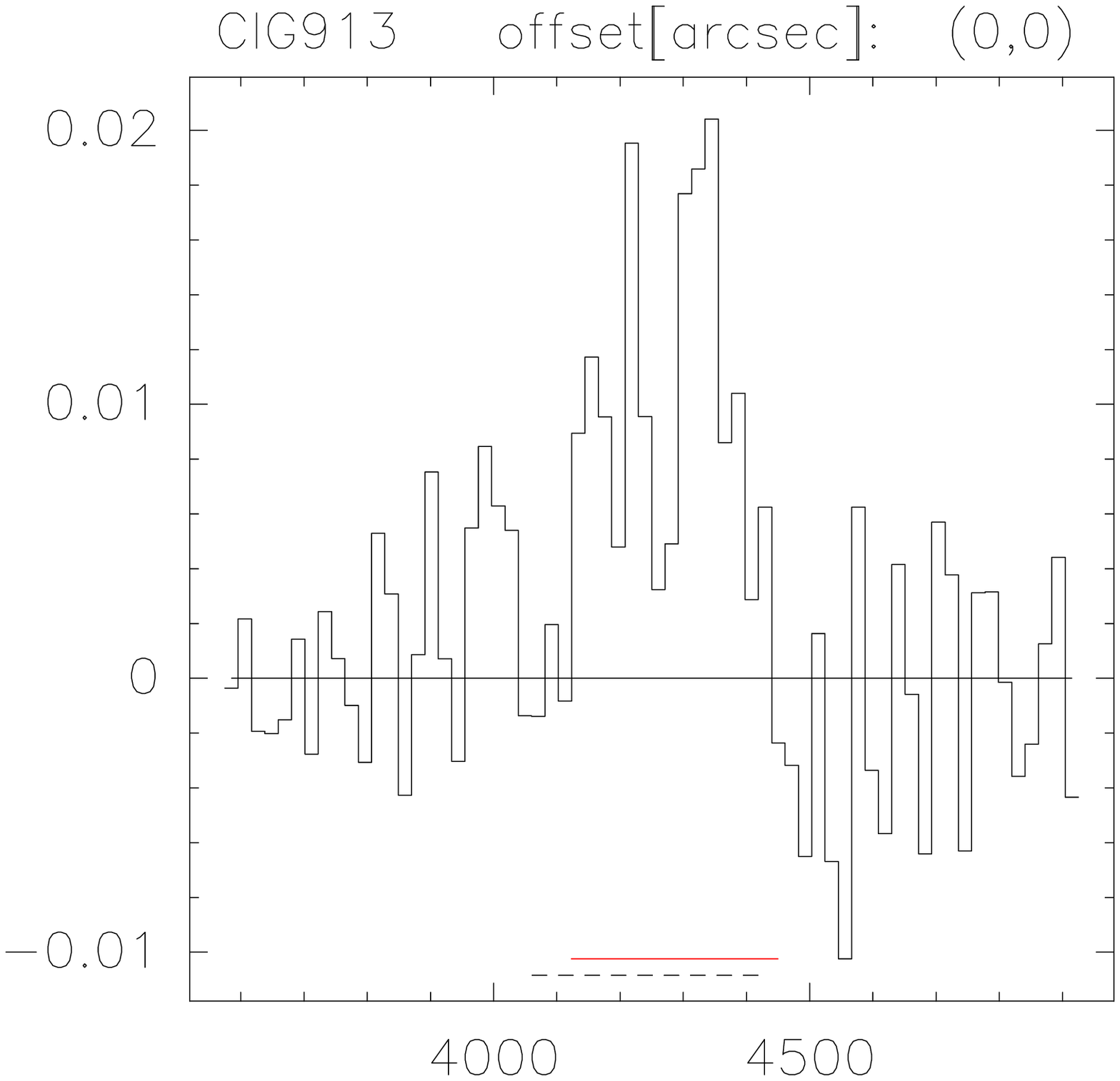}\quad 
\includegraphics[width=3cm,angle=270]{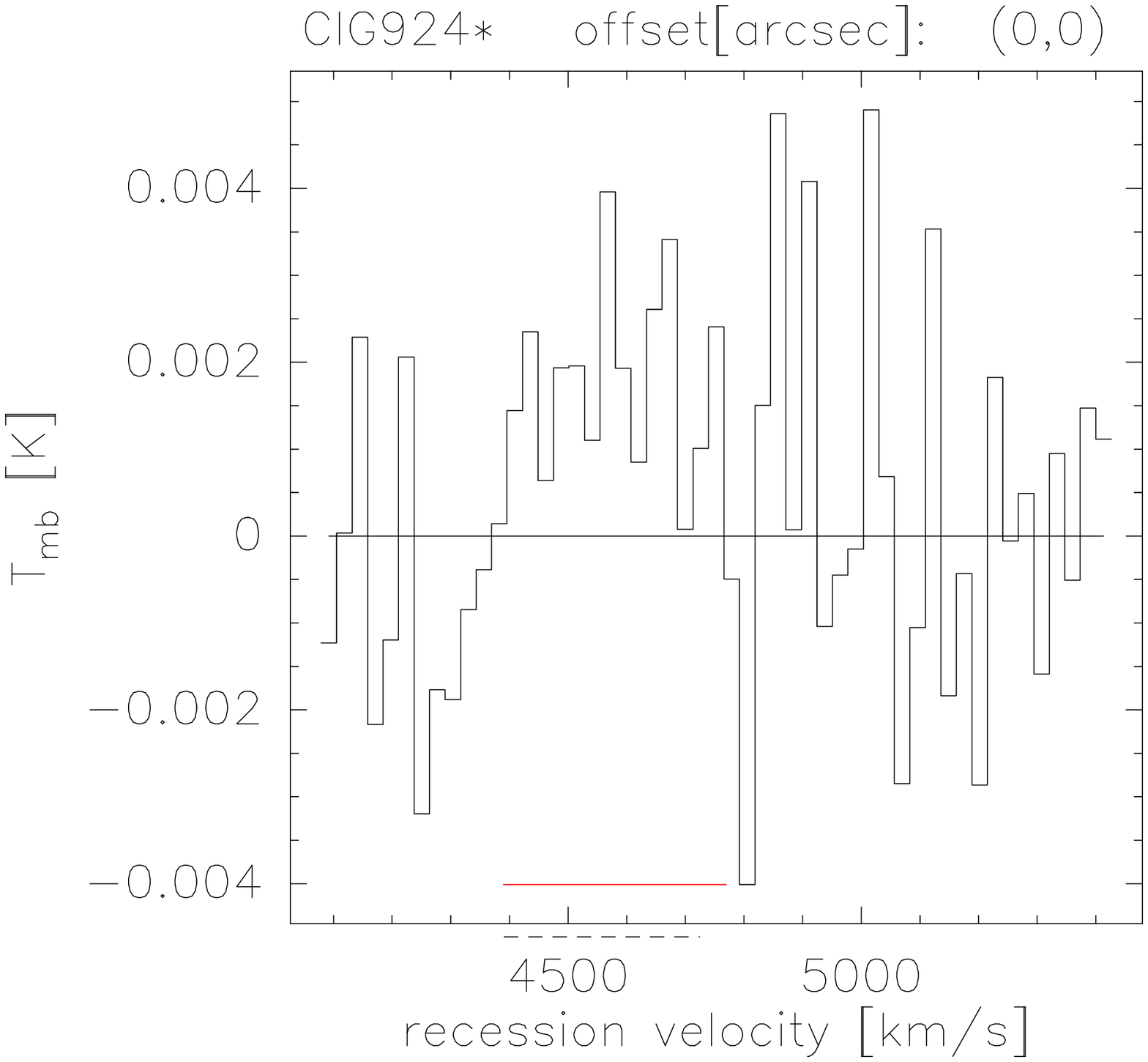}\quad 
\includegraphics[width=3cm,angle=270]{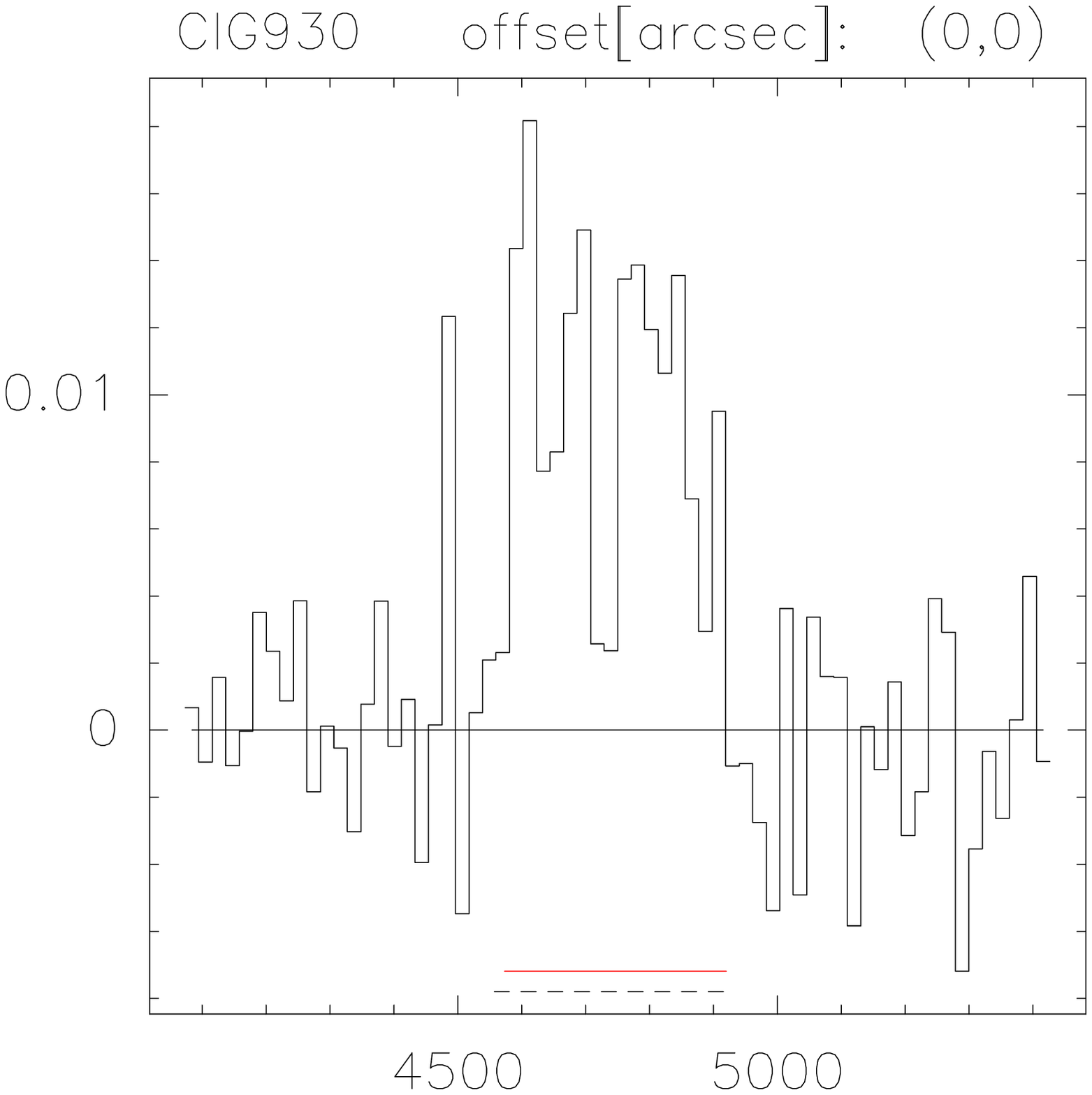}} 
\centerline{\includegraphics[width=3cm,angle=270]{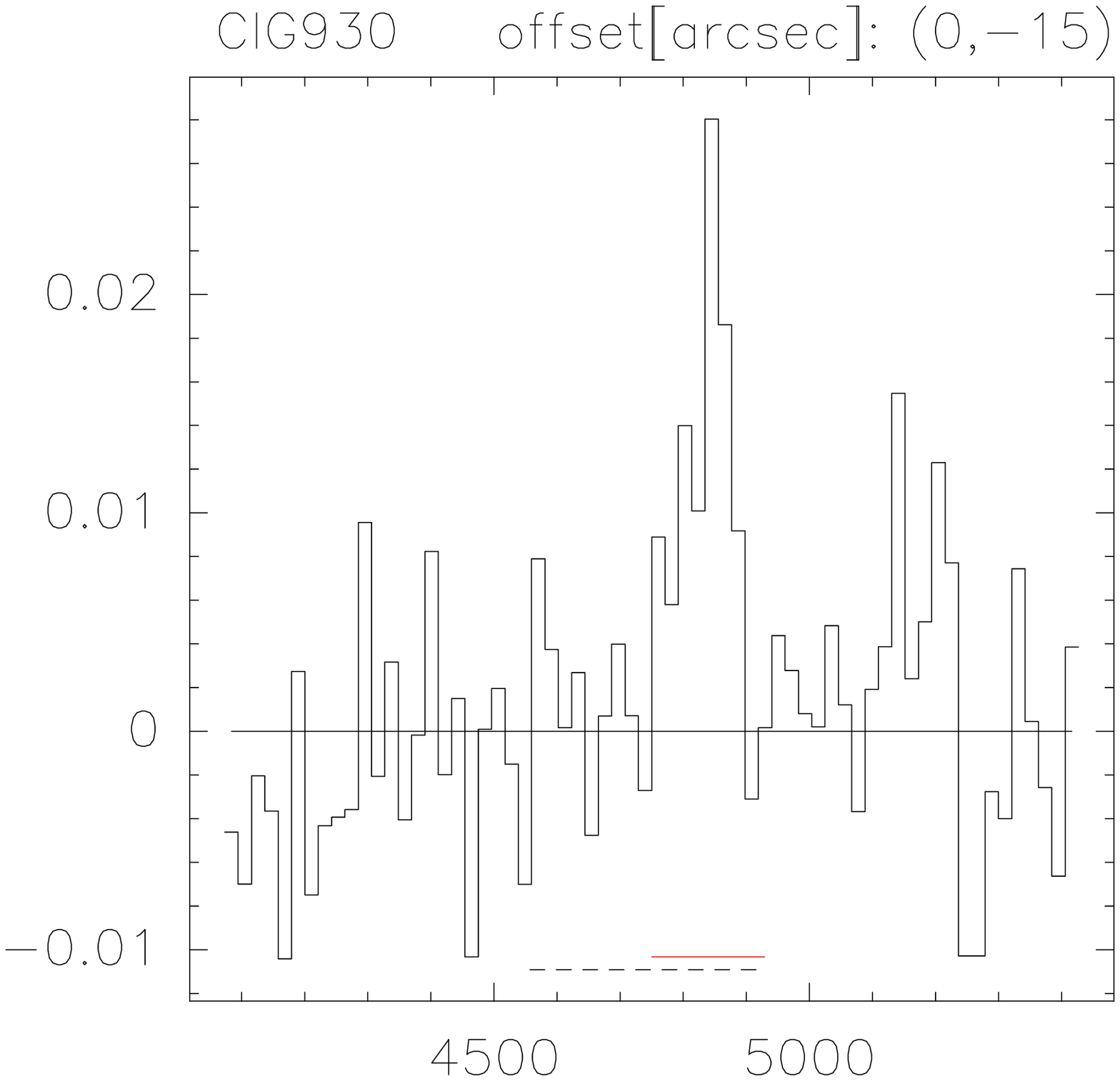} \quad 
\includegraphics[width=3cm,angle=270]{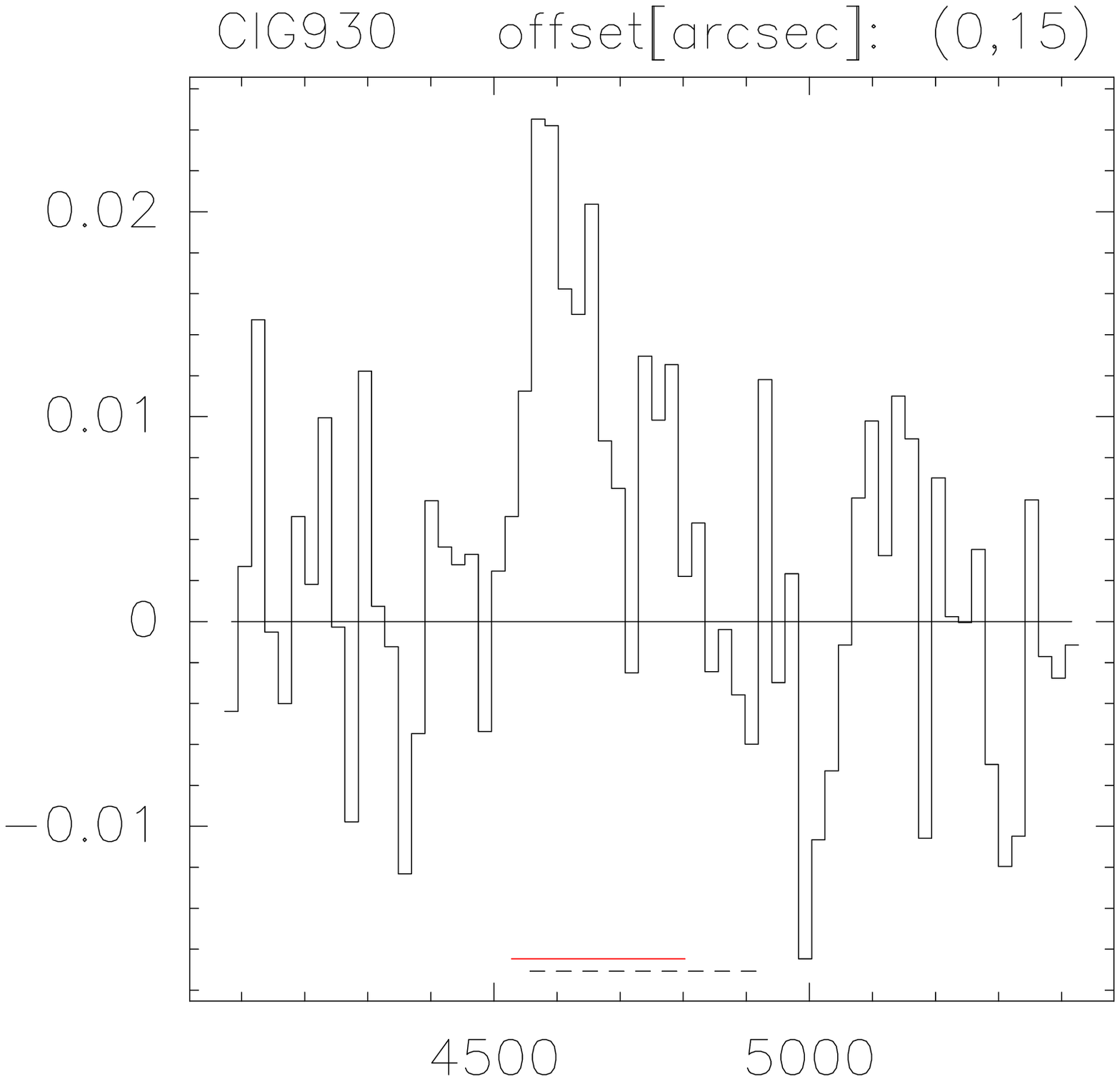}\quad 
\includegraphics[width=3cm,angle=270]{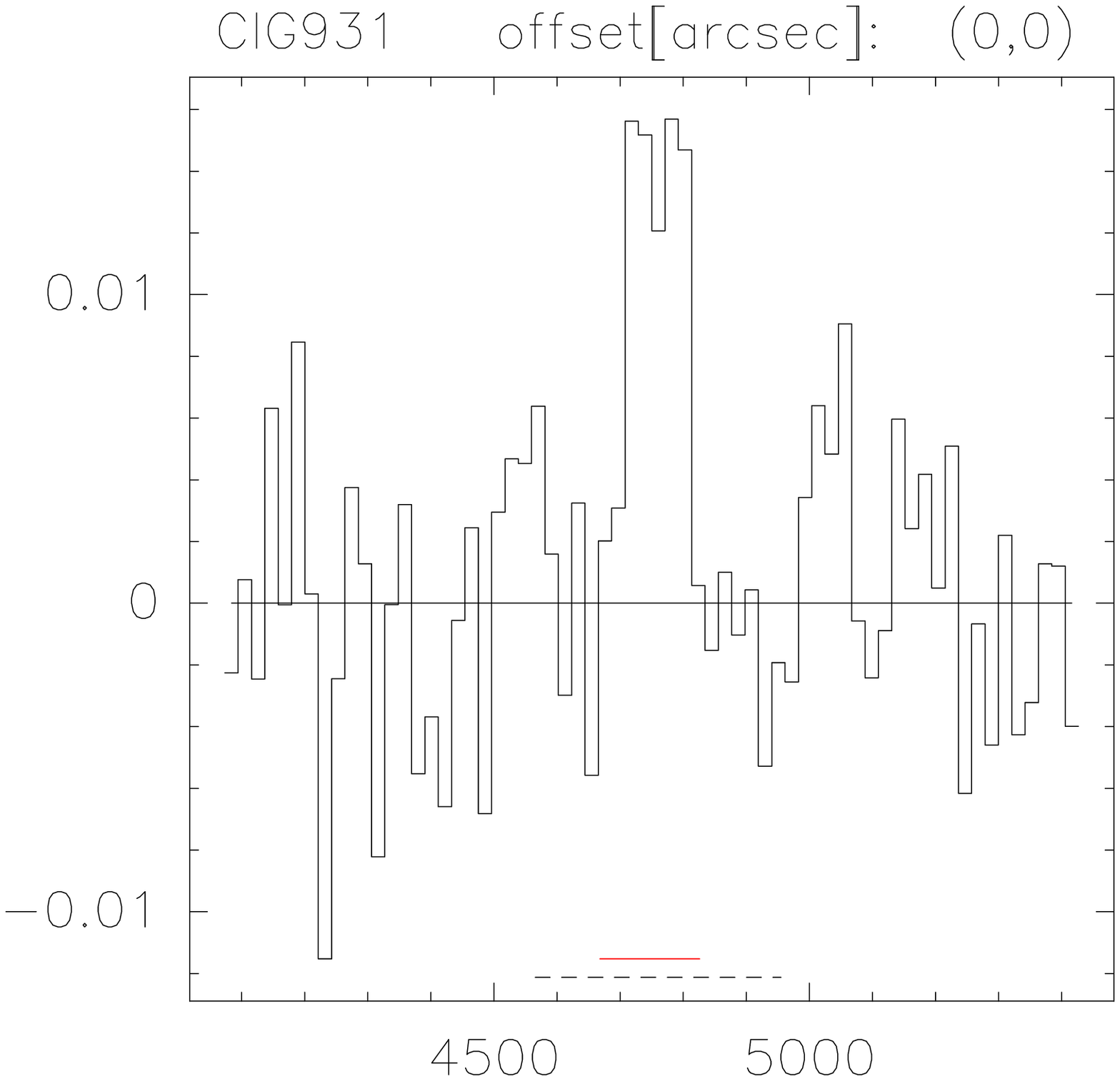}\quad 
\includegraphics[width=3cm,angle=270]{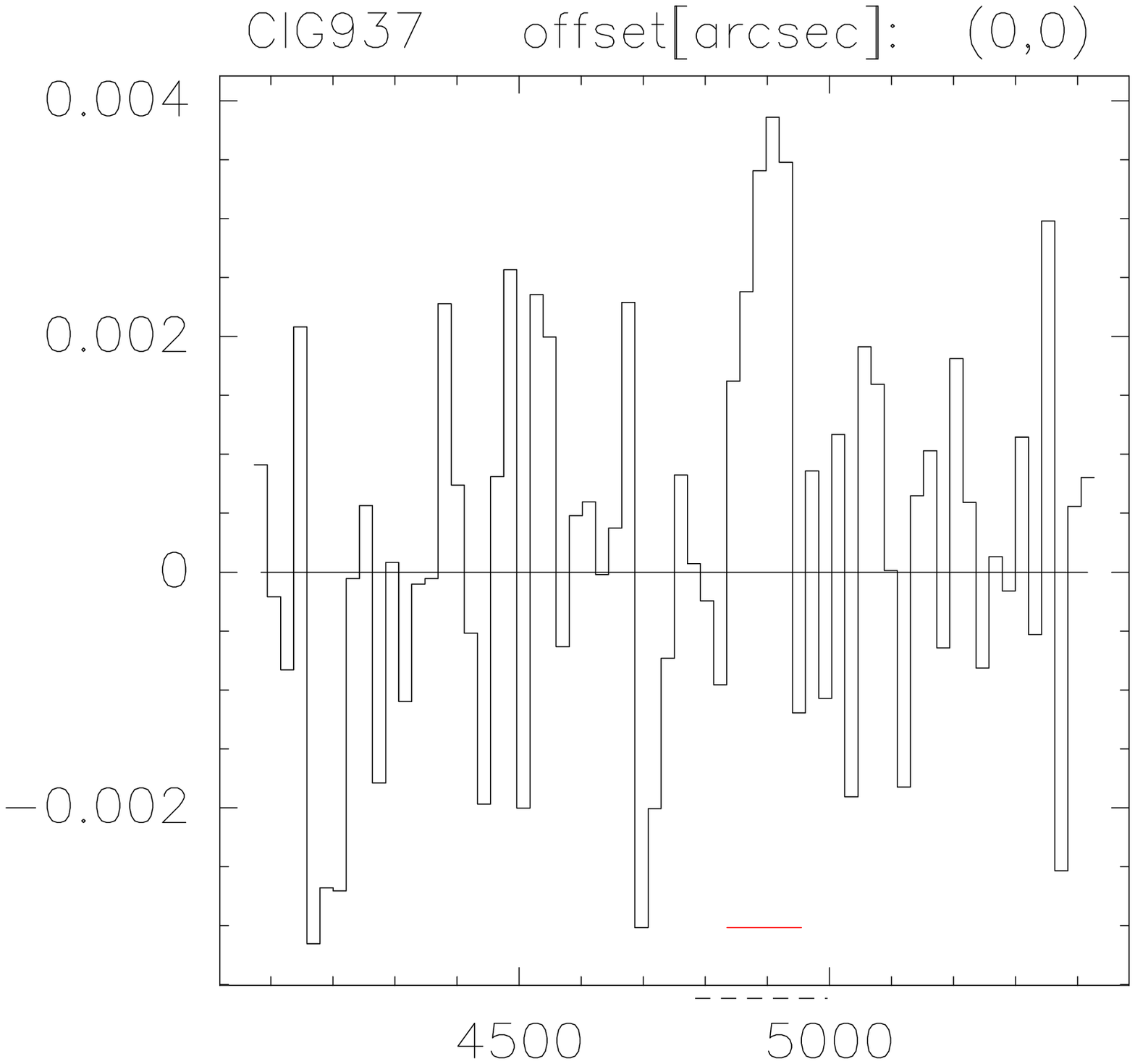}\quad 
\includegraphics[width=3cm,angle=270]{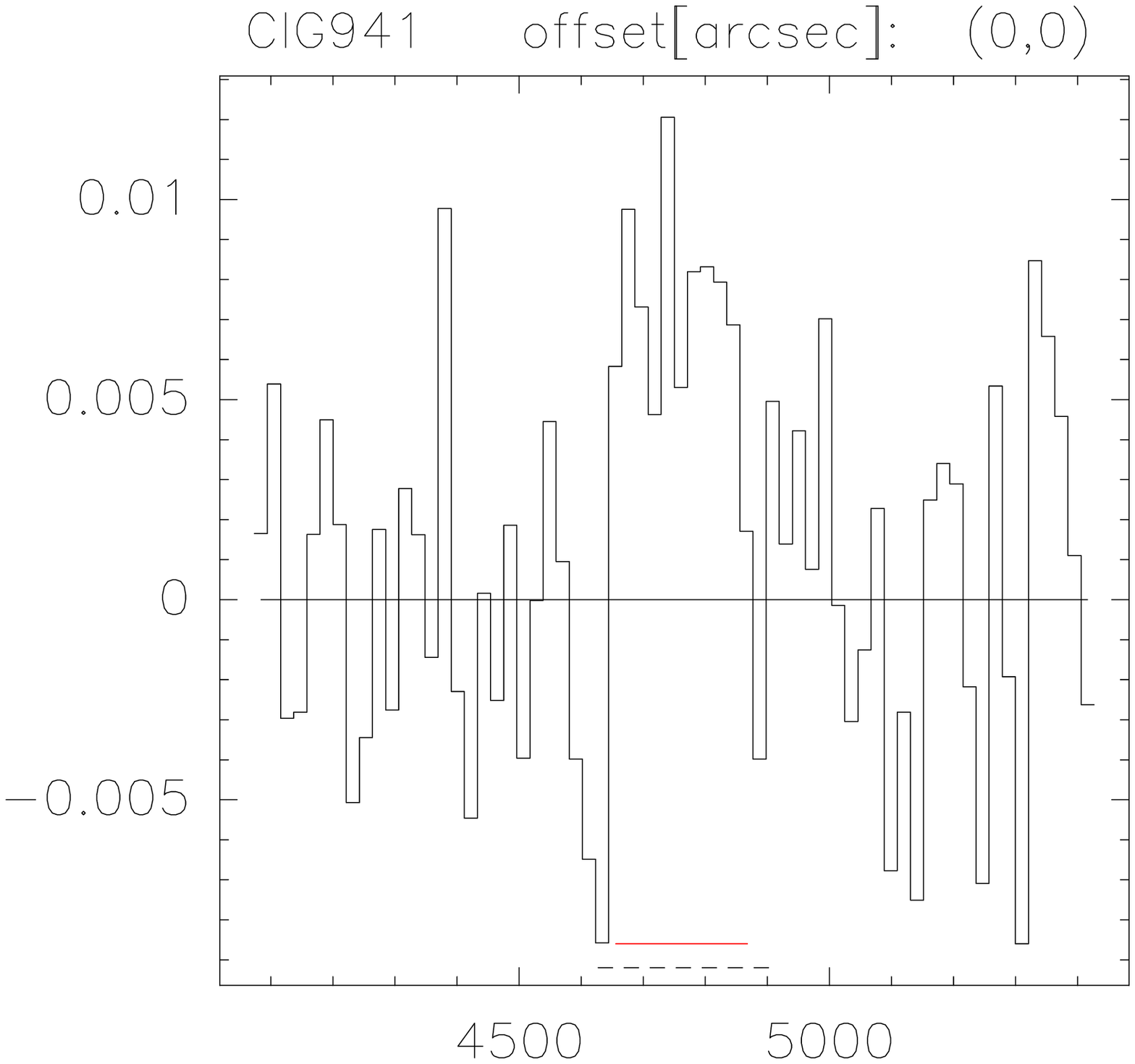}} 
\centerline{\includegraphics[width=3cm,angle=270]{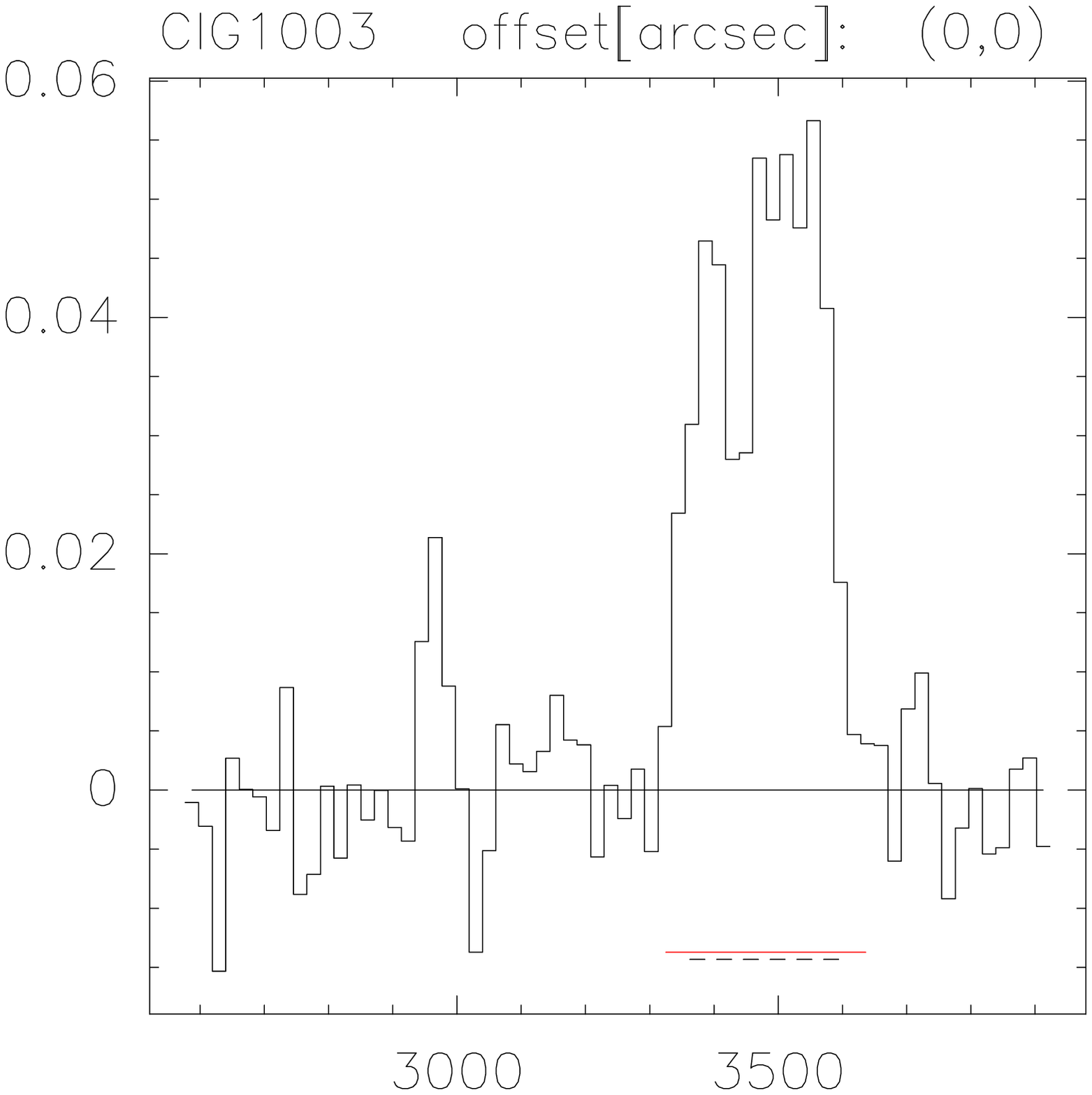} \quad 
\includegraphics[width=3cm,angle=270]{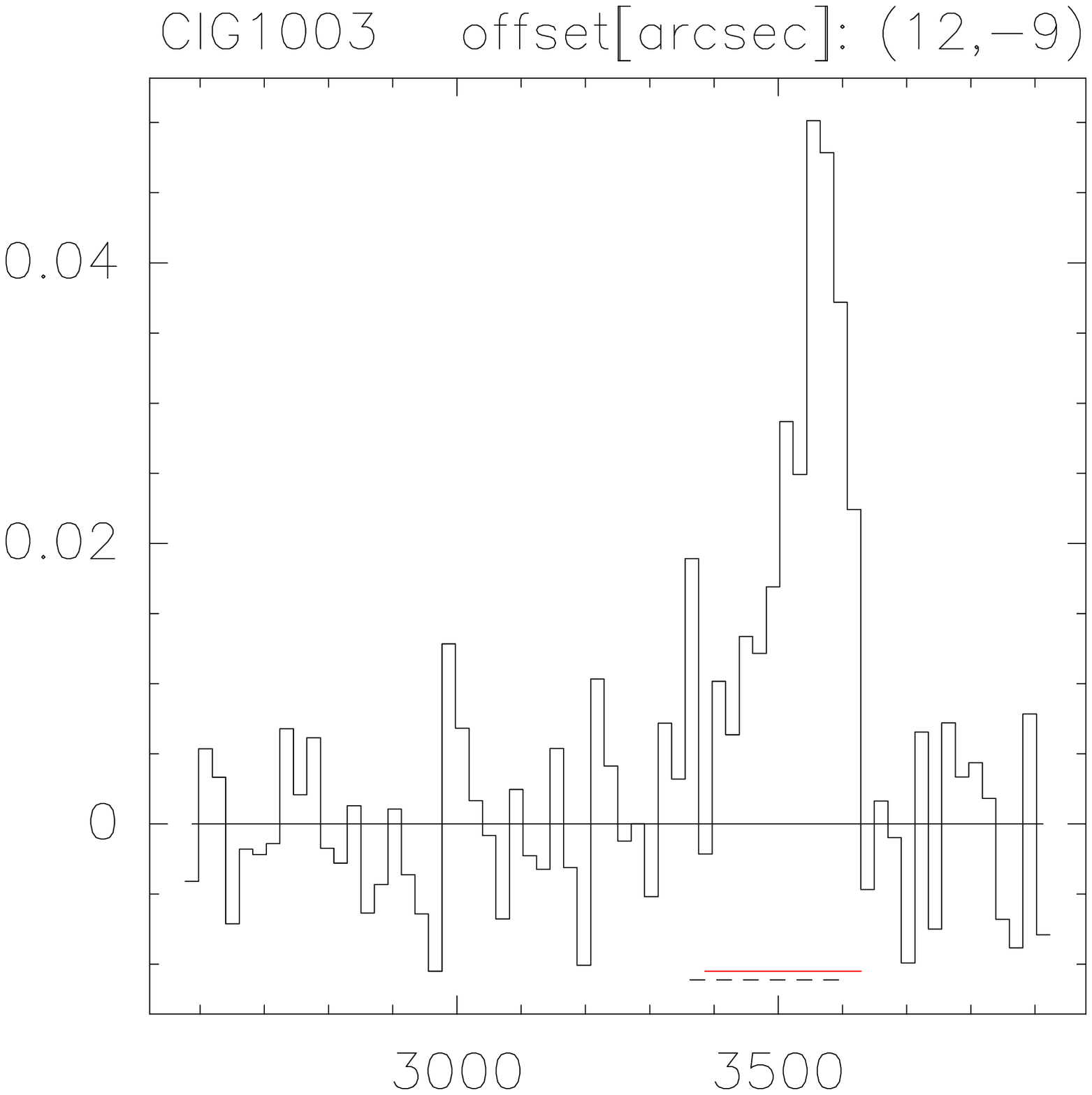}\quad 
\includegraphics[width=3cm,angle=270]{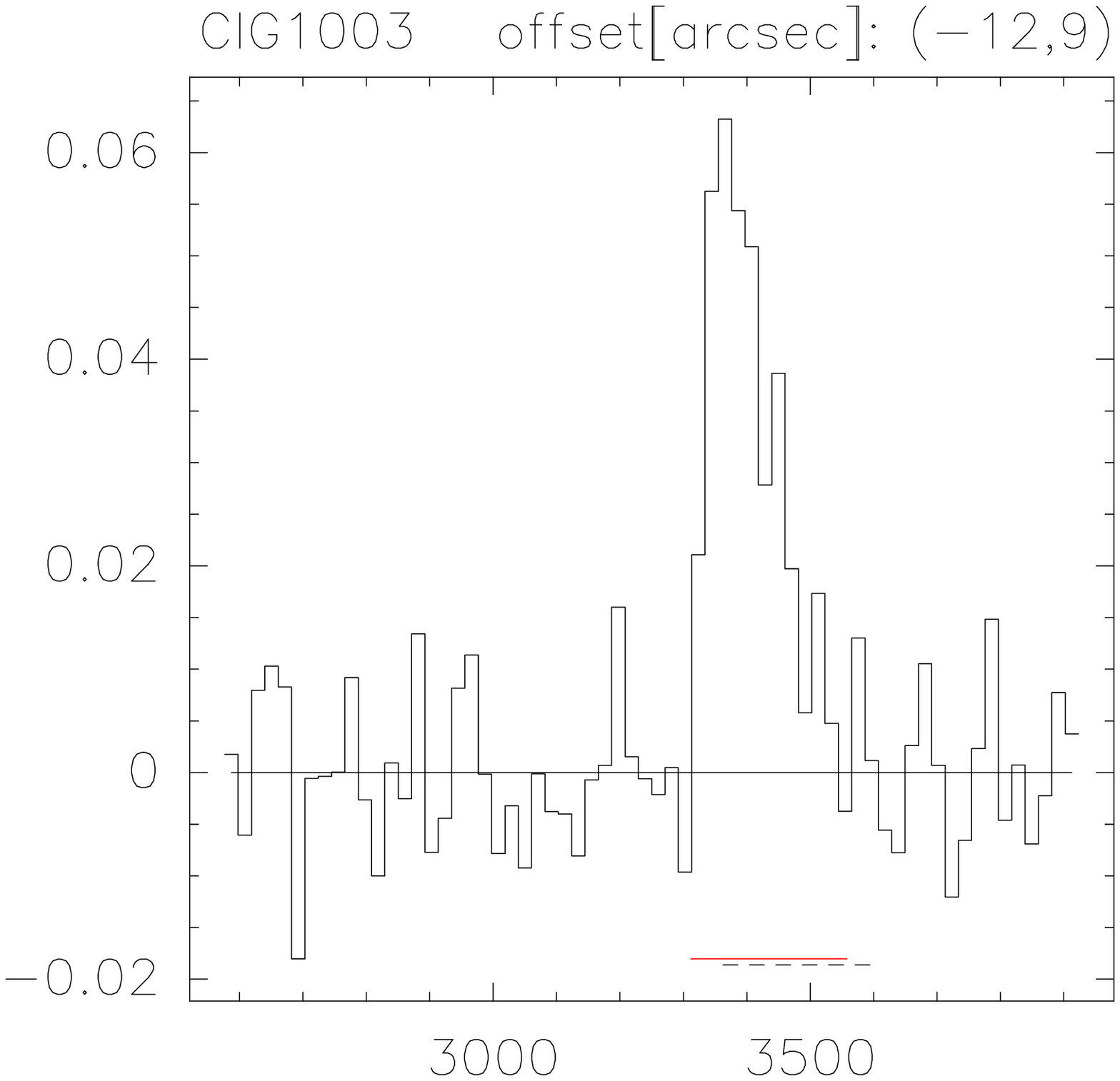}\quad 
\includegraphics[width=3cm,angle=270]{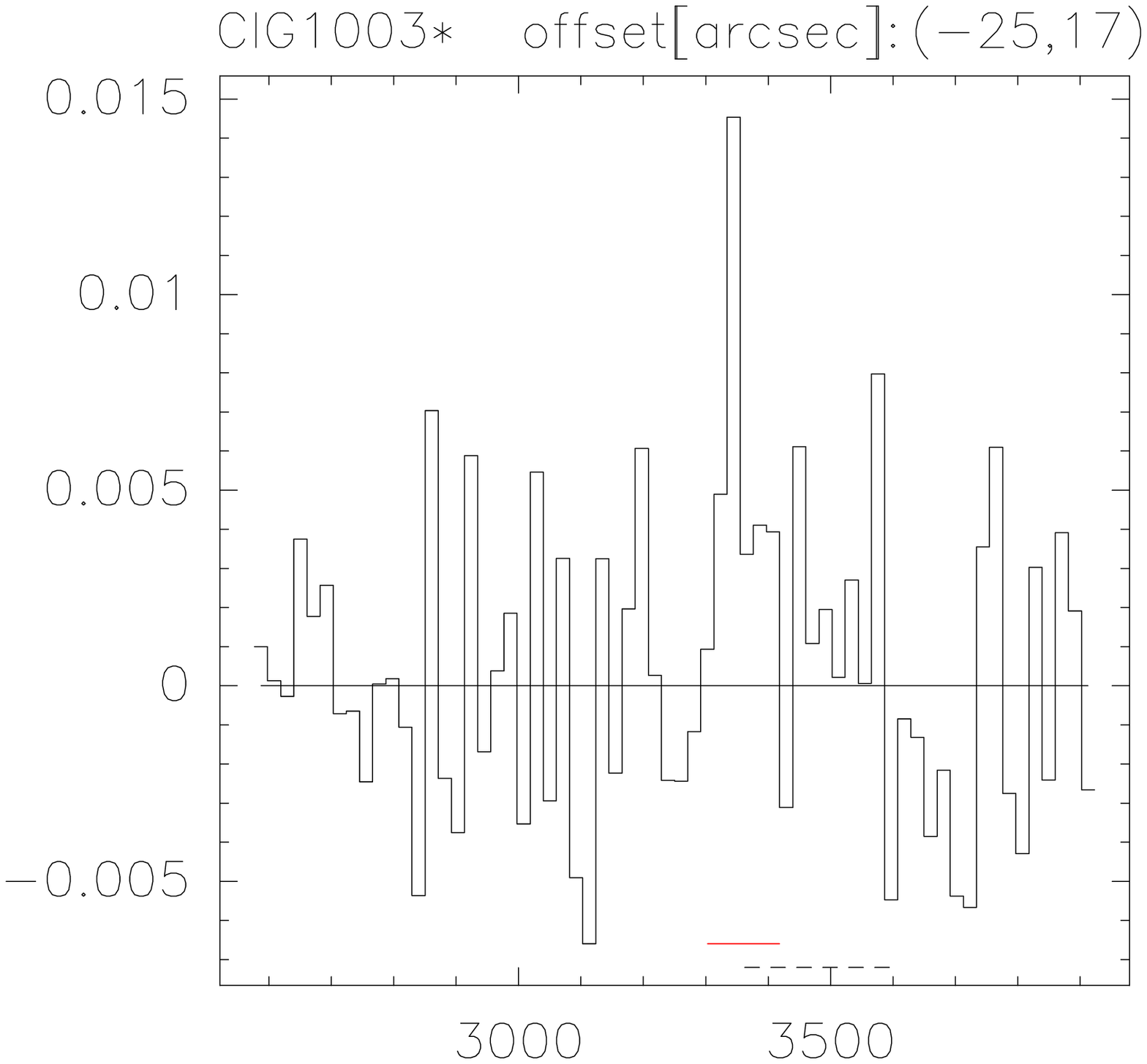}\quad 
\includegraphics[width=3cm,angle=270]{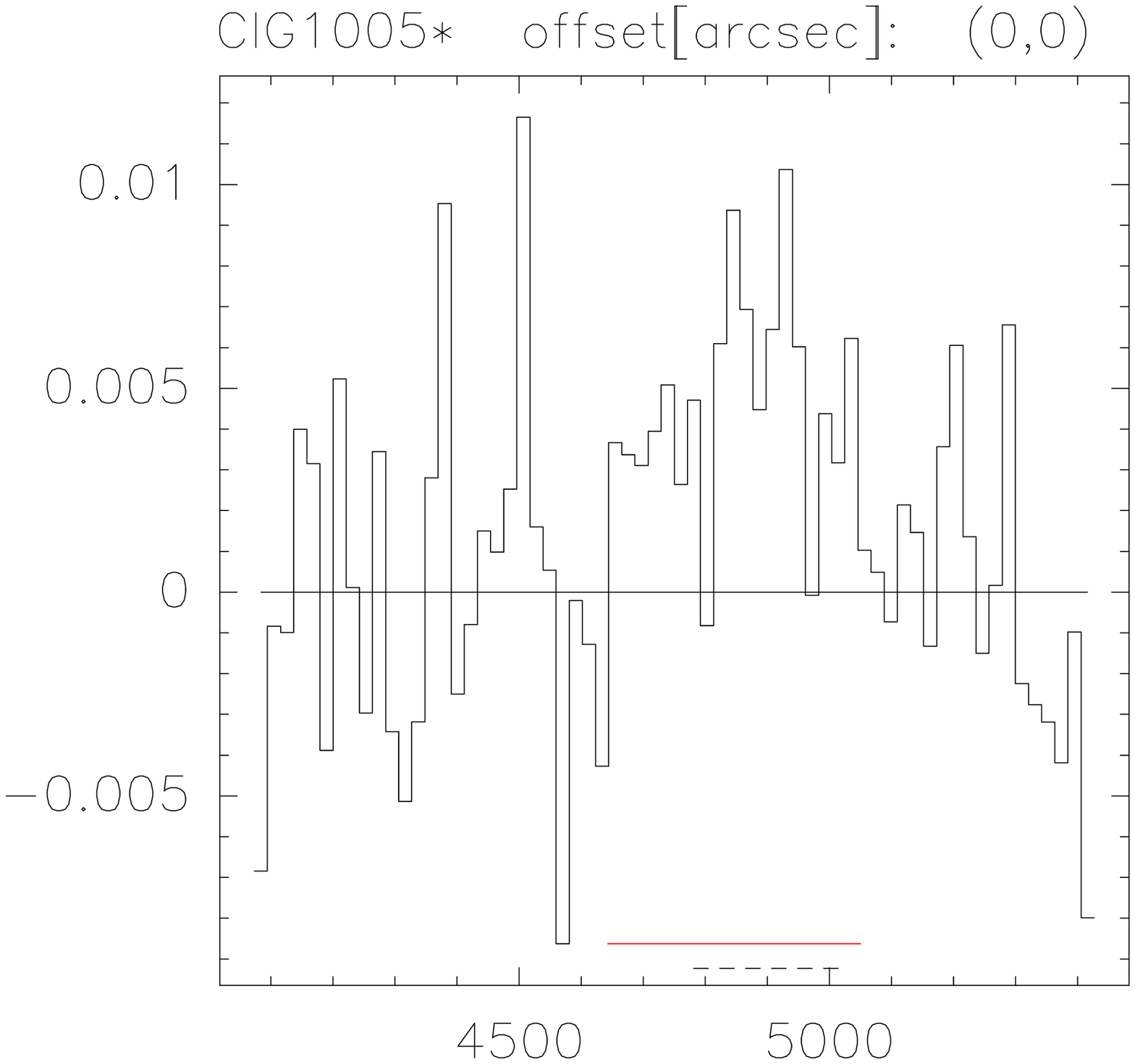}} 
\centerline{\includegraphics[width=3cm,angle=270]{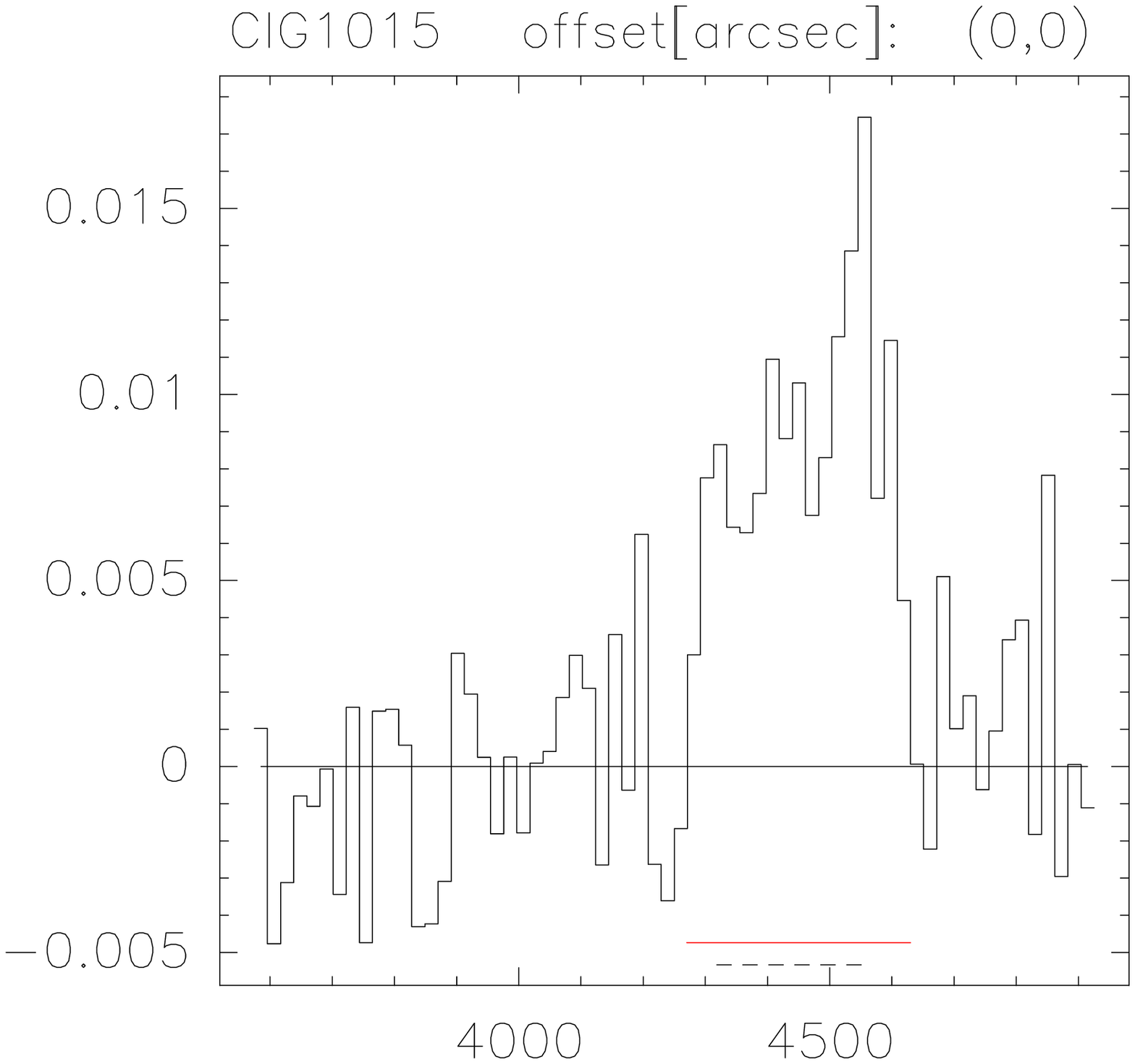} \quad 
\includegraphics[width=3cm,angle=270]{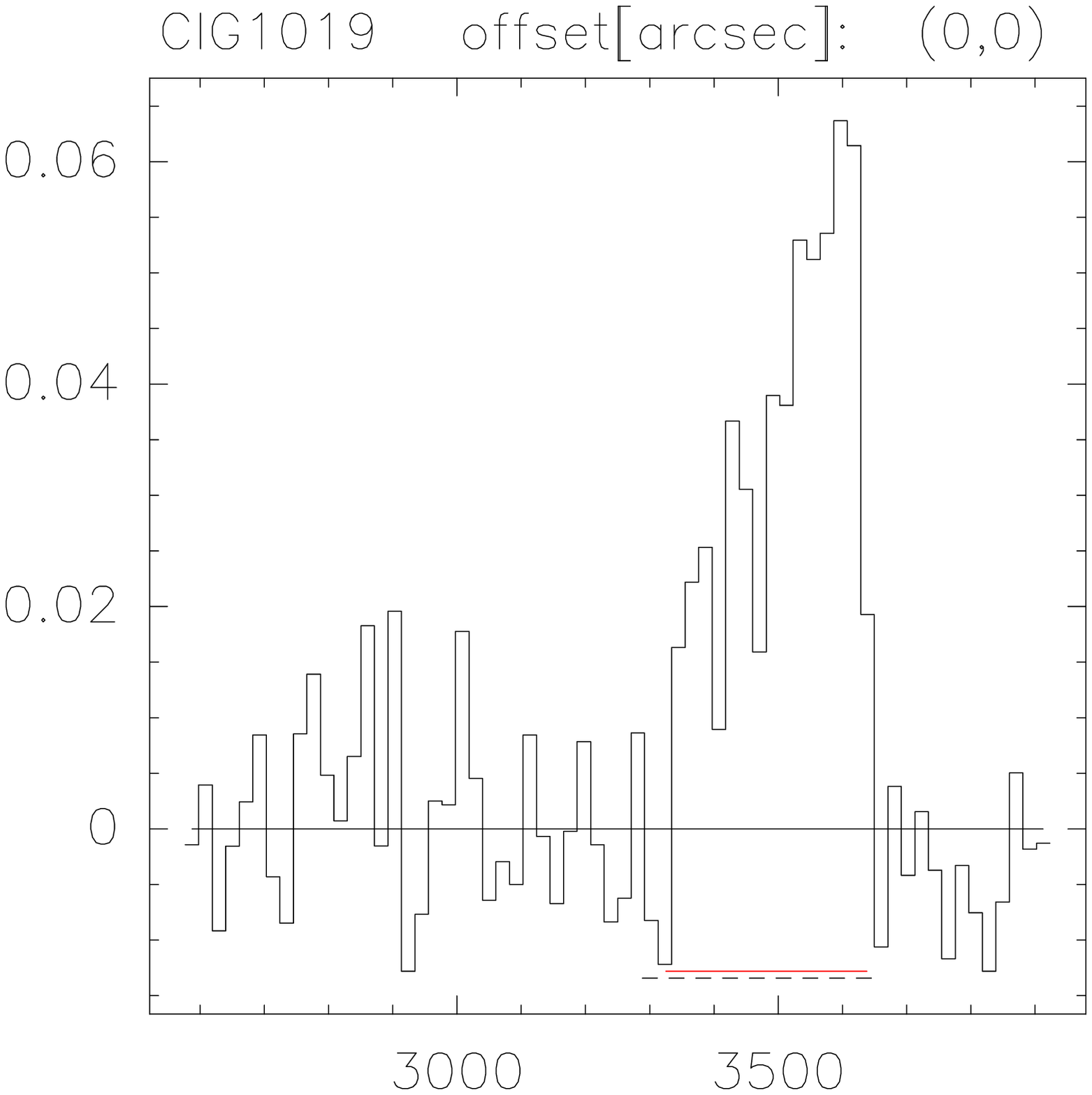}\quad 
\includegraphics[width=3cm,angle=270]{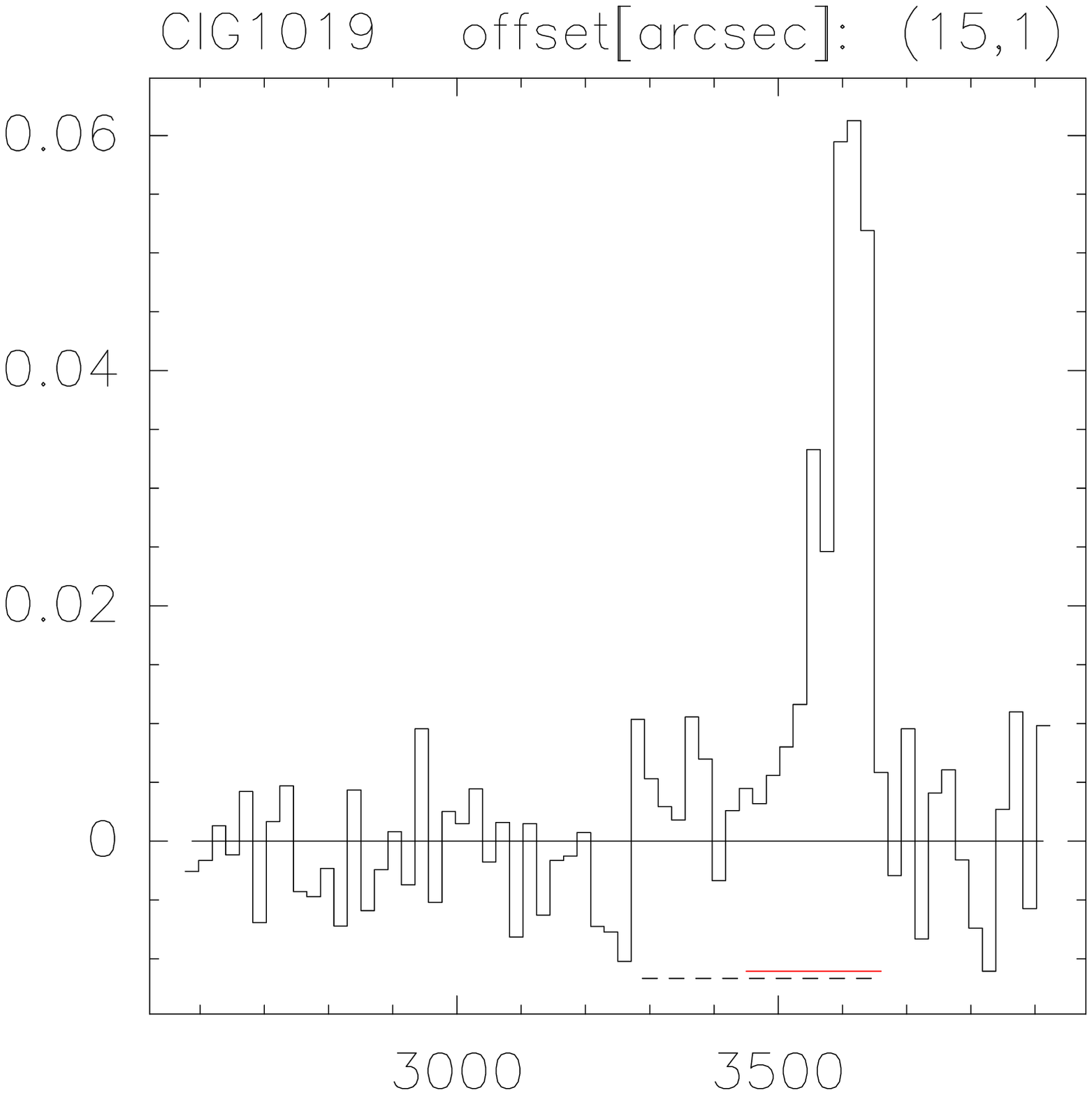}\quad 
\includegraphics[width=3cm,angle=270]{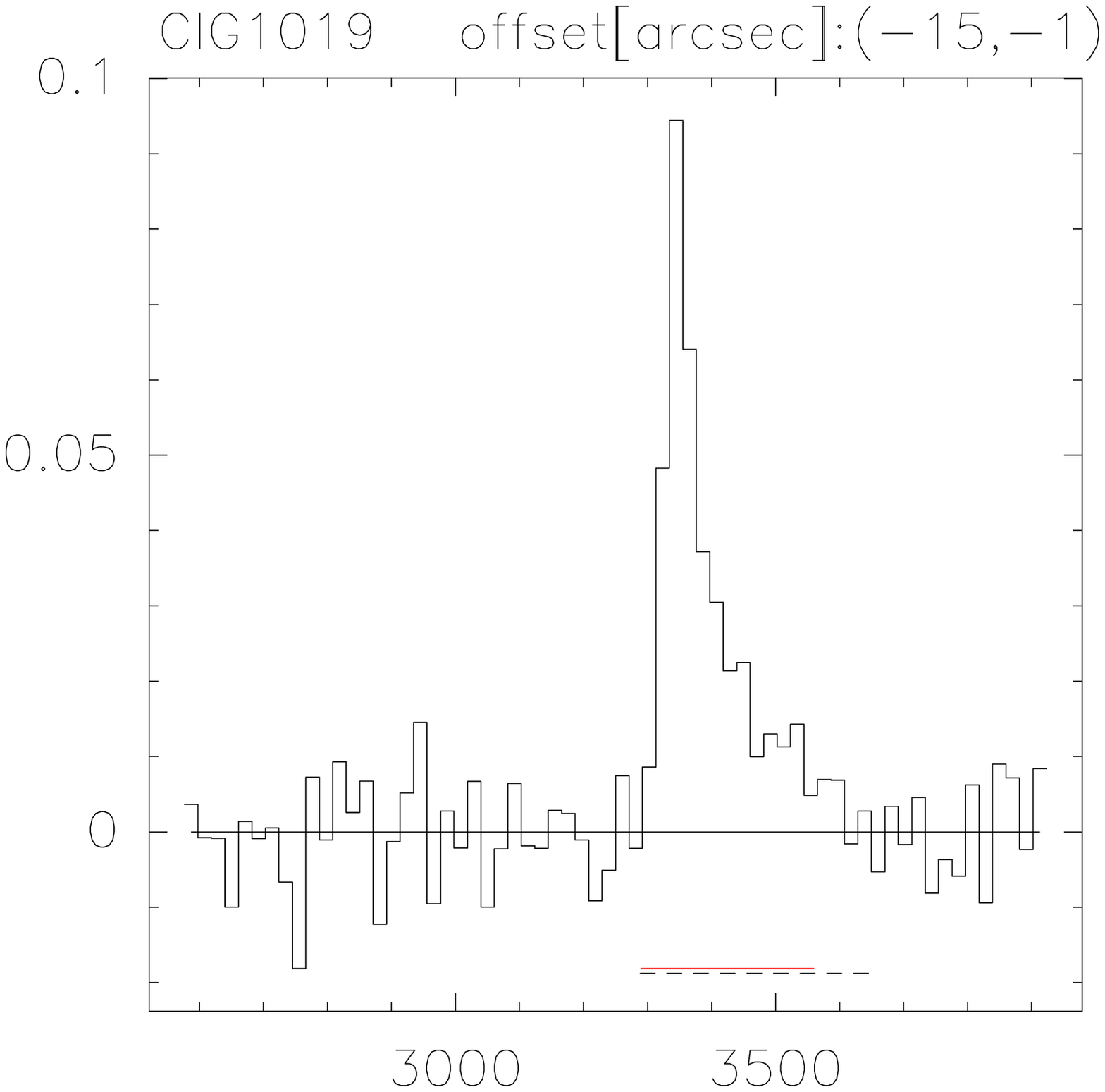}\quad 
\includegraphics[width=3cm,angle=270]{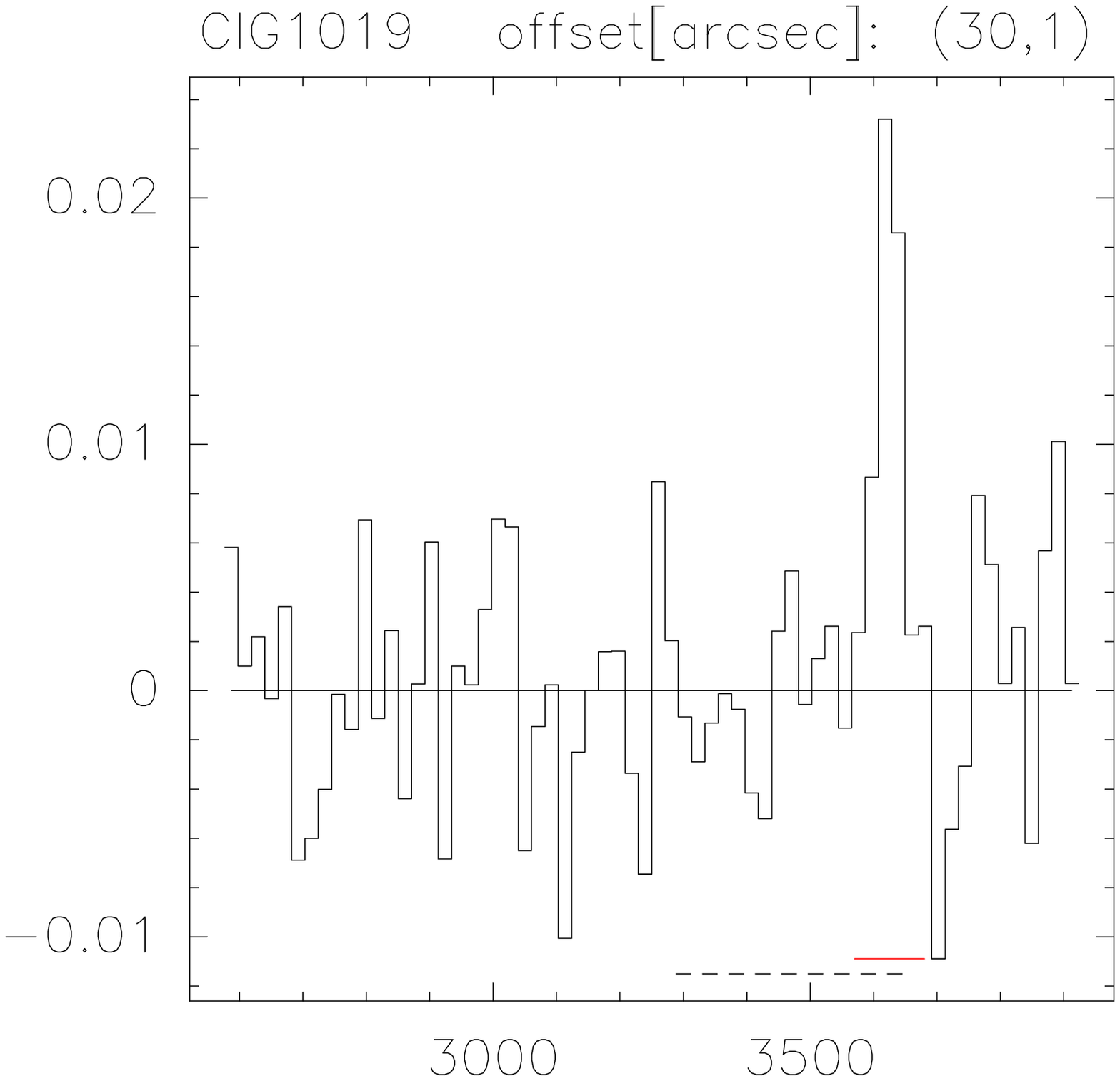}} 
\centerline{\includegraphics[width=3.3cm,angle=270]{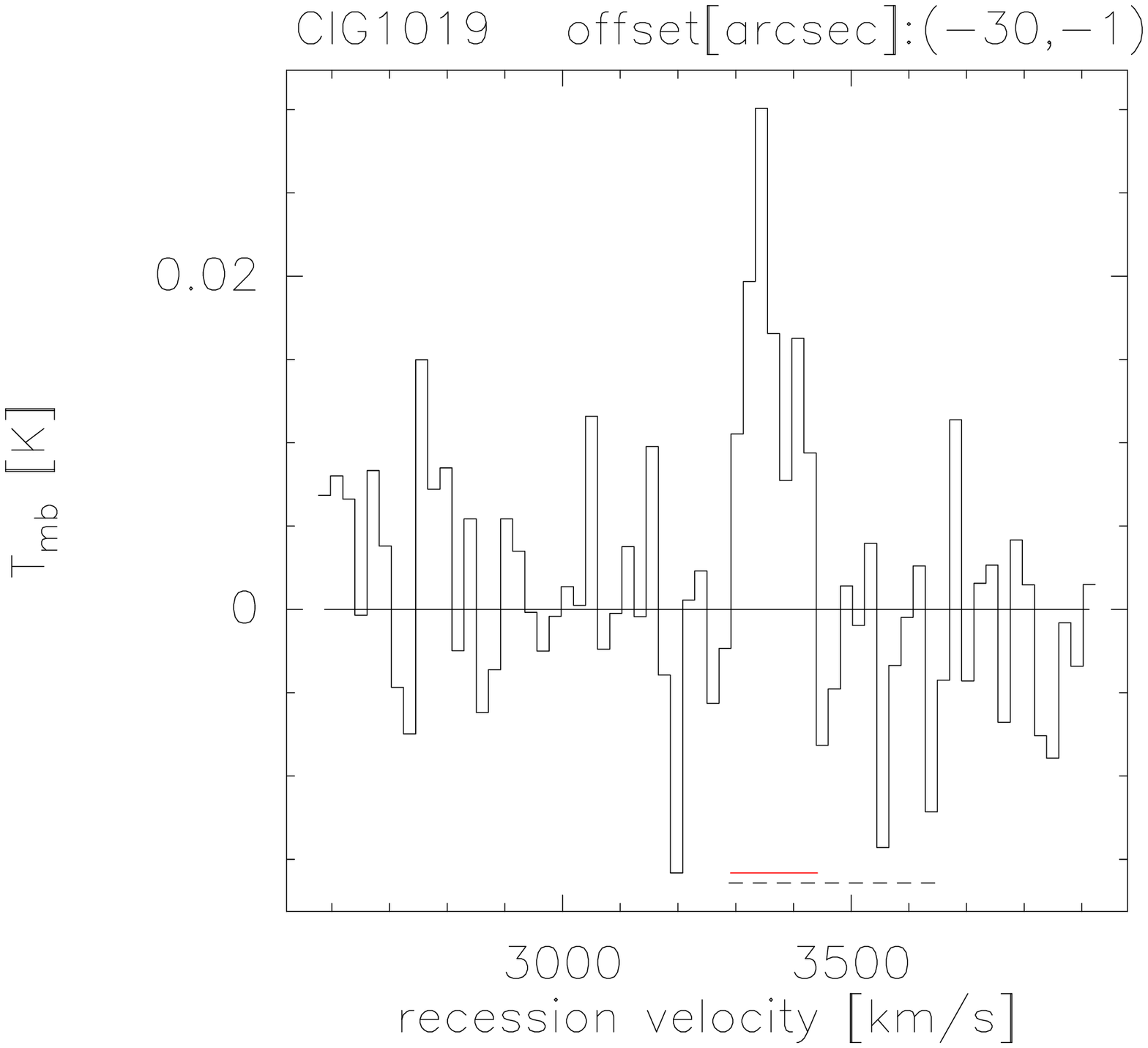} \quad 
\includegraphics[width=3cm,angle=270]{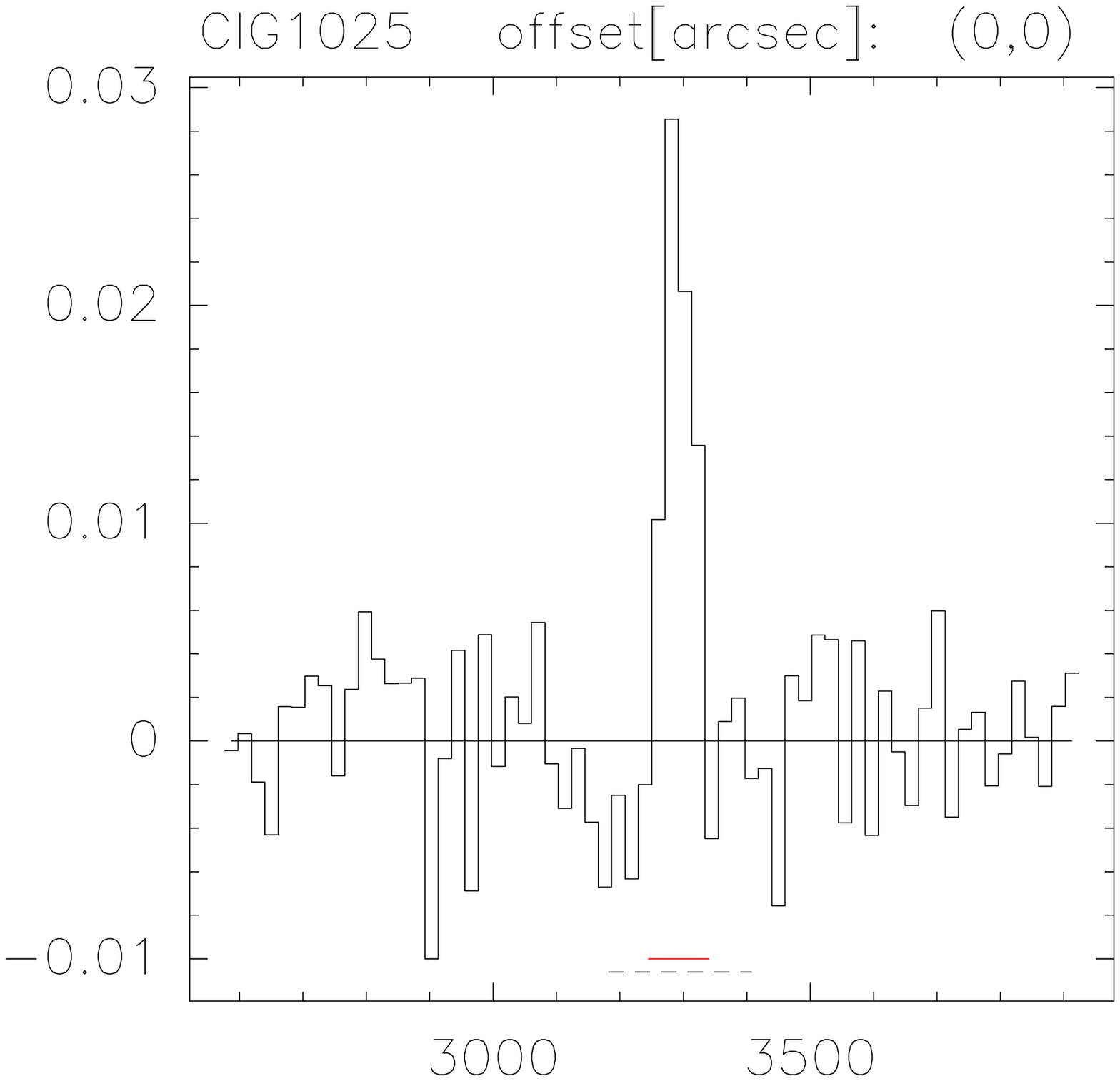}\quad 
\includegraphics[width=3cm,angle=270]{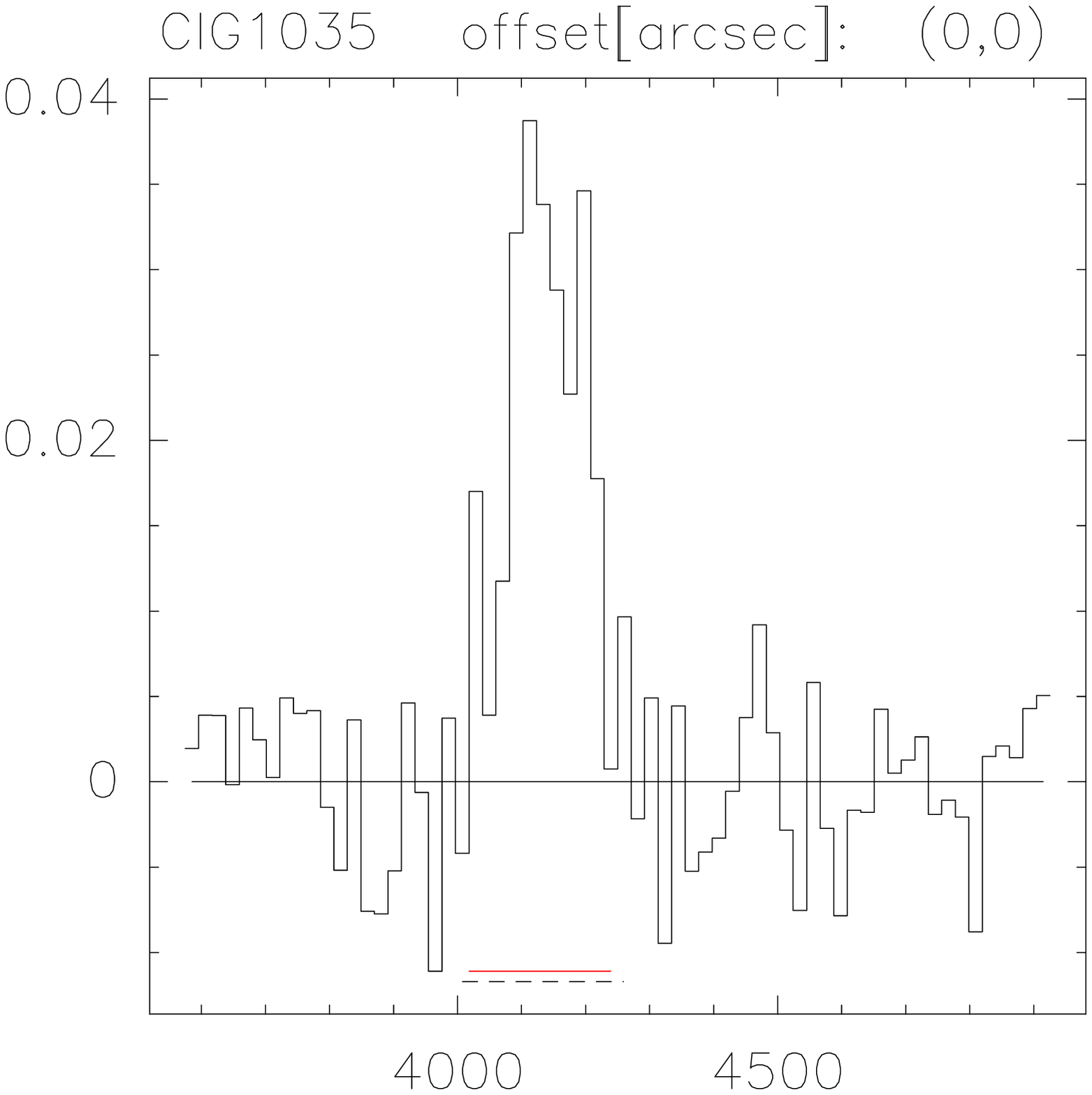}\quad 
\includegraphics[width=3cm,angle=270]{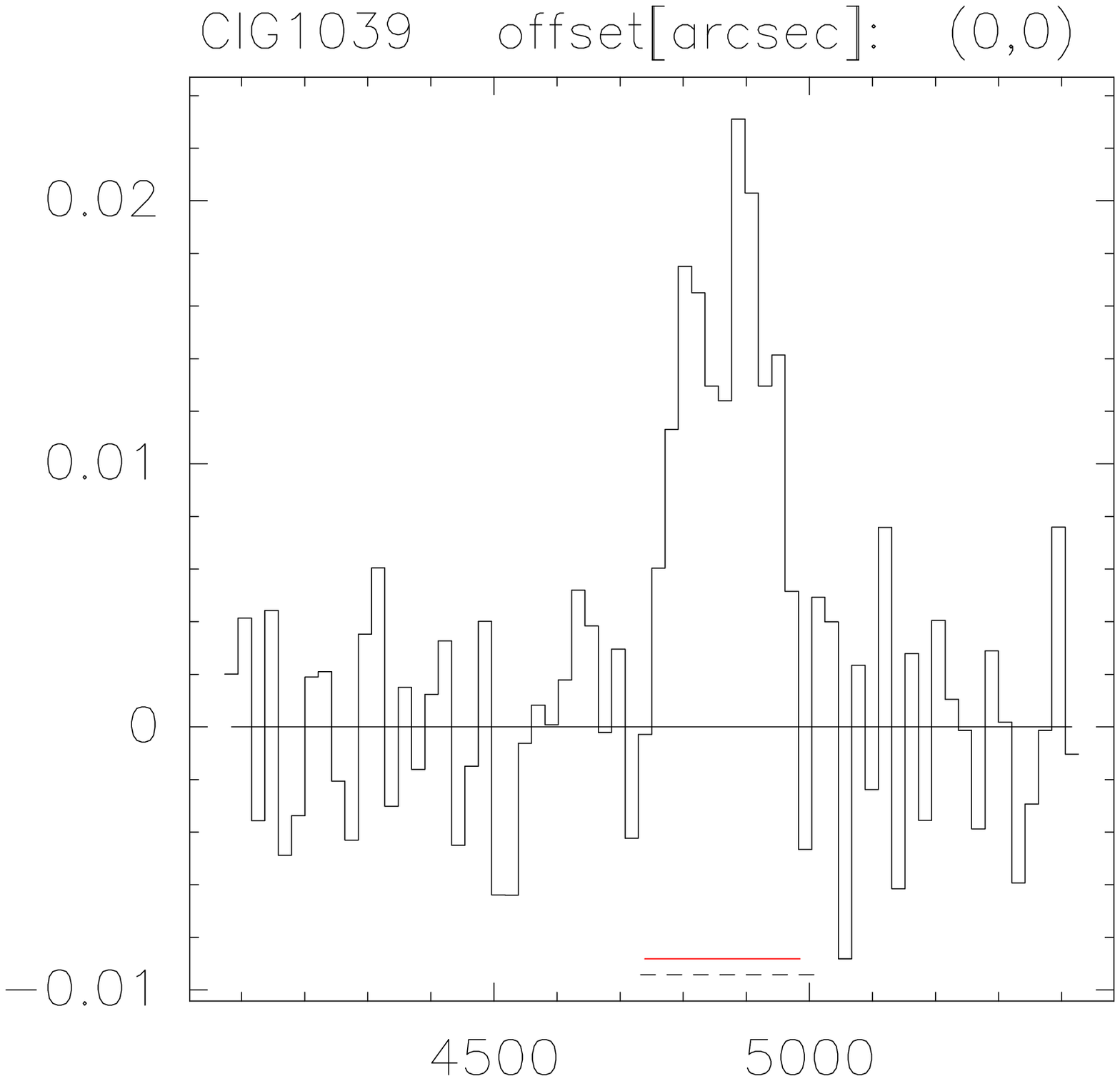}}
\addtocounter{figure}{-1} 
\caption{(continued)} 
\end{figure*} 

\clearpage
%%%%% FCRAO spectra
  
\begin{figure*} 
\centerline{\includegraphics[width=3cm,angle=270]{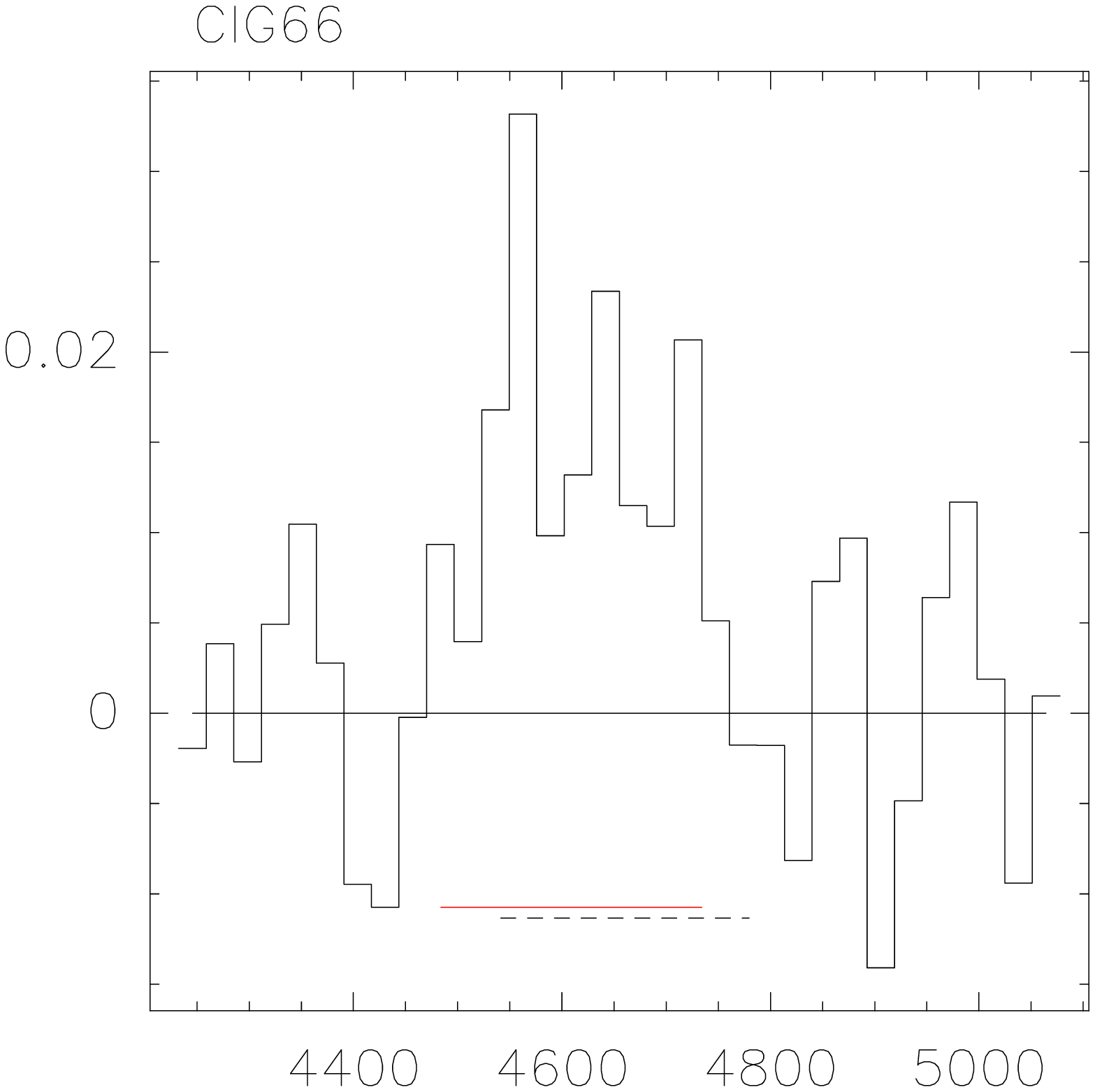} \quad 
\includegraphics[width=3cm,angle=270]{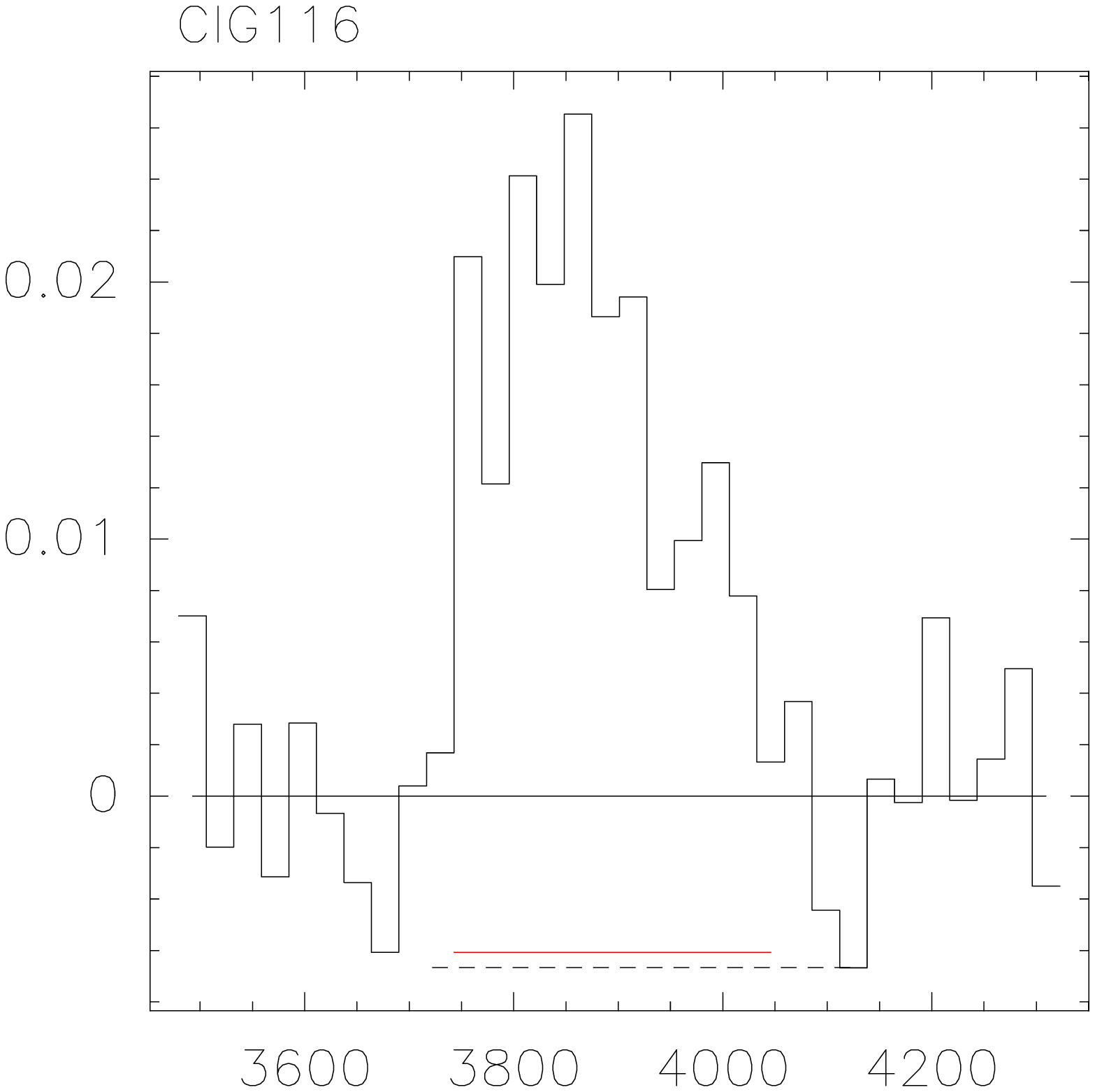}\quad 
\includegraphics[width=3cm,angle=270]{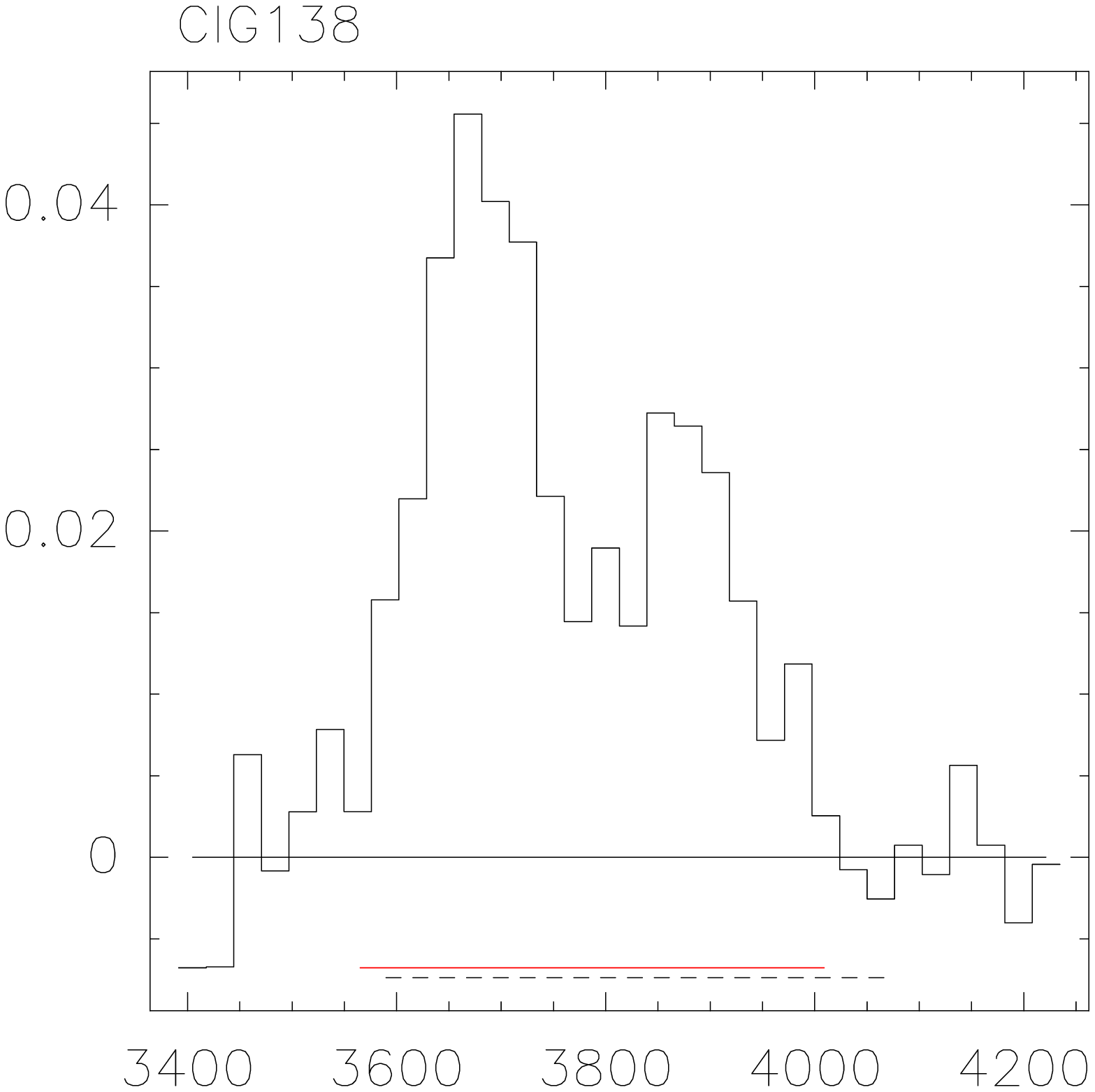}\quad 
\includegraphics[width=3cm,angle=270]{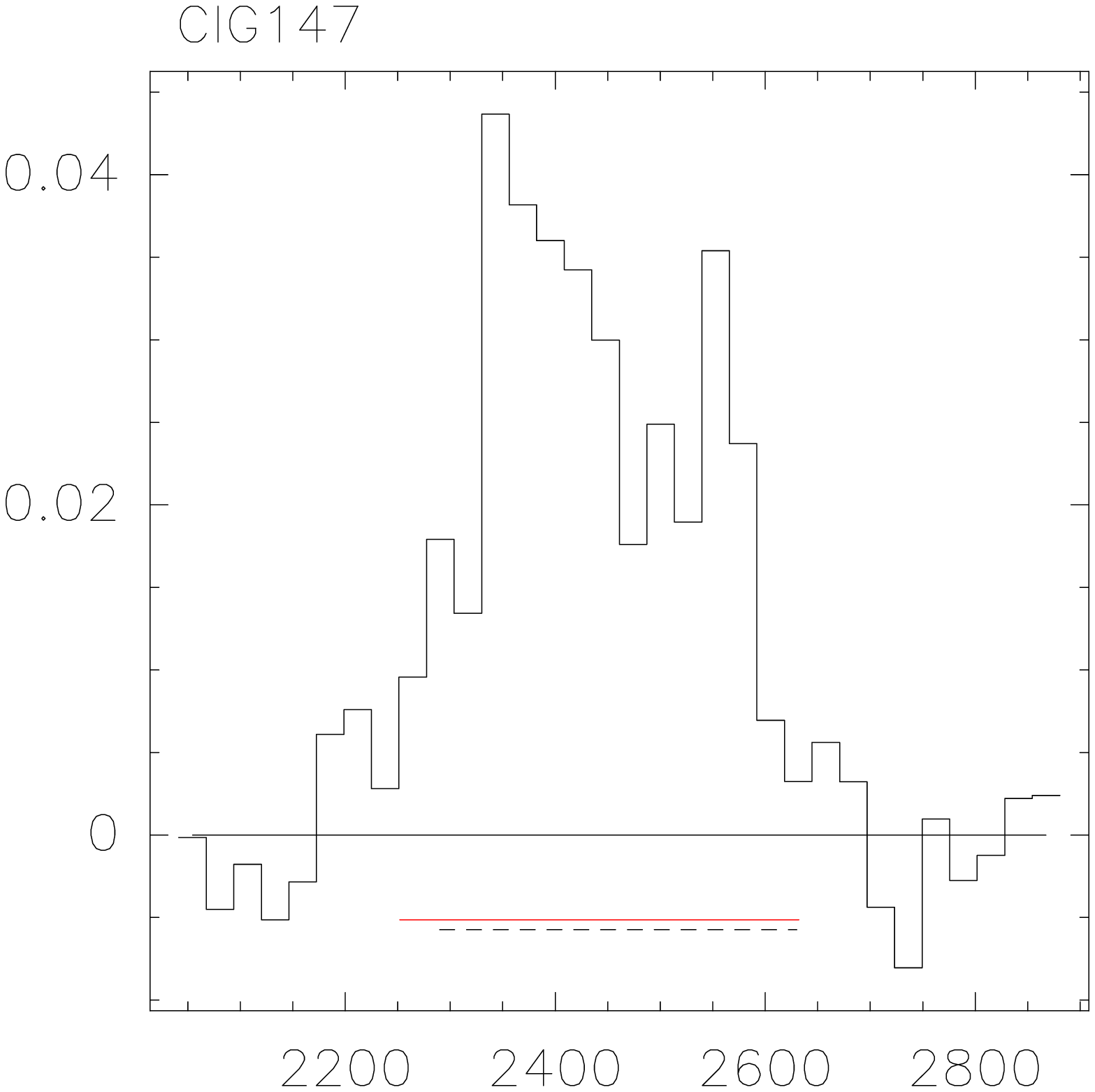}\quad 
\includegraphics[width=3cm,angle=270]{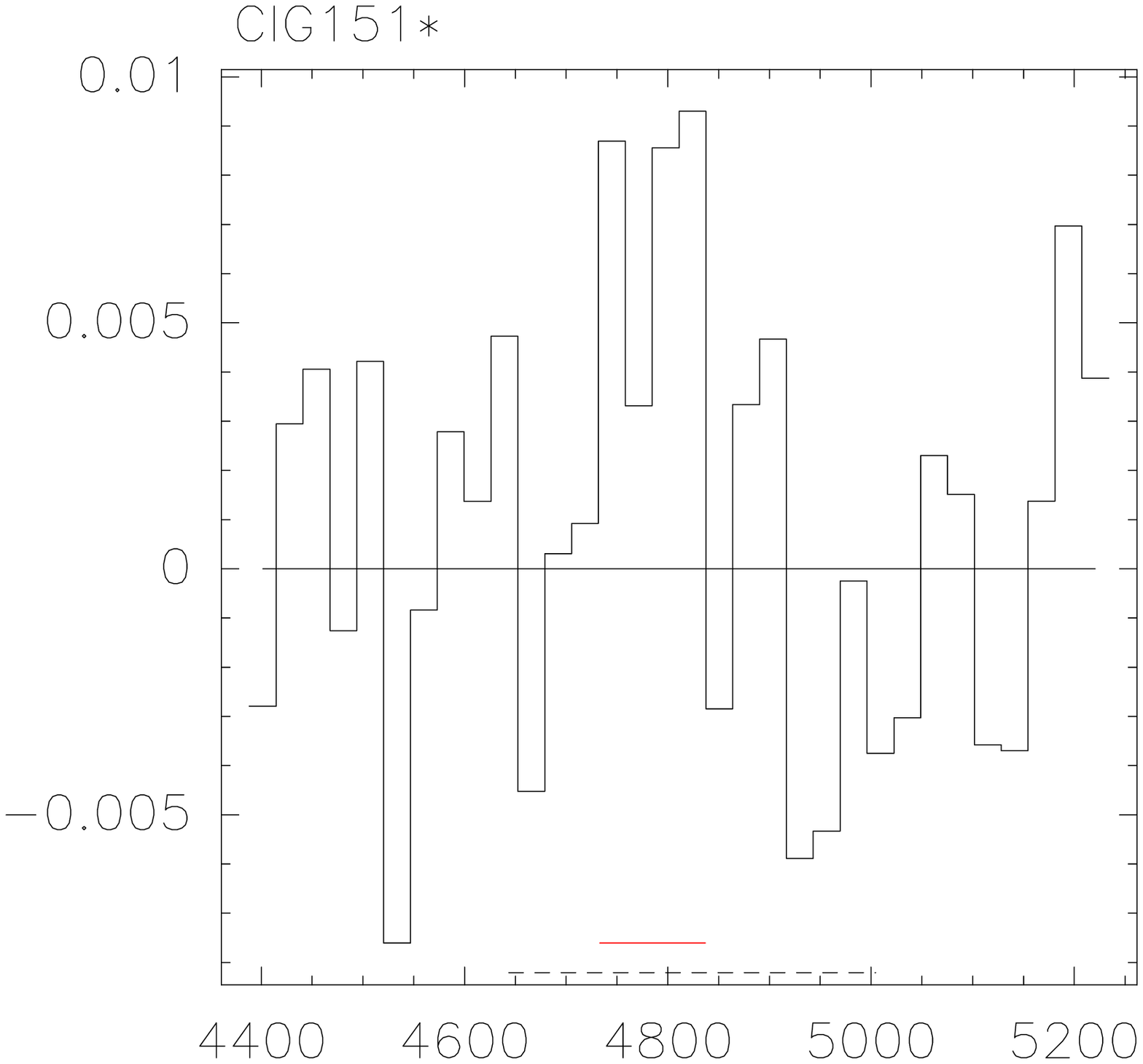}} 
\centerline{\includegraphics[width=3cm,angle=270]{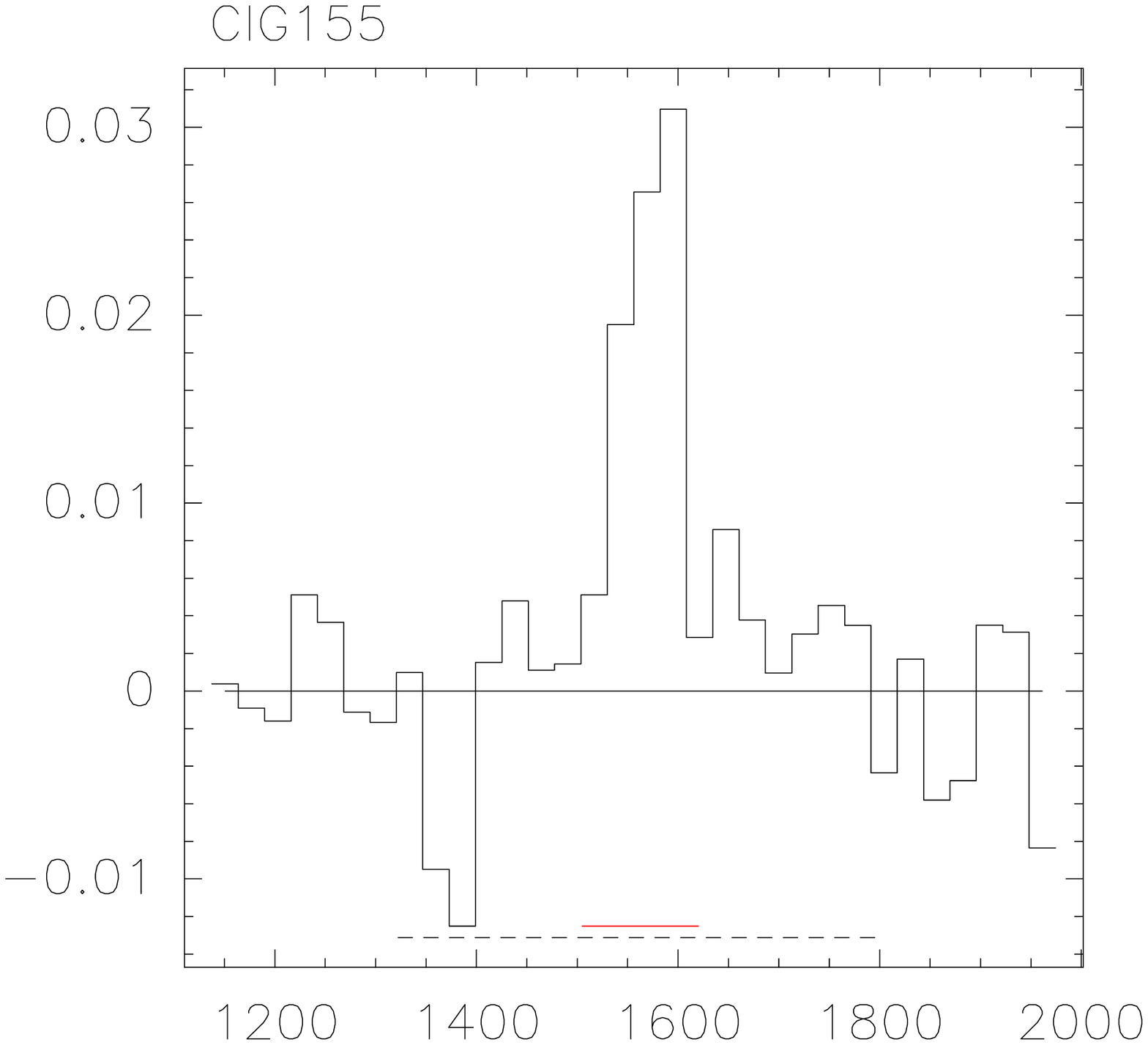} \quad 
\includegraphics[width=3cm,angle=270]{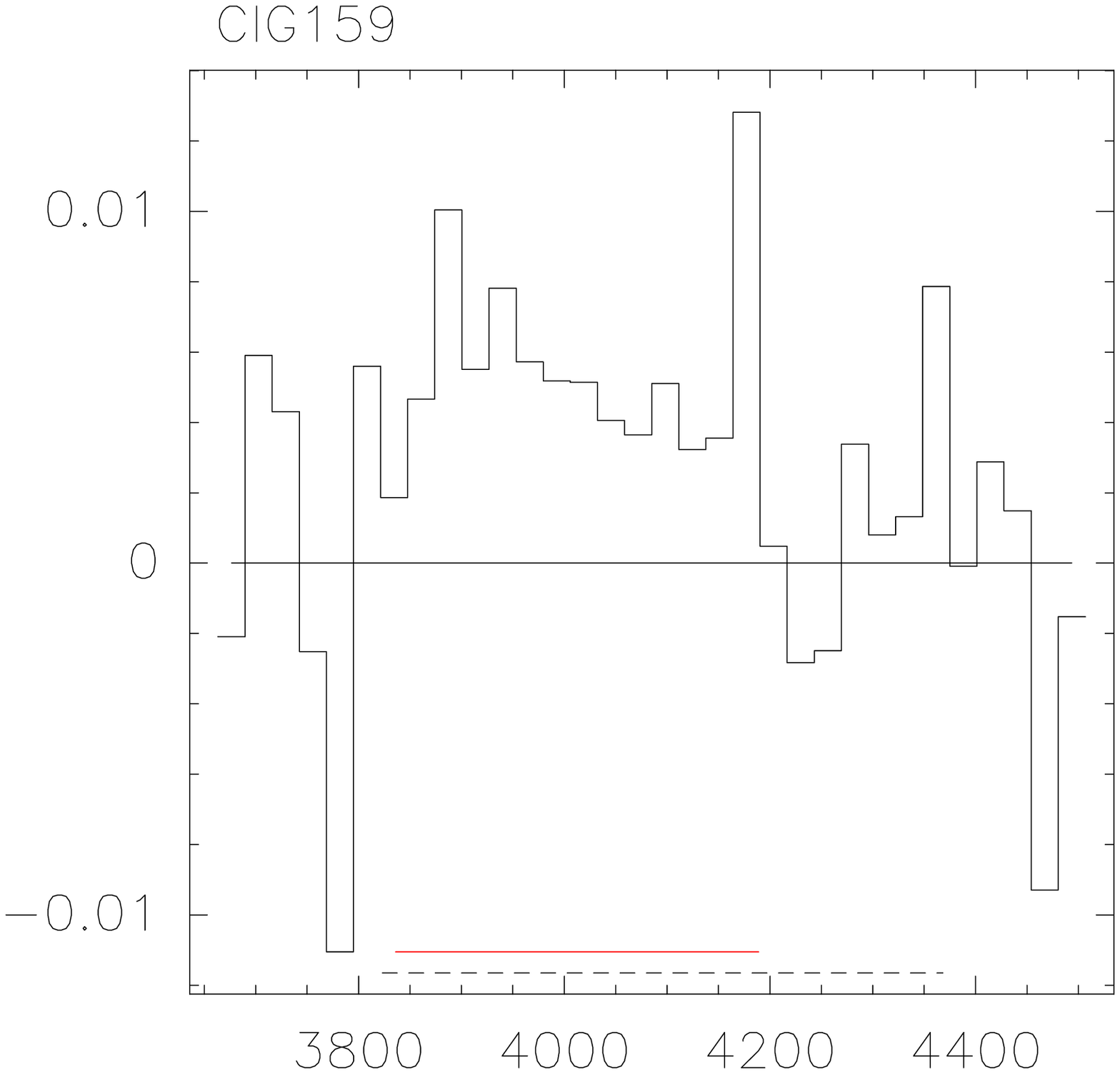}\quad 
\includegraphics[width=3cm,angle=270]{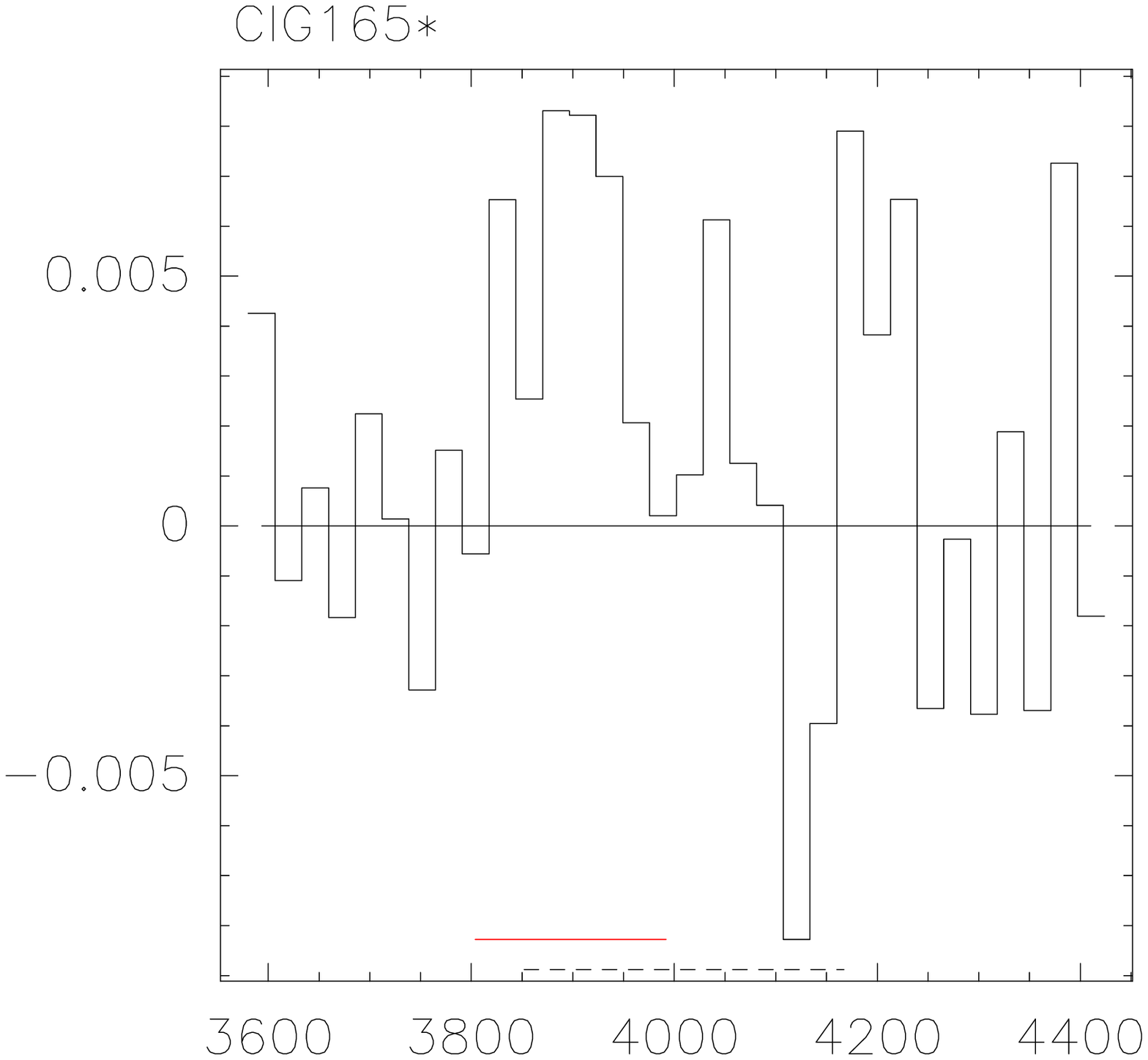}\quad 
\includegraphics[width=3cm,angle=270]{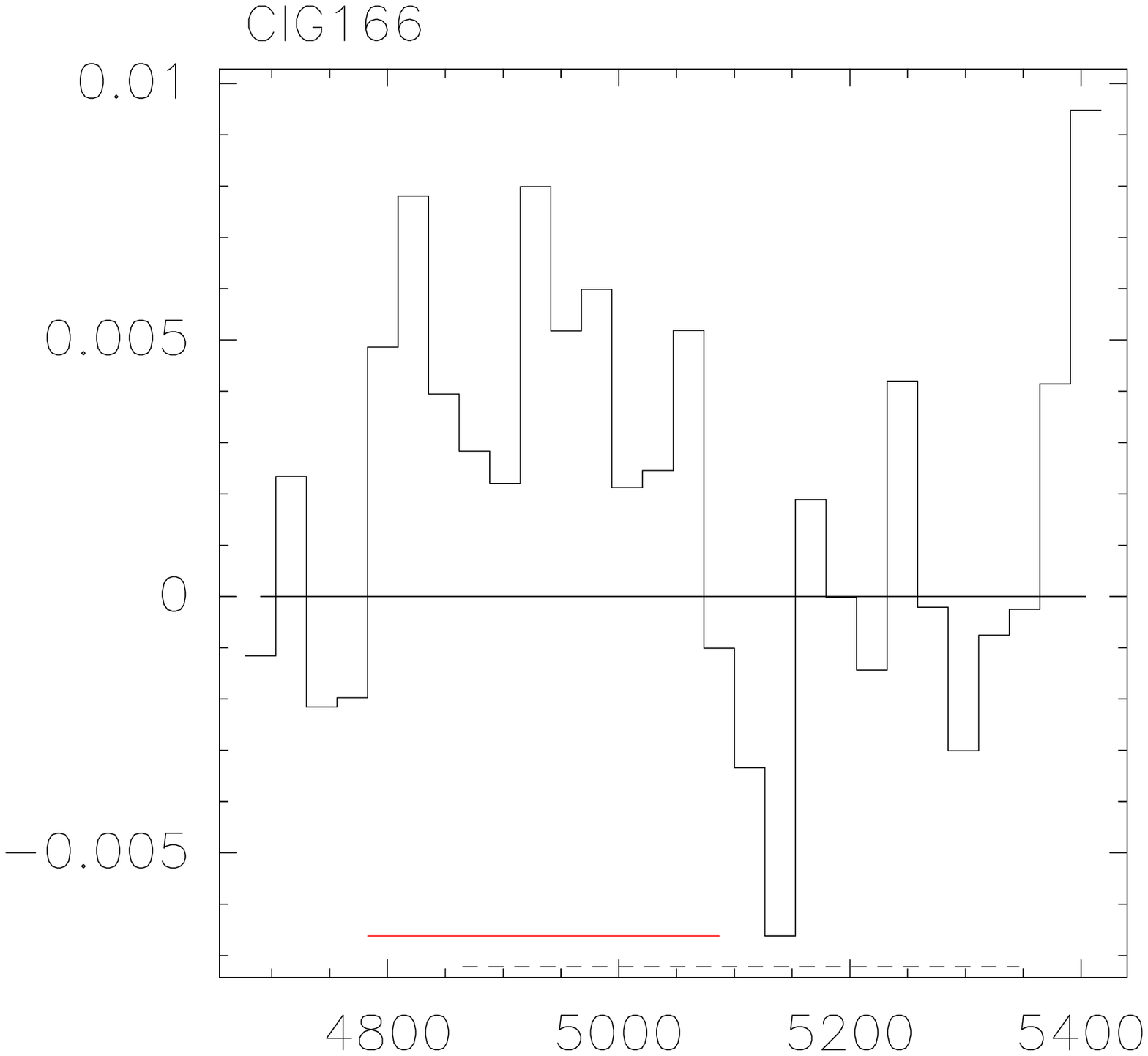}\quad 
\includegraphics[width=3cm,angle=270]{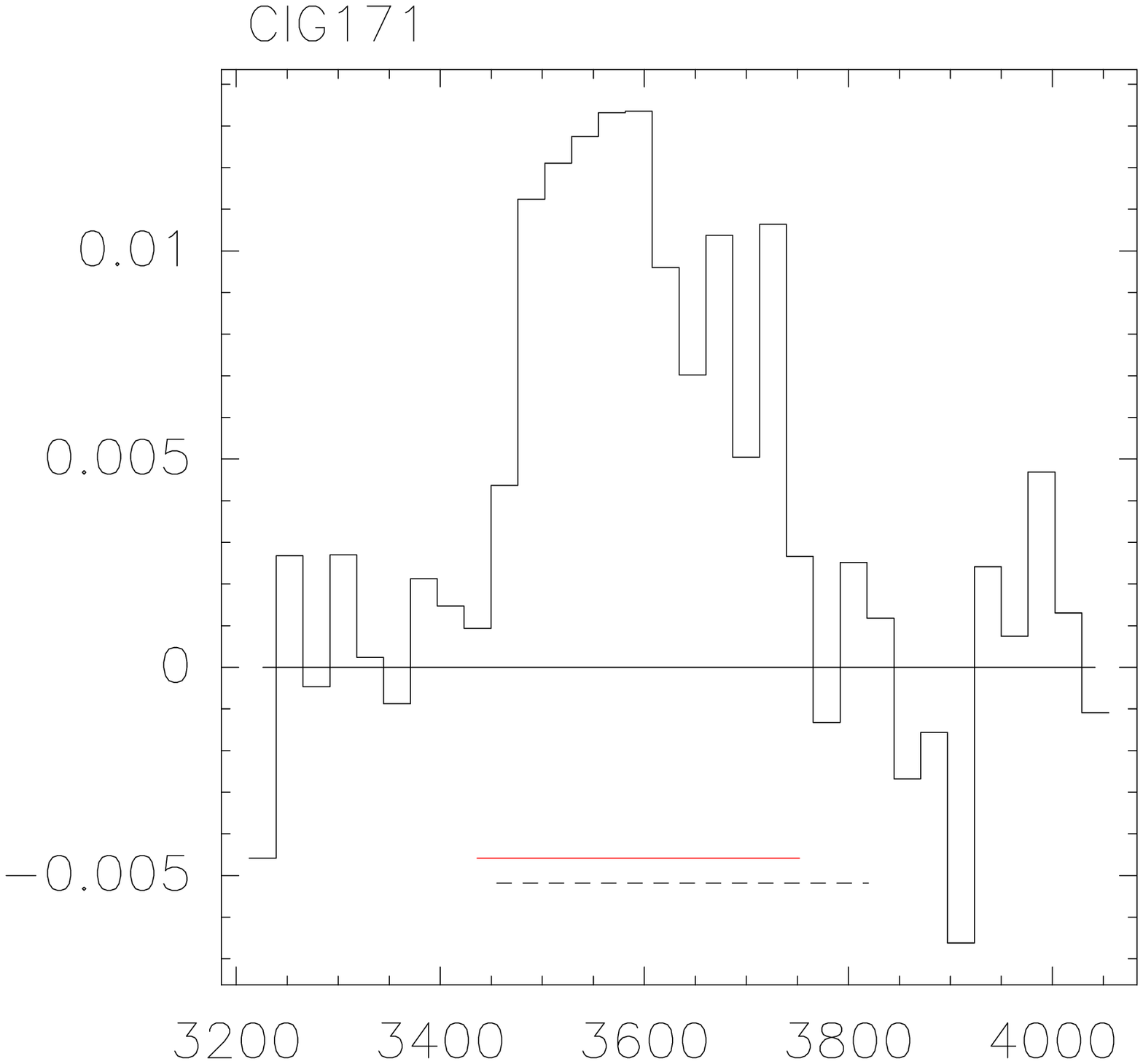}} 
\centerline{\includegraphics[width=3cm,angle=270]{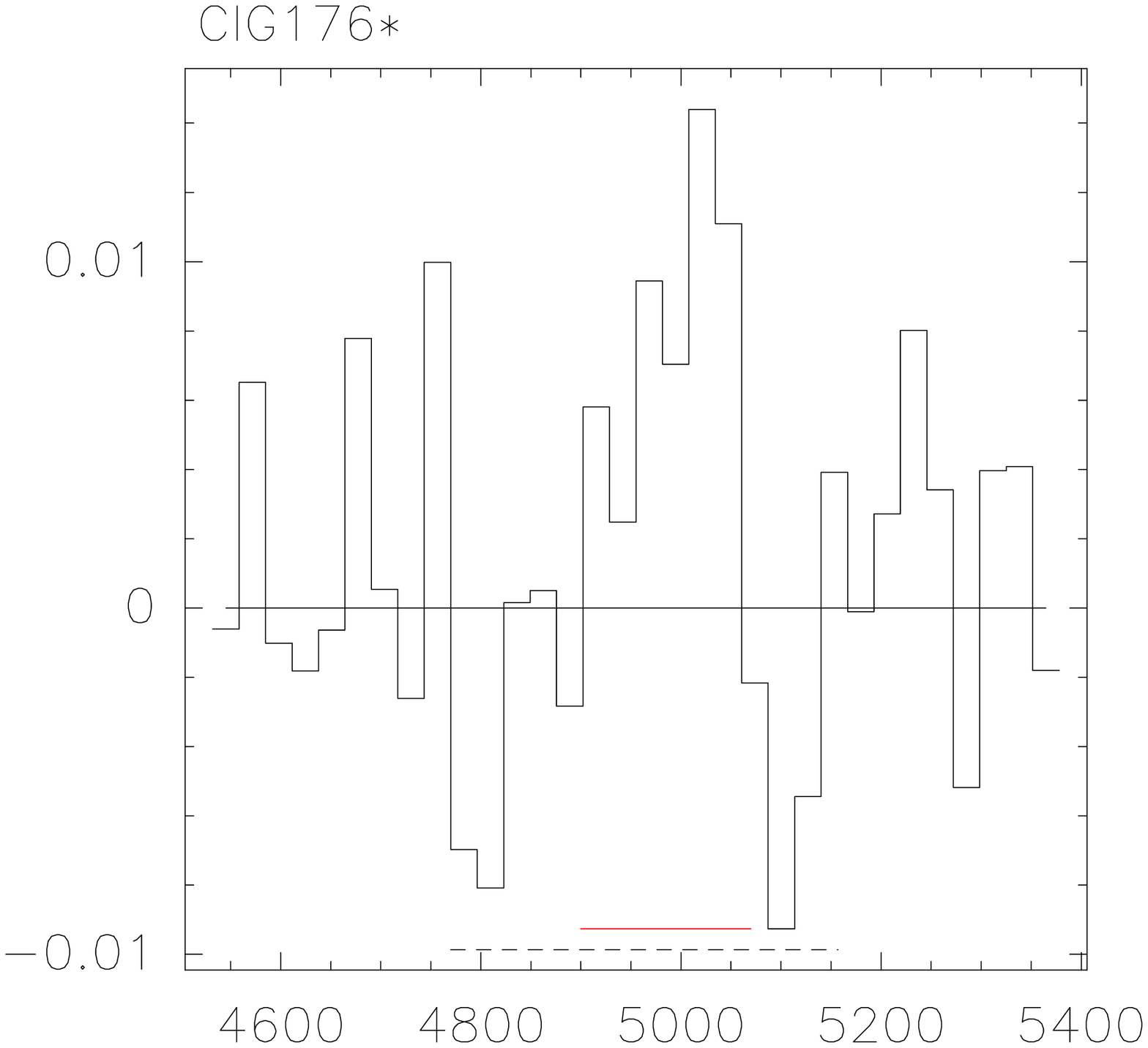} \quad 
\includegraphics[width=3cm,angle=270]{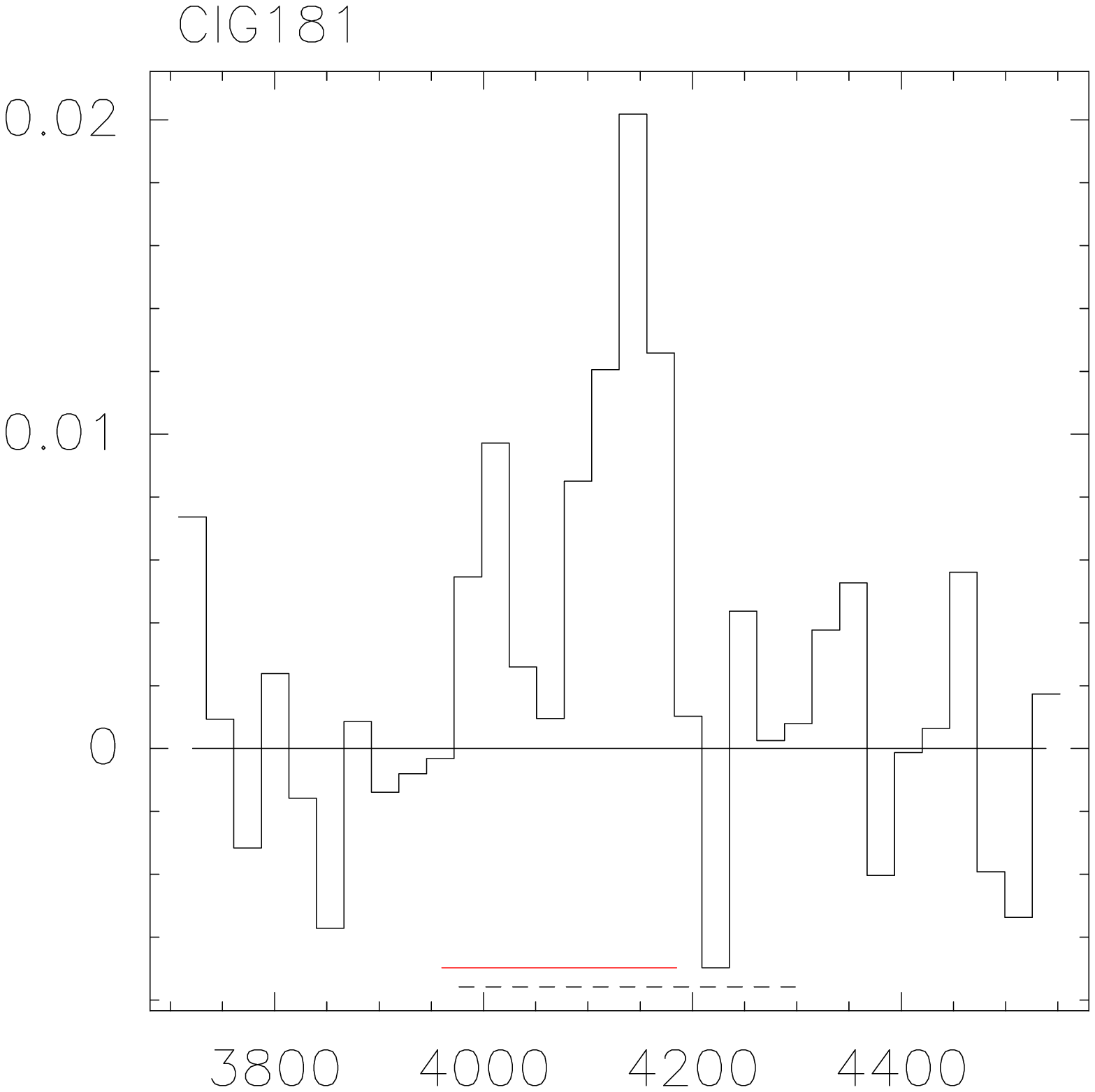}\quad 
\includegraphics[width=3cm,angle=270]{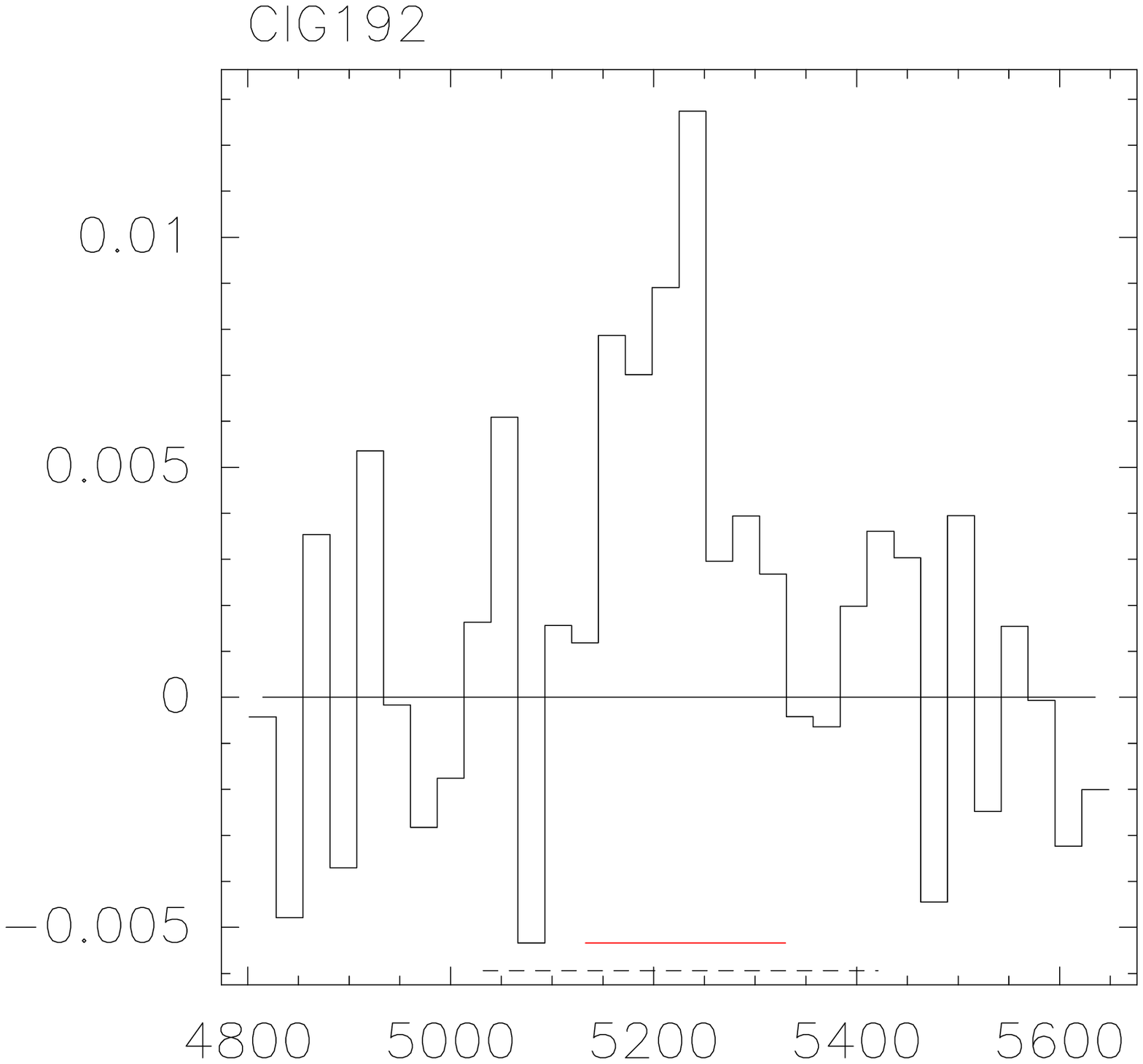}\quad 
\includegraphics[width=3cm,angle=270]{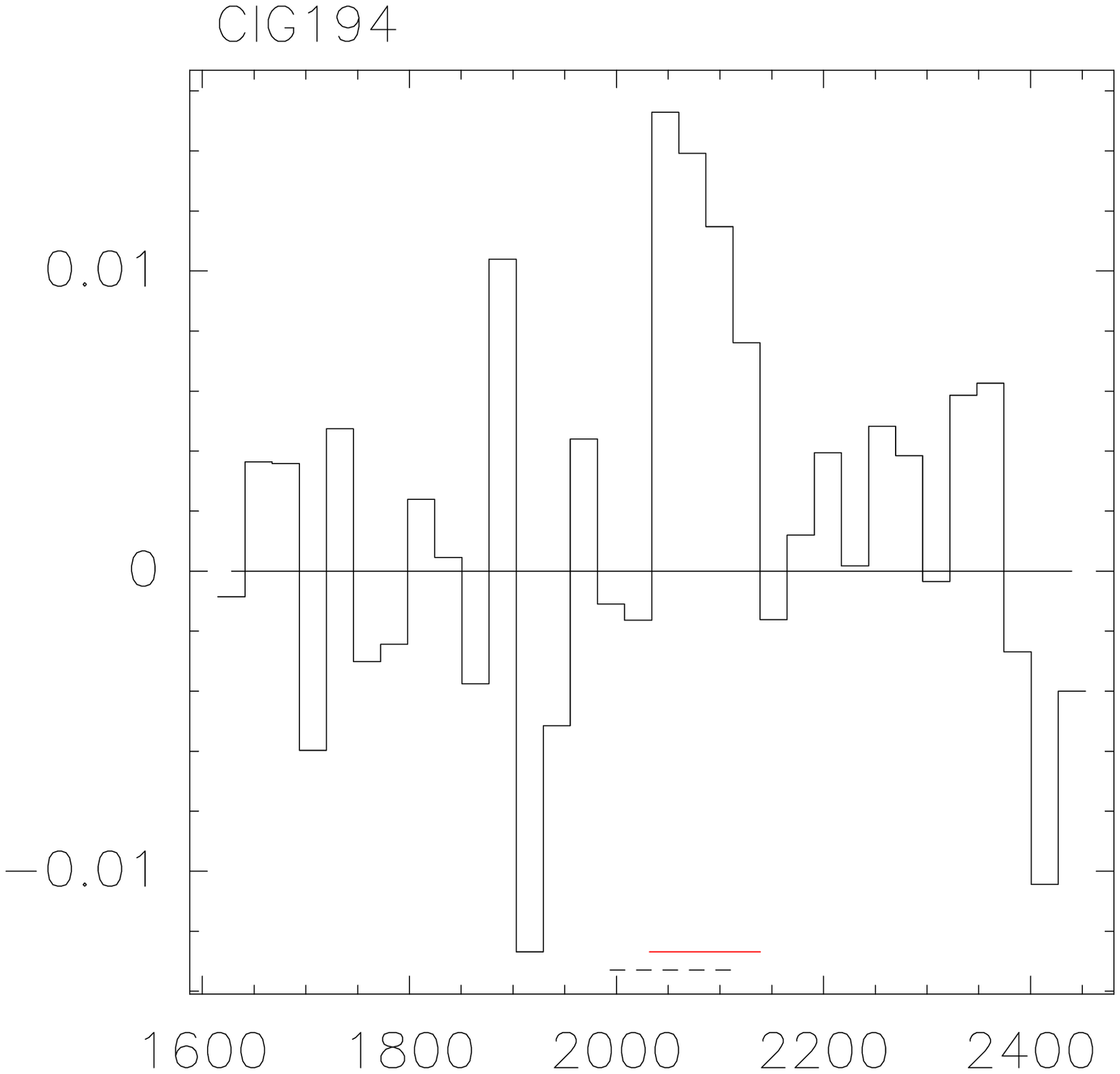}\quad 
\includegraphics[width=3cm,angle=270]{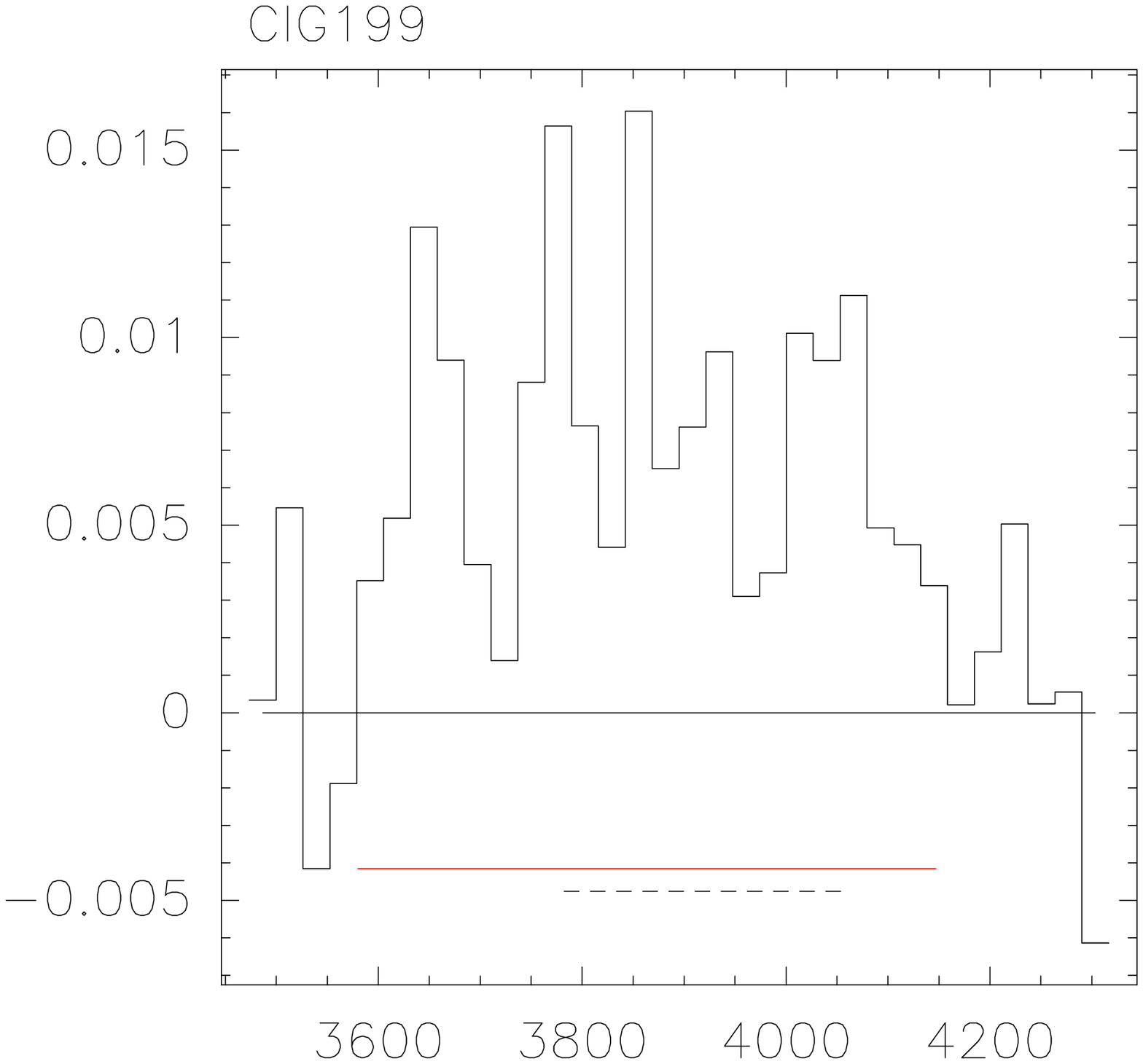}} 
\centerline{\includegraphics[width=3cm,angle=270]{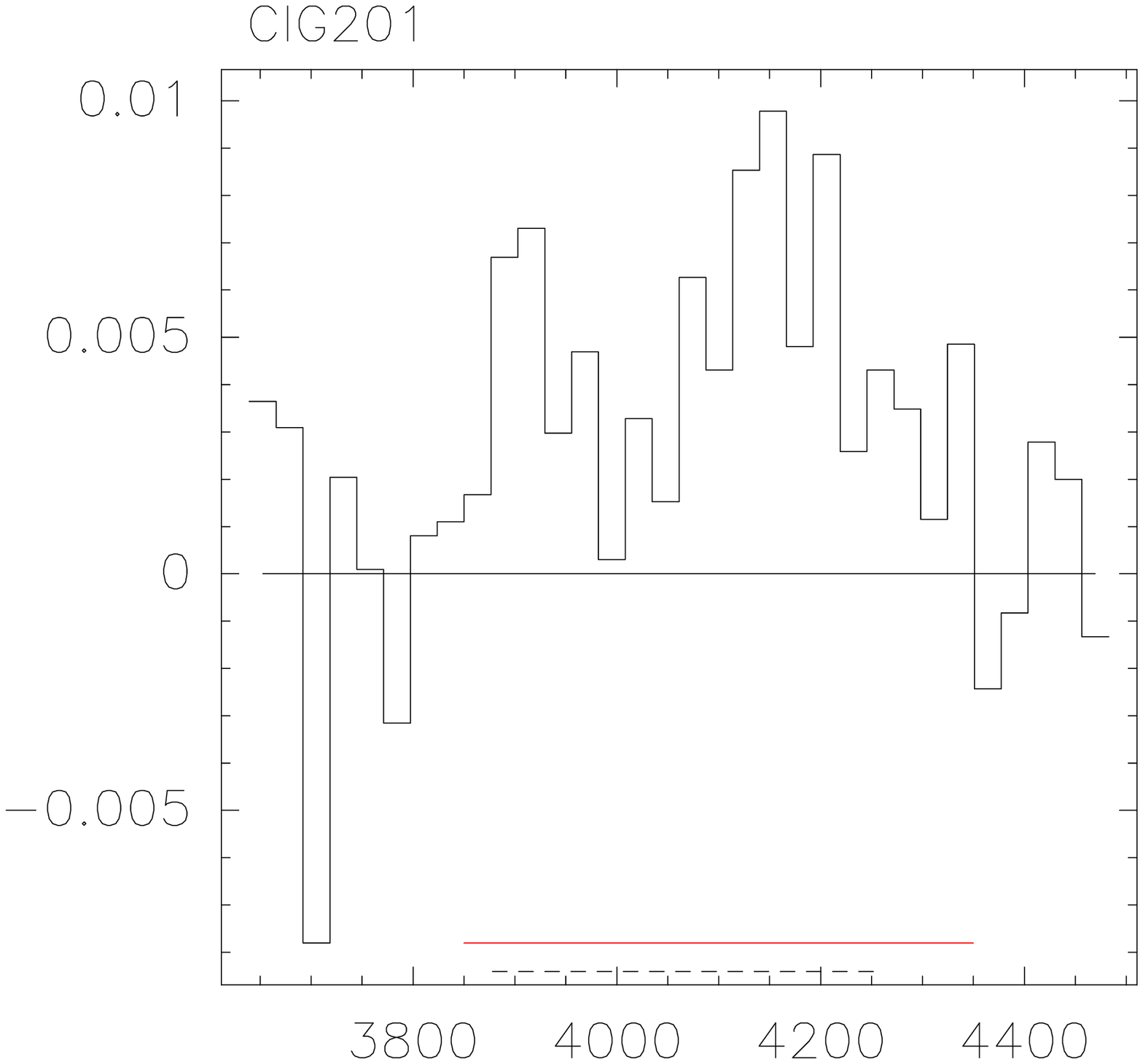} \quad 
\includegraphics[width=3cm,angle=270]{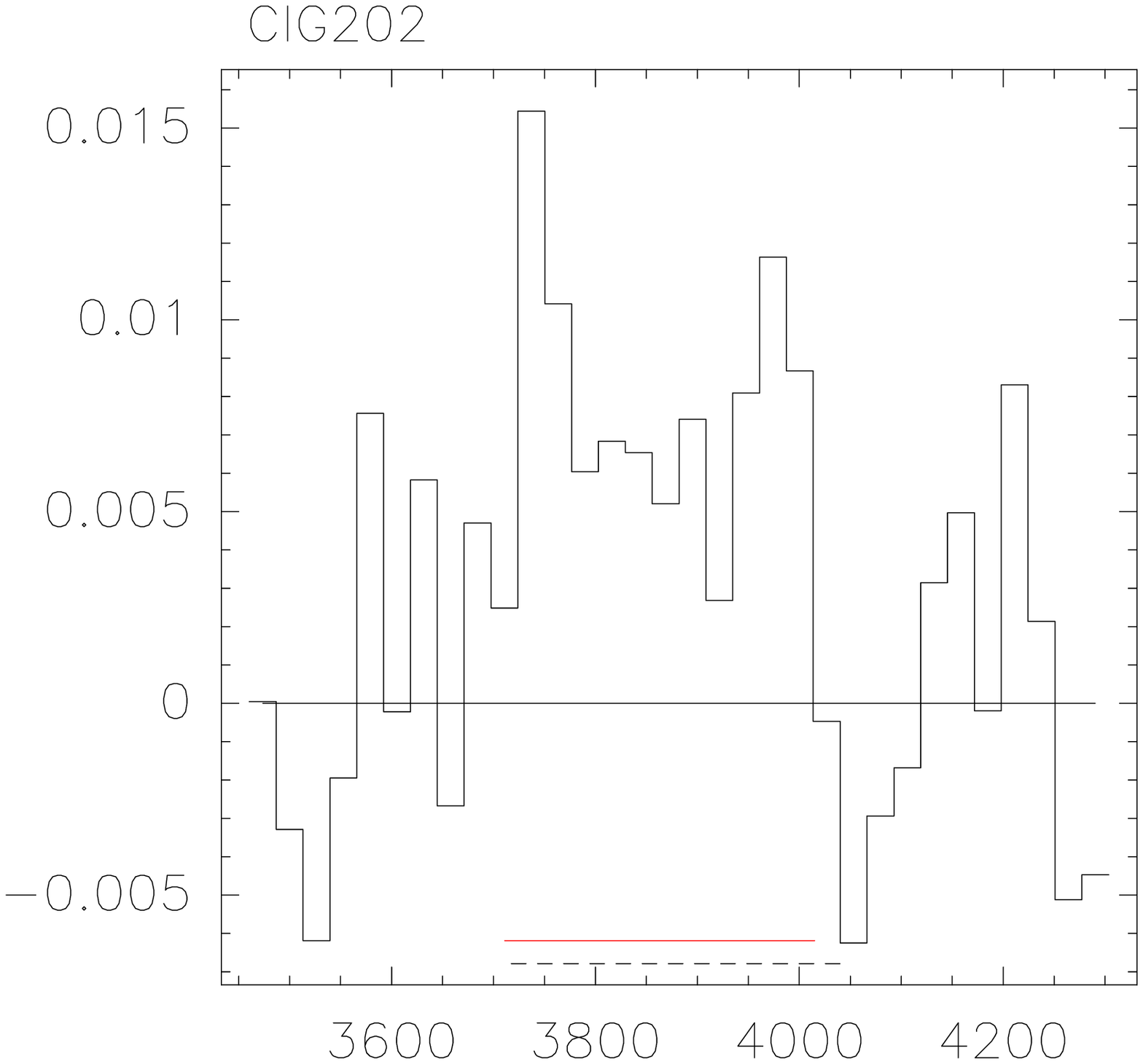}\quad 
\includegraphics[width=3cm,angle=270]{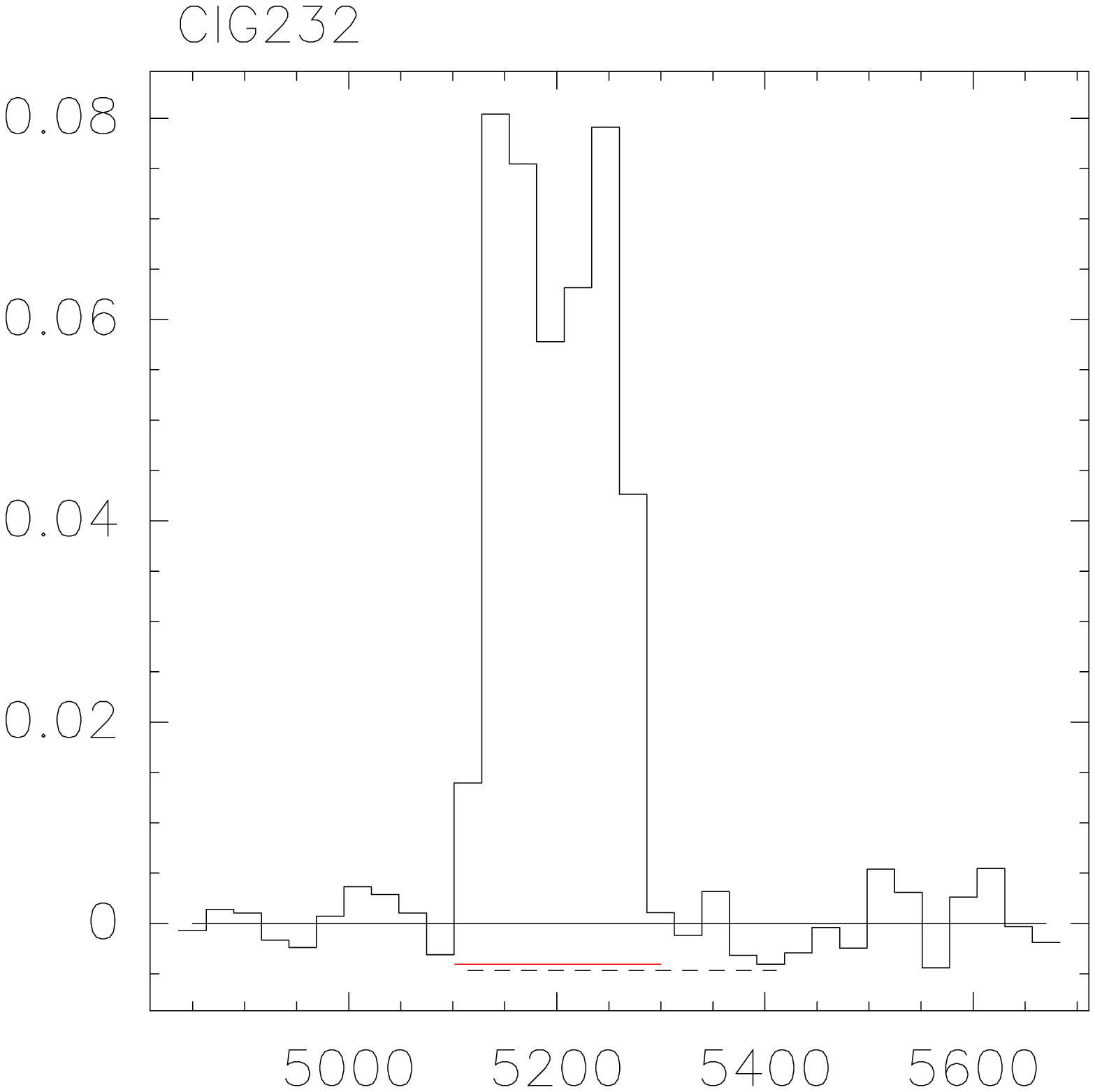}\quad 
\includegraphics[width=3cm,angle=270]{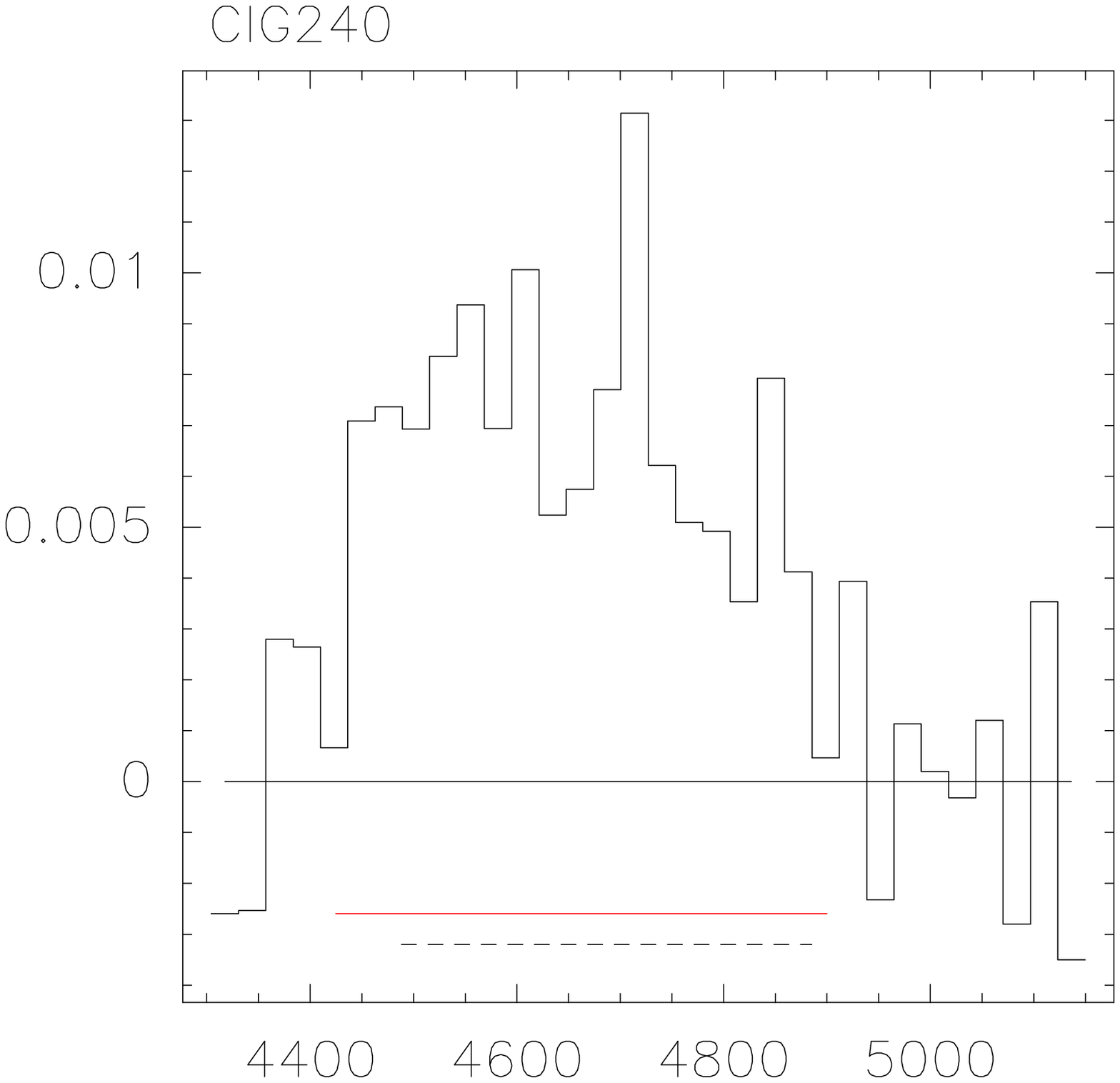}\quad 
\includegraphics[width=3cm,angle=270]{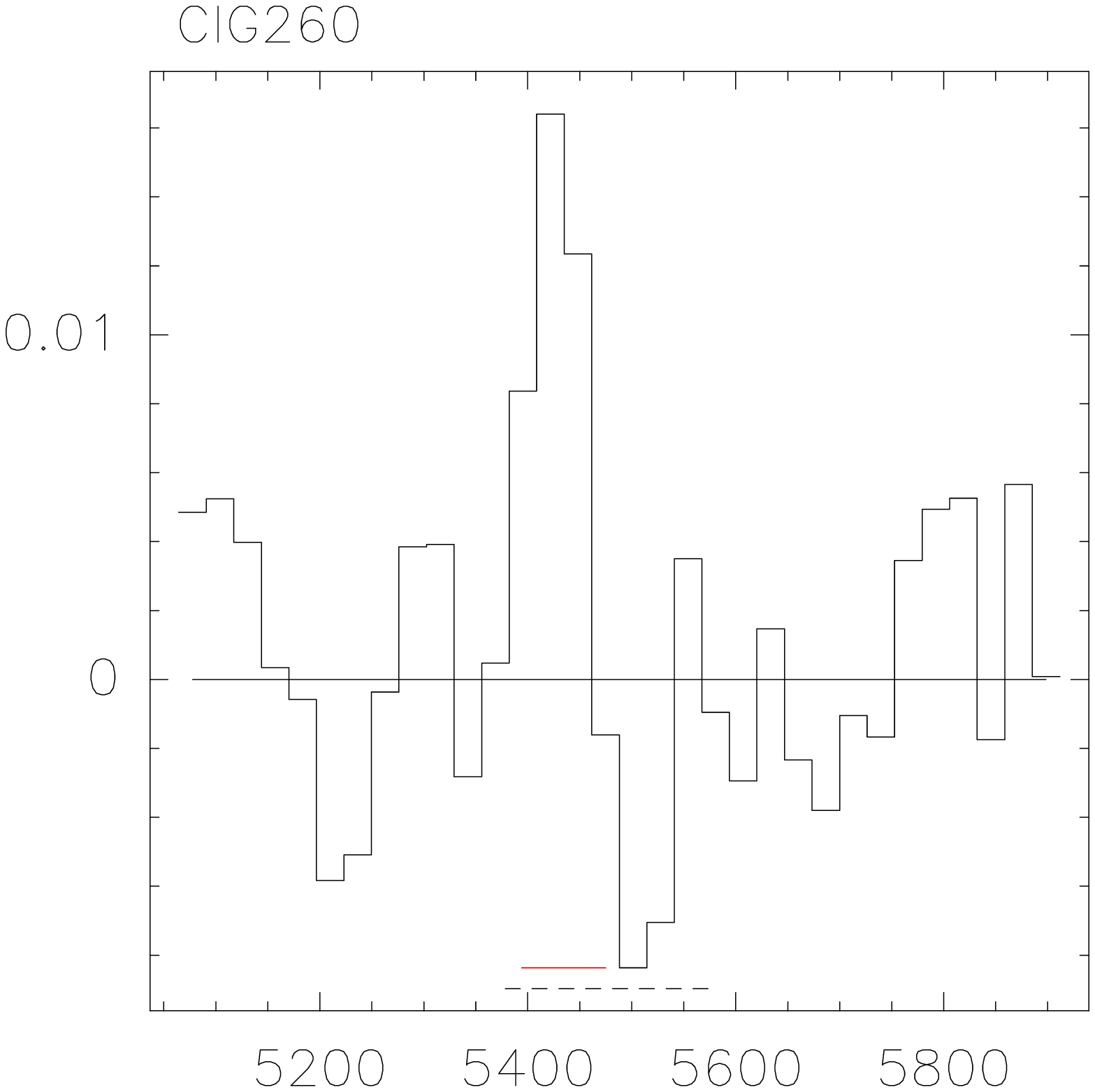}} 
\centerline{\includegraphics[width=3cm,angle=270]{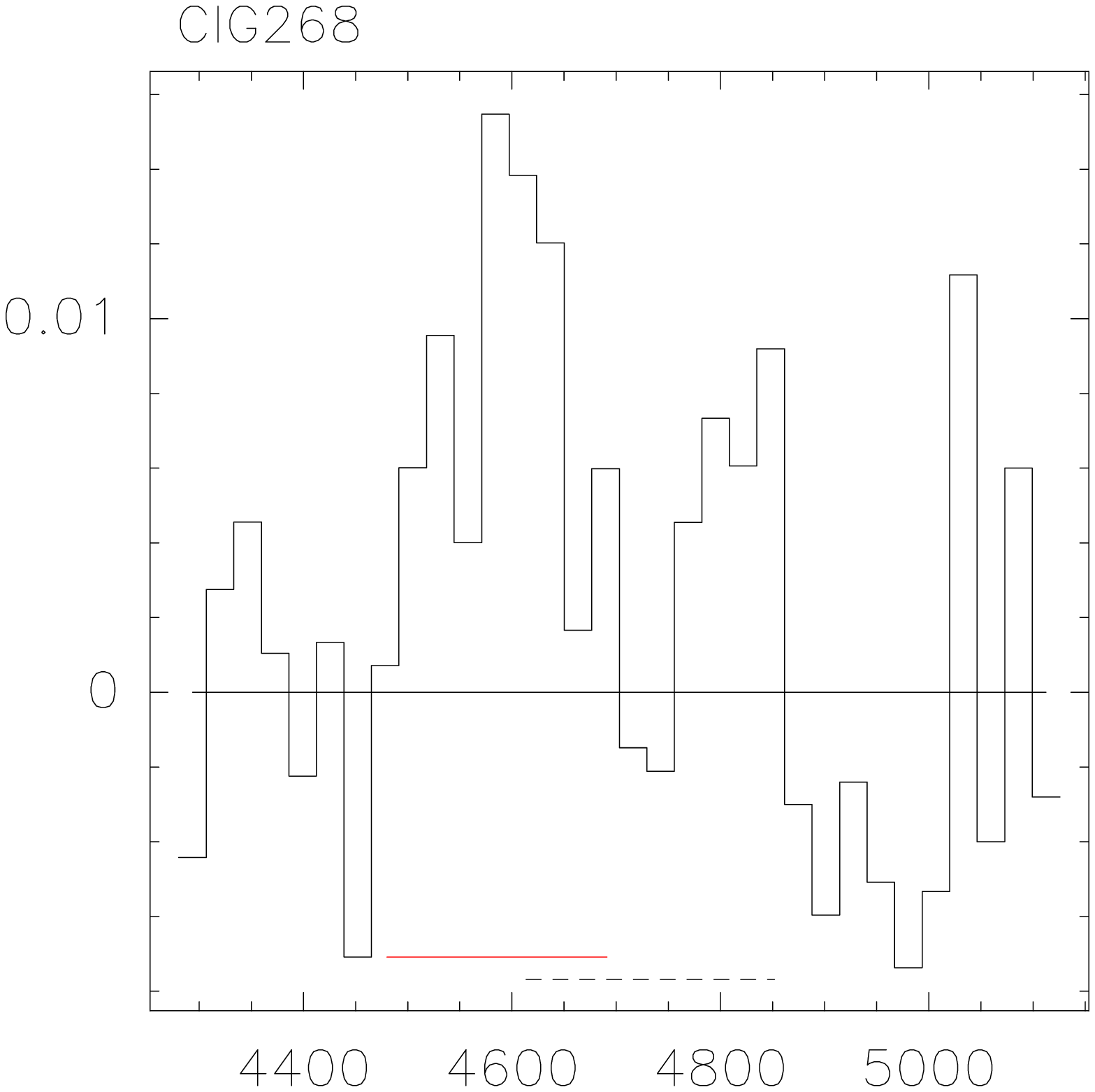} \quad 
\includegraphics[width=3cm,angle=270]{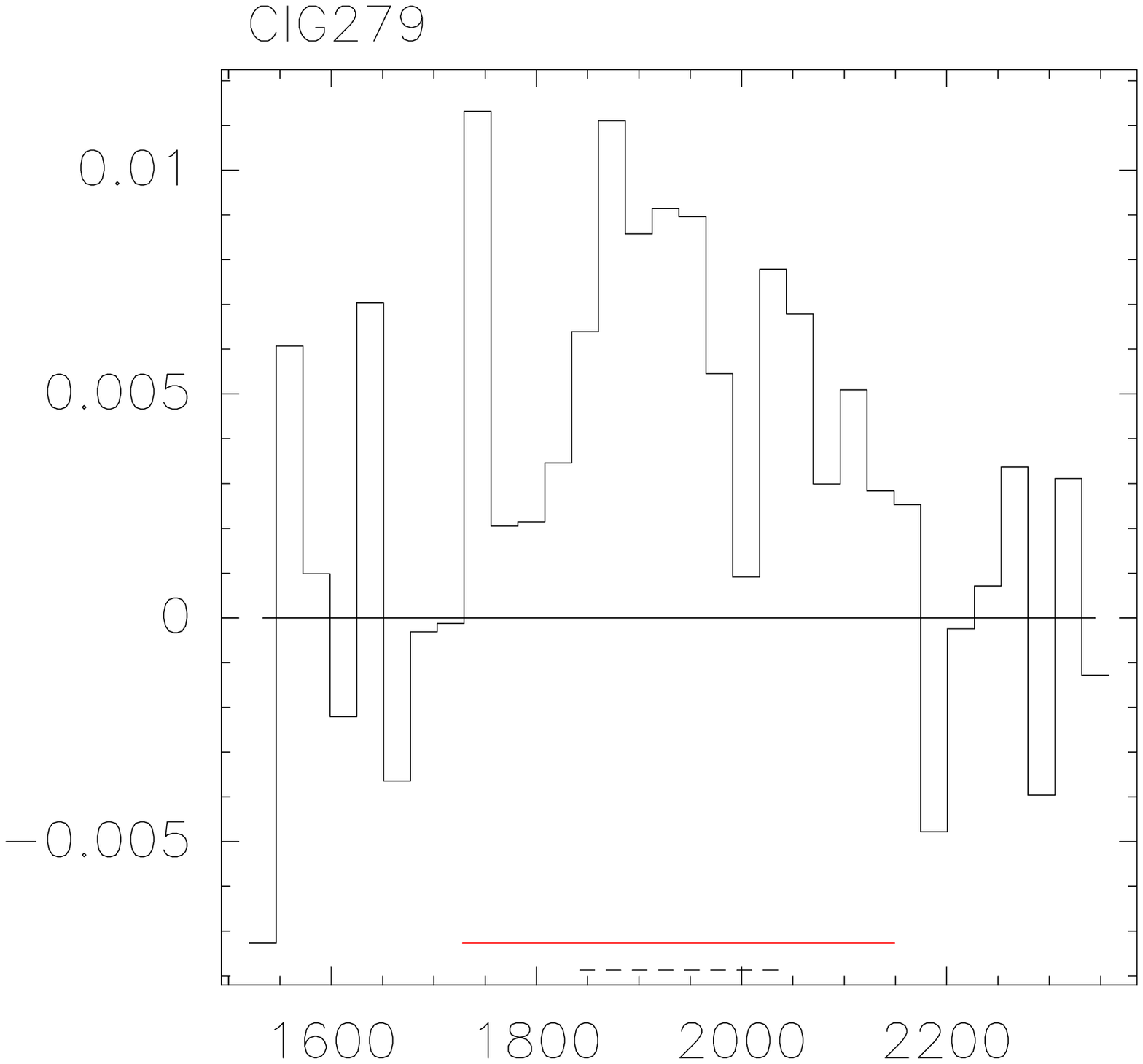}\quad 
\includegraphics[width=3cm,angle=270]{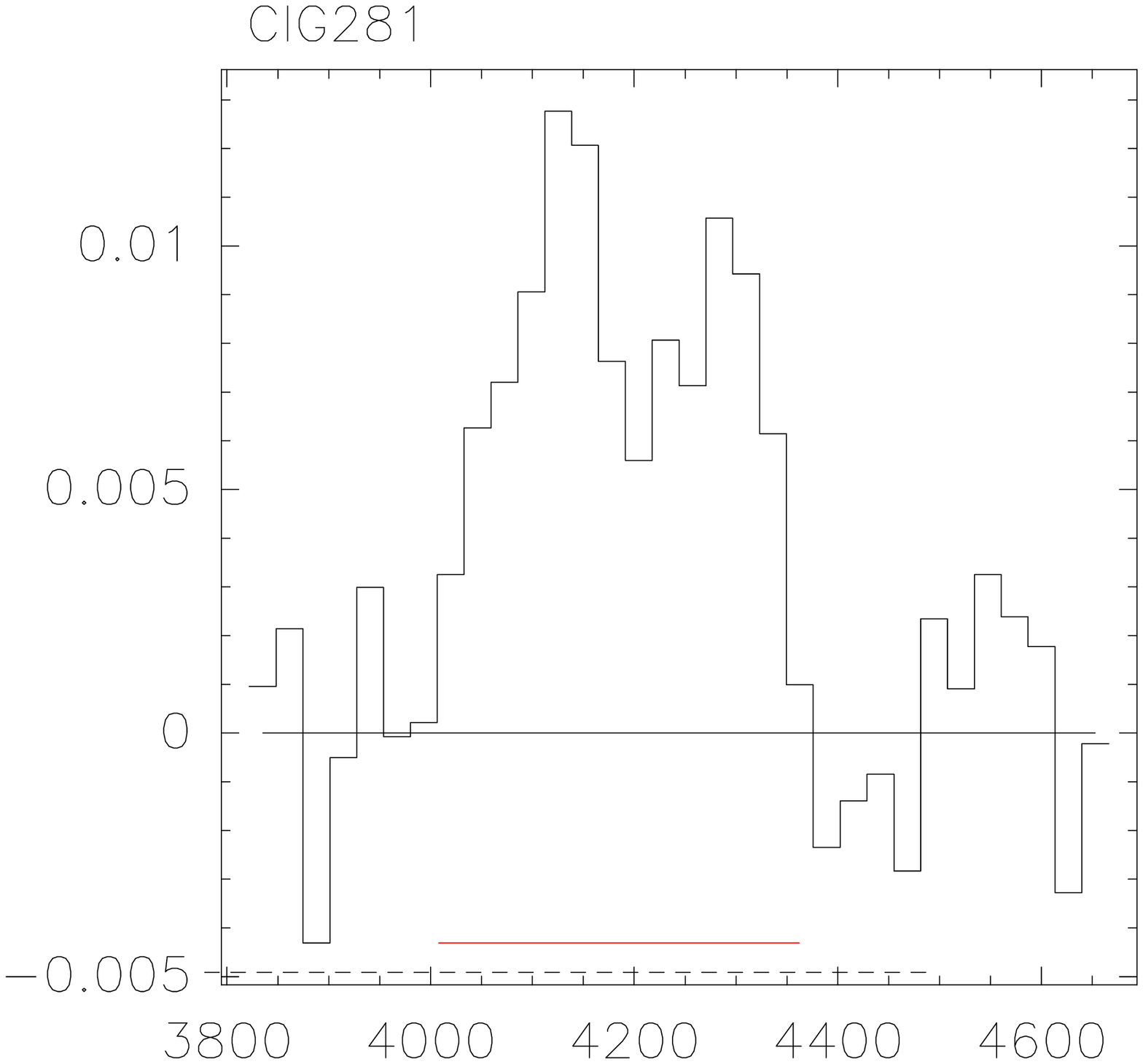}\quad 
\includegraphics[width=3cm,angle=270]{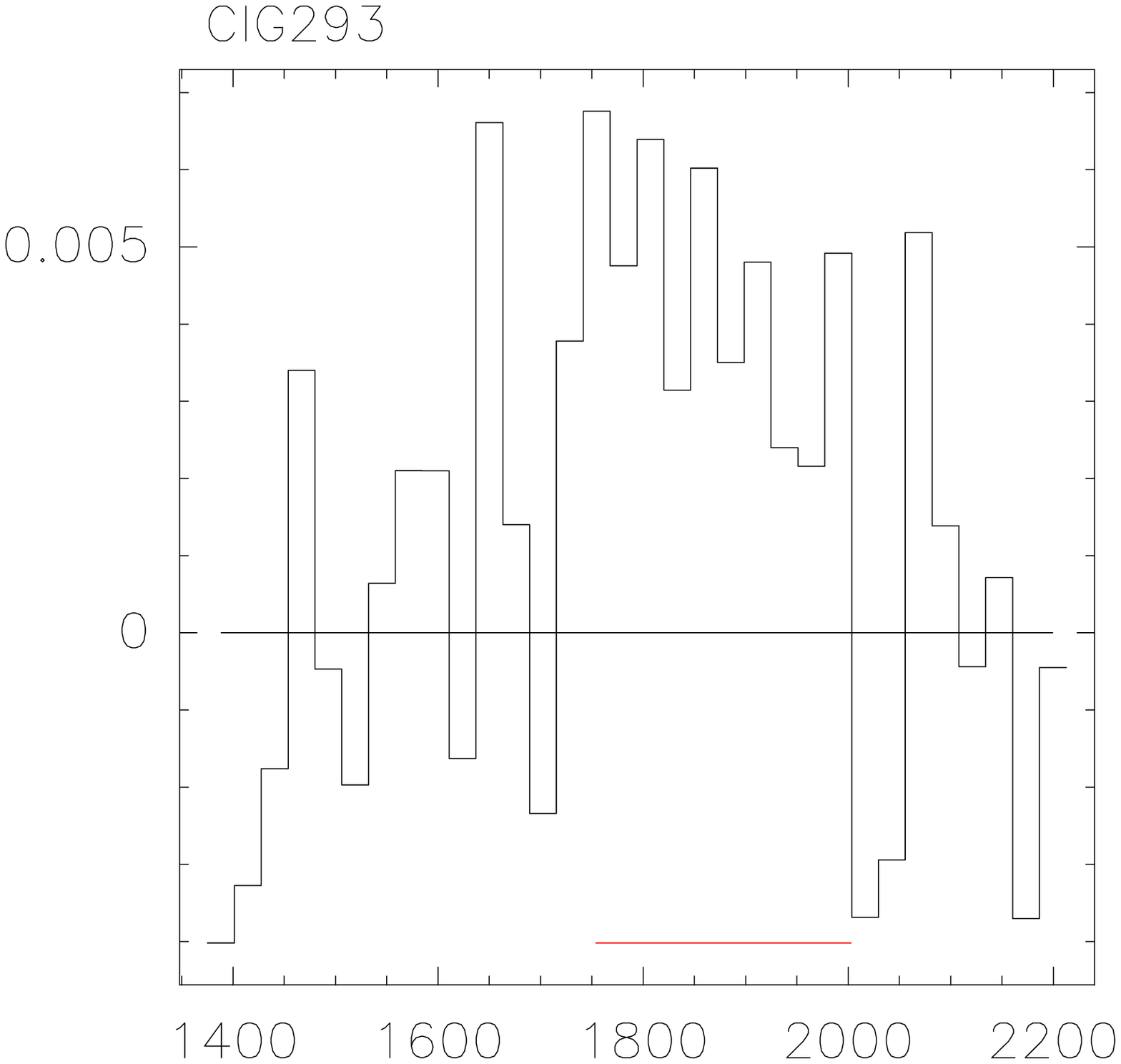}\quad 
\includegraphics[width=3cm,angle=270]{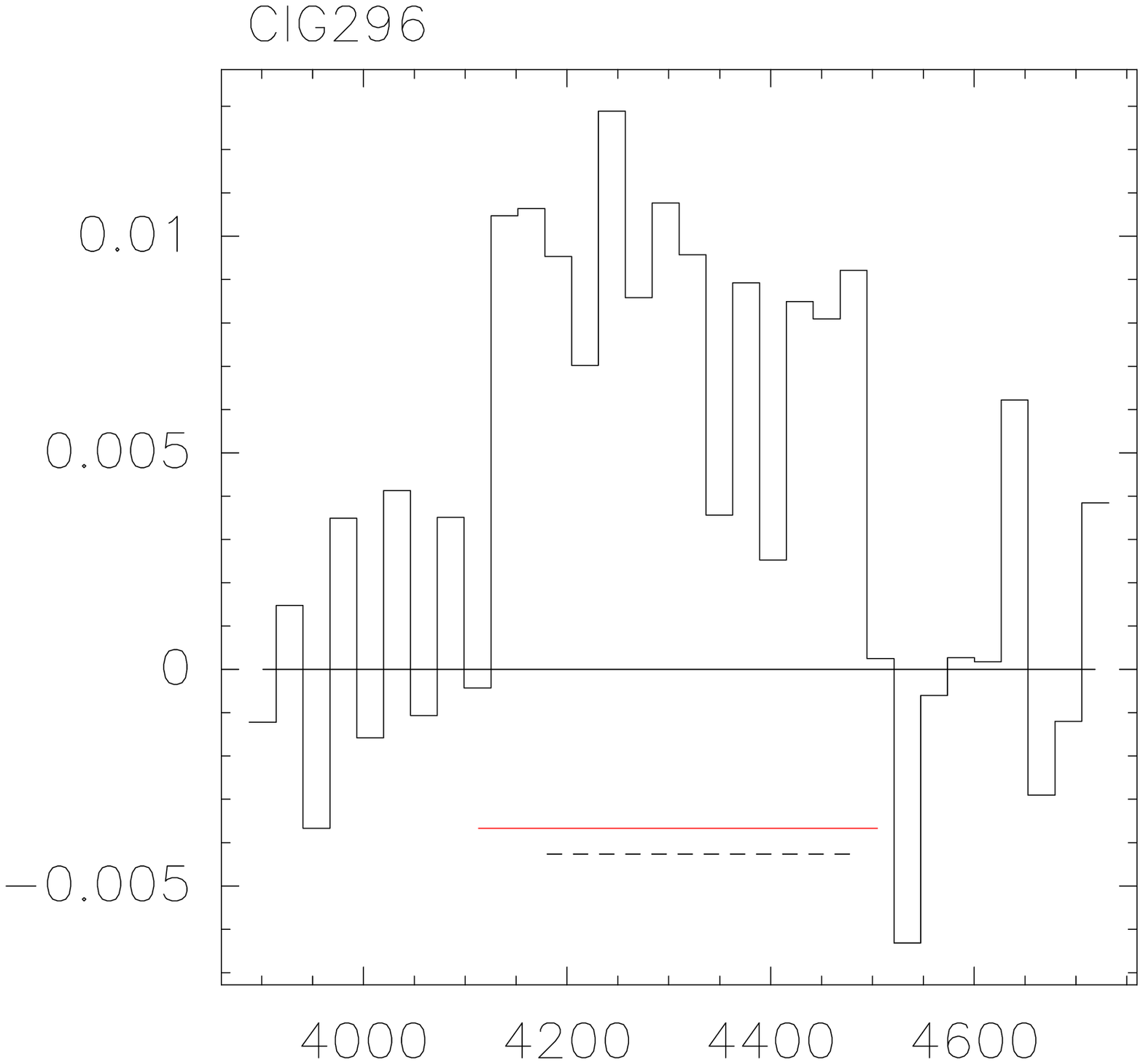}} 
\centerline{\includegraphics[width=3cm,angle=270]{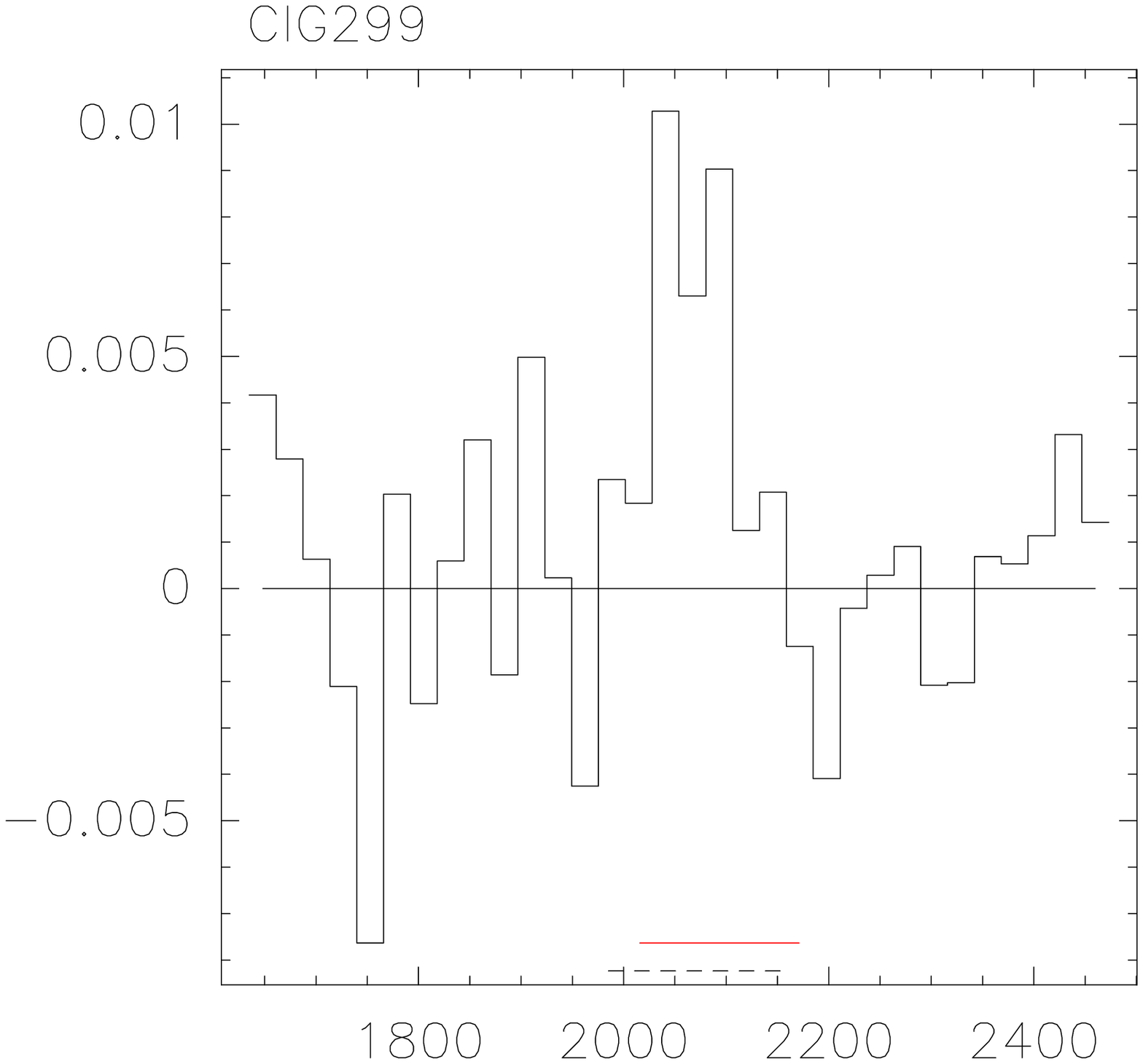} \quad 
\includegraphics[width=3cm,angle=270]{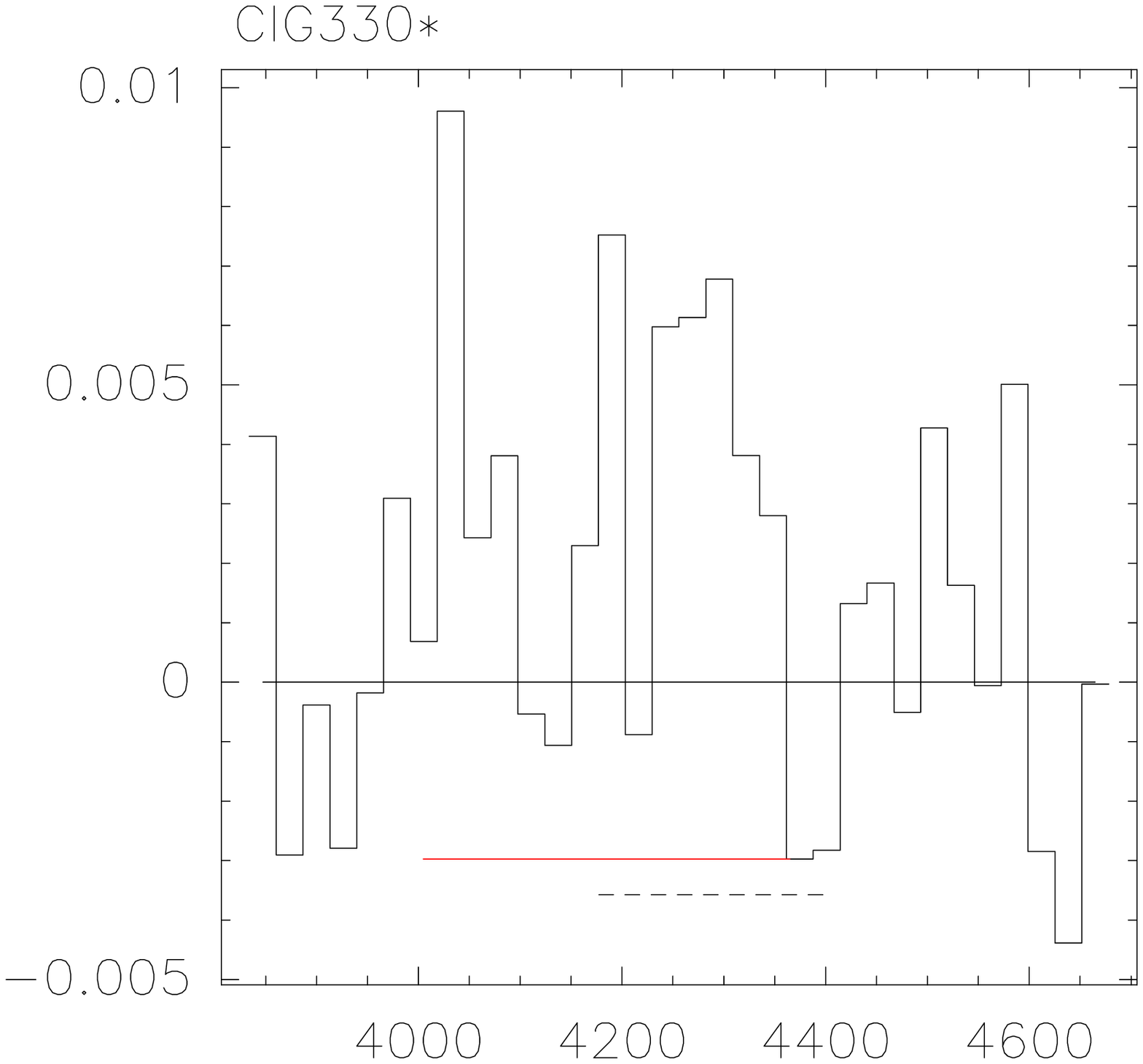}\quad 
\includegraphics[width=3cm,angle=270]{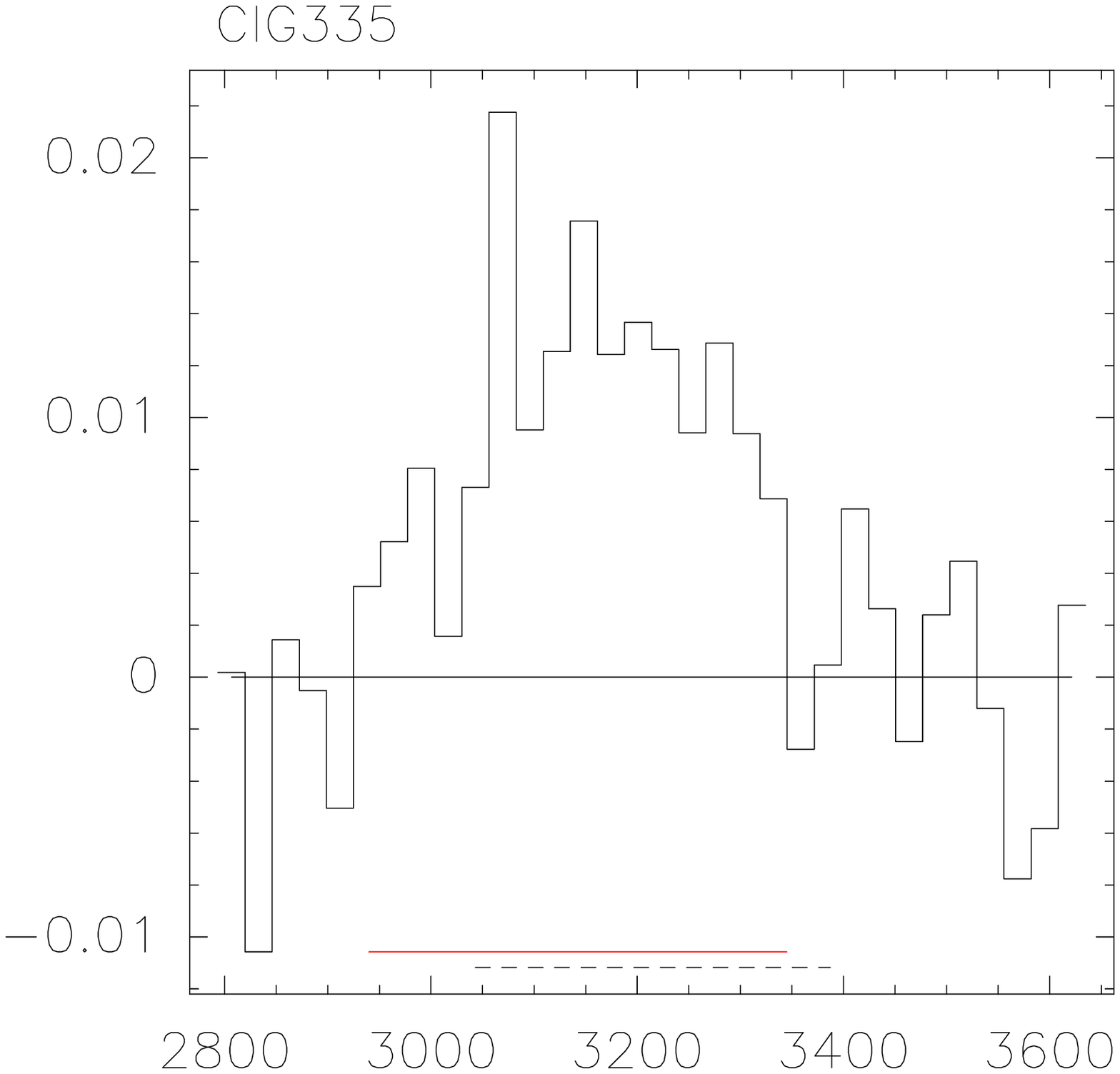}\quad 
\includegraphics[width=3cm,angle=270]{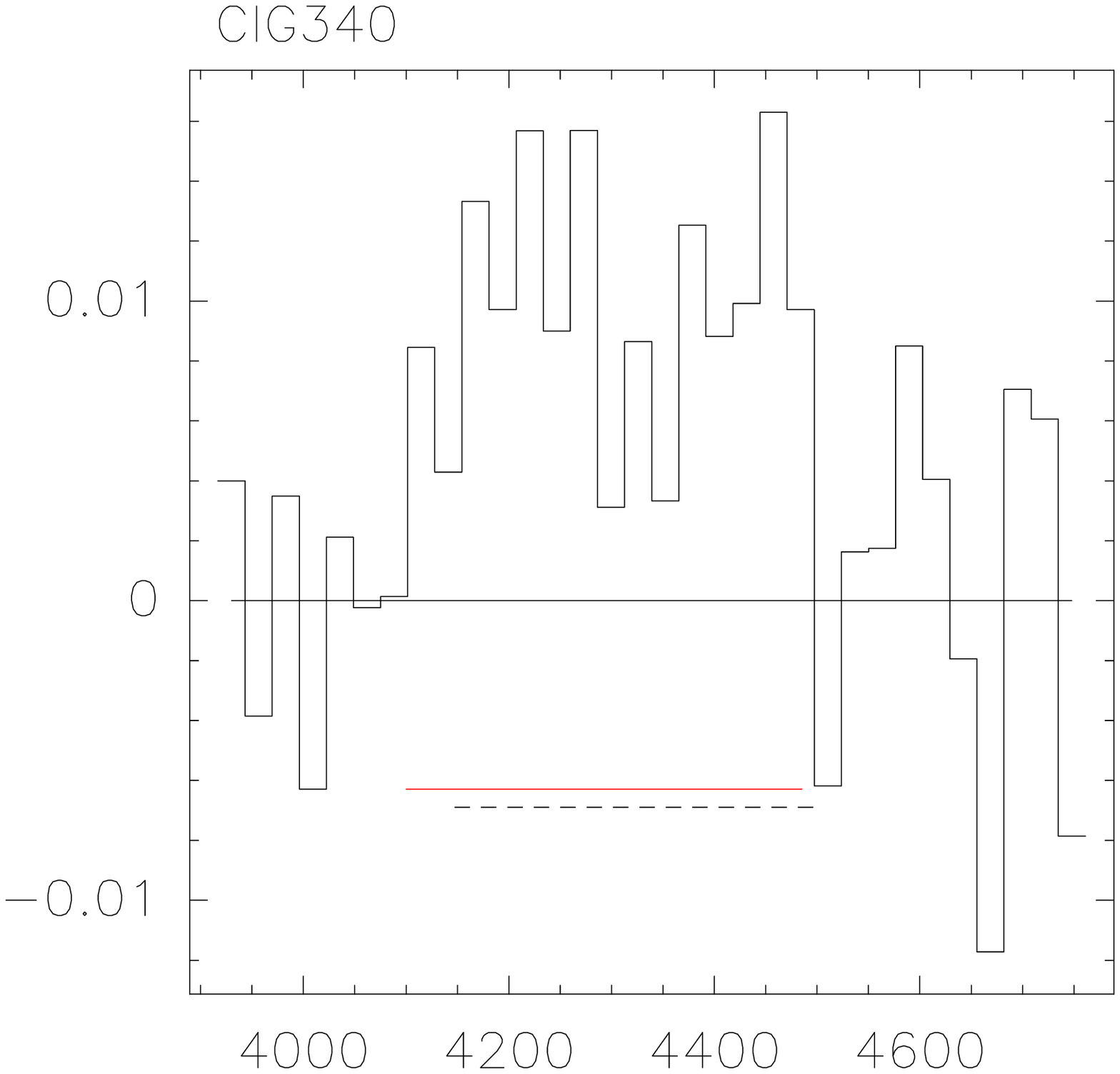}\quad 
\includegraphics[width=3cm,angle=270]{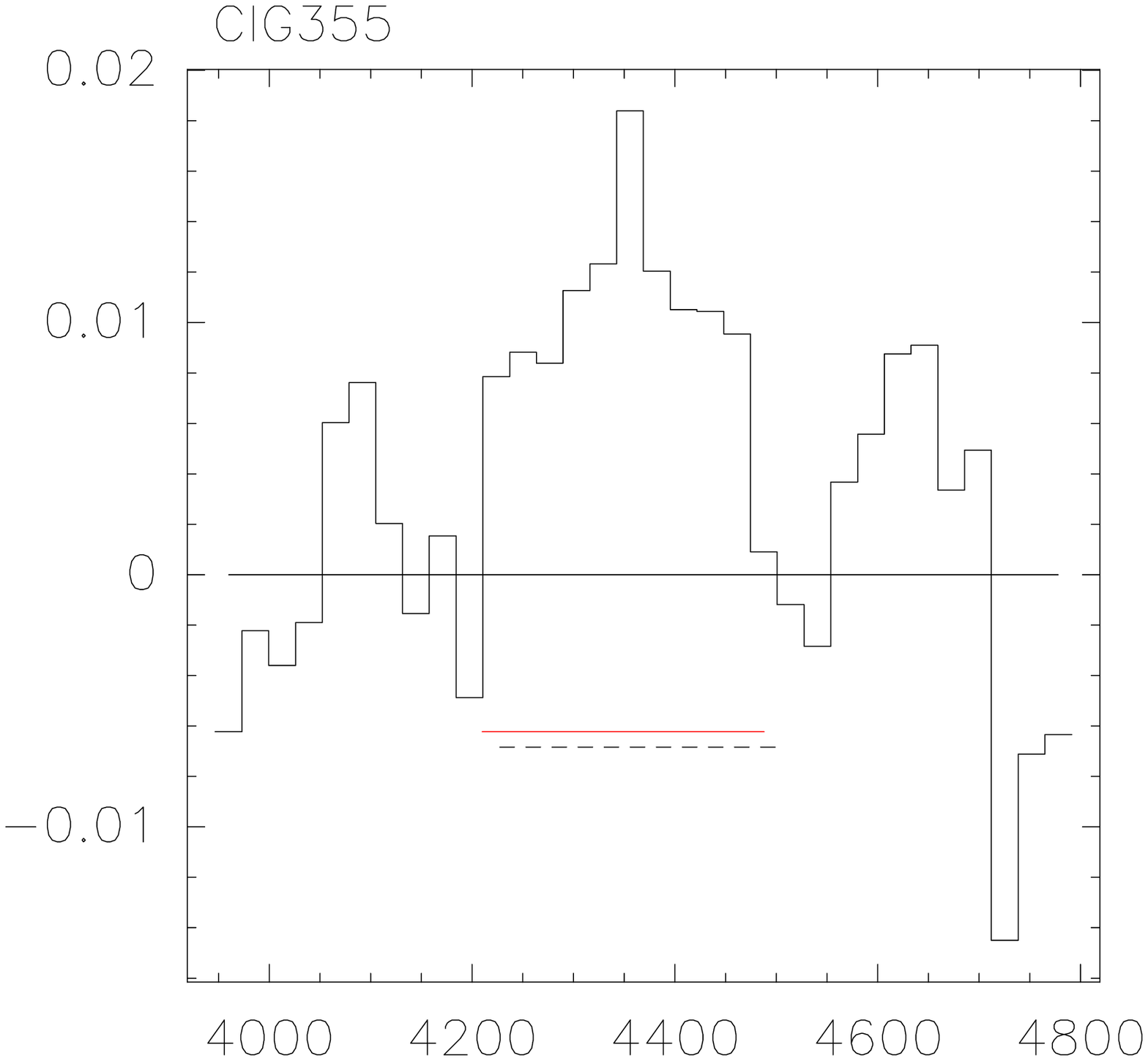}} 
\centerline{\includegraphics[width=4.2cm,angle=270]{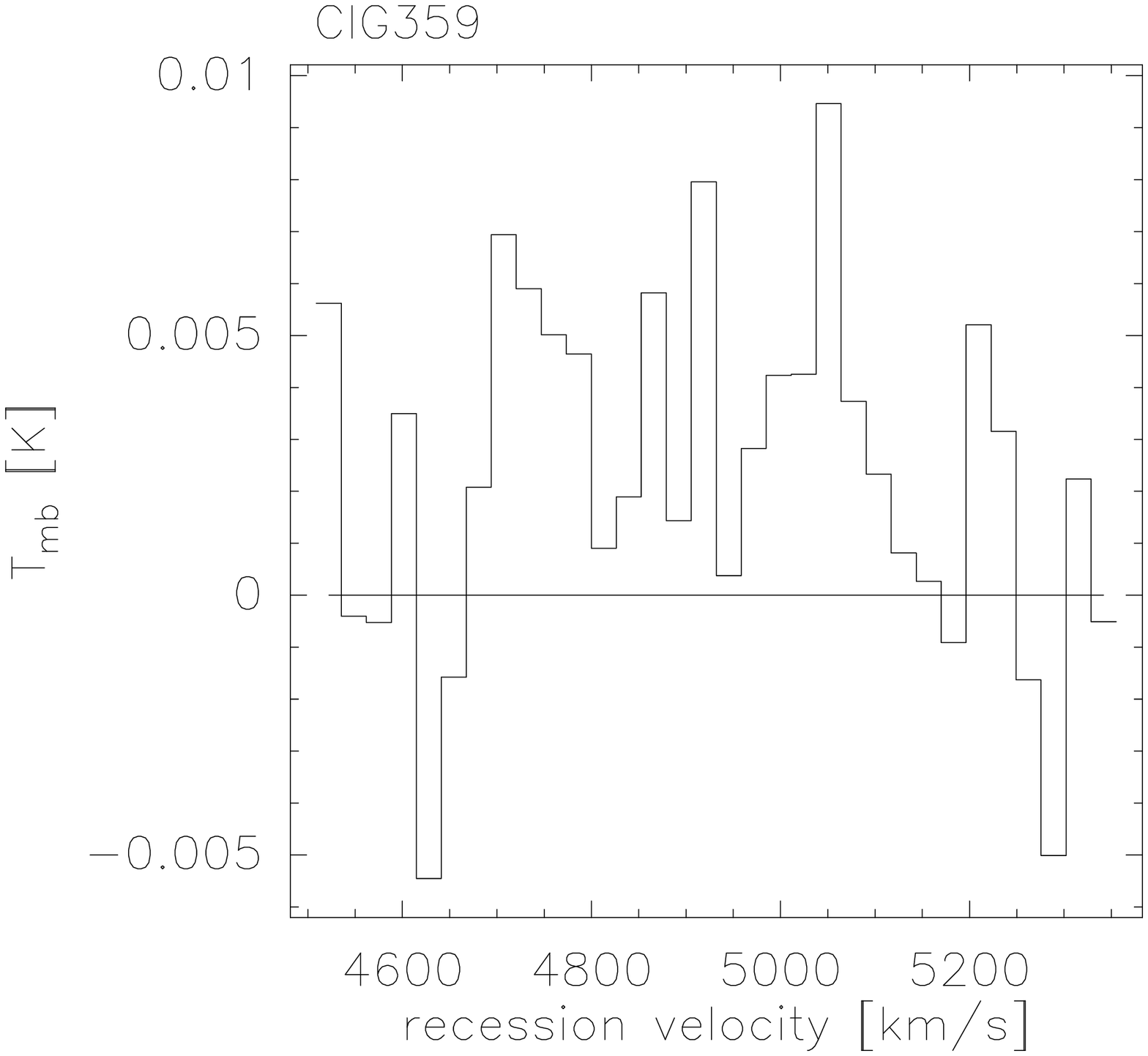} \quad 
\includegraphics[width=3cm,angle=270]{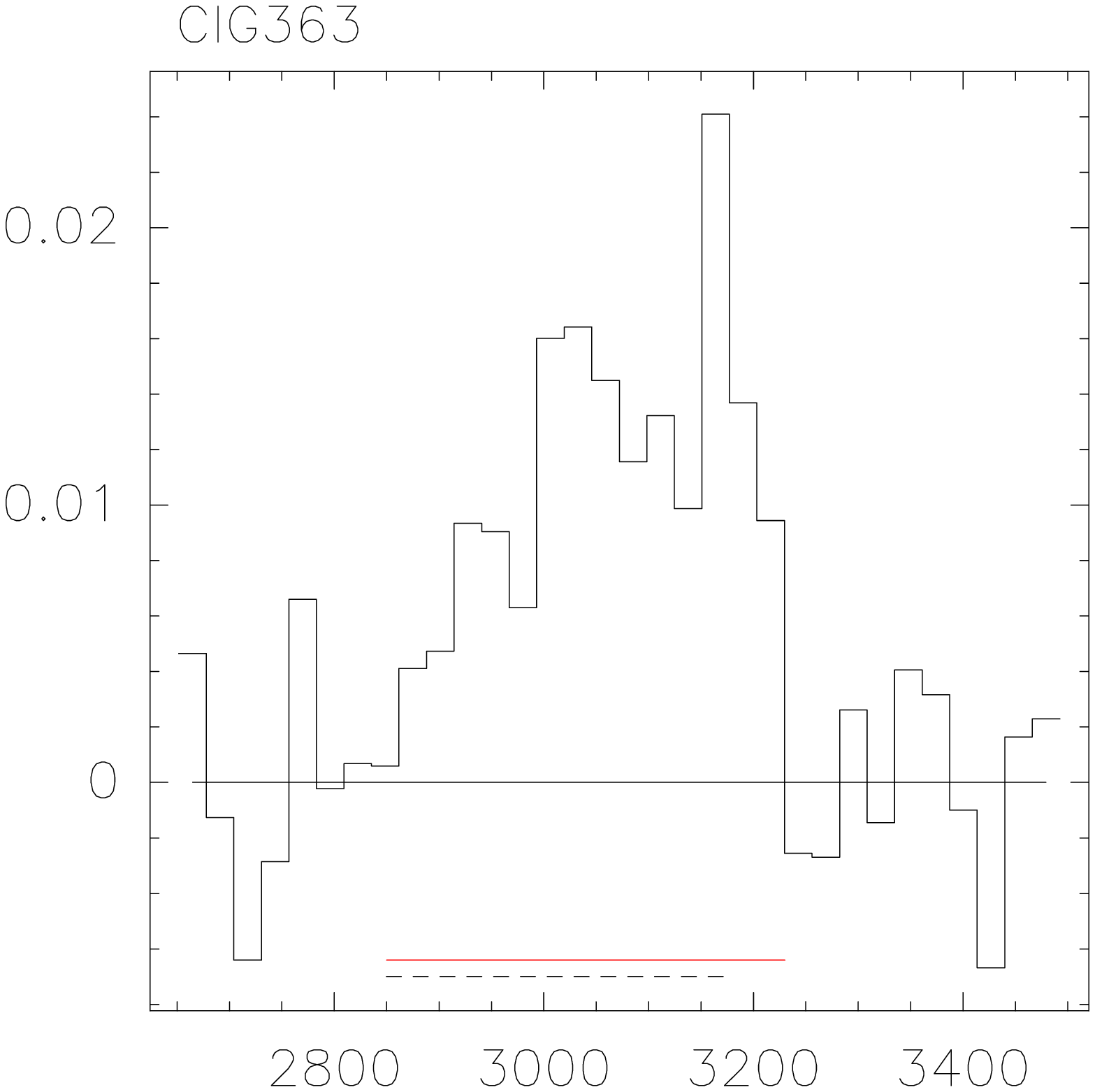}\quad 
\includegraphics[width=3cm,angle=270]{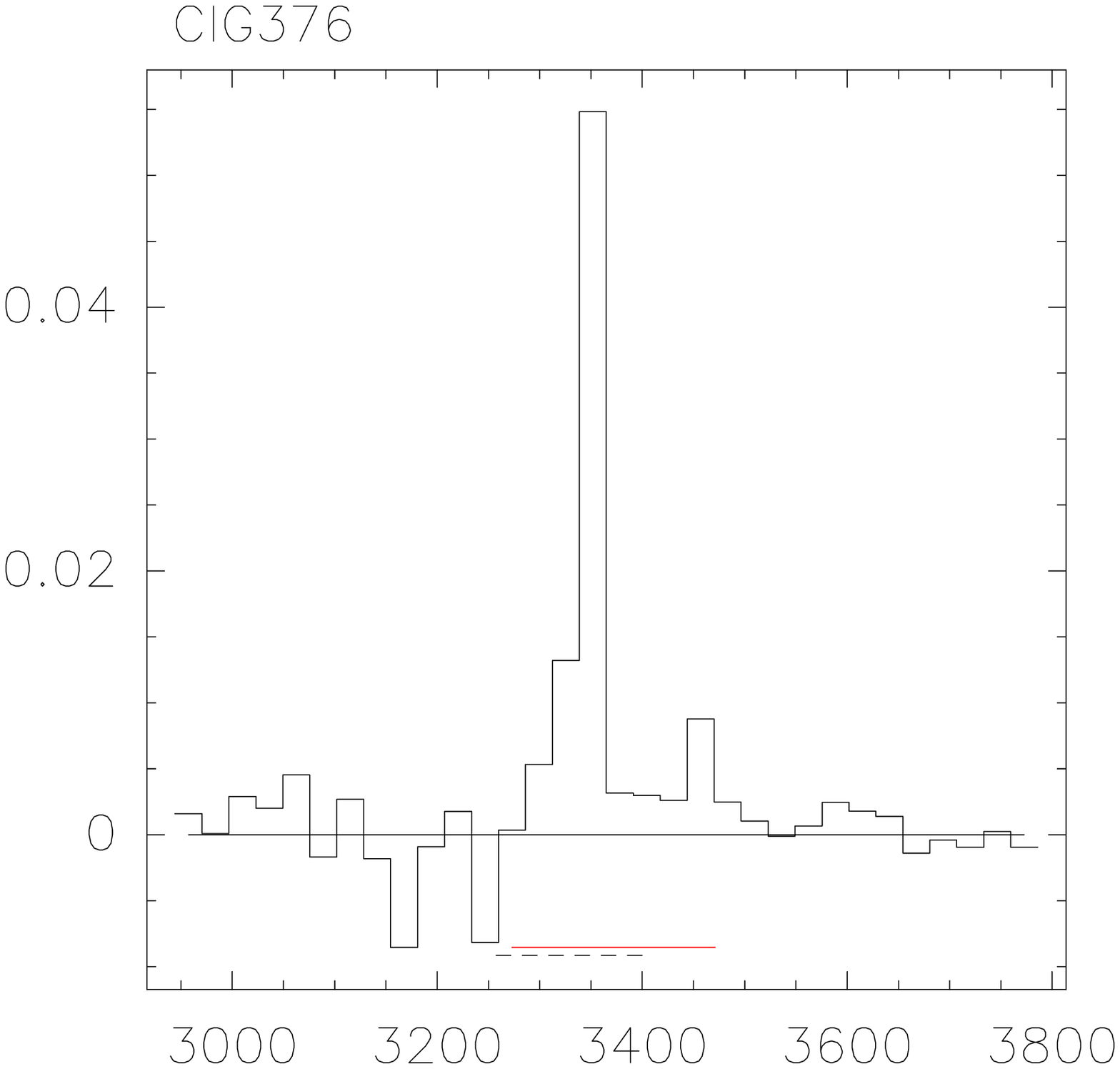}\quad 
\includegraphics[width=3cm,angle=270]{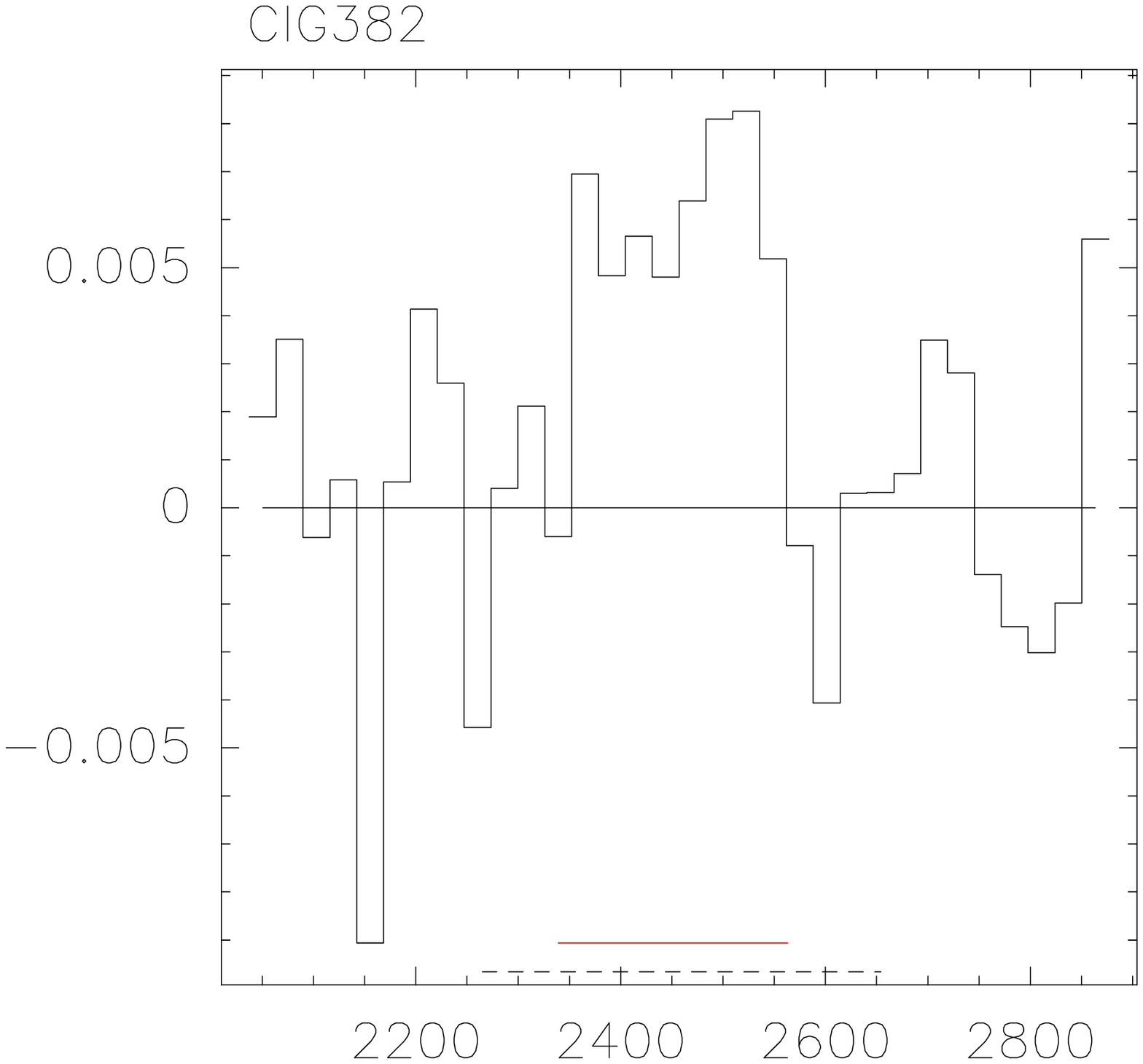}\quad 
\includegraphics[width=3cm,angle=270]{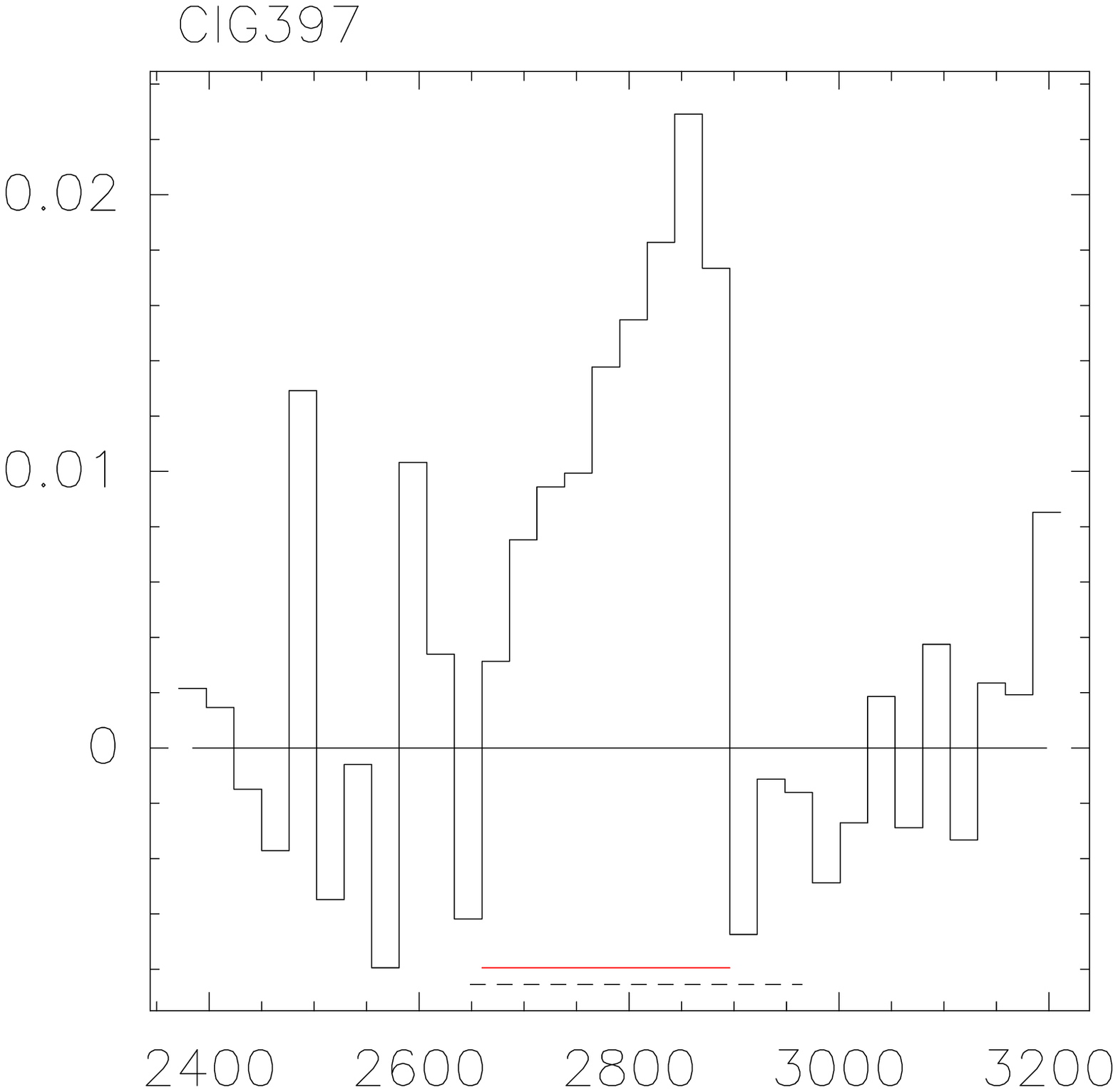}} 
\caption{CO(1-0) spectra of the galaxies detected  with the FCRAO telescope.
The spectral resolution is 13.1  or 26 \kms , depending on the spectral shape and rms noise.
The x-axis represents
 the recession velocity in \kms\ and the y-axis the intensity in K in the $T_{mb}$ scale. 
The full (red) line segment shows the line width of the CO line adopted for the 
determination of the velocity  integrated intensity. The dashed (black) line segment
is the HI  line width at 20\% peak level, W$_{20}$.
An asterisk next to the name indicates a marginal detection.} 
\label{spectra_fcrao}
\end{figure*} 
  
\begin{figure*} 
\centerline{\includegraphics[width=3cm,angle=270]{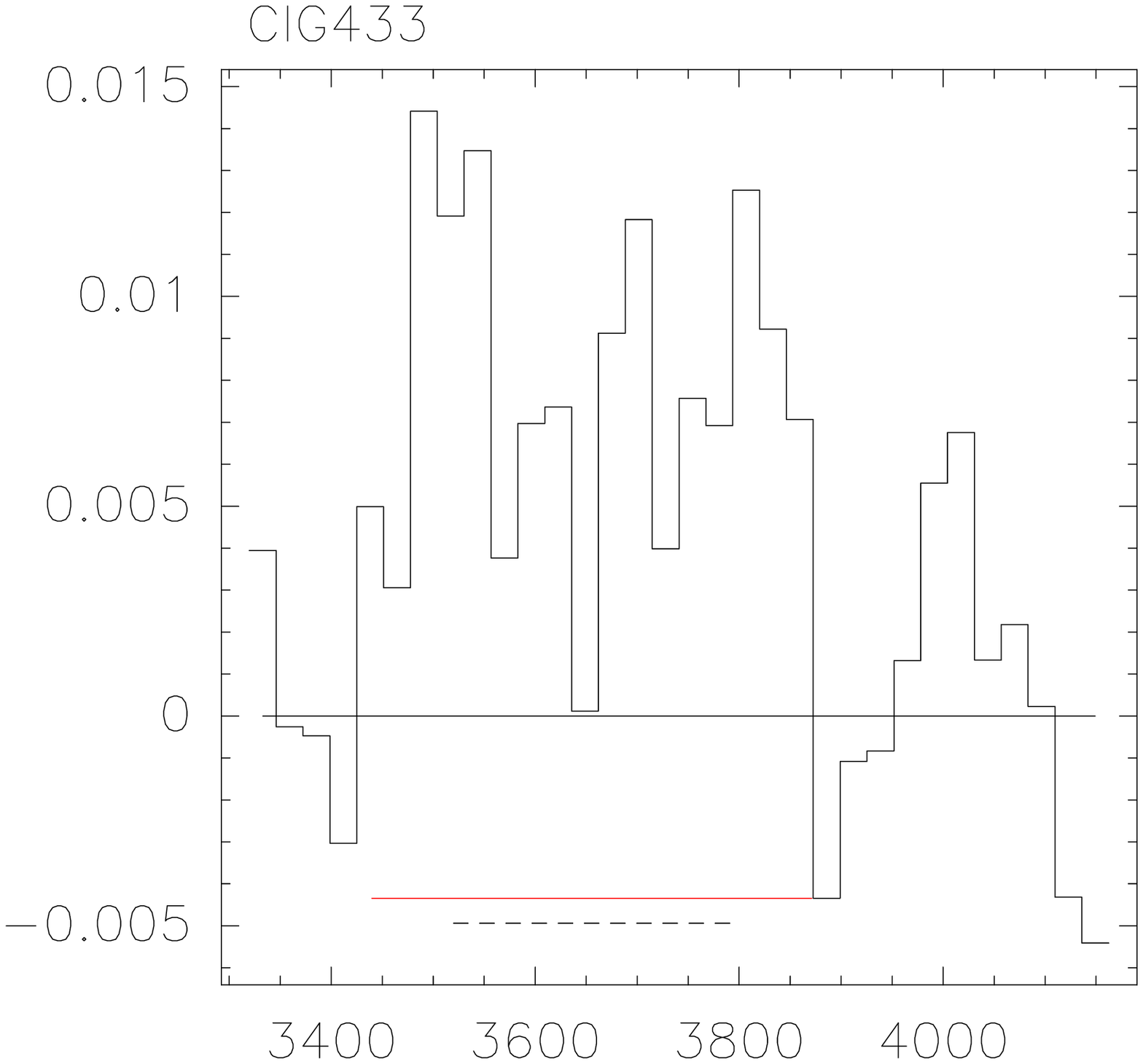} \quad 
\includegraphics[width=3cm,angle=270]{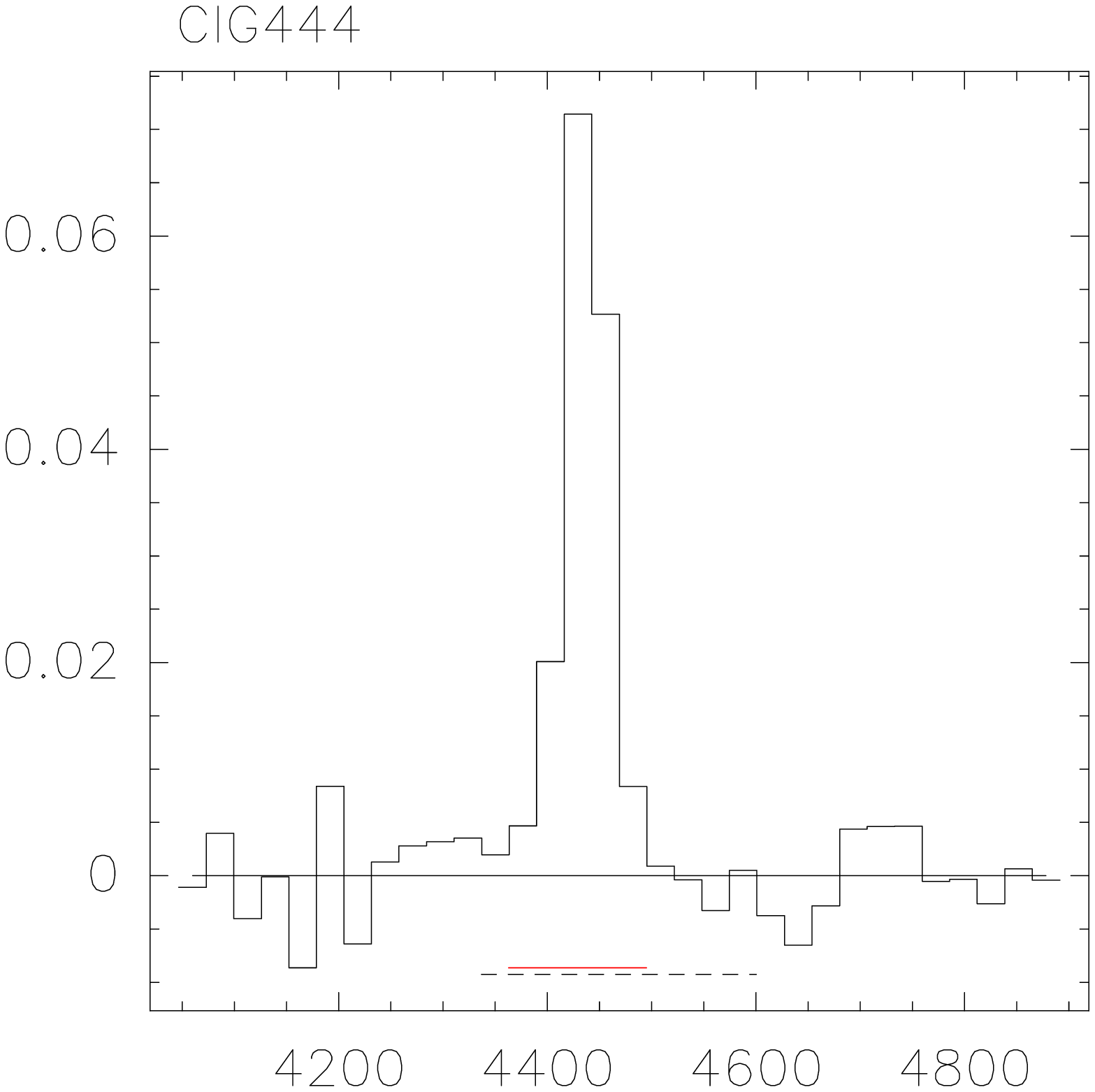}\quad 
\includegraphics[width=3cm,angle=270]{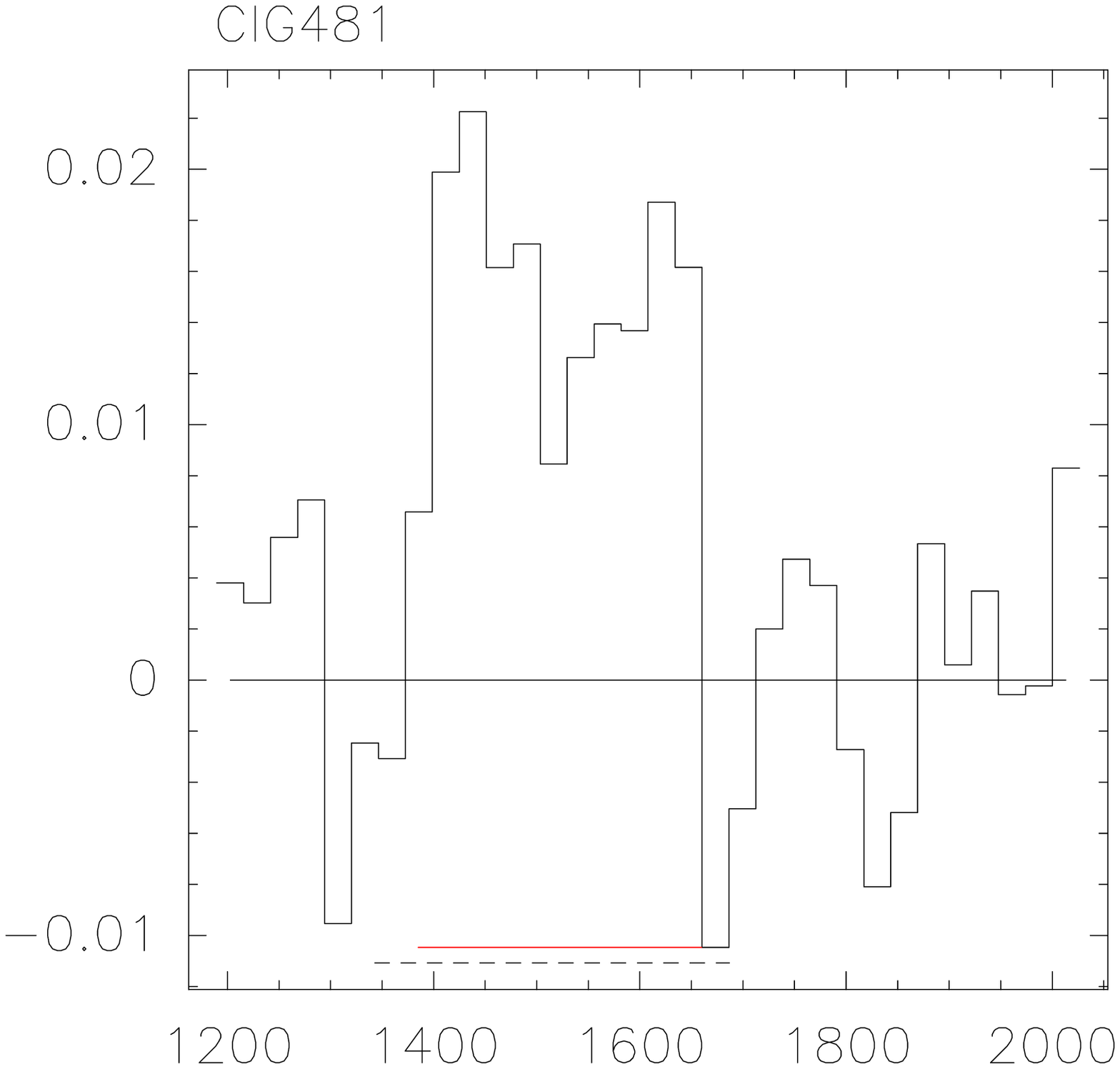}\quad 
\includegraphics[width=3cm,angle=270]{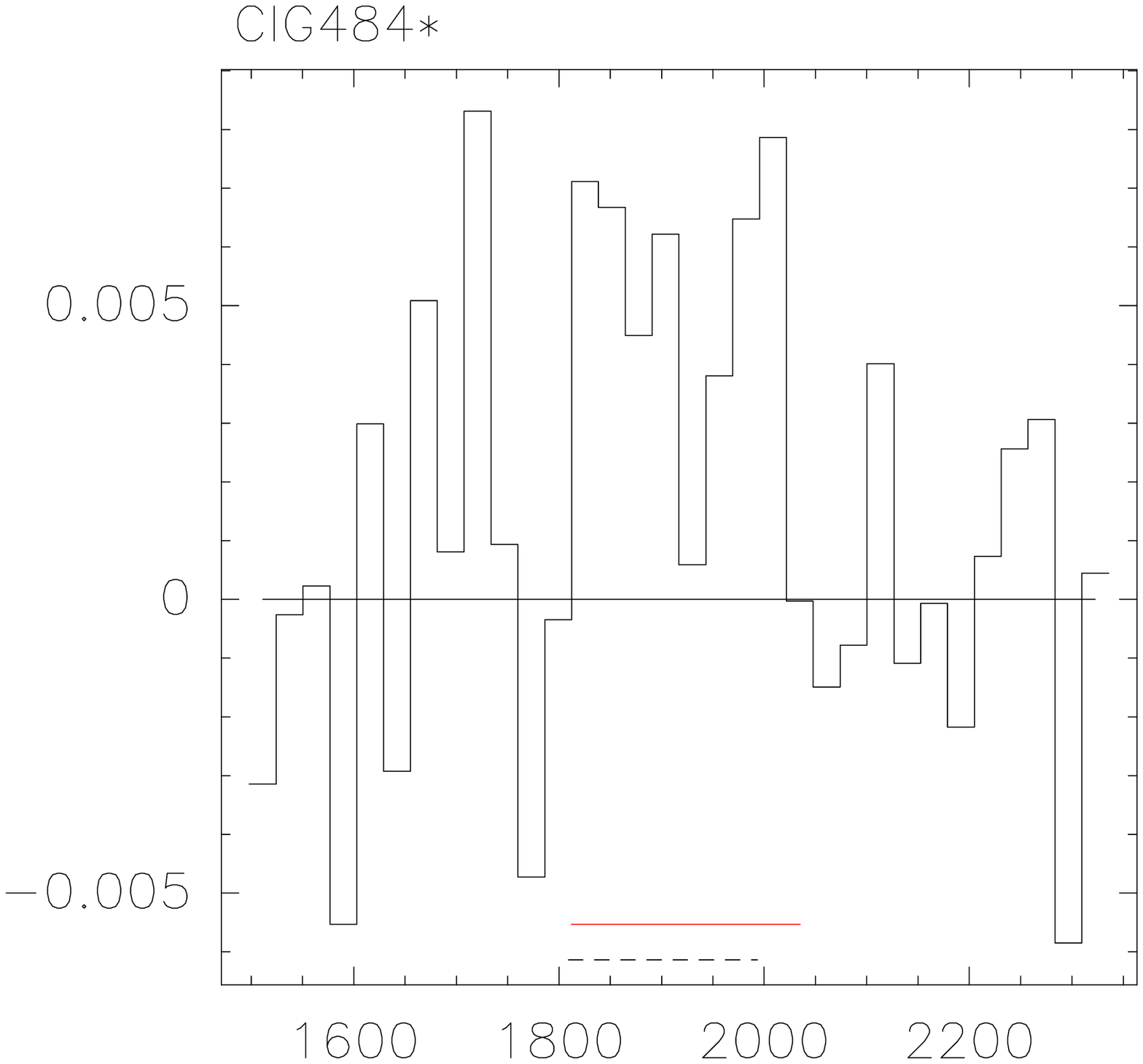}\quad 
\includegraphics[width=3cm,angle=270]{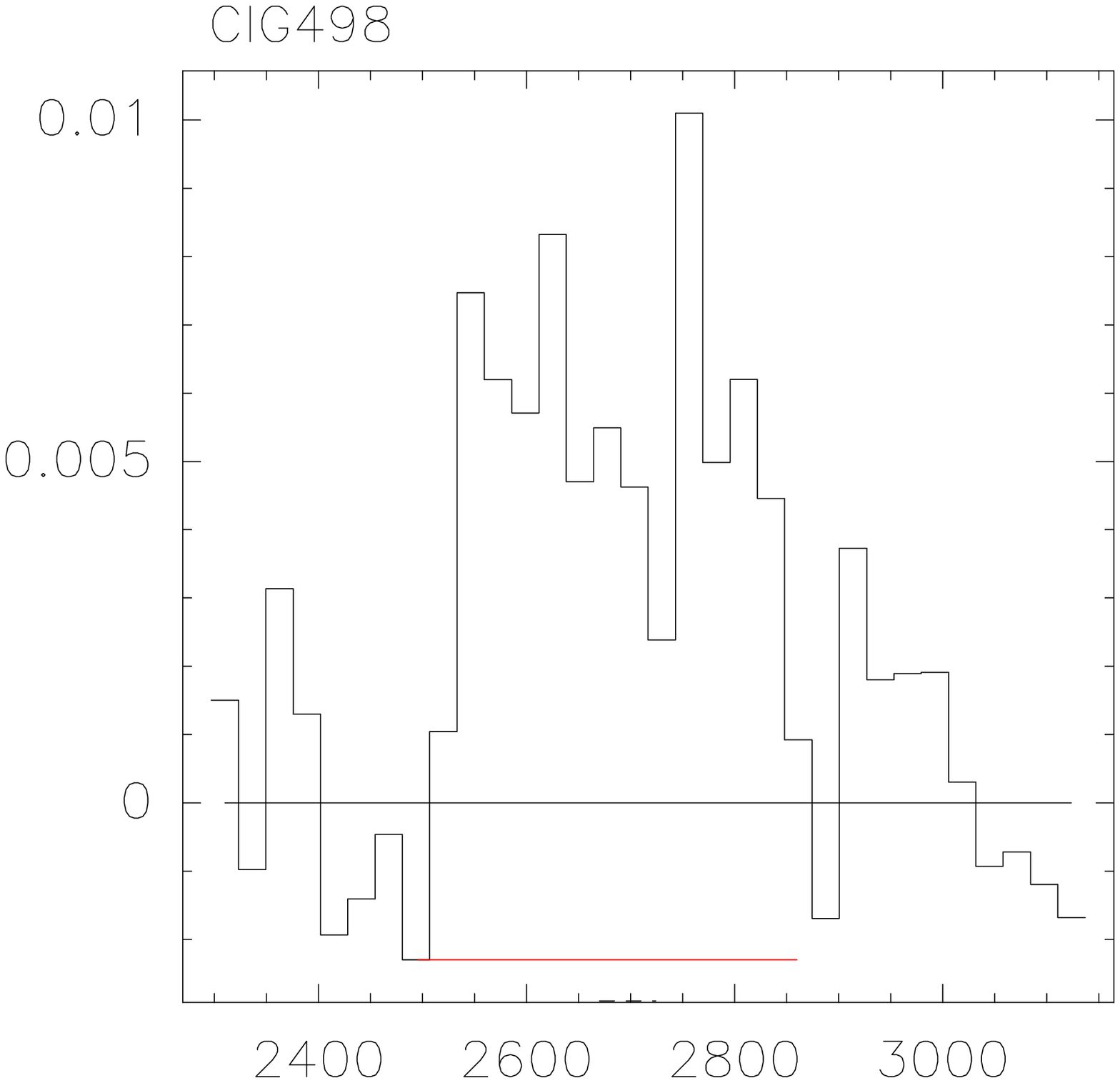}} 
\centerline{\includegraphics[width=3cm,angle=270]{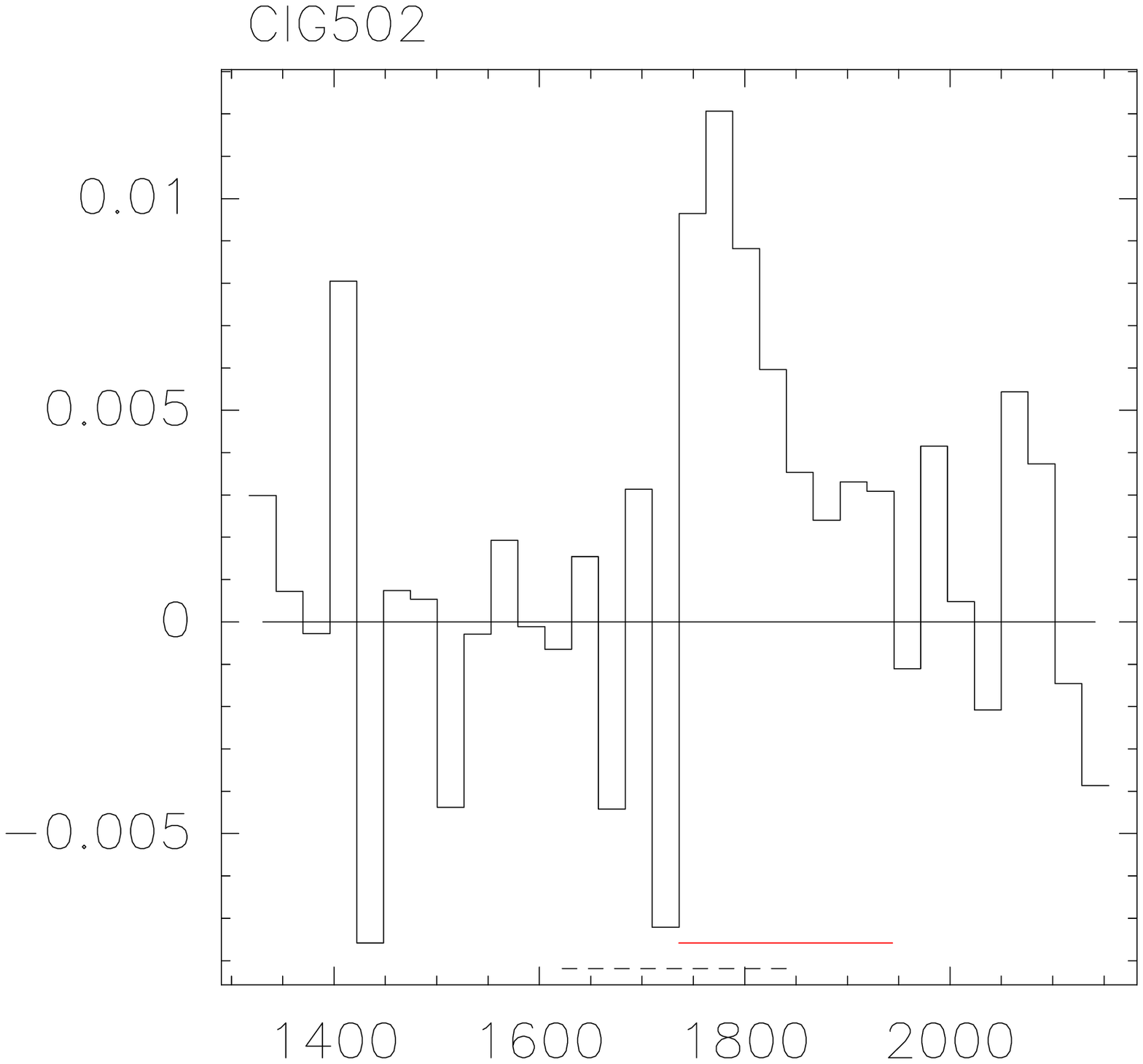} \quad 
\includegraphics[width=3cm,angle=270]{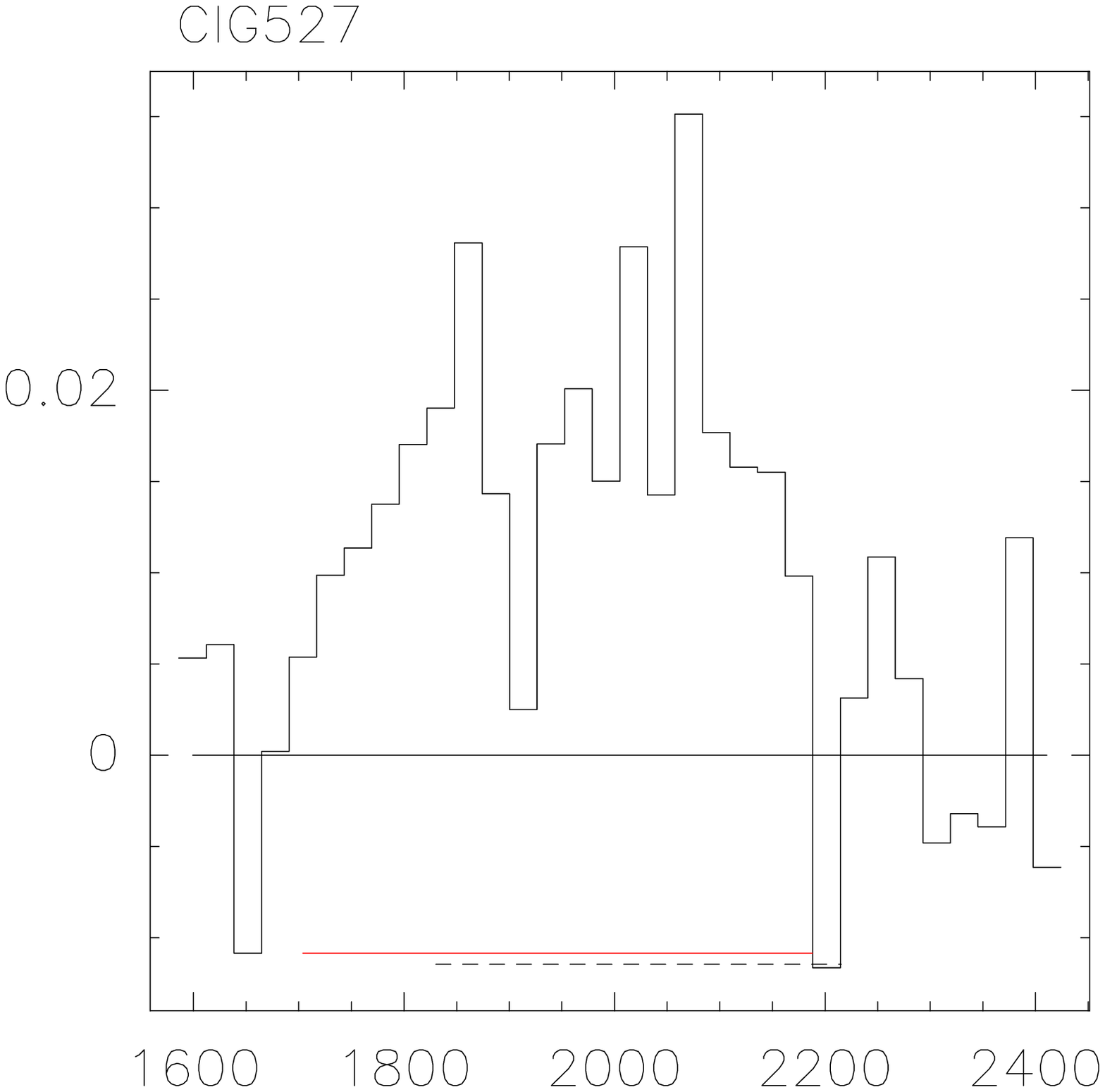}\quad 
\includegraphics[width=3cm,angle=270]{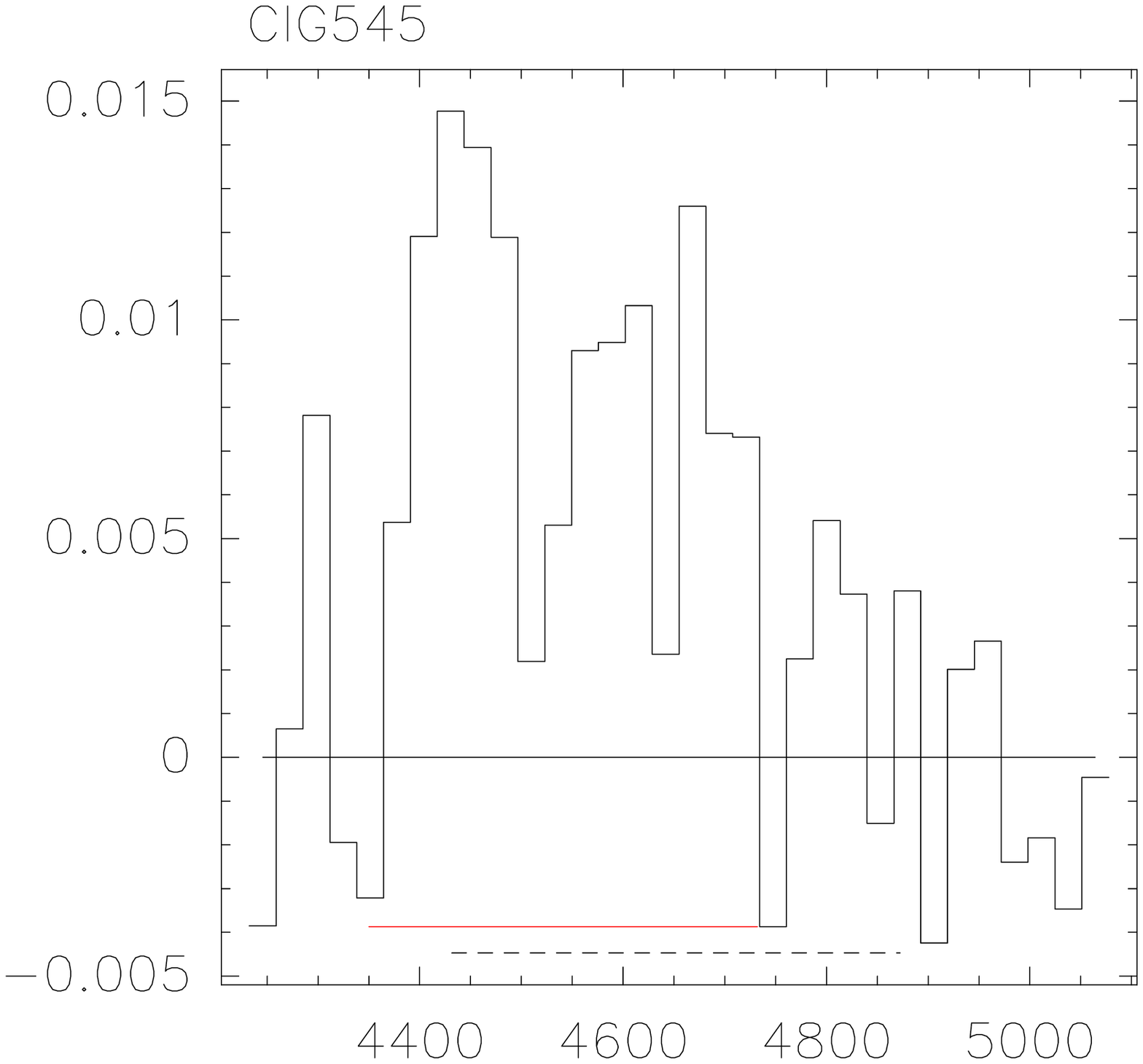}\quad 
\includegraphics[width=3cm,angle=270]{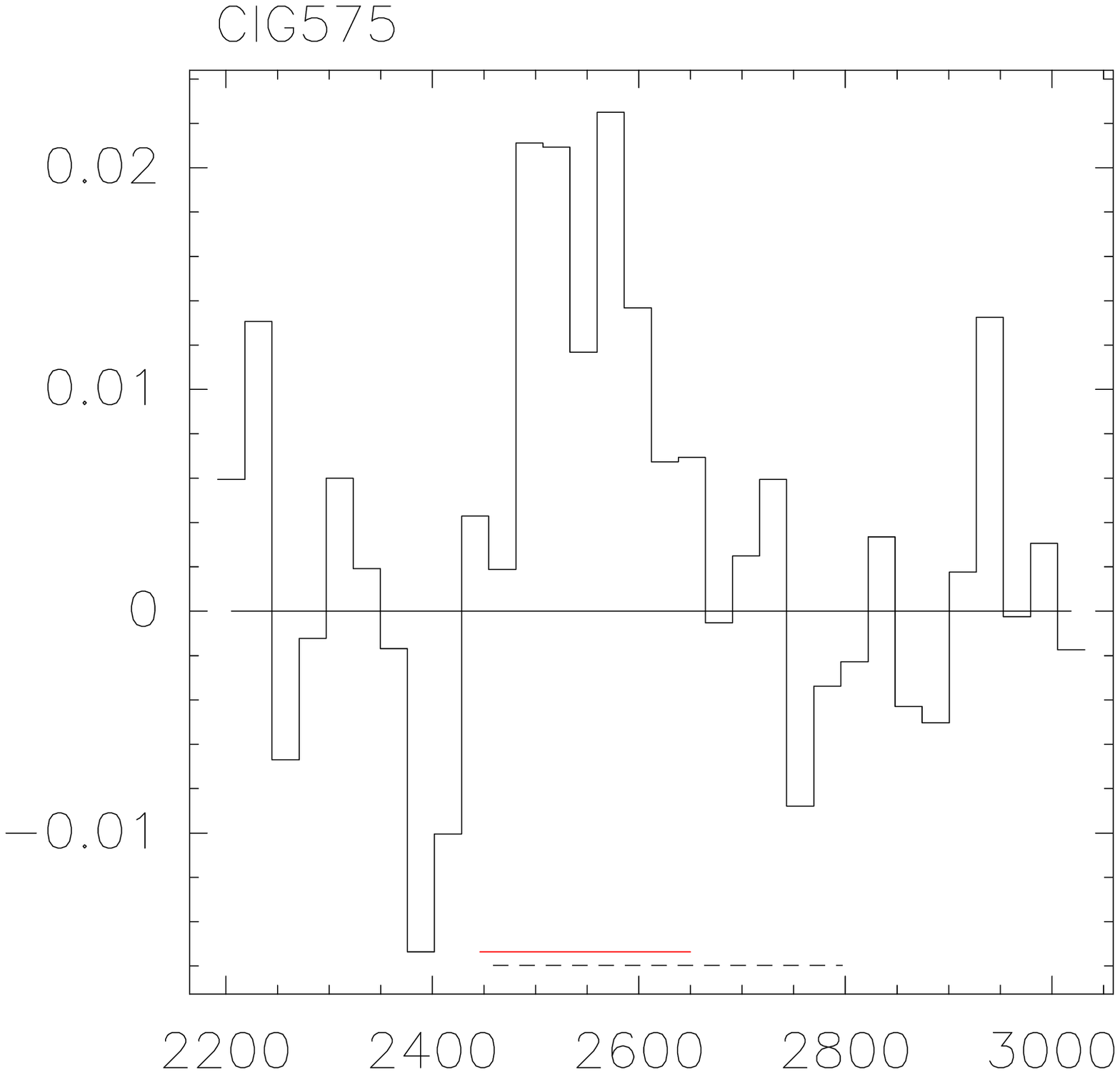}\quad 
\includegraphics[width=3cm,angle=270]{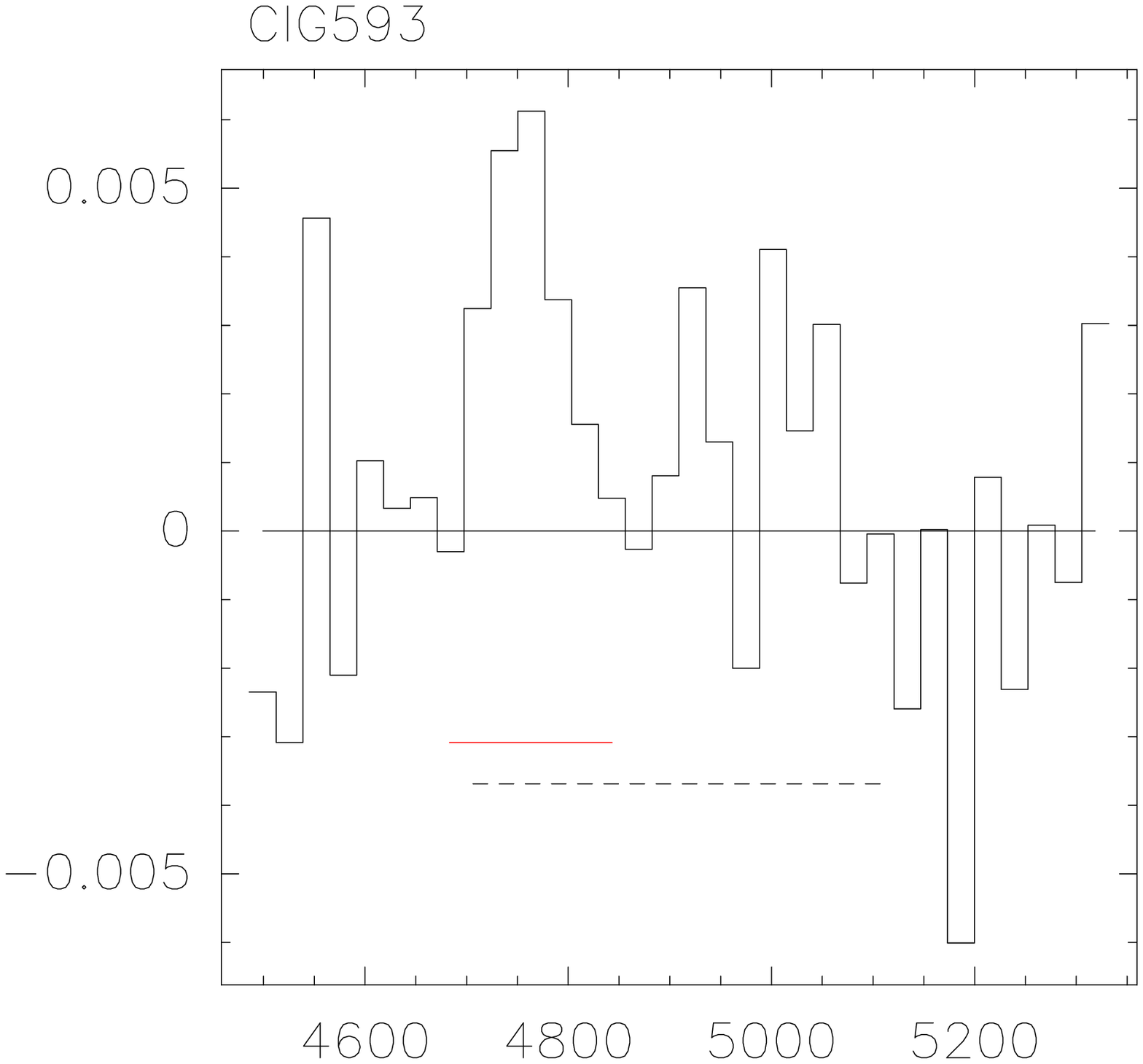}} 
\centerline{\includegraphics[width=3cm,angle=270]{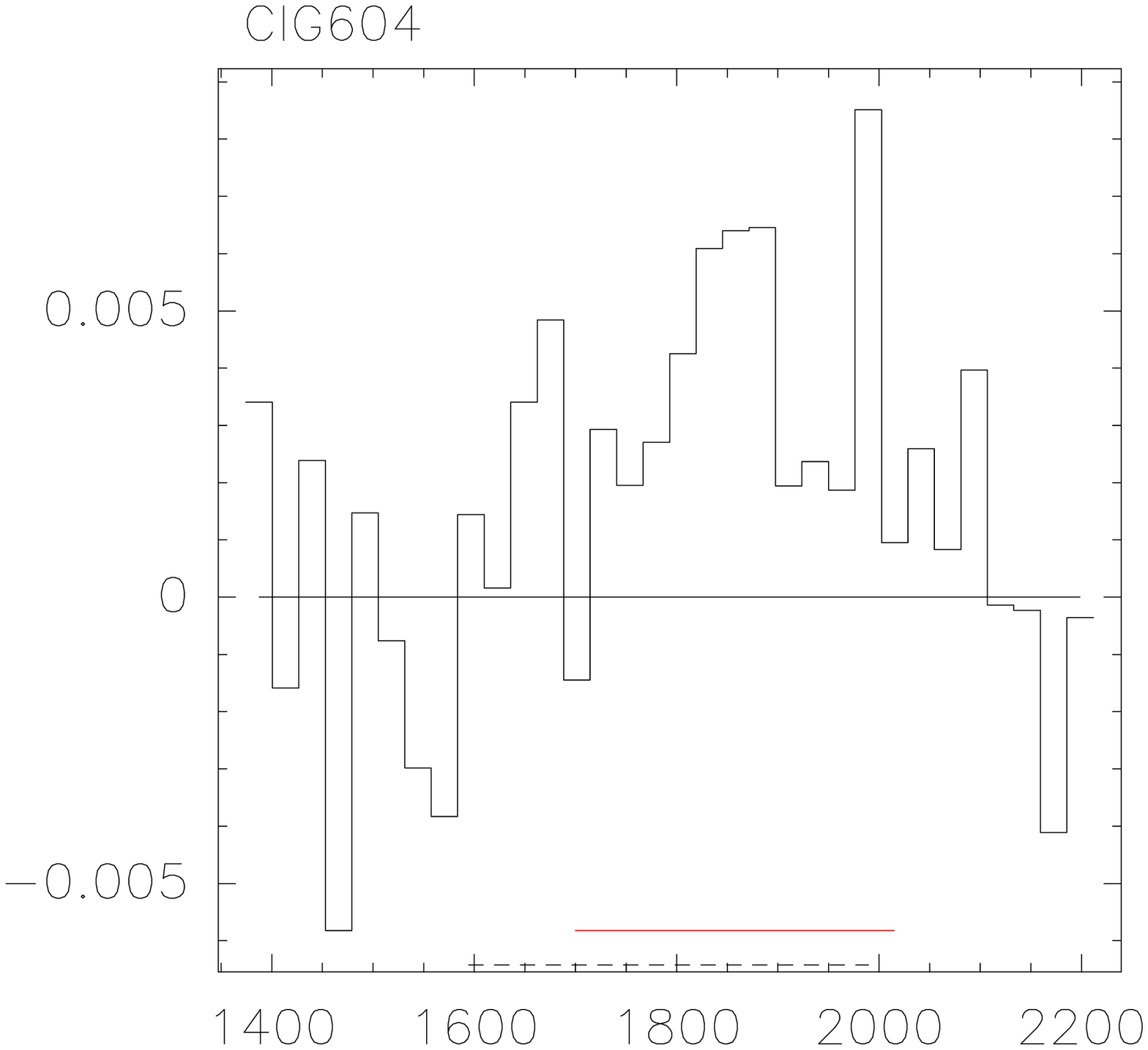} \quad 
\includegraphics[width=3cm,angle=270]{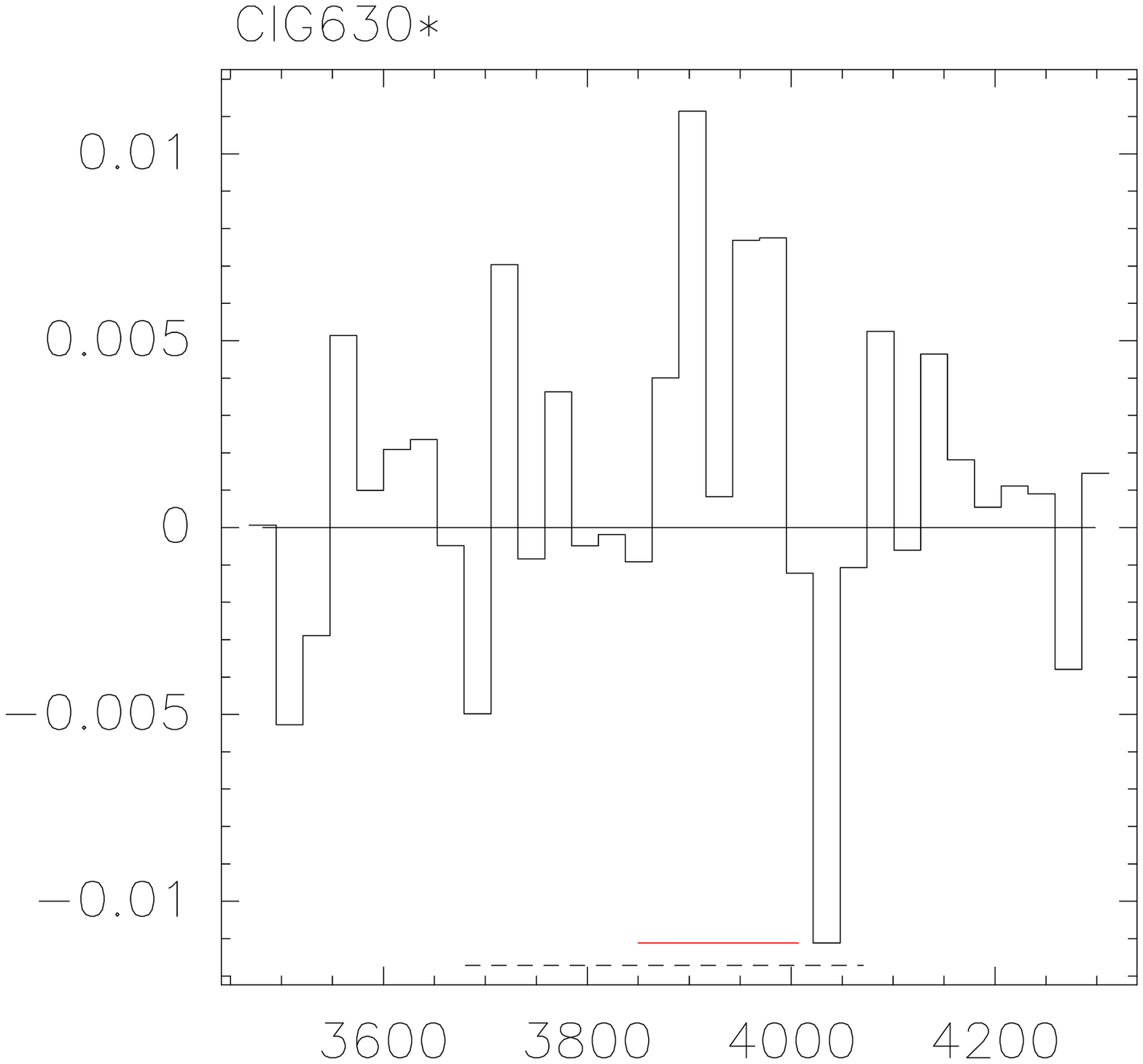}\quad 
\includegraphics[width=3cm,angle=270]{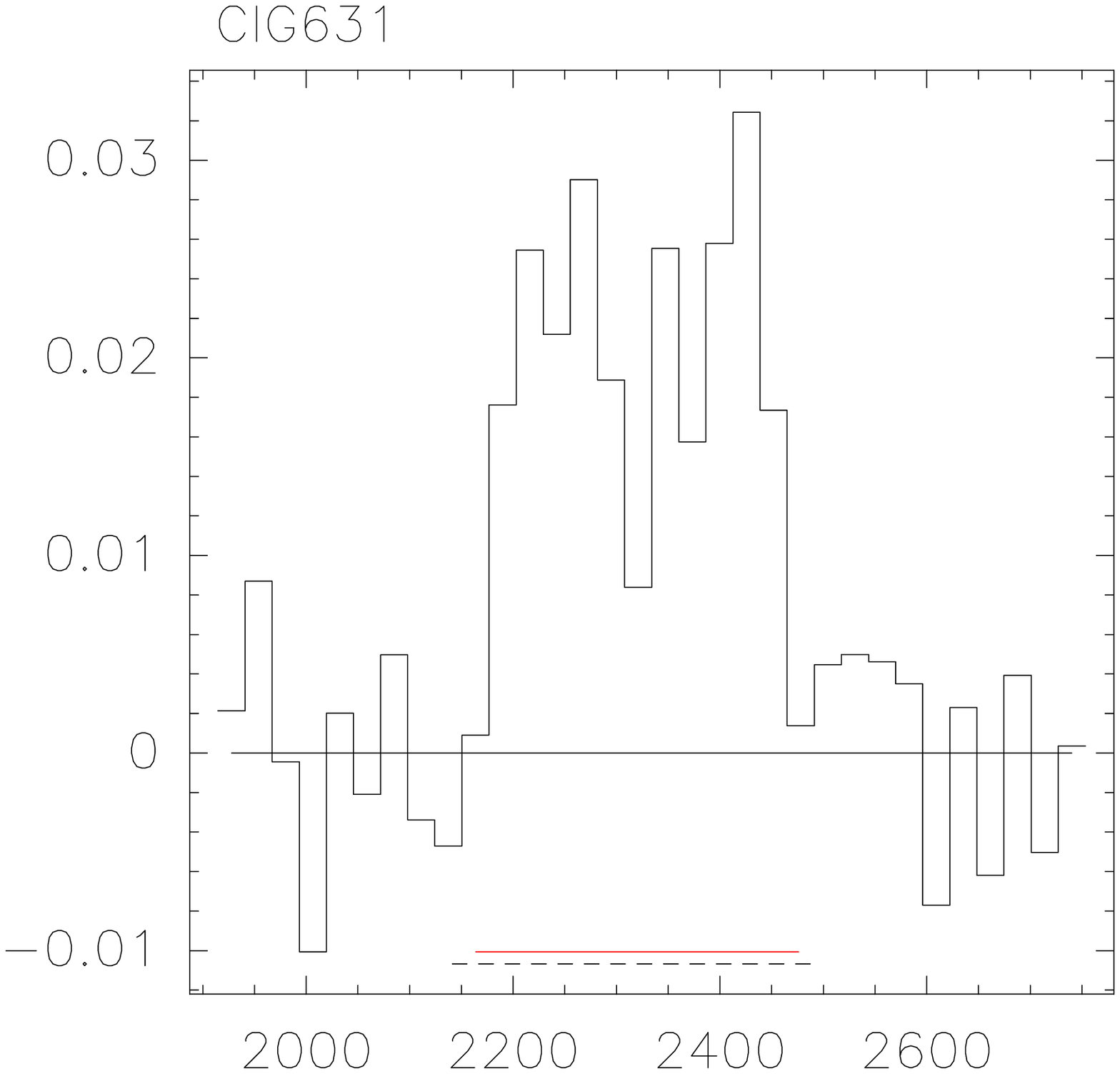}\quad 
\includegraphics[width=3cm,angle=270]{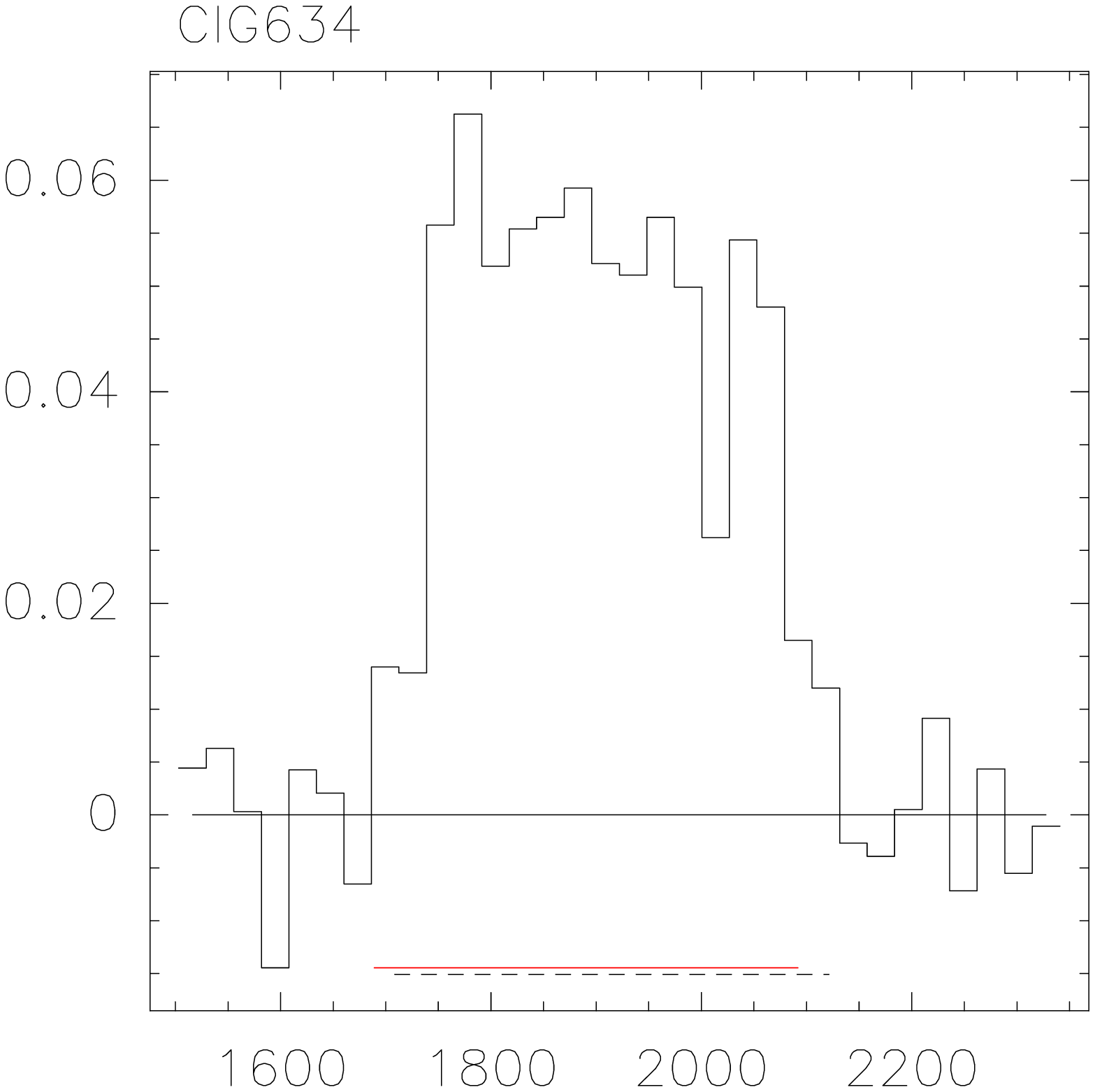}\quad 
\includegraphics[width=3cm,angle=270]{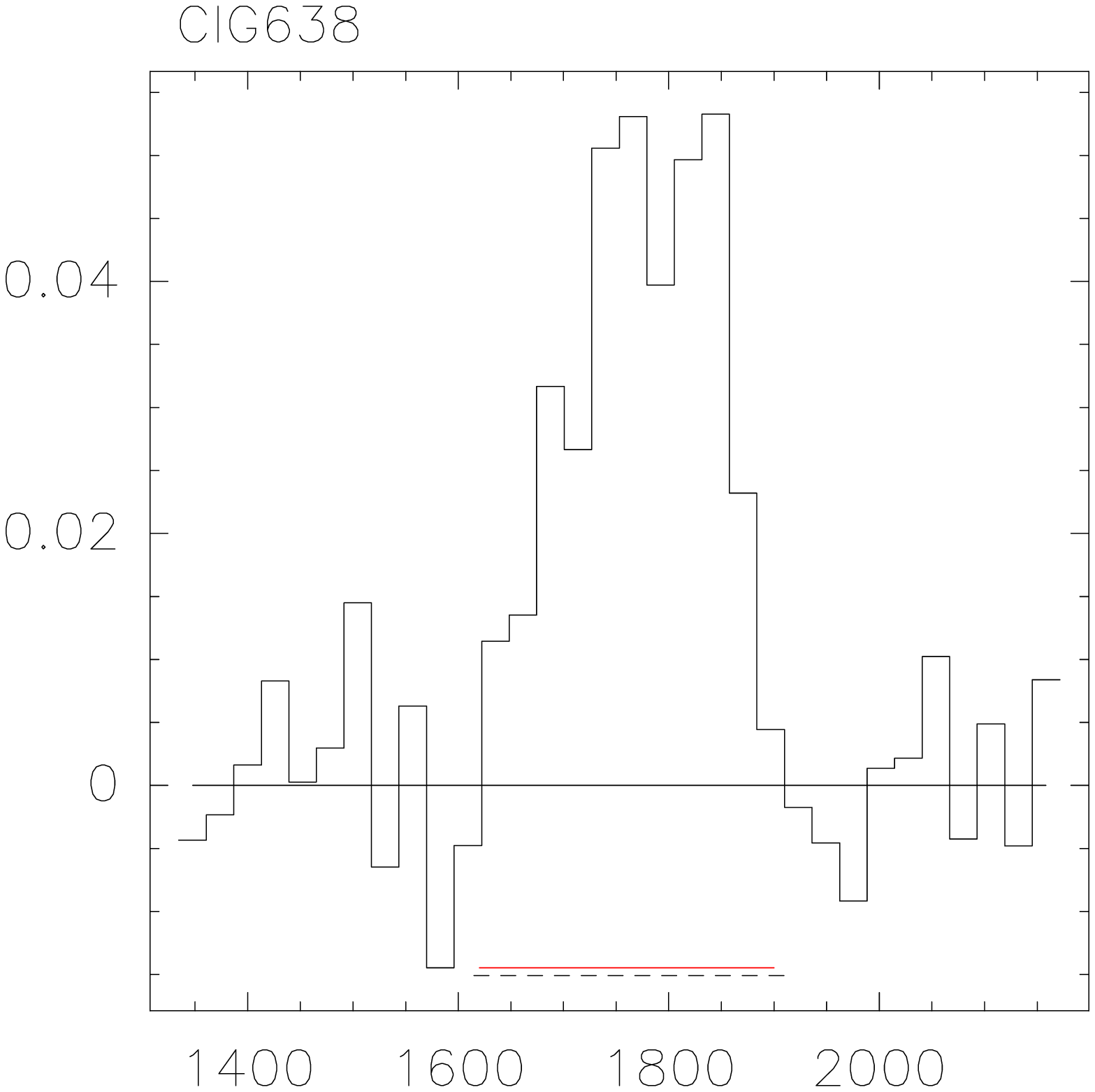}} 
\centerline{\includegraphics[width=3cm,angle=270]{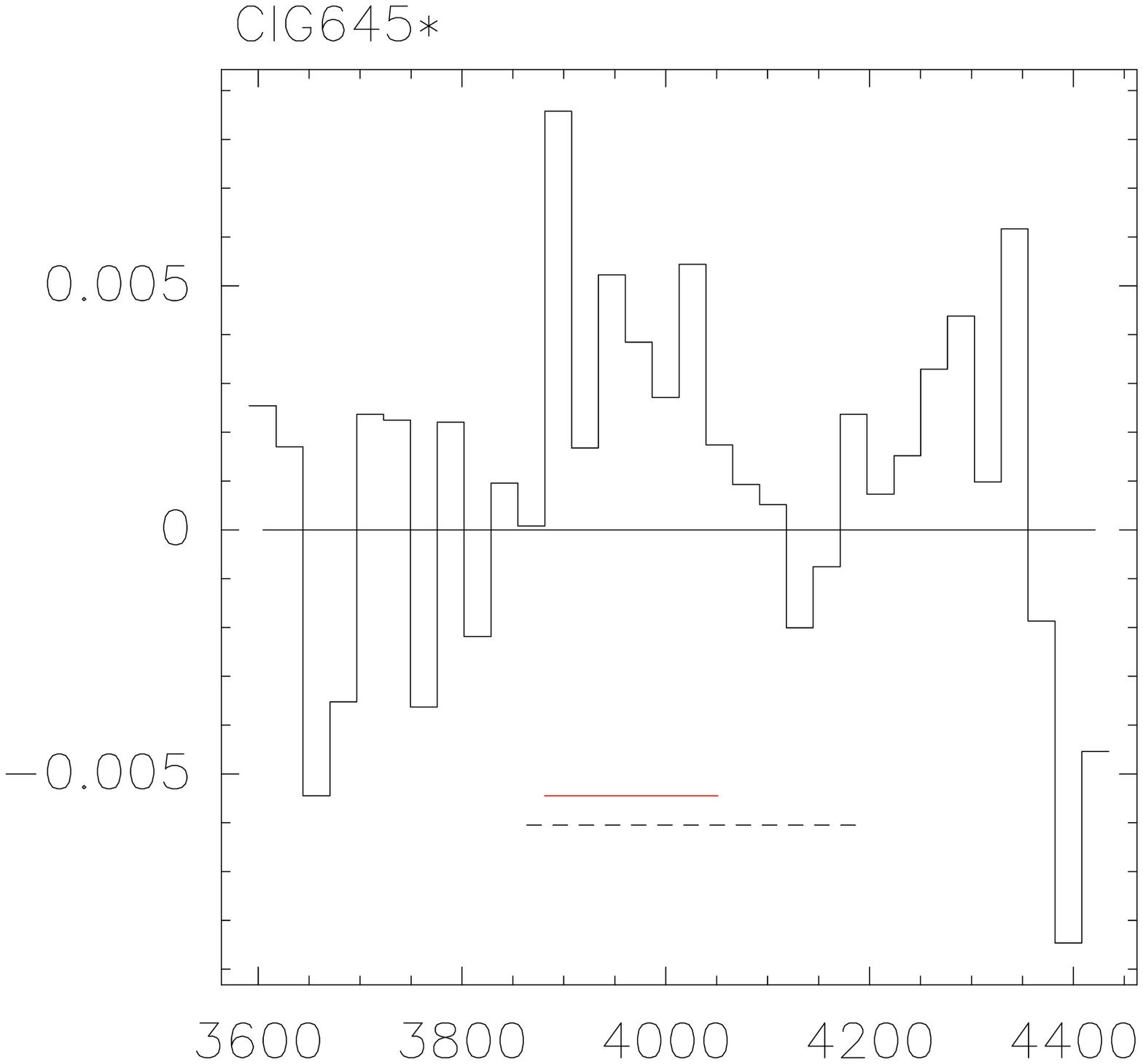} \quad 
\includegraphics[width=3cm,angle=270]{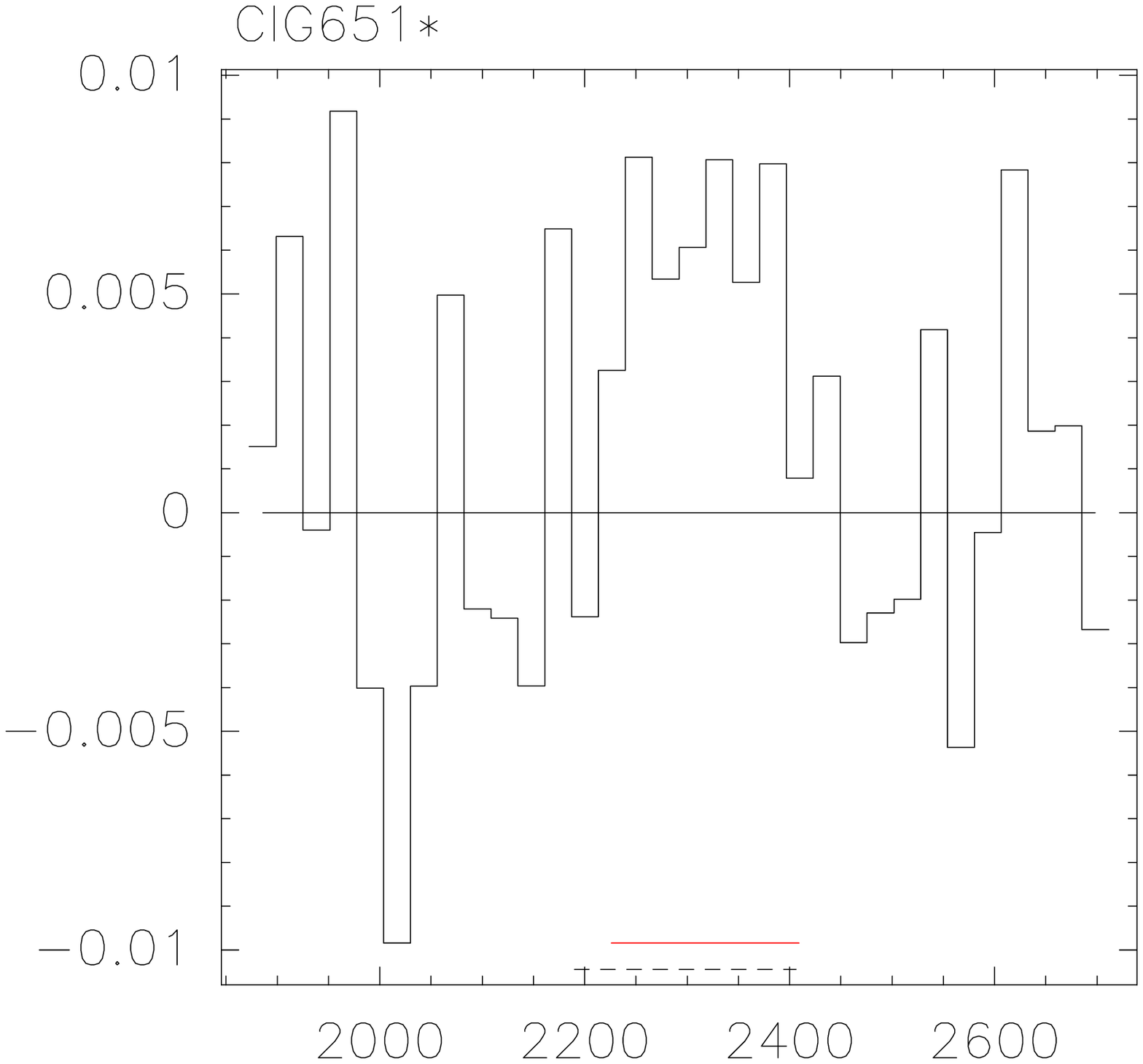}\quad 
\includegraphics[width=3cm,angle=270]{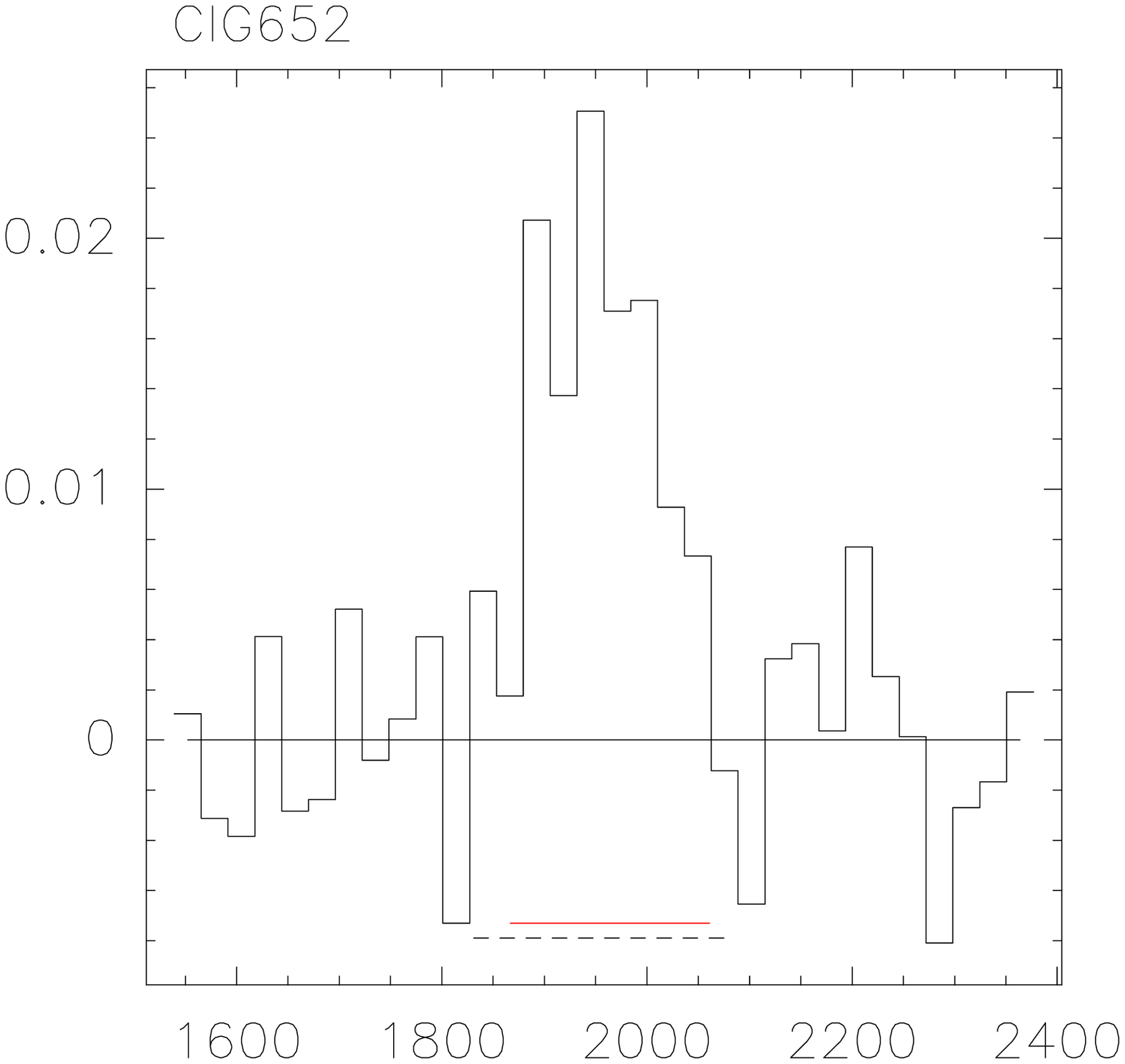}\quad 
\includegraphics[width=3cm,angle=270]{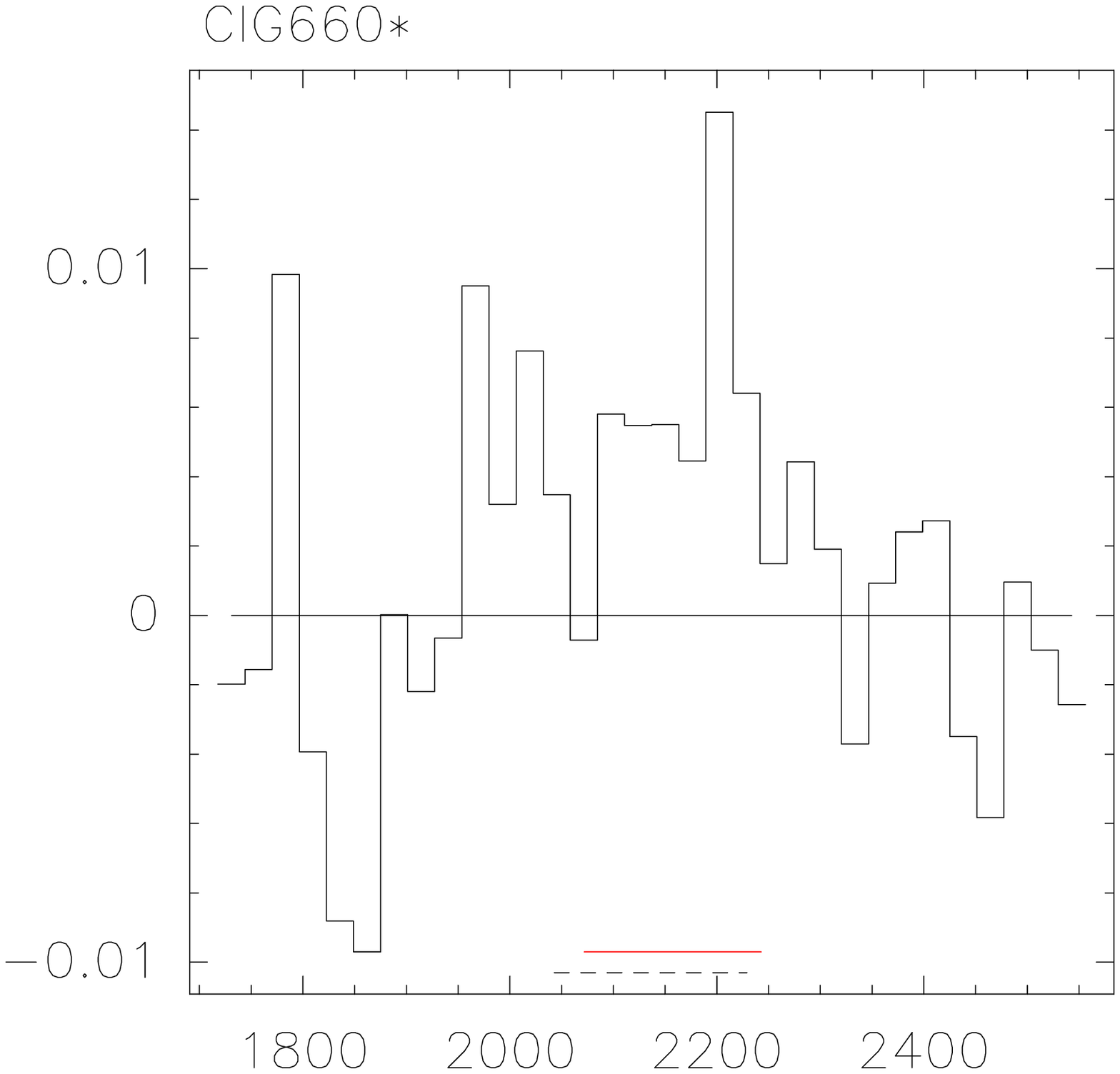}\quad 
\includegraphics[width=3cm,angle=270]{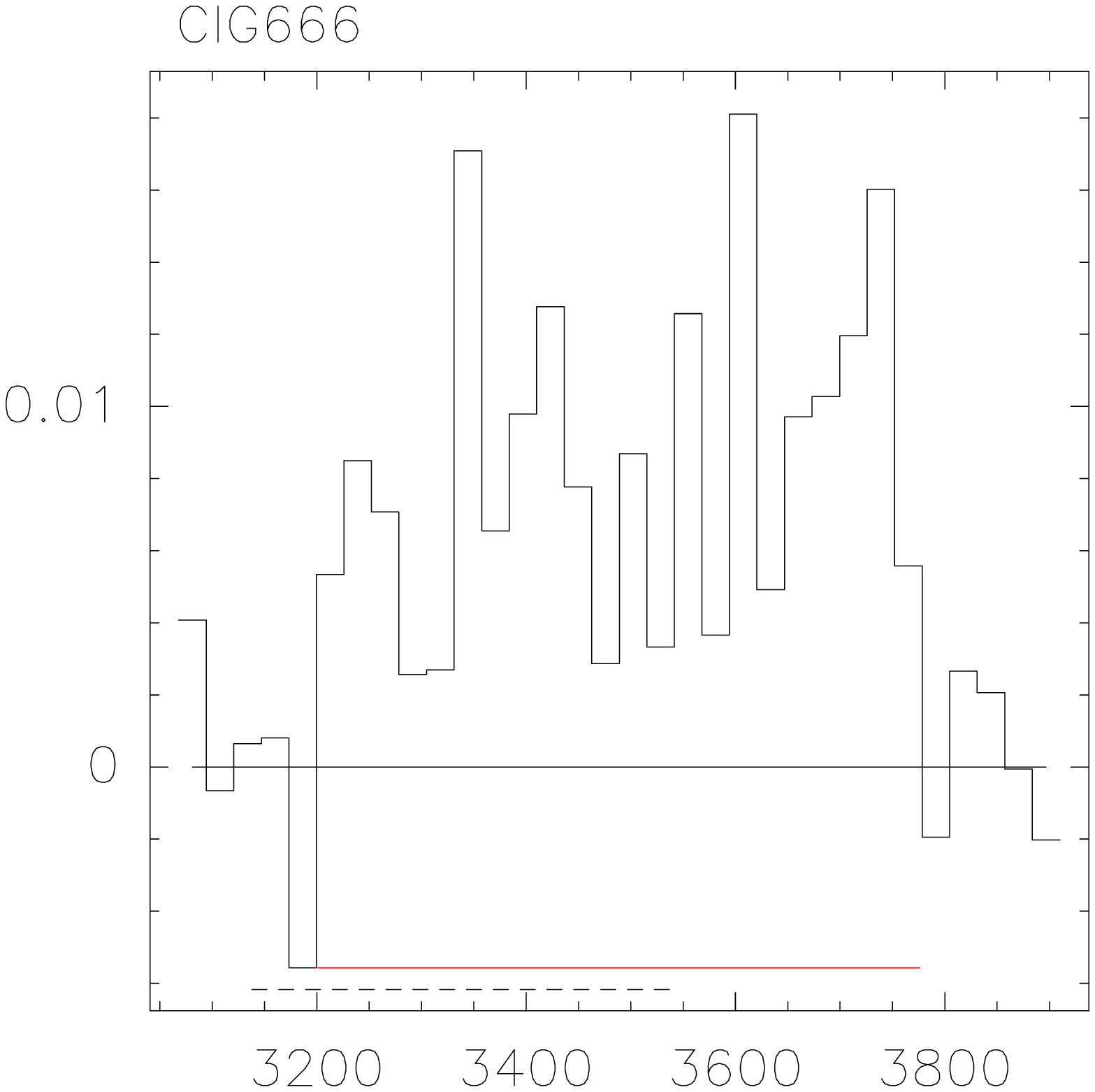}} 
\centerline{\includegraphics[width=3cm,angle=270]{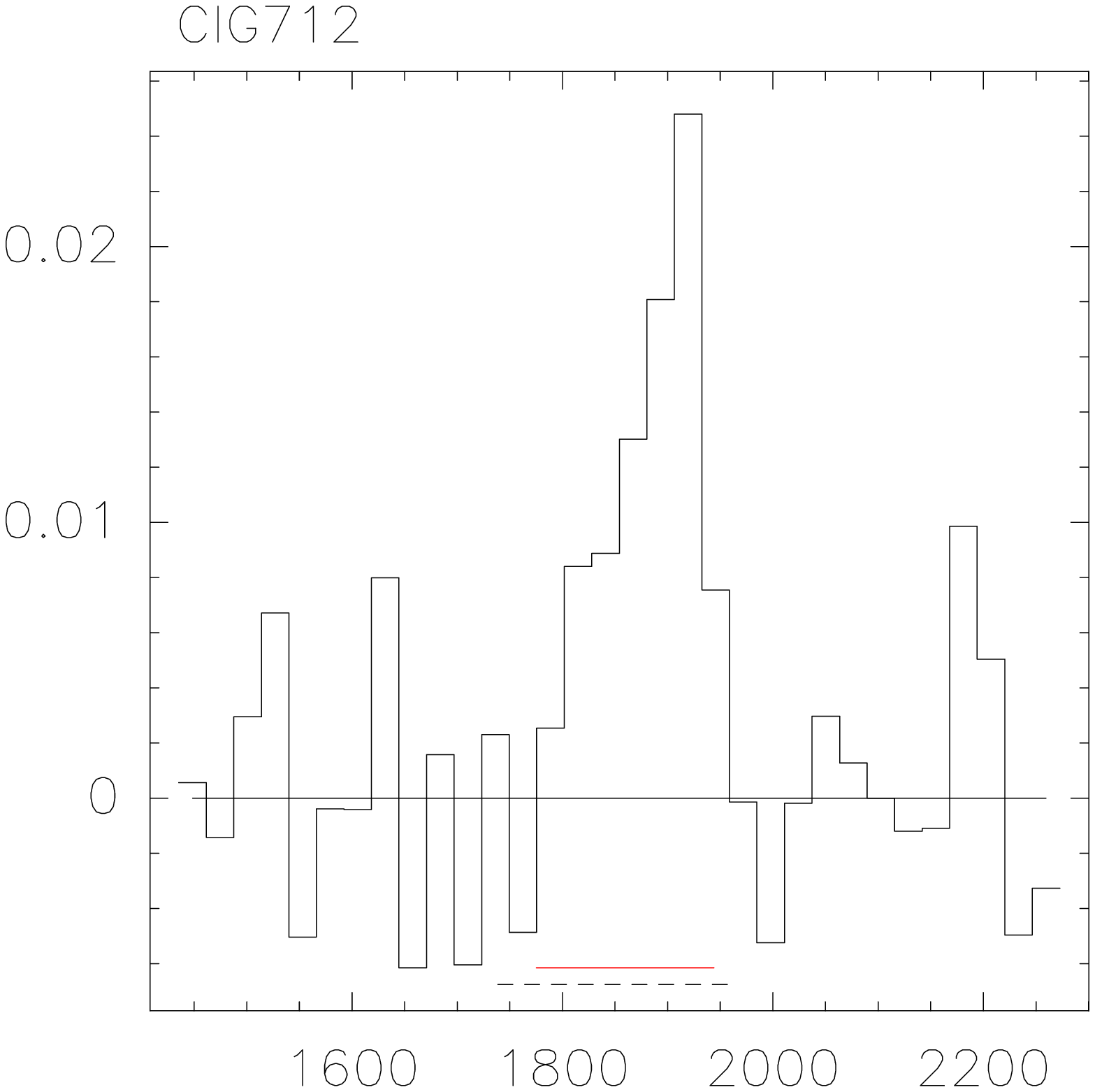} \quad 
\includegraphics[width=3cm,angle=270]{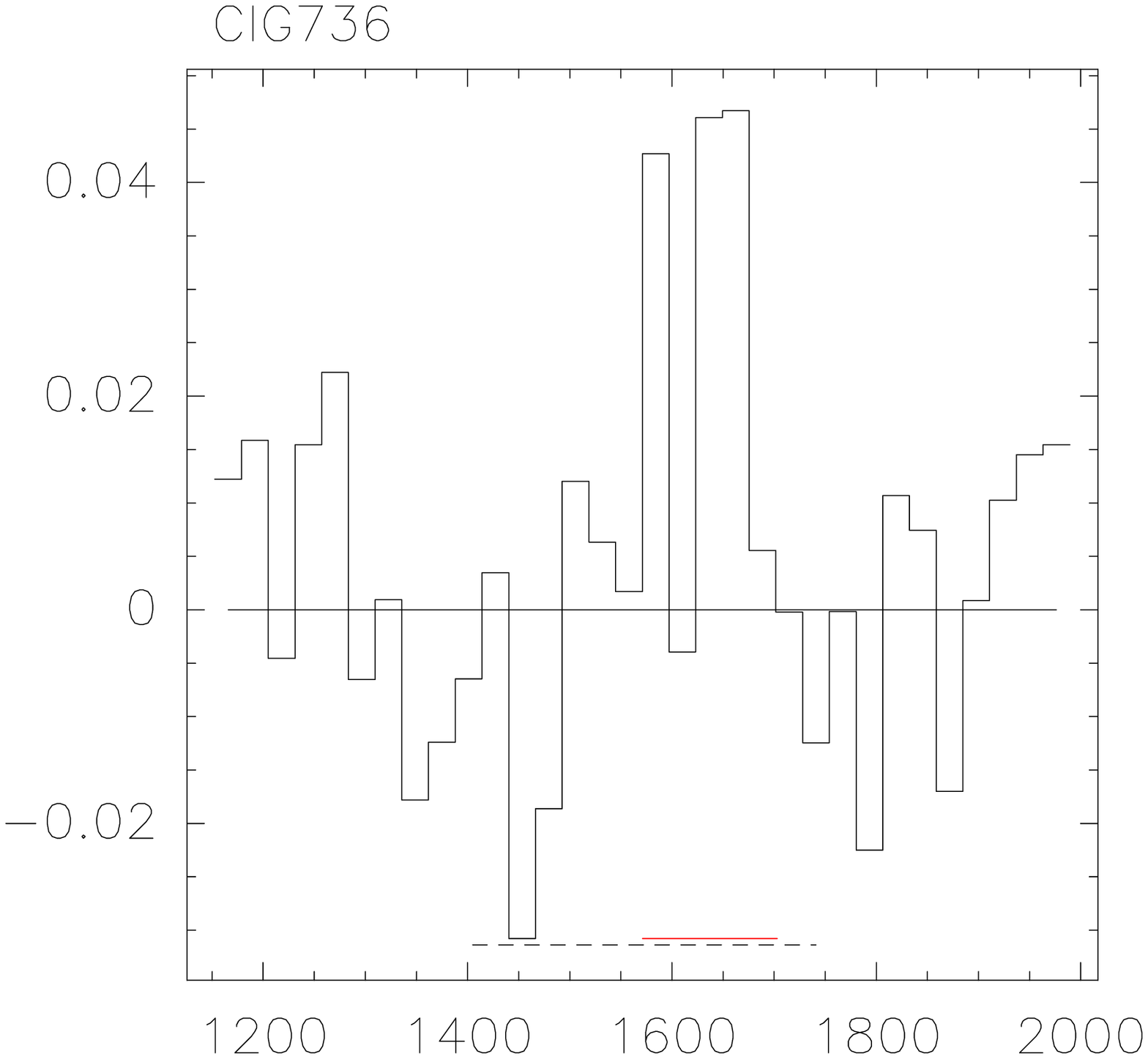}\quad 
\includegraphics[width=3cm,angle=270]{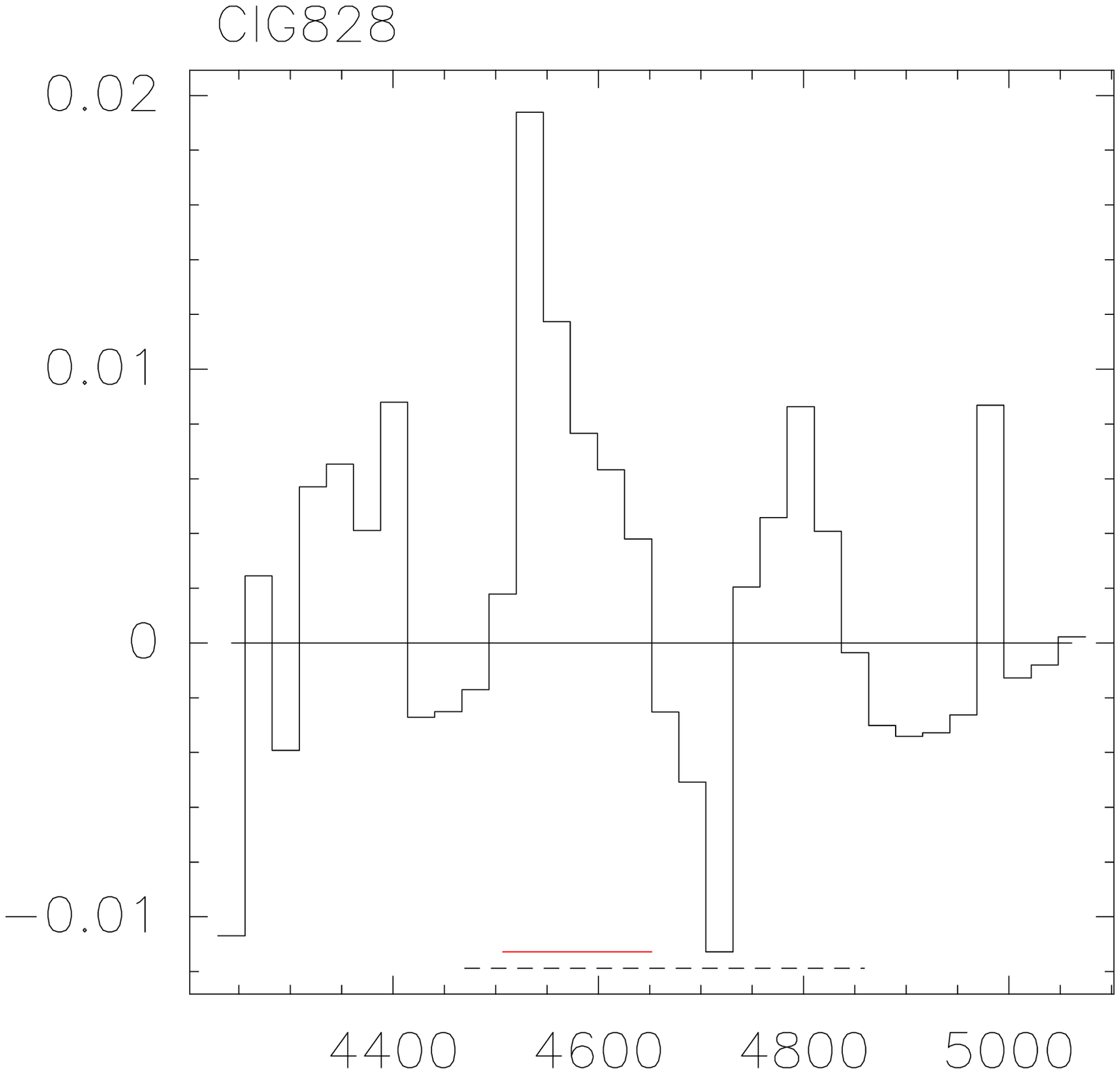}\quad 
\includegraphics[width=3cm,angle=270]{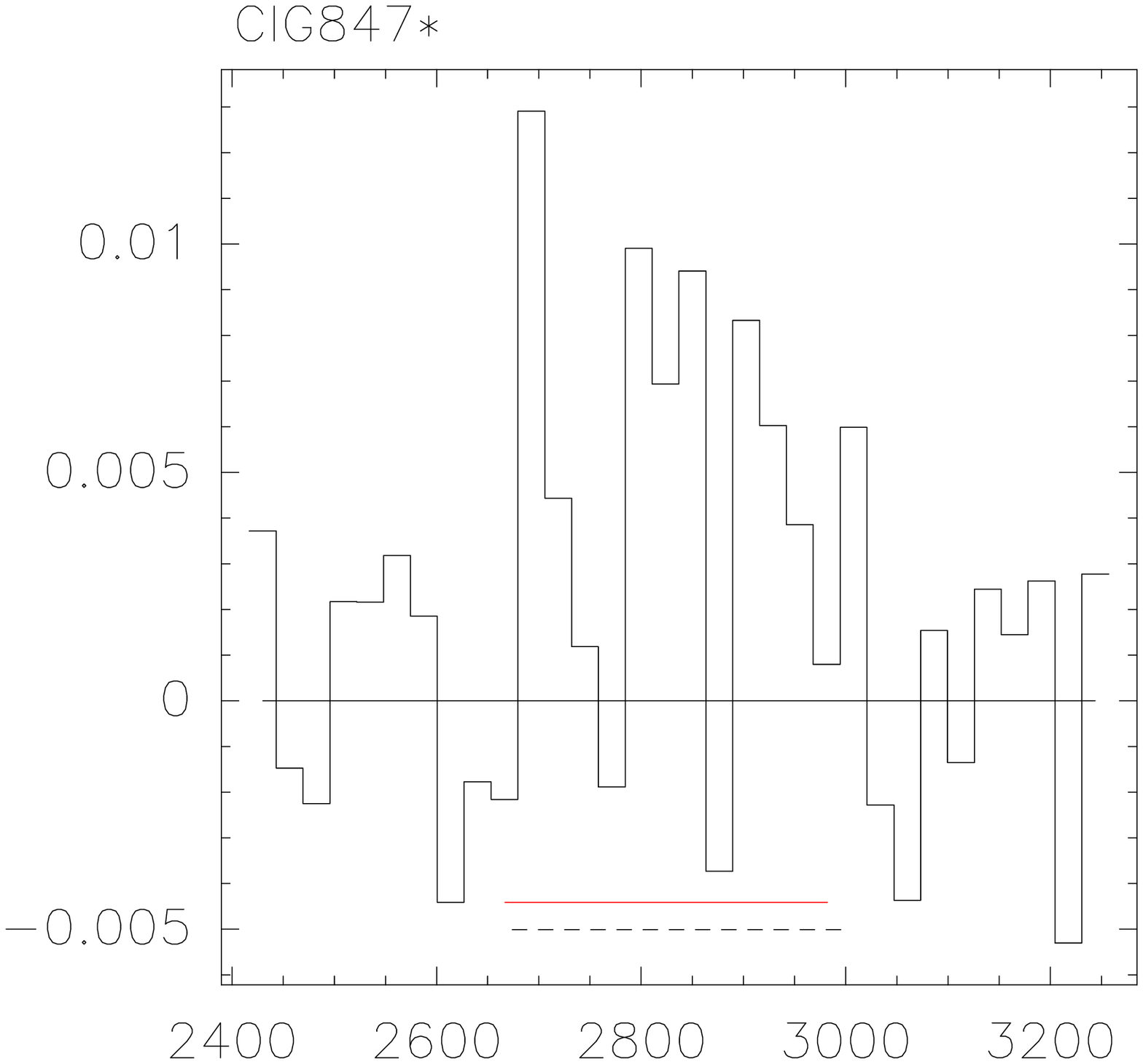}\quad 
\includegraphics[width=3cm,angle=270]{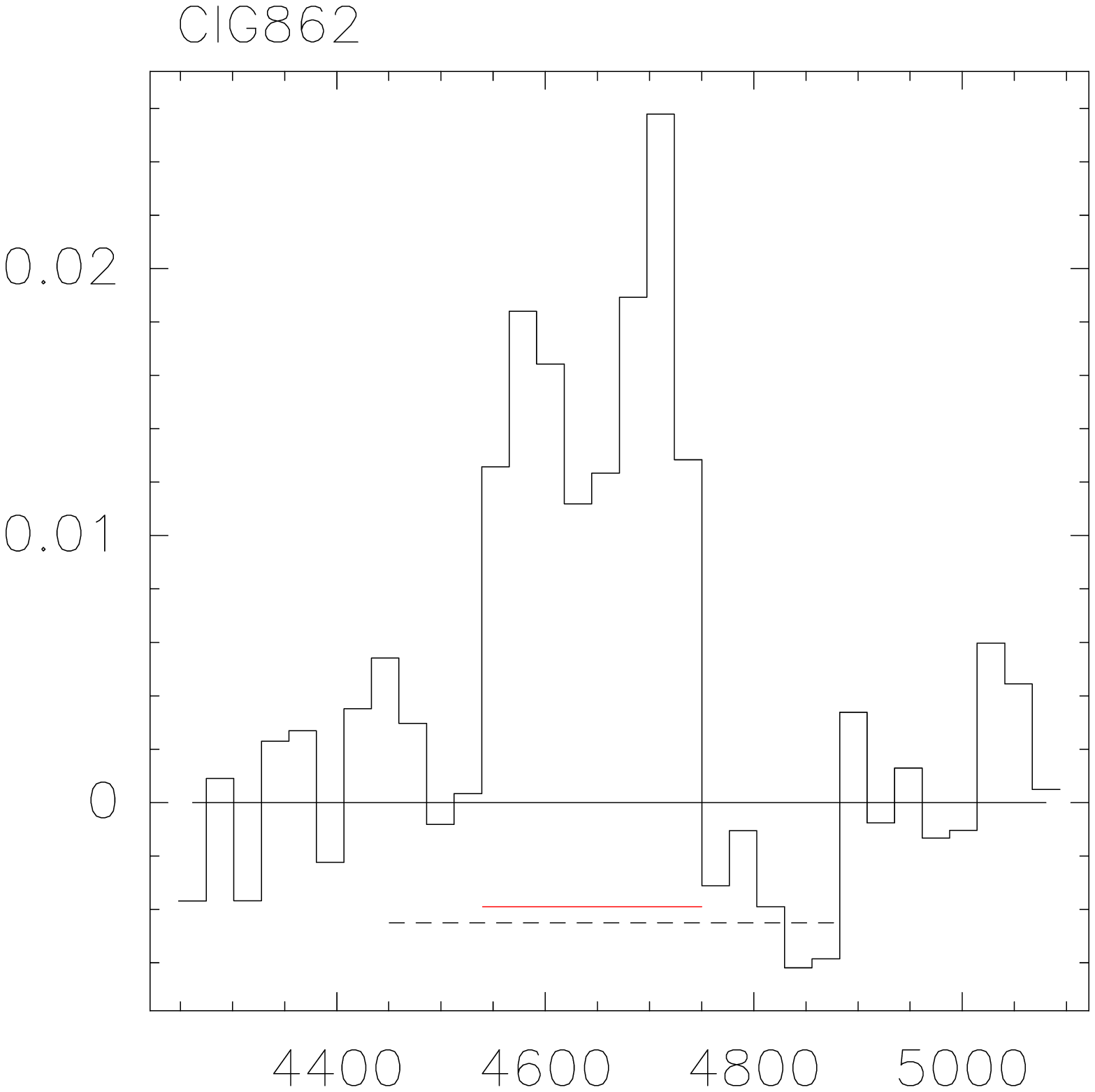}} 
\centerline{\includegraphics[width=3.3cm,angle=270]{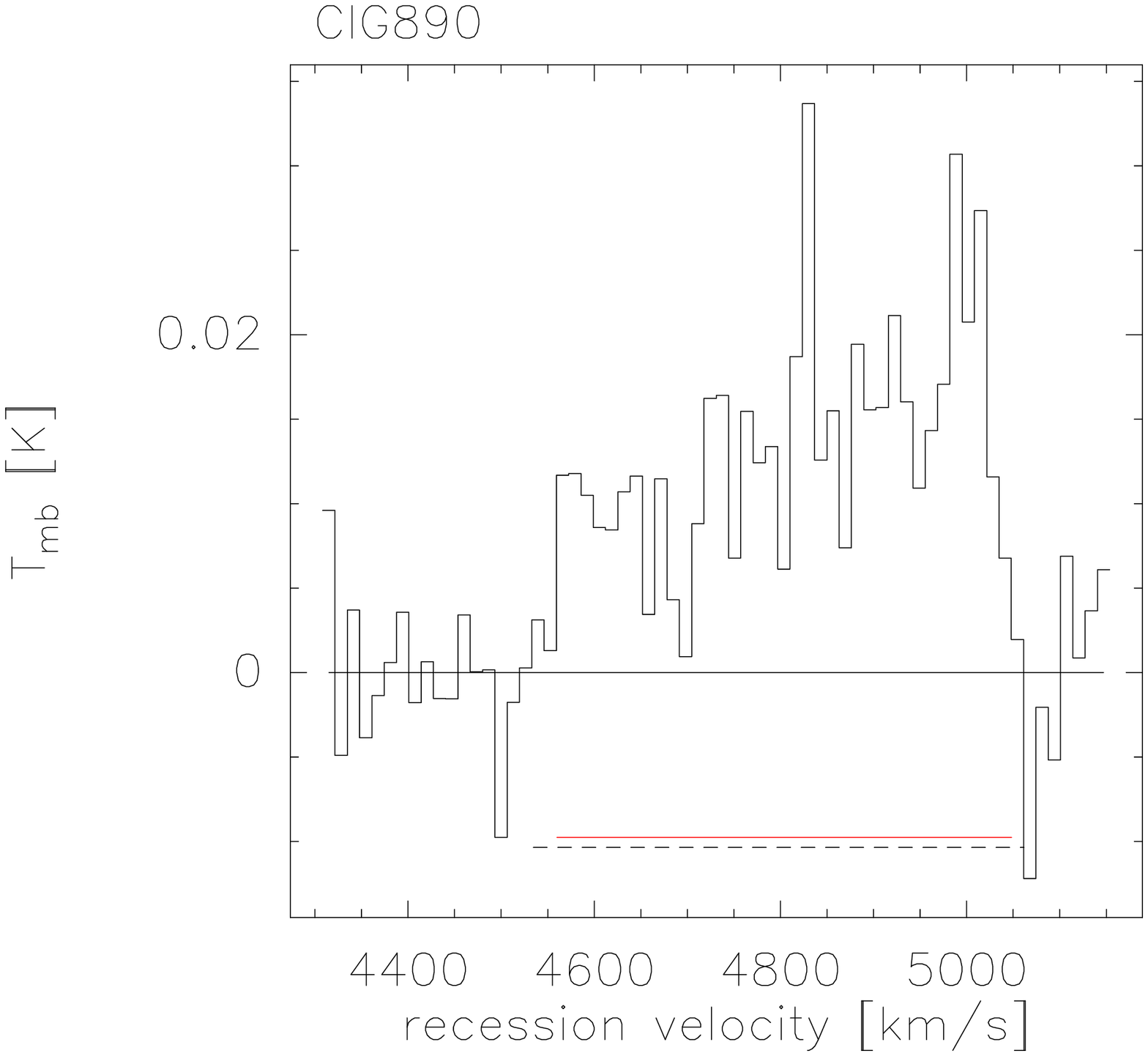} \quad 
\includegraphics[width=3cm,angle=270]{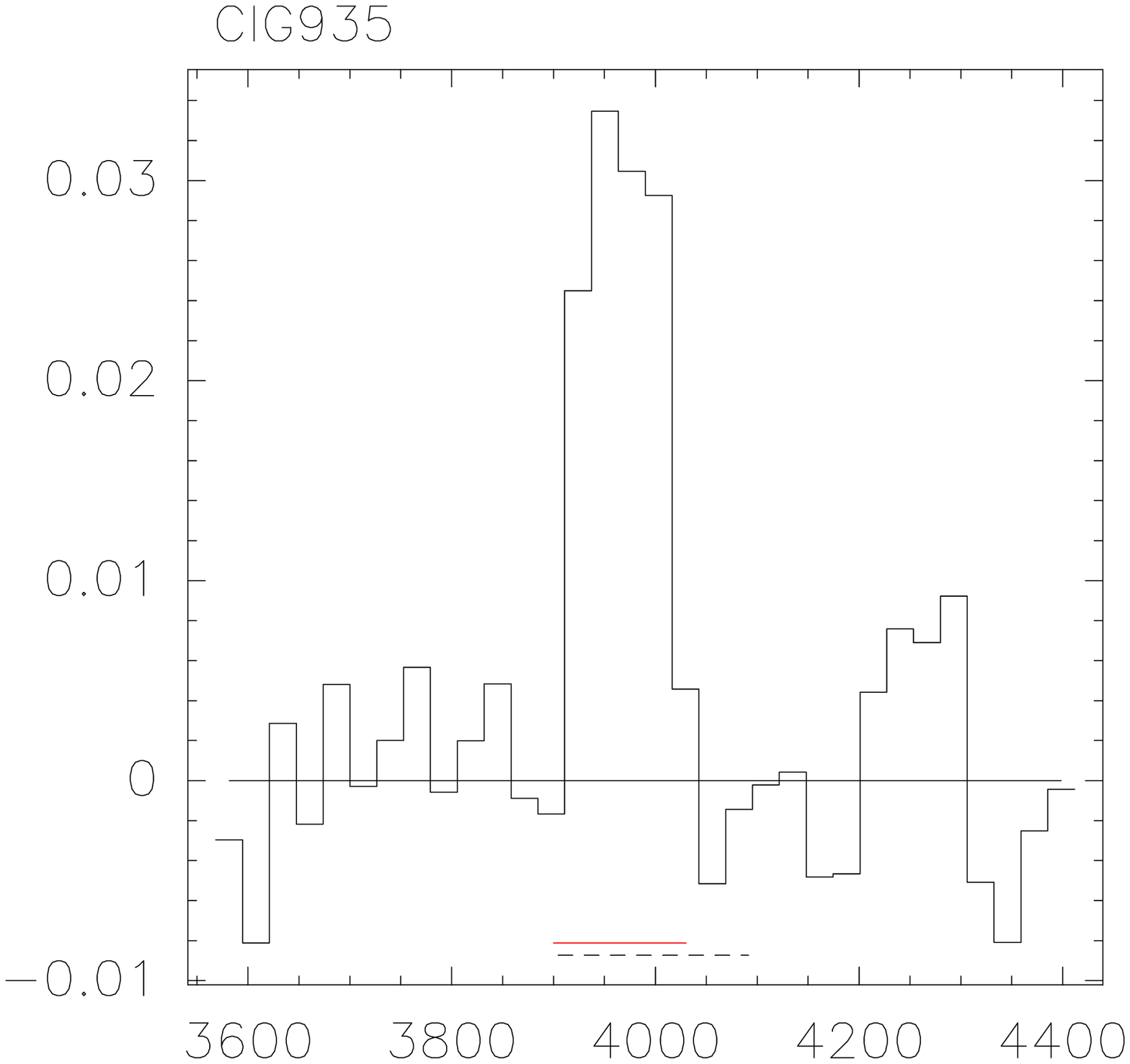}\quad 
\includegraphics[width=3cm,angle=270]{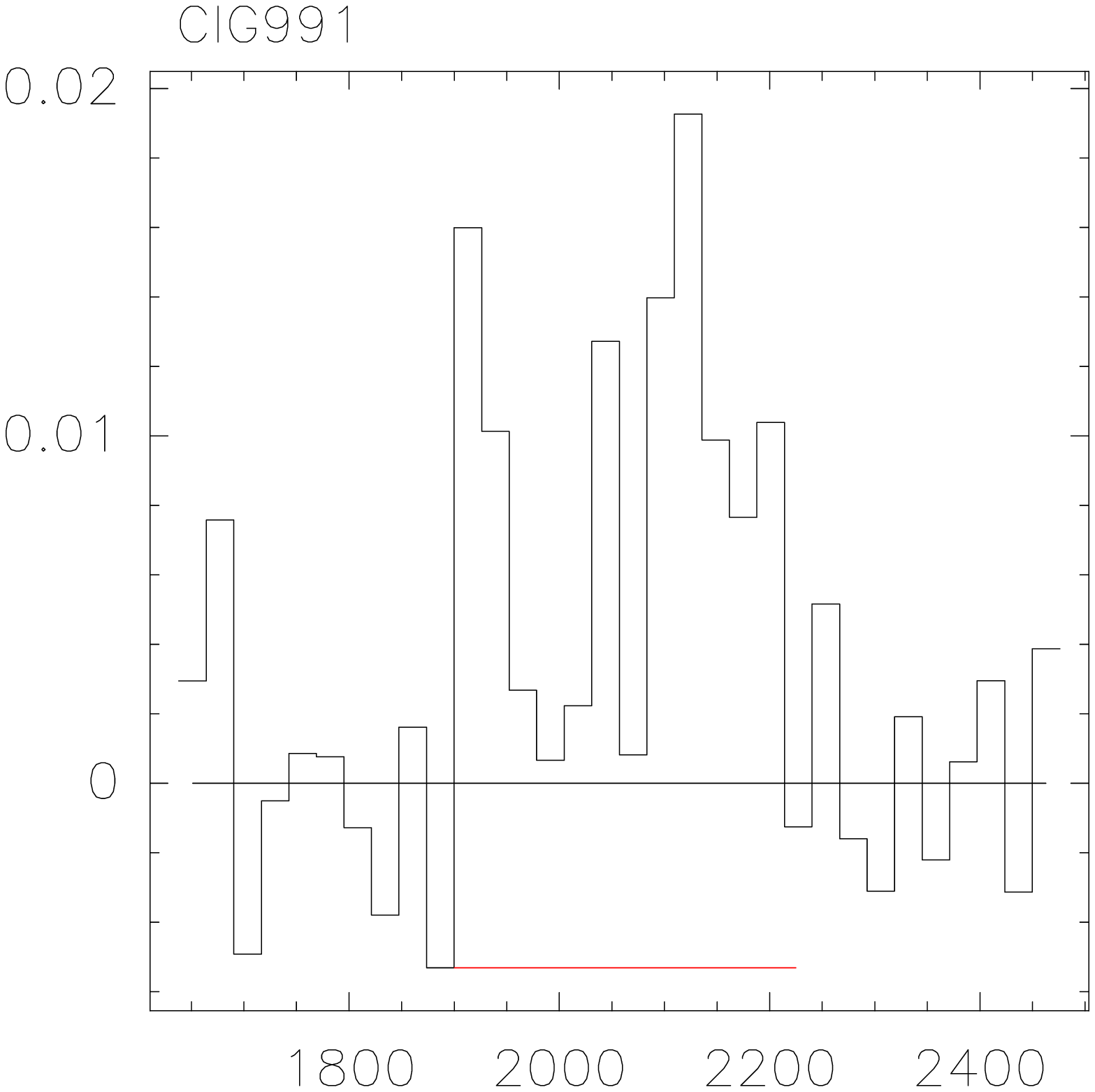}}
\addtocounter{figure}{-1} 
\caption{(continued)} 
\end{figure*}

\end{document}